\documentclass[aps, rmp, reprint, twocolumn, notitlepage, showpacs, floatfix, superscriptaddress, longbibliography]{revtex4-1}

\usepackage{silence}
\usepackage{amsthm,amsmath,amsfonts,amssymb,verbatim, color}
\usepackage{graphicx}
\usepackage[percent]{overpic}
\usepackage{epsfig,slashed}
\usepackage{epstopdf}
\usepackage{lipsum}
\usepackage{float}
\usepackage{mathtools}
\usepackage{natbib}
\usepackage{subfigure}
\usepackage[mathscr]{euscript}
\usepackage[export]{adjustbox}
\usepackage[bottom]{footmisc}
\usepackage{textcomp}

\makeatletter
\newsavebox\myboxA
\newsavebox\myboxB
\newlength\mylenA

\newcommand*\xoverline[2][0.75]{%
    \sbox{\myboxA}{$\m@th#2$}%
    \setbox\myboxB\null
    \ht\myboxB=\ht\myboxA%
    \dp\myboxB=\dp\myboxA%
    \wd\myboxB=#1\wd\myboxA
    \sbox\myboxB{$\m@th\overline{\copy\myboxB}$}
    \setlength\mylenA{\the\wd\myboxA}
    \addtolength\mylenA{-\the\wd\myboxB}%
    \ifdim\wd\myboxB<\wd\myboxA%
       \rlap{\hskip 0.5\mylenA\usebox\myboxB}{\usebox\myboxA}%
    \else
        \hskip -0.5\mylenA\rlap{\usebox\myboxA}{\hskip 0.5\mylenA\usebox\myboxB}%
    \fi}
\makeatother

\begin{document}

\newtheorem{theorem}{Theorem}
\newtheorem{property}{Property}
\newcommand{\tr}{\mathop{\mathrm{Tr}}}
\newcommand{\bsigma}{\boldsymbol{\sigma}}
\newcommand{\re}{\mathop{\mathrm{Re}}}
\newcommand{\im}{\mathop{\mathrm{Im}}}
\newcommand{\diag}{\mathrm{diag}}
\newcommand{\sign}{\mathrm{sign}}
\newcommand{\sgn}{\mathop{\mathrm{sgn}}}
\newcommand{\mb}{\bm}
\newcommand{\ua}{\uparrow}
\newcommand{\da}{\downarrow}
\newcommand{\ra}{\rightarrow}
\newcommand{\la}{\leftarrow}
\newcommand{\mc}{\mathcal}
\newcommand{\bs}{\boldsymbol}
\newcommand{\lra}{\leftrightarrow}
\newcommand{\nn}{\nonumber}
\newcommand{\half}{{\textstyle{\frac{1}{2}}}}
\newcommand{\mf}{\mathfrak}
\newcommand{\MF}{\text{MF}}
\newcommand{\IR}{\text{IR}}
\newcommand{\UV}{\text{UV}}

\renewcommand{\i}{\mathop{\mathrm{i}}}
\renewcommand{\b}[1]{{\boldsymbol{#1}}}

\def\II{\hbox{$1\hskip -1.2pt\vrule depth 0pt height 1.6ex width 0.7pt\vrule depth 0pt height 0.3pt width 0.12em$}}

\DeclareGraphicsExtensions{.png}

\title{Nodal portraits of quantum billiards: Domains, lines, and statistics }

\author{Sudhir R. Jain}
\affiliation{$\mbox{Nuclear Physics Division, Bhabha Atomic Research Centre, Mumbai 400085, India}$}
\affiliation{$\mbox{Homi Bhabha National Institute, Training School Complex, Anushakti Nagar, Mumbai 400094, India}$}

\author{Rhine Samajdar}
\affiliation{$\mbox{Department of Physics, Harvard University, Cambridge, MA 02138, USA}$}

\begin{abstract}
We present a comprehensive review of the nodal domains and lines of quantum billiards, emphasizing a quantitative comparison of theoretical findings to experiments.  The nodal statistics are shown to distinguish not only between regular and chaotic classical dynamics but also between different geometric shapes of the billiard system itself. We discuss, in particular, how a random superposition of plane waves can model chaotic eigenfunctions and highlight the connections of the complex morphology of the nodal lines thereof to percolation theory and Schramm-Loewner evolution. Various approaches to counting the nodal domains---using trace formulae, graph theory, and difference equations---are also illustrated with examples. The nodal patterns addressed pertain to waves on vibrating plates and membranes, acoustic and electromagnetic modes, wavefunctions of a ``particle in a box'' as well as to percolating clusters, and domains in ferromagnets, thus underlining the diversity---and far-reaching implications---of the problem.  
\end{abstract}
\maketitle
\tableofcontents

\hypersetup{linkcolor=blue}

\section{Introduction}

\begin{figure}[htb]
\includegraphics[trim={0cm 0 0 -1cm}, clip, width=\linewidth]{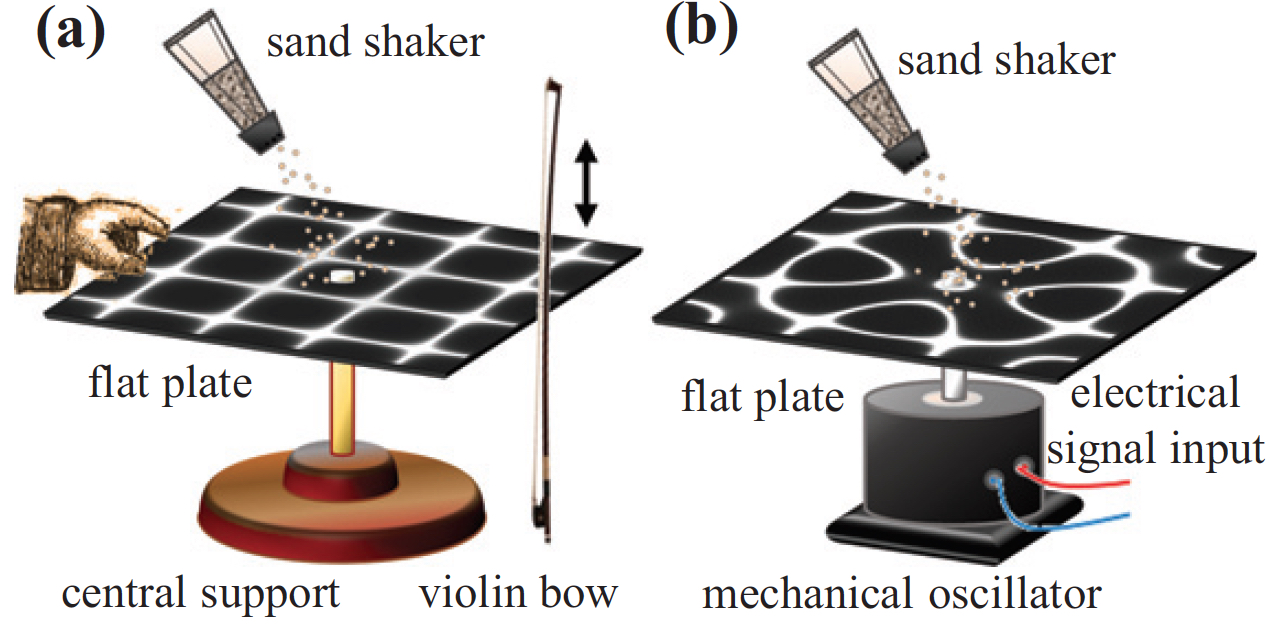}
\caption{\label{fig:modern}Experimental approaches for generating (a) classical and (b) modern Chladni figures. The resonant oscillations excited by the bow, in the original scheme, are eigenmodes of the plate. In modern experiments \cite{waller1937production, waller1938vibrations, waller1940vibrations, jensen1955production}, the resonant modes are locally excited by driving the plate at varying frequencies with an electronically-controlled mechanical oscillator. From \citet{tuan2015resolving}.} 
\end{figure}

In February 1809, Napoleon Bonaparte invited a certain Dr.~Chladni of Wittenberg to his court at the Tuileries Palace in Paris. The Emperor of the French people had been greatly enthused by the prospect of a firsthand demonstration of the ``sound patterns'' that Ernst Florens Friedrich \citet{chladni1802akustik} had documented in his book, \textsl{Die Akustik}, some seven years earlier. The path leading up to this meeting had been paved by Chladni's longstanding scholarly interest in the vibrations of bars and plates. Inspired by Lichtenberg's experiments on using sulphur and minium powder to visualize electric discharges in insulators, he decided to replicate the procedure with vibrating brass plates (see Fig.~\ref{fig:modern}). The rest, as they say, is history. \citet{stockmann2007chladni} narrates: ``He spread sand on the plate, stroke[d] it with the violin bow, and within a few seconds the sand brought about the shape of a star with ten rays.'' Thus, Chladni's sound figures were born. The study of these vibrational patterns soon generated such intense interest that it prompted no less a man than Michael \citet{faraday1859experimental} to remark, ``The beautiful series of forms assumed by sand, filings, or other grains, when lying upon vibrating plates, discovered and developed by Chladni, are so striking as to be recalled to the minds of those who have seen them by the slightest reference. They indicate the quiescent parts of the plates, and visibly figure out what are called the nodal lines.'' Over the last quarter of the $18^\mathrm{th}$ century, \citet{chladni1787entdeckungen} systematically probed the sound patterns of circular, quadratic, and rectangular plates, by fixing them with his fingers at different points, thereby enforcing the occurrence of nodal lines \cite{ullmann2007life}. In fact, his investigations---coupled with pure empirical reasoning---led to the discovery of the relation between the frequency $f$ of a vibrating circular plate and the number of its diametric ($m$) and radial ($n$) nodal lines:
\begin{equation}
f \sim (m + 2 n)^2,
\end{equation}
which, today, is better known as Chladni's law. 

\begin{figure}[htb]
\includegraphics[width=\linewidth]{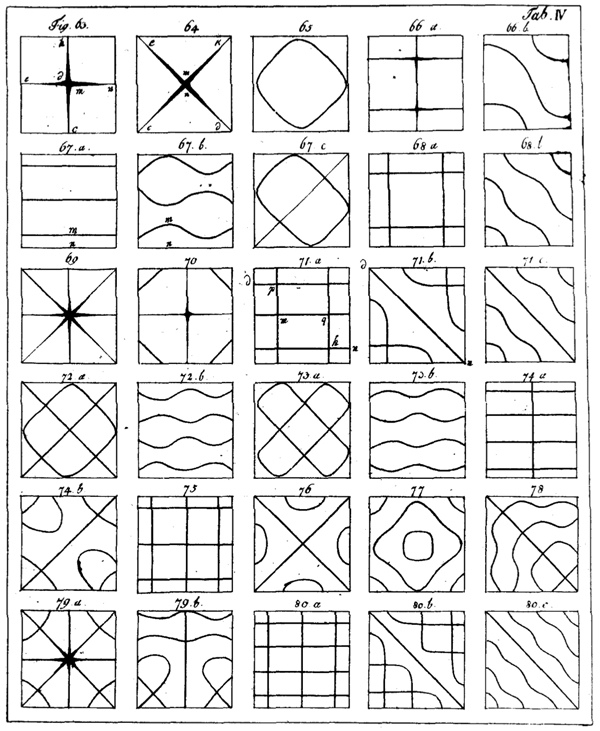}
\caption{\label{fig:chladni}Sound figures of a square plate. From \citet{chladni1802akustik}.} 
\end{figure}

Underlying the patterns traced by the tiny sand particles on an excited plate---and dictating the genesis of the nodal lines---is the fundamental question of symmetry. If a plate has $N$ lines of symmetry (each of which divides its surface into two equal parts that are mirror images of each other), \citet{waller1957interpreting} observed that all the normal modes of vibration supported by the plate can be classified according to the relation $c = 2 F$, where $c$ denotes the number of symmetry classes and $F$, the number of factors of $N$. For instance, for a square plate, there are ($N$ =) 4 lines of symmetry, the factors being $1, 2, 4$ ($F = 3$); the number of classes is then 6. Such taxonomical endeavors were of great personal interest to Chladni.  In \textsl{Die Akustik}, he categorizes the patterns observed for rectangular plates according to the number of nodal lines parallel to both sides, meticulously sketching each individual excitation (Fig.~\ref{fig:chladni}) at assigned frequencies. These bicentenarian figures bespeak a few general features of note. Suppose a square plate, driven into vibration by a violin bow at one corner, is pinned by, say, a finger, at the midpoint of one of the edges. The resultant nodal pattern partitions the square into four (Fig.~\ref{fig:chladni}: $1^{\mathrm{st}}$ row, $1^{\mathrm{st}}$ column), each nodal line being the mean level, having one side of it a rising and on the other a falling surface \cite{rossiter1871}. Alternatively, if the positions of driving and pinning are exchanged, the lines of least agitation span across the diagonals, dividing the plate into four equal triangles (Fig.~\ref{fig:chladni}: $1^{\mathrm{st}}$ row, $2^{\mathrm{nd}}$ column). The nodal portraits therefore strongly suggest that Chladni figures always conform with the symmetry of the geometrical surfaces upon which they are produced \cite{waller1952vibrations}. Any disturbance, in the slightest, suffices to change from one standing wave to another among the normal modes with the same frequency, but not to a combination thereof \cite{waller1954symmetry}. The regions that the nodal contours divide the plate into---where the vibrations are out of phase---are called nodal domains. Understanding these patterns (and counting the number of such domains) has remained a problem generating great intellectual curiosity since the time it was first treated mathematically by Sophie Germain and Gustav Robert Kirchhoff. This question has today become a part of a more general class of problems which include percolation theory, where percolation clusters appear, ferromagnetism, where magnetic domains appear, and microwave cavities, where domains of electromagnetic modes appear---all sharing certain features, and subtle differences.

Meanwhile, across the English channel, the stage was set for the birth of the physics of billiards on April 27, 1900. Lest we sound facetious in making so sweeping a statement, let us clarify the historical context in which it is made. On that fateful day, Lord Kelvin delivered a lecture entitled ``Nineteenth century clouds over the dynamical theory of heat and light'' at the the Royal Institution of Great Britain. One of the portentous clouds that he had been alluding to was the breakdown of the ergodicity hypothesis---the assumption that the phase-space average of a physical quantity should accord with its time average, taken over sufficiently long times. The setup that \citet{kelvin1901nineteenth} considered consisted of a billiard table---``a finite area of plane or curved surface, bounded by \ldots walls, from which impinging particles are reflected at angles equal to the angles of incidence''---and an ``ideal, perfectly smooth, non-rotating billiard ball, moving in straight lines except when it strikes the boundary.'' Examining the motion of a point particle bouncing off the hard walls of a a scalene-triangular billiard (now known to be pseudointegrable) and a flower-like billiard (nonintegrable), he was able to explicitly demonstrate a contradiction to the equipartition theorem \cite{nakamura2004quantum}. Actually, in these experiments, Kelvin approximated the flower-like billiard by a polygonal one, substituting for each arc of the flower (the semicircular corrugations of the circular boundary) its chord. Little did he know that this anodyne approximation would be \textsl{the} reason for the violation of the ergodicity hypothesis. Nonetheless, his investigation set in motion studies on the nonlinear dynamics of classical billiard systems, and in particular, on nonintegrable and chaotic billiards \cite{sinaui1976introduction, sinai1970dynamical, bunimovich1981statistical, arnold1967theorie, krylov1979works}.

So far, the protagonists in our narrative have been two disparate physical systems---quivering plates and dynamical billiards---that, seemingly, lack any connection. The unifying link between the two was provided by the advent of quantum mechanics, and the subsequent inception of \textsl{quantum} billiards. The quantum analogue to a classical billiard is governed by the stationary Schr{\" o}dinger equation \begin{equation}
\label{eq:schr}
\Delta\,\psi_j \equiv \bigg ( \frac{\partial^2}{\partial\,x^2} + \frac{\partial^2}{\partial\,y^2}  \bigg) \psi_j = - k_j^2\,\psi_j
\end{equation}
for a ``particle in a box''. Information about the walls of the billiard table appear only through the boundary conditions enforced by, say, a confining potential in the Hamiltonian. Eq.~\eqref{eq:schr} is nothing but a simple time-independent wave equation---the Helmholtz equation. If one identifies the wavefunction $\psi$ as the amplitude of a wave field (sound waves, to be concrete), it is evident that the wavefunctions of quantum billiards allow essentially the same description as the Chladni figures of yore.\footnote{In actuality, the amplitude of flexural vibrations of stiff acoustic plates are known \cite{landau1959course} to be described by 
\begin{equation}
\label{eq:chladniVib}
\Delta^2\,\psi_j \equiv \bigg ( \frac{\partial^2}{\partial\,x^2} + \frac{\partial^2}{\partial\,y^2}  \bigg)^2 \psi_j = k_j^4\,\psi_j
\end{equation}
for the $j^\mathrm{th}$ resonance, or, with driving, by the Kirchhoff-Love equation:
\begin{equation}
\left( D\,\nabla ^4 + \rho\, h\,\frac{\partial ^2}{\partial t^2}\right) \psi\, (x, y, t) = F\,(x, y, t),
\end{equation}
where $D$ is the flexural rigidity, $\rho$ the mass density, $h$ the thickness, and $F(x, y, t)$ the effective force function. As opposed to the vibrations of membranes without internal stiffness, it is the square of the Laplace operator that figures in Eq.~\eqref{eq:chladniVib}.} This opens up new avenues to address questions and theories, which were originally prompted by quantum mechanics, by means of classical wave experiments on acoustics \cite{weaver1989spectral, ellegaard1995spectral, tanner2007wave}, water waves \cite{blumel1992quasilinear}, light \cite{huang2002observation, chen2006manifestation, chen2012generation}, and microwave networks \cite{hul2004experimental, hul2012scattering, lawniczak2014resonances}. In the course of the last two decades, quantum billiards have been experimentally realized in gated, mesoscopic GaAs tables \cite{berry1994influence}, microwave cavities \cite{stockmann1990quantum, richter1999playing} and ultracold atom traps \cite{milner2001optical, andersen2006decay, friedman2001observation, montangero2009quantum}. The eigenfunctions of these planar quantum billiards once again organize themselves into domains, with positive and negative signs, often in remarkably complicated geometric shapes. Formally, such nodal domains may be defined as the maximally connected regions wherein the wavefunction does not change sign. Unfortunately, quantifying the nodal patterns is a major challenge since it is extremely hard to discern any order when the wavefunctions are arranged in ascending order of energy. In principle, the problem seems (deceptively) straightforward---for each billiard of interest, we need only solve the Schr{\"o}dinger equation in appropriate coordinates, and count the domains as a function of the quantum numbers. However, in the absence of an exact solution to the counting problem, the statistics of nodal domains and lines have been studied for regular to chaotic billiards. Besides its connections to an array of different subjects such as the seismic response of sedimentary valleys \cite{flores2007nodal}, violins \cite{gough2007violin}, floaters in surface waves \cite{lukaschuk2007nodal}, evanescent waves in the brain \cite{schnabel2007random}, and turbulence \cite{falkovich2007nodal} to name a few, there are two primary reasons why nodal portraits are of academic interest. Firstly, a host of analytical and numerical evidence \cite{blum2002nodal, toth2009counting} today suggests that the sequence of nodal counts in a billiard encodes the difference between integrability and chaos, i.e., it can shed light on the quantum signatures of classical chaos. Secondly, and equally interestingly, the nodal count also bears additional information about the dynamics and the geometry of the billiard that are not accessible from the spectral statistics alone \cite{gnutzmann2006can}.

This article is organized as follows. We begin with a brief overview of the classical and quantum dynamics of billiard systems in Sec.~\ref{sec:billiards}. The nodal properties of Laplacian eigenfunctions of these billiards constitute a subject that has garnered widespread attention from mathematicians and physicists alike. In Sec.~\ref{math}, we rummage through the motley mathematical literature, sans rigor and jargon, to demonstrate that far from being \textsl{l'art pour l'art}, some crucial results therefrom find direct applicability to quantum billiards. A lot of work has been devoted to chaotic systems owing to Berry's random wave hypothesis, which we examine in Sec.~\ref{sec:chaotic}. This correspondence between nodal domains and random waves also draws us closer to the problems of percolation theory and, relatively more recently, Schramm-Loewner evolution. Proceeding thereafter, in Sec.~\ref{sec:stats}, we explore the various established results on nodal domain and line statistics. As we show, for separable billiards, it is possible to write down exact expressions for certain limiting distributions as also for fully chaotic systems, whereas for integrable but nonseparable, quasi-integrable, and pseudointegrable billiards, only partial results exist. Sec.~\ref{sec:count} brings us to the question of counting nodal domains and, by association, ``the shape of a drum''. For separable systems, a trace formula has been developed in this regard. Likewise, for integrable systems, one can formulate a difference-equation formalism. Other geometries where the problem of counting has been well studied include flat tori, surfaces of revolution, and the boundaries of 2D quantum billiards---we also survey the mathematical developments in these directions. In Sec.~\ref{sec:exp}, we revisit our previously abstract discussions but now, grounded in actual experiments on microwave billiards, which have been instrumental in the investigation of not only nodal statistics, but also of quantum chaos, in general. Finally, we conclude this Review by enlisting some open questions in Sec.~\ref{sec:conclude} and summarizing our perspectives on the prospective directions of this burgeoning field.

\section{Billiards: Classical and quantum dynamics}
\label{sec:billiards}

A point particle moving freely inside an enclosure, reflecting from the (piecewise smooth) boundary in accordance with Snell's law---this is the dynamical system, almost endearing in its simplicity, termed a billiard. Amongst other abstractions, billiards serve as useful models for many-body systems, which can be approximated by a single particle in a mean field, the complexity of the former being simulated by the boundary \cite{sinaui1976introduction, rabouw1981three, glashow1997three, sen1996multispecies, turner1984quantum, krishnamurthy1982exact}. For instance, the Boltzmann-Gibbs gas in statistical mechanics---elastically colliding hard balls in a box---can be exactly mapped onto a billiard \cite{bunimovich2013hard}. The dynamics of the billiard is completely determined by the shape of its boundary, which also defines its symmetries and therefore, the constants of motion. Classifying along these lines, billiards can, by and large, be divided into a few broad categories such as (i) convex with smooth boundaries, (ii) polyhedral or polygonal, and (iii) dispersive or semi-dispersive.  

\begin{figure}[htb]
\includegraphics[width=0.85\linewidth]{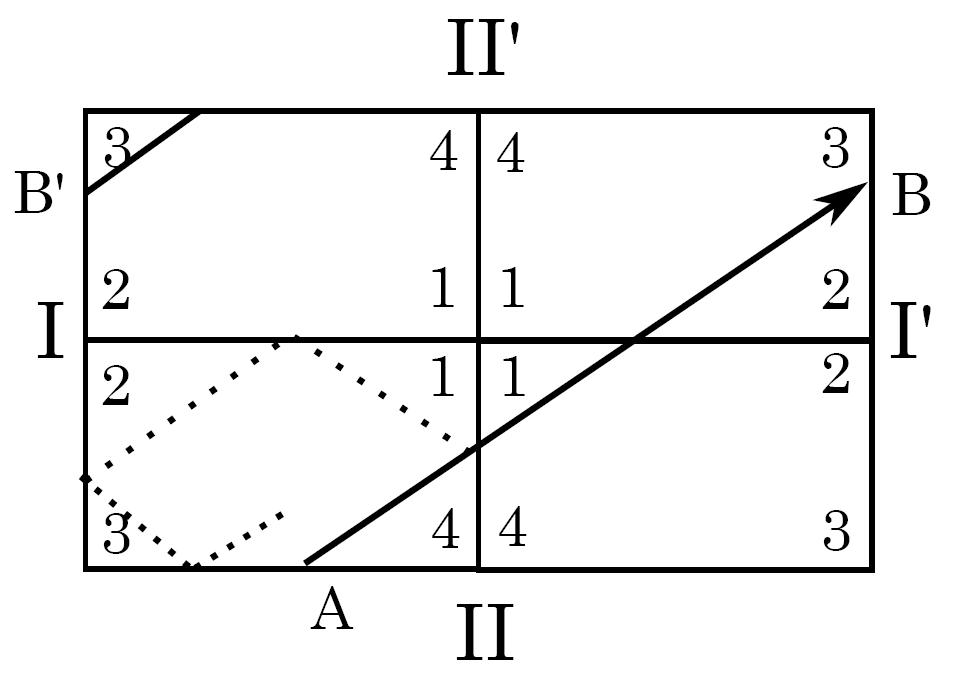}
\caption{\label{fig:rectangle}A trajectory in a rectangular billiard. After the first reflection, it follows the dotted path. Instead, if we reflect the domain, the trajectory straightens into a domain with reflected orientation. On reaching B, the next reflected copy is identical to the one at the top-left, so the trajectory resumes from B$'$, and eventually, from 3--4 in the lower-left copy. For any given initial direction, there are a maximum of four directions generated by successive reflections. Four copies therefore make a fundamental domain, which can be stacked to tessellate the entire plane. Pairwise identification of the sides yields the two-dimensional manifold equivalent to a 2-torus.} 
\end{figure}

A trajectory of a particle undergoing reflection from a surface can be straightened\footnote{In the mathematical literature, the unfolding is a map from the billiard table to a Riemannian surface $X$ endowed with a holomorphic (complex-analytic) 1-form, the latter being useful in assigning coordinates in the complex plane via a holomorphic function \cite{demarco2011conformal}.} by reflecting the domain $\mathcal{D}$ (instead of the particle). For concreteness, consider the rectangular billiard in Fig.~\ref{fig:rectangle}---applying the abovementioned prescription shows that the manifold is a torus. The rational directions, characterized by slopes which are rational multiples of the aspect ratio, correspond to periodic trajectories---the \textsl{periodic orbits}---while the irrational directions fill the torus uniformly. Periodic orbits in polygonal billiards are never isolated. Owing to the lattice structure in two dimensions, the number of periodic orbits of length $\leq  \ell$ is related to counting the number of coprime lattice points inside a circle of radius $\ell$, which grows as $\ell ^2$. It is quite obvious that for billiards with non-symmetrical shapes, the dynamics become unpredictable in the sense that any initial correlations decay after a few reflections from the boundary. Identifying the periodic trajectories for such systems is also a much more complex problem as the number of these trajectories proliferates exponentially \cite{ott2002chaos}. Obviously, in quantum mechanics, these notions break down as the very concept of a trajectory becomes undefined courtesy of the uncertainty principle $\Delta x \, \Delta p \ge \hbar /2$. Contrarily, the correspondence principle necessarily requires quantum mechanics to reduce to the classical description in the semiclassical limit of large quantum numbers, i.e., $E \rightarrow \infty$. This dichotomy is certainly nontrivial and therein lies the tale. However, before we get ahead of ourselves, let us halt to make precise the definitions of integrable and chaotic systems.

\subsection{From integrability to chaos}

\begin{figure}[htb]
\includegraphics[width=\linewidth]{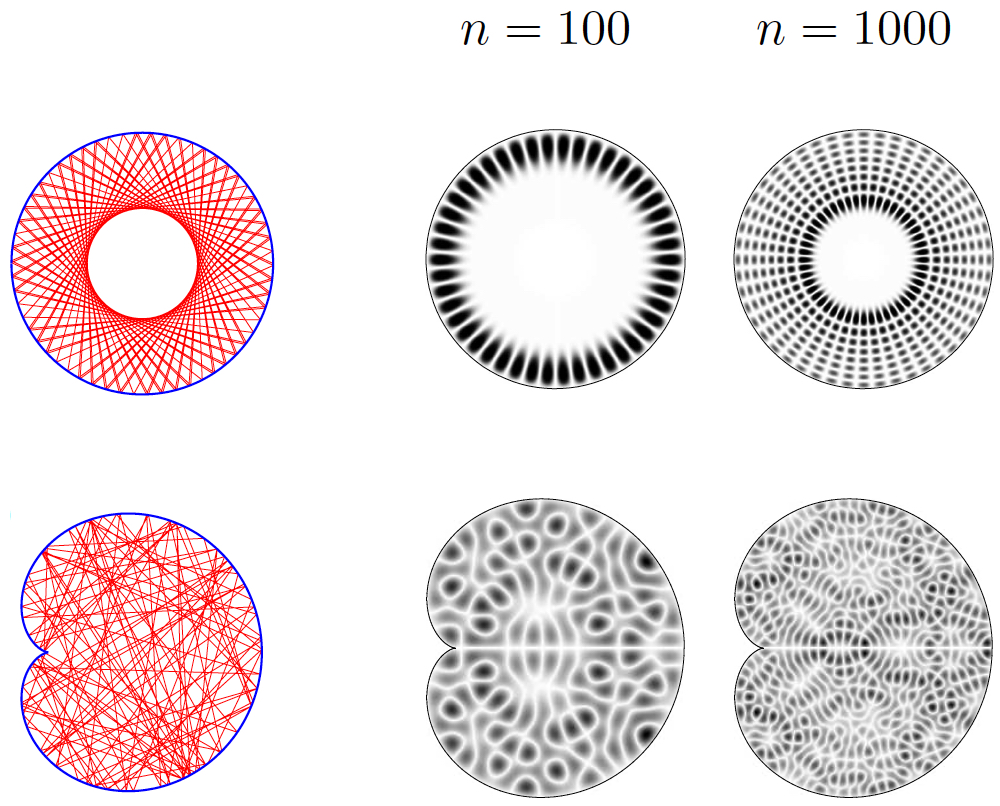}
\caption{\label{fig:classical}The eigenstates of (a) the integrable circular billiard (non-ergodic) and (b) the chaotic cardioid billiard (ergodic) reflect the structure of the corresponding classical dynamics. Shown is a density plot of $\lvert \psi (\mathbf{r}) \rvert^2$ where black corresponds to high probability. Adapted from \citet{arnd_thesis}.}
\end{figure}

A dynamical system is termed as \textsl{integrable} in the sense of Liouville-\citet{arnol2013mathematical} if there  are $f$ constants of motion for a system with $f$ degrees of freedom. These constants must be functionally independent and in involution. Furthermore, the vector fields have to remain regular everywhere in phase space. A circular billiard (Fig.~\ref{fig:classical}, $f = 2$) is an example of this type, where the total energy $E$ and the angular momentum $L$ are conserved. Since the system is integrable, the distance between two ``nearby'' trajectories increases linearly with time.  For any generic system with two degrees of freedom, one can find action-angle variables $(I_1, I_2, \theta _1, \theta _2)$ by employing an appropriate generating function \cite{lichtenberg2013regular}. There are two canonical frequencies, $\omega _i = \mathrm{d}\,\theta _i/\mathrm{d}\,t$, $i = 1, 2$, which are functions of phase-space variables for a nonlinear system. In exceptional situations (called nonlinear resonances), the ratio of these two frequencies becomes a rational number and certain regions in phase space are endowed with a chain of islands. Each island has at its center an elliptic point (stable equilibrium), and a separatrix connecting hyperbolic points. The invariant tori are broken, and new intricate structures are formed (see, for example, \citet{berry1981regularity}). Such systems are said to be \textsl{quasi-integrable}.  However, there exists another qualitatively different class of behavior demonstrated by most billiards. Consider the cardioid billiard of Fig.~\ref{fig:classical}; now, the only constant of motion is the total energy $E$. Classically, the motion on the billiard table is irregular and unpredictable, displaying extreme sensitivity to initial conditions. More importantly, the separation between neighboring trajectories grows exponentially with time with a divergence characterized by the Lyapunov exponent. The cardioid billiard is therefore \textsl{chaotic}. A ``typical'' billiard is expected to be non-integrable \cite{siegel1941ann} and most of these are believed to be chaotic, possessing a positive Kolmogorov-Sinai entropy \cite{bunimovich2007dynamical}. 

Clearly, this trajectory-centric distinction between order and chaos does not hold in the quantum regime. Moreover, since the Schr{\"o}dinger equation is linear, there is no scope for chaos in the conventional sense.\footnote{Hence, the widely prevalent usage of the term ``quantum chaos'', which just connotes the quantum mechanics of classically chaotic systems, should always be taken with a grain of salt.} However, the latent classical dynamical properties do leave their imprints on the statistical behavior of eigenvalues. The statistics of the energy levels of a generic integrable system can be characterized by a Poissonian random process \cite{berry1977level}. Conversely, the energy-level statistics of fully chaotic systems are described by the eigenvalues of random matrices obeying appropriate symmetries \cite{bohigas1984characterization, reichltransition, mehta2004random}. For quasi-integrable systems, the statistics can be modeled as a superposition of Poisson and Wigner distributions with weights proportional to the relative fractions of regular and chaotic subregions \cite{berry1984semiclassical}. Of course, as always, exceptions abound: prominent counterexamples include arithmetic systems \cite{aurich1988periodic, bogomolny1992chaotic, bolte1992arithmetical, bolte1993some, sarnak1995arithmetic} and quantized cat maps \cite{keating1991cat, hannay1980quantization}.

\label{sec:seh}
The classical dynamics of the billiard is also betokened by the structure of the eigenfunctions themselves. The semiclassical eigenfunction hypothesis \cite{voros1976semi, voros1977asymptotic, voros1979semi, berry1977regular, berry1983semiclassical} asserts that the eigenstates should concentrate on those regions which a generic orbit explores in the long--time limit. This statement is best formulated in terms of the Wigner function $W\,(p,q)$, which is a phase-space representation of the wavefunction \cite{wigner1932quantum}. In integrable systems, the motion remains confined to invariant tori in phase space---so $W\,(p,q)$ should localize on these tori in the semiclassical limit---whereas in ergodic systems, the whole energy surface is filled in a uniform manner and $W\,(p,q)$ is expected to semiclassically condense on the energy shell as $W\,(p,q) \sim [1/V (\Sigma_E)]\, \delta (H\,(p,q)-E)$, where $H$ is the Hamiltonian and $V (\Sigma_E)$ is the volume of the shell set by $H\,(p,q)=E$ \cite{backer1998rate}. For the latter class, this was established to be equivalent to the quantum ergodicity theorem \cite{shnirel1974ergodic, zelditch1987uniform, colin1985ergodicite, helffer1987ergodicite, gerard1993ergodic, zelditch1996ergodicity}---which states that ``most'' eigenstates are equidistributed in the semiclassical limit---by \citet{backer1998rate, PhysRevE.58.5192}. Written in position space, this implies 
\begin{equation}
\lim_{j \rightarrow \infty} \int_{\mathscr{R}}\, \lvert \psi_{n_j} (\mathbf{r}) \rvert^2\, \mathrm{d} \mathbf{r} = \frac{\mathrm{vol}\, (\mathscr{R})}{\mathrm{vol}\, (\mathcal{D})}
\end{equation}
for a subsequence $\{  \psi_{n_j} \} \subset \{ \psi_n \}$ of density one. Thus, as $E \rightarrow \infty$, the probability of finding a particle in a certain region of the billiard $\mathscr{R} \subset \mathcal{D}$ is identical to that for the classical system, for almost all eigenfunctions. This consonance is pictorially conveyed by Fig.~\ref{fig:classical}.

The phase space representation of eigenfunctions of integrable systems reveals a direct correspondence with the underlying periodic orbits. As the degree of nonintegrability increases, the eigenfunctions become progressively more structured. For chaotic systems, the eigenfunctions often evince an ``anomalous enhancement or suppression of eigenstate intensity on or near an \textsl{unstable} periodic orbit and its invariant manifolds'' \cite{kaplan1999scars}, over and beyond the statistically expected density. Discovered by \citet{mcdonald1979spectrum} and christened ``scarring'' by \citet{heller1984bound}, this observation is notably distinct from the attraction of eigenstate intensity associated with stable classical orbits, the latter being due to reasons well-understood by a purely semiclassical theory of integrable systems \cite{berry1977calculating, berry1976closed, zelditch1990quantum}. Quantitatively, scarring was partially accounted for on the basis of wavepacket dynamics in Husimi phase space \cite{heller1984bound}, coordinate space \cite{bogomolny1988smoothed} and Wigner phase space \cite{berry1989quantum, berry1991houches}. At a more formal level, a rigorous explanation was tendered by \citet{fishman1996fredholm} in their use of the Fredholm method to derive a semiclassical formula for scar strengths that had previously been obtained using Riemann-Siegel resummation techniques  \cite{agam1994semiclassical, agam1993quantum, agam1995semiclassical}. Although the scars themselves do not disappear in the semiclassical limit \cite{kaplan1998linear}, the scarred area of phase space (surrounding the orbit), and hence the total weight of the scar, vanishes as $\hbar \rightarrow 0$. One could equally well look for ``strong scarring'' \cite{rudnick1994behaviour} in the semiclassical limit---this necessitates the convergence of the total weight of the wavefunction on the unstable orbit \cite{kaplan1999scars}. However, this property has not been found to hold for any physical system thus far and its existence has even been explicitly disproved for certain arithmetic hyperbolic manifolds.\footnote{Moreover, the semiclassical defect measures, to adopt mathematical parlance, cannot be a finite sum of delta functions on closed geodesics on compact Reimannian $C^\infty$ manifolds with Anosov flow and the high-energy eigenfunctions are thus, at least, ``half-delocalized'' \cite{Anantharaman2007, anantharaman2008entropy}.}

Incidentally, addressing quasi-integrable or mixed systems, \citet{keating2001orbit} found that the probability density for perturbed cat maps is localized strongly around the bifurcated periodic orbits---the phenomenon was homologously styled ``superscarring''.

\subsection{Pseudointegrable billiards}

\begin{figure}[htb]
\includegraphics[width=\linewidth]{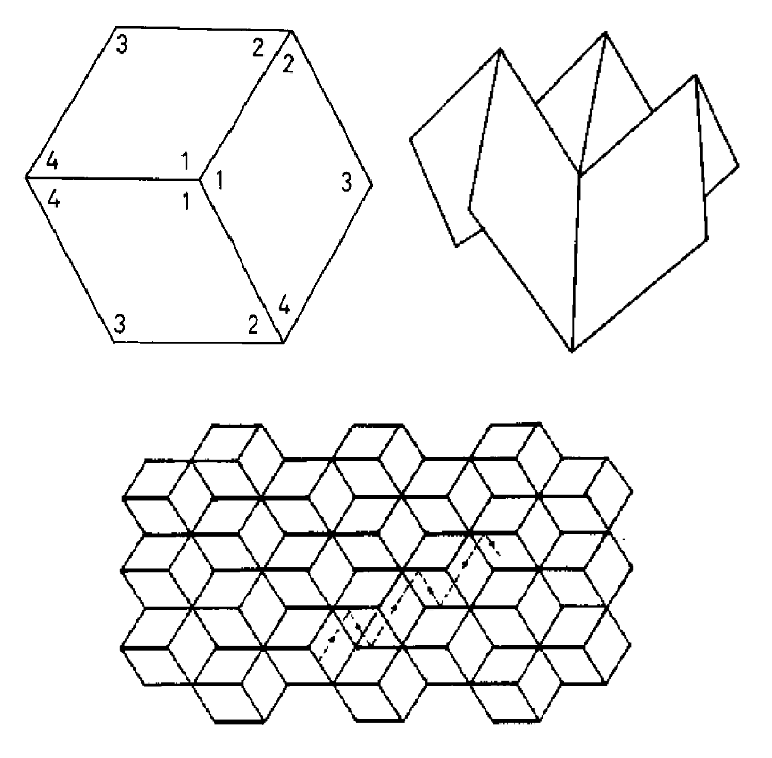}
\caption{\label{fig:rhombus}Upon three subsequent reflections, the rhombus billiard returns onto itself, albeit with a flipped orientation. This double-valuedness is taken care of by constructing a fundamental domain with the six copies required for identification of the sides; the vertex is a ``monkey-saddle'' \cite{eckhardt1984analytically}. The manifold ${\mathcal M}_c$ is thus, topologically, a sphere with two handles or a double torus with genus $g = 2$. Stacking the domains fills the plane, as seen in the lower figure. The bold line segments, although inaccessible, connect two branch points, forming a branch cut. The branch cuts are arranged in a doubly-periodic manner throughout the plane and a classical trajectory is a zig-zag line reflecting from the cuts as the phase changes by $\pi$.}
\end{figure}

Liouville-Arnold integrability is easily broken, when, for instance, the vector fields become singular without disturbing other conditions of integrability. Such dynamical systems are titled \textsl{pseudointegrable}, named thus by \citet{richens1981pseudointegrable}. Nonintegrable polygonal billiards are pseudointegrable; they are also non-chaotic insofar as the Lyapunov exponent is zero.\footnote{A finite-time exponent can also be defined for polygonal billiards as suggested by \citet{moudgalya2015finite}.} In this regard, we now turn to a billiard in a polygon $\mathcal{Q} \subset \mathbb{R}^2$ where all angles are commensurate with $\pi$ \cite{zemlyakov1975topological}. Let us fix a direction $\hat{e}$, say, along one of the sides of ${\mathcal Q}$. One can write the angles between $\hat{e}$ and the remaining sides in the form $\alpha _r = \pi\, m_r/2\,n_r$ with $r = 1, 2, \ldots$. Furthermore, let $N = N({\mathcal Q})$ denote the least common denominator of the fractions $m_r/n_r$. It was observed that the function 
\begin{alignat}{1}\label{eq:const}
F(x, y, p_x, p_y) = F(\phi ) = \lvert \phi \rvert \,\mathrm{mod}\, \frac{\pi}{N}
\end{alignat}
is well-defined on the energy surface ${\mathcal M}$ and invariant with the phase-space flow. For a given $c \in [0,\pi /2N]$, there is an invariant subset ${\mathcal M}_c$ of ${\mathcal M}$. If $c \notin \{0,\, \pi/2N\}$, then the set ${\mathcal M}_c$ is obtained by pairwise identification of the sides of $4N$ non-intersecting copies of the polygon labelled by $\phi _s^{\pm} = \pm c + s\,\pi /N$, with $s = 0, 1, \ldots, (2N-1)$. The identification of the sides is dictated, once again, by Snell's law. \citet{zemlyakov1975topological} proved that ${\mathcal M}_c$ is a two-dimensional manifold of genus $g > 1$ depending on the shape of the polygon. The fact that the motion in phase space is not restricted to a torus as for integrable systems but rather, to a surface with a more complicated topology, is conveyed by the moniker pseudointegrable. A prototypical example is the $(\pi /3, 2\pi /3)$-rhombus billiard, illustrated in Fig.~\ref{fig:rhombus}. 

\begin{figure}[htb]
\includegraphics[width=\linewidth]{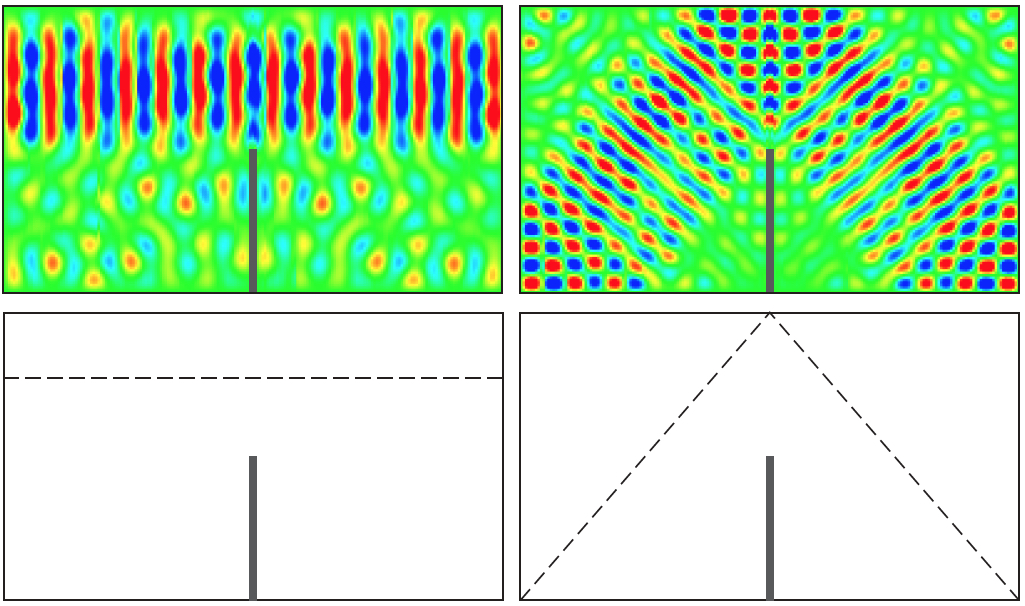}
\caption{\label{fig:superscar}Superscars in a pseudointegrable barrier billiard, experimentally obtained using the perturbing bead method (Sec.~\ref{bead}). Observed is a clear wave function structure connected with the family of classical periodic orbits as well as a distinct localization of excitation strength \cite{richter2008superscars}. The bottom row indicates the corresponding classical orbits (dashed lines). From \citet{bogomolny2006first}.}
\end{figure}

The statistical properties of the eigenvalues of classically pseudointegrable quantum billiards have been numerically established \cite{bogomolny1999models, wiersig2002spectral, biswas1990quantum} to be intermediate between those of regular and chaotic systems. The wavefunctions are also equally intriguing,
demonstrating pronounced scarring behavior that can be related to families of periodic orbits \cite{bogomolny2004structure}. For example, many of the solutions to the Schr\"{o}dinger equation for the rhombus billiard \cite{biswas1990quantum} exhibit significantly enhanced intensities in the close neighborhood of a periodic orbit. However, in contrast to chaotic systems, these scar structures persist even at large quantum numbers, thereby earning the epithet ``superscars''. Experimentally, superscarring (Fig.~\ref{fig:superscar}) has been observed in microwave billiards \cite{bogomolny2006first, richter2008superscars} and LiNC $\leftrightharpoons$ LiCN isomerization reactions \cite{prado2009superscars}. In many-body systems, it is often the case that the amplitude of a collective excitation is spread over states forming the background, well-known examples being the giant dipole resonances in nuclei \cite{bohr1998nuclear, sokolov1997simple} and metallic clusters \cite{brack1993physics}. The distinctly created excitation acts as a ``doorway'' to the background states. Analogously, a superscar of a pseudointegrable billiard also spreads over a large number of non-scarred wavefunctions. This convenient parallel enables the modeling of doorway states in the quantum spectra of nuclei \cite{aaberg2008superscars}.

\subsection{Flows and vortices}
\label{flow}

\begin{figure}[htb]
\includegraphics[width=\linewidth]{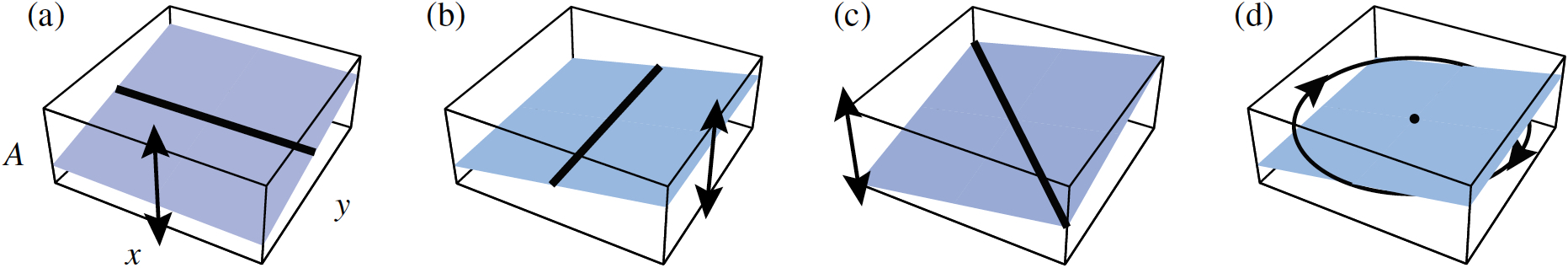}\\
\vspace*{0.25cm}
\includegraphics[width=\linewidth]{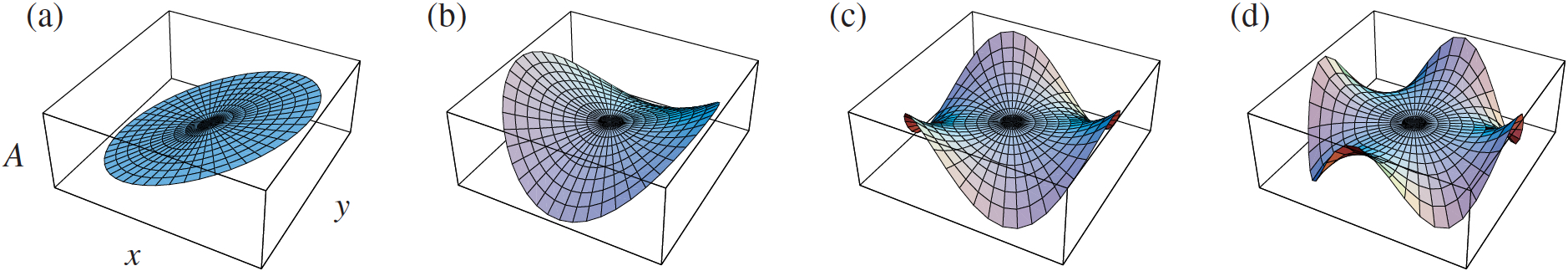}
\caption{\label{fig:Vortex}Chladni lines and vortices. [Top]: Snapshots of the 2-dimensional amplitude field $A(x, y, t)$ that represent nodal lines in the $\hat{x}$ (a) and $\hat{y}$ directions (b) along with an in-phase (c) and a $\pi/2$-out-of-phase (d) superposition, resulting in a realigned nodal line and a vortex, respectively. At the vortex, the amplitude $\sqrt{ x^2 + y^2} = 0$ and the phase of $u$ is indeterminate. [Bottom]: The amplitude fields $A\,(x, y, t)$ representing higher-charged vortices of strengths $Q = 1$ (a) to 4 (d). During one oscillation period, the field rotates through an angle $360^\circ /Q$. From \citet{courtial2007experiments}. With kind permission of The European Physical Journal (EPJ).}
\end{figure}

The nodal lines of Chladni's vibrating plates are a window to a deeper theory---of vortices---that, paradoxically, neither Chladni nor anyone else ever witnessed in the motion of plates \cite{courtial2007experiments}. A direct (theoretical) route to observe the emergence of vortices is to study the displacement of the plate $A\,(x, y, t)$ from its equilibrium position, near the nodal lines. As with any wave phenomenon, it is convenient to regard the actual displacement as the real part of a complex amplitude $u(x, y)\, \exp\,(\mathrm{i}\,\omega\,t)$. In the immediate neighborhood of a nodal line along the $x$-axis, the complex field is $u^{(x)} (x, y) = y$; similarly, $u^{(y)} (x, y) = x$ for a nodal line along $\hat{y}$. The displacements, pictured in Fig.~\ref{fig:Vortex}, pivot about these lines as time evolves. Now, consider two frequency-degenerate eigenmodes with nodal lines in the $\hat{x}$ and $\hat{y}$ directions, which are locally described as above---obviously, these orthogonal lines cross each other at a point, somewhere on the plate. It is interesting to inspect the ensuing oscillation structures in the vicinity of the intersection point for a linear superposition of the eigenstates. If the complex fields are simply summed up as
\begin{equation}
\label{eq:in-phase}
u \,(x,y)= x + y
\end{equation}
the resultant eigenmode displays a nodal line, inclined at 45$^\circ$ to the two axes. Alternatively, if the fields are added out of phase by $\pi/2$,
\begin{equation}
\label{eq:out-phase}
u \,(x,y)= x + \mathrm{i}\,y,
\end{equation}
then the amplitude field $A$ does not merely pivot about a new nodal line but instead, rotates around the crossing point, thus engendering a \textsl{vortex} (at the intersection point). On retracing Chladni's steps, these vortices should, in principle, be visible as fine islands of sand. 

\begin{figure}[htb]
\includegraphics[width = 0.75\linewidth]{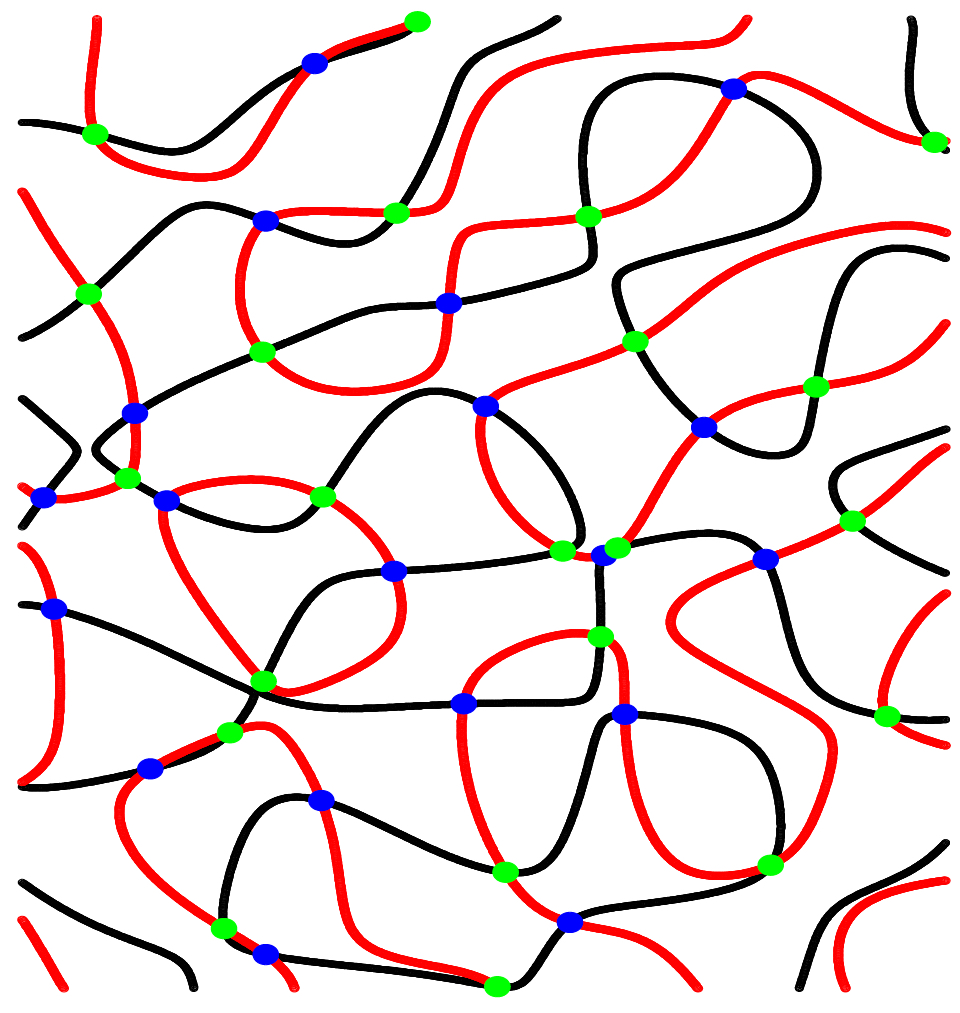}
\caption{\label{fig:uvNL}Typical pattern of the nodal lines $\im\,[ \psi (x, y)] = 0$ (black lines, $\eta$NL) and $\re\, [\psi (x, y)] = 0$ (red lines, $\xi$NL) for a random superposition of plane waves. Nodal lines in each set do not cross. Around the nodal points at which the two sets intersect, there is vortical flow in either clockwise (green dots) or anti-clockwise (blue dots) directions. From \citet{saichev2001distribution}.}
\end{figure}

As early as 1931, \citet{dirac1931} demonstrated that nodal points give rise to current vortices. Vortices, in the most general sense, refer to singular points at which the phase of a field becomes undefined. They are commonplace in physical fields of at least two variables, where the physical quantity signified by the field can be naturally represented in a plane \cite{dennis2009singular}, such as the Argand plane or $\mathbb{R}^2$. In particular, the field could be the wavefunction for an open quantum billiard, or one with broken time-reversal symmetry---in either case, $\psi$ acquires a finite imaginary part. It could also just as well be the complex order parameter of a superconductor or a scalar optical field. Consider a generic complex scalar wave
\begin{equation}
\psi\,(\mathbf{r}, t) = \xi \,(\mathbf{r}, t) + \mathrm{i}\, \eta \,(\mathbf{r}, t)= \rho\,(\mathbf{r}, t) \, \exp\, \left \{\mathrm{i} \,\chi\,(\mathbf{r}, t) \right \}
\end{equation}
with $\xi,\,\eta,\,\rho,\,\chi \in \mathbb{R}$. All is well with this representation, except at the nodal points where $\psi$ (and consequently, the intensity $=  \rho ^2$) is zero. Just as the angle of polar coordinates is ill-defined precisely at the origin, along loci of vanishing $\rho$, the phase $\chi$ is indeterminate. These phase singularities---also called optical vortices---were first illustrated by \citet{nye1974dislocations} as dislocations of wavefronts and later, investigated by \citet{berry1981singularities, berry1998much, nye1999natural}. The dislocations reside on the contour lines of intersection of the two surfaces
\begin{equation}
\label{eq:perp}
\xi \,(\mathbf{r}, t) = 0, \hspace{0.25cm} \eta \,(\mathbf{r}, t) = 0
\end{equation}
as seen in Fig.~\ref{fig:uvNL}; we call these lines $\xi$NL and $\eta$NL, respectively. To borrow \citeauthor{Berry2059}' \citeyearpar{Berry2059} eloquent description, ``in light, they are lines of darkness; in sound, threads of silence.'' Unfortunately, the nodal lines $\xi$NL and $\eta$NL, being dependent on the overall phase of the wavefunction, are hardly uniquely defined. The nodal points, on the contrary, are invariant under a phase transformation of the wave function and remain unmoved in space on multiplying $\psi$ by an (arbitrary) constant phase factor $\exp\, (\mathrm{i} \, \alpha)$ \cite{berggren2002crossover}. This makes them of greater utility when attempting to characterize a complex chaotic wave function.

Around a vortex, the phase assumes all possible values in its $2 \pi$ range, leading to a circulation of the wave energy. Although the change in $\chi$ is usually nonuniform \cite{mondragon1989quantum}, upon traversing a closed circuit $\gamma$ enclosing the phase-singular point, there is a net change of phase
\begin{equation}
\label{eq:Q}
Q = \frac{1}{2\pi} \oint \mathrm{d}\,\mathbf{r} \cdot \nabla \chi = \frac{1}{2 \pi} \oint_\gamma \mathrm{d}\,\chi,
\end{equation}
which is quantized in units of $2 \, \pi$. The integer $Q$ is referred to as the strength or topological
charge \cite{halperin1981} and $\mathscr{S} = \sgn\, (Q)$, which is positive if the phase increases in a right-handed sense, is colloquially termed the sign of the singularity. In the language of Eqs.~(\ref{eq:in-phase},~\ref{eq:out-phase}), the complex field near the center of a canonical charge-$Q$ vortex located at the origin takes the form \cite{molina2001observation}
\begin{equation}
u\,(x, y) = (x + \mathrm{i}\, y)^{\,\lvert Q \rvert} = \exp\, (\mathrm{i}\,Q\,\phi).
\end{equation}
Higher-charge vortices with $Q > 1$ (Fig.~\ref{fig:Vortex}), however, are non-generic in that they are not stable with respect to perturbations \cite{freund1999critical}, which split a charge-$Q$ vortex into $\lvert Q \rvert$ vortices of charge $\pm 1$ \cite{courtial2007experiments}. 

Associated with the field $\psi$ is a current
\begin{equation}
\label{eq:j}
\mathbf{j} = \im\, \left(\psi^*\, \nabla \psi  \right) = \xi \,\nabla \eta - \eta\, \nabla \xi = \rho^2\, \nabla \chi.
\end{equation}
This is the usual quantum-mechanical probability current density. In electrodynamics \cite{jackson1999classical}, $\mathbf{j}$ is the familiar Poynting vector if $\psi$ stands for a polarized (scalar) component of the electric field. Since $\mathbf{j}$ is aligned along the gradient $\nabla \chi$, i.e., in the direction of change of phase, phase singularities constitute vortices of the optical current flow \cite{dennis2009singular}. The corresponding vorticity is defined as
\begin{equation}
\label{eq:vorticity}
\mathbf{\Omega} = \frac{1}{2} \nabla \times \mathbf{j} = \frac{1}{2} \im \, \left(\nabla \psi^* \times \nabla \psi \right) = \nabla \xi \times \nabla \eta.
\end{equation}
The vector $\mathbf{\Omega}$, being perpendicular to the normals to the two surfaces in Eq.~\eqref{eq:perp}, points along the dislocation line. In three dimensions, the total length of dislocation line in a volume $\mathcal{V}$ is
\begin{equation}
L (\mathcal{V}) = \int_{\mathcal{V}} \mathrm{d}\,\mathbf{r}\,\,\delta\,(\xi)\,\delta\,(\eta)\, \lvert \Omega (\mathbf{r}) \rvert,
\end{equation}
whereby we can calculate the dislocation line density. For the time being, however, let us focus on a unit topological charge in one lower dimension. This simplifies life, and Eqs.~(\ref{eq:Q},~\ref{eq:j}), a little, whereupon
\begin{alignat}{1}
Q &= \sgn \,(\mathbf{\Omega} \cdot \hat{z} ) = \sgn \left(\xi_x\,\eta_y - \xi_y\,\eta_x\right),\\
\mathbf{j} &= \left(\xi\,\eta_x - \eta\,\xi_x, \,\xi \,\eta_y - \eta\,\xi_y \right),
\end{alignat}
the subscripts indicating partial derivatives. Note that the strength $Q$ is now identical to the sign $\mathscr{S}$---also called the winding number (WN)---that defines the sense of circulation of the current swirling about the vortex. Neighboring nodal points on the same nodal line (either $\xi$NL or $\eta$NL) always have opposite WNs \cite{berggren2002crossover} and generically, \textsl{any} nearest neighbors are strongly anticorrelated in sign \cite{shvartsman1994vortices}. In fact, the sign of any \textsl{single} vortex in a random Gaussian wave field determines the sign of \textsl{all} other vortices in the field. 

The current is zero at both vortices, where $\psi = \xi = \eta = 0$, and saddle points (stagnation points in the flow), where $\xi / \eta  = \xi_x/ \eta_x= \xi_y/ \eta_y$  \cite{hohmann2009density}. To tell them apart,\footnote{In conventional mathematical nomenclature, vortices (saddles) are the elliptic (hyperbolic) points in the flow.} we scrutinize the Jacobian
\begin{equation}
\mathcal{J} = \partial_x\, j_x\,\, \partial_y\, j_y - \partial_y\, j_x\,\,\partial_x j_y,
\end{equation}
which is positive at vortices but negative at saddles. As a consequence of $\psi$ satisfying the Helmholtz equation, $\mathcal{J}$ neatly separates into two parts
\begin{equation}
\mathcal{J} = \mathcal{J}_v -  \mathcal{J}_s,
\end{equation}
with the individual contributions
\begin{alignat}{1}
\mathcal{J}_v &\equiv \left ( \xi_x\,\eta_y - \xi_y\,\eta_x \right)^2,\\
\nonumber\mathcal{J}_s &\equiv \frac{1}{2} \left[ \left( \xi\,\eta_{xx} - \eta\,\xi_{xx} \right)^2 + \left( \xi\,\eta_{yy} - \eta\,\xi_{yy} \right)^2\right]\\
 &+ \left( \xi\,\eta_{xy} - \eta\,\xi_{xy} \right)^2.
\end{alignat}
From the general density of critical points
\begin{equation}
D_\mathrm{crit} (\mathbf{r}) =\hspace*{-0.25cm} \sum_{\{\mathbf{r}': \,\mathbf{j} (\mathbf{r}') = 0 \}} \hspace*{-0.25cm}\delta^2 (\mathbf{r}-\mathbf{r}') = \delta^2 \left(\mathbf{j}\, (\mathbf{r})\right) \,\lvert \mathcal{J} (\mathbf{r} ) \rvert,
\end{equation}
it is now easy to compute the saddle and vortex densities \cite{Berry2059, saichev2001distribution, dennis2001phase},
\begin{alignat}{1}
\label{eq:disdensity}
D_s (\mathbf{r}) &= \delta^2 \left(\mathbf{j}\, (\mathbf{r})\right) \mathcal{J}_s (\mathbf{r} ),\\
\nonumber D_v (\mathbf{r}) &= \delta^2 \left(\mathbf{j} \,(\mathbf{r})\right) \mathcal{J}_v (\mathbf{r} ) = \delta\,(\xi)\,\delta\,(\eta)\,\lvert \xi_x\,\eta_y - \xi_y\,\eta_x \rvert,
\end{alignat}
respectively.
$D_v$, which can be determined statistically (at least, in the mean), specifies the dislocation point (or vortex) density---the mean number of dislocation lines piercing unit area of a plane \cite{berry1978disruption}. These statistics have been thoroughly studied for quasimonochromatic paraxial waves \cite{baranova1981dislocations}, monochromatic waves in two dimensions \cite{freund1993optical, freund1994wave, freund1997critical, freund1994optical, freund1997parameterization, freund1998critical, shvartsman1994wave}, isotropic random waves \cite{Berry2059}, and random waves subject to the Aharonov-Bohm effect \cite{houston2017random} among others, with novel extensions to the statistics of knotted nodal lines by \citet{taylor2014geometry, taylor2016vortex}.
  
Closer to the theme of our discussion, vortices were predicted in eigenfunctions of highly symmetric 
billiards \cite{chen2002quantum, chen2003vortex, chen2003vortex2} and their telltale fingerprints have been spotted in closed, microwave cavities, coherently excited at multiple points \cite{dembowski2003phase}. Similar predictions \cite{berry1986quantum} and experiments \cite{chibotaru2001vortex, barth2002current, berggren2002crossover, olendski2003localized, sadreev2004current} were proffered for quantum billiards without time-reversal symmetry, and more recently, for open billiards \cite{kuhl2007nodal}. On the optical physics front too, phase singularities have generated considerable enthusiasm, in part due to their connections with beams carrying orbital angular momentum \cite{bliokh2017theory, allen1992orbital, allen1999iv, turnbull1996generation, he1995direct, simpson1996optical, garces2003observation, o2002intrinsic}, which fueled remarkable experiments \cite{karman1997creation, bm_thesis, o2006topology, o2006illustrations} seeking to actively create and annihilate optical dislocations in laser fields. However, as fascinating as the physics thereof may be, we wouldn't want to let our focus stray too far from nodal domains, to which we now thus return.

\section{Mathematical preliminaries}
\label{math}
The classical eigenvalue problem 
\begin{equation}
\label{eq:EigEq}
\Delta\, u\, ({\bf x}) = - \lambda\, u\, ({\bf x})
\end{equation}
for the Laplace operator  
\begin{equation}
\Delta = \partial^2/\partial x_1^2 + \ldots + \partial^2/\partial x_d^2,
\end{equation}
is, to say the least, ubiquitous in all branches of physics. Known as the Helmholtz equation \cite{von1865lehre}, it naturally arises on separating out the time variable from the wave equation. To begin with, let the Laplacian be defined in $d$ spatial dimensions on an open bounded connected domain\footnote{Here, the word ``domain'' is loosely used to signify the enclosed region on which the Laplacian eigenfunctions are defined. This is not to be confused with \textsl{nodal} domains (also called subdomains); the difference should be clear from the context of use.} $\mathcal{D} \in \mathbb{R}^d$ with a piecewise smooth boundary $\partial\, \mathcal{D}$. Although, in the context of quantum billiards, we focus on Euclidean domains in two dimensions ($d = 2$), the discussion can be completely generalized to any-dimensional manifolds.

The actual solutions to Eq.~\eqref{eq:EigEq} are inseparably intertwined with the boundary conditions on $\partial\, \mathcal{D}$. Perhaps the most frequently encountered is the Dirichlet boundary condition where the eigenfunctions are constrained to vanish on the boundary, i.e.,
\begin{equation}
\label{eq:Dirichlet}
u \, ({\bf x}) = 0 \hspace{0.5cm} \textrm{on} \hspace{0.5cm} {\bf x} \in \partial\,\mathcal{D}.
\end{equation}
For instance, borrowing from acoustics, the vibrational modes of a thin membrane (or drum) that is clamped along its boundary are exactly the Dirichlet Laplacian eigenfunctions with the drum frequencies proportional to $\sqrt{\lambda_j}$ \cite{rayleigh1945theory}; a particular vibrational eigenmode can thus be selectively excited at the corresponding frequency \cite{sapoval1991vibrations,sapoval1993vibrations,sapoval1997acoustical}. The combination of Eqs.~\eqref{eq:EigEq} and \eqref{eq:Dirichlet} also describes the propagation of a wave down a waveguide  with cutoff frequency $\sim \sqrt{\lambda_j}$; in this case, the eigenfunctions $u_j$ correspond to the so-called TM-mode \cite{collin1960field}. Other choices of boundary condition include the Neumann condition (which corresponds to the vibration of a free membrane)
\begin{equation}
\frac{\partial\,u}{\partial\,{\rm n}} \, ({\bf x}) = 0 \hspace{0.5cm} \textrm{on} \hspace{0.5cm} {\bf x} \in \partial\,\mathcal{D},
\end{equation}
$\partial / \partial\,{\rm n}$ being the normal derivative pointed outwards from the domain, or the Robin boundary condition
\begin{equation}
\frac{\partial\,u}{\partial\,{\rm n}} \, ({\bf x}) + h\,u \, ({\bf x}) = 0 \hspace{0.5cm} \textrm{on} \hspace{0.5cm} {\bf x} \in \partial\,\mathcal{D}, 
\end{equation}
for some positive constant $h$.

For more than a century, it has been known that the Laplace operator has a discrete spectrum\footnote{This discreteness, however, cannot always be taken for granted as warned by \citet{hempel1991essential} for the spectrum of a bounded domain with Neumann conditions on an irregular boundary.} of infinitely many non-negative eigenvalues \cite{pockels1892uber}. The eigenvalues can be labeled (and arranged in ascending order) by an integer index $j = 1, 2, 3, \ldots$ as
\begin{equation}
0 \le \lambda_1 < \lambda_2 \le \lambda_3 \le \cdots 
\end{equation}
with possible multiplicities (degeneracies). The set of eigenfunctions $\{u_j\, ({\bf x}) \}$ forms a complete basis in the functional space $L_2 \, (\mathcal{D})$ of measurable and square-integrable functions on $\mathcal{D}$ \cite{courant1953methods,simon1979methods}. To avoid any ambiguity about multiplicative factors, the eigenfunctions are conventionally normalized to unit $L_2$-norm
\begin{equation}
\lvert \lvert u_j \rvert \rvert_2 = \lvert \lvert u_j \rvert \rvert_{L_2\, (\mathcal{D})} = \left ( \int_\mathcal{D} \mathrm{d} {\bf x}\, \lvert u_j\, ({\bf x}) \rvert^2 \right)^{1/2} = 1
\end{equation}
and therefore, constitute an orthonormal set:
\begin{alignat}{2}
\textrm{Orthonormality:}& \int_\mathcal{D} \mathrm{d} {\bf x}\, u_i\, ({\bf x})\,u_j\, ({\bf x}) &&= \delta_{i, j},\\
\textrm{Completeness:}& \sum_j\, u_j\, ({\bf x})\,u_j^*\, ({\bf y}) &&= \delta\, ({\bf x} - {\bf y}).
\end{alignat}

The enormity of the literature on the eigenvalue problem is the quintessential paradox of plenty. From the study of elliptic operators to spectral theory and stochastic processes, the properties of eigenvalues and their attendant eigenfunctions have been at the heart of investigations aplenty. Here, we confront and sieve through this vast mathematical scholarship in attempting to bring together various ``facts'' about Laplacian eigenvalues and eigenfunctions (and eventually, their nodal domains) in one consolidated Section. Besides several books on the subject (e.g.,~\citet{bandle1980isoperimetric, arendt2009weyl, chavel1984eigenvalues, davies1996spectral, edmunds1987spectral, hale2005eigenvalues}), our discussion closely follows the review by \citet{Grebenkov2013}, which in turn builds on the comprehensive treatise by \citet{kuttler1984eigenvalues} along the same lines.

\subsection{Eigenvalues: Basic properties}

Even without inspecting the eigenfunctions, important information about the domain $\mathcal{D}$ can be gleaned from the eigenvalues alone as we demonstrate in this subsection. For the sake of brevity and continuity, we refrain from presenting detailed proofs; instead, we simply state the relevant results, the emphasis being on their underlying physics.

A good point to start is the variational formulation of the eigenvalue problem which owes itself to the \textsl{minimax} principle \cite{courant1953methods}
\begin{alignat}{1}
\lambda_j &= \frac{\lvert \lvert \nabla u_j \rvert \rvert^2_{L_2\, (\mathcal{D})} + h\, \lvert \lvert u_j \rvert \rvert^2_{L_2\, (\partial \, \mathcal{D})}} {\lvert \lvert u_j \rvert \rvert^2_{L_2\, (\mathcal{D})}}, \label{eq:Variational}\\
&= \min \max \frac{\lvert \lvert \nabla v \rvert \rvert^2_{L_2\, (\mathcal{D})} + h\, \lvert \lvert v \rvert \rvert^2_{L_2\, (\partial \, \mathcal{D})}} {\lvert \lvert v \rvert \rvert^2_{L_2\, (\mathcal{D})}}, \label{eq:Minimax} \\
\nonumber & = \min \max {\displaystyle \frac{{\displaystyle \int}_\mathcal{D} \left[\bigg({\displaystyle \frac{\partial v}{\partial x}} \bigg)^2 + \bigg({\displaystyle \frac{\partial v}{\partial y}} \bigg)^2 \right]\mathrm{d} {\bf x}\,\,\mathrm{d} {\bf y}}{{\displaystyle\int_\mathcal{D} v^2\,\,\mathrm{d} {\bf x}\,\mathrm{d} {\bf y}}}} \textrm{ for } d = 2,
\end{alignat}
where the maximum is over all linear combinations of the form
\begin{equation*}
v = a_1\, u_1 + a_2\, u_2 + \cdots + a_j\, u_j,
\end{equation*}
and the minimum is over all choices of $j$ linearly independent continuous and piecewise-differentiable functions $u_1,\, u_2,\, \ldots,\,\,u_j$ \cite{henrot2006extremum}; this space of functions is often termed the Sobolev space $H^1 (\mathcal{D})$. While on the subject of terminology, we add that the ratio of quadratic forms in the third equality is known as the Rayleigh quotient. Although Eq.~\eqref{eq:Minimax} looks cumbersome, simplification immediately follows for both Dirichlet and Neumann boundary conditions. It is easy to recognize that the second term in the numerator of Eq.~\eqref{eq:Minimax} vanishes with these conditions; in the former case, $v$ itself is zero on the boundary $\partial \, \mathcal{D}$ whereas for the latter, $h = 0$. In addition to its mathematical elegance, the minimax principle has other important consequences like the Rayleigh-Ritz method for obtaining upper bounds for eigenvalues. The Rayleigh-Ritz method has also been used to estimate the eigenvalues of the Laplacian for rhombical \cite{birkhoff1970accurate,weinstein1966some} and parallelogram \cite{durvasula1968natural} regions. For a more detailed discussion about the numerical aspects of this connection, the curious reader is directed to Ch.~12 of \cite{kuttler1984eigenvalues}. 

\begin{figure}[htb]
\includegraphics[width=0.75\linewidth]{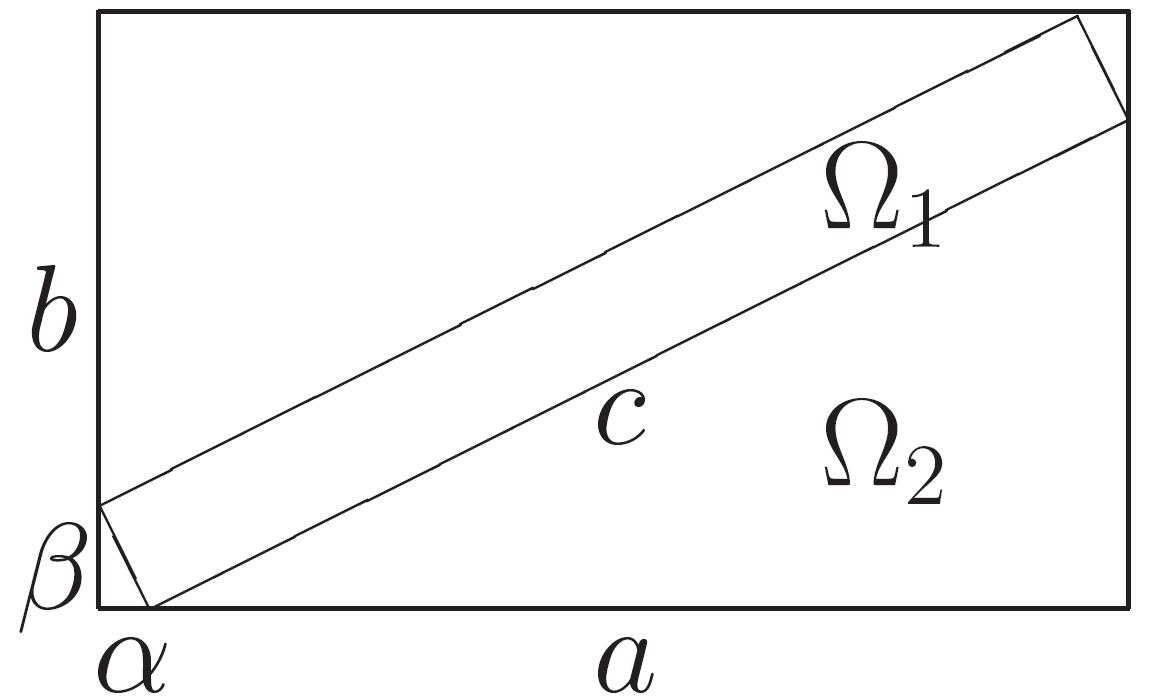}
\caption{\label{fig:Minimax}A counterexample to the property of domain monotonicity for the Neumann boundary condition. Although $\mathcal{D}_1 \subset \mathcal{D}_2$, the second eigenvalue $\lambda_2\, (\mathcal{D}_1)= \pi^2/c^2 < \lambda_2\, (\mathcal{D}_2)= \pi^2/a^2$ (if $a> b$) when ${\displaystyle c = \sqrt{(a-\alpha)^2 +(b-\beta)^2} >a}$. From \citet{Grebenkov2013}; figure by Naoki Saito.}
\end{figure}

Another significant implication of the minimax principle is the property of domain monotonicity for Dirichlet boundary conditions, i.e., eigenvalues monotonously decrease when the domain is enlarged. To see this, consider a region $\mathcal{D}$ that is properly contained in $\mathcal{D}'$; a set of admissible eigenfunctions for $\mathcal{D}'$ is then simply $u_j' ({\bf x}) = u_j$ if ${\bf x} \in \mathcal{D}$ and $0$ otherwise. Applying the minimax principle shows that $\lambda_j (\mathcal{D}) > \lambda_j (\mathcal{D}')$ or in words, the larger the region, the smaller the eigenvalues. This argument breaks down for Neumann or Robin boundary conditions, as illustrated by Fig.~\ref{fig:Minimax}. A word of caution, however: domain monotonicity, in general, is \textsl{not} equivalent to the (trivially true) statement that on magnifying/shrinking a domain by a factor $\alpha$, all the eigenvalues are rescaled by $1/\alpha^2$, which is a much weaker property instead. Another aspect of monotonicity implied by the minimax principle is that the eigenvalues $\lambda_j$ increase with $h$, namely, if $h < h'$, then $\lambda_j (h) \le \lambda_j (h')$; hence, the eigenvalues of the Robin problem are always intermediate between those of the corresponding Neumann and Dirichlet problems. 

The importance of restructuring the problem in variational terms is that Eq.~\eqref{eq:Variational} directly ensures that all eigenvalues are non-negative. In fact, the first eigenvalue $\lambda_1$ is simple (i.e., nondegenerate) and strictly positive for Dirichlet and Robin boundaries but zero with Neumann boundary conditions for which $u_1$ is itself a constant. For the second Dirichlet eigenvalue, \citet{cheng1976eigenfunctions} proved that the multiplicity $M\, (\lambda_2) \le 3$ and the inequality is \textsl{sharp} at that, which means that the equality is actually achieved for a particularly constructed domain. With regard to the higher excited states $j \ge 3$, the best known bound is $M\, (\lambda_j) \le 2j -3$ \cite{hoffmann1999bounds, hoffmann1999multiplicity}. 

The final question that we ask here is how are the eigenvalues affected when the domain $\mathcal{D}$ is perturbed? Obviously, the eigenvalues are invariant under translations and rotations of $\mathcal{D}$, which correspond to area-preserving linear transformations of the coordinate axes. This fact that has been routinely exploited for image recognition and analysis \cite{reuter2006laplace,saito2009analysis}. Ideally, for the sake of numerical computation by finite-element or other approximation methods, one would like $\lambda_j$ to be minimally disturbed by more generic (small) perturbations on $\partial\, \mathcal{D}$ which alter the shape of the domain. Indeed, for Dirichlet boundary conditions, this is indeed the case: the eigenvalues vary continuously under a ``continuous'' perturbation of the domain \cite{courant1953methods}. This statement is unfortunately false, in general, for Neumann boundary conditions in which case, it holds only if a bounded domain with a smooth boundary is deformed by a ``continuously differentiable transformation'' \cite{burenkov2002spectral}.

\subsection{Weyl's law}

The shape of a billiard is intimately related to the properties of the associated eigenvalues and in the low-frequency limit, this connection manifests itself in the form of several isoperimetric inequalities (reviewed in Appendix~\ref{sec:isoperimetric}). Arguably the most famous of the connections between the spectrum and the geometric shape of the domain is Weyl's law \cite{baltes1976spectra,weyl1911asymptotische, Weyl1912}, which asserts that in $d$ dimensions, the asymptotic $(j \rightarrow \infty)$ behavior of the eigenvalues is given by 
\begin{equation}
\label{eq:Weyl}
\lambda_j \propto \frac{4\, \pi^2}{\left(\omega_d\, \mu_d (\mathcal{D})\right)^{2/d}}\, j^{2/d};\quad \omega_d = \frac{\pi^{d/2}}{\Gamma\,(d/2 + 1)},
\end{equation}
where $\mu_d (\mathcal{D})$ is the Lebesgue measure of $\mathcal{D}$ (nothing but the area in two dimensions or the volume in three) and $\omega_d$ is the volume of a $d$-dimensional unit sphere, $\Gamma$ being the Gamma function. Therefore, the area (volume) in two (three) dimensions can be extracted from the slope of $\lambda_j$ graphed against $j^{2/d}$, i.e., by counting how rapidly the eigenvalues grow. Eq.~\eqref{eq:Weyl} can be inverted and rewritten as a formula for the index $j$ instead---this naturally leads us to define a counting function (the number of eigenvalues smaller than $\lambda$) as
\begin{equation}
\label{eq:Counting}
N (\lambda) =  \sum_{j = 1}^{\infty} \Theta (\lambda - \lambda_j ) \propto \frac{\omega_d\, \mu_d (\mathcal{D})}{(2 \pi)^d} \, \lambda^{d/2}\quad (\lambda \rightarrow \infty),
\end{equation}
with $\Theta$ denoting the Heaviside step function.
The description of Eq.~\eqref{eq:Counting} only accounts for the leading-order term in the counting function and there are higher-order corrections that yield information about the boundary of the domain. These corrections, initially proposed by Weyl (and justified later by \citet{Ivrii1980} and \citet{melrose1980weyl} for convex $\mathcal{D}$), are given by
\begin{alignat}{1}
\label{eq:WeylCorr}
N(\lambda) = \begin{cases} \displaystyle \frac{\mathcal{A}}{4 \pi} \lambda \mp \frac{\mathcal{P}}{4 \pi} \sqrt \lambda &\mbox{for } d = 2 \vspace*{0.2cm} \\
\displaystyle \frac{\mathcal{V}}{6 \pi^2} \lambda^{3/2} \mp \frac{\mathcal{S}}{16 \pi} \lambda & \mbox{for } d = 3 \end{cases},
\end{alignat}
where the $- \,(+$) sign is applicable for Dirichlet (Neumann) boundary conditions. Here, and later (unless explicitly stated otherwise), $\mathcal{A}$, $\mathcal{P}$, $\mathcal{V}$ and $\mathcal{S}$ stand for the area, perimeter, volume and surface area of the billiard, respectively, in the appropriate dimensions. \citet{Berry1979,berry1980wavefront} hypothesized that for irregular boundaries, the correction term should be $\lambda^{H/2}$, where $H$ is the Hausdorff dimension of the boundary (rather than $\lambda^{(d-1)/2}$ as Eq.~\eqref{eq:WeylCorr} seems to suggest)---this conjecture was disproved by \citet{Brossard1986} who suggested the use of the Minkowski (or box-counting) dimension instead. The thus modified Weyl-Berry conjecture was proved shortly thereafter for $d = 1$ \cite{lapidus1991fractal,lapidus1993riemann} and falsified for $d > 1$ \cite{lapidus1996counterexamples}. Extensions to domains with fractal \cite{levitin1996spectral} or rough boundaries \cite{netrusov2005weyl} as well as to manifolds and higher-order Laplacians \cite{Desjardins1994, DESJARDINS1998257} soon followed. However, we digress.

\subsection{Eigenfunctions and nodal lines}

After much deliberation on the eigenvalues of the Laplacian in the preceding subsections, we now examine the universal properties of the eigenfunctions themselves as a segue into nodal portraits. Of great import are questions surrounding the smoothness of the eigenfunctions. First of all, the eigenfunctions are infinitely differentiable ($C^\infty$) at any point inside the region $\mathcal{D}$ \cite{bernstein1950existence}. Observe that this $C^\infty$ smoothness is preserved upon reflecting $u_j$ as an odd function across any straight (or more generally, $C^\infty$) portion of the boundary---the resultant function still satisfies the Helmholtz equation locally in a neighborhood of that boundary segment. The fact that such odd reflections can be used to extend the eigenfunctions to larger and larger regions was precisely exploited by \citet{lame1866leccons} to obtain the eigenfunctions and eigenvalues of the equilateral triangle by tiling the plane therewith. Moreover, the eigenfunctions are characterized by the so-called \textsl{unique continuation property} \cite{kuttler1984eigenvalues}: a function $u_j$ satisfying the Helmholtz equation on a domain $\mathcal{D}$ cannot vanish on an open subset $\mathbb{D} \subset \mathcal{D}$ without vanishing identically in the region (i.e., $u_j (\mathbf{r})$ must be zero $\forall\, \mathbf{r} \in \mathcal{D}$). 

To gain some insight into the geometrical structure of the eigenfunctions, it is rewarding to look at the corresponding nodal lines. The nodal set is formally defined as
\begin{equation}
\mathscr{N}_j = \left \{ \mathbf{r} \subset \mathcal{D}\, \vert \, u_j (\mathbf{r}) = 0\right \}.
\end{equation}
Owing to the unique continuation property, the nodal set is also comprised of curves that are $C^\infty$ in $\mathcal{D}$. Where nodal lines cross, they do so at equal angles \cite{courant1953methods}. To put it formally, if $u\, (\mathbf{r}) = 0$, then in any neighborhood of $\mathbf{r}$, the nodal line is either a smooth curve, or an intersection of $n$ smooth curves at equal angles \cite{bers1955local}. This equiangularity also extends to when nodal lines intersect a $C^\infty$ portion of the boundary. Thus, as \citet{kuttler1984eigenvalues} note, ``a single nodal line intersects the $C^\infty$ boundary at right angles, two intersect it at 60$^\circ$ angles, and so forth.'' These nodal lines branch out across $\mathcal{D}$, forming an intricate network that partitions the region into nodal domains. The mathematical corpus on nodal domains is rich and varied with several deep and powerful theorems. Interestingly enough, it is also replete with rather many equally valuable but incorrect assertions and falsified conjectures that strikingly bespeak the circuitous evolution of the subject.

For concreteness, let us illustrate these results in the context of a specific example---the two-dimensional rectangular billiard with (for now) Dirichlet boundary conditions. The solutions for the free vibrations of a thusly-shaped membrane were first studied by \citet{poisson1829memoire} whose analysis, as has been remarked, ``left little to be desired'' \cite{rayleigh1945theory}. For the rectangle defined by $\mathcal{D} = \{ (x, y) \in \mathbb{R}^2 \vert 0 \le x \le a, \, 0 \le y \le b \}$, the Helmholtz equation can be solved quite easily by the method of separation of variables. The eigenfunctions are
\begin{equation}
\label{eq:rect}
u_{m, n} (x, y) = \sqrt{\frac{4} {a \,b}} \sin \left(\frac{m\, \pi \,x}{a} \right) \sin \left(\frac{n\, \pi\, y}{b} \right),
\end{equation}
with eigenvalues 
\begin{equation}
\lambda_{m,n} = \pi^2 \left[ \left(\frac{m}{a} \right)^2 + \left(\frac{n}{b} \right)^2 \right]; \quad m, n = 1, 2,\ldots .
\end{equation}
Since the eigenfunction \eqref{eq:rect} neatly splits into a product of functions that depend on $x$ and $y$ individually (and not on any combination thereof), the rectangle is the prototypical example of a \textsl{separable} billiard. The nodal set is straightforward to visualize: it consists of a grid formed by vertical and horizontal lines at $x = q_1\,a / m$ and $y = q_2\,b/n$, respectively, where $q_1, q_2 \in \mathbb{N}$; $q_1 < m, \, q_2 < n$. Evidently, the total number of nodal domains for the eigenstate $(m, n)$  is $\nu_{m, n} = m\, n$. Instead of the pair of quantum numbers $(m, n)$, the same state can be interchangeably indexed by a label $j$ on the eigenvalues such that
\begin{equation}
0 < \lambda_1 < \lambda_2 \le \ldots \le \lambda_j \le \lambda_{j+1} \le \ldots.
\end{equation}
To highlight some of the nontrivial features that emerge, even in this simple system, we list a sequence of such eigenstates and their associated number of nodal domains for a square billiard of side $\pi$ in Table~\ref{Table:Rectangle}.

\bgroup
\def\arraystretch{1.5}
\begin{table}[htb]
\centering
{
\bgroup
\small
\setlength{\tabcolsep}{15.25pt}
\begin{tabular}{c r c c r} \hline
\multicolumn{1}{c}{$j$} &\multicolumn{1}{c}{$(m,n)$} &\multicolumn{1}{c}{$E_{m,n}$} &\multicolumn{1}{c}{$\nu_{m,n} $} &\multicolumn{1}{c}{$\xi _j$}  \\ \hline
$5$ &$(1,3)$&$10$&$3$&$3/5$ \\
$6$ &$(3,1)$&$10$&$3$&$1/2$ \\
$7$ &$(2,3)$&$13$&$6$&$6/7$ \\
$8$ &$(3,2)$&$13$&$6$&$3/4$ \\
$9$ &$(1,4)$&$17$&$4$&$4/9$ \\
$10$ &$(4,1)$&$17$&$4$&$2/5$ \\
$11$ &$(3,3)$&$18$&$9$&$9/11$ \\
$12$ &$(2,4)$&$20$&$8$&$2/3$ \\ 
$13$ &$(4,2)$&$20$&$8$&$8/13$ \\
$14$ &$(3,4)$&$25$&$12$&$6/7$ \\ \hline
\end{tabular}
}
\egroup
\caption{\label{Table:Rectangle}The eigenstates for a square of side $\pi$ are arranged in increasing order of their eigenvalues $E_j = E_{m,n} = m^2 + n^2$, indexed by the label $j$. For various pairs of quantum numbers, the number of domains $\nu_j = \nu _{m,n}$, and the normalized mode number $\xi _j = \nu_j / j$ are shown here. This tabulation is a direct way to appreciate the complexity and nonmonotonicity of the sequence $\{\xi _j\}$ even for so tractable a system. This makes the pursuit of finding a statistical description for $\xi$ important and worthwhile. }
\end{table}
\egroup

\begin{figure}[htb]
\includegraphics[width=\linewidth]{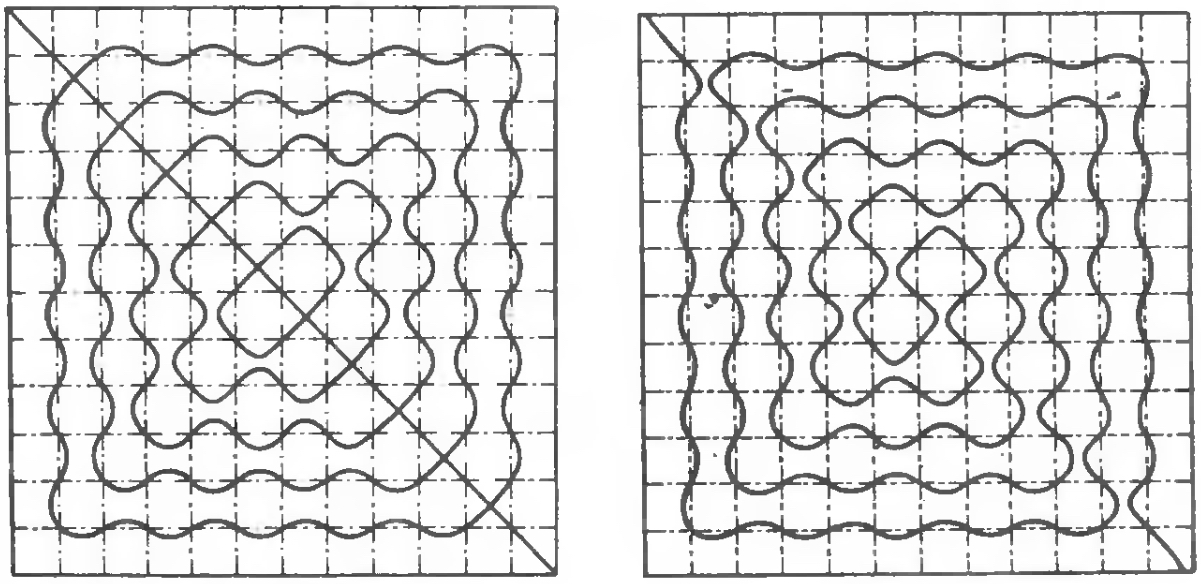}
\caption{\label{fig:RectCH}Consider the family of (linear superpositions of) eigenfunctions of the Laplacian on the square $[0, \pi]^2$
\begin{equation}
\Phi^\theta_{1, 2 r} (x,y) = \cos \theta \sin x \sin (2 r y) + \sin \theta \sin (2 r x) \sin y,
\end{equation}
associated with the Dirichlet eigenvalue $1+4r^2$. For $\theta = \pi/4$ (left), the nodal pattern exhibits 12 domains whereas for $\theta \lesssim \pi/4$ (right), the double points all disappear and the nodal set (a single connected line) divides the square into only two domains \cite{Berard2015}. In fact, \citet{stern1924} established that there are actually infinitely many eigenfunctions having exactly two nodal domains. This example also illustrates why counting nodal domains is such a hard problem, requiring one to resort to powerful numerical techniques. From \citet{courant1953methods}. \textcopyright\, 2004 WILEY-VCH Verlag GmbH \& Co. KGaA.}
\end{figure}

Two particular attributes become apparent upon perusing Table~\ref{Table:Rectangle} for a while. Firstly, for $j >1$, the eigenvalues can be degenerate---this is characteristic of regions with symmetries. Oftentimes, the degeneracy can be lifted by choosing a rectangle with incommensurate side lengths, thereby breaking the aforesaid symmetry. Whenever $m \ne n$, $E$ necessarily has a multiplicity of at least two as it is symmetric under $i \leftrightarrow j$.  Another class of degeneracies stems from purely number-theoretic origins. An example in this category is the fourfold-degenerate eigenvalue $65 = 1^2 + 8^2 = 4^2 + 7^2$. The question therefore is, in how many ways can a given integer $E$ be written as the sum of the squares of two integers? Number theory provides an answer. If $E$ is decomposed into distinct primes as
\begin{equation}
E = 2^\alpha\,p_1^{r_1} \cdots p_k^{r_k} \,q_1^{s_1} \cdots q_l^{s_l},
\end{equation}
where the $p_i$ are of the form $4\,t + 1$ and the $q_i$ of the form $4\,t+3$, then $s_i$ is even $\forall\,i$ and the multiplicity of $E$ is \cite{hardy1979introduction}
\begin{equation}
M  (E) = \prod_{i=1}^k \left (r_i + 1 \right).
\end{equation} 
As expected, a quick check shows that this formula correctly yields the multiplicity of 65 as 4. Accordingly, one can have eigenvalues of arbitrarily large multiplicity. For eigenvalues with multiplicities greater than one, the nodal sets, now formed by a superposition of eigenstates, can be much more exotic as Fig.~\ref{fig:RectCH} illustrates. Referring to \cite{courant1953methods} provides one with ample pictures to argue that the number of nodal domains for a linear combination of two given independent eigenfunctions can be smaller or larger than the number of nodal domains of either \cite{Berard2015}. However, for a generic region, there are still certain constraints on the eigenfunctions. \citet{uhlenbeck1972eigenfunctions, uhlenbeck1976generic} proved that for ``most" regions, the eigenvalues are all simple (nondegenerate), the nodal lines do not intersect, and the eigenfunctions' critical points are either maxima or minima. Given a region $\mathcal{D}$ that violates one or more of these properties, one can always obtain another $\mathcal{D}'$---by arbitrarily small (perhaps symmetry-breaking) perturbations---that indeed satisfies them. Fig.~\ref{fig:RectCH} gives a tangible example of how nodal crossings pull apart under such perturbations. Recognizing the instability of Dirichlet nodal domains under various perturbations due to the ``avoided crossings'', \citet{mcdonald2014neumann} proposed a partition of the domain $\mathcal{D}$ by trajectories of the gradient linking saddle points to extrema. These lines are (misleadingly enough) designated ``Neumann nodal curves'' since their tangent vectors are always parallel to $\nabla u$ and resultantly, $u$ satisfies the Neumann boundary condition along them. Such a construction, as artificial as it may seem, largely eliminates the problem of avoided crossings.

The second observation underscored by the data in Table~\ref{Table:Rectangle} concerns the ratio $\xi_j = \nu_j/j$. The percipient reader would surely have presaged that it is not mere coincidence that $\xi_j$ always happens to be less than unity. In fact, it is not and that brings us to our next set of theorems.

\subsubsection{Courant's nodal domain theorem}

Setting aside billiards and membranes for a moment, let us regress to the simplest possible example that permits a description of nodal sets---a one-dimensional, vibrating string. The nodes, which are now points rather than lines, are the spots on the string that remain stationary at all times. As regards the nodal ``domains'' (more correctly, the nodal intervals that the string is partitioned into), \citet{sturm1836a, sturm1836b} proved the following result, made rigorous by \citet{bocher1898}.

\begin{theorem}{Sturm oscillation theorem \cite{simon2005sturm}:}
\begin{enumerate} 
\item Let $\lambda_j$ be the eigenvalues of $H = -\mathrm{d}^2/\mathrm{d}\, x^2 +V(x)$ with boundary conditions $u_j\,(0) = u_j\,(a) = 0$. Then $u_j (x)$ has exactly $j$ zeros in $(0, a)$.
\item The number of eigenvalues of $H$ strictly below $\Lambda$ is exactly the number of zeros of $u_{j (\Lambda)} (x)$ in $(0,a)$.
\end{enumerate}
\end{theorem}
Sturm, Liouville, and Rayleigh extended this statement to add that a linear combination of $u_m$, $u_{m+1}$, $\ldots$, $u_n$ with constant coefficients has at least $m-1$ and at most $n-1$ zeros in the open interval $(0,a)$ spanned by the string \cite{pleijel1956remarks}. Generalizing results of this sort to higher-dimensional regions dovetails into \citeauthor{courant1923}'s \citeyearpar{courant1923} theorem.  Although the original proof outlined by \citet{courant1953methods} was for planar domains, it has since been adapted to compact Riemannian manifolds \cite{berard1982inegalites}.
\begin{theorem}{Courant's nodal domain theorem:}
\begin{enumerate} 
\label{item1}
\item The first eigenfunction, $u_1, (\mathbf{r})$ corresponding to the smallest eigenvalue, $\lambda_1$, on a domain $\mathcal{D}$ with arbitrary homogeneous boundary conditions does not have any nodes.
\item For $j \ge 2$, $u_j (\mathbf{r})$, corresponding to the $j^{\mathrm{th}}$ eigenvalue of the Laplacian counting multiplicity, divides the region $\mathcal{D}$ into at least two and no more than $j$ domains.
\end{enumerate}
No assumptions are made about the number of independent variables.
\end{theorem}
Even without getting into the details of the derivation, which can be found elsewhere \cite{strausspartial}, we can easily see that assuming Theorem~\ref{item1}.1 is correct, $u_j (\mathbf{r})$, for $j>2$, must divide $\mathcal{D}$ into two domains at the very least. The proof proceeds as follows \cite{saito2007}. Since $u_1$ is orthogonal to $u_{j \ne 1}$, we have
\begin{equation}
\int_\mathcal{D} u_1 (\mathbf{r})\,u_j (\mathbf{r})\,\,\mathrm{d}\, \mathbf{r} = 0,
\end{equation}
and from Theorem~\ref{item1}.1, we also know that $u_1 (\mathbf{r}) > 0$ or $u_1 (\mathbf{r}) < 0$ $\forall \,\mathbf{r} \in \mathcal{D}$. Hence, $u_j (\mathbf{r})$ must necessarily change its sign somewhere in $\mathcal{D}$ and therefore, appealing to the continuity of $u_j (\mathbf{r})$, there exist zeros of $u_j (\mathbf{r})$ in $\mathcal{D}$. These zeros form the nodal set. At the end of Courant's original proof of Theorem~\ref{item1} \cite{courant1953methods}, there appears a rather innocuous (and now notorious) footnote:
\begin{quote}
The theorem just proved may be generalized as follows: Any linear combination of the first $n$ eigenfunctions divides the domain, by means of its nodes, into no more than $n$ subdomains.
\end{quote}
This assertion, attributed to \citet{herrmann1932beitrage, herrmann1936beziehungen}, is---without additional qualifiers---egregiously untrue. Evidence to its falsification was first pointed out by V. I. Arnold.\footnote{\citet{arnold2011topological} reminisces: I wrote a letter to Courant, ``Where can I find this proof now, 40 years after Courant announced the theorem?'' Courant answered that ``one can never trust one's students: to any question they answer either that the problem is too easy to waste time on, or that it is beyond their weak powers.''} He realized that by generalizing Courant's theorem to include Herrmann's, one could arrive at conclusions about the topology of algebraic curves \cite{arnold1973topology} that outright contradicted the known results of quantum field theory. Eventually, putting the final nail in the coffin, \citet{viro1979construction} constructed a real algebraic hypersurface as an explicit counterexample. In general, Herrmann's theorem is valid only under some restrictions on the number of independent variables; in particular, it is false for the Laplacian on $\mathbb{S}^3$ and higher-dimensional spheres \cite{kuznetsov2015delusive}.

\begin{figure}[htb]
\includegraphics[width= 0.24\linewidth]{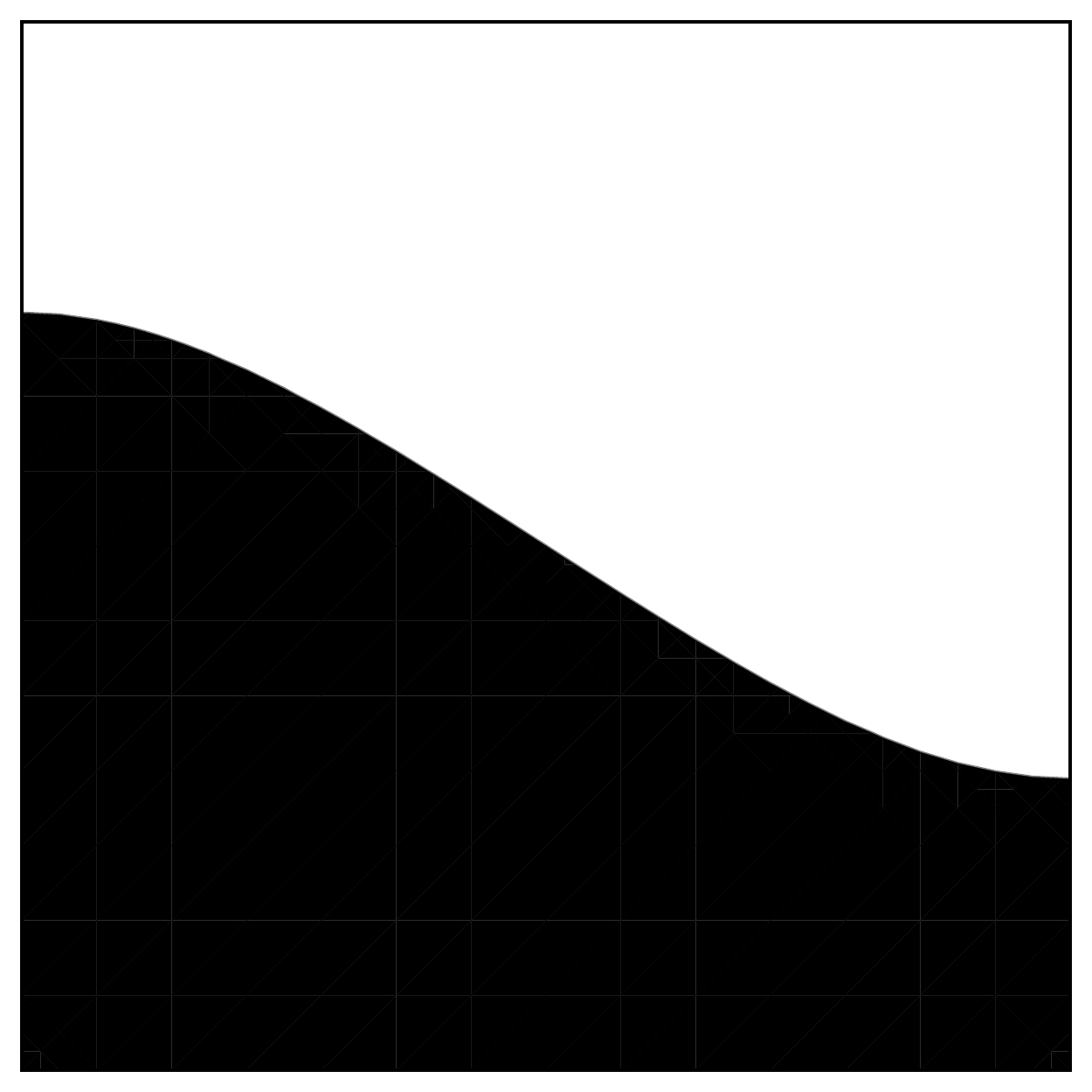}
\includegraphics[width= 0.24\linewidth]{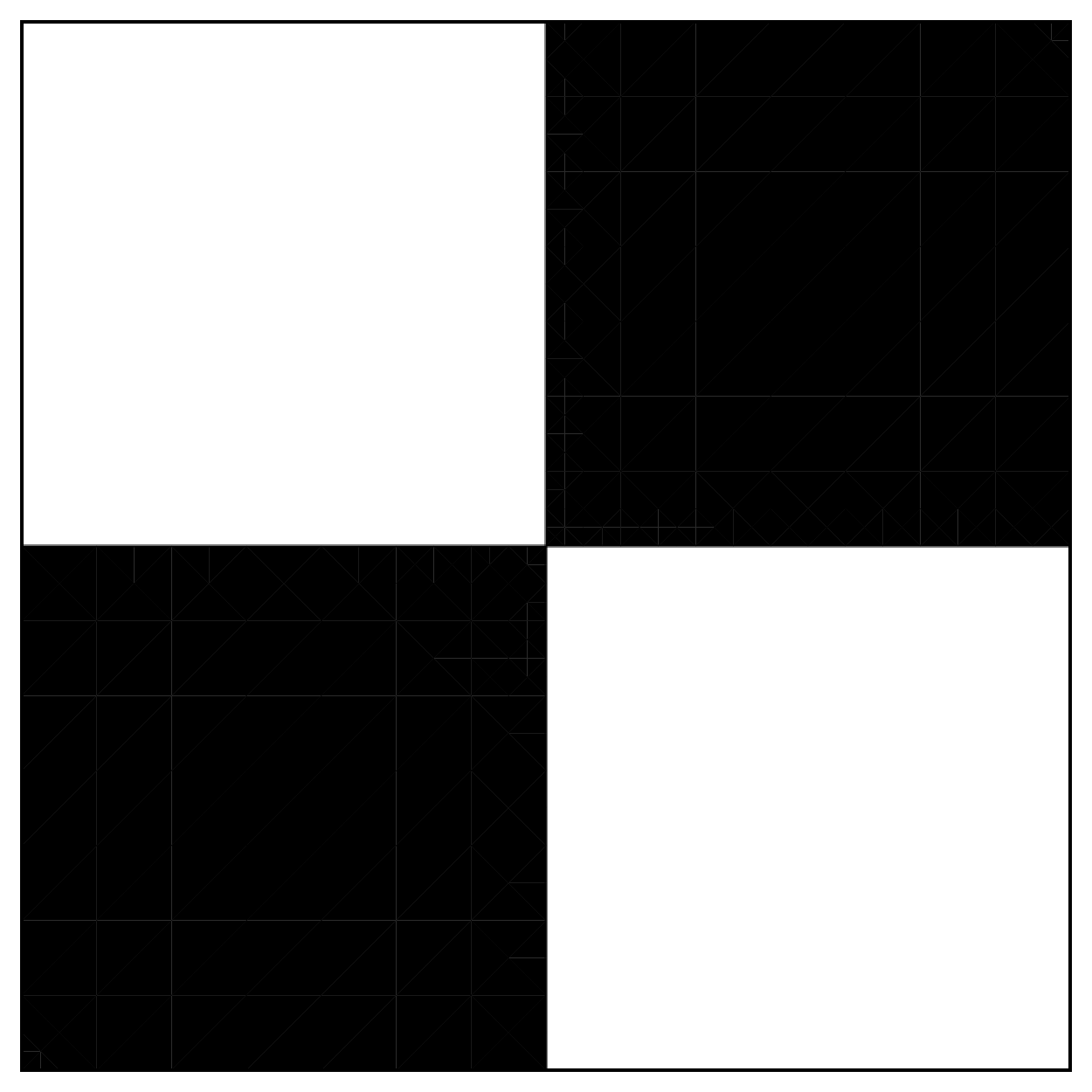}
\includegraphics[width= 0.24\linewidth]{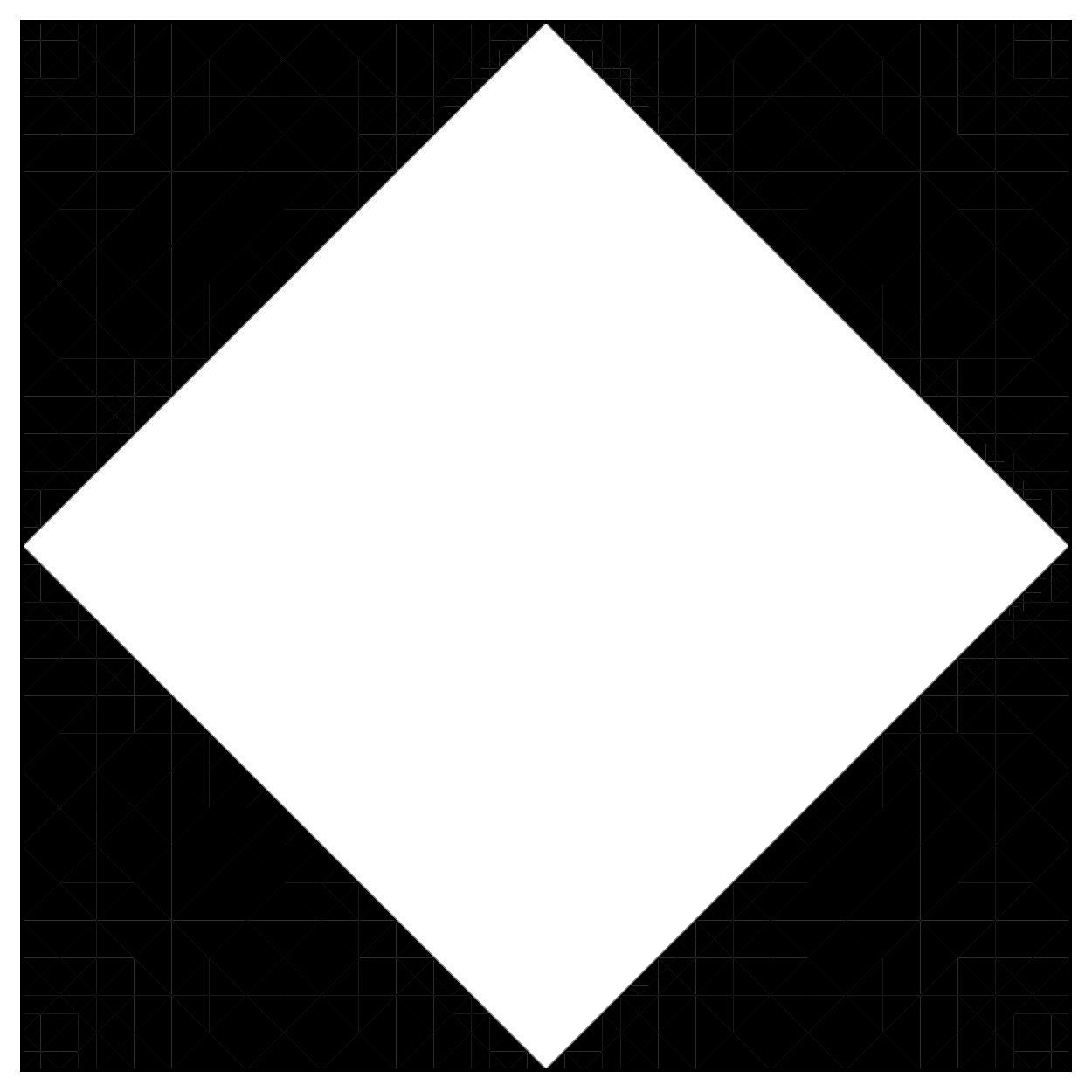}
\includegraphics[width= 0.24\linewidth]{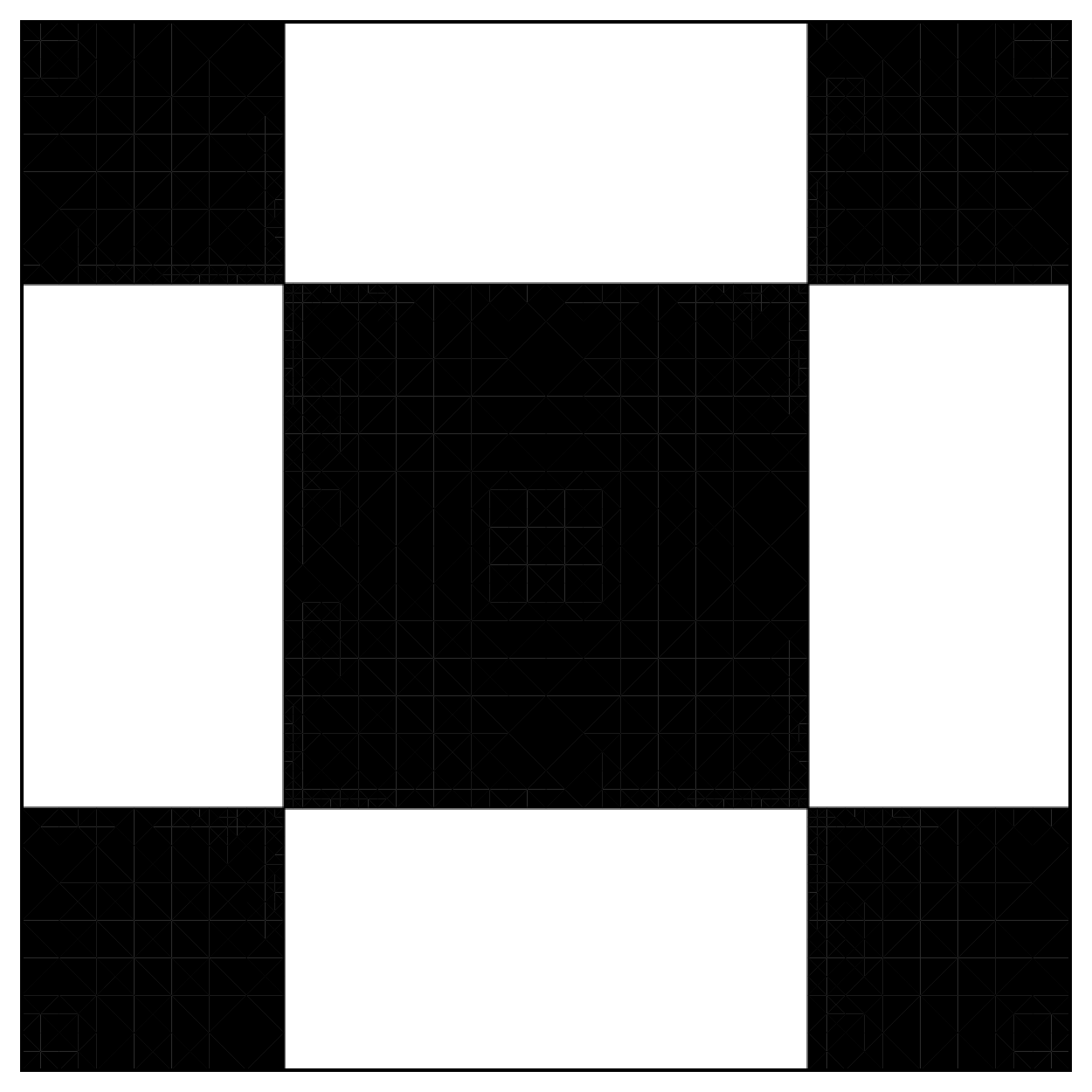}
\caption{\label{fig:neumann}Nodal patterns of Neumann eigenfunctions in the square. The nodal sets are depicted for the four nontrivial Courant-sharp cases, $\lambda_2$, $\lambda_4$, $\lambda_5$, and $\lambda_9$, corresponding to wavefunctions $\cos \theta \cos x + \sin \theta \cos y\, (\theta = 1$ here), $\cos x \cos y$, $\cos 2x + \cos 2y$, and $\cos 2x \cos 2y$, respectively.  Regions where the wavefunction is positive (negative) are painted black (white).}
\end{figure}

On the contrary, there do exist other modifications to Courant's proposition that are actually correct. \citet{pleijel1956remarks} showed that for planar domains with Dirichlet boundary conditions, Courant's bound can be asymptotically improved, proving that for an infinitely long sequence,
\begin{equation}
\label{eq:pleijel}
\lim_{j \to \infty} \sup \frac{\nu _j}{j} \leq \left( \frac{2}{\mathscr{J}_{0,1}} \right)^2 \approx 0.691,
\end{equation}
where $\mathscr{J}_{\,\upsilon,1}$ is the first positive zero of the Bessel function $J_\upsilon (z)$. Let us briefly recall Pleijel's argument here, not only because the proof is enlightening but also because it gives us the chance to justify our compilation of the much-vaunted isoperimetric inequalities in Appendix~\ref{sec:isoperimetric}. Denote by $\omega_1, \omega_2, \ldots, \omega_\nu$ the nodal domains of the $j^\mathrm{th}$ eigenfunction of the Dirichlet Laplacian, $u_j$. Inside each domain $\omega_i$, the function $u_j$ is nonzero; hence, $\lambda_j$ is the first eigenvalue (with Dirichlet boundary conditions along the perimeter) of the region $\omega_i$. One can then apply the Faber-Krahn inequality (Eq.~\ref{eq:fk}) individually to each $\omega_i$ and sum over $i$ to get
\begin{equation}
\label{eq:plProof1}
\sum_i \frac{\mathcal{A} \left( \omega_i \right)}{\pi\,\mathscr{J}_{0,1}^2} = \frac{\mathcal{A} }{\pi\,\mathscr{J}_{0,1}^2} \ge \sum_i \frac{1}{\lambda_j} = \frac{\nu_j}{\lambda_j},
\end{equation}
following which, we can decompose $\nu_j / \lambda_j$ as $\nu_j / j \times j / \lambda_j$. According to Weyl's law (Eq.~\ref{eq:Weyl}), in the limit $j \rightarrow \infty$, $j/\lambda_j \propto \mathcal{A}/4\pi$. Making this substitution in the second term of the product, we obtain Eq.~\eqref{eq:pleijel}.
To attain the maximal upper bound of $\nu_j = j$ in Courant's inequality, a necessary condition is that
\begin{equation}
\label{eq:plProof2}
\frac{\mathcal{A} }{\pi\,\mathscr{J}_{0,1}^2} \ge \frac{j}{\lambda_j}.
\end{equation}
Suppose  $\nu_j = j$ for infinitely many values of $j$. Reusing Weyl's law, the right-hand side above can be replaced as:
\begin{equation}
\frac{\mathcal{A} }{\pi\,\mathscr{J}_{0,1}^2} \ge \frac{\mathcal{A} }{4\, \pi},
\end{equation}
which is blatantly false because $\mathscr{J}_{0,1}  \approx 2.4048 > 2$. This led \citet{pleijel1956remarks} to conclude that for the eigenfunctions $u_j$ of a membrane with a fixed boundary, the maximum $j$ of the number of nodal domains is attained only for a \textsl{finite} number of eigenvalues.  For the square membrane, we can actually identify the eigenvalues that are Courant sharp (i.e., $\lambda_j$ such that $\nu_j = j$). In this case, \citet{pleijel1956remarks} derived the identity
\begin{equation}
j > \frac{\pi}{4} \lambda_j - 2 \sqrt{\lambda_j} + 2,
\end{equation}
which, combined with Eq.~\eqref{eq:plProof2} (now in the form $j < 0.54323 \,\lambda_j$ for $\mathcal{A} = \pi^2$), can be effectively rearranged to $\lambda_j< 51$. Manually examining the spectral sequence (and deferring the missing analysis of $\lambda_5, \lambda_7$, and $\lambda_9$ in Pleijel's proof to \citet{Berard2015}) we find that $\lambda_j$ is Courant sharp only for $j = 1, 2$, and $4$. Pleijel's theorem was generalized to surfaces by \citet{peetre1957generalization} and therefore, the sphere too has only finitely many Courant-sharp eigenvalues. Additionally, the estimate \eqref{eq:pleijel} also holds for a piecewise real analytic domain with Neumann boundary conditions \cite{polterovich2009pleijel}, for which the only Courant-sharp eigenvalues are \cite{helffer2015nodal} $\lambda_j\mbox{ with } j \in \{1, 2, 4, 5, 9 \}$ (Fig.~\ref{fig:neumann}).

Further generalizations of Courant's nodal domain theorem include extensions to nonlinear eigenvalue problems for the $p$-Laplacian $\Delta_p$ \cite{drabek2002generalization, cuesta2000nodal}, selfadjoint second-order elliptic operators \cite{alessandrini1998courant}, and inequalities on spectral counting functions \cite{ancona2004nodal}. Specifically worth mentioning in this list is the important conjecture for the biharmonic eigenvalue problem: 
\begin{equation}
\begin{aligned}
\label{eq:biharmonic}
\Delta^2 u &= \lambda\, u & &\mbox{ in } \mathcal{D}\\
u &= \frac{\partial}{\partial {\mathrm{n}}} u = 0\quad & &\mbox{ on } \partial\, \mathcal{D}.
\end{aligned}
\end{equation}
\citet{szego1950membranes} hypothesized that if­ $\mathcal{D} \in \mathbb{R}^2$ is a ``nice'' domain (i.e., $\partial\,\mathcal{D}$ is an analytic curve), then $u_1$ for Eq.~\eqref{eq:biharmonic} does not change its sign \cite{sweers2001first}. However, surprisingly, the conjecture is not even true for the first eigenfunction \cite{coffman1980structure, coffman1982structure, duffin1948question, garabedian1950partial, kozlov1990sign, charles1953generation, shapiro1994elementary}.

\subsubsection{Nodal line conjecture}

Another richly-debated conjecture that sparked intense discussion in the field was the proposition by \citet{payne} that the second Dirichlet eigenfunction $u_2$ does not have a closed interior nodal line in a bounded planar domain $\mathcal{D} \subset \mathbb{R}^2$ of \textsl{any} arbitrary shape. This longstanding question was also concurrently posed by \citet{yau1982survey}. \citet{payne1973two} himself verified the conjecture under the added condition that $\mathcal{D}$ is symmetric with respect to a line and convex with respect to the direction vertical to it. Thereafter, \citet{lin1987second} ratified it for a smooth, convex domain invariant under rotations by $2 \pi p/q$; $p, q \in \mathbb{Z}^+$; a few years later, \citet{jerison1991first} did so for long, thin convex sets. Finally, \citet{melas1992nodal}, furnishing the first fully-general result, certified that ``if $\mathcal{D} \subset \mathbb{R}^2$ is a bounded, convex domain with $C^\infty$ boundary, then the nodal line $\mathscr{N}$ of any second eigenfunction $u_2$ must intersect the boundary $\partial\,\mathcal{D}$ at exactly two points''---the analogous statement for simply-connected concave domains was proved by \citet{yang2013nodal}. Since Courant's theorem ensures that $u_2$ can have at most two nodal domains, the (only) nodal line must connect these intersection points.\footnote{In fact, when the eccentricity of $\mathcal{D}$ is sufficiently large, $\mathscr{N}$ is close to a straight line in the sense that the width of the nodal line $< \mathcal{C} / r_\mathcal{D}$, where $\mathcal{C}$ is a constant and $r_\mathcal{D}$ is the inradius of $\mathcal{D}$ (the radius of the largest ball that can be inscribed in $\mathcal{D}$) \cite{jerison1995diameter, grieser1996asymptotics}.} The impossibility of a closed nodal curve thus follows. This argument was extended upon by \citet{alessandrini1994nodal} by removing the requirement of $C^\infty$ smoothness on the boundary. The higher-dimensional generalization of this conjecture is due to a theorem by \citet{liboff1994nodal}: the nodal surface of the first excited state of a three-dimensional convex domain intersects its boundary in a single simple closed curve \cite{Grebenkov2013}.

Noteworthily, Payne's surmise does not pass muster for nonconvex domains; the nodal line of the second eigenfunction of the Laplacian can be closed in $\mathbb{R}^d$ \cite{fournais2001nodal} and, in particular, in $\mathbb{R}^2$ with Dirichlet \cite{hoffmann1997nodal} and Robin \cite{kennedy2011nodal} boundary conditions. It has also been explicitly disproved for the eigenvalue problem where the Schr{\" o}dinger operator has a potential $V \ne 0$ in addition to the Laplacian $\Delta$ \cite{lin1988counterexample}.

\subsubsection{Geometry of nodal sets}

Before concluding this Section, we quickly expound on two other results of a purely geometric nature on the size---or volume---of nodal domains. To characterize the asymptotic geometry of the domains, we quantify the size by the inradius. Let $r_\lambda$ be the inradius of a nodal domain for the eigenfunction with eigenvalue $\lambda$. On a closed Riemannian manifold of dimension $d \ge 3$, we then have \cite{mangoubi2008inner}
\begin{equation}
\frac{\mathcal{C}_1}{\lambda^{d^2 - 15\, d/8 + 1/4}\, (\log \lambda)^{2 d - 4}} \le r_\lambda \le  \frac{\mathcal{C}_2}{\sqrt{\lambda}}. 
\end{equation}
In two dimensions also, there is a sharp bound \cite{donnelly1988nodal, donnelly1990nodal, 0036-0279-43-4-L23, Zelditch_2013}:
\begin{equation}
\frac{\mathcal{C}_3}{\sqrt{\lambda}}  \le r_\lambda \le  \frac{\mathcal{C}_4}{\sqrt{\lambda}}. 
\end{equation}
After some jugglery and gentle coaxing, this inequality offers a crude but utilitarian estimate for the number of nodal domains:
\begin{equation}
\label{eq:juggle}
\frac{\mathcal{A}\,\lambda_j}{\pi\,\mathcal{C}_4^2}  \le \nu_j \approx \frac{\mathcal{A}}{\pi\,r_j^2} \le  \frac{\mathcal{A}\,\lambda_j}{\pi\,\mathcal{C}_3^2}. 
\end{equation}
These relations reappear in different guises when we talk about arithmetic random waves and difference equations later in this review. In the meantime, the more mathematically-oriented reader can find a delightful survey of results on the geometric properties of eigenfunctions in \cite{jakobson2001geometric}.


\section{Nodal sets of chaotic billiards}
\label{sec:chaotic}

\subsection{The random wave model---An introduction}

\begin{figure}[htb]
  \includegraphics[width=\linewidth]{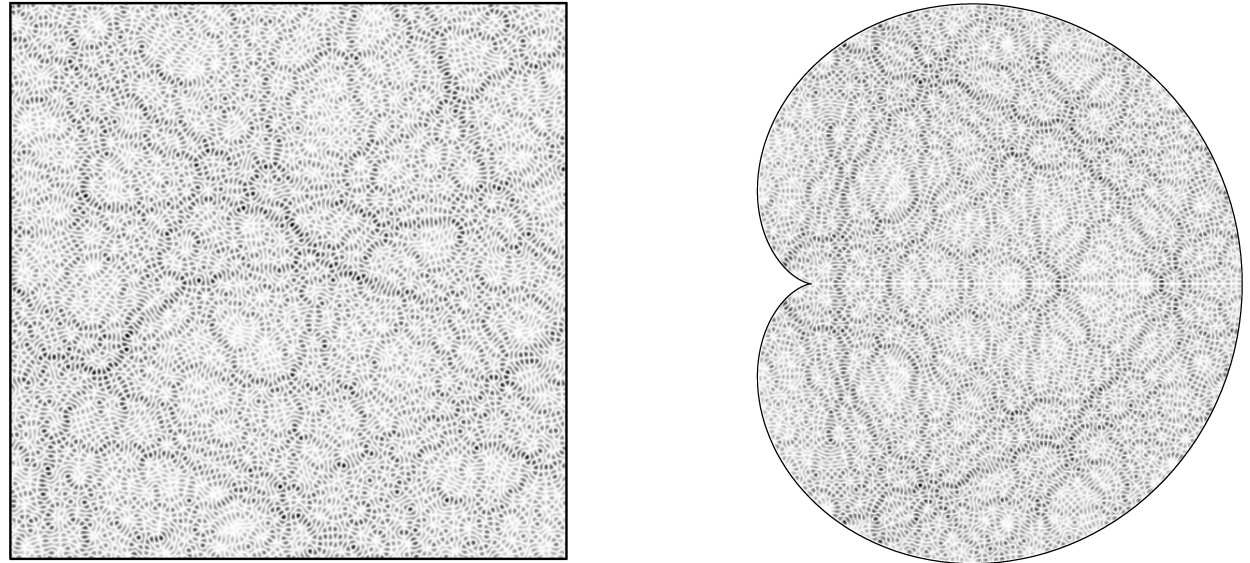}
  \caption{\label{fig:cardioid}Example of a random wave (Eq.~\ref{eq:RWM}) in comparison to the $6000^{\textrm{th}}$ eigenfunction of the chaotic cardioid billiard (of odd symmetry). Locally the appearance of the states is practically indistinguishable. \citet{o1988quantum} investigated the properties of the eigenfunctions constructed by such random superpositions and revealed the existence of structures resembling precursors of periodic orbit scar localization \cite{heller1984bound, kaplan1999measuring}. From \citet{arnd_thesis}.}
\end{figure}

The statistical properties of wavefunctions of classically chaotic systems are well-described by a surprisingly simple model, first conjectured by \citet{berry1977regular}. Known as the random wave model (RWM), it proposes that the eigenfunctions of strongly chaotic systems ``behave'' like a random superposition of plane waves of fixed wavevector magnitude. Even today, nearly half a century after Berry's initial proposition, there is no formal proof of this statement but rather, only heuristic justifications. The starting point for any such argument is the semiclassical eigenfunction hypothesis that we have seen previously in Sec.~\ref{sec:seh}. Then, the model can be motivated by the underlying (classical) chaotic (also called ``irregular'') dynamics of the quantum billiard where a typical trajectory gets arbitrarily close to every point in position space with apparently random directions and random phases (corresponding to the length of trajectory segments) \cite{backer2007random}. Since the whole energy surface is filled uniformly for ergodic systems, the probability density of finding the particle somewhere has, on average, a uniform distribution over the full billiard \cite{arnd_thesis}. In other words, as \citet{urbina2007random} note, ``in the semiclassical regime, the eigenfunctions should appear isotropic, structureless and roughly homogeneous owing to the lack of structure of the classical phase space.'' Hence, a random superposition of plane waves suffices for a reasonably good description of the system as Fig.~\ref{fig:cardioid} indeed demonstrates. 

The most intriguing aspect is the universality of this deceptively simple model, which finds applications in diverse fields ranging from optics \cite{ Berry2059} to wave mechanics in disordered media \cite{barth2002current, mirlin2000statistics}. In studies of mesoscopic systems as well, the RWM has proven to be immensely successful in describing conductance fluctuations in quantum dots \cite{beenakker1997random, alhassid2000statistical, baranger1994mesoscopic, jalabert1994universal}, especially in the Coulomb blockade (CB) regime \cite{beenakker1991theory, patel1998changing, aleiner2002quantum} where it relates the distribution of tunneling amplitudes to the statistics of CB peak heights \cite{jalabert1992statistical, alhassid1997signatures, ullmo2008many}. This robustness has prompted observers to regard the RWM as a litmus test for wave signatures of classically chaotic dynamics \cite{blum2002nodal}. Further applications and theoretical details of the RWM are discussed by by \citet{urbina2013random} in their state-of-the-art review.

\begin{figure*}[htb]
\includegraphics[width=0.6\textwidth, height = 5cm, keepaspectratio]{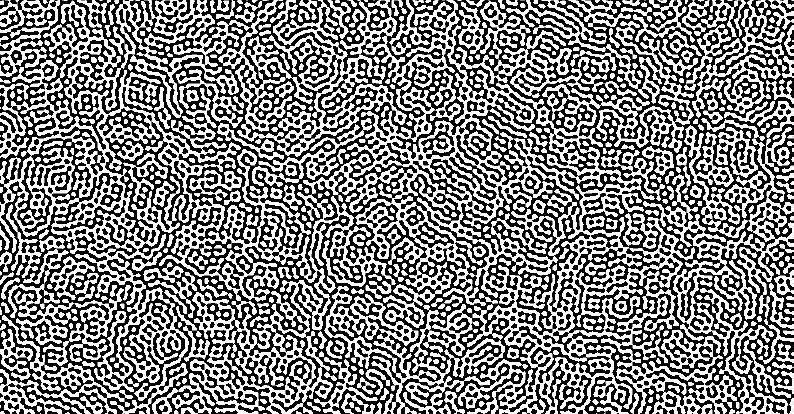}\quad \quad
  \includegraphics[width=0.3\textwidth,height = 5cm, keepaspectratio]{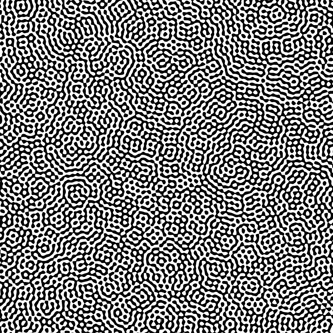}
  \caption{\label{fig:RWM}[Left]: Nodal domains of the eigenfunction of a quarter of the stadium billiard with area $4 \pi$ and energy $E = 10092.029$. [Right]: Nodal domains of a random wavefunction (Eq.~\ref{eq:Gaussian}) with $k = 100$. As before,, black (white) regions represent nodal domains where the function is positive (negative). The two figures look very similar. From \citet{bogomolnyRWM}. \textcopyright\, IOP Publishing. Reproduced with permission. All rights reserved.}
\end{figure*}

That the model works so well is no coincidence as reasoned succinctly by \citet{urbina2003supporting}. Their reasoning hinges on the two fundamental arguments at play here. First, the RWM wavefunction is, roughly speaking, a function taking random values at each point (formally called a \textsl{stationary} random process) \cite{goodman2015}; in addition, it is also Gaussian, which implies that it can be uniquely characterized by a two-point correlation function that encodes the appropriate symmetries. It is this generality of the random wave two-point correlation that accounts for the effectiveness of the theory when neglecting boundary effects, as for bulk properties. A cautionary remark, however: despite its spectacular triumphs, one would do well to remember that the random wave model is precisely just that---namely, a model and that too, of sufficiently excited states. For instance, it must be emphasized that  the ground-state eigenfunction for a quantum chaotic billiard, in general, would be unassumingly well-behaved, not resembling anything that looks like a random superposition of plane waves \cite{jain2002quantum,jain2009ground}. Nonetheless, for the object of our interest, namely, the nodal set, it works very well, numerically (see Fig.~\ref{fig:RWM}).

Now, to the mathematics. In two Euclidean dimensions, the random superposition of plane waves on a region $\mathcal{D} \subset \mathbb{R}^2$ may be written as
\begin{equation}
\label{eq:RWM}
\psi_\textsl{\textsc{n}} (\mathbf{r}) = \sqrt{\frac{2}{\mathrm{vol}\,(\mathcal{D})\, N}} \sum_{i=1}^{N} \, a_i\,\cos(\mathbf{k}_i.\mathbf{r} + \phi _i) 
\end{equation}
where $a_i \in \mathbb{R} $ are independent Gaussian random variables with zero mean and unit variance, $\phi_i$ are uniformly-distributed random variables on $[0, 2 \pi)$, and the momenta $\mathbf{k}_i \in \mathbb{R}^2$ are randomly equidistributed, lying on the circle of radius $\sqrt{E}$. Note that the factor of $1/\sqrt{N}$ up front takes care of the normalization in the limit $N \rightarrow \infty$. A property of random waves of this type is the universality of the spatial autocorrelation function
\begin{alignat}{1}
\label{eq:Berry}
\nonumber C (\mathbf{r}, \delta \mathbf{r}) &=  \langle \psi_{\textsc{rwm}} (\mathbf{r} - \delta \mathbf{r} /2)\, \psi_{\textsc{rwm}} (\mathbf{r} + \delta \mathbf{r} /2) \rangle\\
&= \frac{1}{\textrm{vol}\, (\mathcal{D})}\, J_0 (\sqrt{E} \, \lvert \delta \mathbf{r} \rvert).
\end{alignat}
One should be wary that Eq.~\eqref{eq:Berry} is not a consequence of the randomness but rather, of the dispersion relation $\lvert\, \mathbf{k} \, \rvert = \sqrt{E}$, and more directly, of the quantum ergodicity theorem. Although semiclassical techniques cannot be directly used to calculate higher-order correlations, Berry went on to conjecture that \textsl{all} the statistical properties of the spatial fluctuations of the eigenfunctions of a chaotic systems are described by a superposition of waves with fixed wavenumber and random phases. This description is reminiscent of the ergodicity hypothesis in statistical physics in its assertion that the spatial average in Eq.~\eqref{eq:Berry} is essentially equivalent to averaging over the random phases of Eq.~\eqref{eq:RWM}. Such universal spatial fluctuations have also been identified in systems with nontrivial spin dynamics such a confined two-dimensional electron gas in the presence of spin-orbit interaction \cite{urbina2013universal}. 

\begin{figure}[htb]
\includegraphics[width=\linewidth]{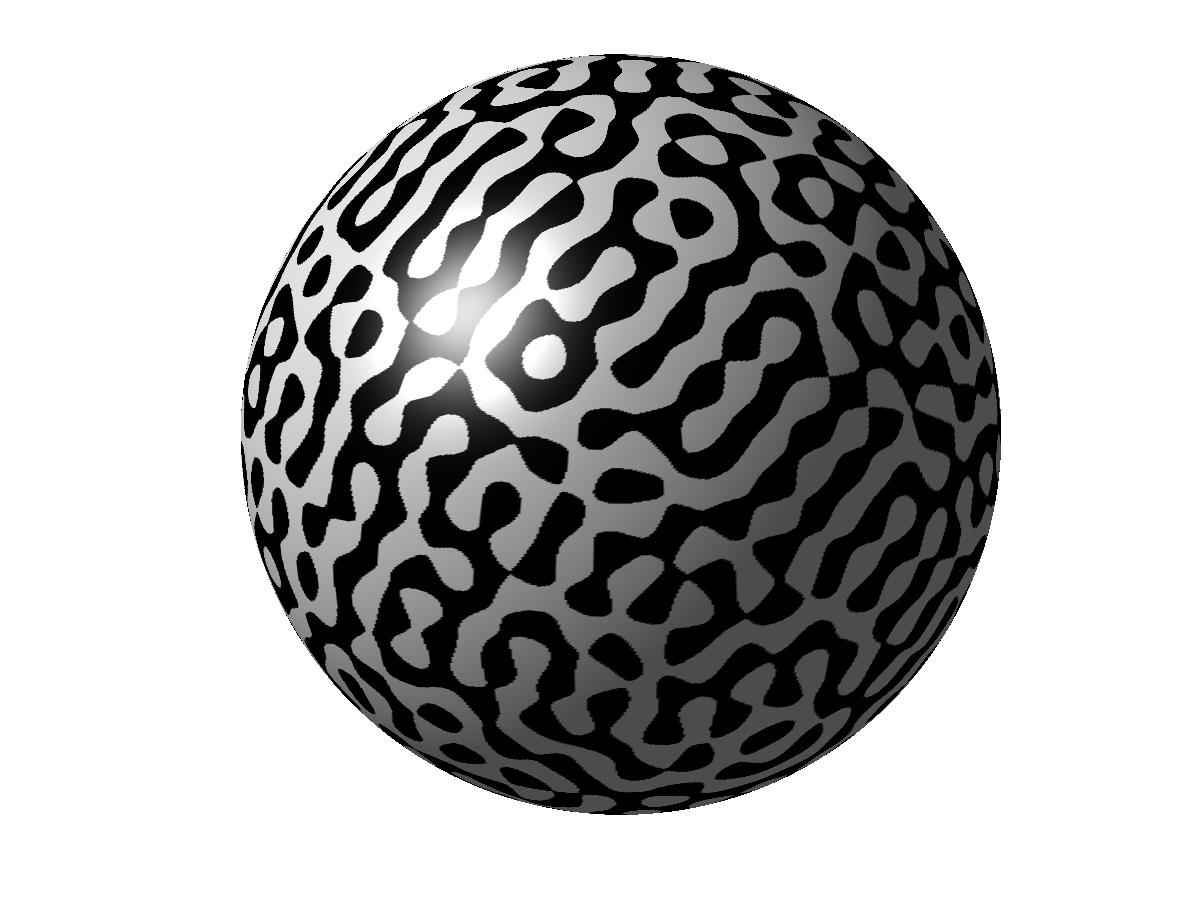}
\caption{\label{fig:GaussianHarmonic}Nodal portrait of a Gaussian spherical harmonic of degree 40. Figure prepared by Alex Barnett.}
\end{figure}

More rigorously, \textsl{any} wavefunction of a two-dimensional billiard obeying the Helmholtz equation with energy $E = \mathbf{k}^2$ can be written as the superposition (upto normalization)
\begin{equation}
\label{eq:Gaussian}
\psi\, (r, \theta) = \sum_{n\, \in\, \mathbb{Z}} C_n\, J_{\lvert n \rvert} (r, \theta)\, \mathrm{e}^{\mathrm{i}\,n \, \theta},
\end{equation}
where $J_n$ are Bessel functions and the coefficients satisfy $C_n = C^{*}_{-n}$ if the wavefunction is real. Berry's conjecture contends that in the semiclassical limit $k \rightarrow \infty$, for all statistical purposes, the coefficients $C_n$ could be taken as independent Gaussian random variables with $\langle C_n \rangle = 0$ and $\langle C_n\, C_m^{*} \rangle = \sigma^2\, \delta_{m,n}$. However, perhaps the most general way to think about a random plane wave would be to regard it as the 2D Fourier transform of white noise\footnote{We use the term in the generic sense of a random signal that has equal intensity at all frequencies and thus, a flat power spectrum.} on the unit circle. In this context, let $L_{\textrm{sym}}^2 (\mathbb{T})$ denote the Hilbert space of square-integrable ($L^2$) functions on the unit circle with the symmetry $f (-z) = f^{*} (z)$. The Fourier image of the space $L_{\textrm{sym}}^2$ is simply the space of real analytic functions, $\mathcal{H}$, satisfying the Helmholtz equation. A random plane wave is then $F = \sum_n C_n \Phi_n$ where the $C_n$ are independent Gaussians and $\{ \Phi_n \}$ is any orthonormal basis in $\mathcal{H}$; this has a covariance function $\mathrm{Cov}\, \{F(x), F(y)\} = J_0 (\lvert x - y \rvert)$. It may seem a bit pompous to replace the simple sums over exponentials and Bessel functions with a Gaussian field in its full glory but this is, in many ways, the more natural representation to use since the first two definitions can always be extracted from it. The raison d'\^{e}tre for this excursion into overtly mathematical terrain is that it enables us to link our discussion to the Gaussian spherical harmonics (Fig.~\ref{fig:GaussianHarmonic}) and tap into the treasure trove of already well-known results \cite{wigman2009} on nodal sets thereof. The connection is simple: the Gaussian plane wave is a large $n$ limit of the Gaussian spherical harmonic of degree $n$.

To be precise, consider the eigenfunctions of the Laplacian $\Delta$ on the $m$-dimensional unit sphere $\mathbb{S}^m$. Given an eigenvalue $E_n \, (n= 1, 2, \ldots)$, the corresponding eigenspace is the space $\mathcal{H}_n$ of the spherical harmonics of degree $n$, which form a $2n + 1$ dimensional space of eigenfunctions of the Laplacian on $\mathbb{S}^m$; let $g_n \in \mathcal{H}_n$. Using integral formulae due to Poincar{\'e} and Kac-Rice (see, for example, \citet{cramer2013stationary}), it is possible to obtain some universal estimates such as the following:
\begin{itemize}
\item For each $g_n$, every nodal domain of $g_n$ contains a disk of radius $c\,n^{-1}$, where $c$ is an absolute constant; this tells us about the minimum size of a domain.

\item Let $\mathcal{Z}(g_n)$ be the `volume' of the nodal set (in two dimensions, the total length of the nodal lines). \citet{yau1982survey, yau1993differential} conjectured that for any smooth metric on a manifold $\mathcal{M}$, there exist constants $c\, (\mathcal{M})$, $\mathcal{C} (\mathcal{M})$ such that for every $g_n$, $c\,\sqrt{E_n} < \mathcal{Z}(g_n) <  \mathcal{C}\,\sqrt{E_n}$; the upper bound was also proposed by \citet{cheng1976eigenfunctions}. The lower bound has been proved for the planar case \cite{bruning1972lange, bruning1978knoten} and for smooth metrics \cite{logunov2016nodal1, logunov2016nodal2}. We refer to \cite{savo2000eigenvalue} and \cite{logunov2016nodal3} for estimates of the constants $c$ and $\mathcal{C}$, respectively. Although \citet{donnelly1988nodal} proved Yau's conjecture for real-analytic metrics, the problem, in its full generality, still remains open.

\item For Gaussian spherical harmonics $g_n$, the expectation value $\mathbb{E}\, \left[\mathcal{Z}(g_n)\right] = \pi \sqrt{2 E_n} = \sqrt{2} \pi\, n + \mathcal{O}(1)$ \cite{berardvolume, zelditch2009real}. Subsequently, it was proved \cite{wigman2010fluctuations, wigman2012erratum} that 
\begin{equation}
\label{eq:GSHVar}
\mathrm{Var}\, \mathcal{Z}(g_n) = \frac{1}{32} \log n + \mathcal{O}(1),
\end{equation} 
which is much smaller than the prior $\mathcal{O}\,(n)$ estimate \cite{wigman2009} on account of Berry's cancellation phenomenon \cite{berry2002statistics}.

\item By Courant's nodal domain theorem, and Pleijel's refinement, $\nu\,(g_n) < 0.69\, n^2$. Since there are spherical harmonics with one or two nodal domains, there is no nontrivial deterministic lower bound \cite{lewy1977mininum}. A sharp upper bound for the number of nodal domains of spherical harmonics, for the first six eigenvalues, is given by \cite{leydold1996number}. On the other side of the coin, it is still an open problem whether---for a general surface (or higher-dimensional Riemannian manifold)---there exists a sequence of eigenfunctions for which the number of nodal domains tends to infinity with the eigenvalue. This question was answered in the affirmative for nodal domains of Maass forms \cite{ghosh2013nodal, ghosh2015nodal} and non-positively curved surfaces with concave boundaries, i.e., generalized Sinai or Lorentz billiards \cite{jung2016number}.
\end{itemize}
Specializing to two dimensions, $m=2$, we can exploit all of the above results on $\mathbb{S}^2$ for billiards.

After all this song and dance, it becomes imperative to address the elephant in the room: what are the quantitative predictions (if any) of the random wave model? One (and the simplest) of the many answers to this query concerns the amplitude distribution of the wavefunction. Using the central limit theorem one immediately obtains that random waves exhibit a Gaussian distribution of eigenfunction amplitudes
\begin{equation}
P (\psi) = \frac{1}{\sqrt{2\pi}\,\sigma} \exp \left(- \frac{\psi^2}{2 \,\sigma^2} \right),
\end{equation}
where $\sigma^2 = 1 / \textrm{vol} (\Omega)$. This prediction was borne out by several numerical studies \cite{mcdonald1988wave, o1988quantum, aurich1993statistical, li1994statistical}.

Buoyed by this preliminary success, we now direct our attention to explicitly constructing random wave models for billiards with the intent to eventually study their nodal structure. The first hurdle that we run into is that the theory we have looked at so far is isotropic. For a billiard, with well-defined boundaries, we need a nonisotropic RWM constructed from a random superposition of waves satisfying both the Schr\"{o}dinger equation and the boundary conditions---this problem, regrettably enough, turns out to be at least as difficult as solving the full quantum mechanical problem employing standard techniques \cite{urbina2003supporting}. The nontrivial deviations from the isotropic case owing to finite-size effects \cite{ullmo2001interactions} emphasize the relevance of extending this approach to include arbitrary boundaries.

In order to incorporate boundaries into the model, our desideratum is an ensemble of random functions, constructed so as to respect the boundary conditions of the billiard. This patently calls for a departure from the spatial averaging prescribed in the original RWM, such as in Eq.~\eqref{eq:Berry}, as it would destroy any information about the boundary. The way out of this conundrum is to substitute the spatial average with a spectral one. Instead of dealing with a single eigenfunction, we now evaluate the average for fixed sets of positions (without any spatial integration) over a set of normalized solutions $\psi_j (\mathbf{r})$ of the Schr\"{o}dinger equation with nondegenerate eigenvalues $E_j$ lying in the interval $w = [e - \delta \,e/2, e + \delta\, e/2]$. The principal idea is that given a functional $F[\psi] \equiv F(\psi(\mathbf{r}_1), \ldots , \psi(\mathbf{r}_N))$, we can always define the spectral average of the functional $F$ around energy $e$ as \cite{urbina2006statistical}
\begin{equation}
\mathcal{F} = \frac{1}{\rho_w (e)} \,\sum_j w(e-E_j)\, F[\psi_n], 
\end{equation}
where $w(x)$ represents a normalized window function around $x = 0$ and $\rho_w(e)$ is the density of states smoothed over the window $w$. The two-point correlation function associated with this spectral average is
\begin{equation}
R_w (\mathbf{r}_1, \mathbf{r}_2; e) =  \frac{1}{\rho_w (e)} \,\sum_j w(e-E_j)\, \psi_n (\mathbf{r}_1)\, \psi_n (\mathbf{r}_2), 
\end{equation}
which exactly satisfies the boundary conditions. This correlation function can actually be calculated using semiclassical expansions of the propagator and represented in a multiple reflection expansion \cite{hortikar1998,urbina2004semiclassical} of the form:
\begin{alignat}{1}
R_w (\mathbf{r}_i, \mathbf{r}_j, e) &=  \frac{1}{\mathcal{A}\, (\mathcal{D})} J_0\,(k(e)\, \lvert \mathbf{r}_i - \mathbf{r}_j \rvert) \\
\nonumber &+ \mbox{sum over reflections at the boundary}.
\end{alignat}
Thus, the prescription most amenable to generalization is to work directly with the two-point correlation function that assimilates all nonuniversal effects stemming from boundary constraints.

To make the above notions precise, let us demonstrate this approach in the context of computing the following  one-point averages that find use in the nodal counting statistics:
\begin{alignat*}{3}
B\, (\mathbf{r}) & \equiv \left \langle \psi\, (\mathbf{r})^2 \right \rangle, \quad &&K_y\, (\mathbf{r}) &&\equiv \left \langle \psi\, (\mathbf{r})\, \frac{\partial\, \psi\, (\mathbf{r}) }{\partial\, y} \right\rangle,\\
D_x\, (\mathbf{r}) & \equiv \left \langle \left ( \frac{\partial\, \psi\, (\mathbf{r}) }{\partial\, x} \right)^2 \right \rangle, \quad &&D_y\, (\mathbf{r}) && \equiv \left \langle \left ( \frac{\partial\, \psi\, (\mathbf{r}) }{\partial\, y} \right)^2 \right \rangle,
\end{alignat*}
written in the notation of \cite{berry2002statistics}. For billiard systems, the isotropic RWM is defined by the ensemble
\begin{equation}
\label{eq:isotropic}
\psi^i \,(\mathbf{r}) = {\displaystyle \sqrt{\frac{2}{J}}}\sum_{j=1}^N \cos \left(k\, x\, \cos \theta_j + k\, y\, \sin \theta_j + \phi_j \right)
\end{equation}
with $\theta_j = 2 \pi j/J$ and the average $\langle \cdots \rangle$ being determined by integration over a set of independent random phases $\phi_j \in (0, 2 \pi]$ (formally, the limit $J\rightarrow\infty$ is taken \textsl{after} averaging). Explicit calculation yields \cite{Berry2059}
\begin{equation}
\label{eq:corrFn}
B^{i}\,(\mathbf{r}) = 1, \,\, K^i_y \,(\mathbf{r}) = 0, \,\, D^i_x\,(\mathbf{r}) = D^i_y\,(\mathbf{r}) = \frac{k^2}{2} .
\end{equation}
Since boundary effects are neglected in this calculation, these results only represent bulk approximations to the system. In a first attempt to overcome this limitation, \citet{berry2002statistics} introduced the following ensemble of nonisotropic superpositions of random waves for an idealized system with an infinite straight wall at $y = y_0$, on which the wavefunction satisfies Dirichlet (D) or Neumann (N) boundary conditions:
\begin{equation}
\label{eq:heller}
\psi^{(\mathrm{D,N})} = {\displaystyle \sqrt{\frac{4}{J}}}\sum_{j=1}^N (\sin, \cos)(k (y-y_0))\cos \left(k\, x\, \cos \theta_j + \phi_j \right).
\end{equation}
The nonisotropic analogues of Eq.~\eqref{eq:corrFn} can again be worked out in this ensemble. For a more generic situation, where the confining potential is smooth, one can construct an ensemble of random Airy functions $Ai(\mathbf{r})$ to locally satisfy the Schr\"{o}dinger equation, as demonstrated by \citet{bies2002nodal} for a linear ramp potential $V (x, y) = Vy$. Another (and, to the best of our knowledge, the only other) boundary to which a nonisotropic RWM has been adapted is the edge between two infinite lines angled at a rational multiple of $\pi$ \cite{bies2003}.

The modus operandi for actually calculating the above-defined averages is to evaluate the two-point correlation function defining the nonisotropic and finite-size RWM
\begin{alignat}{1}
\nonumber &R \,(\mathbf{r}_1, \mathbf{r}_2) \equiv \langle \psi(\mathbf{r}_1)\, \psi^{*} (\mathbf{r}_2) \rangle = \frac{1}{N} \sum_{E_j \in w} \psi_j(\mathbf{r}_1)\, \psi_j^{*} (\mathbf{r}_2)
\end{alignat}
and then take derivatives. The correlation function is more conveniently expressed in terms of the Green's function of the system
\begin{alignat}{1}
&G (\mathbf{r}_1, \mathbf{r}_2, E+\mathrm{i} 0^{+}) = \sum_{j=1}^{\infty} \frac{\psi_j(\mathbf{r}_1)\, \psi_j^{*} (\mathbf{r}_2)}{E - E_j +\mathrm{i}\, 0^{+}}, \mbox{ as }\\
\nonumber &\displaystyle F = \frac{\Delta (e)}{2 \pi \mathrm{i}\, \delta e} \int_w \left(G^* (\mathbf{r}_1, \mathbf{r}_2, E+\mathrm{i} 0^{+})- G (\mathbf{r}_2, \mathbf{r}_1, E+\mathrm{i} 0^{+})\right) \mathrm{d}E,
\end{alignat}
\noindent upon converting the sum to an integral by introducing the (approximately constant) mean level spacing $\Delta (e)$. This form of the correlation function---in analogy with the partition function in statistical mechanics---can be used to compute the relevant averages by differentiation. For instance,
\begin{alignat}{1}
D_x (\mathbf{r}) &= \left[\frac{\partial^2 F(\mathbf{r}_1, \mathbf{r}_2)}{\partial x_1 \,\partial x_2 } \right]_{\mathbf{r}_1 = \mathbf{r}_2 = \mathbf{r}}, \\
K_y (\mathbf{r}) &= \left[\frac{1}{2} \left(\frac{\partial}{\partial y_1} + \frac{\partial}{\partial y_2} \right) F(\mathbf{r}_1, \mathbf{r}_2) \right]_{\mathbf{r}_1 = \mathbf{r}_2 = \mathbf{r}}.
\end{alignat}
For billiards, the bulk results are obtained by replacing the exact Green's function with the free propagator given by the Hankel function \cite{heller2007statistical}
\begin{equation}
\label{eq:Green}
G^0 (\mathbf{r}_2, \mathbf{r}_1, E+\mathrm{i} 0^{+}) = \frac{\mathrm{i}}{4 \pi} H_0^{(1)} \left(\frac{\sqrt{E}}{\hbar} \lvert \mathbf{r}_1 - \mathbf{r}_2 \rvert  \right).
\end{equation}
The (bulk) contribution to the two-point correlation is 
\begin{equation}
\label{eq:FiniteRWM}
F^b (\mathbf{r}_1, \mathbf{r}_2) = \frac{1}{\mathcal{A}\, \delta e} \int_{e - \frac{\delta e}{2}}^{e + \frac{\delta e}{2}} J_0 \left(\frac{\sqrt{E}}{\hbar} \lvert \mathbf{r}_1 - \mathbf{r}_2 \rvert  \right) \mathrm{d}E.
\end{equation}
This not only agrees with the previous calculation of Eq.~\eqref{eq:corrFn} upon differentiation but also reduces to Berry's result, Eq.~\eqref{eq:Berry} (identifying $\mathrm{vol}\,(\mathcal{D}) = \mathcal{A}$), when $\lvert \mathbf{r}_1 - \mathbf{r}_2 \rvert \ll \sqrt{4 \mathcal{A} /\pi}$. In the opposite limit, the correlation decays much faster, as long as $\delta e \geq \hbar \sqrt{\pi e/4 \mathcal{A}}$. Recognizing $\sqrt{4 \mathcal{A} /\pi}$ as a length scale that sets the average linear size of the system, $L$, it is apparent that the RWM defined by Eq.~\eqref{eq:FiniteRWM} subsumes finite size effects when scales of averaging are larger than the ballistic Thouless energy, i.e., $\delta\, e \ge e_{\textrm{Th}} = \hbar \sqrt{e} /L$ \cite{urbina2003supporting}. Similarly, one can find the two-point correlation in the case of an infinite (smooth or straight) barrier and then take the appropriate limits (either short distances or infinite system size) to recover the known RWM results. The salient attribute worth noting in this elegant calculation is that despite Berry's original conjecture being an inherently statistical statement, the two-point correlation function defining the RWM, Eq.~\eqref{eq:FiniteRWM}, is derived from purely quantum mechanical expressions without reference to any statistical assumptions about the wavefunction and is thus independent of the character of the classical system. An instructive summary of the different approaches we have discussed here can be espied in the example of the wedge billiard, reviewed by \citet{heller2007statistical}.

\subsubsection{The normalization problem}

Before moving on, we would like to stress a subtlety about normalization first presented by \citet{mirlin2000statistics} and explored by \citet{narimanov2001semiclassical}. The snake lurking in the grass is the assumption of a chaotic wavefunction being a Gaussian process, which apparently contradicts the normalization condition for the wavefunction \cite{gornyi2002wave}; more specifically, the existence of finite boundaries is incompatible with Gaussian statistics. This disagreement can be seen as follows. Consider the functional
\begin{equation}
\eta\, [\psi] = \int \lvert\, \psi (\mathbf{r})\,\rvert^2\, \mathrm{d}\,\mathbf{r},
\end{equation}
where $\psi (\mathbf{r})$ belongs to the ensemble chosen to describe the statistical properties of the wavefunction. The normalization of \textsl{all} $\psi (\mathbf{r})$ in the ensemble imposes the constraint that the ensemble variance $\mathrm{Var}\, (\eta) = \langle (\eta\, [\psi])^2 \rangle - (\langle \eta\, [\psi] \rangle )^2 = 0$ since $\eta\,[\psi] = 1$ over the set of normalized eigenfunctions. However, with a Gaussian distribution of wavefunction amplitudes, we find
\begin{equation}
\mathrm{Var}\, (\eta) = 2 \int \int \lvert \langle \psi(\mathbf{r}_1)\, \psi^{*} (\mathbf{r}_2) \rangle \rvert^2\, \mathrm{d}\,\mathbf{r}_1\, \mathrm{d}\,\mathbf{r}_2 \ne 0.
\end{equation}
Physically, the ensuing implication is that the Gaussian distribution, owing to fluctuations of the normalization integral, would produce spurious contributions to any statistics beyond the two-point correlation function, or, for that matter, to the spectral average of any functional of order higher than two. This ostensible discrepancy was partially resolved by \citet{urbina2003supporting}, who explicitly proved that $\mathrm{Var}\, (\eta) \sim \mathcal{O} (1/N)$ and hence, goes to zero as $N \rightarrow \infty$, thereby avoiding any conflict. Nonetheless, in the presence of boundaries, the Gaussian conjecture for the fluctuations of irregular eigenfunctions must be modified---it turns out that the Principle of Maximum Entropy \cite{grandy1987} selects a particular kind of distribution, known as the ``Gaussian Projected Ensemble'' \cite{goldstein2006distribution}. This leads to yet another alternate formulation of Berry's conjecture: to quote \citet{urbina2007random}, ``in systems with classically chaotic dynamics, spectral averages of functionals defined over the set of eigenfunctions are given by the corresponding average over the Gaussian Projected Ensemble with fixed system-dependent covariance matrix.''

\subsubsection{Systems with mixed phase space}

\begin{figure}[htb]
\includegraphics[width=\linewidth]{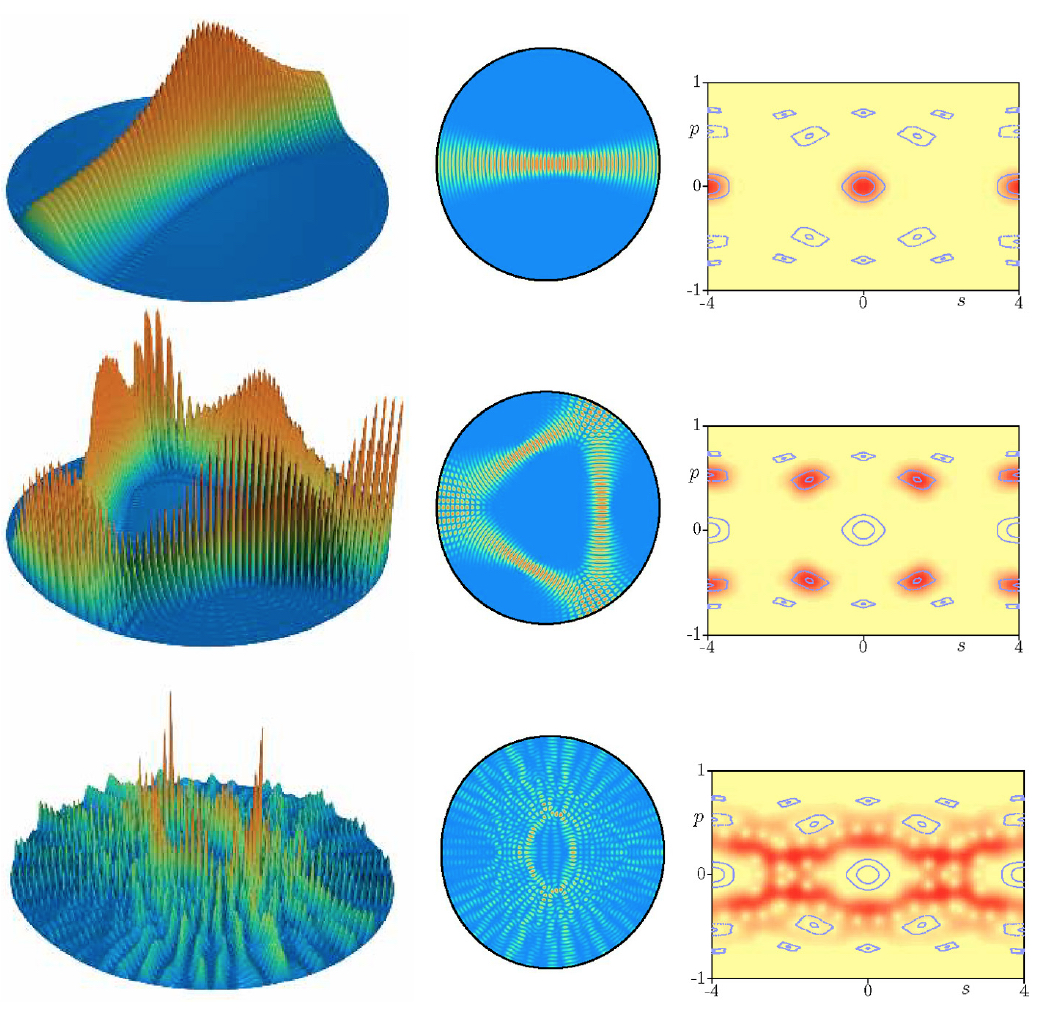}
\caption{\label{fig:Mixed}Eigenstates of the lima{\c c}on billiard, formed by deforming the circular billiard \cite{robnik1983classical}, with its boundary defined in polar coordinates by $r (\phi) = 1 + \varepsilon \, \cos \phi$. This is an example of a system with a mixed phase space for $\varepsilon > 0$. For $\varepsilon = 0.3$ as pictured, the eigenstates either concentrate in the regular islands, or extend over the chaotic region. [Left to right]: $\lvert \psi_n (\mathbf{r})\rvert^2$, density plot, and quantum Poincar{\'e}-Husimi representation (embedding the wavefunctions into the classical phase space). From \citet{backer2007random}. With kind permission of The European Physical Journal (EPJ).}
\end{figure}

A generic system, in contrast to our discussion heretofore, is neither fully chaotic nor regular; instead typically, systems would have a \textsl{mixed} phase space characterized by the coexistence of both regular and chaotic motions. \citet{percival1973regular} conjectured that for such mixed systems, like a mushroom billiard \cite{bunimovich2001mushrooms, gomes2015percival}, a full density subsequence of a complete set of eigenfunctions divides into two disjoint subsets, one corresponding to the ergodic and completely integrable regions of phase space each. The transition of the system from integrable to mixed dynamics brought about by small perturbations is described by KAM theory \cite{kolmogorov1954preservation, arnold1963proof, moser1962invariant}. Broadly speaking, the KAM theorem propounds that if the system is subjected to a weak nonlinear perturbation, some of the invariant tori (that satisfy the non-resonance condition of having ``sufficiently irrational'' frequencies \cite{casati1979stochastic}) are deformed but survive nonetheless, while others are destroyed and become invariant Cantor sets \cite{percival1979variational}. The fingerprints of this behavior can be discerned in a Poincar{\'e} section of the billiard flow, which exhibits a prominent irregular component---the chaotic sea---along with regular islands accompanying the stable periodic orbits. The same structure is manifest in the quantum states as well (see Fig.~\ref{fig:Mixed}).\footnote{To be fair, chaotic states can actually extend into the region of the regular islands when far away from the semiclassical limit \cite{backer2005flooding}.}

The organization of the eigenstates prompts one to believe that if we consider irregular eigenfunctions that are concentrated on a region $\mathscr{R}$ in phase space, the statistical properties should once again be described by a superposition of plane waves but with wave vectors of the same lengths and directions distributed uniformly on $\mathscr{R}$ alone this time (as opposed to the whole phase space). Indeed, this belief is not misplaced. Thus, one obtains the \textsl{restricted} random wave model \cite{arnd2002amplitude}:
\begin{equation*}
\psi_{\textsc{rrwm}, \mathscr{R}} (\mathbf{r}) = \sqrt{\frac{4 \,\pi}{\mathrm{vol}\,(\mathscr{R})\, N}} \sum_{i=1}^{N} \, \chi_\mathscr{R}(\hat{\mathbf{k}}_i,\mathbf{r})\,\cos(\mathbf{k}_i\,\mathbf{r} + \varepsilon_i), 
\end{equation*}
where the characteristic function $\chi_\mathscr{R} (\mathbf{k})$, which is one if $\mathbf{k}\, \in\, \mathscr{R}$ and zero otherwise, ensures the localization on $\mathscr{R}$. The amplitude distribution is locally Gaussian
\begin{equation}
P_{\mathbf{r}} (\psi) = \frac{1}{\sqrt{2\pi\,\sigma^2 (\mathbf{r})}} \exp \left(- \frac{\psi^2}{2 \,\sigma^2 (\mathbf{r})} \right),
\end{equation}
but with a position-dependent variance $\sigma^2 (\mathbf{r})$. Since $\sigma^2 (\mathbf{r}) = (\mathrm{volume})^{-1}$ for an ergodic system, this calculation is compatible with the previous expectation of a Gaussian distribution. On the other hand, if the variance bears some explicit position dependence, significant deviations from the Gaussian are possible. For more detailed discussions on nonisotropic random waves, we refer the interested reader to \cite{berry2002nodal,bies2003,urbina2003supporting,urbina2007random}.

\subsection{Random wavefunctions and percolation}

Building upon our analysis of the nodal structure of chaotic wavefunctions in the previous subsections, we pick up the thread that runs through this entire review, namely, the problem of enumerating the nodal domains. However, analytical results for the actual number of nodal domains, alas, are relatively scarce. The main difficulty in counting stems from non-locality: local observation of the nodal curves alone does not permit one to arrive at conclusions about the number of connected components. A major breakthrough in such enumerative pursuits was the introduction of the percolation model for nodal domains of chaotic wave functions by \citet{bogomolny2002percolation}. This model not only facilitated the analytical calculation of several physical quantities but also enabled one to exploit the link with percolation theory to gain a foothold on the structure of chaotic wavefunctions. Today, such borrowed ideas regularly find applications in quantum chaos and varied problems where nodal domains of random functions are of importance (see, for example, \citet{berk1987scattering}). Given the overarching reach of percolation in diverse contexts such as clustering, diffusion, fractals, phase transitions, and disordered systems, as well as its recurrent presence in the rest of this review, this would be a good point to quickly recapitulate the basic ideas behind the phenomenon. 

\subsubsection{Percolation in statistical physics}

\begin{figure}[htb]
\includegraphics[width=\linewidth]{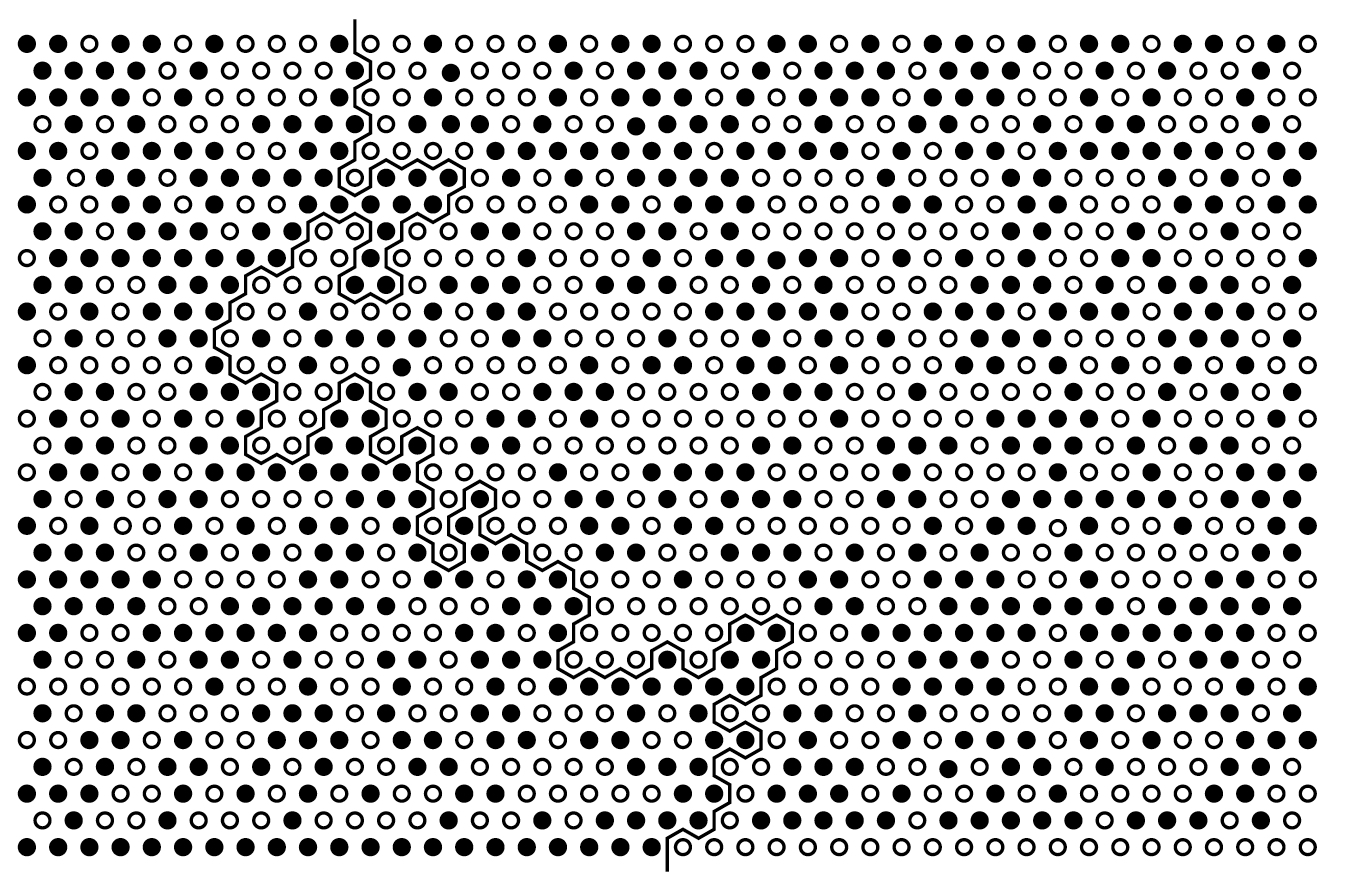}
\caption{\label{fig:Percolation}The percolation process on the triangular lattice at $p = 1/2$. Figure prepared by Geoffrey Grimmett. First published as Fig.~6 by \citet{lawler2009conformal}. \textcopyright\, 2008 American Mathematical Society (AMS); used by permission of the AMS. }
\end{figure}

A simple model for the percolation process can be described as follows. Consider a triangular lattice where every lattice point in the upper half-plane is colored white or black, independently, with probability $p$ and $1-p$, respectively. It is perhaps easiest to envisage a white vertex as being an ``open'' site, which permits the flow of a liquid through it; alternatively, one can think of the black (white) vertices as being the (un)occupied sites of a given lattice. Consequently, percolation may be reckoned as a model of the permeability of a material, regulated by the value of $p$. The question of interest concerns the existence of an infinite and connected collection of such open sites; in the language of our fluid flow analogy, this asks whether there exists a continuous ``pipe'' through the extent of the lattice. In statistical physics, we are chiefly concerned with critical phenomena, i.e., the study of systems at or near the point where a phase transition occurs. At a critical concentration $p_{\textsc{c}}$, one finds that an infinite cluster of white sites, embedded in the black background, extends across the lattice. For site percolation on a triangular lattice, this so-called percolation threshold is known to be $p_{\textsc{c}} = 1/2$ \cite{fogedby2012stochastic}. The point $p = 1/2$ is ``critical'' in the sense that for $p > 1/2$, there will always be an infinite connected cluster of white sites whereas this is certainly not the case for $p < 1/2$. A realization of this system at criticality is drawn in Fig.~\ref{fig:Percolation} with the sole difference of a boundary condition on the bottom row that segregates black and white sites on different sides. The rest of the upper half-plane is colored according as the chosen value of $p$. Once all the sites have been filled in, there exists a unique curve commencing at the bottom row such that it has only white vertices on its one side and all black vertices on the other. Thus, by imposing an appropriate boundary condition, we have induced a domain wall that meanders across the system.

\begin{figure}[htb]
\includegraphics[width=0.85\linewidth]{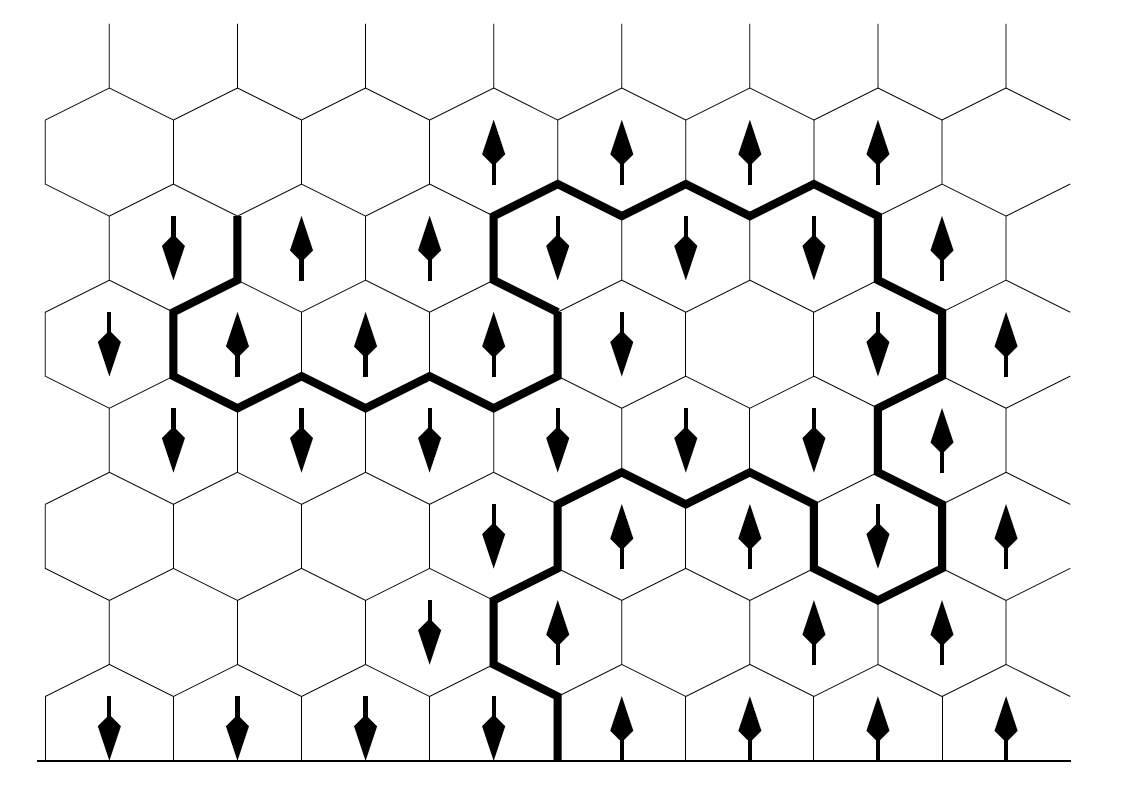}
\caption{\label{fig:Ising} The exploration process for the Ising model. At each step the walk turns left or right according as the value of the spin in front of it. The relative probabilities are determined by the expectation value of this spin given the fixed spins either side of the walk up to this time. The walk never crosses itself and never gets trapped. From \citet{cardy2005sle}, with permission from Elsevier.}
\end{figure}

Closely related to the problem of site percolation is the Ising model, which is, in all likelihood, the simplest interacting many-particle system in statistical physics \cite{binney1992theory, stanley1987introduction}. Each lattice site $i$ is occupied by an Ising ``spin'' $\sigma_i = \pm 1$ (or more generally, a single degree of freedom), which points either up or down. As a model of a ferromagnet, the spins interact via a short-range exchange interaction $J$ and are described by the nearest-neighbor coupled Hamiltonian
$H = - J\, \sum_{\langle i, j \rangle} \sigma_i\, \sigma_j$,
which favors parallel alignment of adjacent spins for $J>0$. The statistical weight assigned to each spin configuration is given by the conventional Boltzmann factor $\exp \,(-\beta \, H) $ and the partition function assumes the form
\begin{equation}
Z = \sum_{\{ \sigma_i\}} \, \exp(-H/ k_B\, T).
\end{equation}
In the limit of small $\beta \ll 1/J$ (or high ``temperature'') the system is disordered: the spin correlations are localized and decay exponentially fast, which means that spins separated by a large distance are almost independent of one another. On the other hand, when $\beta \gg 1/J$ at low temperatures, the system possesses long-range order. The Ising model thus exhibits a phase transition, at a critical temperature, from a disordered paramagnetic phase (at $T > T_c$) to a ferromagnetic phase (for $T < T_c$) with a non-zero order parameter $m \ne 0$ (the magnetization). Near the critical point, the order parameter, the correlation length, and the correlation function scale as
\begin{alignat}{2}
m &\sim \lvert T - T_c \rvert^{\,\beta};\hspace*{0.2cm} &&(T \rightarrow T_c^{\,-})\\
\xi &\sim \lvert T - T_c \rvert^{- \upsilon}; \hspace*{0.2cm} &&(T \rightarrow T_c^{\,+})\\
\langle \sigma_i\, \sigma_j \rangle &\sim \frac{1}{\lvert i - j \rvert^\eta}\,\exp\, &&\left(-\lvert i - j \rvert\, /\,\xi\right),
\end{alignat}
respectively, with critical exponents $\beta$, $\upsilon$ and $\eta$. It has long been known \cite{fisher1967} that the phase transition at $T = T_c$ is signaled by the divergence of the correlation length $\xi$ as the system becomes scale invariant. The intimate connection of the Ising model to our previous description (in Fig.~\ref{fig:Percolation}) is evinced by the duality between site percolation and domain wall formation, which can be seen as follows. Instead of selecting a configuration of all the spins $\sigma_i$ first and then identifying the domain wall, the curve can be generated by sequential steps in what is known as an exploration process, as demonstrated by Fig.~\ref{fig:Ising} for a honeycomb lattice. This exploration path \textsl{is} the interface for the statistical mechanics of percolation \cite{fogedby2012stochastic}. Another interesting parallel between the two descriptions is that the lattice dual to the honeycomb \textsl{is} the triangular lattice, whose sites are positioned at the centers of the hexagons of the former.

That is all we have to say about the percolation process for now. We will come back to it shortly in Sec.~\ref{SLE} when we take a closer look at conformal invariance and scaling limits in the context of Schramm-Loewner Evolution (SLE). At present, however, we return to the description of nodal lines afforded by the original model of \citet{bogomolny2002percolation}.

\subsubsection{The Bogomolny-Schmit percolation model}

\begin{figure}[htb]
\includegraphics[width=\linewidth]{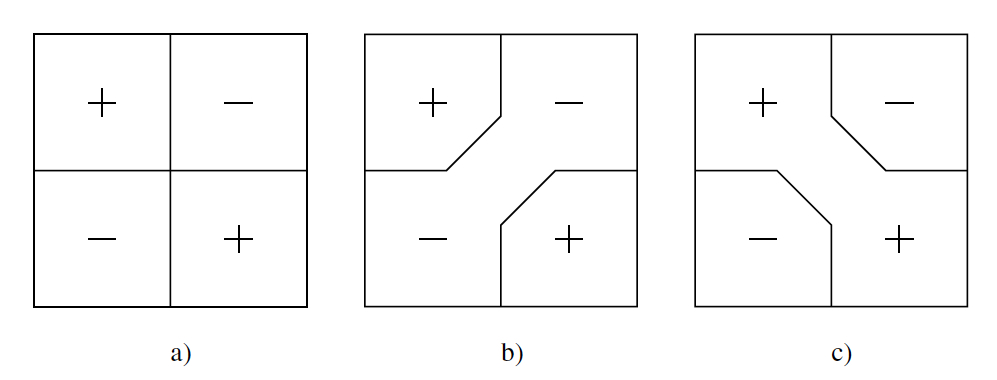}
\caption{\label{fig:Crossing}(a) True nodal crossing. (b) and (c) Avoided nodal crossings. Typically, nodal lines do not intersect as this would imply that $\psi$, $\psi_x$ and $\psi_y$ are simultaneously zero at some point. These three functions being independent Gaussians, the probability of this event is nearly zero. From \citet{bogomolny2002percolation}.}
\end{figure}

Consider the nodal portrait of a random function; we came across one in Fig.~\ref{fig:RWM}. The mean number of zeroes (or nodes) of a wavefunction of this type along a given straight line (without loss of generality, let's say the vertical one) can be estimated from the approximate quantization condition $\bar{k}_y\,L_y \approx \pi\,m$, where $m \in \mathbb{Z}$ and $\bar{k}_y= \mathbf{k}^2/2$ is the mean-square momentum along the $y$-axis. The mean density of nodal lines is roughly
\begin{equation}
\xoverline{\rho}_y = \frac{m}{L_y} = \frac{k}{\pi\, \sqrt{2}}.
\end{equation}
Hence, the nodal lines of random functions form a rectangular grid in the \textsl{mean} and the total number of sites in the resultant lattice is asymptotically
\begin{equation}
N_{\mathrm{tot}}\, =\, \left(\xoverline{\rho}_y\right)^2\,\mathcal{A}\, =\, k^2\,\frac{\mathcal{A}}{2\,\pi^2}\, =\, \frac{2}{\pi} \xoverline{N} \, (E),
\end{equation}
where $\xoverline{N} \, (E) = \mathcal{A} \, E / 4\,\pi$, in accordance with the Weyl formula, is the mean number of levels below energy $E$. In principle, this result can be derived rigorously using the methods of \citet{bogomolny1996quantum}. This back-of-the-envelope calculation, albeit appealing in its simplicity, comes with a potential pitfall. At any point inside the billiard, the actual wavefunction can be written as the sum of its average and a small correction as $\psi\, (x, y) = \xoverline{\psi}\, (x, y) + \delta \psi\, (x, y) $. Now, even if the average wavefunction forges a checkerboard nodal picture, the addition of the correction term recasts the crossing of nodal lines into one of the two possible avoided crossings in Fig.~\ref{fig:Crossing}. The sign of the critical point between two maxima or minima determines whether the positive or negative nodal components connect.

\begin{figure}[htb]
\begin{overpic}[width= \linewidth]{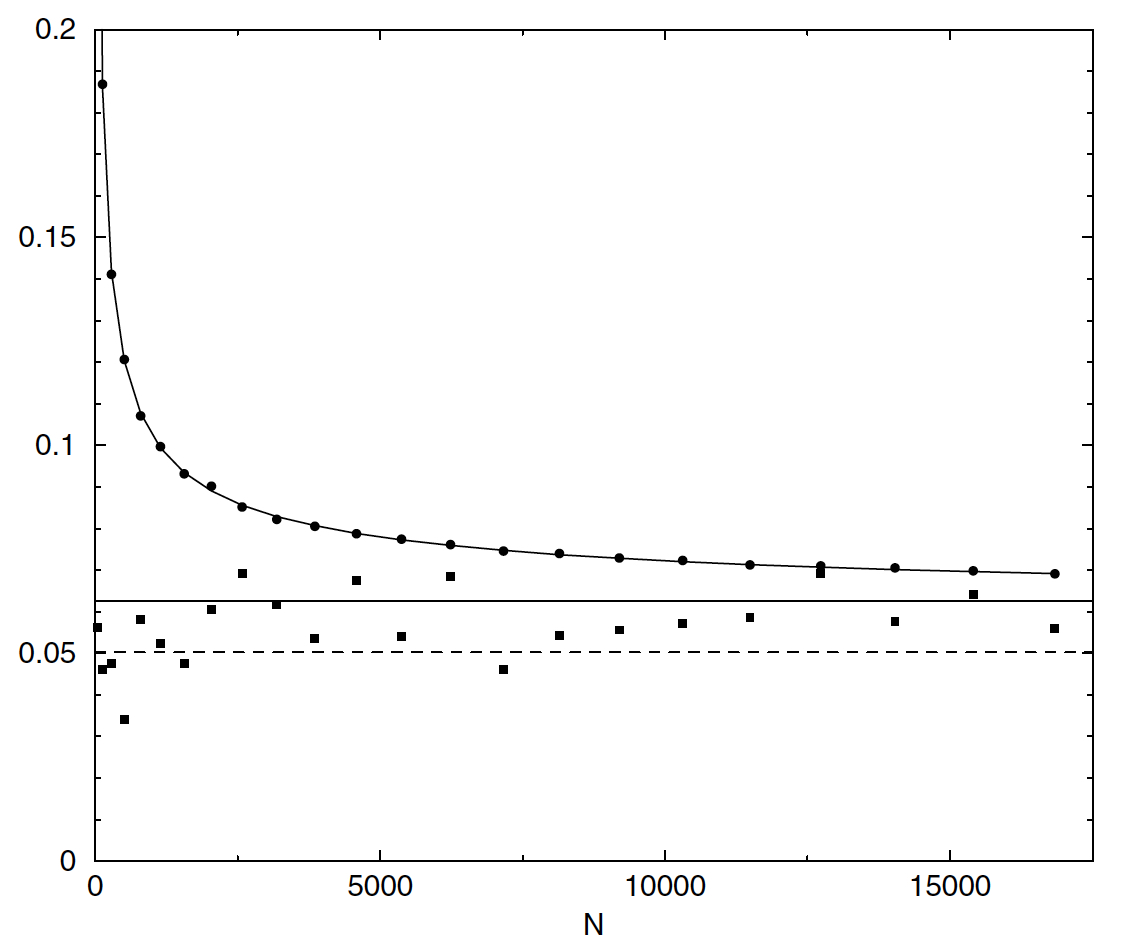}
     \put(35,39){\includegraphics[width= 0.6\linewidth]{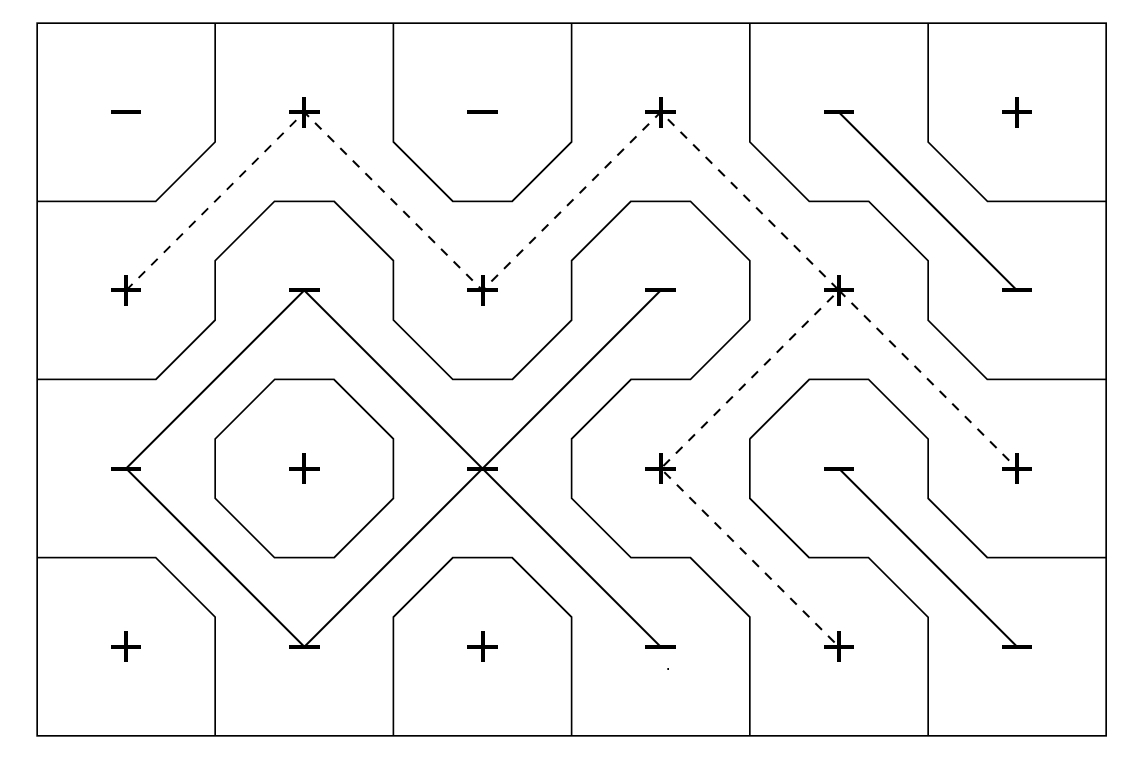}}
\end{overpic}
\caption{\label{fig:NDomains}Mean values of the nodal domains (dots) and their variances (squares) for random functions, normalized by and plotted against $N \equiv \xoverline{N}\,(E)$. The solid and dashed horizontal lines represent theoretical predictions of Eqs.~\eqref{eq:BS1} and \eqref{eq:BS2}, respectively. [Inset]: A realization of a random percolation-like process. The centers of the positive (negative) regions, denoted by $+$ ($-$), constitute two dual lattices. Dashed (solid) lines indicate graphs for the positive (negative) dual lattice. Adapted from \citet{bogomolny2002percolation}.}
\end{figure}

This observation lies at the crux of the Bogomolny-Schmit conjecture, which posits that the distribution of nodal domains for random functions is the same as that for a specific percolation-like process. Starting with a rectangular lattice where each of the $N_\mathrm{tot}$ sites represents a saddle point\footnote{Strictly speaking, this picture is not accurate as the function $\psi$ and its gradient $\nabla \,\psi$ cannot vanish simultaneously.} (with zero saddle height) akin to Fig.~\ref{fig:Crossing}~(a), each nodal crossing is amended to one of the permitted avoided intersections. As previously, the bond between two neighboring maxima is set to be ``open'' if the saddle height is positive and ``closed'' otherwise. The only stipulation governing this percolation-like process is that the saddle heights are uncorrelated and have equal probabilities of being positive or negative. Hence, the process, although random, is well-defined and corresponds to critical bond percolation. A particular instance is to be found in the inset of Fig.~\ref{fig:NDomains}. It is not difficult to discern that the original lattice can be decomposed into two dual lattices (of size $a = 2 \pi / k$, the de Broglie wavelength) with sites at the centers of the regions where the wavefunction is positive or negative. Two vertices on a dual lattice are connected by an edge if and only if the corresponding cells of the grid belong to the same nodal domain of the random function \cite{sodin2016lectures}. \textsl{Any} realization of the aforementioned random process therefore uniquely delineates two graphs on these lattices and conversely, there is a one-to-one correspondence between a graph on a dual lattice and an allowed realization. \citet{bogomolny2002percolation} cognized that the number of connected nodal domains equals the sum of the numbers of different components of both the positive and negative graphs, $n_\pm$, and bond percolation on the square lattice thus provides a good description of the nodal domains.

Utilizing the correspondence between the generating function
\begin{equation}
Z\,(x) = \sum_{\mathrm{realizations}}\,x^{n_{+} + n_{-}}
\end{equation}
and the partition sum of the Potts model \cite{wu1982potts}, it was shown that the total number of nodal domains for a random function has a universal Gaussian distribution with mean  $\bar{n}\,(E)$ and variance $\sigma^2\,(E)$ given by
\begin{alignat}{1}
\label{eq:BS1}\frac{\bar{n}\,(E)}{\xoverline{N}\,(E)} &= \frac{3\,\sqrt{3} - 5}{\pi} \approx 0.0624,\\
\label{eq:BS2}\frac{\sigma^2\,(E)}{\xoverline{N}\,(E)} &= \frac{18}{\pi^2} + \frac{4\,\sqrt{3}}{\pi}- \frac{25}{2\,\pi} \approx 0.0502.
\end{alignat}
Similar formulae were obtained by \citet{ziff1997universality} from Monte-Carlo simulations for the two-dimensional percolation problem. The asymptotic predictions of Eqs.~(\ref{eq:BS1},~\ref{eq:BS2}) are compared with numerical calculations of the mean value and the variance for several random functions in Fig.~\ref{fig:NDomains}, which demonstrates reasonable agreement, within statistical errors. Deviations from Eq.~\eqref{eq:BS1} larger than that sanctioned by Eq.~\eqref{eq:BS2} can be attributed to the existence of wavefunction scars. Regions of quasi-integrable behavior have considerably fewer nodal domains than chaotic regions, which accounts for the origin of large fluctuations \cite{bogomolnyRWM}.

Additionally, percolation theory prognosticates that the distribution of the areas $s$ of clusters (the connected nodal domains), $n (s)$, should follow a power-law behavior
\begin{equation}
\label{eq:fisher}
P_A (s) \propto \left(\frac{s}{s_{\min}} \right)^{-\tau}
\end{equation}
where $\tau = 187/91$ is the Fisher exponent  \cite{stauffer1994introduction}. The constant $s_{\min} = \pi\,\left(\mathscr{J}_{0,1}/k \right)^2$ is the smallest possible area for a fixed wavenumber $\mathbf{k}$ (Eq.~\ref{eq:fk}). The nodal domain perimeters $l$ should also exhibit a similar dependence $n_l \propto l^{-\tau'}$ with the scaling exponent $\tau' = 15/7$ \cite{ziff1986test}. Another result that was originally derived in the context of percolation and can be directly carried over is the fractal dimension of the nodal domains. Within the framework of the percolation model, this is given by the fractal dimension of critical percolation clusters, $D = 91/48$ \cite{stauffer1994introduction}. Numerically, the fractal dimension of a domain can be extracted by juxtaposing it onto a grid of squares (say, of side $R$) and counting the number of crossings of the region with the grid, which is expected to scale as
\begin{equation}
n_{\mathrm{crossing}} \propto R^{-D}
\end{equation}
whenever $a = 2\pi/k \ll R \ll l$ (the size of the domain).

Thus began the steady influx of ideas from percolation into the study of nodal domains, which soon proved to be a rather fruitful intellectual exercise. However, this celebrated conjecture has not been without controversy and has received its fair share of criticism and defense alike. At first sight, the manifest connection between critical percolation and the random plane wave model should itself come as a surprise \cite{foltin2004morphology}. In the standard percolation scheme, each site/edge of a lattice is set to be occupied/positive with probability $p$; this assignment has the important property that the occupation fraction (or concentrations) at different points are independent random variables. In contrast, for a Gaussian random function, the probability of its values having the same sign at two points, far from being uncorrelated, is
\begin{equation}
\label{eq:qccorr}
P\,(\lvert \mathbf{r} - \mathbf{r}' \rvert) = \frac{1}{2} + \frac{1}{\pi}\,\arcsin\,G(\lvert \mathbf{r} - \mathbf{r}' \rvert),
\end{equation}
which, assuming that the two-point correlation function is normalized as $G(0) = 1$, is nearly one for nearby pairs of points. It is only in the limit of large distances when $G(r) \rightarrow 0$ that the probabilities of finding a point positive and negative equalize. For a random superposition of plane waves of the form \eqref{eq:Gaussian}, it follows that\footnote{This correlator holds whenever the assumption that wavefunctions have quantum chaotic correlations does. For instance, applied to fluctuations of the transmission phase in interacting quantum dots, Eq.~\eqref{eq:qccorr} predicts large universal sequences of resonances and transmission zeros \cite{molina2012scattering}.}
\begin{equation}
\label{eq:CorrDecay}
\langle \psi\,(\mathbf{r})\,\psi\,(\mathbf{r}') \rangle = J_0 (k\, \lvert \mathbf{r} - \mathbf{r}' \rvert) \sim \frac{\cos\, (k\, \lvert \mathbf{r} - \mathbf{r}' \rvert - \pi/4)}{\sqrt{k\, \lvert \mathbf{r} - \mathbf{r}' \rvert}}
\end{equation}
and the random wavefunction correlation $G(r)$ decays rather slowly ($\sim r^{-1/2}$). In fact, even for the density of the random field,
\begin{equation}
\chi = -\sgn (\psi) \,\Theta \left(- \det \left(\partial_x\partial_y\psi \right) \right)\,\det \left(\partial_x\partial_y \psi \right)\,\delta^2(\nabla \psi),
\end{equation}
the correlations decay just as slowly:
\begin{alignat}{2}
\langle \chi\,(\mathbf{r})\,\chi\,(0) \rangle &= \frac{1}{72\,\pi^3}\,J_0 (r) &&+ \mathcal{O}\left(J_0^3(r) \right)\\ 
\nonumber &\sim \frac{1}{\sqrt{r}}\,\cos\, (2\pi r) &&+ \mathcal{O} \left(r^{-3/2}\right),
\end{alignat}
and consequently, are long-ranged \cite{foltin2003}. It is not unreasonable to expect such long-range correlations to undermine the validity of percolation theory.\footnote{Obversely, if one constructs a Gaussian ensemble of random functions characterized by the correlation function $G_0 (r) = \exp\,(- k^2\, r^2 /4)$, the applicability of the critical (short-range) percolation picture is almost self-evident \cite{zallen1971percolation, weinrib1982percolation}.} This contention was addressed by \citet{bogomolnyRWM} by invoking what is known as Harris' criterion \cite{harris1974effect} to argue that the specious problem of the slow decay of Eq.~\eqref{eq:CorrDecay} is effectively circumvented. To see this, consider a critical percolation problem where the concentrations at different sites, denoted by $\psi \,(x)$, are correlated. The mean concentration at each site, of course, equals the critical value $p_{\textsc{c}}$. Let $p_{\textsc{v}}$ represent the average concentration in a (finite) volume $V$:
\begin{equation}
p_{\textsc{v}} = \frac{1}{V}\,\sum_{\mathbf{r}\,\in\, V} \psi\,(\mathbf{r});
\end{equation}
obviously, $\langle p_{\textsc{v}} \rangle = p_{\textsc{c}}$. Furthermore, we assume, on grounds of translational invariance, that the \textsl{connected} correlation function depends not on $\mathbf{r}$ and $\mathbf{r}'$ individually, but rather only on the difference between them, i.e., 
\begin{equation}
\langle (\psi\,(\mathbf{r}) - p_{\textsc{c}})\,(\psi\,(\mathbf{r}')-p_{\textsc{c}}) \rangle = G(\lvert \mathbf{r} - \mathbf{r}' \rvert). 
\end{equation}
Briefly, Harris' eponymous criterion, later extended by \citet{weinrib1984long} to correlated percolation, can be formulated as follows. If the variance of $p_{\textsc{v}}$,
\begin{alignat}{1}
\Delta &\equiv \langle (p_{\textsc{v}} - p_{\textsc{c}})^2 \rangle \\
\nonumber &= \frac{1}{V^2}\sum_{\mathbf{r}, \mathbf{r}'\,\in\, V} \langle (\psi\,(\mathbf{r}) - p_{\textsc{c}})\,(\psi\,(\mathbf{r}')-p_{\textsc{c}}) \rangle\\
\nonumber &\approx \frac{1}{V^2} \int_{\mathbf{r}\,\in\, V}\int_{\mathbf{r}'\,\in\, V} G(\lvert \mathbf{r} - \mathbf{r}' \rvert)\,\mathrm{d}\,\mathbf{r}\,\, \mathrm{d}\,\mathbf{r}',
\end{alignat}
is small in the sense that $\Delta \ll \lvert p_{\textsc{v}} - p_{\textsc{c}} \rvert^2$, then correlations are unessential and all critical quantities are the same as for the standard uncorrelated percolation \cite{bogomolnyRWM}. For random wavefunctions, $\Delta \sim \xi^{-3}$ and $\xi \sim \lvert p_{\textsc{v}} - p_{\textsc{c}} \rvert^{-\upsilon}$, where $\upsilon = 4/3$ is a critical index for two-dimensional percolation. Hence, Harris' criterion is satisfied\footnote{Actually, we have been a little too quick with this calculation. A pivotal role is played here by the oscillating nature of the correlation function, which is responsible for strong cancellations. The fact that $G\,(r)$ is not always positive, coupled with the requirement of $\Delta$ being non-negative (by definition), saves the day.} and the system belongs to the short-range percolation universality class. For the sake of completeness, let us mention that the Bogomolny-Schmit conjecture can be generalized along these lines to level domains, i.e., regions where $\psi\,(x, y) > \varepsilon$ for some fixed $\varepsilon \ne 0$. Level domains are also described by percolation theory but by \textsl{noncritical} percolation wherein the deviation from criticality $p- p_{\textsc{c}} \sim \varepsilon$ \cite{bogomolnyRWM}.

\begin{figure}[htb]
\subfigure[]{\includegraphics[width=0.49\linewidth, trim ={2cm 1cm 2cm 0.75cm}, clip]{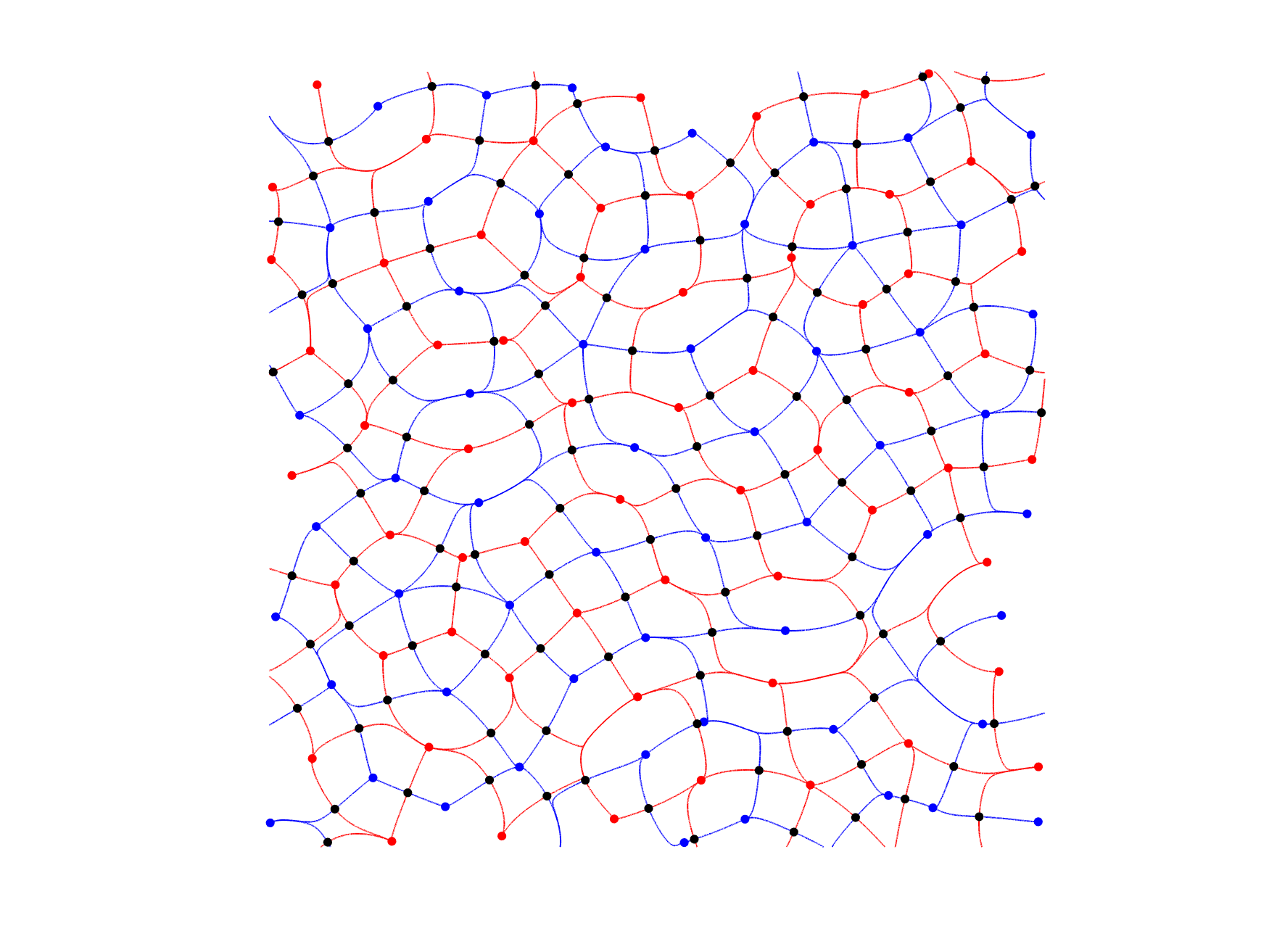}}
\subfigure[]{\includegraphics[width=0.49\linewidth, trim ={2cm 1cm 2cm 0.75cm}, clip]{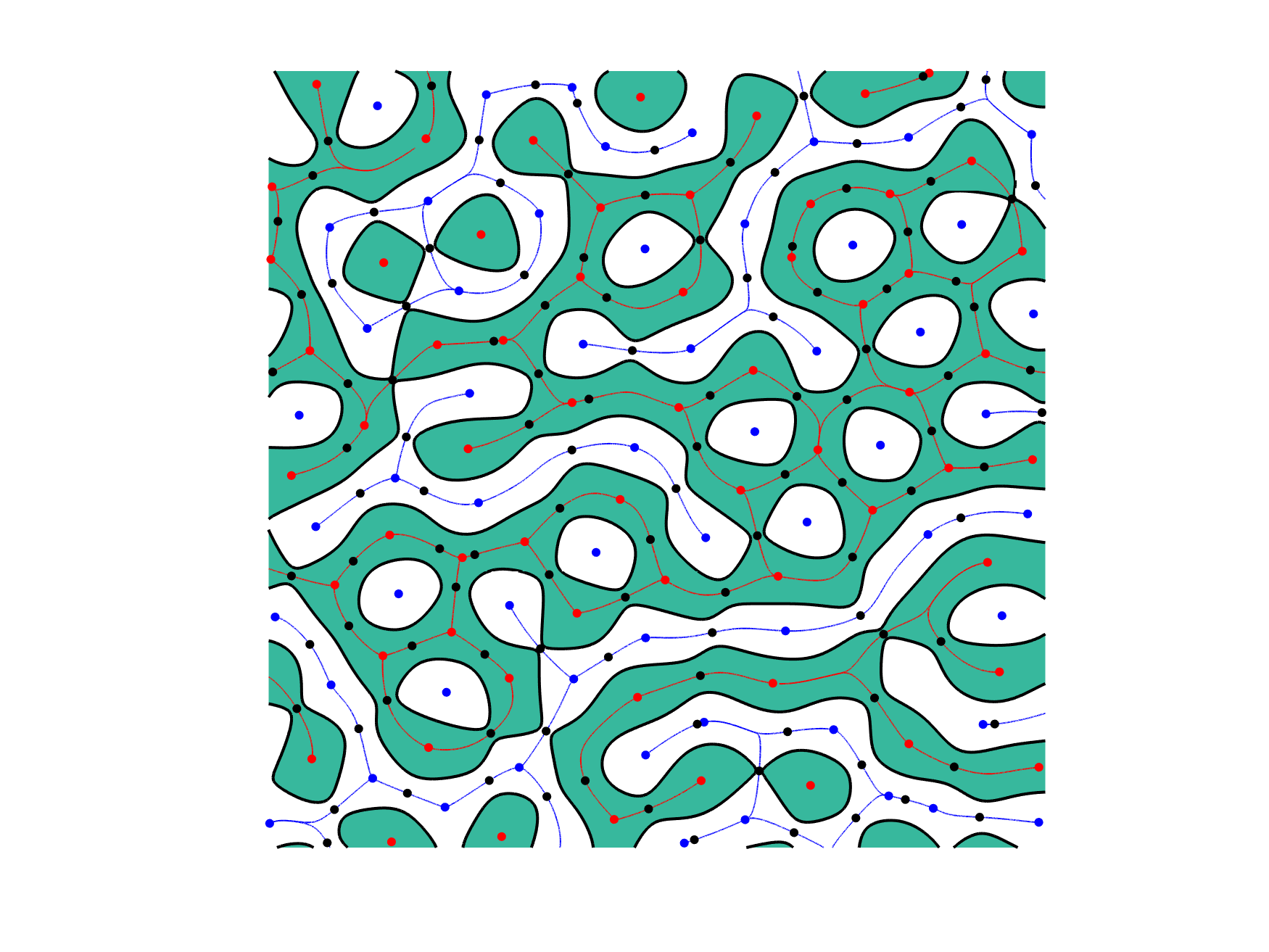}}
\caption{\label{fig:NewPerc}Bond percolation, with probability $p = 1/2$, on a graph generated by the random plane wave. (a) The nodes of the graph (red) are local maxima and the edges are gradient streamlines passing through saddles. The dual graph (blue) is formed by local minima. (b) The percolation model is critical and clusters represent the connected nodal components. Figures courtesy Dmitry Belyaev.}
\end{figure}

Disconcertingly enough, more serious objections to the conjecture have come to the forefront in recent times, fueled by several high-precision numerical studies. The first questions were raised by \citet{nastasescu2011}, who computed the density of the mean number of nodal domains for random spherical harmonics to be $0.0598\, \pm\, 0.0003$, nearly $5\, \sigma$ away from the theorized value. Initially, this discrepancy was brushed away by attribution to finite size and curvature effects. Adding to the growing unease, \citet{konrad2012} repeated the calculation for plane waves  to obtain a density of $0.0589\, \pm\, 0.000142$, which is $6\%$ below the prediction. The general belief is that the normalized number of nodal domains should behave like $a + b/k$, for constants $a$ and $b$. While \citet{konrad2012} found a best fit of $0.0589 + 4.6209/k$, simulations at higher energies \cite{beliaev2013bogomolny} determined the fitting parameters to be $a = 0.0589$ and $b = 4.717$. Puzzlingly, the latter group also observed that the crossing probabilities for a nodal line of a random plane wave to connect the sides of a box $\mathcal{D}$ converge to their percolation counterparts \cite{cardy1992critical, watts1996crossing}. That the probabilities are macroscopic observables and hence, universal from a percolation perspective, reaffirms the suspicion that this is no coincidence. Seeking better agreement with the numerics, \citet{beliaev2013bogomolny} proposed an alternative normalization scheme in which the number of vertices of the square lattice chosen is the same as the number of local maxima of the random plane wave (see Fig.~\ref{fig:NewPerc}). Pursuant to this prescription, the technical details of which are to be found in \cite{kereta2012}, the average density of the critical points can be computed using Gaussian integrals to be $0.0919 \,k^2$ (a quarter of which are maxima) and the number of domains is estimated as $\bar{n}\,(E) / \xoverline{N}\,(E) = 0.0566$. It is important to emphasize that the choice of normalization condition should not affect the density of nodal domains, which is believed to be a universal quantity in the following sense. Given Laplacian eigenfunctions $\psi_n$, we can define 
\begin{equation}
f_n = \sum_{k=n}^{n+C\sqrt{n}} c_k\,\psi_k,
\end{equation} 
where $c_k$ are i.i.d. normal variables and $C$ is a (large) constant. Then, the properly rescaled number of nodal domains of $f_n$ (asymptotically) has the same density as the random plane wave \cite{beliaev2013bogomolny}. The same cannot be said, however, for the number of clusters per vertex, which, being a nonuniversal quantity in percolation theory, is strongly dependent on the lattice structure.

All things considered, at present, we are far from understanding ``a hidden universality law'' of sorts that would provide a rigorous foundation for the Bogomolnny-Schmit conjecture. One of the strongest known results in this direction pertains to the number of nodal domains $\nu\, (g_n)$ for a two-dimensional Gaussian spherical harmonic of degree $n$ . \citet{nazarov2009number} proved that for every $\varepsilon > 0$, there exist positive constants $\mathcal{C} (\varepsilon)$ and $c\, (\varepsilon)$ such that the probability tail
\begin{equation}
P\, \bigg \{ \left \lvert\, \frac{\nu \,(g_n)}{n^2} - \nu_{\textsc{ns}} \,\right \rvert  > \varepsilon \bigg \} \le \mathcal{C} (\varepsilon) \,\mathrm{e}^{- c \,(\varepsilon)\,n}.
\end{equation} 
In addition, the expected number of nodal domains is asymptotic to
\begin{equation}
\label{eq:NS}
\mathbb{E}\, \left[\nu \,(g_n) \right] = \nu_{\textsc{ns}}\,n^2 + \mathcal{O} \left( n^2 \right).
\end{equation}
This theorem, which is based on the Gaussian isoperimetric inequality \cite{borell1975brunn, sudakov1978extremal}, implies that the normalized number of nodal domains  is exponentially concentrated around some strictly positive constant, which, in this case, happens to be $\nu_{\textsc{ns}} =  (3\,\sqrt{3} - 5) / \pi$. The constant $\nu_{\textsc{ns}}$ is certainly not universal and genuinely depends on the underlying random Gaussian field \cite{kurlberg2015non}. Recently, \citet{rozenshein2016number} extended these results to random eigenfunctions of the Laplacian on the torus $\mathbb{T}^2 = \mathbb{R}^2/\mathbb{Z}^2$, showing that the number of nodal domains $\nu_n$ localizes around its median, mean, and limiting mean, exponentially, with an optimal lower bound \cite{buckley2016number}. The asymptotic law for the expectation of $\nu \,(g_n)$, Eq.~ \eqref{eq:NS}, is actually much more general as proved by  \citet{nazarov2016asymptotic} in a setting of ensembles of Gaussian functions on Riemannian manifolds. These and other methods have been systematically exploited to obtain distributions of nodal volumes \cite{wigman2009, rudnick2008volume, beliaev2016volume} and topological invariants \cite{gayet2014betti, gayet2014lower, gayet2015expected, canzani2014topology, canzani2016topology}.

\subsection{Morphology of nodal lines: Schramm-Loewner Evolution}
\label{SLE}

One of the oldest tricks in the playbook of critical phenomena is to define a model on a finite subset of a lattice and then ask what happens as the lattice size is allowed to grow. Ceteris paribus, one could also work in a bounded region and increase the resolution of the grid thereon by considering finer and finer lattices. Either way, the objective of such endeavors is to determine the scaling (continuum) limit of the system, if it exists, and understand its geometric and fractal properties. Many years ago, \citet{belavin1984infinite, belavin1984infinite2} postulated that several two-dimensional systems had scaling limits at criticality that were, loosely speaking, conformally invariant. Although not rigorous, their predictions were consistent with numerical simulations and hinted at some deeper physics beneath the surface.

\subsubsection{Scaling and conformal invariance}
As is probably evident from our preceding remarks, the recurrent motifs of this narrative are the conjoint ideas of scaling and conformal invariance. Scale invariance is ubiquitous in nature; its most notable application in physics is perhaps the renormalization group flow \cite{wilson1974renormalization}. Mathematically, a function (or scaling operator) $\Phi\, (\mathbf{r})$ is said to be scale invariant if
\begin{equation}
\label{eq:scale}
\Phi\,(\lambda\,\mathbf{r}) = \lambda^{\Delta}\, \Phi\, (\mathbf{r})
\end{equation}
holds with the same exponent $\Delta$ for all rescaling factors $\lambda$ (which generate the dilatation $\mathbf{r} \mapsto \mathbf{r}' = \lambda\,\mathbf{r}$); in other words, if $\Phi$ is a generalized homogeneous function. Discrete scale invariance is better recognized as self-similarity: for instance, the famous Koch curve scales with $\Delta = 1$, but only for values of $\lambda = 1/3^n;\, n \in \mathbb{Z}$. If the rescaling factors are permitted to be space-dependent $\lambda = \lambda(\mathbf{r})$, a natural generalization of global scale-invariance, Eq.~\eqref{eq:scale}, is
\begin{equation}
\Phi\,(\mathbf{r}) \mapsto \Phi'\,(\mathbf{r}) = J\, (\mathbf{r})^{\,x/d}\, \Phi \left(\mathbf{r}/\lambda(\mathbf{r})\right),
\end{equation}
where $J (\mathbf{r})$ is the Jacobian of the transformation $\mathbf{r} \mapsto \mathbf{r}' = \mathbf{r}/\lambda (\mathbf{r})$ in $d$ spatial dimensions and $x$ the scaling dimension of $\Phi$. Restricting to those coordinate transforms that additionally conserve angles, one arrives at conformal transformations \cite{henkel2012conformal}. Visualized on an
elastic medium, conformal transformations represent deformations
without shear \cite{landau1959course}. In two dimensions, these concepts can easily be extended to the complex plane by bartering a vector $\mathbf{r} = (r_1, r_2)$ for a complex number $z = r_1 + \mathrm{i}\,r_2$. Then, Riemann's mapping theorem \cite{ahlfors2010conformal} ensures that we can map any simply connected domain $\mathcal{D}$ (topologically equivalent to a disk) to another, $\mathcal{D}'$, i.e., there exists an invertible holomorphic (complex-analytic) map $g$ between them. In quotidian dealings, both $\mathcal{D}$ and $\mathcal{D}'$ are customarily the upper half-plane $\mathbb{H}$. Moreover, any analytic or anti-analytic coordinate transformation, $z \mapsto g\,(z)$ or $\bar{z} \mapsto \bar{g}\, (\bar{z})$, is always conformal. The bourne of the mathematicians to whom SLE owes its present form was to make these notions of conformal invariance precise and introduce the rigor that field theorists had previously glossed over.

\begin{figure}[htb]
\includegraphics[width=\linewidth]{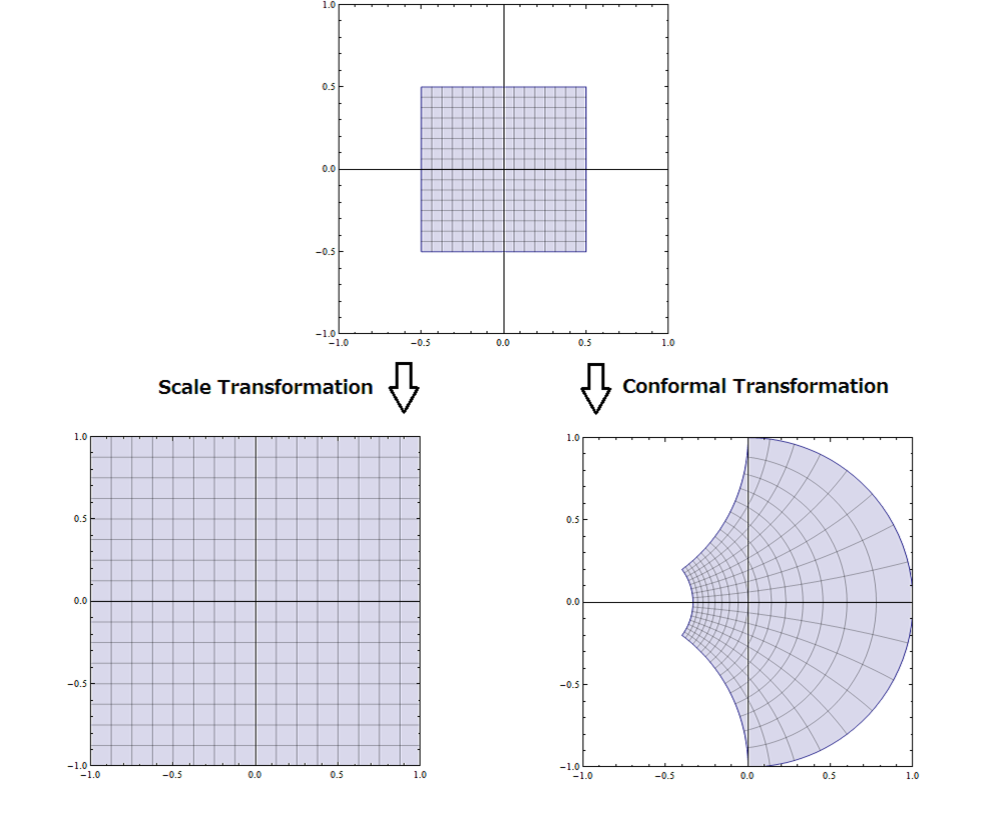}
\caption{\label{fig:Conformal}Comparison between scale and conformal transformations. The latter are also angle-preserving and basically correspond to a combination of a local rotation, local translation, and local dilatation. There is no a priori reason why a scale-invariant system must also be conformally invariant although it is often found to be so. From \citet{nakayama2015scale}, with permission from Elsevier.}
\end{figure}

Before diving into the details, we must concede that it is nigh impossible to do justice to this incredibly rich field in just a few pages and hence it is obligatory for us to point the interested reader to further references. Besides the original set of articles by \citet{lsw2001-1, lsw2001-2, lsw2002} relating SLE to various aspects of Brownian motion, mention must be made of the excellent albeit mathematically intensive reviews by \citet{lawler2001vienna, werner2004random, kager2004guide, lawler2008conformally}. For the reader disinclined towards the jargon of mathematicians, there are also a few semi-pedagogical (and more accessible) introductions written for physicists \cite{fogedby2012stochastic, cardy2005sle, lawler2009conformal}, from which we mainly source our discussion. An exhaustive bibliography up to 2003 can be found in \cite{gruzberg2004loewner}.

\subsubsection{Critical interfaces in 2D: Lattice models}
Our spotlight now falls on models that describe random non-intersecting paths, which define the boundaries of clusters on a lattice, with the hope that a picture of their continuum limits might be rendered by SLE. Motivated by analogous ideas in two-dimensional critical behavior such as the Coulomb gas approach \cite{nienhuis1987}, the central question that arises is, what are the properties of the measure on such curves as the lattice spacing tends to zero \cite{cardy2005sle}? The simplest, if not the oldest, example is the random walk, which has been extensively studied over the decades \cite{ash2000probability, feder1988fractals, reichl2016modern}. We begin with an unbiased simple random walk (SRW) consisting of exactly $n$ steps on a two-dimensional plane. The $i^{\mathrm{th}}$ step, represented by the vector $\mathbf{s}_i$ (of magnitude equal to a constant step size $S$), is random, isotropic and uncorrelated, i.e., $\langle \mathbf{s}_i \rangle = 0$, and $\langle s^{\,\alpha}_i\,s_j^\beta \rangle \propto \delta_{i,j}\,\delta_{\alpha,\beta}$. The net displacement during the drunkard's excursion is just $\mathbf{x} = \sum_{i=1}^n \mathbf{s}_i$ and the corresponding end-to-end (Euclidean) distance traversed is $R = \sqrt{\langle \mathbf{x}^2\rangle} \sim \sqrt{n} \sim t^{1/2}$, characteristic of diffusive motion. Calculating the fractal dimension\footnote{Crudely, a subset of $\mathbb{Z}^d$ is said to have fractal dimension $D$ if the number of points in a disk of radius $R$ grows as $R^D$.} of a typical path using the box-counting procedure \cite{feder1988fractals, mandelbrot1983fractal} yields $D =2$, implying that the walk fills the plane modulo the lattice spacing. As a consequence, the scaling exponent, defined by the relation $R \sim n^{\upsilon}$, is found to be $\upsilon = D^{-1} = 1/2$.

To formally characterize this process, let us represent a walk by a sequence of vertices $\omega = [ \omega_1, \omega_2, \ldots, \omega_n]$ with $\lvert \lvert \omega_j - \omega_{j-1} \rvert \rvert = S$ for each $j$. Each such walk $\omega$ is assigned a probability $P_{\omega}(n) = \exp\, (- \beta\, n)$ resembling the familiar Boltzmann factor, whereby all walks of the same length have the same weight. The sum of the weights of all possible paths is recognized as the partition function
\begin{equation}
\label{eq:partition}
Z_{\beta} = \sum_{\omega} \mathrm{e}^{-\beta\, \lvert \omega \rvert},
\end{equation}
where the length (number of edges) $\lvert \omega \rvert$ can be arbitrarily large. At, and only at, a critical value of $\beta = \beta_c = \log 4$ ($\because$ the number of SRWs of length $n$ is $4^n$ in 2D), this sum neither grows nor decays exponentially with the number of lattice sites (or equivalently, the finesse of the mesh), scaling as a power law instead. Said otherwise, it becomes scale invariant! The scaling limit of an unbiased random walk is actually Brownian motion \cite{ash2000probability}, denoted $B_t$. This limit is obtained as $S \rightarrow 0$ but with the number of steps $n$ simultaneously scaled up, so as to keep the size $R \sim \sqrt{n}\,S$ constant. The resultant Brownian path is a continuous non-differentiable random curve, once again plane-filling with fractal dimension $D = 2$. The most significant departure from our previous discourse is that the notion of the probability density of a walk no longer makes sense. Indeed, what do the probabilities $P_{\omega}(n)$ that we allotted so insouciantly even mean when $n \rightarrow \infty$? Nevertheless, this is not an insurmountable hurdle as we can make the appropriate replacements with the more general concept of a measure. The partition function in Eq.~\eqref{eq:partition} is now identified with what mathematicians would call the total \textsl{mass} of this measure. For concreteness, consider a Brownian excursion on a domain $\mathcal{D}$, traced by a curve $\gamma$ that commences at $\mathbf{r}_1$ and concludes at $\mathbf{r}_2$. This acquires a measure $\mu \,(\gamma;\mathcal{D}, \mathbf{r}_1, \mathbf{r}_2)$, which is conformally invariant \cite{levy1965}. The proof, which follows directly from It{\^o}'s lemma \cite{ito1944109}, hinges on the premise that if $B_t$ is a complex Brownian motion and $g$ a conformal mapping between simply connected domains $\mathcal{D} \rightarrow \mathcal{D}'$, then $g\,(B_t)$, modulo a reparametrization of time, is also a Brownian motion. Consequently, we have
\begin{property}{Conformal Invariance:}
\label{prop1}
\begin{equation}
(g \circ\mu) \,(\gamma;\mathcal{D}, \mathbf{r}_1, \mathbf{r}_2) = \mu \,\left(g(\gamma);\mathcal{D}', \mathbf{r}_1', \mathbf{r}_2' \right).
\end{equation}
\end{property}
\noindent Although Brownian motion does not fall within the purview of SLE owing to self-crossings that annul Riemann's mapping theorem, its perimeter (which has a fractal dimension of $D = 4/3$) does \cite{lawler2000dimension}, as do several other variations upon it.

\begin{figure}[htb]
\subfigure[]{\includegraphics[width=0.49\linewidth]{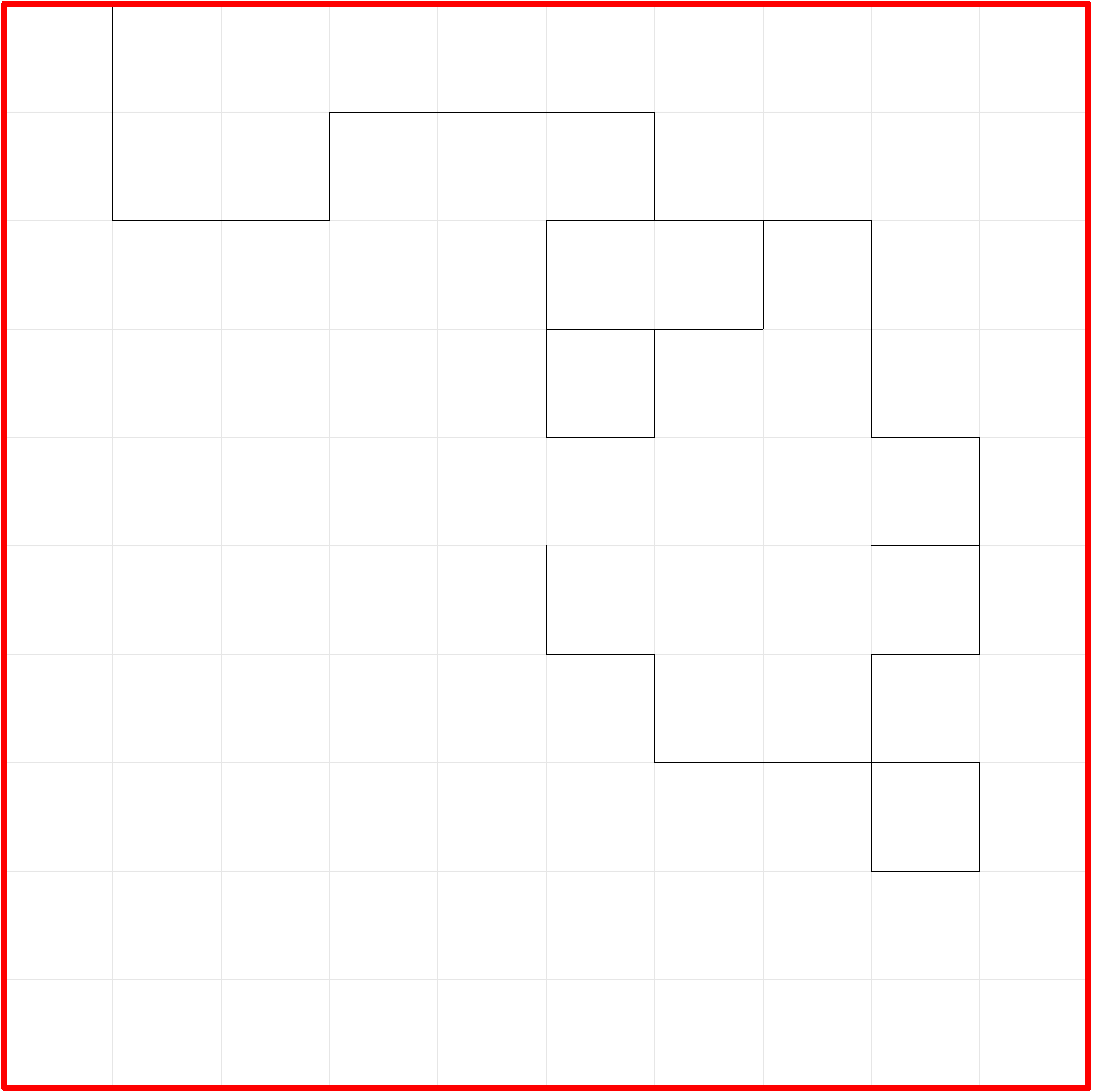}}
\subfigure[]{\includegraphics[width=0.49\linewidth]{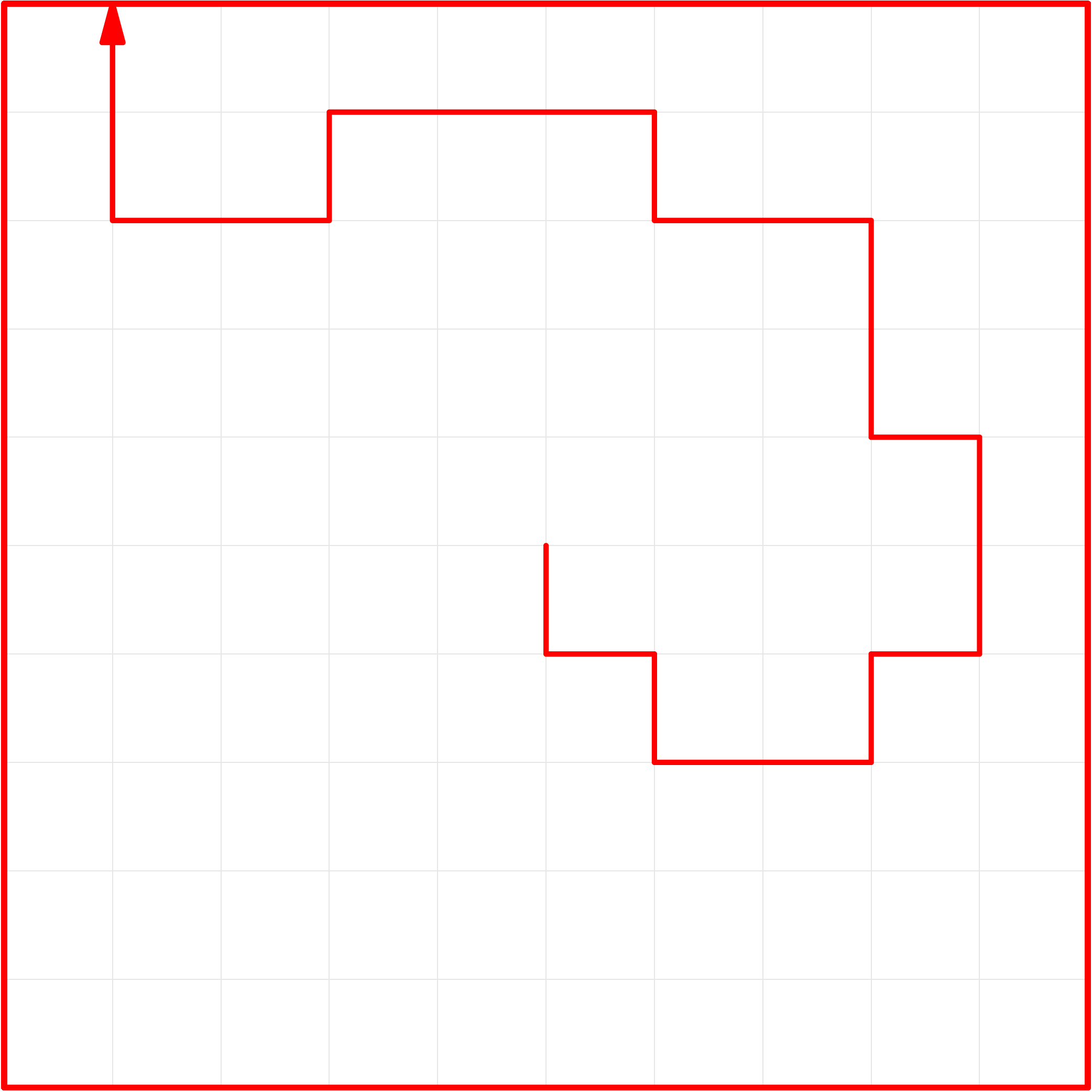}}
\subfigure[]{\includegraphics[width=0.49\linewidth]{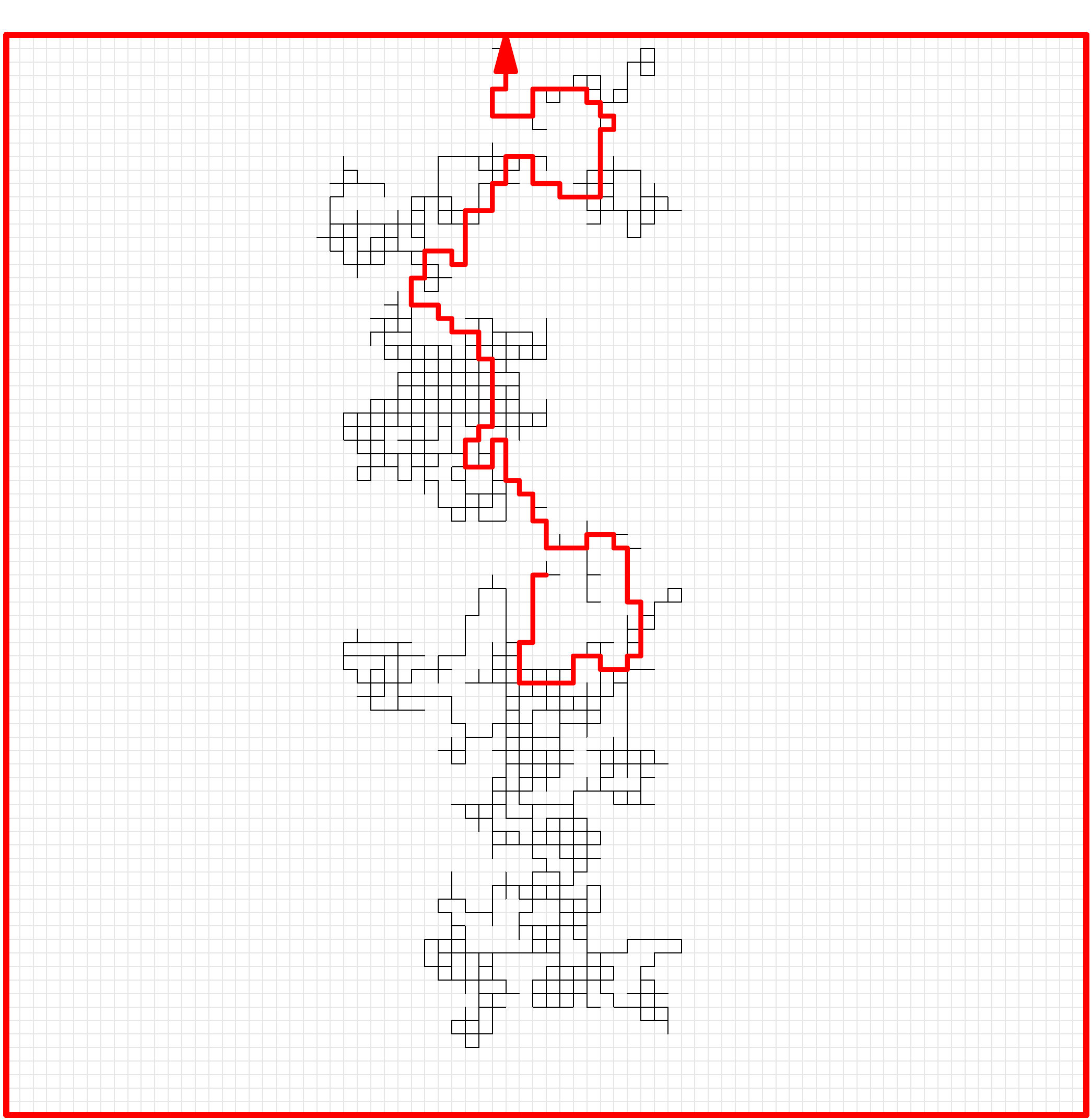}}
\subfigure[]{\includegraphics[width=0.4901\linewidth]{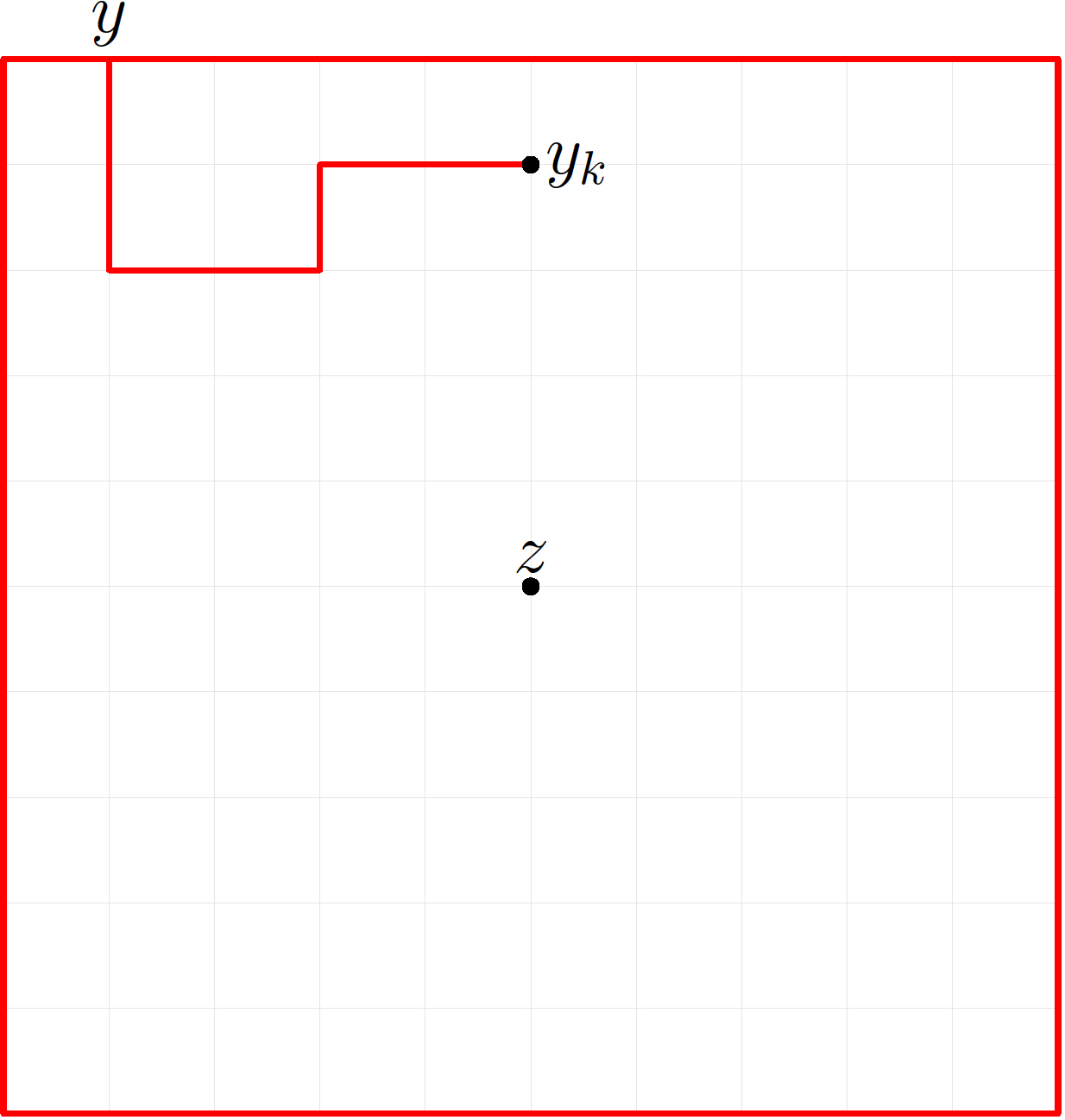}}
\caption{\label{fig:LERW}(a) An unbiased random walk on a domain $\mathcal{D}$ (a $20 \times 20$ square lattice) that starts in the interior and terminates when it encounters the boundary. (b) The same, after loop erasure. (c) LERW on a $80 \times 80$ grid (in red) together with its \textsl{shadow}. (d) Construction of the domain Markov property. Figures courtesy Tom \citet{alberts2008slides}.}
\end{figure}

A common variant is the loop-erased random walk (LERW), which was studied by \citet{schramm2000scaling}. This process, a subset of a broader (and more intractable) class of self-avoiding random walks, describes systems like polymers that possess strong tendencies to be self-avoiding. The prescription, outlined in Fig.~\ref{fig:LERW}, is to sample an unconstrained random walk and then, chronologically erase the loops along the way, in the order in which they are encountered; the resultant path, by construction, does not cross itself. The bad news is that LERWs forfeit a salient property of the SRW---Markovian (memoryless) time evolution. Since the LERW curve cannot cross itself, the future path, at any point, is always highly correlated with and dependent on its past. To make things explicit, consider Fig.~\ref{fig:LERW}~(b), which depicts a LERW, specifically conditioned to exit $\mathcal{D}$ at a fixed point on the boundary, $y \in \partial \,\mathcal{D}$. We call this law\footnote{We will precisely define what ``law'' means in a minute. For now, we put the cart before the horse and treat it as a placeholder for ``the rule/equation governing the evolution of $\gamma$."} $LERW_{S}\,(\gamma;\, \mathcal{D}, z, y)$. Let us trace out this path in reverse: having observed the first $k$ steps of the walk (Fig.~\ref{fig:LERW}~d), it turns out that the law of the remaining path is  nothing but $LERW_{S}\,(\gamma;\, \mathcal{D}\backslash [y, y_k], z, y_k)$. Thus we recover the original law itself but this time, on a truncated slit domain, $\mathbb{D}_k \equiv  \mathcal{D}\backslash [y, y_k]$. One can therefore still salvage a Markov property in the LERW process, albeit in the evolution of the domain rather than the curve. Formally \cite{cardy2005sle}, 
\begin{property}{Domain Markovian condition:}
\label{prop2}
If the curve $\gamma$ is divided into two disjoint parts: $\gamma_1$ from $\mathbf{r}_1$ to $\tau$, and $\gamma_2$ from $\tau$ to $\mathbf{r}_2$, then the conditional measure satisfies
\begin{equation}
\mu\,(\gamma_2 \vert \gamma_1;\mathcal{D}, \mathbf{r}_1, \mathbf{r}_2) = \mu\,(\gamma_2;\mathcal{D} \backslash \gamma_1, \mathbf{r}_1, \mathbf{r}_2). 
\end{equation}
\end{property} 
\noindent The LERW measure of each self-avoiding path $\omega$ is the total measure of all the different SRWs that can be loop erased to $\omega$. This weight can be written as
\begin{equation}
4^{-\lvert \omega \rvert}\,\mathrm{e}^{\,\Lambda \,(\omega)},
\end{equation}
where $\Lambda \,(\omega)$ is a measure of the number of loops in the domain that intersect $\Lambda \,(\omega)$ \cite{lawler2009conformal}. Properties~\ref{prop1} and \ref{prop2}, in combination with Loewner evolution, are sufficient to determine the measures in the scaling limit \cite{fogedby2012stochastic}. LERW, which satisfies both, was proved to have a conformally-invariant scaling limit that can be accessed by SLE \cite{lawler2004conformal}. By the way, the growth of the curve $\gamma$ is also an example of an exploration process. This should immediately remind us of our old friend, percolation. It is easily seen that the exploration curve in Fig.~\ref{fig:Ising} exhibits the same domain Markov property as LERW. This correspondence can be taken a step further. Indeed, in the scaling limit, which does manifest conformal invariance, the critical percolation cluster has a fractal boundary described by SLE. Interestingly, the model also bears an additional conformal invariant, which was predicted by \citet{cardy1984conformal, cardy1992critical}.

\subsubsection{The Loewner differential equation}

\begin{figure}[htb]
\includegraphics[width=\linewidth]{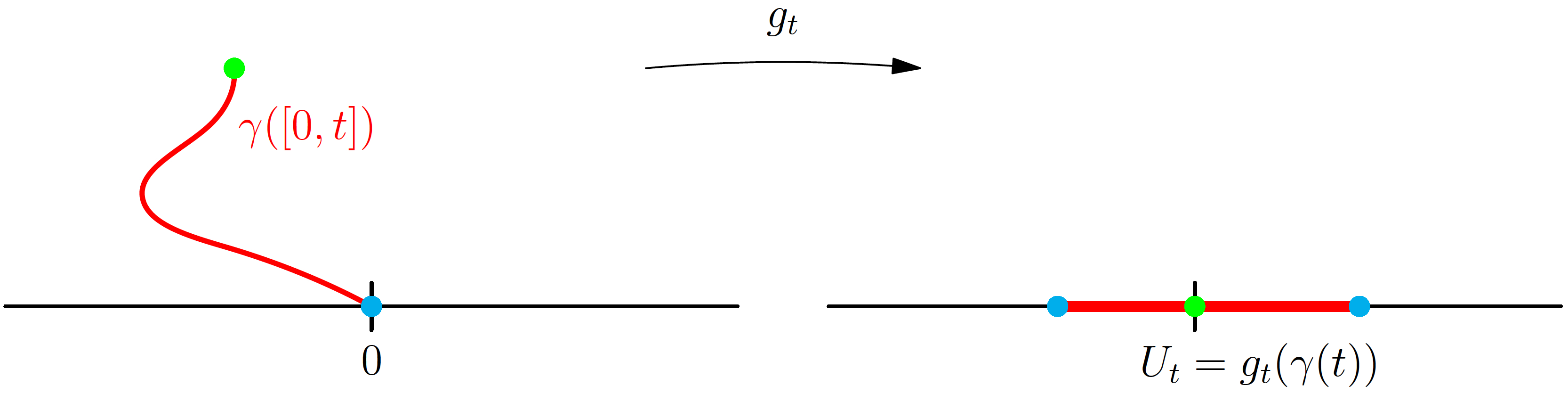}
\includegraphics[width= \linewidth]{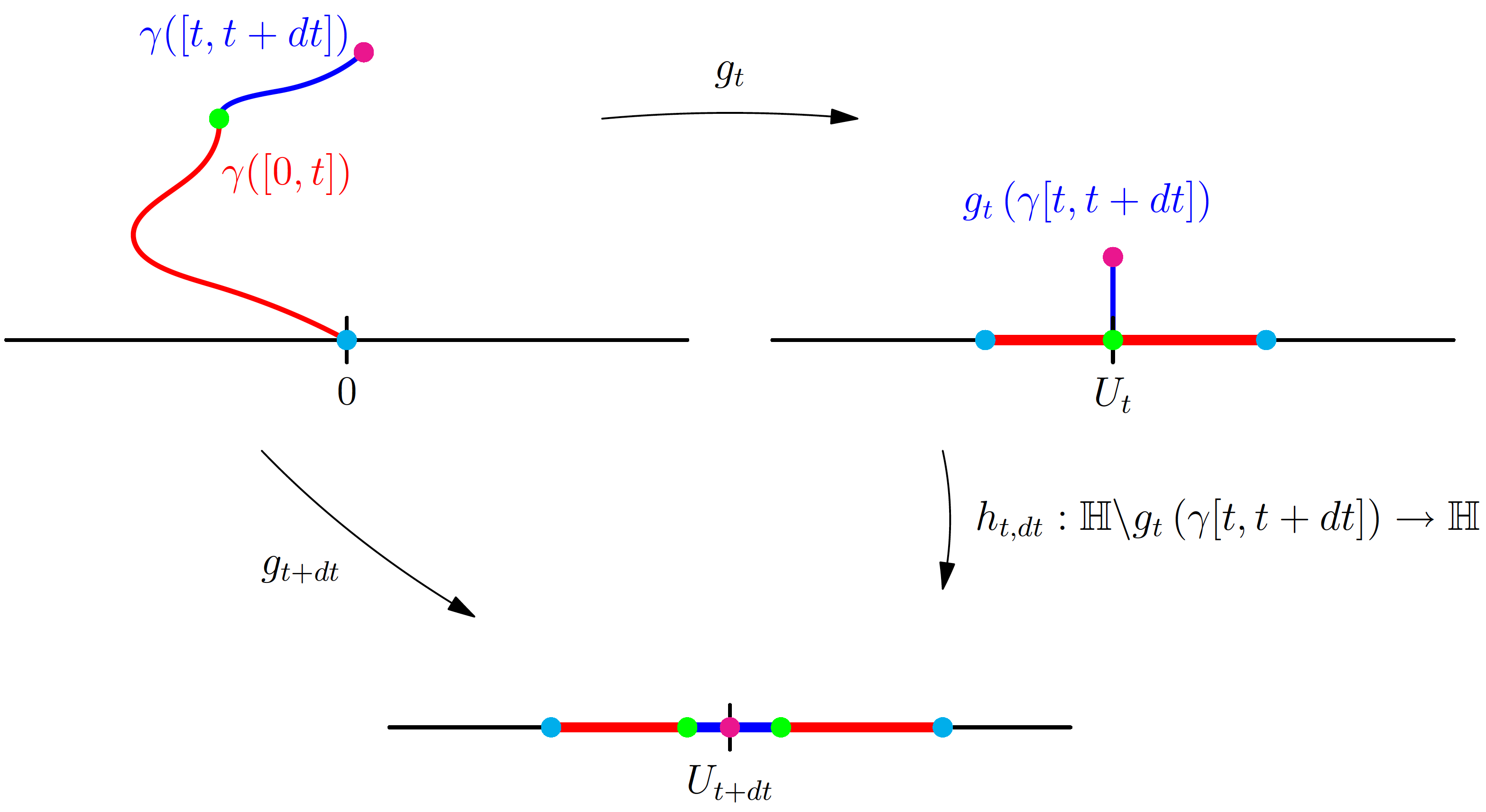}
\caption{\label{fig:Loewner}Loewner evolution (topmost panel) and the schematic outline for the derivation of the Loewner differential equation, Eq.~\eqref{eq:LDE}. Figure courtesy Tom \citet{alberts2008slides}.}
\end{figure}

Armed with this background, we can now begin the story of Loewner evolution, which originally unfolded in the context of the Bieberbach conjecture. Proposed in 1916, the latter surmised that if $f (z) = z + \sum_{n \ge 2} a_n\,z^n$ is a holomorphic, injective (one-to-one) function on the unit disc $\mathbb{D} = \{z \in \mathbb{C};\, \lvert z \rvert < 1\}$, then $\lvert a_n \rvert \le n$ for $n \ge 2$ \cite{bieberbach1916, gong1999bieberbach}. In 1912, Bieberbach proved $\lvert a_2 \rvert \le 2$ and later, \citet{lowner1923untersuchungen} considered the dynamics of the coefficients $a_n$ to prove $\lvert a_3 \rvert \le 3$. It then took a further 62 years for conclusive progress before the conjecture was finally proved by \citet{de1985proof}. L\"{o}wner's ideas in this regard can be stated quite simply; we closely follow the discussion by \citet{alberts2008slides}. Consider a self-avoiding curve $\gamma: [0, \infty) \rightarrow \mathbb{H}$ such that $\gamma\, (0) = 0$ and $\gamma\, (\infty) = \infty$. It follows that as the curve grows, $\forall\, t \ge 0$, $\mathbb{H} \backslash \gamma\,([0, t])$ is a simply connected domain.\footnote{If $\gamma$ touches itself, there may be regions (enclosed by loops) that cannot be accessed without crossing the curve. The union of the set of such points with $\gamma$ is called the \textsl{hull} $\mathbb{K}_t$. In less simplistic descriptions, it is $\mathbb{H}\backslash \mathbb{K}_t$ that is taken to be simply connected.} SLE catalogs (local) growth processes of this type for which the resulting set $\gamma$ is eventually a continuous curve \cite{henkel2012conformal}. Riemann's mapping theorem avers that there exists a `time-dependent' conformal transformation $g_t: \mathbb{H} \backslash \gamma\,[0, t] \rightarrow \xoverline{\mathbb{H}}$ as Fig.~\ref{fig:Loewner} limns. The map $g_t$ is certainly not unique and most generally, will have three real degrees of freedom. For instance, $\mathbb{H}$ is transformed to itself by a three-parameter group of fractional transformations 
\begin{equation}
g_t (z) = \frac{a\,z + b}{c\, z + d},
\end{equation}
for $a, b, c, d\,\in\,\mathbb{R}$ \cite{bogomolny2006sle, siegel1969topics}. Firstly, two degrees of freedom can be absorbed by imposing the hydrodynamic normalization which constrains the behavior out at infinity: $g_t (\infty) = \infty$, $g_t' \,(\infty) = 1$. The Laurent expansion for $\lvert z \rvert \rightarrow \infty$ must therefore resemble 
\begin{equation}
\label{eq:Laurent}
g_t (z) = z + a_0 + \frac{b_1}{z} + \frac{b_2}{z^2} + \ldots\,.
\end{equation}
Specifying $a_0 = 0$ fixes all the other coefficients uniquely. The coefficient  $b_1$ (which is also a function of $t$) is, for reasons historical and arcane, called the half-plane capacity of $\gamma\,[0, t]$, and notationally denoted by $a\, (\gamma\,[0, t])$. The fact that the capacity is additive, i.e.,
\begin{equation}
a\, (\gamma\,[0, t+s]) = a\, (\gamma\,[0, t]) + a\, (g_t(\gamma\,[t, t+s]))
\end{equation}
lends credence to our interpretation of $t$ as a ``conformal time''. Furthermore, since $a (\gamma\,[0, t])$ is continuous and increasing, one can reparametrize the curve such that $a\, (\gamma\,[0, t]) = 2\,t$ (again, by convention) and Eq.~\eqref{eq:Laurent} is recast as
\begin{equation}
g_t (z) = z + \frac{a\, (\gamma\,[0, t])}{z} + \ldots =  z + \frac{2\,t}{z} + \ldots\,.
\end{equation}
An instructive example is when $\gamma$ is a straight line, growing vertically upwards in the upper half-plane. 
If $(\zeta, h)$ are the coordinates of the tip of the slit, then $z = \zeta + \mathrm{i}\,h$ and the required transformation is
\begin{equation}
\label{eq:slit}
g_t (z) =  \zeta + \sqrt{(z- \zeta)^2 + 4\,t},
\end{equation}
with $h = \sqrt{4\,t}$ being the parametric equation of the growing slit. In fact, $g_t \left(\gamma (t)\right) \propto \sqrt{t}$ is a signature of a straight line growing at a fixed angle to the real axis \cite{cardy2005sle}. More involved deterministic examples can be found in \cite{kager2004exact}.

This nifty calculation enables us to derive the Loewner differential equation (LDE) that describes the evolution of the maps $g_t$ with the growth of the curve  $\gamma\,([0, t])$. One starts by asking if the map $g_{t+\mathrm{d}t}$ could be determined just from knowledge of the one infinitesimally earlier, $g_t$. The procedure to do so is sketched in Fig.~\ref{fig:Loewner}. In particular, we can explicitly compute the map $h_{t, \mathrm{d}t}$, which, by Eq.~\eqref{eq:slit}, is
\begin{alignat}{1}
\nonumber h_{t, \mathrm{d}t}\, (w) &= U_{t+\mathrm{d}t} + \sqrt{(w - U_t)^2 + 2\, \mathrm{d}t}\\
&\approx w + \frac{2\, \mathrm{d}t}{w - U_t},
\end{alignat}
where $U_t = g_t \left(\gamma (t)\right)$, called the ``driving'' or ``forcing'' function, encodes all the topological properties of the curve $\gamma\,([0, t])$. Recognizing that
\begin{equation}
g_{t+\mathrm{d}t} (z) = h_{t, \mathrm{d}t}\, \left(g_t (z) \right) \approx\, g_t(z) +\frac{2}{g_t(z) - U_t},
\end{equation}
we are led to the celebrated (and till now, elusive) LDE:
\begin{equation}
\label{eq:LDE}
\partial_t\,g_t(z) = \frac{2}{g_t(z) - U_t};\,\, g_0 (z) = z.
\end{equation}
Given a curve $\gamma$, the maps $g_t$ must necessarily satisfy Eq.~\eqref{eq:LDE}. The converse is also possible: given a driving function $U_t : [0,\infty) \rightarrow \mathbb{R}$, the LDE can be solved for $g_t$, which then determines $\gamma\,([0, t])$. When $U_t$ itself is a continuous random function, Eq.~\eqref{eq:LDE} yields a stochastic equation of motion and consequently, a stochastic map $g_t\,(z)$. The growing tip then traces out a random curve, determined by \cite{bauer2003discrete, kennedy2007fast}
\begin{equation}
\gamma\, (t) = g_t^{-1}\, \left(U_t\right).
\end{equation}
A sufficient condition for generating such a curve is that the driving function $U_t$ be H\"{o}lder continuous\footnote{A real or complex-valued function $f$ is said to be H\"{o}lder continuous if $\exists$ non-negative real constants $M$, $\alpha$, such that $\lvert f(x) - f(y) \rvert \le M\, \lvert \lvert x - y \rvert \rvert^{\,\alpha}\, \forall \,x, y \,\in$ the domain of $f$.} with exponent $\alpha > 1/2$. By a deep general result \cite{henkel2012conformal}, any random process $\zeta\, (t)$ with continuous samples and independent identically distributed increments must necessarily be of the form $\zeta\, (t) = \sigma\, B_t + \rho\,t$ for some $\sigma >0$ and $\rho \, \in\,\mathbb{R}$. In his seminal work, Oded \citet{schramm2000scaling} connected the dots to show that conformal invariance and the domain Markov property together conspire to impose this very form on $U_t$, which is thereby restricted to standard Brownian motions with drift. In addition, invariance under reflections about the imaginary axis requires $\rho = 0$ with the outcome 
\begin{alignat}{1}
U_t = \sqrt{\kappa}\,B_t; \mbox{ with }\,\langle B_t \rangle = 0, \,\langle B_t\, B_s \rangle = \min\, (t, s),
\end{alignat}
where $\kappa$ is the diffusion constant. To be precise, $U_t$ is distributed as a Gaussian
\begin{equation}
P\,(U, t) = \frac{1}{\sqrt{2\pi\, t}}\exp \left(-\frac{U^2}{2\, \kappa\, t}\right)
\end{equation}
with correlations $\langle\, \left(U_t - U_s\right)^2\, \rangle = \kappa\, \lvert t - s \rvert$ \cite{fogedby2012stochastic}. We \textsl{define} (chordal) $\mathrm{SLE}_{\,\kappa}$ (from $0$ to $\infty$ in $\mathbb{H}$) as the random collection of conformal maps $g_t$ obtained by solving Eq.~\eqref{eq:LDE} with $U_t =\sqrt{\kappa}\,B_t$; the curve itself is called the $\mathrm{SLE}_{\,\kappa}$ trace. On another simply connected domain $\mathbb{D}$ with distinct boundary points $w, \,z$ (such as in the percolation exploration process), $\mathrm{SLE}_{\,\kappa}$ from $z$ to $w$ is directly obtained by a conformal transformation \cite{lawler2009conformal} mapping $z$ to the origin and $w$ to infinity. Likewise, one can define radial $\mathrm{SLE}_{\,\kappa}$ for paths from a boundary point to one in the interior (as for LERW)---this is the ``law'' that we had blithely anticipated earlier.

\begin{figure}[htb]
\includegraphics[width= \linewidth]{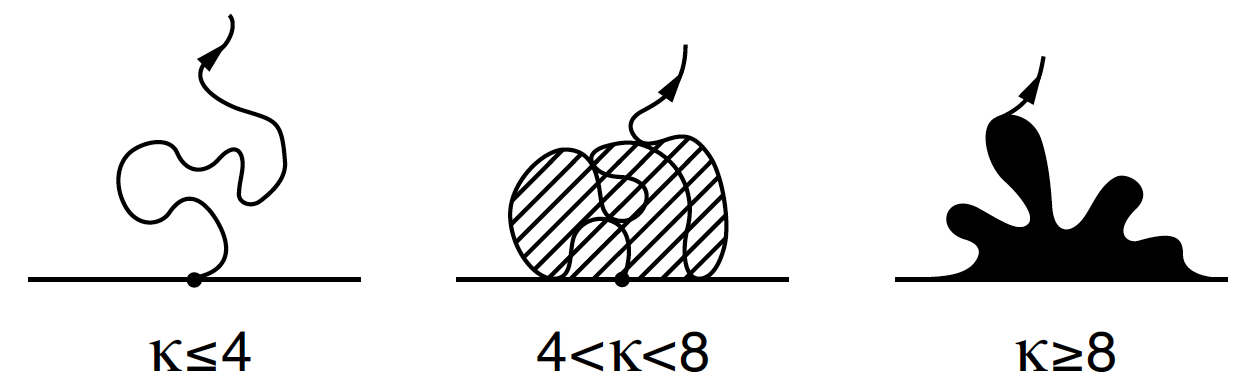}
\caption{\label{fig:SLEPhases}The phases of $\mathrm{SLE}_{\,\kappa}$. For $\kappa \le 4$, the trace is a simple non-intersecting scale invariant random curve from the origin to $\textsl{infty}$ that does not intersect the real line. When $4 <\kappa < 8$, the curve touches (but does not cross) itself and intersects part of the real line i.e $\gamma \cap \mathbb{R} \subsetneq \mathbb{R} $. The trace is plane-filling and self-osculating for $\kappa \ge 8$. From \citet{fogedby2012stochastic}. With permission of Springer Nature.}
\end{figure}

The only caveat is that $B_t$ is not H\"{o}lder-$1/2$ continuous, which insinuates that the credibility of the entire approach might be resting on tenuous grounds. Fortunately, \citet{schramm2005basic} proved that although it is not immediately obvious, and perhaps, even questionable, that a curve should always be produced for $\mathrm{SLE}_{\,\kappa}$, in practice, it is indeed the case. The value of $\kappa$ determines the universality class of critical behavior, thereby establishing a well-defined classification into different phases as shown in Fig.~\ref{fig:SLEPhases}. This innocuous parameter has yet another significance: it is related to the Hausdorff dimension of the $\mathrm{SLE}_{\,\kappa}$ trace \cite{beffara2004hausdorff, beffara2008dimension} as
\begin{equation}
D_{\,\kappa} \equiv \dim \big(\gamma\,[0,t] \big) = \min \bigg(2,\,1 + \frac{\kappa}{8}\bigg).
\end{equation}
Thus a typical LERW path, which corresponds to $\mathrm{SLE}_{2}$, has a fractal dimension of $5/4$ in the continuum limit and accordingly, the average number of steps in a LERW across a $N \times N$ grid is $N^{\,5/4}$ \cite{kenyon2000}. Other familiar lattice models that are known to have a SLE scaling limit include the self-avoiding walk [$\kappa = 8/3$ \cite{lawler2004scaling}] as well as cluster boundaries in the Ising model [$\kappa = 3$] and in percolation [$\kappa = 6$ \cite{smirnov2001critical}].

\subsubsection{SLE$_6$ description of nodal lines}

After having perhaps tested the patience of the reader who would not be unreasonable to wonder what the preceding mathematical palaver might possibly have to do with nodal lines, we are finally ready to explore the connections between the two. We know that the boundaries of percolation clusters are generated by $\mathrm{SLE}_{6}$ traces, as proved by \citet{smirnov2001critical} for critical percolation on the triangular
lattice. We have also seen that the nodal portraits of random wavefunctions are adequately captured by a critical percolation model. Putting two and two together, we conclude that the nodal lines of chaotic billiards should be described by $\mathrm{SLE}_{6}$ curves. This correspondence was examined in detail by \citet{bogomolny2006sle} with the aid of numerical calculations on a semicircular region of area $4 \pi$. For each of $N = 2248$ realizations of the random wavefunction \eqref{eq:Gaussian}, they inspected the longest nodal line stretching from the origin to the boundary and the statistical properties of its forcing function $U_t$, determined using the geodesic algorithm \cite{marshall2006convergence}. Strictly speaking, SLE predicts that
\begin{alignat}{1}
\xoverline{U}_t &= \frac{1}{N}\,\sum_{j=1}^N\,U_t (j) = 0,\\
\label{eq:VarSLE}\sigma^2(t) &= \frac{1}{N}\,\sum_{j=1}^N\,\left(U_t (j)-\xoverline{U}_t \right)^2 \approx\bigg(\kappa \pm \kappa \sqrt{\frac{2}{N}} \bigg)\, t. 
\end{alignat}
The reality, however, differs from this simple picture. Contrary to expectations, the dependence of the calculated variance $\sigma^2$ on $t$ is not linear. At best, the initial (approximately) straight segment, which constitutes only $\sim 1/4$ of the total curve, can be fit to a quadratic equation as 
\begin{equation}
\sigma^2(t) = -0.003 + 6.05\,t - 10.0\,t^2.
\end{equation}
Noteworthily, the slope of the linear term, which yields $\kappa = 6.05$ (as opposed to $6$ for pure percolation) is within the confidence interval $6 \pm 0.18$ set by Eq.~\eqref{eq:VarSLE}.

\begin{figure}
\includegraphics[width= \linewidth]{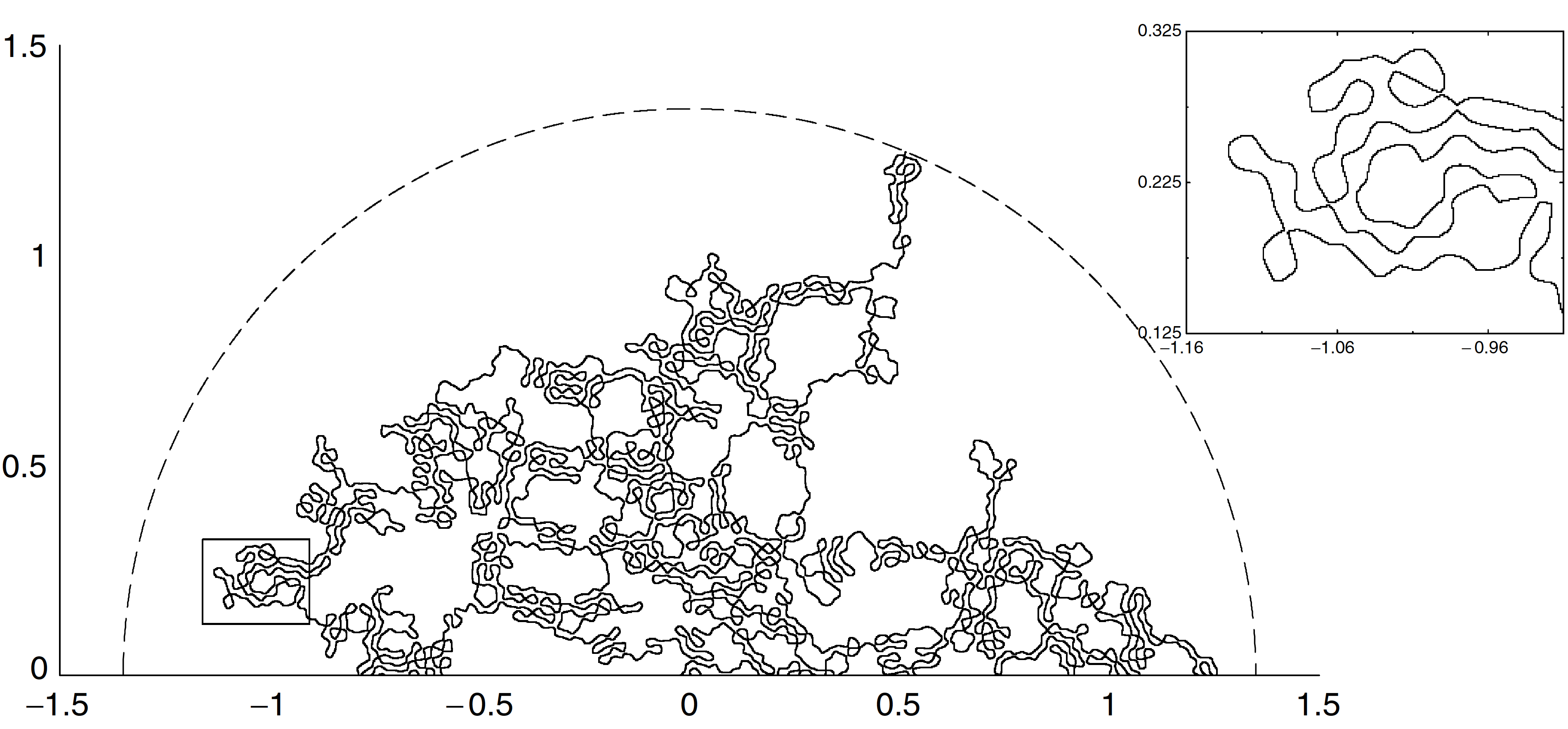}
\includegraphics[width= \linewidth]{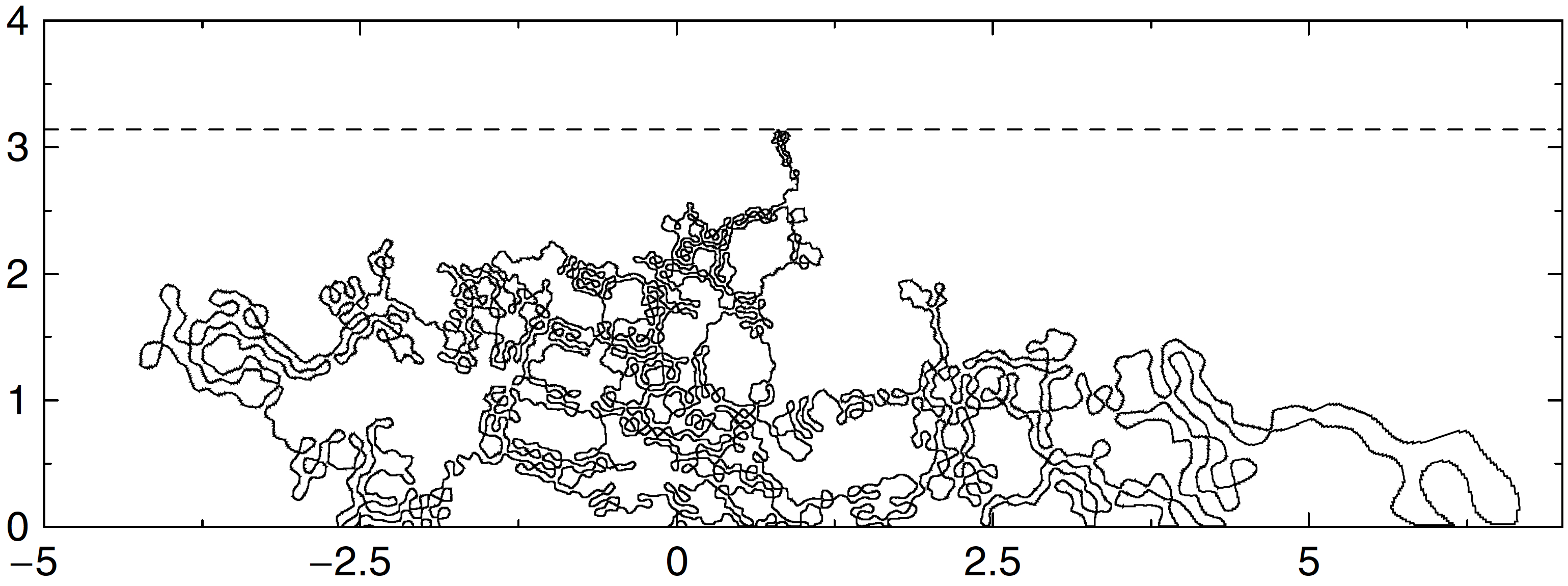}
\caption{\label{fig:dipolar}[Top]: A nodal line of a random wavefunction. [Bottom]: Image of the same nodal line under the map \eqref{eq:mapSLE}. In both figures, the dashed line indicates the absorbing boundary. From \citet{bogomolny2006sle}. \textcopyright\, IOP Publishing. Reproduced with permission. All rights reserved.}
\end{figure}

These aberrations can be ascribed to the inexorable finite-size effects concomitant with numerical computations. Whilst the parallel between percolation boundaries and SLE$_6$, technically, only holds for infinite curves, realistically, one is always restricted to curves of finite size.  \citet{bogomolny2006sle} reasoned that ideally, by dint of the locality of SLE$_6$ \cite{lsw2001-1, lawler2003conformal}, the trace should not ``feel'' the boundary until it actually encounters a point thereon.  This argument breaks down because even for infinitesimally small capacities (or time scales), there exist traces that can go arbitrarily far from the origin \cite{bogomolny2006sle}. The statistical properties of such curves are bound to be affected by the presence of boundaries, which therefore, must be taken into account.  To this end, one employs an amended version of the traditional formalism, known as dipolar SLE \cite{bauer2005dipolar}. Consider a region with two boundary arcs capped by points $z_+$ and $z_{-}$. One of the arcs is assumed to be an absorbing (Dirichlet) boundary and the other reflecting (Neumann). The origin of the random curves, which meander all the way only to be rudely stopped by the absorbing arc, is a point $z_0$ on the reflecting boundary. The guile lies in that the reflecting boundary, such as the real axis in Fig.~\ref{fig:dipolar}, now thwarts the ambitions of the more adventurous curves desirous of a quick stroll outside the domain. This setup can be conformally mapped to the strip
\begin{equation}
\mathbb{S} = \{z\,\in\,\mathbb{C},\,0< \im z < \pi \}
\end{equation}
such that the points $z_-$, $z_+$, and $z_0$ are mapped to $-\infty$, $\infty$, and $0$, respectively. For the semicircle of radius $L$, this is achieved by the map 
\begin{equation}
\label{eq:mapSLE}
F\,(z)= \log\,\left(\,\frac{(L+z)^2}{(L-z)^2}\, \right),
\end{equation}
whereas for a rectangle $[-L/2, L/2] \times [0, l]$, the same is accomplished  by
\begin{equation}
F\,(z)= \log\,\left(\frac{\wp\,(z+L/2) - \wp\, (L)}{\wp\,(L/2) - \wp\,(L)} \right),
\end{equation}
where $\wp\,(z)$ is the Weierstrass elliptic function with periods $2L$ and $2\mathrm{i}\,l$.  The dipolar $\mathrm{SLE}_{\,\kappa}$ process is effected by the Loewner-type equation:
\begin{equation}
\partial_t\,g_t(z) = \frac{2}{\tanh\,\left(g_t(z) - U_t \right)}; \quad g_0 (z) = z,
\end{equation}
whereupon subsequent calculations refine the quadratic fitting to 
\begin{equation}
\label{eq:refine}
\sigma^2(t) = 5.92\,t - 0.103\,t^2
\end{equation}
Eq.~\eqref{eq:refine}, if not a better estimate of $\kappa$, at least minimizes the coefficient of  the quadratic term compared to our previous estimate in Eq.~\eqref{eq:VarSLE}. The remaining mismatches are vestiges of discretization errors that fade away upon including more points along a trace and do not bear any new physical information. In the same vein, \citet{keating2006nodal, keating2008nodal} corroborated that the nodal lines for a perturbed quantum cat map are described by SLE$_6$. The nodal lines of the vorticity field in two-dimensional turbulence are also known to bear resemblance to SLE$_6$ curves \cite{bernard2006conformal, cardy2006turbulence}.

From a broader viewpoint, the universally accredited importance of SLE is due in no small part to its intimate and profound connections to conformal field theories \cite{bauer2002sle, bauer2003conformal, cardy2003stochastic} and the flavor it lends to general ideas such as criticality and dualities \cite{beffara2004hausdorff, duplantier2000conformally}. In all fairness, one could go on all day about SLE and its myriad applications. However, to prevent this from devolving into a full-blown review about SLE alone, we have to draw the line somewhere and this seems to be as good a place as any.

\section{Statistical measures}
\label{sec:stats}

\subsection{Nodal domain statistics}

The nodal domain statistics of quantum billiards, as promised, can be used to distinguish between quantum systems with integrable and chaotic classical dynamics; the limiting distributions of the same are believed to be universal (system-independent). This yields a criterion for quantum chaos, which is complementary to that established based on spectral statistics. In this subsection, we examine each of these statistical measures individually.

\subsubsection{Limiting distributions of nodal counts}

Bearing Courant's nodal domain theorem and Eq.~\eqref{eq:pleijel} in mind, \citet{blum2002nodal} defined the normalized number of nodal domains as
\begin{equation}
\xi _j = \frac{\nu _j}{j}, \qquad 0 < \xi _j \leq 1. 
\end{equation}
To extract the universal features thereof, it was proposed that a limiting distribution be constructed as
\begin{equation}
\label{eq:limiting}
P\,(\xi ) = \lim_{E \to \infty} P \left (\xi , I_g(E) \right)
\end{equation}
by considering the energy levels in an interval $I_g(E) = [E, E+gE]$, $g > 0$. With the number of eigenvalues in $I_g(E)$ given by the Weyl formula, the distribution of $\xi$ associated with $I_g(E)$ is 
\begin{equation}\label{eq:dist_int}
P\,(\xi , I_g(E)) = \frac{1}{N_I} \sum_{E_j \in I_g(E)} \delta \left(\xi - \frac{\nu _j}{j} \right).
\end{equation}

\paragraph{Separable, integrable billiards}

\begin{figure}[htb]
\includegraphics[width=\linewidth]{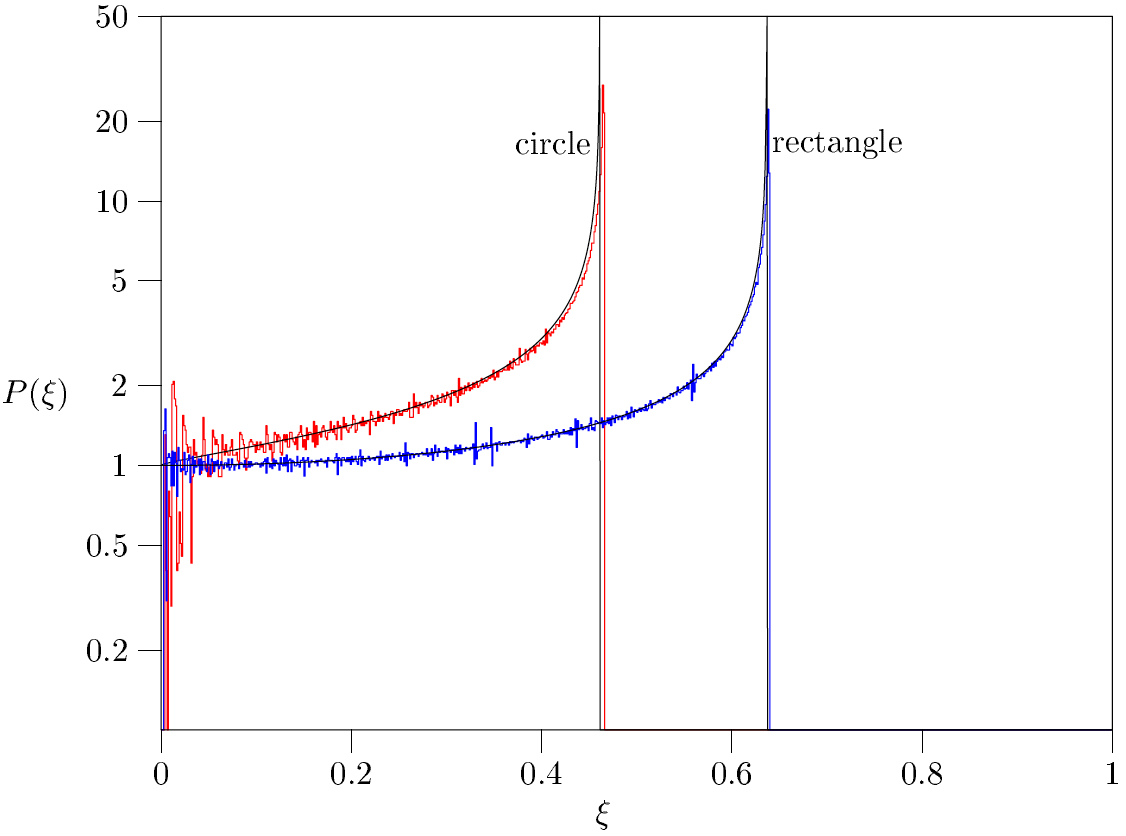}
\caption{\label{fig:ld}Numerically calculated limiting distributions of the nodal domain number for rectangular (blue) and circular (red) billiards in the spectral intervals $62500 \le j \le 125000$ and $30000 \le j \le60000$, respectively. For the rectangle, the limiting distribution (smooth line) coincides with the analytical result  of Eq.~\eqref{eq:ld_rect}. The singularities seen in the numerical data mark the contributions of periodic orbits. From \citet{blum2002nodal}.}
\end{figure}

For integrable (separable) systems, the Hamiltonian can be expressed in terms of action-angle variables, $H \,(I_1, I_2)$---this is a homogeneous function of degree two. The Einstein-Brillouin-Keller quantization \cite{bleher1994distribution} of such systems gives the energy levels $E_{n_1, n_2} = H(n_1+\alpha _1, n_2+\alpha _2) + {\cal O}(\sqrt{E})$, $n_1, n_2 \in \mathbb{Z}$; $\alpha _1, \alpha _2$ being Maslov indices. Separability of the system implies that the number of domains of the eigenfunction with quantum numbers, $n_1, n_2$ is $\nu _{n_1, n_2} = n_1n_2 + {\mathcal O}(1)$. The nodal domain number is thus $\xi _{n_1, n_2} = \nu _{n_1, n_2}/N(E_{n_1, n_2})$, where $N(E) = \mathcal{A}\,E\,(1 + {\mathcal O}(E^{-1/2}))$. Eq. \eqref{eq:dist_int} can be rewritten as
\begin{alignat}{1}
\label{eq:pxi1}
P\,(\xi , I_g(E)) &= \frac{1}{g\,E\,\mathcal{A}} \int_{H \in I_g} \mathrm{d}I_1\, \mathrm{d}I_2\, \delta \left( \xi - \frac{I_1I_2}{\mathcal{A}\,H(I_1, I_2)} \right) \nonumber \\ 
&+ {\mathcal O}(E^{-1/2}).
\end{alignat}
The homogeneity of $H$ leads to an expression for $P$ that is independent of $g$ and $E$. Employing a change of variables $(I_1, I_2) \to ({\cal E}, s)$, where ${\cal E} = H(I_1, I_2)$ and $s$ is the constant-energy curve, Eq.~\eqref{eq:pxi1} becomes an integral
\begin{equation}\label{eq:lt}
P\,(\xi ) = \frac{1}{\mathcal{A}} \int_{\Gamma} \mathrm{d}\,s\,\,\delta \left( \xi - \frac{I_1(s)I_2(s)}{\mathcal{A}} \right). 
\end{equation}
over the line $\Gamma$ defined by $H\,(I_1, I_2) = 1$. The limiting distributions for rectangular and circular billiards, calculated numerically according to Eq.~\eqref{eq:lt}, are seen in Fig.~\ref{fig:ld}.  

For the rectangular billiard, we elucidate the calculations in more detail now, following \citet{smilansky2005nodal}. Let the lengths of the sides of the rectangle be $L_x$ (set equal to $\pi$) and $L_y = L_x/\alpha$ where $0 < \alpha < 1$. The energy spectrum is proportional to $n^2 + \alpha\, ^2m^2$. Weyl's law gives us the cumulative density of energy levels:
\begin{equation}
N(E) \simeq \frac{\pi E}{4\alpha} \left( 1 - \frac{2}{\pi} \frac{1+\alpha}{E} \right).
\end{equation}
The simple form of the energy level sequence admits easy parametrization in terms of the continuous variables
\begin{equation}
n(E, \theta ) = \sqrt{E} \cos \theta, \qquad m(E, \theta ) = \sqrt{E} \sin \theta /\alpha,
\end{equation}
whereupon the normalized nodal domain number for the $j^\mathrm{th}$ state is
\begin{equation}
\xi _j = \frac{2}{\pi} \sin\, 2\theta\,\left[ 1 - \frac{2}{\pi} \frac{1 + \alpha }{\sqrt{E}} \right]^{-1}.
\end{equation}
The distribution \eqref{eq:dist_int} can be expressed in terms of $E, \theta$:
\begin{equation}
P(\xi, I) \simeq \frac{1}{2\alpha N_I}\int_{E_0}^{E_1}\hspace*{-0.2cm} \int_{0}^{\pi /2} \delta \left[\xi - \xi _j(E, \theta)\right]\, \mathrm{d}\,E\,\, \mathrm{d}\,\theta .
\end{equation}
$N_I$ gives the number of energy levels between $E_0$ and $E_1$. We first perform the $\theta$-integral by taking advantage of the Dirac delta function; with some convenient definitions, this can be easily achieved so that 
\begin{equation}
\label{eq:pxi_before}
P\,(\xi, I) \simeq \frac{E_0}{\alpha\, N_I}\int_{1}^{\ell} x \left[\frac{2\,\cos 2\theta _0}{\pi\, (1 - \varepsilon /x)}\right]^{-1} \mathrm{d}\,x,
\end{equation}
where $x = \sqrt{E/E_0}$, $\varepsilon\, (\alpha) = 2\,(1+\alpha )/(\pi \sqrt{E_0})$, $\sin 2\theta _0 = (\pi\, \xi/2) (1 - \varepsilon /x)$, and 
\begin{alignat}{1}
\ell = 
\begin{cases}
\mathcal{G}  = \sqrt{E_1/E_0},&\mbox{if~} \xi < 2/\pi ,\nonumber \\
\mbox{min}\left[\mathcal{G},\, {\displaystyle \frac{\varepsilon\, \pi\, \xi}{2}} \left( {\displaystyle\frac{\pi\, \xi}{2}} - 1 \right)^{-1}\right], &\mbox{if}~ {\displaystyle\frac{2}{\pi}} < \xi \leq {\displaystyle \frac{2}{\pi (1-\varepsilon )}}.
\end{cases}
\end{alignat}
$P\,(\xi , I)$ is zero for $\xi > 2/ [\pi\, (1- \varepsilon )]$. In this manner, the aspect ratio $\alpha$ of a rectangular domain (with Dirichlet boundary conditions) can be retrieved by counting its nodal domains. The limiting distribution is 
\begin{alignat}{1}
\label{eq:ld_rect}
P\,(\xi ) = 
\begin{cases}
{\displaystyle \frac{1}{\sqrt{1 - (\pi\, \xi /2)^2}}}, &\mbox{for } \xi < 2/\pi ,\\
 0, &\mbox{for } \xi > \xi_{\mathrm{max}} = 2/\pi .  
\end{cases}
\end{alignat}
In Fig.~\ref{fig:ld}, the peak around $\xi \approx 0.64$ exactly coincides with the value of $2/\pi$ in Eq.~\eqref{eq:ld_rect}. 

\paragraph{Surfaces of revolution}

The Schr\"{o}dinger equation can be solved for certain surfaces of revolution in $\mathbb{R}^3$. \citet{karageorge2008counting} presented the first results for ellipsoids with different eccentricities and proved that ``the nodal sequence of a  mirror-symmetric surface is sufficient to uniquely determine its shape (modulo
scaling).'' The form of the limiting distribution is nearly identical to that for separable billiards (Fig.~\ref{fig:ld}) but, of course, with differing values of $\xi_{\mathrm{max}}$ (that actually decrease with the eccentricity $\varepsilon$); specifically, for the sphere, $\varepsilon = 1$ and $\xi_{\mathrm{max}} = 0.5$. In general, for separable systems in $d$ dimensions,
\begin{equation}
P\,(\xi) \approx \left( 1 - \xi/\xi_\mathrm{max} \right)^{(d-3)/2}.
\end{equation}

\paragraph{Nonseparable, integrable billiards}
The two chief billiards of interest in this category are the right-angled isosceles triangle and the equilateral triangle, the nodal domain distributions for both of which have been studied numerically.

\begin{figure}[htb]
\includegraphics[width=\linewidth]{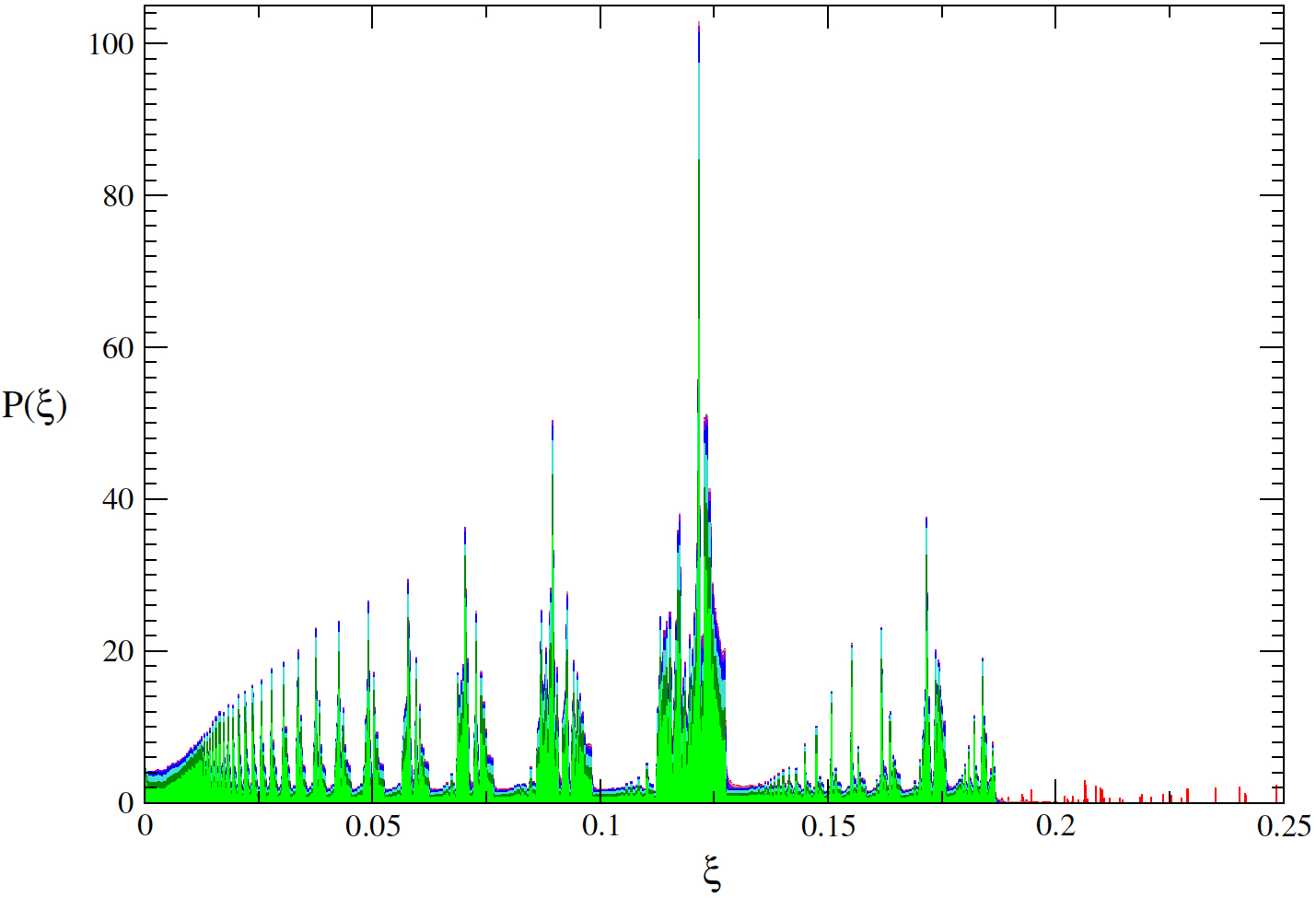}
\caption{\label{fig:pxi} The nodal count distribution for the right-isosceles triangle, considering the energy interval $[9000^2, 2 \times 9000^2]$. The colors represents the proportions of wavefunctions with specific tiling behaviors; in particular, light green denotes no tiling. The structure of the distribution remains invariant on inclusion of the data from tiling eigenfunctions. From \citet{aronovitch2012nodal}. \textcopyright\, IOP Publishing. Reproduced with permission. All rights reserved.}
\end{figure}

First, let us consider the right-isosceles triangle. For simplicity, let the length of the equal legs be $\pi$; the area is just $\pi ^2/2$. The unnormalized solutions of the Schr\"{o}dinger problem in the interior of the triangle, with Dirichlet conditions on the boundary, have the form
\begin{alignat}{1}
\label{eq:wfun_iso}
\psi _{mn}(x, y) = \sin (m\,x) \sin (n\,y) - \sin (n\,x) \sin (m\,y),
\end{alignat}
and the spectrum of eigenvalues is given by 
\begin{equation}
E_{m,n} = m^2 + n^2, \qquad m, n \in {\mathbb{N}}, ~m > n.
\end{equation}
When $\gcd\,(m, n) = d >1$ for $(m > n)$, the nodal set is composed of $d^2$ identical nodal patterns (``tiles''), each contained within a subtriangle. The nodal domains of the billiard were counted using the Euler formula for graphs by \citet{aronovitch2012nodal} (refer to Sec.~\ref{sec:graph} for details). Fig.~\ref{fig:pxi}  shows the corresponding distribution of the nodal domain counts. One observes a large number of peaks, each seemingly converging to a Dirac delta function.

\begin{figure}[htb]
\subfigure[]{\includegraphics[width=0.32\linewidth]{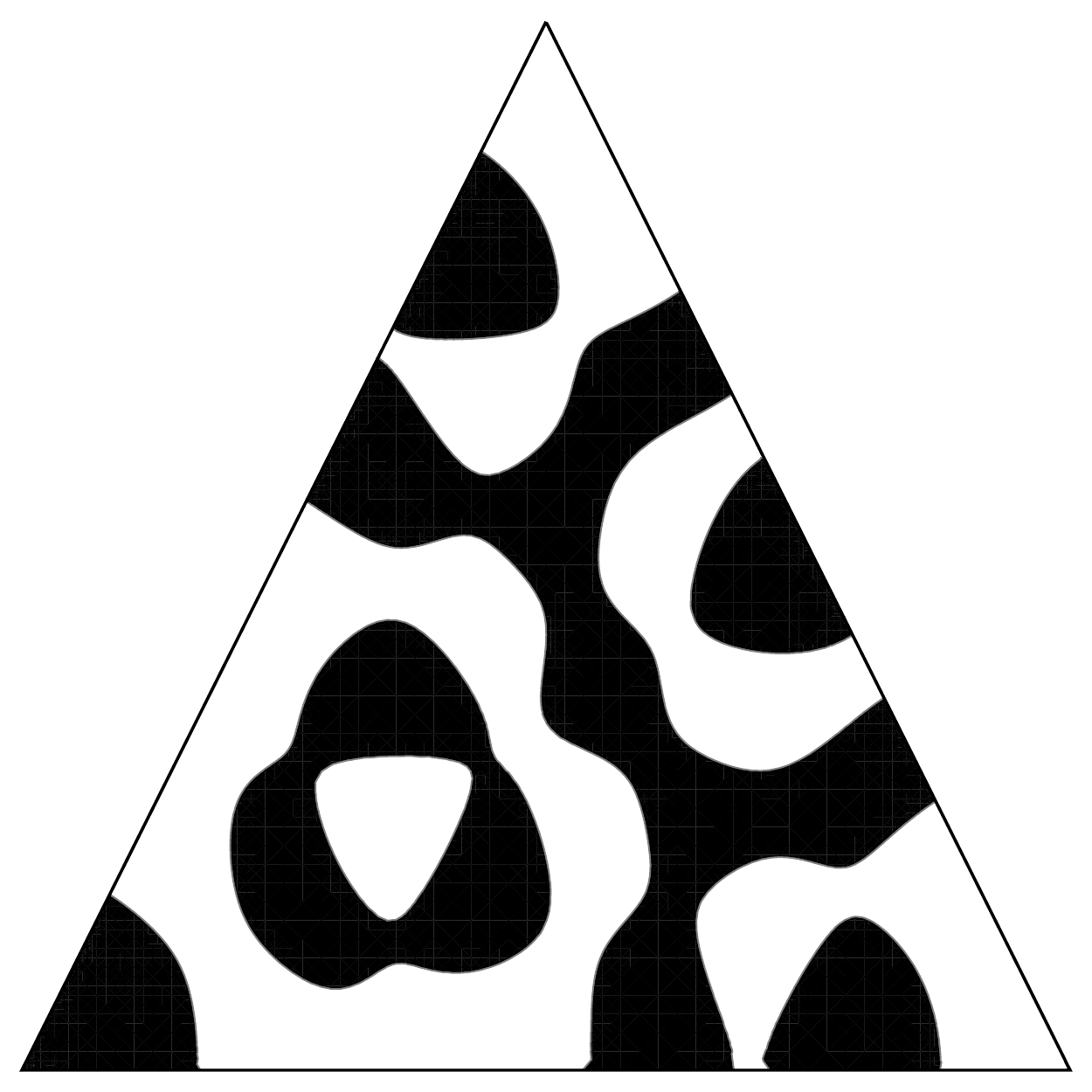}}
\subfigure[]{\includegraphics[width=0.32\linewidth]{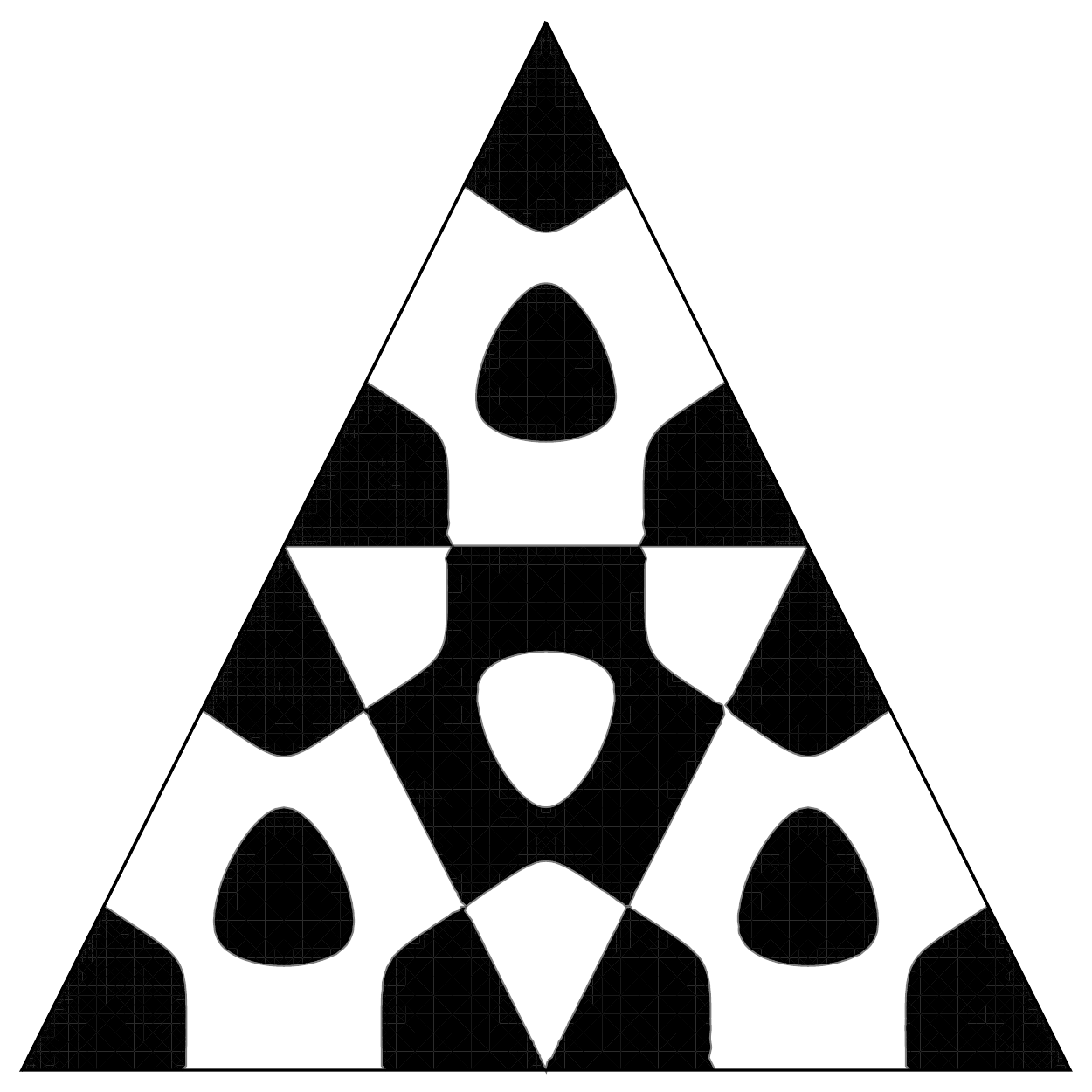}}
\subfigure[]{\includegraphics[width=0.32\linewidth]{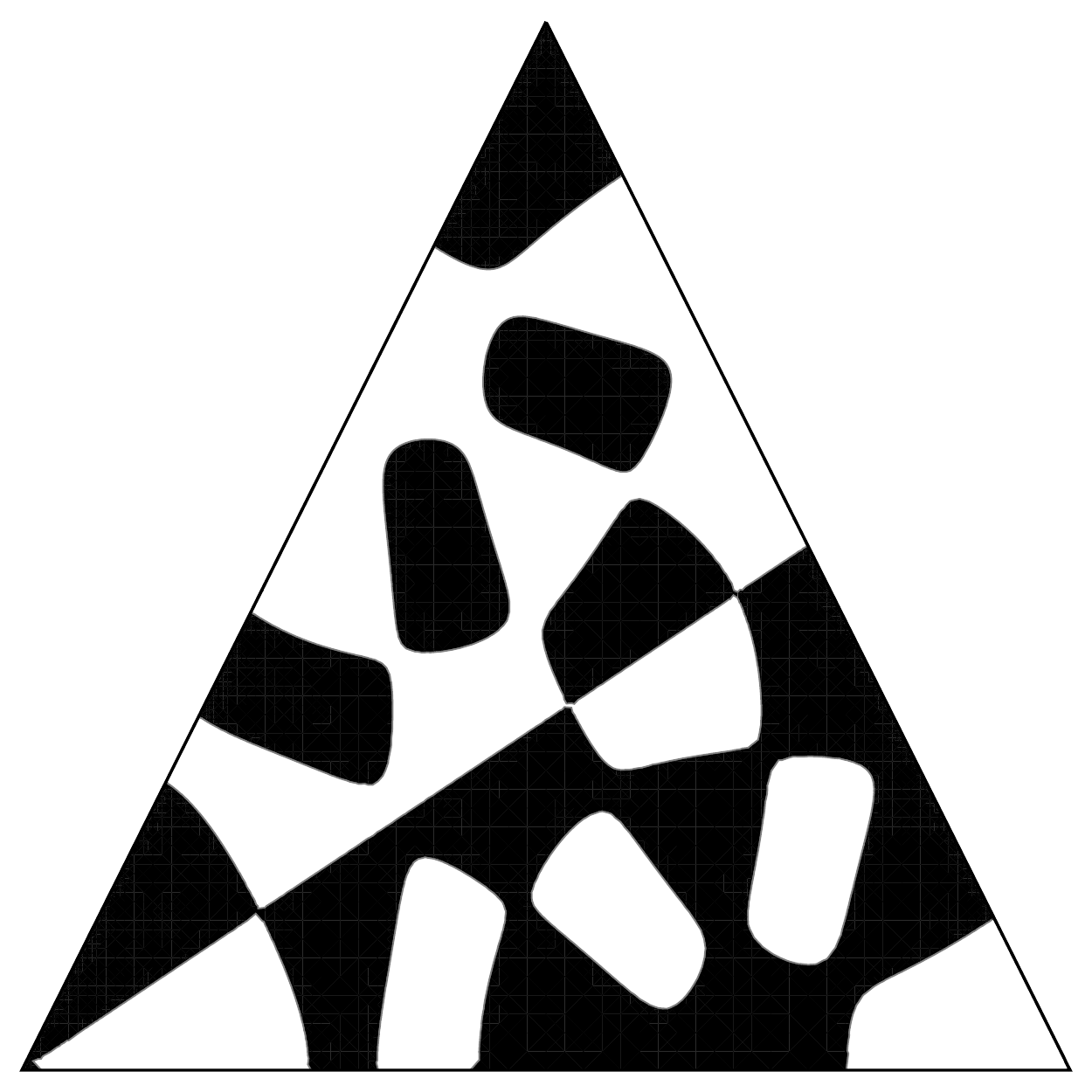}}
\caption{\label{fig:eq} (a) Nodal domains for the cosine combination of the eigenfunctions of the equilateral triangle for $(m, n) = (10, 3)$. (b) The tiling pattern of the domains (with 4 tiles) is seen for $(m, n) = (10, 2)$. (c) The antisymmetric (sine) combination  for $(m, n) = (10, 3)$. The eigenfunctions are positive (negative) in the white (black) regions.}
\end{figure}

\begin{figure}[htb]
\begin{overpic}[width= \linewidth]{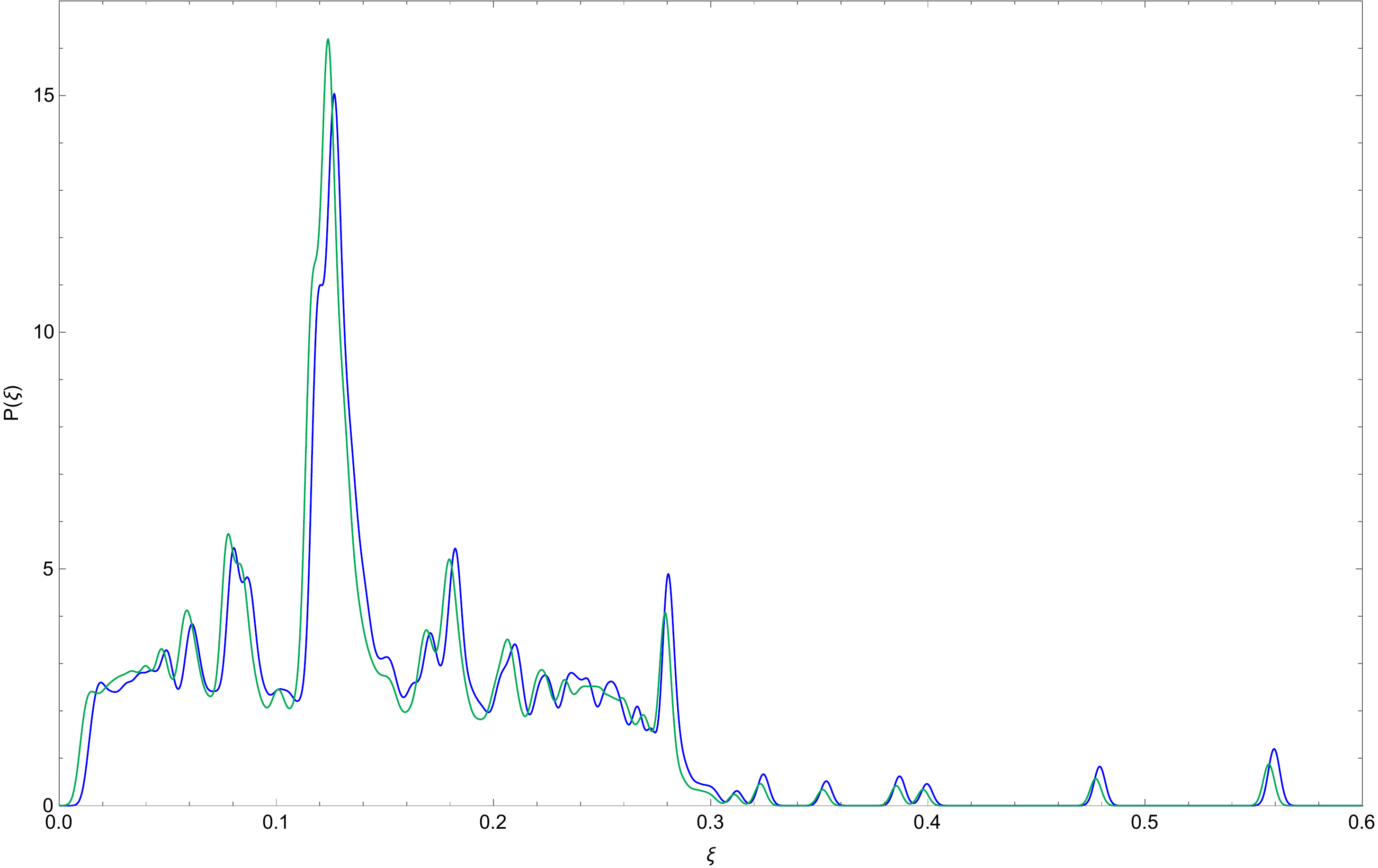}
     \put(36, 22.5){\includegraphics[width= 0.6\linewidth]{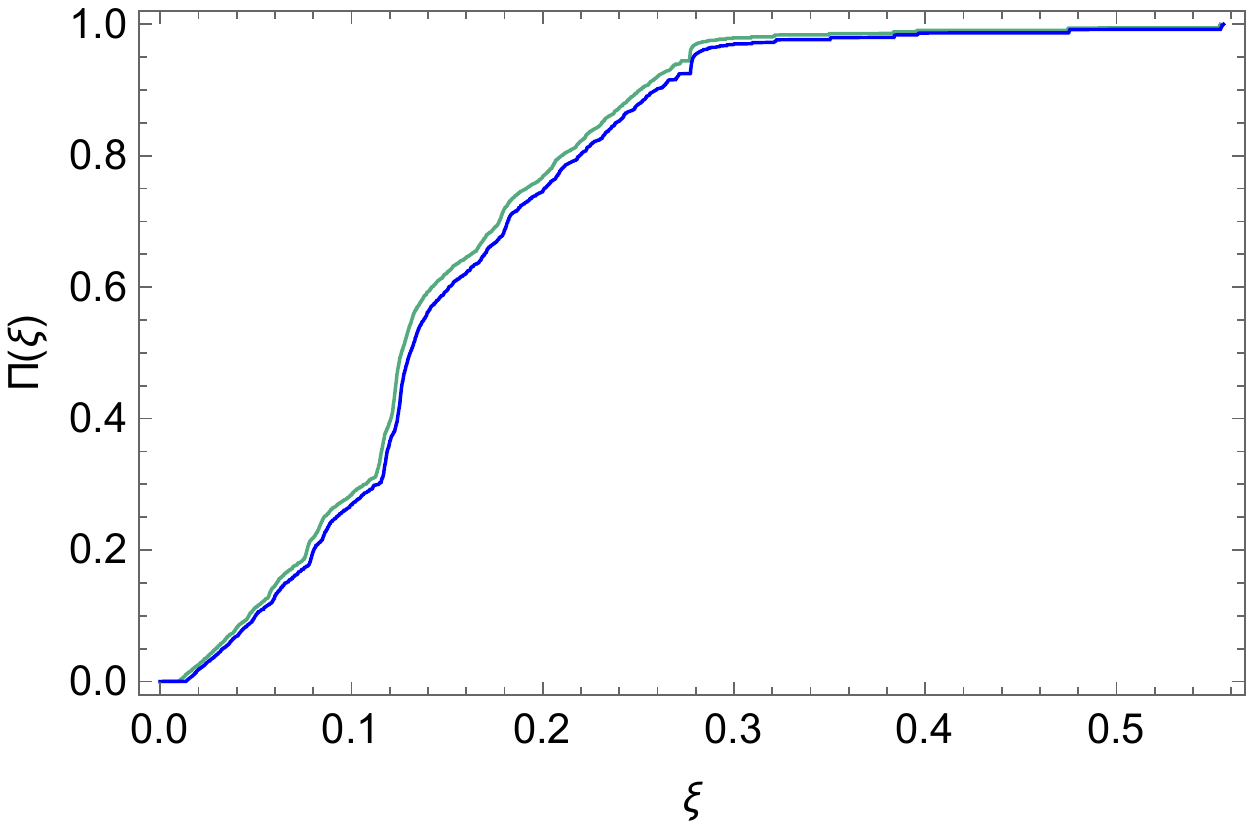}}
\end{overpic}
\caption{\label{fig:pxi_eq} The probability distribution and the integrated distribution (inset) of the nodal domain number for the eigenfunctions of an equilateral triangle billiard corresponding to two spectral intervals, [10000, 20000] (blue) and [20000, 40000] (green) containing 3014 and 6028 eigenfunctions, respectively.}
\end{figure}

The analysis for the equilateral triangle proceeds analogously. An equilateral-triangular domain of side $L = \pi$ and area $\mathcal{A} = \sqrt{3}\,\pi^2/4$ is  
\begin{alignat}{1}
{\cal D} &= \bigg\{(x,y) \in \bigg[0,\frac{\pi}{2}\bigg] \times \bigg[0,\frac{\sqrt{3}\pi}{2}\bigg] : y\le\sqrt{3}x\bigg\} 
\\
\nonumber &\cup \bigg\{(x,y) \in \bigg[\frac{\pi}{2},\pi \bigg] \times \bigg[0,\frac{\sqrt{3}\pi}{2}\bigg] : y\le\sqrt{3}(\pi-x)\bigg\}. 
\end{alignat}
The Dirichlet eigenfunctions, which form a complete orthogonal basis, are \cite{brack2003semiclassical} 
\begin{alignat}{1}
\label{eq:wf}
\psi_{m,n}^{c,s} (x,y) =  \sin{\bigg(n\frac{2\pi}{\sqrt{3}L}y\bigg)} &(\cos,\sin)\bigg[ (2m-n)\frac{2\pi}{3L}x \bigg]
\nonumber \\
-\sin{\bigg(m\frac{2\pi}{\sqrt{3}L}y\bigg)}  &   (\cos,\sin)\bigg[(2n-m)\frac{2\pi}{3L}x\bigg]
\nonumber \\
+ \sin{\bigg[(m-n)\frac{2\pi}{\sqrt{3}L}y\bigg]}&(\cos,\sin)\bigg[-(m+n)\frac{2\pi}{3L}x\bigg], 
\end{alignat}
where $m$ and $n$ are integer quantum numbers with the restriction $m\geq2n$ and $m,n>0$. The eigenfunctions $\psi_{m,n}^c$ and $\psi_{m,n}^s$ correspond to the symmetric and antisymmetric modes respectively \cite{mccartin2003eigenstructure}. The eigenenergies of the Hamiltonian for the system is
\begin{equation}
E_{m,n} = \frac{16}{9}\frac{\pi^2\hbar^2}{2\,\mathfrak{m}\, L^2}(m^2+n^2-mn)
\end{equation}
for a particle of mass $\mathfrak{m}$. This spectrum possesses interesting and deep number-theoretic properties as shown by \citet{itzykson1986arithmetical}. The nodal patterns of the eigenfunctions of the equilateral triangle also exhibit certain symmetry relations, including, once again, a tiling structure of the nodal lines: for $m \ge 2n$ and $\gcd\, (m,n) = d > 1$, $\psi_{m,n}$ is tiled by $\psi_{m',n'}$ with $m' = m/d$ and $n' = n/d$. This arrangement, which follows directly from Eq.~\eqref{eq:wf}, is illustrated by the examples in Fig.~\ref{fig:eq}. Counting the nodal domains for this billiard with the Hoshen-Kopelman algorithm \cite{hoshen1976percolation}, one arrives at the distribution $P \,(\xi)$, displayed for two spectral intervals in Fig.~\ref{fig:pxi_eq}.

\subsubsection{Distribution of boundary intersections}

For bounded domains, one can also study the statistics of the number of nodal intersections with the boundary of the billiard, $\tilde{\nu}_j$---this is exactly the number of times the normal derivative at the boundary vanishes. The appropriate normalized parameter is now $\eta_{j} = \tilde{\nu}_{j}/\sqrt{j}$. Even with Neumann boundary conditions, $\tilde{\nu}_j \sim \mathcal{O}\,(\lambda_j)$ \cite{toth2009}, so $\eta_{j}$ is correctly normalized either way. The distribution of $\eta$ for the interval $I_{g}(E)$, which is a characteristic of the system, is
\begin{equation}
P\,[\eta, I_{g}(E)] = \frac{1}{N_{I}} \sum_{E_{j} \in I_{g}(E)} \delta (\eta -
\eta_{j}).
\label{eq: eta}
\end{equation}
Hence, the limiting distribution for the system is defined, in exact correspondence to Eq.~\eqref{eq:limiting}, as
\begin{equation}
P\,(\eta) = \lim_{E \rightarrow \infty} P\,[\eta, I_{g}(E)].
\end{equation}

\begin{figure}[htb]
\includegraphics[width=\linewidth]{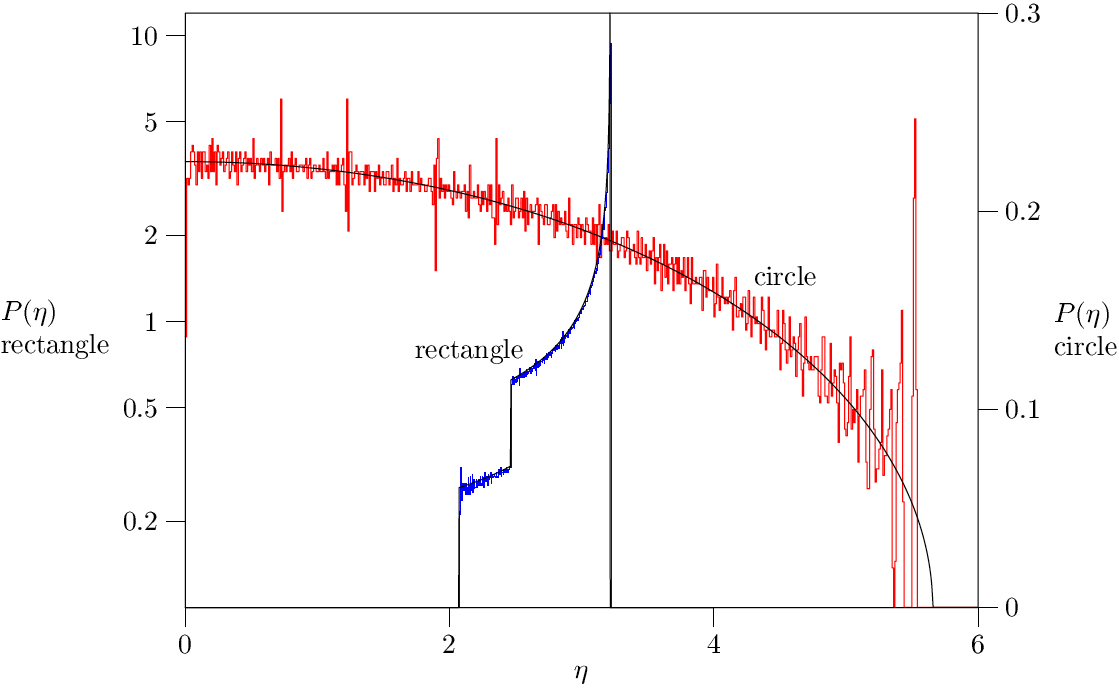}
\caption{\label{fig:etad}The distribution of the normalized numbers of boundary intersections for the rectangular and circular billiards, in the same spectral intervals as in Fig.~\ref{fig:ld}. The smooth lines indicate the limiting distributions. From \citet{blum2002nodal}.}
\end{figure}

\paragraph{Separable, integrable billiards}
In terms of the action variables $I_{1,2}$, the analogue of Eq.~\eqref{eq:ld_rect} for the boundary intersections was calculated by \citet{blum2002nodal} to be
\begin{alignat*}{1}
\label{eq:ld_boundary}
P\,(\eta ) = 
\begin{cases}
{\displaystyle \frac{1}{4 \sqrt{\mathcal{A}}} \left(I_2 (I_1) - I_2' (I_1)\, I_1 \right) \bigg \vert_ {I_1 =\frac{ \eta \sqrt{\mathcal{A}}}{2}} }, &\eta <  \eta_m ,\\
 0, &\eta \ge \eta_m, 
\end{cases}
\end{alignat*}
where the maximum value of $\eta$, $\eta_m = 2\,I_{1, m}/\sqrt{\mathcal{A}}$, is determined by the intersection point, $I_{1, m}$, of the $I_1$-axis with the contour line $\Gamma$. The numerics (Fig.~\ref{fig:etad}) are consistent with this prediction.

\begin{figure} [htb]
\includegraphics[width=\linewidth]{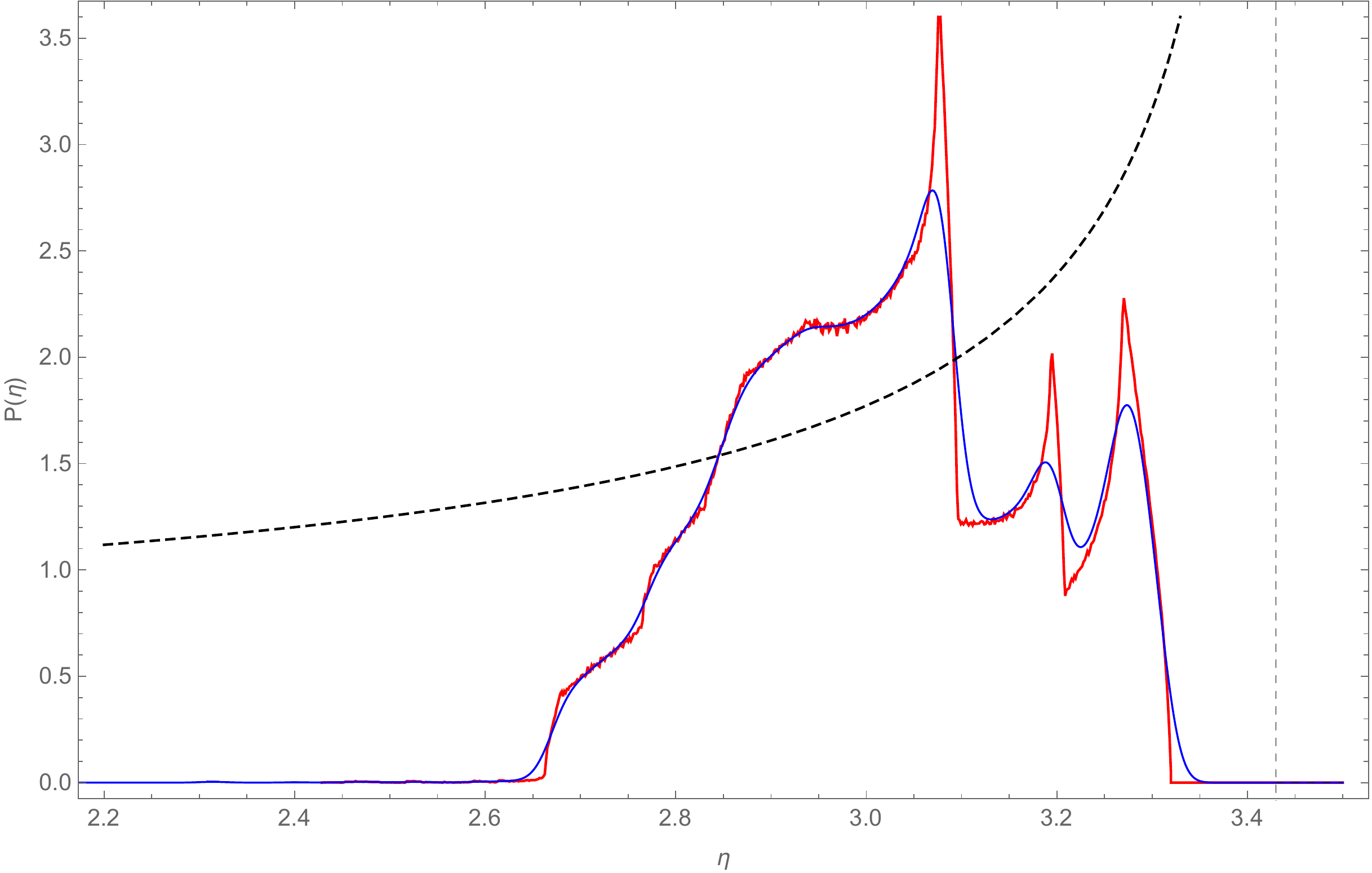}
\caption{\label {fig:PEta}The distribution $P\,(\eta)$ of the normalised number of boundary intersections for the equilateral triangle, evaluated over 882455 wavefunctions in the spectral interval $[2000^{2}, 2 \times 2000^{2}]$. The red curve shows the exact distribution as a function of $\eta$ whereas the blue curve depicts the smoothened histogram representation of the same. The dotted line marks the prediction of Eq.~\eqref{eq:PEta_analytic}. }
\end{figure}

\paragraph{Nonseparable, integrable billiards} 
We revert to our old example of the equilateral triangle as a prototype for this class of systems. As seen in Fig. \ref{fig:PEta}, $P\,(\eta)$ consists of multiple dominant peaks of different strengths at certain characteristic values of $\eta $. \citet{samajdar2014JPA} obtained an analytical form for this distribution for non-tiling wavefunctions. The number of eigenfunctions in a generic spectral interval $I = [E_0,E_1]$ is set by Weyl's law:
\begin{equation}
N_I = \displaystyle{\frac{\sqrt{3}\pi}{16}} \bigg[ \big(E_1-E_0\big) - \displaystyle{\frac{4\sqrt{3}}{\pi}}\big(\sqrt{E_1}-\sqrt{E_0}\big)\bigg].
\end{equation}
Introducing the variable $\varepsilon = 4\,\sqrt{3}/(\pi\sqrt{E_0})$ and retaining $x, \mathcal{G}$ from Eq.~\eqref{eq:pxi_before}, the distribution can be expressed as
\begin{alignat}{2}
\label{eq:PEta_analytic}
P(\eta, I) 
&= \displaystyle{\frac{E_0\sqrt{\pi\sqrt{3}}}{4\,N_I}} \int\limits_{1}^{\ell} \displaystyle{\frac{x \, (1-\varepsilon/x)^{1/2}}{\sqrt{1-\{f(x)\}^2}}}\, \mathrm {d}\, x, &&\mbox{ where}\\ 
f(x) &= \displaystyle{\frac{16+\sqrt{\pi E_0\sqrt{3}}\:\eta\:(x^2-\varepsilon \,x)^{1/2}}{8\,x\,\sqrt{E_0}}}, &&\mbox{ and}
\end{alignat}
\begin{equation}
\ell = \begin{cases}
\mathcal{G}, & \hspace{0.5cm}\mbox{ if }\, \eta < \varphi_1, \\ 
\min \, \big[\mathcal{G}, X_{\mathrm{max}}\big], & \hspace{0.5cm} \mbox{ if }\, \eta > \varphi_1,
\end{cases}
\mbox{ with }
\end{equation}
\begin{alignat*}{1}
\varphi_1 = \begin{cases}
\displaystyle{\frac{8}{3}}\sqrt{6-\sqrt{3}\,\pi}, \quad &\mbox{if } \,0<\sqrt{E_0} \le \displaystyle{\frac{2\sqrt{3}}{\pi-\sqrt{3}}},\\
\displaystyle{\frac{8}{3^{1/4}}}\displaystyle{\frac{\sqrt{E_0}-2}{\sqrt{ \pi\, E_0-4\,\sqrt {3\,E_0}}}}, &\mbox{otherwise}.
\end{cases}
\end{alignat*}
Furthermore, $P(\eta, I) = 0$ for all $\eta > \varphi_2$ specified by
\begin{alignat}{1}
\varphi_2 = 
\begin{cases}
\nonumber\displaystyle{\frac{8}{3^{1/4}}}\displaystyle{\frac{\sqrt{E_0}-2}{\sqrt{ \pi E_0-4\sqrt {3E_0}}}}, \hspace*{-0.25cm}&\mbox{if }\, \displaystyle{\frac{4\sqrt{3}}{\pi}}<\sqrt{E_0} \le \displaystyle{\frac{\pi}{\pi-\sqrt{3}}},
\\
\displaystyle{\frac{8}{3^{1/4}\sqrt{\pi}}}, &\mbox{ if }\, \sqrt{E_0} >{\displaystyle \frac{\pi}{\pi-\sqrt{3}}}.
\end{cases}
\end{alignat}
However, there does not exist any such upper bound when $\sqrt{E_0} \le 4\sqrt{3}/\pi$. $X_{\mathrm{max}}$, appearing in the definition of $\ell$ above, is the maximum permissible value of $x$ as regulated by the inequality
\begin{equation*} 
0<\varphi_1 \le \displaystyle{\frac{8\,x \,\sqrt{E_0} - 16}{\sqrt{\pi\, E_0\, \sqrt 3\, (x^2-\varepsilon\, x) }}} \le \varphi_2.
\end{equation*}
Evaluating $X_{max}$ for the most general case (considering sufficiently excited states such that $E_0>[2\sqrt 3/(\pi-\sqrt{3})]^2$), the integral for $P(\eta,I)$ can be easily computed numerically for any $E_0$. The theoretical estimate of $\varphi_2 = 3.43$ for $E_0 = 2000^2$ is in close agreement with the numerical result (Fig. \ref{fig:PEta}), which suggests that the distribution $P(\eta ) $ is zero beyond, approximately, 3.35.

\paragraph{Chaotic billiards}

\begin{figure}[htb]
\includegraphics[width=\linewidth]{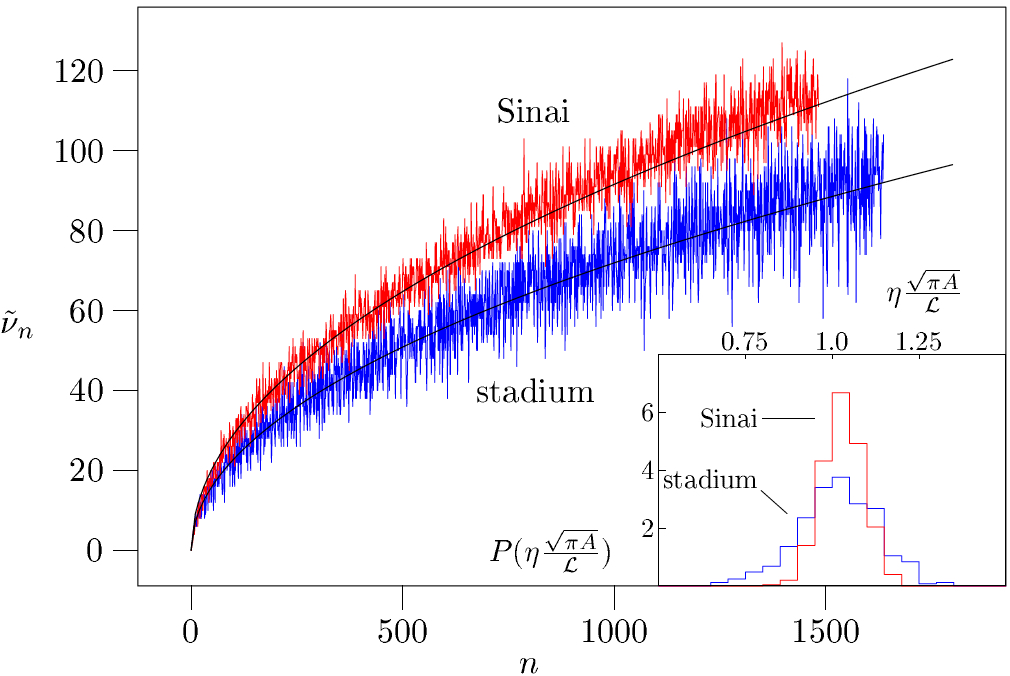}
\caption{\label{fig:pxi_ch} The number of boundary intersections $\tilde{\nu}_n$ for chaotic systems follows Eq.~\eqref{eq:boun_int_ch}, where the numerical results are based on 1637 (1483) eigenfunctions of the stadium (Sinai) billiard. [Inset]: The distribution $P\,(\eta \sqrt{\pi\, \mathcal{A}}/\mathcal{P})$. From \citet{blum2002nodal}.}
\end{figure}

For the chaotic Sinai and stadium billiards, the nodal domain statistics are usually analyzed in the random wave approximation. Given a wavefunction of the type \eqref{eq:Gaussian}, the number of zeros of its normal derivative $u \,(\theta) = \partial_r \psi (r, \theta) \vert_{r=R}$, on a circle of radius $R$, is \cite{kac1959probability}
\begin{equation*}
\tilde{\nu}_u = \int_0^{2\pi}  \hspace*{-0.1cm} \mathrm{d}\,\theta \int_{-\infty}^{\infty} \int_{-\infty}^{\infty} \frac{\mathrm{d}\,\xi\,\,\mathrm{d} \,\eta}{2\, (\pi\,\eta)^2} \,\mathrm{e}^{\mathrm{i}\,\xi\,u (\theta)} \left(1- \mathrm{e}^{\mathrm{i}\,\eta\, \dot{u}\, (\theta)} \right),
\end{equation*}
where $\dot{u} \,(\theta) = \mathrm{d}_\theta u \,(\theta)$. The mean and variance of $\tilde{\nu}_j$ are thus \cite{blum2002nodal}
\begin{alignat}{1}\label{eq:boun_int_ch}
\langle \tilde{\nu } \rangle &= k\,R \,\sqrt{1 + \left( \frac{2}{k\,R} \right)^2} \approx k \,L\approx \frac{k\,\mathcal{P}}{2\,\pi}, \\
\mathrm{Var}\, (\tilde{\nu}) &\approx 0.0769\, k\,\mathcal{P}.
\end{alignat}
This implies that the scaled number of nodal intersections tends to a Dirac delta function, centered at $\eta \sqrt{\pi\, \mathcal{A}}/\mathcal{P} = 1$ (seen in the inset of Fig.~\ref{fig:pxi_ch}).

\subsubsection{Geometric characterization of nodal domains}

\paragraph{Area-to-perimeter ratio}

\begin{figure} [htb]
\includegraphics[width=\linewidth, trim={-0.45cm 0 0 0},clip]{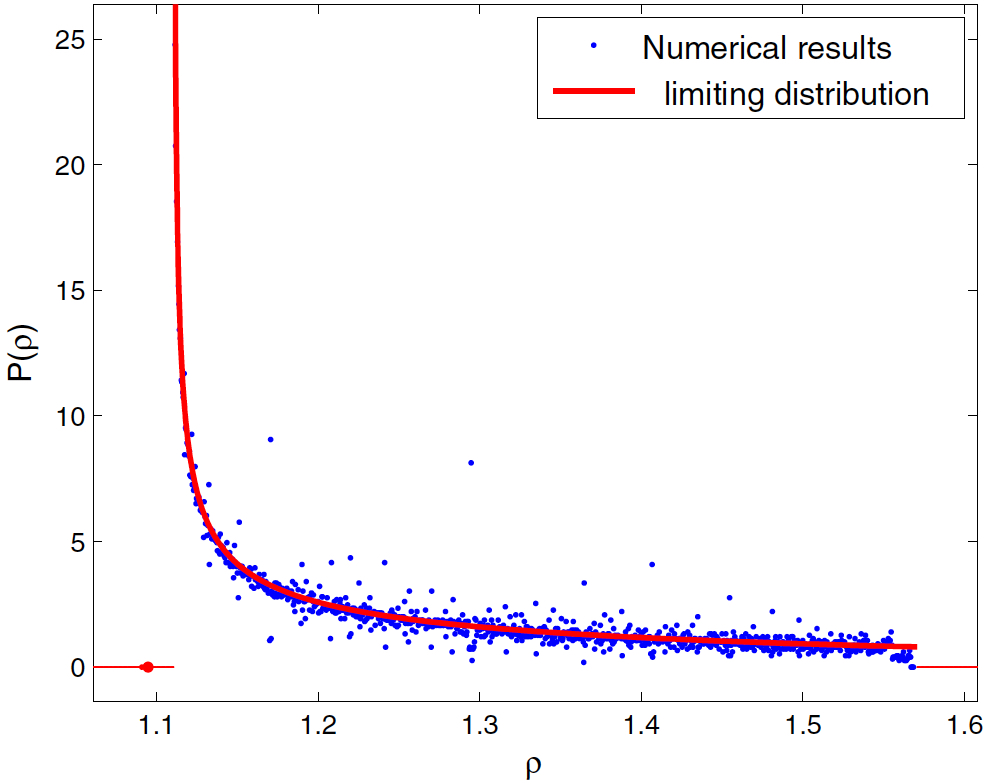}
\includegraphics[width=\linewidth, trim={0.2cm 0 0 0},clip]{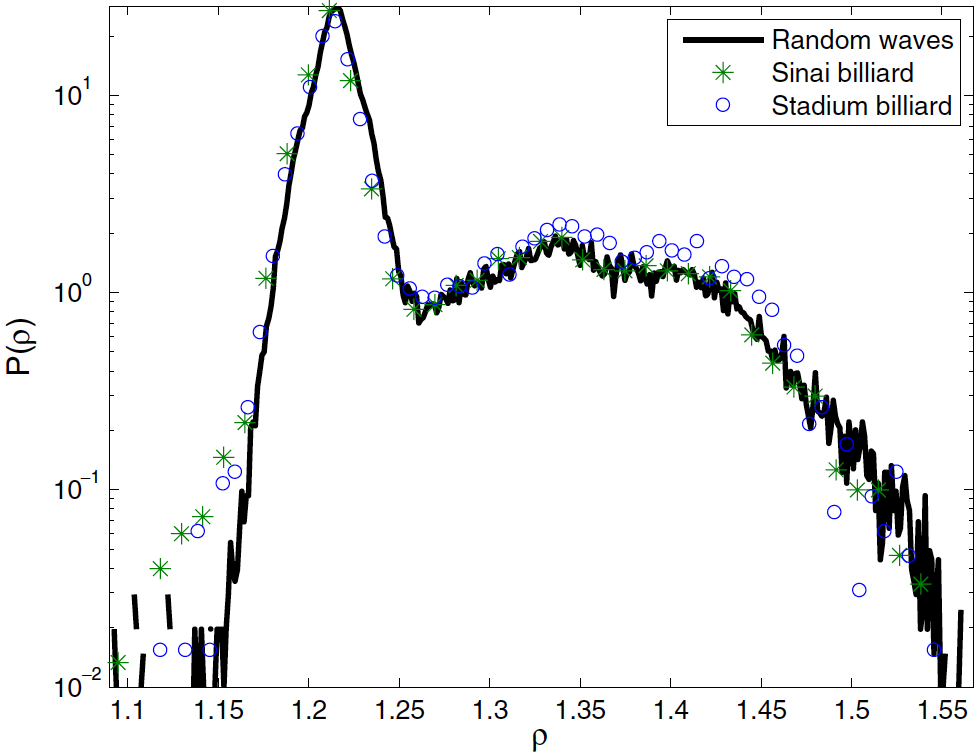}
\caption{\label {fig:A2P}[Top]: The limiting distribution of $P\,(\rho)$ for a rectangular billiard, Eq.~\eqref{eq:ratiorect}, compared to the calculated distribution for eigenfunctions with $E\cdot\mathcal{A}< 10^5$. [Bottom]: A comparison between the distribution function $P\,(\rho)$ calculated for the random-wave ensemble and for the inner domains of a Sinai and a stadium billiard. From \citet{elon2007geometric}. \textcopyright\, IOP Publishing. Reproduced with permission. All rights reserved.}
\end{figure}

A nodal domain of an eigenfunction, being a bounded region in space itself, has obvious geometric characteristics like a well-defined perimeter and area. The ratio of these two quantities turns out to be another statistically significant tool to sniff out the underlying dynamics of the system. To interrogate the morphology of the nodal lines, \citet{elon2007geometric} considered the set of nodal domains of the $j^{\rm th}$ eigenfunction of a billiard in a domain ${\mathcal D}$; this can be represented as the sequence $\{ \omega _j^{(m)} \}$, $m = 1, 2, \ldots, \nu _j$. One can then define the ratio
\begin{equation}
\label{eq:ratio}
\rho _j^{(m)} = \frac{{\mathcal A}_j^{(m)}\sqrt{E_j}}{L_j^{(m)}},
\end{equation}
where ${\mathcal A}_j^{(m)}$, $L_j^{(m)}$ denote the area and perimeter of the nodal domain; the factor of the energy eigenvalue ensures the correct scaling. As in Eq.~\eqref{eq:limiting}, we inspect the probability measure
\begin{equation}
\label{eq:measure}
P_{\mathcal D}(\rho, E, g) = \frac{1}{N_I} \sum_{E_j \in I} \frac{1}{\nu _j} \sum_{m=1}^{\nu _j} \delta \left(\rho - \rho _j^{(m)}\right),
\end{equation}
which again tends to a limiting distribution in the same fashion as previously. For a rectangular billiard, the distribution is of the form
\begin{alignat}{1}
\label{eq:ratiorect}
P_{\rm rectangle}(\rho ) =
\begin{cases}
{\displaystyle \frac{4}{\rho \sqrt{8\rho ^2 - \pi ^2}}}, \,\,&{\displaystyle \frac{\pi}{\sqrt{8}} \leq \rho \leq \frac{\pi}{2}}, \\
0, \,\, &\mbox{otherwise}.
\end{cases}
\end{alignat}
This function (Fig.~\ref{fig:A2P}), in addition to being independent of the aspect ratio of the billiard, is analytic and monotonically decreasing in the compact interval $[\pi/\sqrt{8}, \pi/2]$ but is discontinuous at the endpoints. All of these properties, including the support, are believed to be universal features for all two-dimensional separable surfaces. Explicit derivations of the limiting distributions for the family of simple surfaces of revolution and the disc billiard \cite{elon2007geometric} further lend weight to this hypothesis. However, for integrable but non-separable and pseudointegrable billiards, the form of $P\,(\rho)$ is unknown. For chaotic Sinai and stadium billiards, numerics suggest a universal limiting distribution $P\,(\rho)$, which converges to that for the random-wave ensemble. This agreement can also be demonstrated for finite energies, as in Fig.~\ref{fig:A2P}, by considering only the inner nodal domains (away from the billiard's boundary).

\paragraph{Signed area distribution}

\begin{figure} [htb]
\begin{overpic}[width= \linewidth]{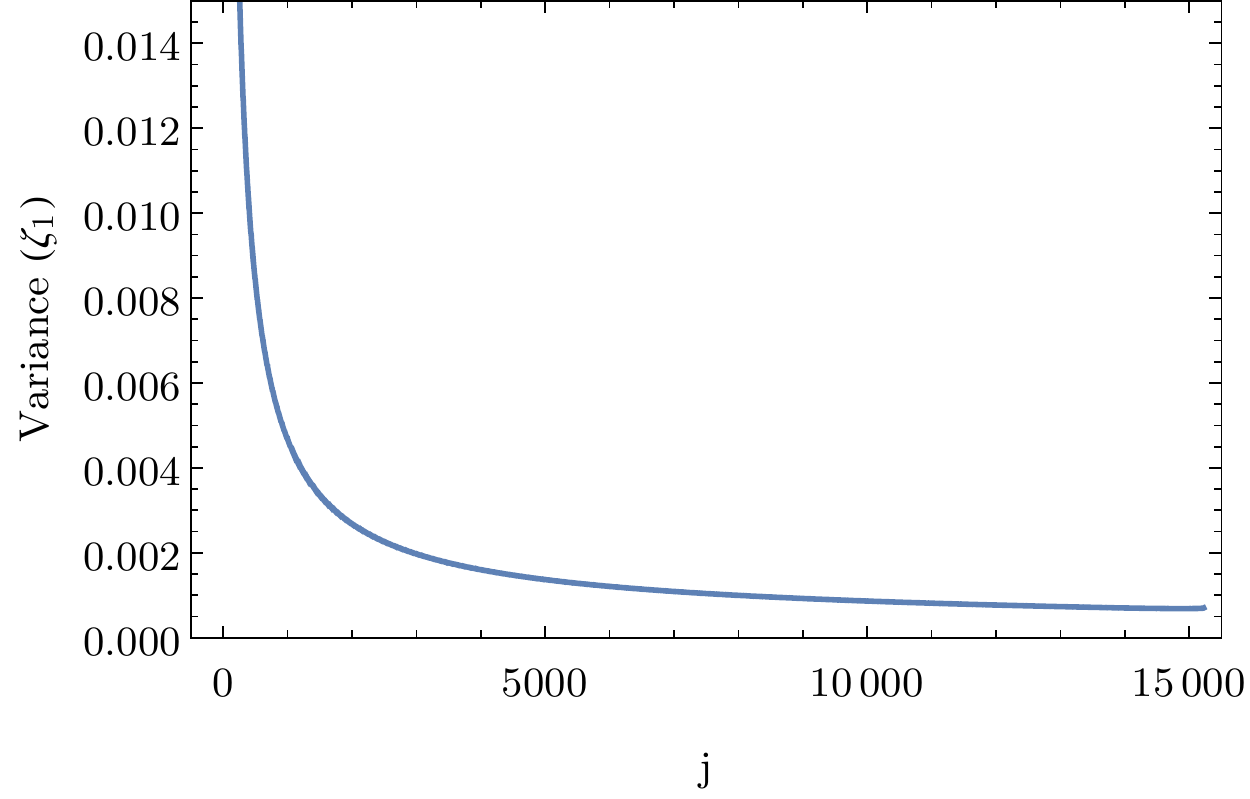}
     \put(37,19){\includegraphics[width= 0.6\linewidth]{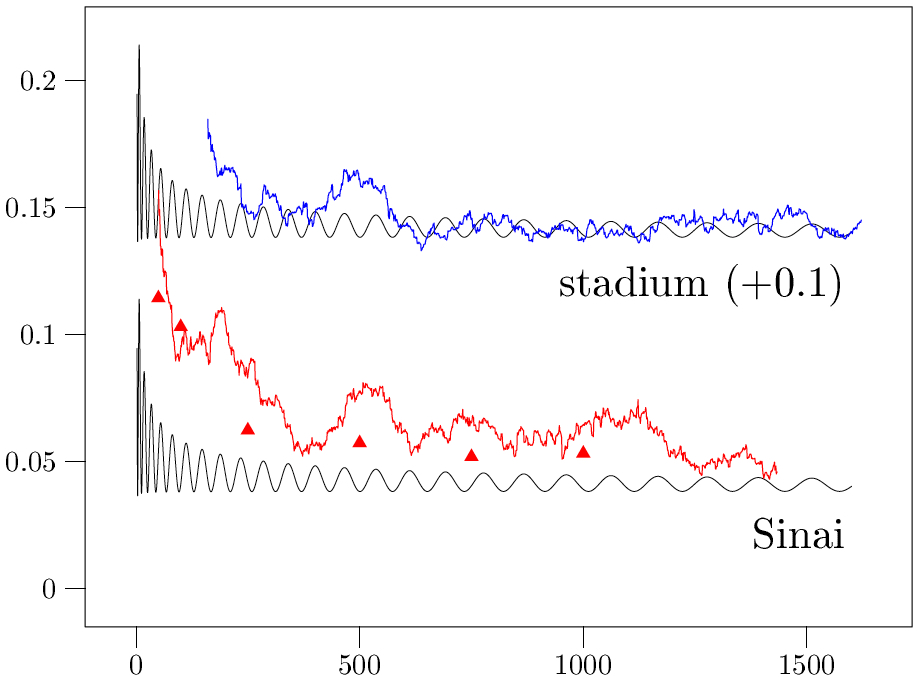}}
\end{overpic}
\caption{\label {fig:SignedArea}The normalized signed area variance for the equilateral triangle. [Inset]: The same for the stadium and Sinai billiards but plotted as a function of $\textsc{n}$. The analytical expression and numerical simulation for random waves are depicted by smooth curves and triangles, respectively. Adapted from \citet{blum2002nodal}.}
\end{figure}

Instead of scrutinizing the areas of individual nodal domains, one may alternatively peruse the collective statistics of the \textsl{total} area where the wavefunction is positive (negative), denoted hereafter by $\lvert\mathfrak{A}\rvert_\pm$; clearly, $\langle \lvert\mathfrak{A}\rvert_\pm \rangle = \mathcal{A}/2$.
Exploiting the identity
\begin{equation}
\lvert\mathfrak{A}\rvert_\pm = \frac{1}{2\,\pi\, \mathrm{i}} \lim_{\epsilon \rightarrow 0^+} \int_{\mathcal{A}} \mathrm{d}\, \mathbf{r} \int_{-\infty}^{\infty} \mathrm{d}\,\xi \,\frac{\mathrm{e}^{\,\pm \mathrm{i}\,\xi\,\psi\,(\mathbf{r})}}{\xi - \mathrm{i}\,\epsilon},
\end{equation}
\citet{blum2002nodal} reckoned the signed area variance 
\begin{equation}
\frac{ \langle (\lvert\mathfrak{A}\rvert_+ -  (\lvert\mathfrak{A}\rvert_- )^2 \rangle}{ \mathcal{A}^2} \approx \frac{0.0386}{(R\, k/2)^2} \equiv 0.0386 \,\textsc{n}^{-1}
\end{equation}
for the random wave model on a circle of radius $R$. Fig.~\ref{fig:SignedArea} affirms the convergence of the variance to this asymptotic limit for the stadium and the Sinai billiards but also proclaims the qualitatively different---non-oscillatory---behavior of the data for the equilateral triangle billiard.

\subsubsection{Nodal volume statistics}

The nodal volume is the hypersurface volume of the nodal set of the $j^{\rm th}$ eigenfunction, denoted by ${\mathcal H}_j$. In order to facilitate comparison of nodal volumes of eigenfunctions at different energies, it is prudent to scale the volumes with the typical wavelength, $\sqrt{E_j}$, $E_j > 0$, and then define the rescaled dimensionless variable, $\sigma _j = {\mathcal H}_j/{\mathcal V}\sqrt{E_j}$, where ${\mathcal V}$ is the volume of the manifold ${\mathcal M}$. It is this rescaled quantity on which Yau's conjecture places the bounds $c_1 \leq \sigma _j \leq c_2$ for $j \geq 2$, where $c_1, c_2$ depend only on the manifold and the metric. We now present some recent results  \cite{gnutzmann2014remarks} on the nodal volume statistics for an $s$-dimensional cuboid---a paradigm of regular classical dynamics---and for boundary-adapted planar random waves---an established model for chaotic wave functions---in irregular shapes \cite{berry2002statistics}.  

\paragraph{$s$-dimensional cuboid}

The normalized eigenfunctions of an $s$-dimensional cuboidal Dirichlet billiard with sides of lengths $\{ a_{\ell},\, \ell = 1, 2, \ldots, s \}$ and volume ${\mathcal V} = \prod_{\ell = 1}^{s} a_{\ell}$ are
\begin{equation}
\label{eq:wfun-cuboid}
\psi _{\bf n}({\bf q}) = \frac{(2\pi)^{s/2}}{{\mathcal V}^{1/2}} \prod_{\ell = 1}^{s} \sin \left( \frac{\pi \,n_{\ell}\,q_{\ell}}{a_{\ell}} \right), 
\end{equation}
$\{n_{\ell}\}$ being positive integers. The corresponding energies and rescaled nodal volumes are
\begin{equation}
\label{eq:en-cuboid}
E_{\bf n} = \pi^2 \sum_{\ell = 1}^{s} \frac{n_{\ell}^2}{a_{\ell}^2},\quad
\sigma _{\bf n} = \frac{1}{\sqrt{E_{\bf n}}}\sum_{\ell = 1}^{s} \frac{n_{\ell} - 1}{a_{\ell}}.
\end{equation}
Let us define an asymptotic mean value of $\sigma _{\bf n}$ in a spectral interval $E_{\bf n} \in [E, E+\Delta E]$ of width $\Delta E$ near $E$,
\begin{equation*}
\langle \sigma _{\bf n} \rangle_{[E, E+\Delta E]} = \frac{1}{N_{[E, E+\Delta E]}}\sum_{{\bf n} \in \mathbb{N}^s} \sigma _{\bf n} \chi _{[E, E+\Delta E]} (E_{\bf n}),
\end{equation*}
where $\chi$ is the characteristic function on the interval and $N_{[E, E+\Delta E]}$ is the number of eigenfunctions with energies in $[E, E+\Delta E]$. For the asymptotic behavior, $\Delta E$ can be chosen to be $g\,E^{1/4}$, $g > 0$, without loss of generality. Weyl's law for the cumulative level density, adapted for an $s$-dimensional cuboid, is repackaged to
\begin{equation}
\label{eq:weyl_cuboid}
N_{\rm Weyl}(E) = \frac{\zeta _s{\mathcal V}}{2^s\pi ^s} E^{s/2} - \frac{\zeta _{s-1}{\mathcal S}}{2^{s+1}\pi ^{s-1}}E^{(s-1)/2} + {\mathcal O}(E^{(s-2)/2}),
\end{equation}
where $\zeta _s = \pi^{s/2}/\Gamma (1 + s/2)$ is the volume of an $s$-dimensional unit sphere and ${\mathcal S} = 2{\mathcal V}\sum_{\ell = 1}^{s} a_{\ell}^{-1}$ is the $(s-1)$-dimensional volume of the surface of the $s$-cuboid. Using Eq.~\eqref{eq:weyl_cuboid} to obtain $N_{[E, E+\Delta E]}$ and employing the Poisson summation formula, the mean value is found to be \cite{gnutzmann2014remarks}
\begin{alignat}{1}
\label{eq:mean-series}
\nonumber\langle &\sigma _{\bf n} \rangle_{[E, E+\Delta E]} = \frac{2\,\zeta_{s-1}}{\pi \,\zeta _{s}}\left( 1 - \beta _s\frac{\mathcal S}{\mathcal V}E^{-1/2} + {\mathcal O}(E^{-3/4}) \right),\\
&\beta _s = \frac{\pi \,(s-1)\, \zeta_{s-2}}{2\,s\,\zeta_{s-1}} + \frac{\pi\, \zeta_s}{4\zeta_{s-1}} - \frac{\pi\, (s-1)\,\zeta_{s-1}}{2\,s\,\zeta_s}.
\end{alignat}
Similarly, the variance of $\sigma _{\bf n}$ can be expressed in an asymptotic series for large $E$ as 
\begin{equation*}
\mathrm{Var}\,(\sigma _{\bf n}) = \frac{1}{\pi^2} + \frac{4\,(s-1)\,\zeta_{s-2}}{s\,\pi^2\,\zeta_s} - \frac{4\,\zeta^2_{s-1}}{\pi^2\,\zeta_s^2} + \mathcal{O} (E^{-1/2}). 
\end{equation*}
Utilizing the higher moments, the limiting distribution
\begin{equation}
\label{eq:moments}
P_s(\sigma) = \lim_{E \to \infty} \langle \delta \,(\sigma - \sigma _{\bf n}) \rangle_{[E, E+\Delta E]} 
\end{equation}
can be calculated for any $s$. The limiting distributions thus evaluated by \citet{gnutzmann2014remarks} are non-zero only over a finite interval. For instance, $P_2(\sigma )$ is non-zero only over $[1/\pi , \sqrt{2}/\pi]$ wherein it varies as $4/\sqrt{2-\pi^2\sigma^2}$, bringing to mind Eq.~\eqref{eq:ld_rect}. Importantly, this observation is also in line with Yau's conjecture. 

\paragraph{Random wave model}

On a different note from the integrable cuboid, we can repeat the above-defined procedure for the eigenfunctions of a chaotic billiard modeled according to boundary-adapted RWM. In $s$ dimensions, the mean of $\sigma$ is \cite{gnutzmann2014remarks}:
\begin{alignat*}{1}
\label{eq:sigma-RWM}
\langle \sigma \rangle_G &= \rho _{\rm bulk} \left( 1 - \frac{\mathcal S}{\mathcal V} \frac{\log k}{32\,\pi\, k} + {\mathcal O}(k^{-1})\right), \qquad s = 2, \nonumber \\
&= \rho _{\rm bulk} \left( 1 - \frac{\mathcal S}{\mathcal V} \frac{I_s}{32\,\pi\, k} + {\mathcal O}(k^{-1}) \right), \qquad s \geq 3,
\end{alignat*}
where $\rho_{\rm bulk} = \Gamma ((s+1)/2)/[\sqrt{\pi \,s}\, \Gamma (s/2)] $ is the constant nodal density of the standard RWM without boundaries and $I_s$ are constants ($I_3 \simeq 0.758,\, I_4 = 0.645$). The limiting distribution of nodal volumes is now sharply peaked for a finite energy interval and converges to $P (\sigma) = \delta\,(\sigma - \rho_{\rm bulk})$. This is to be contrasted with the finite support for the cuboid's distribution---the distinct characters of $P\,(\sigma)$ therefore differentiate between chaotic and regular manifolds.
Moreover, the variance decreases with increasing energies for irregularly-shaped billiards whereas for separable shapes, it remains finite and bounded. Additionally, the boundary corrections to expected rescaled nodal volumes---$1/k$ for the cuboid and $k^{-1}\log k$ for chaotic shapes---also herald this distinction.

\subsection{Nodal line statistics}
The sinuous nature of the nodal curves for classically chaotic systems, particularly for the excited states, renders their study challenging. However, the wealth of statistical information borne by the nodal lines also makes the problem equally rewarding.

\subsubsection{Length fluctuations}

The total length of the nodal curve of a real wavefunction $u({\bf r})$, ${\bf r} = (x, y)$, is proffered by elementary calculus:
\begin{equation}
\label{eq:length}
L = \int\hspace{-0.2cm} \int_{\cal A} \mathrm{d}\,{\bf r}\,\, \delta \left(u({\bf r})\right)\,\lvert\nabla\, u({\bf r})\rvert,
\end{equation}
where the integral runs over the area of the billiard enclosure ${\cal A}$. Although, on paper, Eq.~\eqref{eq:length} offers a deterministic expression for the nodal length, the practical drawback is that exact analytical results for eigenfunctions of chaotic billiards are extremely rare and only known for a few low-lying states \cite{jain2002quantum}. For the excited states of these systems, the eigenfunctions display intricate scaling properties, reminiscent of multifractal objects, but the exact forms thereof remain an open question. Thus, foiled, we resort to the random wave model, consoling ourselves that since the RWM describes the high-energy eigenvalues, the nodal lines of random waves should also model the nodal lines of honest eigenfunctions \cite{wigman2012nodal}. To adapt Eq.~\eqref{eq:isotropic} to the boundary conditions, we switch to coordinates in which, at any point on the billiard's boundary, $\hat{x}$ is along $\partial\, \mathcal{D}$ (traversing counterclockwise) and $\hat{y}$ points along the normal to it, inwards (much like $\hat{\theta}$ and $-\hat{r}$, respectively, in polar coordinates). For real eigenfunctions, the superpositions can be written as (cf. Eq.~\ref{eq:heller})
\begin{eqnarray}
u_D({\bf R}) &=& \sqrt{\frac{2}{N}} \sum_{j=1}^{N} \sin \,(Y \sin \theta _j)\cos \,(X\cos \theta _j + \phi _j); \nonumber \\ & &(u = 0 ~{\rm for}~y = 0), \\
u_N({\bf R}) &=& \sqrt{\frac{2}{N}} \sum_{j=1}^{N} \cos\, (Y \sin \theta _j)\cos\, (X\cos \theta _j + \phi _j);\nonumber \\  & &(\partial u/\partial y = 0 ~{\rm for}~y = 0),
\end{eqnarray}
where ${\bf R} = (k\,x, k\,y)  \equiv (X, Y)$ introduces a convenient set of dimensionless coordinates. On replacing cosines by sines in these equations, we obtain the corresponding expressions for complex waves. To perform the averages, we integrate over the random phases $\phi_j$ and directions $\theta_j$, both of which are uniformly distributed in $[0, 2\pi]$.\\

The average total length of nodal lines $\langle L \rangle$ can be calculated on knowing the mean density of nodal length as a function of $y$ \cite{berry2002statistics}. From Eq.~\eqref{eq:length}, this is
\begin{equation}\label{eq:av_line}
\langle L(k)\rangle = \frac{k}{2\sqrt{2}} \int\hspace{-0.2cm} \int_{\cal A} \mathrm{d}\,{\bf r}\,\,\rho _L(Y);
\end{equation}
the prefactor ensures that $\rho _L(Y)$ tends to unity as $Y \to \infty$. The mean density is 
\begin{equation}
\label{eq:density}
\rho _L(Y) = 2\sqrt{2}\, \langle \delta (u)\,\lvert \nabla _R\,u \rvert \, \rangle,
\end{equation}
which requires information about the distribution functions of $u_X$ and $u_Y$. These distributions are Gaussian,
\begin{alignat}{1}
P(u_X) &= \frac{1}{\sqrt{2\pi\, D_X}}\exp \left[ -\frac{u_X^2}{2\,D_X}\right], \\
\nonumber P(u=0, u_Y) &= \frac{1}{2\pi\, \sqrt{B\,D_Y - K^2}}\exp \left[ -\frac{B\,u_Y^2}{2\,(B\,D_Y - K^2)}\right],
\end{alignat}
with the parameters expressible in terms of various cylindrical Bessel functions as $D_X=(1\mp J_0(2\,Y) \mp J_2(2\,Y))/2$, $D_Y=(1\pm J_0(2\,Y) \mp J_2(2\,Y))/2$, $B=1 \mp J_0(2\,Y)$, and $K = \pm J_1(2\,Y)$ for Dirichlet and Neumann boundary conditions, respectively. Employing these distributions, the density reduces to
\begin{alignat}{1}
\rho _L(Y) &= \frac{2\sqrt{2}}{\pi}D_X\,(B\,D_Y-K^2) \nonumber \\
&\times \int_{0}^{\pi /2} \frac{d\theta}{[B\,D_X\cos ^2\theta + (B\,D_Y-K^2)\sin ^2\theta]^{3/2}} \nonumber \\
&= \frac{2\sqrt{2}}{\pi B}\sqrt{B\,D_Y - K^2}\,\, \mathscr{E}\left[ 1 + \frac{B\,D_X}{K^2 - B\,D_Y} \right],
\end{alignat}
$\mathscr{E}$ being the complete elliptic integral \cite{whittaker1996course}. 

\begin{figure}[htb]
\includegraphics[width=\linewidth]{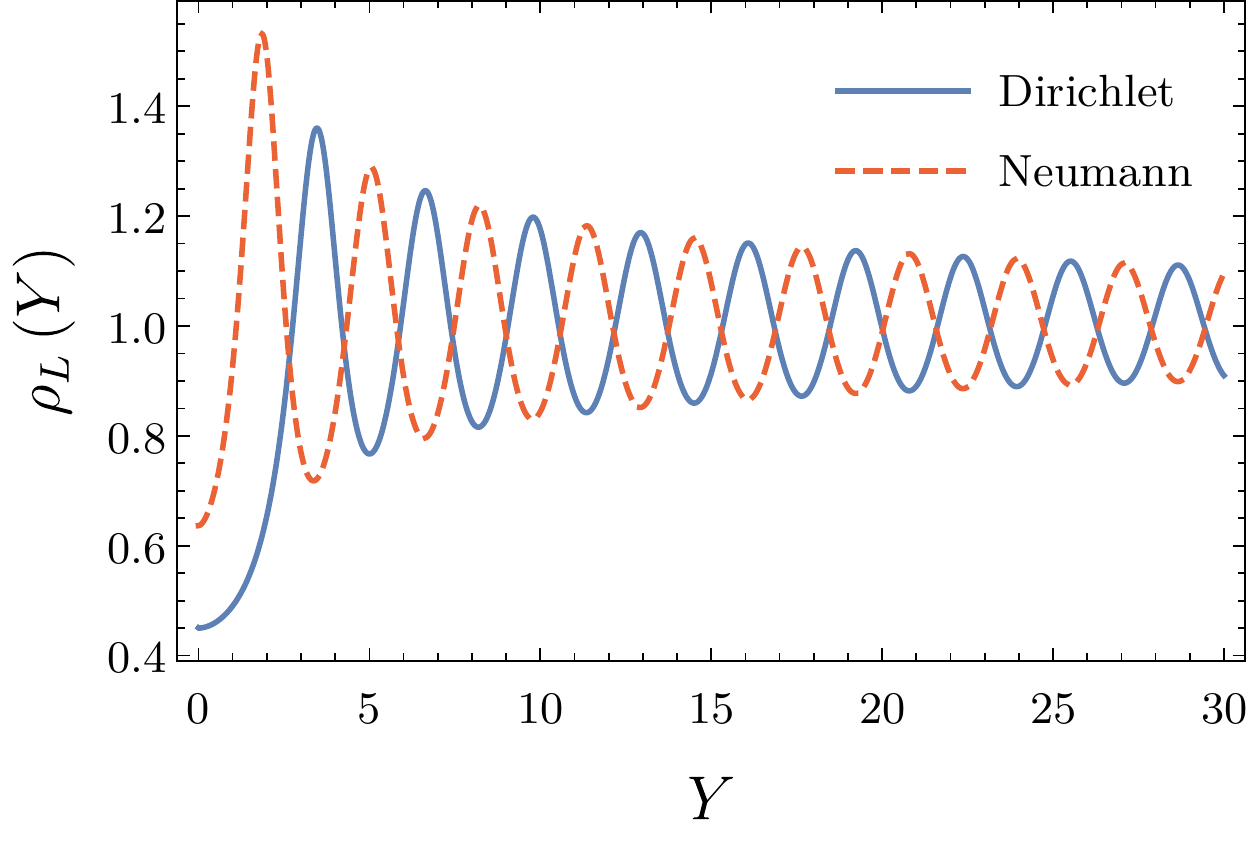}
\caption{\label{fig:line_density}Mean nodal line density $\rho_L\,(Y )$, for Dirichlet (thick curve) and Neumann (thin curve) boundary conditions.}
\end{figure}

The behavior of $\rho _L(Y)$ for Dirichlet (Neumann) boundary conditions in Fig. \ref{fig:line_density} shows a stronger (weaker) repulsion from the boundary, along with an oscillatory asymptotic behaviour for large $Y$, eventually settling at unity. Eqs.~(\ref{eq:av_line},~\ref{eq:density}) now tell us that
\begin{equation}
\label{eq:ch1}
\frac{\langle L \rangle}{\sqrt{\cal A}} = \sqrt{\frac{\pi \,j}{2}} - \frac{\cal P}{128 \,\pi\, \sqrt{2\,{\cal A}}}\log j + \mathcal{O}(1).
\end{equation}
The first piece is  an ``area term'', originating from the $k{\cal A}/2\sqrt{2}$ factor, whereas the second term stems from the boundary. More generally, it is conjectured \cite{nonnenmacher2013} that on any chaotic surface $\mathbb{M}$, such as an ergodic billiard, the nodal lines are asymptotically equidistributed in $\mathbb{M}$ \cite{wigman2012nodal}. In particular, for a discrete spectrum of eigenvalues $\{ E_j \}$, the nodal length of the $j^{\mathrm{th}}$ eigenfunction is asymptotically shaped to obey 
\begin{equation}
\label{eq:ch2}
\langle L_j \rangle \sim c_{\,\mathbb{M}}\cdot \sqrt{E_j}
\end{equation}
for some constant $c_{\,\mathbb{M}} > 0 $.\\

It is quite obvious that the nodal lines' length depends on the state; correspondingly, there are fluctuations over different eigenstates. These fluctuations are quantified by the variance
\begin{equation}
(\delta L)^2 \equiv \left(\sqrt{\langle L^2 - \langle L \rangle ^2 \rangle}\right)^2 \approx \frac{\cal A}{256\pi }\log\, (k\,\sqrt{\cal A}).
\end{equation}
Written in the dimensionless form (for level number $j$)
\begin{equation}
\frac{\delta L}{\sqrt{\mathcal{A}}} \approx \frac{1}{16} \sqrt{\frac{\log j}{2 \pi}},
\end{equation}
this estimate \cite{berry2002statistics} is palpably smaller than the boundary term, thereby fanning the expectation that the boundary corrections could be gauged without the necessity of averaging over long sequences of states. Although, these results were obtained for a straight-line boundary, the average nodal line length has the same leading-order logarithmic boundary term for a convex circular boundary as well \cite{wheeler2005curved}. \\

While chaotic billiards were the cynosure of the preceding exercise, studying the nodal lines of the purportedly ``simpler'' integrable systems also provides unique insights. An example that has recently spurred great inquiry in mathematical circles is the square, but with periodic boundary conditions, which reduces it to the standard 2-torus $\mathbb{T}^2 = \mathbb{R}^2/\mathbb{Z}^2$ by sewing together opposite sides. The energy spectrum is $E_j = 4 \pi^2\,j$ with $j\,\in\,\mathbb{N}$, as we have seen earlier, and the cardinality $N_j$ of the set of frequencies
\begin{equation}
\Lambda_j = \{ \lambda = (m,n) \,\in\,\mathbb{Z}^2: \,m^2 + n^2 = j \}
\end{equation}
is just the degeneracy of the level, which grows on average as $\sqrt {\log j}$ \cite{landau1909}. The eigenspace is spanned by the complex exponentials $e_\lambda \,(\mathbf{r}) \equiv \exp\, (2\pi\, \mathrm{i}\, \lambda\cdot\mathbf{r})$ or their trigonometric equivalents. For this system (or any other with a checkerboard nodal pattern), the total nodal length of a ``pure'' (non-superposed) eigenstate is trivial; in a square of side $\mathcal{L}$, it equals $(m-1)\,(n-1)\,\mathcal{L}$. It is much more interesting to look at the nodal lengths and their fluctuations when the state is instead a linear superposition of such individual eigenfunctions. In this context, we consider arithmetic random waves (also called random Gaussian toral Laplacian eigenfunctions), which are the random fields
\begin{equation}
T_j (\mathbf{r}) = \frac{1}{N_j} \,\sum_{\lambda\, \in\, \Lambda_j}\,a_\lambda\,e_\lambda (\mathbf{r});\hspace{0.5cm} \mathbf{r} \,\in \,\mathbb{T},
\end{equation}
where the coefficients $a_\lambda$ are independent standard complex-Gaussian random variables save for the relations $a_{-\lambda} = \bar{a}_\lambda$. The expected total nodal length of the random eigenfunctions is \cite{rudnick2008volume}
\begin{equation}
\label{eq:int1}
\mathbb{E}\,\left[L_j\right] = \frac{1}{2\sqrt{2}}\,\sqrt{E_j}, 
\end{equation}
in consistence with Yau's conjecture \cite{yau1982survey, donnelly1988nodal}. The corresponding variance was calculated by \citet{krishnapur2013nodal} as
\begin{equation}
\label{eq:int2}
\mathrm{Var}\, (L_j) = c_j\, \frac{E_j}{N_j^2}\,\left[ 1 + \mathcal{O}_{N_j \rightarrow \infty} (1) \right],
\end{equation}
where $c_j \,\in\,[1/512, 1/256]$. This expression is to be juxtaposed with Eq.~\eqref{eq:GSHVar} for the Gaussian spherical harmonic, which, in addition, is garnished by the peculiar factor of $1/32$ arising from the nontrivial local geometry of the sphere. In this fashion, Eqs.~\eqref{eq:int1} and \eqref{eq:int2} define the integrable analogues to the chaotic versions, Eqs.~\eqref{eq:ch1} and \eqref{eq:ch2}, respectively.  The fine asymptotic behavior can be extracted by examining the distributions of the sequence of normalized random variables
\begin{equation}
\tilde{L}_j \equiv \frac{L_j - \mathbb{E}\,\left[L_j\right]}{\sqrt{\mathrm{Var}\, (L_j)}};\hspace{0.5cm} j\,\in\,\mathbb{N}.
\end{equation}
For $\eta \,\in [0,1]$, we also introduce the random variable
\begin{equation}
\mathrm{M}_\eta = \frac{1}{2\,\sqrt{1+\eta^2}}\,\left(2 - (1+\eta)\,\chi_{1}^2 - (1-\eta)\,\chi_2^2 \right),
\end{equation}
$\chi_1$ and $\chi_2$ being standard independent Gaussians. As proved by \citet{marinucci2016non}, the sequence $\tilde{L}_j$ converges to $\mathrm{M}_\eta $ under certain technical measure-theoretic conditions determining $\eta$, which we will not get into except to note that $\tilde{L}_j$ does \textsl{not} converge in distribution for $N_j \rightarrow \infty$ \cite{kurlberg2016probability}.

\subsubsection{Curvature distributions}

\begin{figure}[htb]
\includegraphics[width=\linewidth]{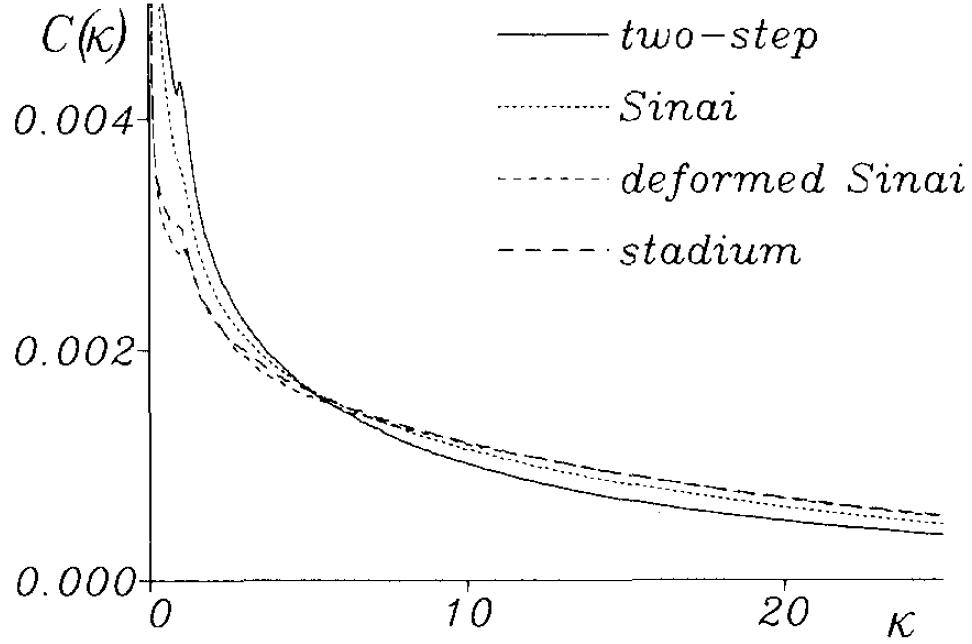}
\caption{\label{fig:curvature} Nodal line curvature distribution, $C(\kappa )$, over 201--600 eigenfunctions for pseudointegrable (two-step) and chaotic (Sinai, deformed Sinai, Stadium) billiards show comparable trends. From \citet{simmel1996statistical}, with permission from Elsevier.}
\end{figure}

The avoidance of the labyrinthine nodal curves for nonseparable and chaotic systems suggests studying their curvature. Separable billiards can hold eigenfunctions with nodal curves of zero curvature. Let us denote the local curvature by $\kappa = 1/r$ ($r$ being the radius of curvature) and the length of nodal curves with curvature $\leq \kappa $ by $\ell (\kappa )$. A differential measure of curvature can be defined as \cite{simmel1996statistical}

\begin{equation}
C(\kappa ) = \frac{1}{\ell (\infty )}\frac{\mathrm{d}\, \ell (\kappa )}{\mathrm{d}\, \kappa }
\end{equation}
with $\int_{0}^{\infty } C(\kappa )\,\,\mathrm{d}\,\kappa = 1$. For separable billiards, due to the checkerboard arrangement of the nodal lines, $C(\kappa )$ is trivially $\delta (\kappa )$. Numerical studies on pseudointegrable and chaotic billiards reveal that the curvature distribution, averaged over the eigenstates, is similar (Fig. \ref{fig:curvature}).   

The nodal lines of the eigenfunctions of nonseparable systems avoid intersections. To illustrate this, let us expand a plane wave in cylindrical coordinates, following \citet{courant1953methods}
\begin{equation}\label{eq:random_wave}
\psi({\bf r}) = \sum_{l=-\infty}^{\infty} \alpha _l\,J_l(k\,r)\,\mathrm{e}^{\mathrm{i}\,l\,\theta}, 
\end{equation}
where $\alpha _l*=(-1)^l \alpha _{-l}$. The origin can be made arbitrary by shifting it by $\rho $, whereupon the wavefunction can be rewritten using the addition theorem of Bessel functions \cite{watson1995treatise}:
\begin{equation}
\psi ({\bf r} + \rho ) = \sum_{m} \alpha _{m}({\bf r})\,J_m(k\,\rho)\,\mathrm{e}^{\mathrm{i}\,m\,\phi},
\end{equation}
where $\phi$ is measured from the direction defined by ${\bf r}$. Here, $\alpha _m = \beta _m + \mathrm{i} \gamma _m$ is given by 
\begin{equation}
\alpha _{m}({\bf r}) = \sum_{l=-\infty}^{\infty} \alpha _l(0)\,J_{l-m}(k\,r)\,\mathrm{e}^{\mathrm{i}\,l\,\phi}. 
\end{equation}
The real and imaginary parts of $\alpha _m$ are related to the wavefunction and its derivatives at ${\bf r}$; for instance, \cite{monastra2003avoided}
\begin{alignat}{1}
\beta _0({\bf r}) &= \psi ({\bf r}), \qquad \beta _1({\bf r})=\frac{1}{k}\frac{\partial \,\psi}{\partial\, r}, \nonumber \\
\beta _2({\bf r}) &= \psi ({\bf r}) + \frac{2}{k^2}\frac{\partial ^2\,\psi}{\partial\, r^2},\nonumber \\
\gamma _1({\bf r}) &= -\frac{1}{k\,r}\frac{\partial\, \psi}{\partial\, \theta}, \qquad \gamma _2({\bf r})=\frac{2}{(k\,r)^2}\left(\frac{\partial \,\psi}{\partial\, \theta} - r\frac{\partial ^2\psi}{\partial\, r\,\partial\, \theta}\right), \nonumber \\
\frac{\partial ^2 \psi }{\partial\, \theta ^2} &= -k\,r\beta _1({\bf r}) - \frac{(k\,r)^2}{2}(\beta _2({\bf r}) + \beta _0({\bf r})).
\end{alignat}
For $k\,\rho < 1$, the wavefunction to second order in $k\,\rho$ is
\begin{eqnarray}\label{eq:zero_set}
\psi ({\bf r}+\rho) &=& \beta _0({\bf r})\,\bigg[ 1 - \left( \frac{(k\,\rho)^2}{2}\right) \bigg] \nonumber \\ &+& |\alpha _1({\bf r})|\left( \frac{(k\,\rho)}{2}\right) \cos (\phi + \phi _1) \nonumber \\  
&+& \frac{1}{2}|\alpha _2({\bf r})|\left( \frac{(k\,\rho)^2}{2} \right)\cos [2(\phi + \phi _2)],
\end{eqnarray}
taking $\phi _{1,2}$ to be the phases of $\alpha _{1,2}$. Two nodal lines intersect at $\mathbf{r}$ if $\beta_0 = 0$, $\alpha _1=0$, and $\alpha _2 \neq 0$. The intersection of the nodal lines is at right angles since $\cos 2(\phi + \phi _2) = 0$ along two perpendicular lines intersecting at ${\bf r}$. If more than two nodal lines intersect such that the first non-vanishing coefficient at ${\bf r}$ is $\alpha _q$, then ${\bf r}$ is a nodal point of order $q$. At this point, $q$ nodal lines intersect at angles $\pi /q$. For higher-order $q$, more conditions need to be satisfied by the coefficients. Thus the intersections become rarer, which is the essential result of Uhlenbeck's theorem.

An avoided crossing of two nodal lines occurs at ${\bf r}$ if ${\bf r}$ is a saddle point of the wavefunction. In terms of the local coordinates of $\rho = (\xi, \eta)$, Eq.~\eqref{eq:zero_set} takes the form
\begin{equation}
\left(1-\frac{|\alpha _2({\bf r})|}{\beta _0({\bf r})}\right) \xi ^2 + \left(1 + \frac{|\alpha _2({\bf r})|}{\beta _0({\bf r})} \right) \eta ^2 = 1.
\end{equation}
This is the equation of a hyperbola or an ellipse according as whether $|\alpha _2({\bf r})|$ is larger or smaller than $|\beta _0({\bf r})|$. The coefficients, $\alpha _2({\bf r})$ and $\beta _0({\bf r})$, are, of course, related to the wavefunction and its derivatives. One can quantify the scaled distance between the two branches by defining an avoidance range associated with the avoided crossing at ${\bf r}$ \cite{monastra2003avoided}:
\begin{equation}
z({\bf r}) = \left[ \frac{16\,k^2 \,\lvert \psi \rvert}{k^2\,\lvert \psi \rvert + \sqrt{4\,\psi _{xy}^2 + (\psi _{xx} - \psi _{yy})^2}}  \bigg| _{\bf r} \right]^{1/2},
\end{equation}
the subscripts denoting partial derivatives. Clearly, $z$ is zero at an intersection of nodal lines. If the intersection is at a saddle point, then $|\alpha _2({\bf r})| > |\beta _0({\bf r})|$, which puts a bound on $z$ so that it can range only between $0$ and $2\sqrt{2}$. 

\begin{figure}[htb]
\includegraphics[width=\linewidth]{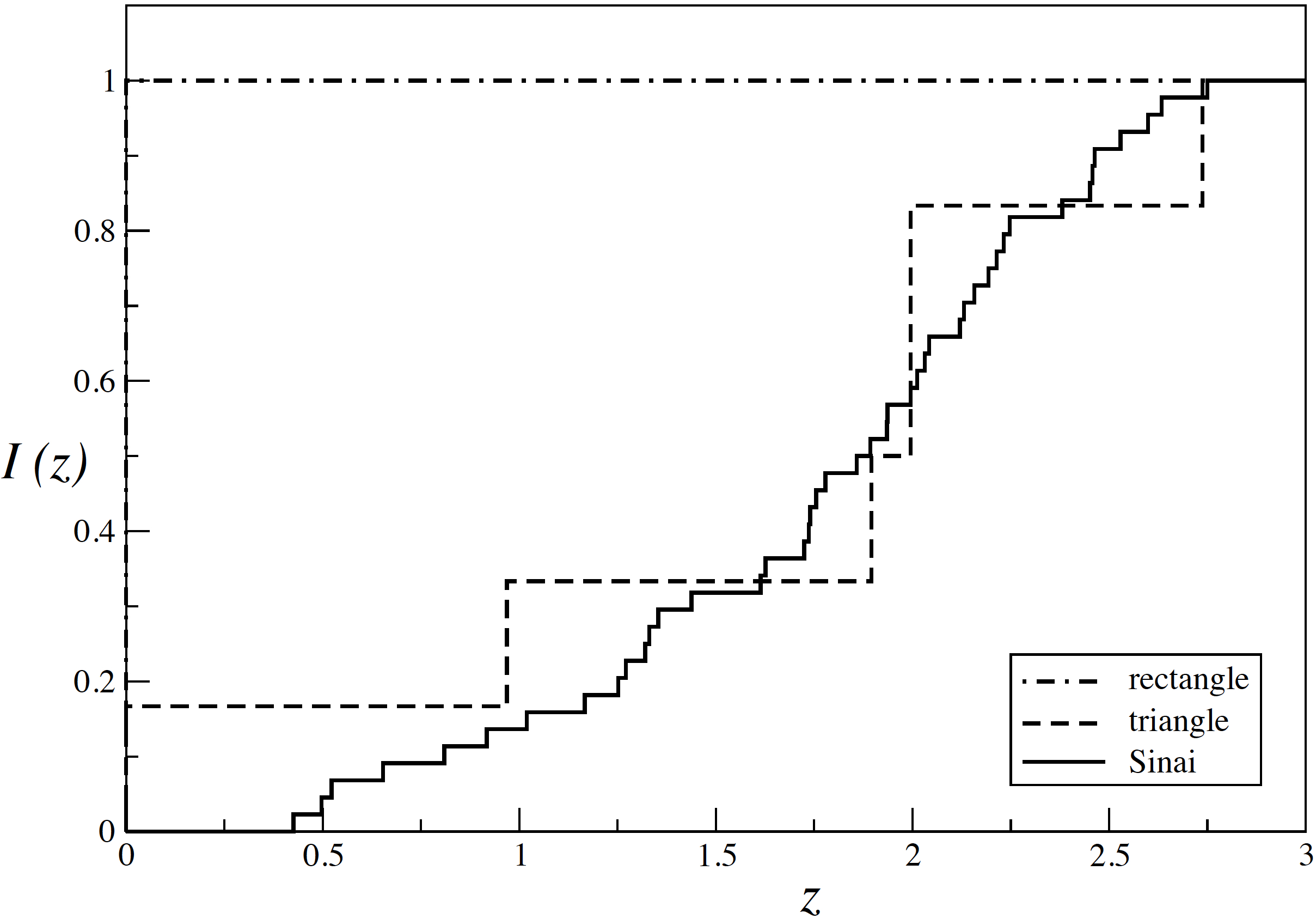}
\caption{\label{fig:avoidance} The number of avoided crossings normalized with the number of saddle points gives the cumulative distribution of the avoidance range, $I(z)$. The case of the rectangular billiard is trivial with $I(z) = 1$. The nodal lines of the equilateral triangle do display avoided crossings as it is a nonseparable billiard; the histogram thereof follows a trend matched closely by that for the chaotic Sinai billiard. A continuous curve drawn through the histogram for Sinai billiard is well approximated by the result obtained for the random wave model.  From \citet{monastra2002avoided}. Refer to \cite{monastra2003avoided} for further details.}
\end{figure}

The distribution function of $z$ is expected to capture the classical nature of the billiards under consideration. The normalized avoidance range distribution, $I(z)$, is simply the ratio of the number of avoided crossings $\leq $ $z$, $\tilde{I}(z)$, to the number of saddle points, $N_S$. The number of saddle points can be counted by recalling that $|\alpha _2({\bf r})|^2 - |\beta _0({\bf r})|^2 > 0$ and integrating over the domain $\mathcal{D}$ to obtain
\begin{eqnarray}
N_S &=& \frac{k^2}{4} \int_{\mathcal{D}} r\,\,\mathrm{d}\,r\,\,\mathrm{d}\,\theta\,\, \delta (\beta _1({\bf r}))\,\delta (\gamma _1({\bf r}))\,\left(\lvert \alpha _2({\bf r})\rvert ^2 - \beta^2 _0({\bf r})\right) \nonumber \\ &~& \times \Theta (|\alpha _2({\bf r})|^2 - \beta^2 _0({\bf r})).
\end{eqnarray}
Similarly, 
 \begin{eqnarray}
\tilde{I}(z) &=& \frac{k^2}{4} \int_{\mathcal{D}} r\,\mathrm{d}\,r\,\mathrm{d}\,\theta\,\, \delta (\beta _1({\bf r}))\,\delta (\gamma _1({\bf r}))(\lvert \alpha _2({\bf r})\rvert^2 - \beta^2 _0({\bf r})) \nonumber \\ &~& \times \Theta (|\alpha _2({\bf r})|^2 - \beta^2 _0({\bf r})) \Theta (z-z({\bf r})).
\end{eqnarray}
Equipped with the above, $I(z) = \tilde{I}(z)/N_S$ can be studied for different billiards. 

Closed-form expressions for $I(z)$ can be found if the eigenfunction is assumed to be drawn from a random wave ensemble. In this case, Gaussian integrations over the parameters $\beta _0, \beta _1, \beta _2, \gamma _1, \gamma _2 $ give the averaged values, $\langle N_S \rangle $ and $\langle \tilde{I}(z)\rangle $, leading to
\begin{equation}\label{eq:avoidance}
I_{\textsc{rwm}} (z) = \frac{3\sqrt{3}\,z^2\, (16 - z^2)^2}{(512 - 64\,z^2 + 3\,z^4)^{3/2}};\,\, 0 < z < 2\sqrt{2}.
\end{equation}
The trend displayed by the Sinai billiard in Fig. \ref{fig:avoidance} is closely mimicked by Eq.~\eqref{eq:avoidance}. The probability distribution of the avoidance $P(z) = \mathrm{d}\,I(z)/\mathrm{d}\,z$ exhibits linear repulsion. However, the proportionality constant is different from the case of the Gaussian Orthogonal Ensemble of random matrices \cite{mehta2004random}. The analogy with random matrix theory (RMT) is unsurprising given the results of \citet{johansson2002non} who considered the trajectories of $N$ non-intersecting Brownian particles on a line starting from $t=0$ and returning at $t = 2T$ to show that the distribution function of their nearest-neighbor spacings is the same as that for the eigenvalues of random matrices.  

More formally, the curvature of $u(r)$ can be defined  \cite{struik2012lectures} as the rate of turning of the
tangent of a contour line of $u$ at a point $\mathbf{r}$ \cite{berry2002statistics}, i.e.,
\begin{equation}
\kappa (r) = \frac{u_x^2\,u_{xx} + u_y^2\,u_{yy} - 2\,u_x\,u_y\,u_{xy}}{|\nabla u|^3}.
\end{equation}
The probability distribution of $\kappa $ is procured by appropriately weighting it over the nodal line length:
\begin{equation}
P(\kappa) = \frac{\langle \delta (u)\lvert\nabla u\rvert\delta (\kappa - \kappa (r))\rangle}{\langle \delta (u) \lvert\nabla u\rvert \rangle}.
\end{equation}
On the nodal curves, since the wavefunction is zero, $u_{xx} + u_{yy} = 0$. This fact, in combination with the polar representation of $\nabla u$ for Gaussian random waves, Eq.~\eqref{eq:isotropic}, yields the  distribution \cite{berry2002statistics}
\begin{equation}
P\left( \frac{\kappa}{k} \right) = \frac{4}{\pi \left[1 + \left( {\displaystyle \frac{\kappa}{k}}\right)^2 \right]^2}.
\end{equation}
This implies
\begin{equation}
\langle \lvert \kappa \rvert \rangle = \frac{k}{\pi} = \frac{2}{\lambda}; \qquad \sqrt{\left\langle \kappa ^2 \right\rangle} = \frac{k}{2} = \frac{\pi }{\lambda},
\end{equation}
which is a physically appealing result---the radii of curvature of the nodal lines are of the same order as the wavelength. Higher moments of $\kappa$ are divergent and correspond to regions where the nodal lines are strongly curved. 

\subsubsection{Complexity of the network of nodal lines}

\begin{figure}[htb]
\includegraphics[width=\linewidth]{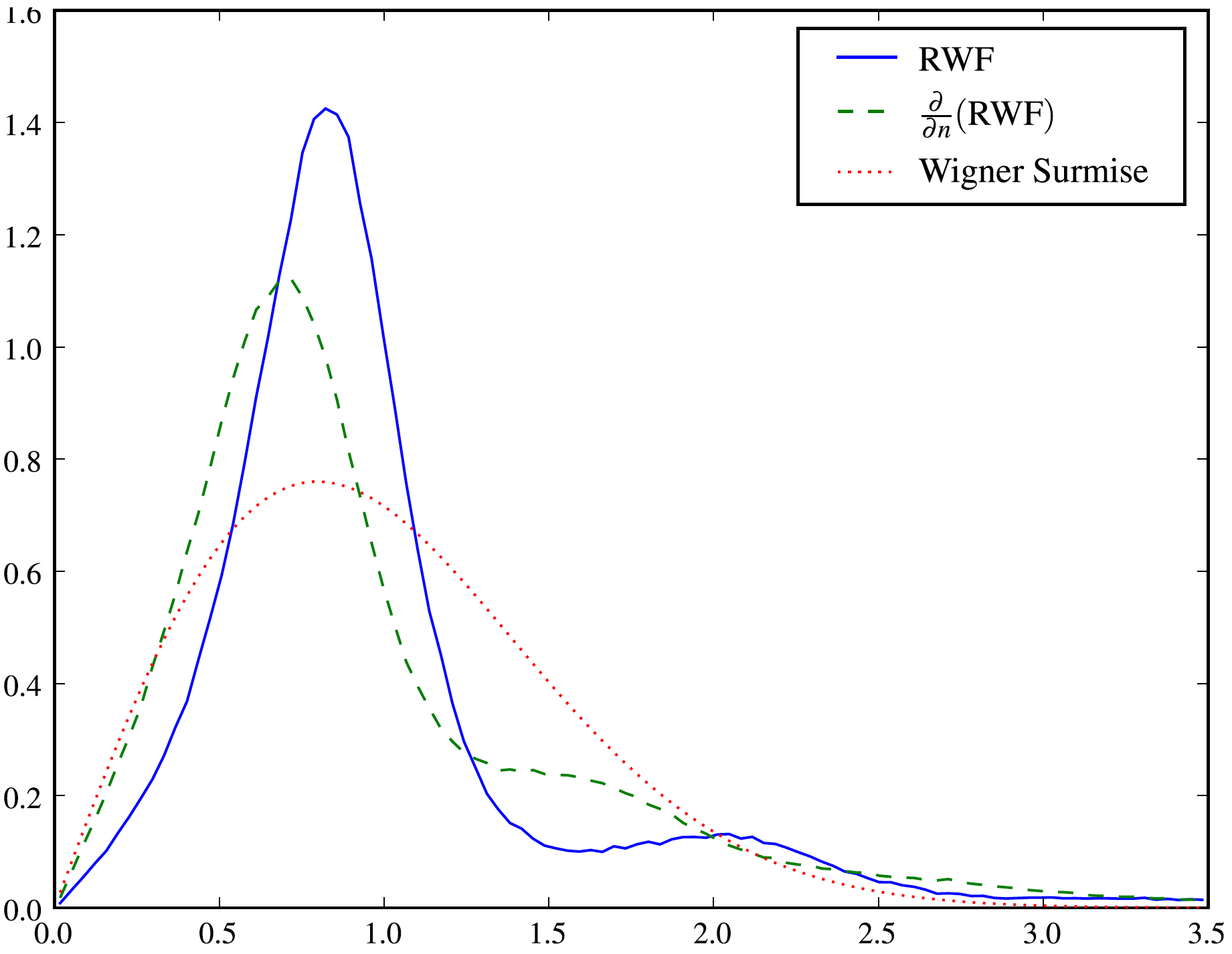}
\caption{\label{fig:intersections} Density $p\,(s)$ of the nearest-neighbour spacings $s$ between the points where a reference curve intersects a network of nodal lines created by a random wave (RW), the normal derivative of a random wave (NRW), or the Wigner surmise of RMT. For small distances, $p\,(s)$ grows linearly. On closer inspection, one can observe persistent oscillations relative to the mean decaying curve for the RW and the NRW. From \citet{aronovitch2006slides}.}
\end{figure}

Nodal lines form a serpentine network in nonseparable and nonintegrable billiards. One of the ways of quantifying such a network is to examine its intersection with a reference curve \cite{aronovitch2007statistics}. Interestingly, the first efforts to do so were made in the context of studying a moving surface of the sea in terms of the zeros of its wave pattern along a straight line \cite{longuet1957statistical}. In order to characterize the complex wave pattern belonging to a random wave ensemble, consider the points generated by the intersection of a reference curve with the nodal lines. The nearest-neighbor spacing statistics for these points can be found numerically but a naive comparison with the spacing statistics for the Gaussian Orthogonal or Unitary ensembles of RMT fails miserably. Fig.~\ref{fig:intersections} clearly illustrates the difference with the corresponding results for the GOE of RMT \cite{mehta2004random}. 

We can write the density of the points of intersection as the density of zeros of a function $f$:
\begin{equation}
\rho\, (t) = \sum_{i} \delta\, (t - t_i) = \sum_{i} \frac{\delta\, [f(t)]}{\lvert\partial\, f/\partial\, t\rvert},
\end{equation}
where $f(t_i) = 0$. Kac's Fourier representation \cite{kac1959probability} gives the mean density function:
\begin{equation}
\label{eq:kac}
\langle \rho (t) \rangle = \frac{1}{2\,\pi ^2}\int\hspace*{-0.2cm} \int_{-\infty}^{\infty} \frac{\mathrm{d}\,\xi\,\, \mathrm{d}\,\eta }{\eta ^2} \left \langle \mathrm{e}^{\mathrm{i}\,\xi\, f(t)} (1 - \mathrm{e}^{\mathrm{i}\,\eta\, \partial_t f(t)}) \right\rangle .
\end{equation}
This can be easily evaluated for Gaussian $f(t)$ and used to calculate the two-point correlation function:
\begin{eqnarray}
R(t, t') &=&\left \langle \sum_{i\ne j} \delta \,(t - t_i)\,\,\delta\, (t' - t_j)\right \rangle  \nonumber \\
&=& \langle \rho (t) \,\rho (t')\rangle - \delta (t - t')\,\langle \rho (t)\rangle .
\end{eqnarray}

\begin{figure}[htb]
\includegraphics[width=\linewidth]{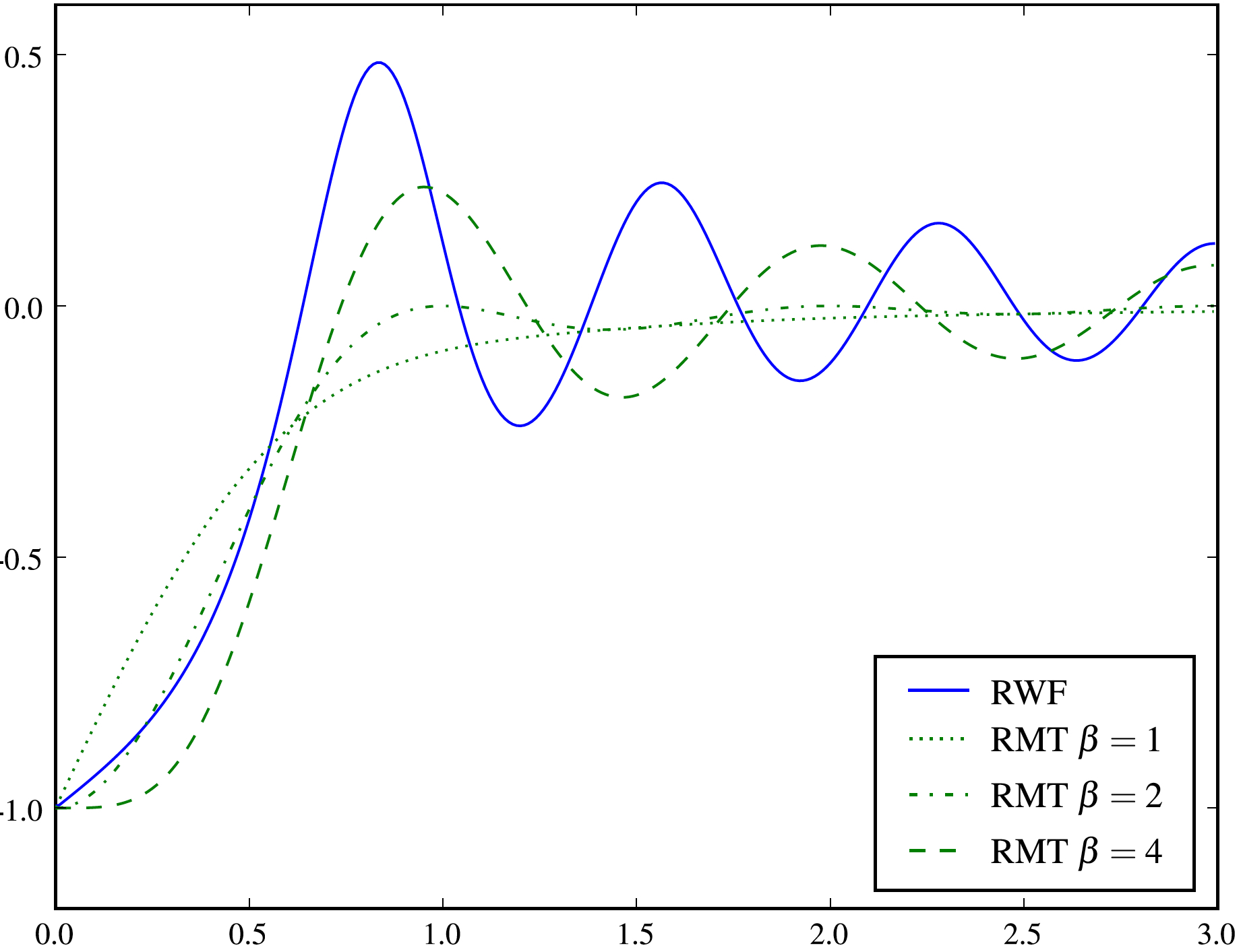}
\caption{\label{fig:correl}Normalized correlation: nodal intersections of RW versus RMT level spacings---GOE ($\beta = 1$), GUE ($\beta = 2$), and GSE ($\beta = 4$). From \citet{aronovitch2006slides}.}
\end{figure}

The random wave model is appropriate for excited states satisfying $k \gg \kappa$, where $\kappa$ is now the curvature of the reference curve. In this limit, the first such reference curve of relevance is a straight line. Restricting our discussion to this example, the average density is $\langle \rho \rangle = k/(\sqrt{2}\, \pi )$. 
In terms of $s = (t' - t)\,\langle \rho \rangle $, the normalized correlation function, ${\mathcal R}(s) = R/\langle \rho \rangle ^2 - 1$ exhibits the following behaviour \cite{aronovitch2007statistics}:
\begin{eqnarray}
{\mathcal R}(s) &\sim & -1 + \frac{\pi ^2}{16}s + \frac{37 \pi ^4}{2304} s^3 + \frac{\pi^4}{1296\sqrt{2}} s^4 +\ldots;\,s \to 0, \nonumber \\
&\sim & \frac{1}{2\sqrt{2}\,\pi^2\,s} \left[ 1 + 9\,\sin\, (2\sqrt{2}\,\pi\, s) \right];\, s \to \infty . 
\end{eqnarray}
Fig.~\ref{fig:correl} presents a comparison of these asymptotics along with the results expected from the three random matrix ensembles---the agreement is only on average.  

When working with chaotic billiards such as the (desymmetrized) Sinai and stadium billiards, one can choose a reference curve inside the domain and inspect the nodal intersections with this curve. The near-perfect agreement of the spacing statistics of nodal intersections between RWM and chaotic billiards \footnote{This was verified by simulations \cite{aronovitch2007statistics} based on 1500 (10000) eigenfunctions for the stadium (Sinai) billiard in the range of wavenumbers 110--165 (350--500); the numerically-obtained densities all lie precisely on the RWM (blue) curve in Fig.~\ref{fig:intersections}.} demonstrates the success of Berry's conjecture. In fact, for random Gaussian toral eigenfunctions in two and three dimensions, the expected intersection number (against any smooth curve) is \textsl{universally} proportional to the length of the reference curve times the wavenumber, but independent of the geometry \cite{maffucci2016nodal, rudnick2015nodal, rudnick2016nodal}.

\subsubsection{Density of line shapes}

One of the ways to find the distribution of the shapes of nodal lines is to compute the probability that a nodal line matches a given reference curve ${\bf r}\,(s)$ (parametrized by the arclength $s$ in the plane) within a certain precision $\varepsilon$ \cite{foltin2004morphology}. Of interest is the integral of the square of the amplitude of a random function $\psi ({\bf r})$,
\begin{equation}
X = \frac{1}{2} \int \mathrm{d}\,s \,\,\psi ^2\,({\bf r}\,(s));
\end{equation}
$X$ is also a random variable itself. When $\psi$ has a nodal line close to ${\bf r}\,(s)$, $X$ is small. \citet{foltin2004morphology} calculated the distribution of $X$ and its cumulants for a circular reference curve, ${\cal C}$, for the random wave and short-range ensembles (corresponding to critical percolation). The hope was that the scaling properties of these moments as a function of the radius (size) of ${\bf r}\,(s)$ would detect the long-range correlations in $\psi$. An approximate expression for the probability that a nodal line falls inside a tube of width $\varepsilon$ is thus considered. 

\begin{figure}[htb]
\includegraphics[width=0.75\linewidth]{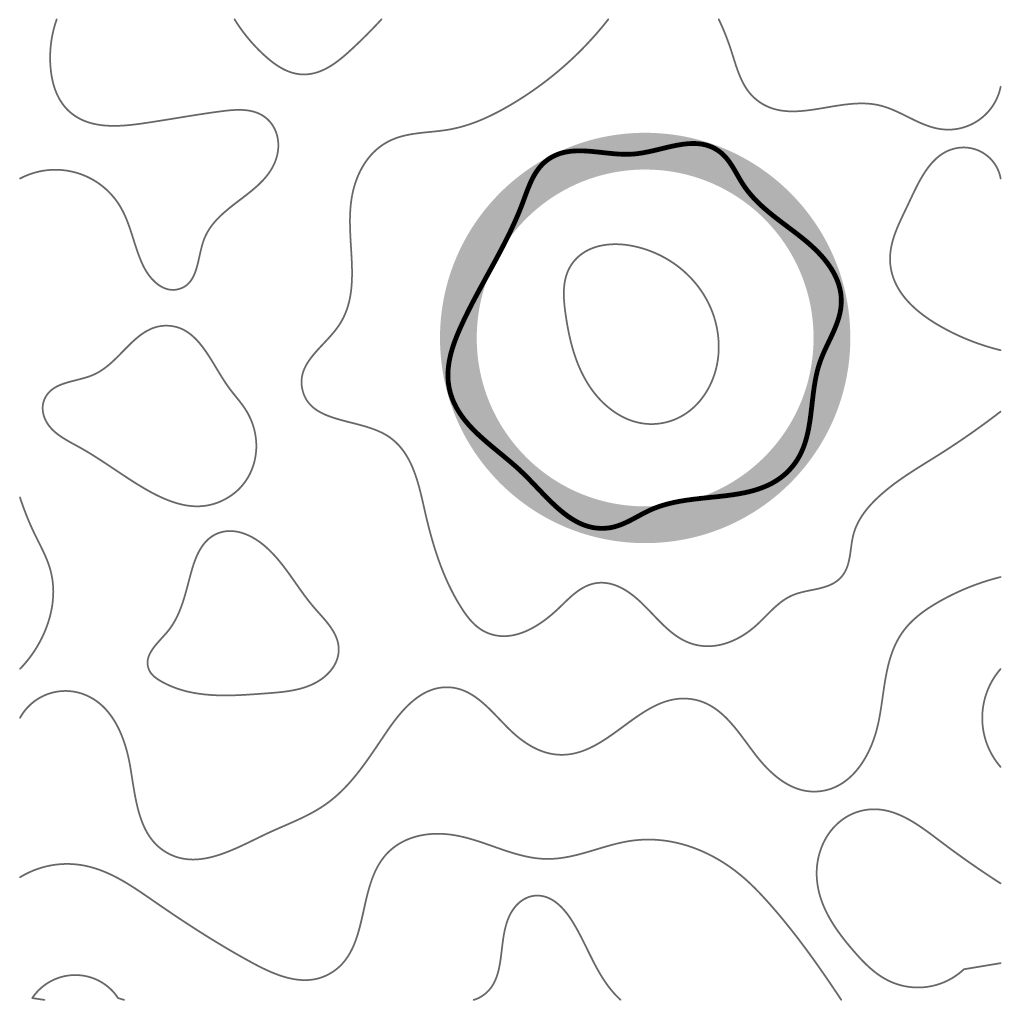}
\caption{\label{fig:tube}A section of the nodal set of a random wavefunction. One of its nodal lines lies within the prescribed thin circular tube, i.e., this configuration contributes to the density $\rho$. From \citet{foltin2004arxiv}. Refer to \cite{foltin2004morphology} for further details.}
\end{figure}

Assuming that $\psi ({\bf r})$ has a nodal line close to ${\cal C}$, the normed distance $\zeta\, (s)$ of the nodal line from the reference curve can be obtained by linearization
\begin{alignat}{1}
\psi\, ({\bf r} + \zeta\, {\rm n}) & \simeq \zeta\, \frac{\partial\, \psi \,({\bf r})}{\partial \,\mathrm{n}} + \psi \,({\bf r}) = 0; \,\mbox{ wherefore, } \\ 
\zeta &= - \frac{\psi \,({\bf r})}{\partial \,\psi\, ({\bf r})/\partial \,\mathrm{n}},
\end{alignat}
$\hat{\rm n}\,(s)$ being the unit vector normal to ${\cal C}$. Finding the probability that a nodal line lies within a tube $\lvert \zeta\, (s) \rvert < d$ by analytical means alone is difficult. We thus diffuse the boundary of the tube and consider instead the expectation value
\begin{equation}
P_{\varepsilon} = \bigg{\langle} \exp \left( -\frac{1}{2\varepsilon} \int \mathrm{d}\,s~ \zeta ^2 \right) \bigg{\rangle}
\end{equation}
where the line integral is along ${\cal C}$ and $\varepsilon \sim d^3$. However, we need to simplify even further as the numerator and denominator of $\zeta $ are not independent, implying that $\eta$ itself need not be Gaussian distributed. Thus, we consider an isotropic, mean-field expression for the shape probability which can be evaluated:
\begin{alignat}{1}\label{eq:approx_prob}
P_{\varepsilon} &= \bigg{\langle} \exp \left( -\frac{1}{\varepsilon} \int \mathrm{d}\,s~ \frac{\psi ^2}{\langle (\nabla \psi)^2 \rangle} \right) \bigg{\rangle} \nonumber \\ 
&= \det \left( 1 + \frac{\hat{B}}{\varepsilon\, \langle (\nabla \psi)^2 \rangle /2} \right)^{-1/2}.
\end{alignat}
$\hat{B}$ is an integral operator with symmetric kernel, which gives the correlation function of $\psi ({\bf r}(s))$,
\begin{equation}
B(s, s') = \langle \psi ({\bf r}(s))\,\psi ({\bf r}(s')) \rangle = G\,(\lvert{\bf r}\,(s) - {\bf r}\,(s')\rvert).
\end{equation}
Since the operator is positive semi-definite with a finite trace, $\int \mathrm{d} s\, B\,(s,s) = L$, its eigenvalues $\beta _{\mu} \geq 0$ have an accumulation point at zero. Thus we arrive at the final form of the generating function,
\begin{alignat}{1}
F(\varepsilon) &\equiv \log P_{\varepsilon} = -\frac{1}{2}\sum_{\mu} \log \left( 1 + \frac{\beta _{\mu}}{\varepsilon\, \langle (\nabla \psi)^2 \rangle /2} \right), \nonumber \\
&= \log\, \langle \exp(-X/\tilde{\varepsilon})\rangle,  
\end{alignat}
which entails all the cumulants upon expansion in powers of $\tilde{\varepsilon} = \varepsilon\, \langle (\nabla \psi)^2 \rangle /2$.

It is well-known \cite{berry1977regular,hortikar1998} that for a random wave ensemble (RWE), the correlation function is $G_{RW}\,(r) = \langle \psi ({\bf r})\, \psi (0)\rangle = J_0(k\,r)$ whereas for the short-range ensemble (SRE), it is $\exp (-k^2r^2/4)$ \cite{foltin2004morphology}. The probability \eqref{eq:approx_prob} can now be calculated for a circle of radius $R$ for both. The kernel of the integral operator corresponding to the correlation function $G_{RW}\,(r)$ is $B(\theta - \theta ') = J_0\,(2\,k\,R\,\sin (\theta - \theta ')/2)$ for two positions $\theta , \theta '$ on the circle. The eigenfunctions of $\hat{B}$ are $\exp (\mathrm{i}\,m\,\theta)$; $m = 0, \pm 1, \pm 2, \ldots$, with eigenvalues
\begin{alignat*}{1}
\beta _m = R\,\int_{0}^{2\pi} \mathrm{d}\,\theta\,\, J_0(2\,k\,R\,\sin (\theta /2))\,\mathrm{e}^{\mathrm{i}\,m\,\theta} = 2\pi\, R\,J_m^2(k\,R). 
\end{alignat*}
This long-range character stands in contrast to the situation with SRE where the eigenvalues are
\begin{alignat*}{1}
\beta _m = 2\pi\, R\,\mathrm{e}^{-k^2\,R^2/2}I_m(k^2R^2/2) \approx \frac{2\sqrt{\pi}}{k}\mathrm{e}^{-m^2/(k\,R)^2}.
\end{alignat*}
The spectrum is seen to behave smoothly for SRE whereas there are strong fluctuations for RWE. 

The generating function $F(\varepsilon )$ can now be expressed for different regions where $m < k\,R$ or $m \approx k\,R$. Let us scrutinize RWE first: the asymptotic expansions of the Bessel functions in both regimes can be combined into a single scaling law with a universal scaling function $f(x)$ \cite{foltin2004morphology} as
\begin{equation}
\lvert J_m(k\,R) \rvert \sim (k\,R)^{-1/3}f\left( \frac{m^2 - (k\,R)^2}{(k\,R)^{4/3}} \right). 
\end{equation}
Thus, the eigenvalues of the operator scale as
\begin{equation}
\beta _m = \frac{2\pi}{k}(k\,R)^{1/3}\left[f\left( \frac{m^2 - (k\,R)^2}{(k\,R)^{4/3}} \right)\right]^2.
\end{equation}
The leading behavior of the $\upsilon^\mathrm{th}$ cumulant of $X$, which is proportional to the trace of the $\upsilon^\mathrm{th}$ power of $\hat{B}$, can be simplified as a function of $kR$ (to first-order therein):
\begin{alignat}{1}
\label{eq:scaling}
\langle X^{\upsilon} \rangle _c \sim k^{-\upsilon} \times 
\begin{cases}
k\,R, \qquad \qquad  &\upsilon < 2,\\
k\,R\, \log\, (k\,R), \qquad &\upsilon = 2,\\
(k\,R)^{(1+\upsilon )/3}, \qquad &\upsilon > 2.
\end{cases}
\end{alignat}
For the SRE, the scaling behavior is \cite{foltin2004morphology}
\begin{equation}
\langle X^{\upsilon} \rangle _c \sim k^{-\upsilon}\,k\,R; \qquad \forall\,\, \upsilon > 0.
\end{equation}
Clearly, below the critical exponent $\upsilon = 2$, the two results agree for large $k\,R$. For $\upsilon > 2$, Eq.~\eqref{eq:scaling} suggests a rather nontrivial scaling of the cumulants for the RWE. Numerical computations for the case of $\upsilon = 3$ also corroborate the scaling relation, with an exponent $4/3$.

\section{Counting nodal domains}
\label{sec:count}

\subsection{Can one count the shape of a drum?}

By now, we can all unanimously agree that the shape of a domain (the geometry of its boundary) is wedded to and determines its Laplacian eigenspectrum. One can turn the tables to pose the question in reverse: does the set of eigenvalues (or emitted frequencies, if we are talking about a vibrating drum) uniquely identify the domain? Or as \citet{kac1966can} put it more colorfully, ``can one hear the shape of a drum?'' A short answer is yes; a shorter answer is no.

The connection between isometry and isospectrality is a question that has both perplexed and fascinated physicists and mathematicians alike. Upon a moment's thought, the nontriviality of the problem becomes apparent. Most definitely, certain geometrical and topological constants associated with the domain, say, $\mathcal{D}$, can indeed be drawn from the spectrum. For instance, \citet{kac1966can} himself conjectured the asymptotic relation for the heat trace, also called the spectral function, 
\begin{alignat}{1}
\label{eq:heat}
\nonumber H (t) &= \mathrm{Tr} \left (\mathrm{e}^{t\,\Delta_\mathcal{D}} \right) = \sum_{j=1}^\infty \mathrm{e}^{- \lambda_j t}\\
&\sim \frac{\mathcal{A}}{4 \pi\, t} -  \frac{\mathcal{P}}{8 \sqrt{\pi\, t}} + \frac{1}{6} (1 - h) \quad \mbox{ as } t \rightarrow 0,
\end{alignat}
where $h$ is the number of holes (genus) in $\mathcal{D}$. The heat trace is thus a spectral invariant. The first term in Eq.~\eqref{eq:heat} is essentially Weyl's law \cite{vaa2005weyl} whereas the second and third components follows from the results of \citet{pleijel1954study} and \citet{mckean1967curvature}, respectively. Similarly, \citet{van1988heat} found that for any polygon in $\mathbb{R}^2$ with angles $\alpha_ i$, the heat trace can be written as
\begin{equation}
\sum_{j=1}^\infty \mathrm{e}^{- \lambda_j t} \sim \frac{\mathcal{A}}{4 \pi\, t} -  \frac{\mathcal{P}}{8 \sqrt{\pi\, t}} + \frac{1}{24} \sum_i \left(\frac{\pi}{\alpha_i} - \frac{\alpha_i}{\pi}\right) + \mathcal{O} \left(\mathrm{e}^{-\frac{\mathcal{C}}{t}} \right).
\end{equation}
However, besides information about $\mathcal{P}$ and $\mathcal{A}$, both Eq.~\eqref{eq:Weyl} and Eq.~\eqref{eq:kac} are silent about retrieving the shape itself. Moreover, unlike with isoperimetric inequalities, here, we are grappling with the whole spectrum rather than individual eigenvalues, which renders the problem fundamentally different from any we have encountered up to this point.

\begin{figure}[htb]
\includegraphics[width=0.45\linewidth, trim={5cm 8cm 4.5cm 8cm}, clip]{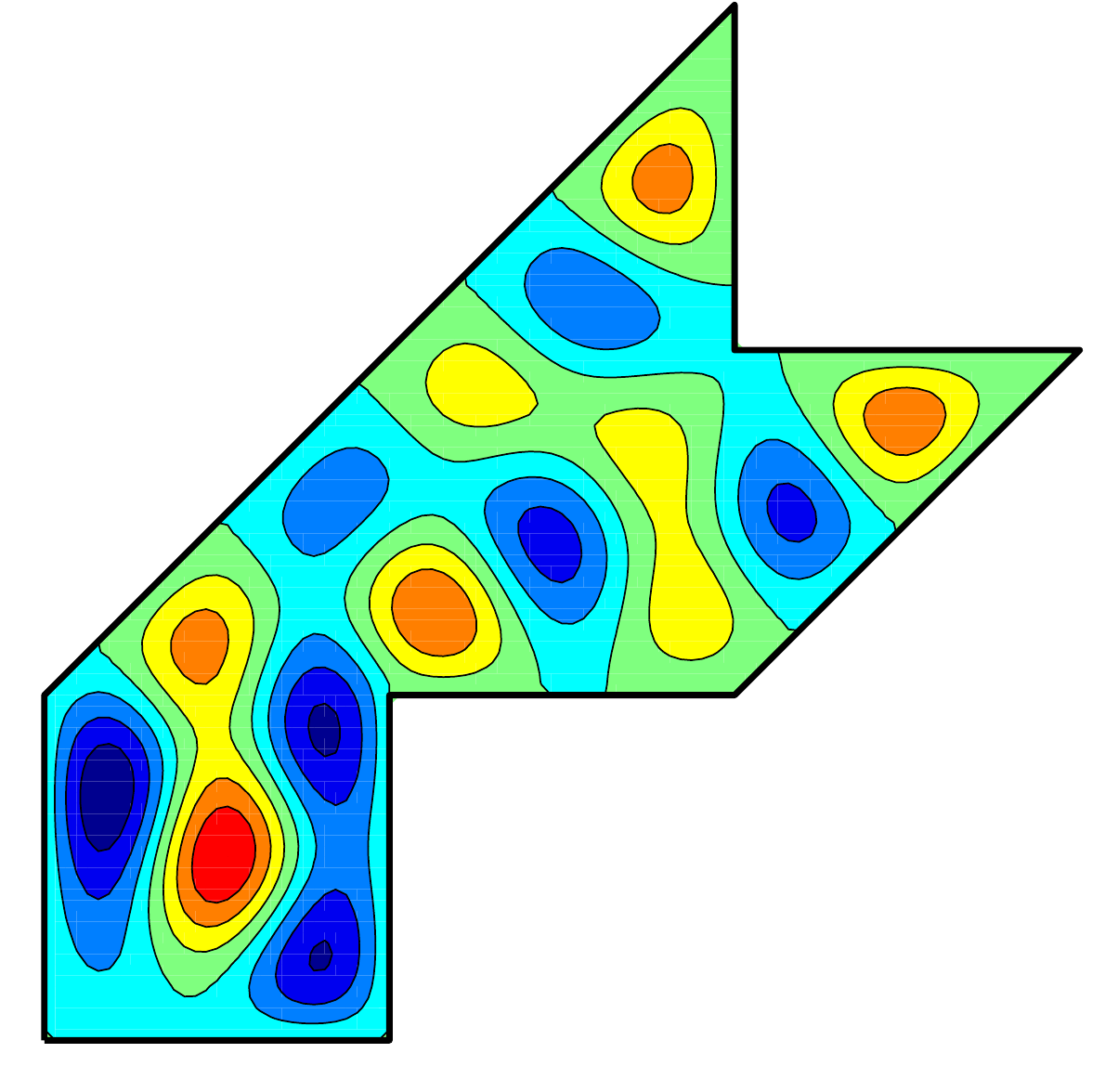}
\includegraphics[width=0.45\linewidth, trim={5cm 8cm 4.5cm 8cm}, clip]{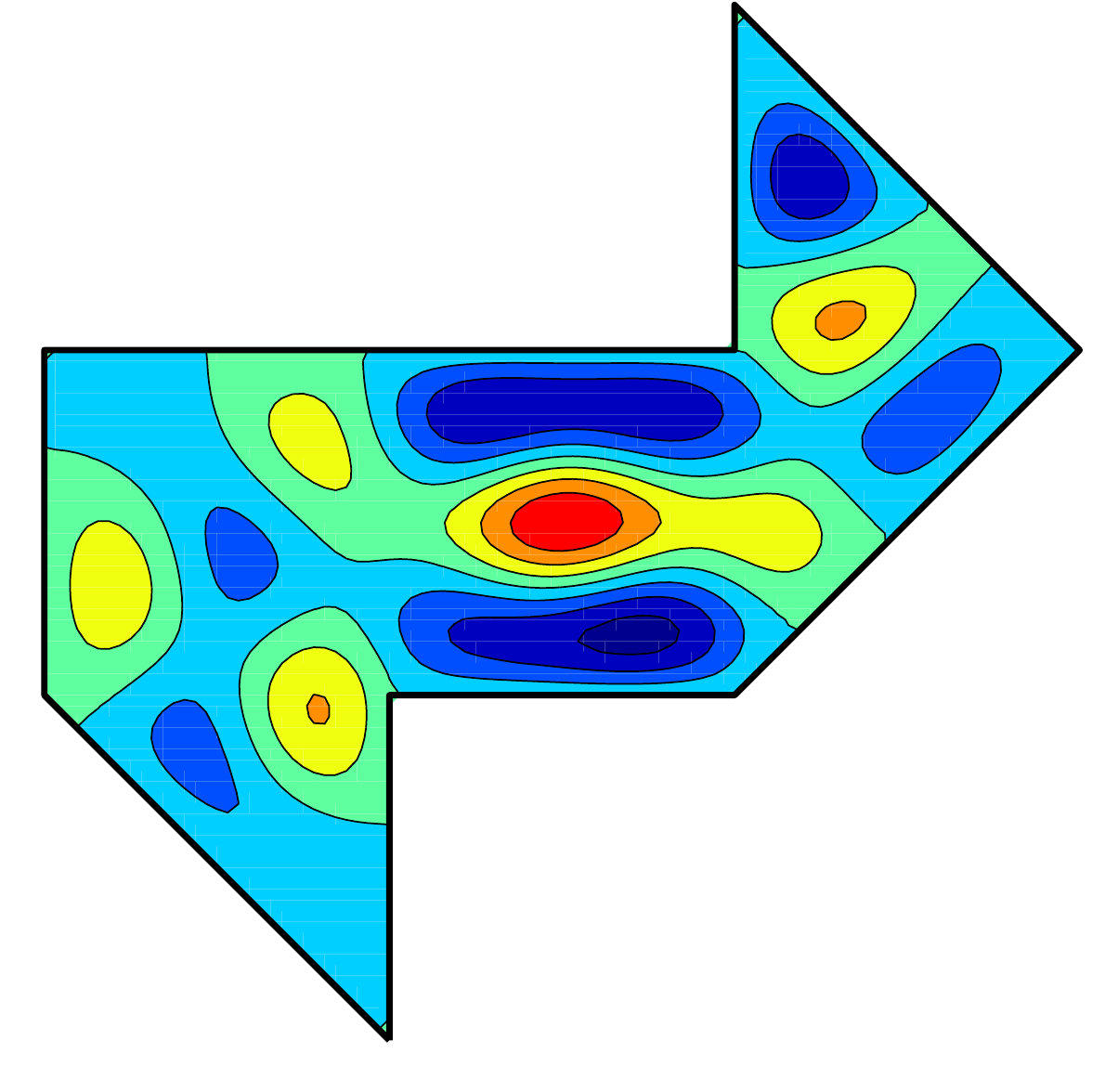}
\includegraphics[width=0.45\linewidth, trim={5cm 8cm 4.5cm 8cm}, clip]{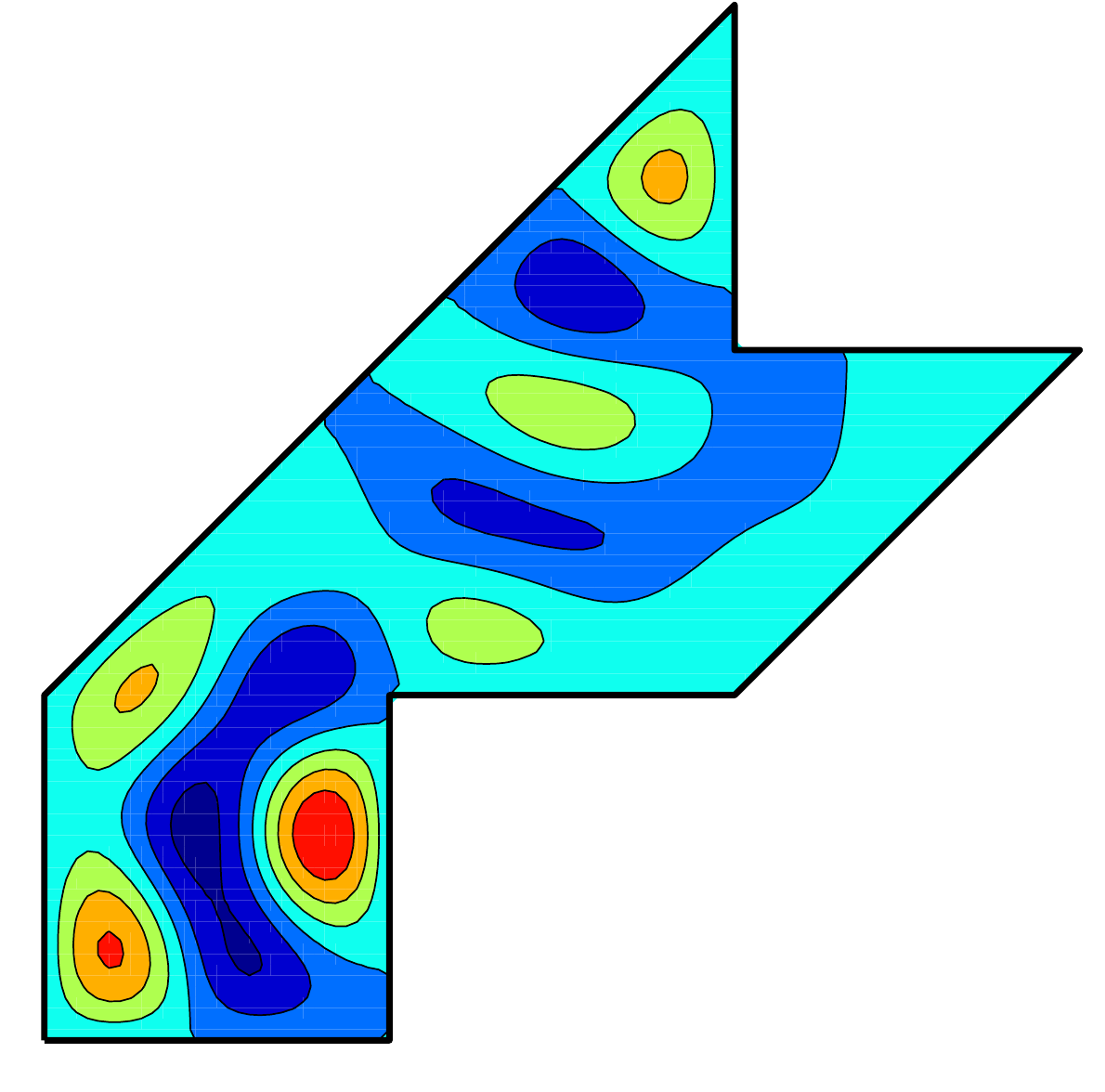}
\includegraphics[width=0.45\linewidth, trim={5cm 8cm 4.5cm 8cm}, clip]{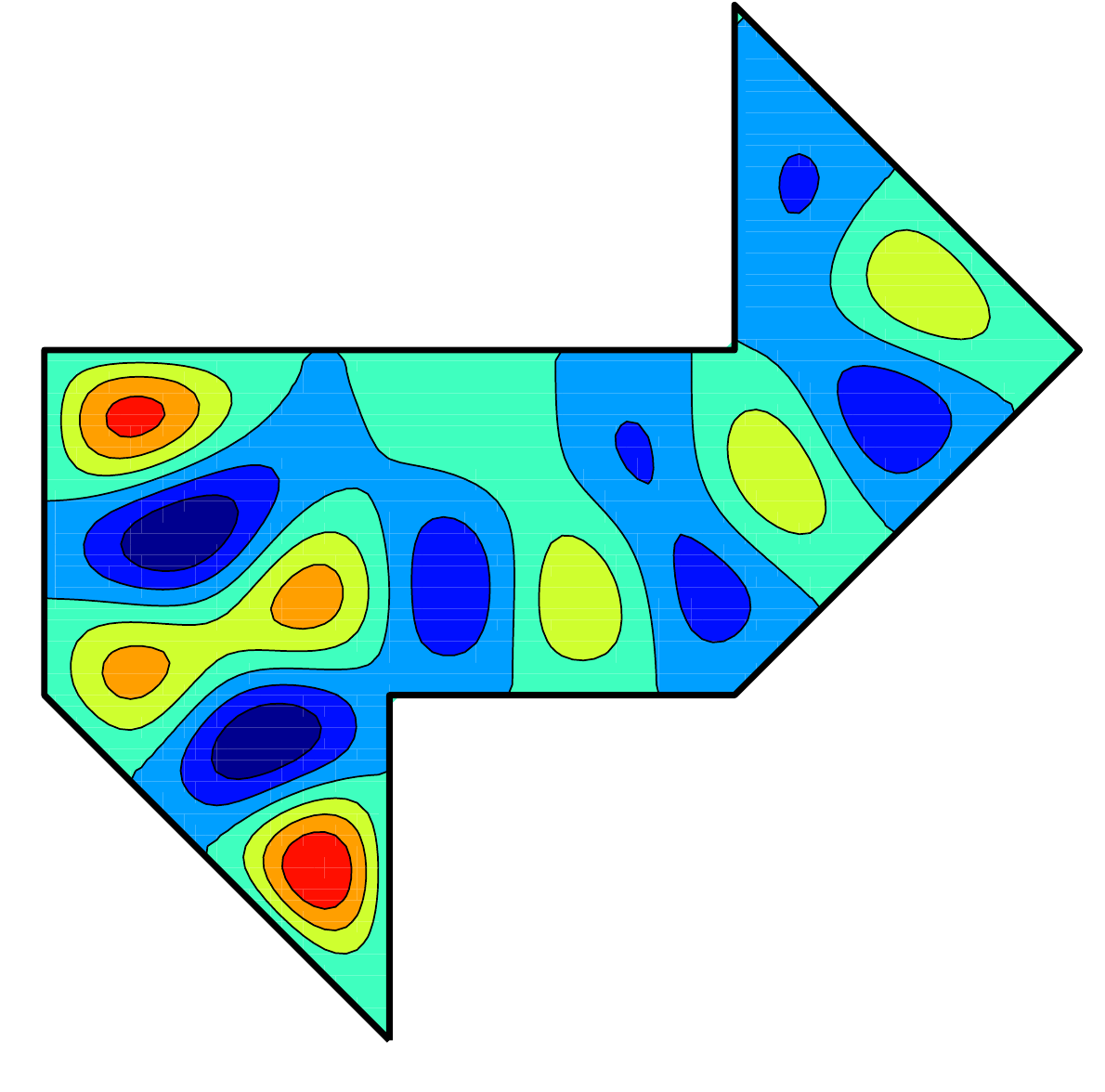}
\caption{\label{fig:count}Two domains that are isospectral but not isometric in the sense of geometric congruence. The $19^\mathrm{th}$ (top) and $20^\mathrm{th}$ (bottom) eigenfunctions are shown. Isospectraity can be proved by the transplantation technique \cite{berard1992transplantation}. Note that the isospectral billiards have the same area and perimeter. Figure generated from code by \citet{moler2012}.}
\end{figure}

Without further ado, let us illustrate why, formally, the answer to Kac's question is in the negative. After a string of early counterexamples \cite{protter1987can, milnor1964eigenvalues, ikeda1980lens, vigneras1980varietes, urakawa1982bounded} in $\mathbb{R}^n$, $n \ge 4$, the fate of the conjecture for planar domains was finally sealed by the discovery of a pair of two-dimensional polygonal billiards \cite{gordon1992isospectral, gordon1992one} having exactly the same eigenspectrum (Fig.~\ref{fig:count}), for both Dirichlet and Neumann boundary conditions. This seemingly serendipitous finding was actually based on \citeauthor{sunada1985riemannian}'s \citeyearpar{sunada1985riemannian} systematic ``paper-folding'' construction. Many other instances of isospectral domains were soon provided by \citet{buser1994some, chapman1995drums}, including experimental realizations by \citet{sridhar1994experiments} (see also \citet{cipra1992you}). Nonetheless, only 17 families of examples that say no to Kac's question were found in a 40-year period \cite{giraud2010hearing}. All these specimens, however, were either disjoint or nonconvex. For $d \ge 4$, \citet{gordon1994isospectral} put this objection to rest by showing the existence of isospectral convex connected domains but the question remains open for billiards in $d=2$.

At the very onset, we had, somewhat tantalizingly, hinted that Kac's question might also have an affirmative answer. This is indeed true for certain classes of domains. \citet{zelditch2000spectral} proved that  for domains possessing elliptical symmetry, the spectrum of the Dirichlet Laplacian uniquely determines the region, subject to the satisfaction of some generic conditions on the boundary. Analogous results hold for real analytic, planar domains with only one symmetry \cite{zelditch2004inverse, zelditch2009inverse}. Moreover, for rectangular or triangular billiards, a finite number of eigenvalues suffice to completely specify the shape \cite{chang1989hearing}. In fact, two isospectral triangles must necessarily be isometric \cite{grieser2013hearing}. The prevailing belief is that a smooth boundary on the domain $\mathcal{D}$ is a sufficient condition for a positive answer to Kac's question; unfortunately this is yet to be proved.

Even if, disappointingly enough, one cannot ``hear'' the shape of a drum, one can still ``count'' the shape to address the question of isospectrality. This scheme was first developed by \citet{gnutzmann2005resolving, gnutzmann2006can} who provided a combination of heuristic arguments and numerical simulations to support the proposition that sequences of nodal counts store information on the geometry (metric) of the domain where the wave equation is considered. To put it explicitly, the information contained in the nodal count sequence is \textsl{different} from that borne by the spectral sequence. Thus the spectral ambiguity---and its associations with isometry---can be resolved by comparing nodal sequences, as we discuss in the following subsection.

\subsection{Trace formula approach}

The nodal count sequence for separable Laplacians can be described by a semiclassical trace formula \cite{gnutzmann2005resolving, gnutzmann2006can, gnutzmann2007trace}. This is analogous to the \citet{gutzwiller1990chaos} spectral trace formula, which relates the quantum density of states, $g\,(E)$, to its semiclassical counterpart, $g_\textsc{sc}$,
\begin{equation*}
g \,(E) = \sum_{j=0}^\infty \delta(E - E_j) \leftrightarrow g_\textsc{sc} (E) = g_0 (E) + \sum_{\alpha \,\in\, po} A_\alpha \mathrm{e}^{\mathrm{i} \,k\,L_{\alpha}},
\end{equation*}
as a sum over all the periodic orbits (of length $L_\alpha$) of the corresponding classical system. Likewise, the trace formula for the nodal counts can be expressed as sums over closed ray trajectories on the
manifold where each term bears geometric information about the orbit \cite{aronovitch2012nodal}. The approach is particularly simple for flat tori (in $\mathbb{R}^2$ and $\mathbb{R}^4$) and surfaces of revolution.

\subsubsection{Flat tori}

The simple 2-dimensional torus can be represented as a rectangular billiard in $\mathbb{R}^2$ with length $a$ and breadth $b$ ($\tau = a/b \notin \mathbb{Q}$) with periodic boundary conditions. The number of nodal domains is, trivially,
\begin{equation}
\nu _{m,n} = (2 \lvert n\rvert + \delta _{n,0})\,(2 \lvert m \rvert + \delta _{m,0}).
\end{equation}
The cumulative density of energy levels affords an asymptotic expression in terms of classical periodic orbits, obtained from saddle-point approximations of all oscillatory integrals:
\begin{alignat}{1}
\label{eq:trace_rect}
\mathcal{N} (E) 
& = \sum_{m, n = -\infty}^{\infty} \Theta \,\left( E - E_{m,n} \right)\\
&= {\mathcal A}\,E + \sqrt{\frac{8}{\pi}}{\mathcal A}\,E^{1/4}\sum_{po} \frac{\sin (L_{po}\sqrt{E} - \pi /4)}{L_{po}^{3/2}}\nonumber \\ &+ {\mathcal O}\,(E^{-3/4}),
\end{alignat}
where ${\mathcal A} = a\,b/4\pi$ and $L_{po}= \sqrt{(Na)^2 + (Mb)^2}; N, M \in \mathbb{Z}$, is the length of a periodic orbit. Similarly, the cumulative nodal count
\begin{equation}
C(K) = \sum_{j = 1}^{\lfloor K \rfloor} \nu _j, \mbox{ for } K > 0,
\end{equation}
after modification to set a unique order within the degenerate states, can be brought to the form
\begin{alignat}{1}\label{eq:ctilde_rect}
&\tilde{c}\,(E) = \sum_{j=1}^{\infty} \nu _j\, \Theta (E - E_j) \nonumber \\
&= \frac{2{\mathcal A}^2}{\pi ^2} E^2 + \frac{2^{11/2}{\mathcal A}^3}{\pi ^{1/2}} E^{5/4} \sum_{po} \frac{\lvert MN\rvert}{L_{po}^{7/2}} \sin \left( L_{po}\sqrt{E} - \frac{\pi}{4} \right) \nonumber \\ &+ {\mathcal O}\,(E).
\end{alignat}
The next step is to invert $\mathcal{N}(E) = K$ with $\mathcal{N}(E)$ from Eq.~ \eqref{eq:trace_rect}, thereby yielding, to leading order,
\begin{equation}
E\,(K) = \frac{K}{\mathcal{A}} - K^{1/4} \frac{2^{3/2}}{\mathcal{A} \sqrt{\pi}} \sum_{po} {\displaystyle \frac{\sin \left(L_{po} \sqrt{\frac{K}{\mathcal{A}}} - \frac{\pi}{4}\right) }{(L_{po}/\sqrt{\mathcal{A}})^{3/2}} },
\end{equation}
which we can now substitute in Eq.~\eqref{eq:ctilde_rect}. Thus, the cumulative counting function $c\,(K) \equiv \tilde{c}\,(E(K))$ is formulated as the sum of an average part, $c_{\rm av}(K)$, and an oscillating part, $c_{\rm osc}(K)$, where
\begin{alignat}{1}\label{eq:trace_tori}
c_{\rm av}(K) &= \frac{2}{\pi ^2} K^2 + {\mathcal O}(K), \nonumber \\ 
c_{\rm osc}(K) &= \sum_{po} \frac{1}{L_{po}^{3/2}}\left( \frac{4\pi^2 \lvert NM \rvert }{L_{po}^2/{\mathcal A}} - 1\right) \sin \left( L_{po}\sqrt{\frac{K}{A}} - \frac{\pi}{4}\right)\nonumber\\
&\times K^{5/4}\sqrt{\frac{2^7{\mathcal A}^{3/2}}{\pi ^5 }}  + {\mathcal O}\,(K). 
\end{alignat}
The oscillatory part of $c\,(K)$ is especially interesting as its Fourier transform with respect to $\sqrt{K}$ entails the length spectrum of the periodic orbits in contrast to the smooth part, which is independent of the geometry of the torus. In fact,  $c_{\rm osc}(K)$ explicitly depends on the aspect ratio $\tau = a/b$, through $L_{po}$, and can therefore resolve different geometries. A kindred calculation in $\mathbb{R}^4$ \cite{gnutzmann2005resolving} insinuates that ``one can count the shape of a drum (if it is designed as a flat torus in four dimensions).''

\subsubsection{Surfaces of revolution}

A surface of revolution ${\mathcal M}$ is the collection of points traced out on rotating a curve $y = f(x)$ for $x \in [-1, 1]$ about the $x$-axis. We concern ourselves only with surfaces which are smooth and convex and, in particular, focus on mild deformations of ellipsoids of revolution, in essence, imposing that $\mathcal{M}$ has no boundary and $f(x)$ possesses a single maximum at $x_{\rm max}$. 

\begin{figure}[htb]
\includegraphics[width=\linewidth]{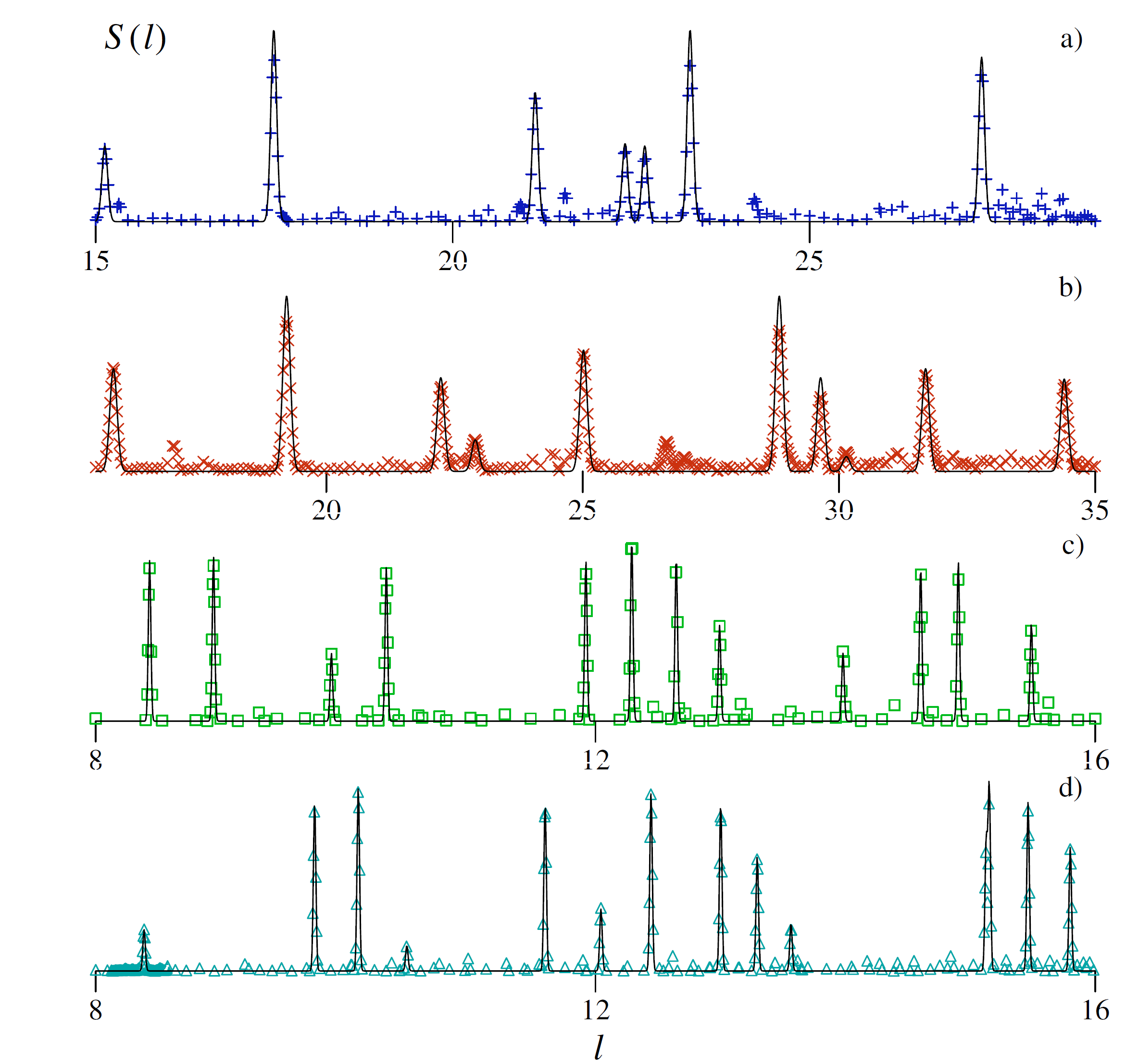}
\caption{\label{fig:length} The absolute value of the length spectra for ellipsoids and flat tori. The ellipsoids correspond to surfaces of revolution for $f(x) = R\sqrt{1 - x^2}$ with $R = 2$ (a) and $1/2$ (b). For the tori chosen, $\tau^2 = 2$ (c) and $\sqrt{2}$ (d). Points represent the numerical data of the length spectra. The continuous line is obtained by the Fourier transform of the trace formulae, Eqs.~\eqref{eq:trace_tori} and \eqref{eq:c}. From \citet{gnutzmann2007trace}. With kind permission of The European Physical Journal (EPJ).}
\end{figure}

The Euclidean metric in $\mathbb{R}^3$ induces a metric on the surface
\begin{equation}\label{eq:metric}
\mathrm{d}\,s^2 = \left[ 1 + \left\{ \frac{\partial f}{\partial\, x}^2 \right\} \right]\, \mathrm{d}\,x^2 + [f(x)]^2 \,\mathrm{d}\,\theta ^2;
\end{equation}
$\theta$ is the azimuthal angle. The wave equation on the surface of revolution is simply $(\nabla _{\mathcal M}^2 + E)\, \psi (x, \theta) = 0$ where 
\begin{equation}
\nabla _{\mathcal M}^2 = \frac{1}{f(x)\,\sigma (x)}\frac{\partial}{\partial \,x} \frac{f(x)}{\sigma (x)}\frac{\partial}{\partial\, x} + \frac{1}{[f(x)]^2}\frac{\partial ^2}{\partial\, \theta ^2}
\end{equation}
with $\sigma (x) = \sqrt{1 + (\partial f/\partial x)^2}$. The wave equation is separable with solutions
\begin{alignat}{1}
\psi\, (x, \theta) = 
\begin{cases}
\cos (m\,\theta )\, \phi _m(x), \qquad &m \geq 0,\\
\sin (m\,\theta )\, \phi _m(x), \qquad &m < 0,
\end{cases}
\end{alignat}
in which $\phi _m(x)$ satisfies the Sturm-Liouville equation:
\begin{equation*}
-\frac{1}{f(x)\,\sigma (x)}\frac{\mathrm{d}}{\mathrm{d}\,x}\frac{f(x)}{\sigma (x)}\frac{\mathrm{d}\,\phi _m(x)}{\mathrm{d}\,x} + \frac{m^2}{[f(x)]^2}\phi _m(x) = E\,\phi _m(x).
\end{equation*}
For each $m$, one has a sequence of solutions $\phi _{n,m}$, $n = 0, 1, 2, \ldots$, with eigenvalues $E_{n,m}$. The nodal domains, of which there are
\begin{equation}
\nu _{n,m} = (n+1)\,(2 \lvert m\rvert  + \delta _{m,0}),
\end{equation}
are arranged in a checkerboard pattern. A semiclassical treatment leads to an asymptotic expression (large-wavenumber expansion) for the cumulative density of energy levels in terms of the classical periodic orbits formed by a rational ratio of windings $(M, N)$ along the $(\theta , x)$ directions \cite{gnutzmann2007trace}:
\begin{equation}
\mathcal{N}(E) = {\mathcal A}\,E + E^{1/4}\hspace*{-0.4cm}\sum_{(M, N) \neq (0, 0)}\hspace*{-0.4cm} \mathcal{N}_{M,N}(E);
\end{equation}
note that the area of the surface is now $4\pi\, {\mathcal A}$. In the stationary phase approximation \cite{bleher1994distribution}
\begin{equation}
\label{eq:N_MN}
\mathcal{N}_{M,N}(E) = (-1)^N \frac{\sin (L_{M,N}\sqrt{E} + \sigma\, \pi /4)}{2 \pi\, \big \lvert N^3 \, n_{M,N}'' \vert_{m = m_{M,N}} \big \rvert^{1/2}} + {\mathcal O}(E^{-1/2}).
\end{equation}
Here, $\sigma = \sgn (n_{M,N}'')$ and the function $n\,(m) \equiv n\,(E = 1, m)$ is itself defined by the action variable
\begin{equation}
n(E, m) = \frac{1}{2\pi} \oint p_x(E, x) \,\,\mathrm{d}\,x.
\end{equation}
Formally, for Eq.~\eqref{eq:N_MN} to remain convergent, $n_{M,N}$ should follow the twist condition, $n_{M,N}'' \neq 0$ for $0 < m \leq f_{\rm max}$. Inverting $\mathcal{N}(E) = K$ now gives
\begin{equation}
E(K) = \frac{K}{\mathcal A} - \left( \frac{K}{\mathcal A} \right)^{1/4} \sum_{M, N} \frac{\mathcal{N}_{M,N}(K/{\mathcal A})}{\mathcal A} + {\mathcal O}(1).
\end{equation}
Accordingly, the cumulative nodal count is
\begin{alignat}{1}
c\,(K) &= \sum_{n = 0}^{\infty} \sum_{m = -\infty}^{\infty} \nu _{m,n} \,\Theta\, [E(K) - E_{m,n}] \nonumber \\
&= \xoverline{c}\,(K) + c_{\rm osc}(K)
\end{alignat}
with 
\begin{alignat}{1}\label{eq:c}
\xoverline{c}(K) &= 2\frac{\xoverline{m\,n}}{\mathcal A}K^2 + \frac{\xoverline{m}}{\sqrt{\mathcal A}}K^{3/2} + {\mathcal O}(K), \\
c_{\rm osc}(K) &= K^{5/4}\sum_{\rm po} a_{po} \sin \left( L_{po}\sqrt{\frac{K}{\mathcal A}} + \sigma \frac{\pi}{4}\right) + {\mathcal O}(K), \nonumber
\end{alignat}
where the amplitudes are given by 
\begin{equation}
a_{po} = (-1)^N \frac{m_{M,N}\,\,n\,(m_{M,N}) - 2\,\xoverline{m\,n}}{\pi\, {\mathcal A}^{5/4}\,|N^3\,n_{M,N}''|^{1/2} }
\end{equation}
and the sum runs over periodic geodesics respecting $(-M/N) \in \mathrm{Range}\,(n'(m)) $, corresponding to the classically accessible domain. The overbars are evaluated as
\begin{equation}
\xoverline{m^pn^q} = \frac{1}{\mathcal A} \int_{E_{mn} < 1} \mathrm{d}\,m \,\,\mathrm{d}\,n\,\, \lvert m \rvert ^p \,n^q.
\end{equation}
That there exists a relation between the nodal count and the periodic geodesics is nothing short of remarkable, even more so given that the cumulative nodal count does not bear any spectral information except the ordering it inherits from the spectrum \cite{gnutzmann2007trace}. Fig.~\ref{fig:length} presents the length spectra of periodic geodesics for some examples of flat tori and surfaces of revolution. The length spectrum $S\,(l)$ is extracted from the Fourier transform of $c_{\rm osc} (K)$ with respect to $\kappa = \sqrt{K}$. 

\subsubsection{Periodic orbits of nonseparable, integrable billiards}

\begin{figure}[htb]
\includegraphics[width=\linewidth]{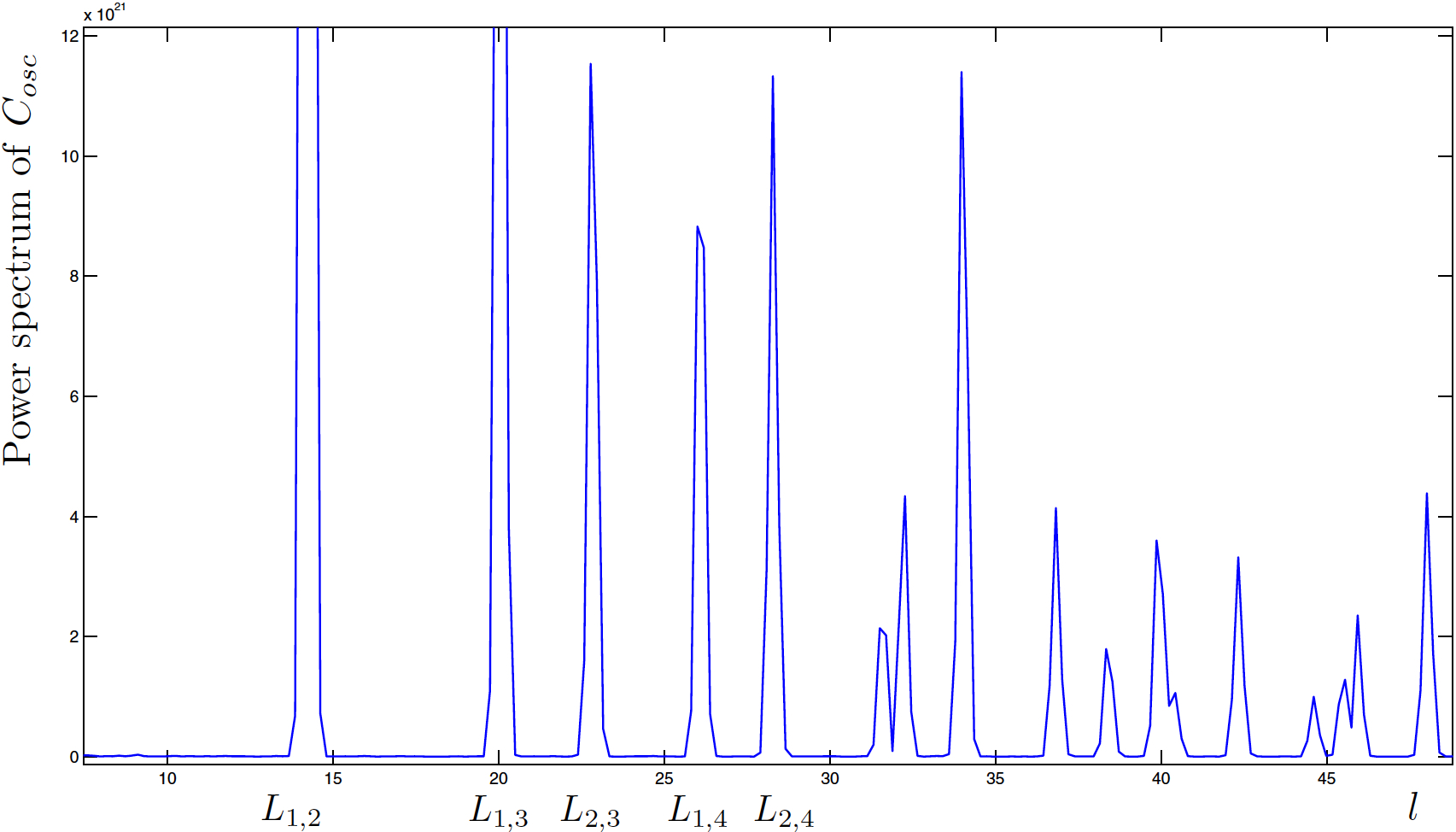}
\includegraphics[width=\linewidth]{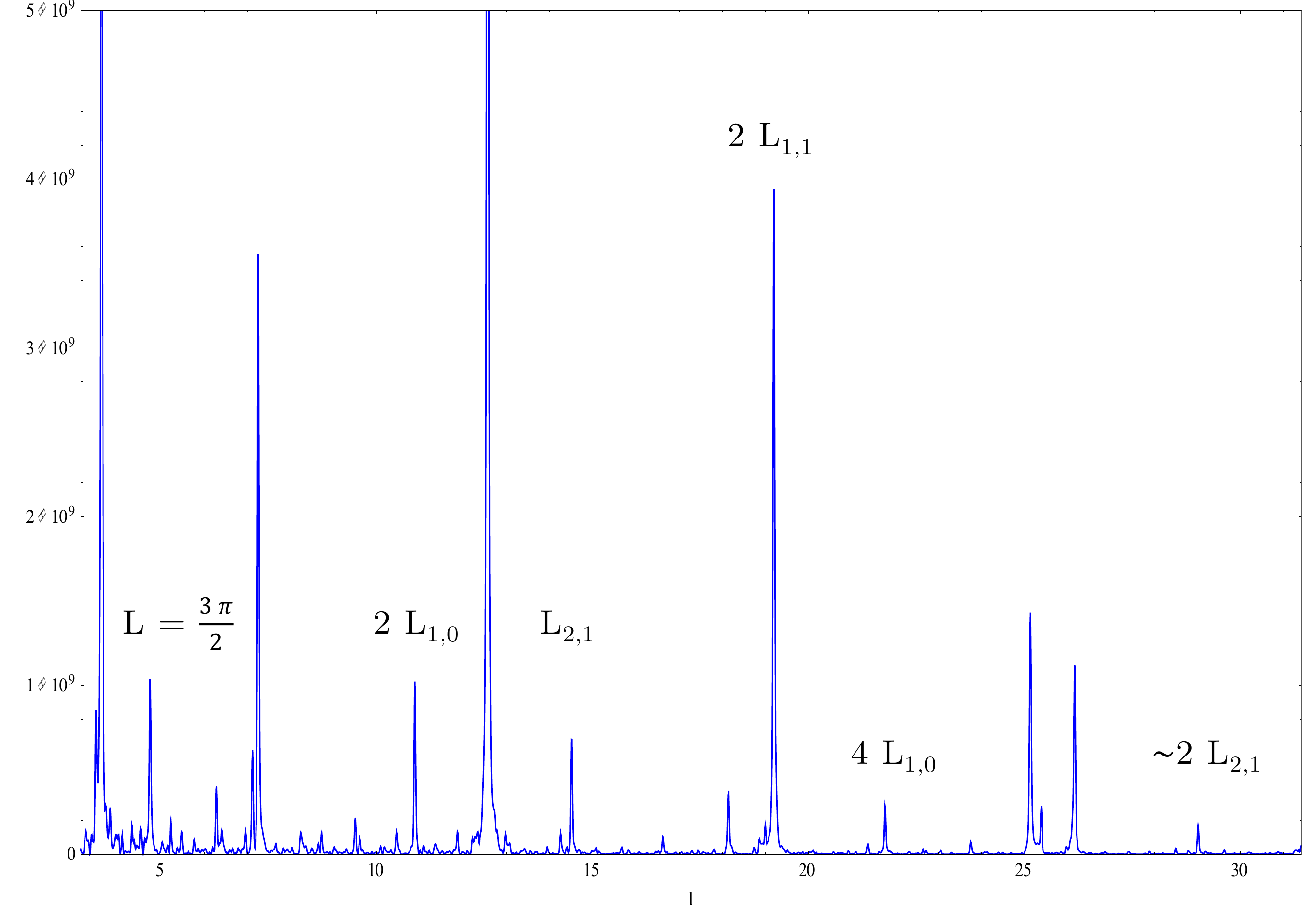}
\caption{\label{fig:length60}The power spectrum of $c_{\rm osc}(K)$ for the right-angled isosceles (top) and equilateral (bottom) triangles. The lengths of some periodic orbits are identified on the $l$ axis. From \citet{aronovitch2012nodal}. \textcopyright\, IOP Publishing. Reproduced with permission. All rights reserved.}
\end{figure}

It is perhaps worth mentioning that the manifestations of periodic orbits in nodal data were also observed for the right-angled isosceles \cite{aronovitch2012nodal} and equilateral \cite{samajdar2014JPA} triangles, which although not separable are at least integrable. For these billiards, the power spectrum of the cumulative count of nodal loops flaunts prominent peaks at the lengths of the periodic orbits (Fig.~\ref{fig:length60}), suggesting forthright the possible existence of a comparable (but yet undetermined) trace formula. 

\paragraph{Inverse nodal problems: Shape analysis}
\label{sec:inverse}
The trace formula approach, although physically appealing, is a little too roundabout when it comes to actually determining the geometry of a billiard. However, there is a more direct way to do so, which is tied to the general inverse nodal problem of determining the metric of a Riemannian manifold using its nodal sequence. Formally, a nodal sequence is defined to be $\mathscr{S} = \{ \nu (\lambda _1), \nu(\lambda _2), \ldots  \}$, arranged in increasing order of $\lambda _j$. To convey the main idea, we concentrate on rectangular billiards  $[0, a] \times [0, b]$ with Dirichlet boundary conditions (in short, Dirichlet rectangles), which are parametrized by the aspect ratio $\tau \leq 1$ such that $\tau\, b = a = 1$. \citet{klawonn2009inverse} proved that the parameter $\tau$ is uniquely determined by the nodal sequence $\mathscr{S}$. Similar results hold for flat tori and Klein bottles, which too can be correspondingly parametrized. Let us now unequivocally outline the methodology for Dirichlet rectangles, $R_{\tau}$, following closely \citet{klawonn2009inverse}. We need to prove that $\tau \neq \tau '$ implies that $\mathscr{S}\,(R_{\tau}) \neq \mathscr{S}\,(R_{\tau}')$ by constructing a sequence limiting to $\tau$. We denote by $\mathbb{P}$ the set of primes and choose $p \in \mathbb{P} \cup \{ 1\}$. For the spectrum of $R_{\tau}$s, there are exactly two positions $i, j$ ($i < j$) with $p \in \nu(\lambda _i), \nu(\lambda _j)$; note that $i$ stands for the pair of quantum numbers $(m,n)$. Then, it is easy to see that
\begin{equation}
\lambda _i = \lambda _{1p} = 1^2 + \tau ^2p^2 \leq p^2 + \tau ^2 = \lambda _{p1} = \lambda _j.
\end{equation}
In general, let us consider a natural number $N$ with prime decomposition, $N = p_1\,p_2 \ldots p_k$; $p_i < p_{i+1}$, $p_i \in \mathbb{P}$. For any such decomposition, one can rewrite $N$ using the permutation $\pi (\{ 1, 2, \ldots , k \}) = \{ i_1, i_2, \ldots , i_k \}$ as the product $p_{i_{1}} \ldots p_{i_{\ell}} \cdot p_{i_{\ell + 1}} \ldots p_{i_{k}} $ with $1 < \ell < k$ such that 
\begin{equation}
N^2 + \tau ^2 > (p_{i_{1}} \ldots p_{i_{\ell}})^2 + \tau ^2 (p_{i_{\ell + 1}} \ldots p_{i_{k}})^2.
\end{equation}
Thus, we can detect the position of every eigenvalue of the forms $N^2 + \tau ^2$ for $N \in \mathbb{N}$ and $1^2 + \tau ^2 p^2$ for $p \in \mathbb{P} \cup \{1\}$. To extract $\tau $ from this data, we construct two sequences, for every $h \in \mathbb{P} \cup \{1\}$, 
\begin{alignat}{1}
h_+ &= \mbox{min}\, \{i \in \mathbb{N}: 1 + h^2\tau ^2 < i^2 + \tau ^2  \}, \nonumber \\
h_- &= \mbox{max}\, \{i \in \mathbb{N}: i^2 + \tau ^2 < 1 + h^2\tau ^2 \},
\end{alignat}
and use the more dense sequence to put bounds on the other. The inequality
\begin{equation}
H_-^h \equiv \frac{h_-^2 - 1}{h^2 - 1} < \tau ^2 < \frac{h_+^2 - 1}{h^2 - 1} \equiv H_+^h, ~\forall\, h \in \mathbb{P}
\end{equation}
is easily verified. Finally, \citet{klawonn2009inverse} illustrates that $\lim_{h\to\infty }H_+^h - H_-^h \to 0$, which implies $H_{\pm}^h \to \tau ^2$, thereby completing the proof. 

The Laplace-Beltrami nodal counts have also been advertised to provide a new signature for 3D shape analysis. Specifically, it was used by \citet{lai2009laplace} to experimentally resolve ambiguities left unaddressed by the ``shape DNA'' (the distribution of eigenvalues). Their method was based on a distance function defined between nodal count sequences ${\mathscr S}$ and ${\mathscr S'}$:
\begin{equation}\label{eq:dist}
D\,({\mathcal S}, \tilde{{\mathcal S}}) = \sqrt{\sum_{n=1}^{\infty} \left( \frac{1}{n^{\alpha}} \right)^2 [{\mathcal N}(\lambda _n) - \tilde{{\mathcal N}}(\lambda _n)]^2};\,\, \alpha > 0. 
\end{equation}
This spectral distance function meets all the conditions of a metric and hence, can be reliably used in shape analysis \cite{lai2010metric}. Pairwise distances, as used in Eq.~\eqref{eq:dist}, are stored in a distance matrix. Further details concerning the usage of this matrix and the multidimensional scaling technique to embed the surfaces into Euclidean space can be found in \cite{lai2009laplace}. Recently, this method has been paired with landmark-based morphometric studies \cite{shi2017conformal} using the neuroimaging data from works on Alzheimer's disease \cite{wang2017towards}.

\subsection{Graph-theoretic analysis}
\label{sec:graph}

\begin{figure}[htb]
\includegraphics[width=\linewidth]{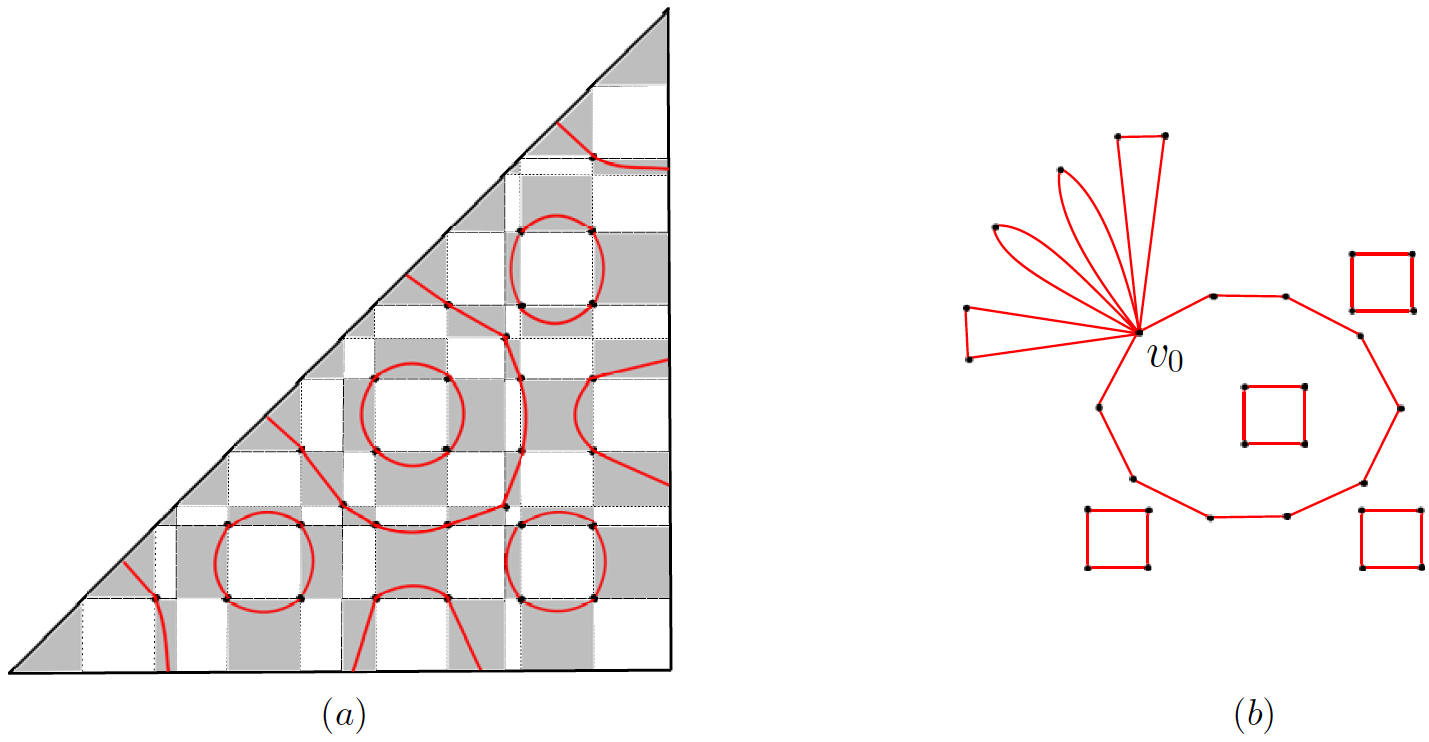}
\caption{\label{fig:graph}(a) The pattern of the nodal set of $\psi_{9,4}$ for the right-isosceles triangle. The shading highlights the cells that the nodal lines pass over; here, $\psi _{mn}^{(1)}$ and $\psi _{mn}^{(2)}$ have the same sign. (b) The graph $G_{9,4}$ produced by the counting algorithm. From \citet{aronovitch2012nodal}. \textcopyright\, IOP Publishing. Reproduced with permission. All rights reserved.}
\end{figure}

The wavefunction for a right-angled isosceles triangle, Eq.~\eqref{eq:wfun_iso}, is easily decomposed into the difference of $\psi _{m,n}^{(1)} = \sin (mx) \sin (ny)$ and $ \psi _{m,n}^{(2)} =\sin (nx) \sin (my)$, which, individually, possess a lattice structure (the length of the legs have been assumed to be $\pi$). Let us denote the nodal sets for the two functions by $N_{m,n}^{(1)}$ and $N_{m,n}^{(2)}$. Their intersection, $N_{m,n}^{(1)} \cap N_{m,n}^{(2)}$, is the set of points
\begin{equation*}
V_{m,n} = \left \{\frac{\pi}{m}(i, j) \vert 0<j<i<m \right\} \cup \left \{\frac{\pi}{n}(i, j) \vert 0<j<i<n \right\}
\end{equation*}
Since $\psi_{m,n}$ necessarily vanishes at these points, the nodal lines pass through them. The union $N_{m,n}^{(1)} \cup N_{m,n}^{(2)}$ divides the triangle $\mathcal{D}$ into cells. \citet{aronovitch2012nodal} translated this to a pictorial representation by constructing a graph $G_{m,n}$ with vertices $V_{m,n}$ and an additional vertex $v_0$ for the boundary of the triangle. The edges of the graph stand for the nodal lines, connecting the vertices. The graph $G_{m,n}$ is endowed, according to certain specified rules, with one, two, or three edge(s) for each (shaded) cell that a nodal line runs over. The number of vertices in a cell determines its connectivity. Once $G_{m,n}$ has been constructed, the number of nodal domains can be counted by the Euler formula for planar graphs:
\begin{alignat}{1}
\nu_{m,n} = 1 + E\,(G_{m,n}) - \lvert V_{m,n}\rvert + c\,(G_{m,n}),
\end{alignat}
where $E\,(G_{m,n}), \lvert \, V_{m,n}\rvert,$ and $c\,(G_{m,n})$ denote the number of edges, vertices, and connected components of $G_{m,n}$, respectively. An example of such a construction is to be seen in Fig.~\ref{fig:graph}.

The nodal domain count can be related to the number of intersections of the nodal lines with the boundary, $\eta _{m,n}$, and the number of nodal loops, $I_{mn}$. For instance, in Fig.~\ref{fig:graph}, $\eta_{\,9,4} = 10$ and $I_{9,4} = 4$. In the generic non-tiling case, the two are linked by 
\begin{alignat}{1}
\nu _{m,n} &= 1 + \frac{1}{2}\eta _{m,n} + I_{m,n},\,\mbox{ where }\\
\eta _{m,n} &= m + n -3.
\end{alignat}
Extensive analysis of the data on the number of domains, obtained numerically using the graph-theoretic algorithm, led \citet{aronovitch2012nodal} to propose that the loop count is given by 
\begin{equation}\label{eq:recursive}
I_{m,n} = \tilde{I}\,(n, (m-n-1)/2, 0)
\end{equation}
where 
\begin{align*}
\tilde{I} (n, k, l) = 
\begin{cases}
\vspace*{0.35cm} 
0; \quad &\hspace*{-2.20cm} n=1 \mbox{ or } k = 0, \nonumber \\
\left\lfloor {\displaystyle \frac{n}{2k+1}} \right\rfloor (l\,k+(2l+1)k^2) \\ 
\vspace*{0.35cm}
+ \tilde{I}\,(n\, \mathrm{mod}\, (2k+1), k, l);  \quad &\hspace*{-2.20cm} 2k + 1 < n, \nonumber \\ 
{\displaystyle \frac{1}{2}\left\lfloor \frac{k}{n} \right \rfloor }(2l + 1)(n^2 - n)  \nonumber \\ 
\vspace*{0.35cm}
+ \tilde{I}\,(n, k\,\mathrm{mod}\, n, l);  \quad &\hspace*{-2.20cm} 2k + 1 > 2n, \nonumber \\
{\displaystyle \left( l + \frac{1}{2} \right) (2k^2 + n^2 - n - 2nk + k) + \frac{k}{2}} \nonumber \\
+ \tilde{I}\,(2k - n +1, n - k - 1, l + 1); \nonumber \\ & \hspace*{-2.20cm} n < (2k + 1) < 2n. 
\end{cases}
\end{align*}
This recursive relation was used, subsequent to verification of its predictions for the first 100,000 non-tiling loop counts, to produce the $P\, (\xi)$ distribution of Fig.~\ref{fig:pxi}.

\subsection{Difference-equation formalism}

For integrable billiards, such as the right-isosceles triangle above, \citet{samajdar2014nodal} put forth an alternative method to count the number of domains. Firstly, the eigenfunctions of all integrable billiards can be classified by an index uniquely specified by the quantum numbers. Within each congruence class, the wavefunctions display exceptional structural similarities and their nodal counts obey certain difference equations. A trivial example is the rectangular billiard; its eigenfunctions are classified by $k = m\, \mathrm{mod}\, n$ and we can set up the relation $\nu_{m+n, n}-\nu_{m,n}= n^2$. Fortunately, the equations for other billiards too turn out to have constant coefficients and can be solved readily given a ``boundary condition'', namely, the number of domains for a low-lying state (which can be counted manually if need be). The solutions thereto  explicitly yield the number of nodal domains for integrable (and importantly, nonseparable) billiards, thus partially solving an otherwise intractable problem. For completeness, we list the analytically-known difference equations below.

\subsubsection{Circles (and annuli/sectors thereof)}

Let $\Delta_{a, \varsigma \,b} \equiv \nu_{a+\varsigma b,b} - \nu_{a,b}$. With this notation, the difference equation for the circle (or a circular annulus) is simply $\Delta_{m, n} = 2\,n^2$ for $m \neq 0$, where $n$ and $m$ are radial and angular quantum numbers, respectively \cite{manjunath2016difference}; this is solved by $\nu _{m,n} = 2\,m\,n$. When $m=0$, the number of domains is just $n$.  For a sector, the domain $\mathcal{D}$ is restricted in radial and angular variables and, like for the rectangular billiard, we find $\nu _{m+n,n} - \nu_{m,n} = n^2$, leading to $\nu _{m,n} = m\,n$. 

\subsubsection{Ellipses and elliptic annuli}
The Helmholtz equation for an elliptical billiard separates into Mathieu equations for the radial and angular components. They can be categorized into symmetry classes by noting whether they are symmetric $(+)$ or antisymmetric $(-)$ with respect to the $x$- and $y$-axes. So, a state designated as $(+-)$ is (anti)symmetric about the $x$ ($y$) axis. The difference equations are: 
\begin{alignat*}{2}
\Delta_{r, l} =
\begin{cases}
4l^2; &(++)\nonumber \\
4l^2+2l; &(+-) \nonumber \\
4l^2+2l; &(-+)  \nonumber \\
4l^2+4l; &(--)
\end{cases},\,
\nu_{r,l}=
\begin{cases}
2l\,(2r+1)+1\\
2\,(2l+1)\,(r+1)\\
(2l+1)\,(2r+1)+1\\
4(l+1)\,(r+1)
\end{cases},
\end{alignat*}
where $l$ ($r$) is the angular (radial) quantum number. For elliptic annuli, the difference equations and number of domains are almost identical:
\begin{alignat*}{2}
\Delta_{r, l} =
\begin{cases}
4l^2; & (++)\\
4l^2+8l; & (+-)\\
4l^2+8l; &(-+)\\
4l^2+4l; &(--)\\
\end{cases},\,
\nu_{r,l}=
\begin{cases}
4l\,(r+l) \nonumber \\
2(2l+1)\,(r+1) \nonumber \\
2(2l+1)\,(r+1) \nonumber \\
4(l+1)\,(r+1)
\end{cases}.
\end{alignat*}   

\subsubsection{Confocal parabolae}
The detailed solutions of the confocal parabolic billiard can be found in \cite{manjunath2016difference}. The Helmholtz equation separates in parabolic coordinates, $(\tau, \sigma)$, related to $(x, y)$ by $x=\tau \sigma$ and $y = (\tau ^2 - \sigma ^2)/2$ with $\sigma \geq 0$. The wavefunctions $\psi \,(\tau, \sigma )$ can be written as a product $S(\sigma)\, T(\tau)$. For Dirichlet boundary conditions, the functions $T$ and $S$ ought to be either both even or both odd. In the former case,
\begin{equation}
\Delta_{m, n} = 2n^2-n, \mbox{ and }\nu_{m,n} = 2\,m\,n - m - n + 1.\, 
\end{equation}
When $S$ and $T$ are odd, $\Delta_{m, n} = 2n^2$ and $\nu_{m,n} = 2\,m\,n$. The parabolic annuli and sectors can be analyzed along the same lines as previously. For all the separable billiards above, the difference equations are but restatements of the exactly attainable nodal counts, which might make the utility of the approach seem dubious. The nonseparable billiards below, hopefully, assuage such fears.

\subsubsection{Right-angled isosceles triangle}
The eigenfunctions of the right-angled isosceles triangle are classified by the index ${\mathcal C}_{2n} = m \mbox{ mod } 2n$. It is sufficient to consider the general non-tiling case as the number of domains with tiling can always be recovered on knowing that for each tile. From extensive numerical analysis, \citet{samajdar2014nodal} found that a simple set of equations hold for the number of nodal domains and loops:   
\begin{alignat}{1}\label{eq:recur_iso}
\Delta_{m, 2n} = \nu_{m+2n,n} - \nu_{m,n} &= \frac{n(n+1)}{2}, \nonumber \\
I_{m+2n,n} - I_{m,n} &= \frac{n(n-1)}{2}.
\end{alignat}
These equations can be retrospectively motivated by succesively inserting $\nu_{m+2n, n}$ and $\nu_{m, n}$ into Eq.~\eqref{eq:juggle}, which thus predicts $\Delta_{m, 2n} \propto n^2 +n$. Let $\zeta _1 = n \mbox{ mod } {\mathcal C}_{2n}$ and $\zeta _2 = n \mbox{ mod } 2\,{\mathcal C}_{2n}$. Taking advantage of Eqs.~\eqref{eq:recursive} and \eqref{eq:recur_iso}, we arrive at an exact solution for the number of domains for even ${\mathcal C}_{2n}$:
\begin{alignat}{1}
\nu_{m,n} &= \frac{m(n+1)+n-2}{4} + \bigg[ -\frac{n^2}{4} + \left( \frac{{\mathcal C}_{2n}}{2} \right)n \nonumber \\ &- \bigg( \frac{{\mathcal C}_{2n}^2 - {\mathcal C}_{2n} - 1}{2} \pm \frac{1}{4}(\zeta _2 - 1)  \bigg) \bigg],
\end{alignat}
with the $+ (-)$ sign applicable when ${\mathcal C}_{2n} < \zeta _2$ (otherwise). For odd ${\mathcal C}_{2n}$:  
\begin{alignat}{1}
\label{eq:iso_fluct}
\nu_{m,n} &= \frac{m(n+1)+n-2}{4} + \bigg[ -\frac{n^2}{4} + \left( \frac{{\mathcal C}_{2n}}{2} \right)n \nonumber \\ &- \bigg( \frac{2\,{\mathcal C}_{2n}^2 - {\mathcal C}_{2n} - 2}{4} + \gamma  \bigg) \bigg].
\end{alignat}
The precise form of $\gamma $ is uncertain, but asymptotically $\lim_{k \to \infty} (\gamma /\nu_{m+kn,n}) = 0$. Since the fluctuations die out as $E \to\infty$, Eq.~\eqref{eq:iso_fluct} lends itself to studying the limiting distributions without problem.

\subsubsection{Equilateral triangle}
\bgroup
\def\arraystretch{1.5}
\begin{table}[htb]
\centering
{
\bgroup
\small
\setlength{\tabcolsep}{9.75pt}
\begin{tabular}{c r c r r r} \hline
\multicolumn{1}{c}{$m$} &\multicolumn{1}{c}{$n$} &\multicolumn{1}{c}{$m \mod 3n$} &\multicolumn{1}{c}{$\nu_{m,n} $} &\multicolumn{1}{c}{$\Delta \nu_{m,n}$} &\multicolumn{1}{c}{$\Delta ^2\nu_{m,n}$}  \\ \hline
$7$ &$2$ &$1$&$6$&$-$&$-$ \\
$13$ &$2$&$1$&$21$&$15$&$-$ \\
$19$ &$2$&$1$&$48$&$27$&$12$ \\
$25$ &$2$&$1$&$87$&$39$&$12$ \\
$31$ &$2$&$1$&$138$&$51$&$12$ \\
$37$ &$2$&$1$&$201$&$63$&$12$ \\
$9$ &$2$&$3$&$10$&$-$&$-$ \\
$15$ &$2$&$3$&$29$&$19$&$-$ \\ 
$21$ &$2$&$3$&$60$&$31$&$12$ \\
$27$ &$2$&$3$&$103$&$43$&$12$ \\ 
$33$ &$2$&$3$&$158$&$55$&$12$ \\
$39$ &$2$&$3$&$225$&$67$&$12$ \\ \hline
\end{tabular}
}
\egroup
\caption{\label{Table:Eq-tr-diff} The second difference of the number of nodal domains for an equilateral-triangular billiard is seen to remain constant if one considers a sequence of wavefunctions differing in $m$ by steps of $3\,n$, for a fixed value of $n$. This defines an equivalence relation, and correspondingly, congruence classes, indexed by $k = m \mbox{ mod } 3\,n$. The second difference $\Delta^2 \nu_{m, n} =  \nu_{m+6n,n} - 2\nu_{m+3n,n} + \nu_{m,n}$ is $12 = 3n^2$ for all the eigenfunctions tabulated.  }
\end{table}
\egroup

The eigenfunctions \eqref{eq:wf} enable classifiication by ${\mathcal C}_{3n} = m \mbox{ mod }3n$ \cite{samajdar2014JPA} as Table~\ref{Table:Eq-tr-diff} explicates. Restricting ourselves to non-tiling patterns, the number of domains and loops satisfy the equations:
\begin{alignat}{1}
\label{eq:count_eq}
\nu_{m+6n,n} - 2\,\nu_{m+3n,n} + \nu_{m,n} &= 3n^2, \nonumber \\
I_{m+6n,n} - 2\,I_{m+3n,n} + I_{m,n} &= 3n^2.
\end{alignat}
The solution to Eq.~\eqref{eq:count_eq}, along with detailed numerical observations, lead to the following formulae for the nodal domain counts:
\begin{alignat}{2}
\label{eq:count}
\nu_{m,n} &= \frac{m^2}{6} - \frac{(4n-3)m}{6} + n^2 -&& \frac{{\mathcal C}_{3n}n-\lambda _1({\mathcal C}_{3n},n)}{3},\nonumber \\ & && (0<{\mathcal C}_{3n}<n), \nonumber \\ 
&= \frac{m^2}{6} - \frac{(4n-3)m}{6} + n^2 -&& \frac{2\,({\mathcal C}_{3n}-n)\,n-\lambda _2({\mathcal C}_{3n},n)}{3} \nonumber \\ 
&~  &&(n<{\mathcal C}_{3n}<3n).
\end{alignat}
$\lambda _1$ and $\lambda _2$ are parameters that contribute to small variations in the count. Their explicit forms are not known analytically although a number of relations satisfied by them are \cite{samajdar2014JPA,samajdar2014nodal}. 

For the closely related hemiequilateral-triangular billiard (a $30^\circ-60^\circ-90^\circ$ scalene triangle), the difference equation turns out to be
\begin{equation}
\label{eq:hemi}
\nu_{m+6n,n} - 2\,\nu_{m+3n,n} + \nu_{m,n} = 0,
\end{equation}
i.e., it is the first rather than the second difference that remains constant. This brings to a close our compilation of the difference equations and domain counts for \textsl{all} planar integrable billiards---both separable and nonseparable. The main advantage of this formalism is that it empowers one to determine $\nu_{m,n}$ for a whole hierarchy of states by starting with the domain count for a simpler wavefunction, in the same congruence class, and ascending the ladder \cite{mandwal2017}.

\subsection{Counting with Potts spins}

Exact counting of the nodal domains of a chaotic billiard is still an outstanding open problem. Within the random wave model, certain correlation functions can be computed, as we have already seen. Here, we discuss a method for calculating the moments of the number of nodal domains with the help of auxiliary Potts spins \cite{foltin2003} that sheds light on the percolative structure of the wavefunction. Suppose the nodal pattern is placed on top of a square lattice of side $a \ll 2\pi/k$. To each node $\mathbf{r}_i$, we assign the variable $\sigma_i = \sgn \,(\psi\,(\mathbf{r}_i))$ and a Potts spin $s_i$ \cite{baxter2007exactly}. The spins can take values $s_i = 1, 2, \ldots , q$ but are constrained to have the same value if they belong to the same nodal domain. Practically, this constraint is implemented as follows. The product over adjacent lattice bonds $\langle ij \rangle $ given by
\begin{alignat}{1}\label{eq:id}
{\cal N}(s_i) &=\prod_{\langle ij \rangle } \left( \frac{1 - \sigma _i\sigma _j}{2} + \frac{1 + \sigma _i\sigma _j}{2}\delta _{s_i,s_j} \right) \nonumber \\ &= \prod_{\langle ij \rangle } \left( 1 - \frac{1 + \sigma _i\sigma _j}{2} (1 - \delta _{s_i,s_j}) \right),
\end{alignat}
is one if and only if $\sigma _i$ and $\sigma _j$ are different, or if $\sigma _i = \sigma _j$ \textsl{and} $s_i = s_j$. Summing over all possible spin configurations, we get \cite{baxter2007exactly}
\begin{equation}\label{eq:sum_potts}
Z(\{\sigma _i\}) = \sum_{s_i} {\cal N}(s_i) = q^C,
\end{equation}
where $C$ is the number of nodal domains. To find the mean number of domains and higher moments, it is expedient to construct a partition function. This is readily defined by averaging over the Gaussian random fields $\psi$ with the correlation function from Eq.~\eqref{eq:CorrDecay}
\begin{alignat}{2}\label{eq:pf}
{\mathcal Z} &= \sum_{\{s _i\}} \left\langle \exp \left( -\beta \sum_{\langle ij \rangle } \frac{1 + \sigma _i\sigma _j}{2} (1 - \delta _{s_i,s_j}) \right)  \right\rangle _{\psi} \nonumber \\ & \rightarrow \langle q^C \rangle _{\psi},\, \mbox{ as } \beta \to \infty. 
\end{alignat}
The mean number of nodal domains is then
\begin{equation}
\langle C \rangle _{\psi} = \frac{\partial \,{\mathcal Z}}{\partial\, \psi } \bigg \vert_{q = 1}\, \mbox{ for } \beta \to \infty.
\end{equation}
As we have endowed each node with a spin, we can also formulate an order parameter $o_i = \delta_{s_i, 1} - 1/q$. The correlation function of $o$ is
\begin{alignat}{1}
G_{k,l} &= \sum_{\{s _i\}} o_k \,o_l\prod_{\langle ij \rangle} \left( 1 - \frac{1 + \sigma _i\sigma _j}{2} (1 - \delta _{s_i,s_j}) \right) \nonumber 
\end{alignat}
and since spins on different domains are independent,
\begin{alignat}{1}
\label{eq:corr_order}
G_{k,l} = 
\begin{cases}
0, &\mbox{ if } k, l \mbox{ are disconnected}, \\ 
q^C - q^{C-1}, &\mbox{ if } k, l \mbox{ are connected}.
\end{cases}
\end{alignat}
The partial derivative of $\langle G_{k,l} \rangle _{\psi}$ with respect to $q$ gives the probability that the nodes $k$ and $l$ are connected.

\section{Experimental realizations}
\label{sec:exp}

For a monograph on a subject that traces its roots to purely experimental origins, it is almost heretical to have a colloquy as unabashedly theoretical as ours thus far. In this Section, we make amends and correct course by delving into some of the pioneering studies on nodal patterns in the laboratory. As \citet{stockmann2006quantum} cedes, prior to the onset of the 1990s, experiments on the quantum mechanics of chaotic systems were few and far between. The initial impetus was borrowed from analyses of nuclear spectra \cite{porter1965statistical}, which, more than half a century later, still continues to reveal new aspects of quantum chaos \cite{dietz2017chaos}. Contemporarily, intriguing investigations were underway on hydrogen atoms in strong microwave \cite{bayfield1974multiphoton} and magnetic fields \cite{holle1986diamagnetism, main1986new}. A completely new direction was lent to the subject by \citet{stockmann1990quantum} with their experiments on irregularly shaped microwave cavities---the microwave billiards, which we turn to discussing first. In Sec.~\ref{sec:expVortex} immediately thereafter, we also detail a few experiments probing the statistics of vortices.

\subsection{Microwave billiards: The physicist's pool table}

\begin{figure}[htb]
\includegraphics[width=\linewidth]{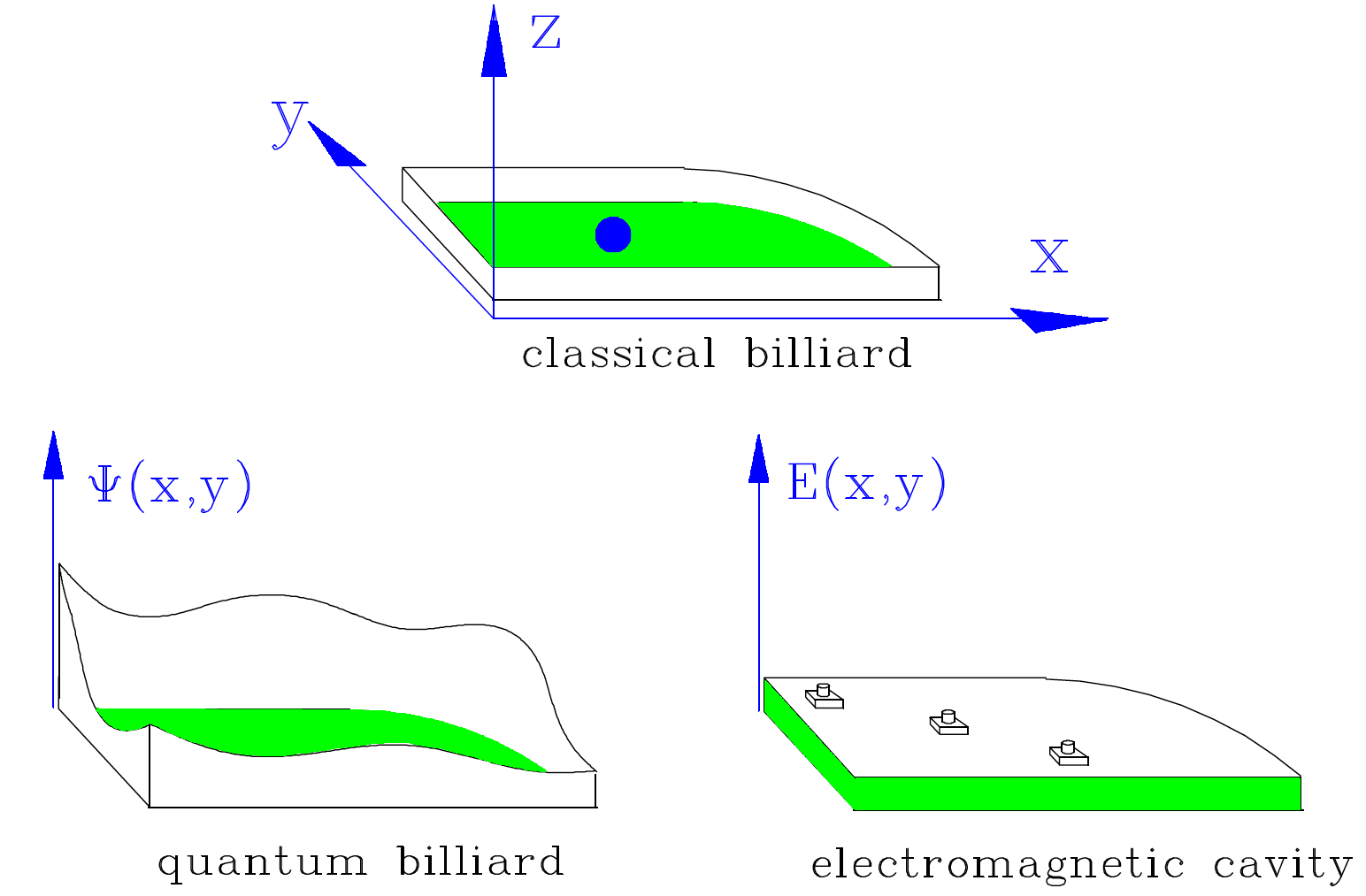}
\caption{\label{fig:analogue}Schematic illustration of a realization of the quantum analogue of a classical billiard through a flat electromagnetic cavity, taking a quarter of the Bunimovich stadium as an example. From \citet{richter1999playing}. With permission of Springer Nature.}
\end{figure}

Playing billiards with microwaves is made feasible by the mathematical analogy between electromagnetism and quantum mechanics in that both are described by linear second-order differential equations. Maxwell's equations, subject to everyday manipulations \cite{jackson1999classical}, can be massaged into the Helmholtz equations for the electric and magnetic fields:
\begin{equation}
(\Delta + k^2) \, \mathbf{E} = 0;\hspace{0.5cm}(\Delta + k^2) \,\mathbf{B} = 0,
\end{equation}
where $k = \omega/c$ is the wavenumber and $\omega$ is the angular frequency. The fields have to further satisfy the boundary conditions 
\begin{equation}
\hat{\mathrm{n}}\, \times\,\mathbf{E} = 0; \hspace{0.5cm} \hat{\mathrm{n}}\cdot \mathbf{B}= 0
\end{equation}
where $\hat{\mathrm{n}}$ is the unit vector normal to the surface. In a sufficiently flat cylindrical resonator $\mathcal{D}$ (of depth $d$), where all the walls are parallel or perpendicular to $\hat{z}$, these assume the form
\begin{equation}
\label{eq:vanish}
\mathbf{E}_z\vert_{\partial\,\mathcal{D}} = 0; \hspace{0.5cm} \nabla_\perp \mathbf{B}_z\vert_{\partial\,\mathcal{D}} = 0.
\end{equation}
The solutions are the familiar transverse magnetic (TM) and transverse electric (TE) modes. The fields belonging to the former category are, generically,
\begin{alignat}{1}
\nonumber \mathbf{E}_z \,(x, y, z) &= \mathscr{E} (x, y)\, \cos \bigg( j\,\frac{\pi\,z}{d} \bigg);  \hspace{0.5cm} j = 0, 1, 2,\ldots,\\
\mathbf{B}_z \,(x, y, z) &= 0,
\end{alignat}
where $\mathscr{E} (x, y)$, which is just a multivariate scalar function, is characterized by the stationary two-dimensional Helmholtz equation
\begin{equation}
\label{eq:Helmholtz}
\bigg[ \Delta + k^2 - \left( j\,\frac{\pi}{d} \right)^2 \bigg] \,\mathscr{E} = 0;  \hspace{0.5cm} \mathscr{E} \,(x, y)\vert_{\partial\,\mathcal{D}} = 0.
\end{equation}
For frequencies below the cutoff $c/2d$ (or for wavenumbers $k < \pi /d$), only TM modes with $j = 0$ are permitted and Eq.~\eqref{eq:Helmholtz} reduces to a more recognizable one, $\left (\Delta + k^2\right)\mathscr{E} = 0$. Endowed with appropriate boundary conditions, this has a discrete spectrum
\begin{equation}
-\bigg ( \frac{\partial^2}{\partial\,x^2} + \frac{\partial^2}{\partial\,y^2}  \bigg)\, \mathscr{E}_{n} = k_n^2\, \mathscr{E}_{n},
\end{equation}
which, fortuitously, bears a striking resemblance to the time-independent Schr\"{o}dinger equation (cf. Eq.~\ref{eq:chladniVib})
\begin{equation}
-\frac{\hbar^2}{2\,\mathrm{m}} \bigg ( \frac{\partial^2}{\partial\,x^2} + \frac{\partial^2}{\partial\,y^2}  \bigg) \,\psi_n = E_n\, \psi_{n}, 
\end{equation}
with $n$ now labelling the energy eigenstates. In fact, there is an exact one-to-one correspondence with the two related as $E \propto k^2$ and $\psi_n \propto \mathscr{E}_{n}$. The analogy can be pushed further. Quantum mechanically, a particle must be trapped within an infinite potential well in order to mimic the hard boundaries of the classical billiard. This infinite barrier exactly translates to the Dirichlet boundary conditions \eqref{eq:vanish} on the surface of the cylinder. Therefore, in a cavity with height $d \le \lambda_{\min}/2 = c/ (2\, f_{\max})$, with $\lambda_{\min}$ being the minimum de Broglie wavelength accessible to experiments, the quantum billiard can be simulated with electromagnetic waves \cite{richter1999playing}, as Fig.~\ref{fig:analogue} outlines. To get a sense of the actual numbers involved, consider a typical ``quasi-two-dimensional'' cavity: with $d = 0.8\, \mathrm{cm}$, we find $\lambda_{\min} = 1.6\,\mathrm{cm}$ and $f_{\max} = 18.75\,\mathrm{GHz}$.

\begin{figure}
\includegraphics[width= \linewidth]{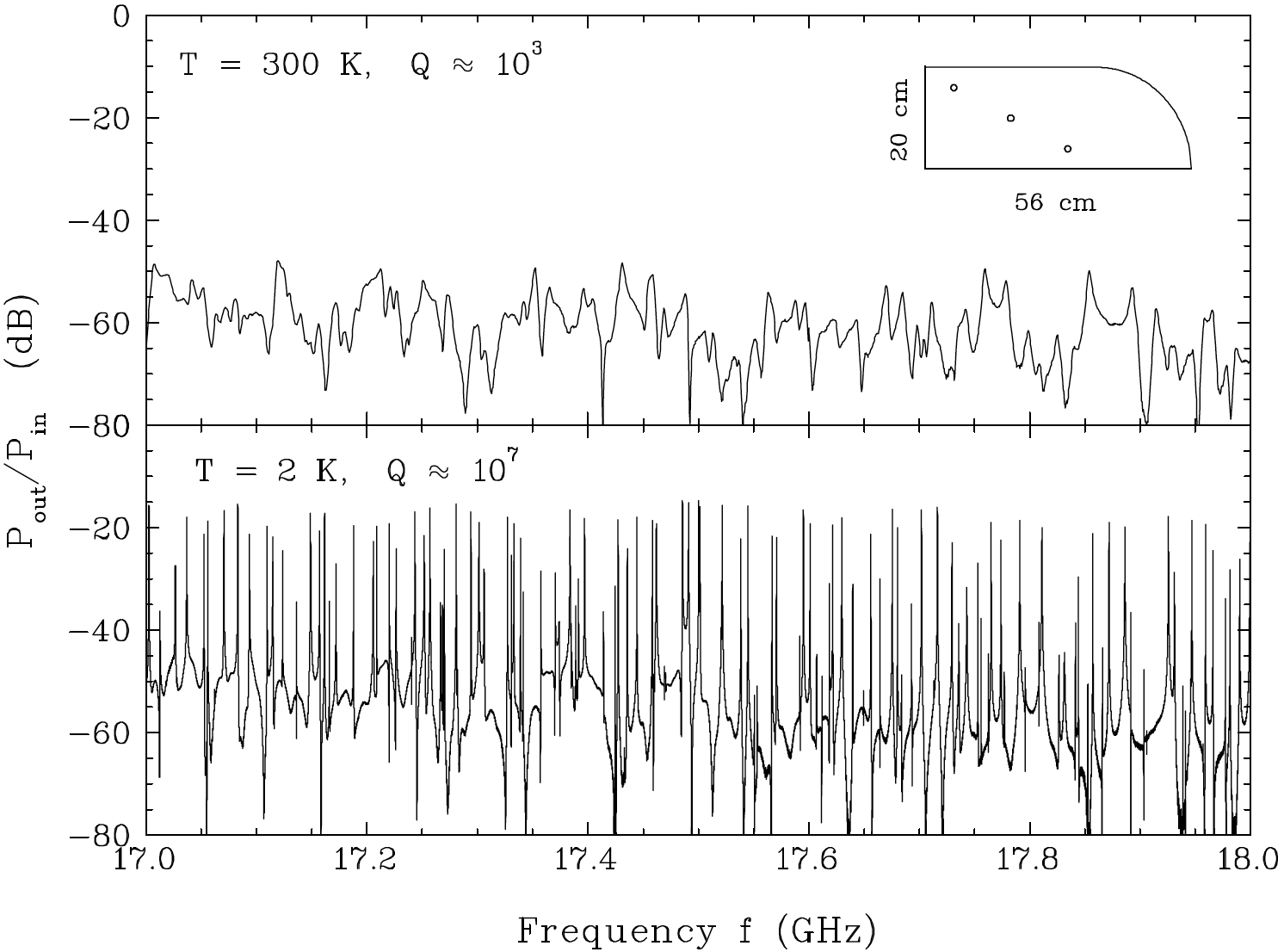}
\caption{\label{fig:spectrum}Measured eigenmode spectrum between 17 and 18 GHz for a cavity in the shape of a quarter stadium billiard. Each maximum in the transmitted microwave power corresponds to an eigenfrequency of the resonator. The upper part is taken at room temperature (normally conducting), the lower part at 2 K (superconducting). [Inset]: The shape of the resonator and the positions of the antennae. From \citet{richter1999playing}. With permission of Springer Nature.}
\end{figure}

Such lines of reasoning are not exactly new. Even earlier, the parallels between microwaves and sound waves were used to simulate the acoustics of enclosed spaces \cite{schroeder1987normal}. Hence, it was no surprise that the first experiments by \citet{stockmann1990quantum} precipitated a flurry of intense investigation into both two- \cite{sridhar1991experimental, stein1992experimental, so1995wave} and three-dimensional \cite{deus1995statistical, alt1996chaotic, alt1997wave} microwave cavities. A time-honored protocol for measurement of the eigenmodes thereof would be as follows. One proceeds by first pumping in microwaves into the resonator with an antenna---this is usually a minuscule wire, no more than a few millimeters in diameter. The antennae are cased in wider semi-rigid leads that are attached to the billiard. To eliminate the possibility of the former influencing the field distributions, they are introduced into the walls of the cavity through small orifices with an abundance of caution to ensure negligible penetration into the resonance chamber itself. Depending on the specifications of the experiment, one could then either record the reflected microwave power at the same antenna as a function of frequency (taking care to to separate the incoming and outgoing waves with a microwave bridge) or measure the transmission between two or more antennae. The entire procedure is repeated with different placements of the antennae so that no electromagnetic mode goes undetected simply due to the unhappy coincidence of it having a node at the address of the recipient antenna. The eigenmode spectrum, a sample of which is captured in Fig.~\ref{fig:spectrum}, can be constructed from the output to input ratios of the microwave power, combined over the different iterations.

The spectrum, even with its ungainly appearance and erratic fluctuations \cite{brink1963widths, dietz2015cross}, is actually a veritable cornucopia of physical information that reveals itself upon systematic inspection. Let us look at the distribution function $P\,(s)$ of the level spacings $s_n = E_n - E_{n-1}$ between adjacent eigenenergies; for convenience, we work in units where the mean spacing $\langle s \rangle$ is normalized to one. This happens to be a particularly convenient quantity to sieve out the underlying classical dynamics from the chaff \cite{mehta2004random}. For integrable dynamics, the spacing is expected to exhibit Poisson statistics
\begin{equation}
P\, (s) = \exp\, (-s),
\end{equation}
whereas it should follow the Wigner distribution
\begin{equation}
P\,(s) = \frac{\pi}{2}s\,\exp \bigg( - \frac{\pi}{4} s^2\bigg)
\end{equation}
for chaotic systems and is widely regarded as a hallmark thereof \cite{haake2013quantum}. The fly in the ointment, however, when working with experimental data, is the uncertainty in counting resonances: if two maxima are separated by a distance smaller than the experimental line width, they are registered as one \cite{stockmann2006quantum}. Such missed eigenmodes could evidently wreak havoc on the spacing distribution. The origin of this nuisance is a straightforward consequence of elementary electromagnetic theory. The microwaves inevitably have a finite skin depth, wherefore they penetrate the walls up to distances on the scale of $\delta = \sqrt{2 / (\mu_0\, \omega\,\sigma)} \sim 1\, \mu$m, determined by the conductivity $\sigma$. Accordingly, the electric fields are damped as $\mathbf{E}\,(t) = \mathbf{E}_0\,\exp(-t/2\tau)\cos\,(\omega_0\,t)$ and the power spectrum
\begin{alignat}{1}
\hat{S} \,(\omega) &= \bigg \lvert \frac{1}{\sqrt{2\pi}} \int_{-\infty}^{\infty} \mathrm{d}\hspace*{0.025cm}t\, \,\mathbf{E}\, (t)\, \exp \,( {\mathrm{i}\,\omega\, t} ) \, \bigg\rvert^2
\end{alignat}
becomes Lorentzian: $[ (\omega - \omega_0 )^2 + ({ {1}/{2\,\tau}} )^2 ]^{-1}$. Careful calculations by \citet{balian1970, balian1971} show that the spectral broadening of the resonances is inversely correlated with the maximum number of resolvable resonances as
\begin{equation}
N_{\max} = \frac{1}{3}Q = \frac{1}{3} \,\frac{\omega_{\max}}{\Delta\,\omega},
\end{equation}
where $\Delta\, \omega = 1/ \tau$ is the full width at half maximum (FWHM) of a resonance peak and $Q$ is the quality factor of the resonator. Typical qualities of normally conducting cavities are in the range of $10^3$--$10^4$ \cite{stockmann2006quantum} and one can distinctly identify abut 1000 resonances through the smudging. The workaround to this unavoidable impediment eventually came from the cryostats of a linear accelerator at Darmstadt, the S-DALINAC \cite{auerhammer1993}. \citet{graf1992distribution} discovered that a superconducting niobium billiard, immersed in a liquid He bath at $4$ K,  could yield $Q \sim 10^5$ to even $10^7$---a remarkable thousandfold improvement upon the experiments of old! This enhancement of the quality factor is not just academic but very much visible to the naked eye as Fig.~\ref{fig:spectrum} seeks to convince. The resultant extraordinarily sharp spectral peaks meant that one could resolve resonances to an unprecedented accuracy of $\Delta\,f < 100$ kHz, nearly two orders of magnitude below the mean level spacing of $17$ MHz. 

\subsection{The S-matrix and transmission measurements}

The unassuming microwave billiard quickly became the workhorse for several ingenious experiments on varied facets of quantum chaos, spanning from tests of random matrix and periodic orbit theory \cite{lewenkopf1992microwave, kudrolli1994signatures, kudrolli1995experimental} to spectral level dynamics \cite{kollmann1994periodic} and scattering matrix approaches \cite{alt1995gaussian, doron1990experimental}. However, our present considerations are of a slightly different nature. The transmission spectrum, invaluable as it is, does not directly tell us about the nodal structure, which must instead be accessed from the electric field distribution inside the cavity. The formalism best equipped to deal with such open systems is that of the S-matrix. The scattering (whence, S) matrix encapsulates the relation between the components of the amplitudes of the waves entering ($a_i$) and departing ($b_i$) through the $i^\mathrm{th}$ channel in a single matrix equation $b = S\,a$. The total number of open channels actually depends on the frequency $f$ because each lead can support $M = \lfloor 2 \, f \,d /c \rfloor$ modes. Using Green's function techniques \cite{stockmann2006quantum}, the S-matrix can be kneaded to the form
\begin{equation}
\label{eq:SMat}
S = 1 - 2 \,\mathrm{i}\,W^\dagger\frac{1}{k^2 - H + \mathrm{i}\,W\,W^\dagger}W,
\end{equation}
where $H = -\Delta$ is the Hamiltonian of the unperturbed system (without channels) and the matrix elements $W_{i, j}$ describe the coupling of the $i^\mathrm{th}$ eigenfunction to the $j^\mathrm{th}$ channel. In the presence of time-reversal symmetry, the scattering matrix is unitary and its off-diagonal terms are not independent: $S^*_{i, j} = S_{j, i}$.  Assuming non-overlapping resonances and point-like coupling \cite{kuhl2007wave}, Eq.~\eqref{eq:SMat} simplifies to
\begin{equation}
S_{i, j} (\mathbf{k}) = \delta_{i, j} -2 \,\mathrm{i}\,\gamma\,\xoverline{G}\left(\mathbf{r}_i, \mathbf{r}_j, \mathbf{k} \right),
\end{equation}
with the modified Green's function \cite{stein1995microwave}
\begin{equation}
\label{eq:breit-wigner}
\xoverline{G}\left(\mathbf{r}_i, \mathbf{r}_j, \mathbf{k} \right) = \sum_n \frac{\psi_n(\mathbf{r}_i) \,\psi_n(\mathbf{r}_j)}{k^2- k_n^2 + \mathrm{i}\,\gamma\,\sum_i \left \lvert \psi_n(\mathbf{r}_i) \right \rvert^2},
\end{equation}
reminiscent of the Breit-Wigner formula in nuclear physics \cite{blatt1979theoretical}. It might be helpful to pause for a moment and survey the barrage of notation introduced by Eq.~\eqref{eq:breit-wigner} in one fell swoop. Here, $\psi_n$ are the real eigenfunctions of the closed systems. The couplings to the antennae (captured by the parameter $\gamma$), in collusion with the summation $\sum_i$ over all open channels, define an effective broadening $\Gamma_n = \gamma\,\sum_i \left \lvert \psi_n(\mathbf{r}_i) \right \rvert^2$ that could also subsume additional effects of absorption in the system \cite{schafer2003correlation, fyodorov2005scattering, kuhl2005direct}. In words, Eq.~\eqref{eq:breit-wigner} claims that $\xoverline{G}\left(\mathbf{r}_i, \mathbf{r}_j, \mathbf{k} \right)$ can be obtained from transmission measurements between two antennae of variable position \cite{kuhl2007wave}. The transmission, in turn, solely depends on the field distributions at the locations of the antennae.\footnote{For instance,  it would be just as difficult, qualitatively speaking, to excite a resonance in the vicinity of a nodal line, as it would be easy in close proximity to a maximum.} Consequently, a two-dimensional scan by a probe/exit antenna (indexed 1), keeping the other entrance antenna(s) stationary, maps out the field $\mathscr{E} (x,y)$ inside the resonator \cite{stockmann2006quantum, stein1995microwave} and with it, a visual image of the wavefunction $\psi\,(x,y)$.

\begin{figure}
\includegraphics[width= \linewidth]{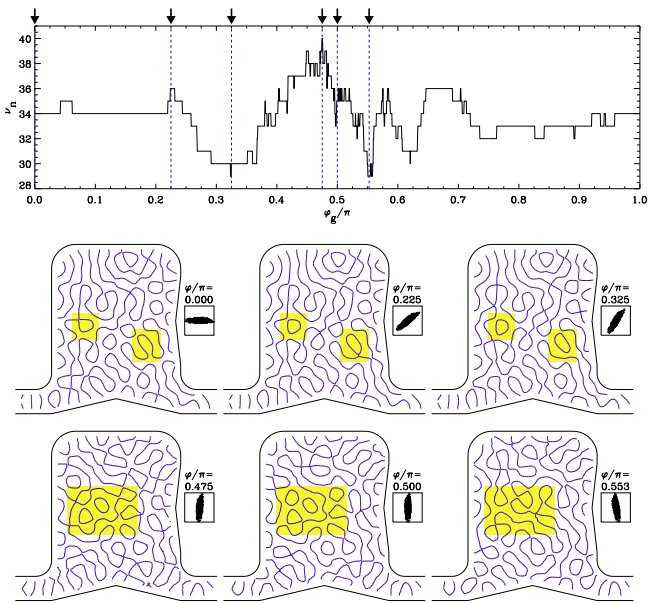}
\caption{\label{fig:phase}Number of nodal domains $\nu_n$ as a function of a global phase $\phi_g$ for
a wave function at $n_\mathrm{Weyl} \approx 223$. Arrows denote the phases $\phi_g$ for which the nodal domains are shown, corresponding to $\phi_g / \pi =$ 0, 0.225, 0.325, 0.475, 0.5, and 0.5525. Areas where appearances and disappearances of nodal domains can be observed are highlighted. The insets show the corresponding plots of $\psi_\textsc{i}$ versus $\psi_\textsc{r}$. From \citet{kuhl2007nodal}.}
\end{figure}

\begin{figure*}[htb]
  \includegraphics[width=0.3075\textwidth, valign=t]{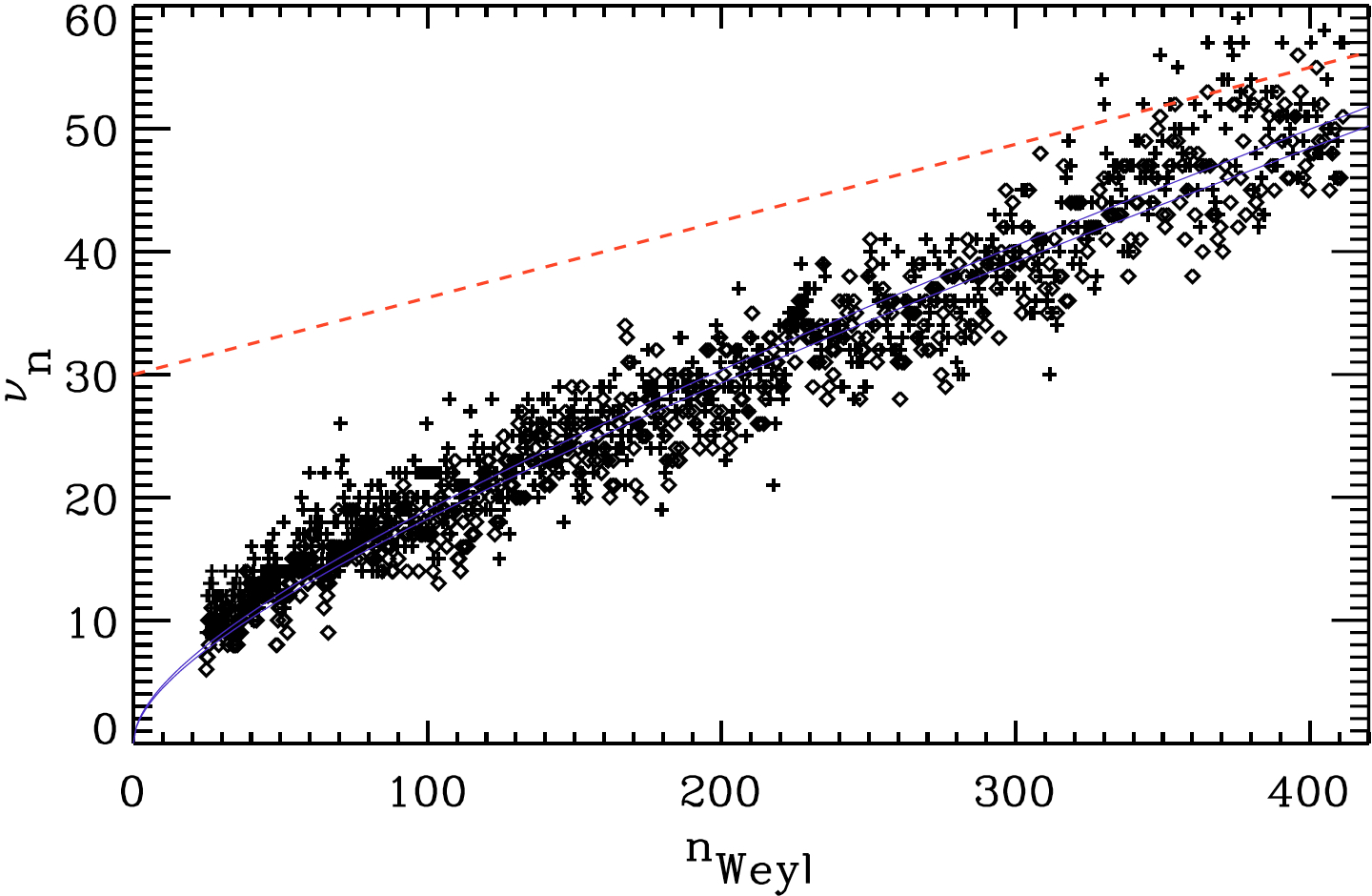}
  \includegraphics[width=0.32\textwidth,valign=t]{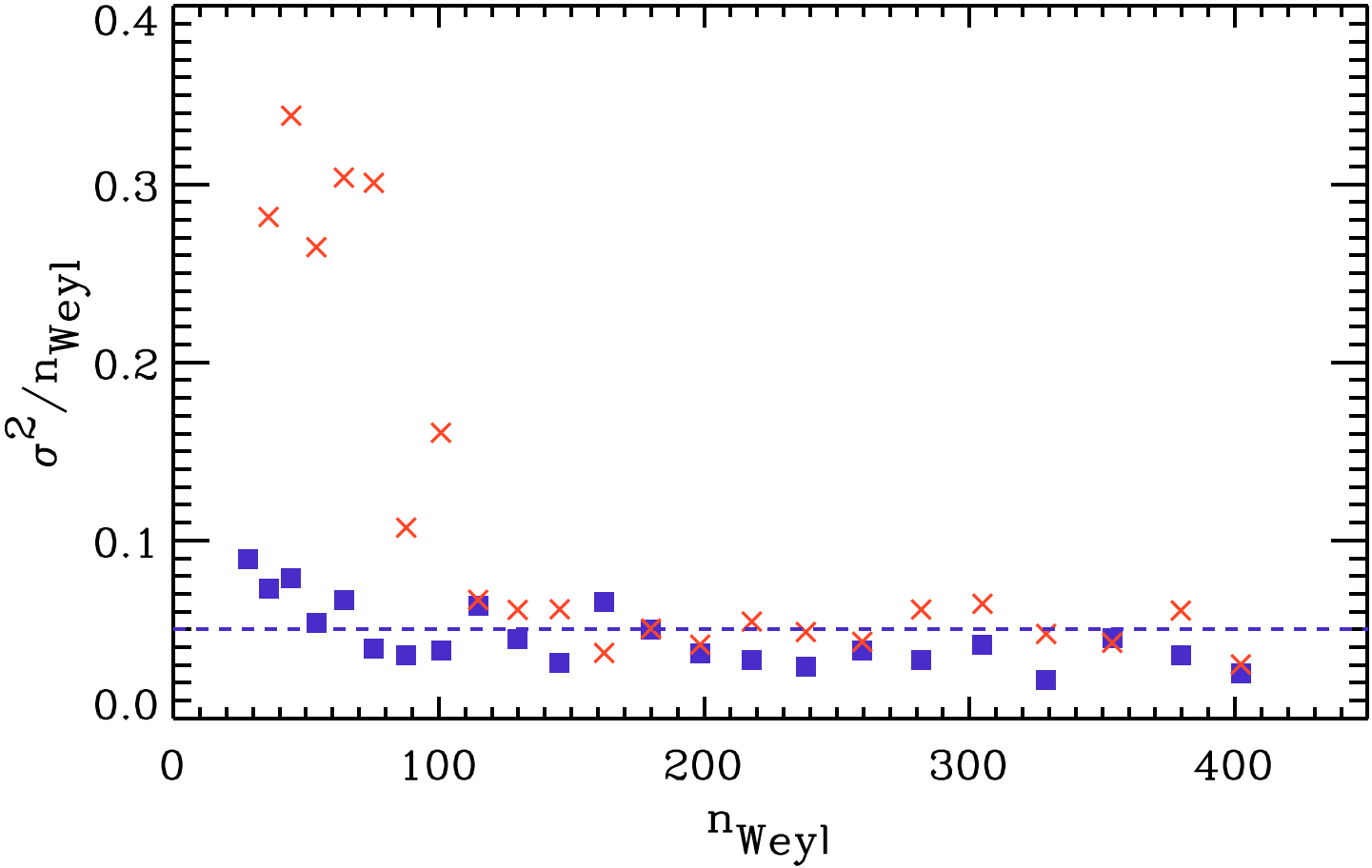} 
  \begin{picture}(0,0)
  \put(-70,-48){\includegraphics[width=2cm]{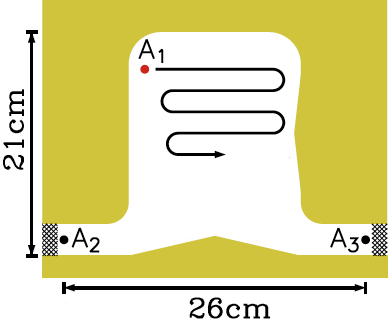}}
  \end{picture}  
  \includegraphics[width=0.335\textwidth,valign=t]{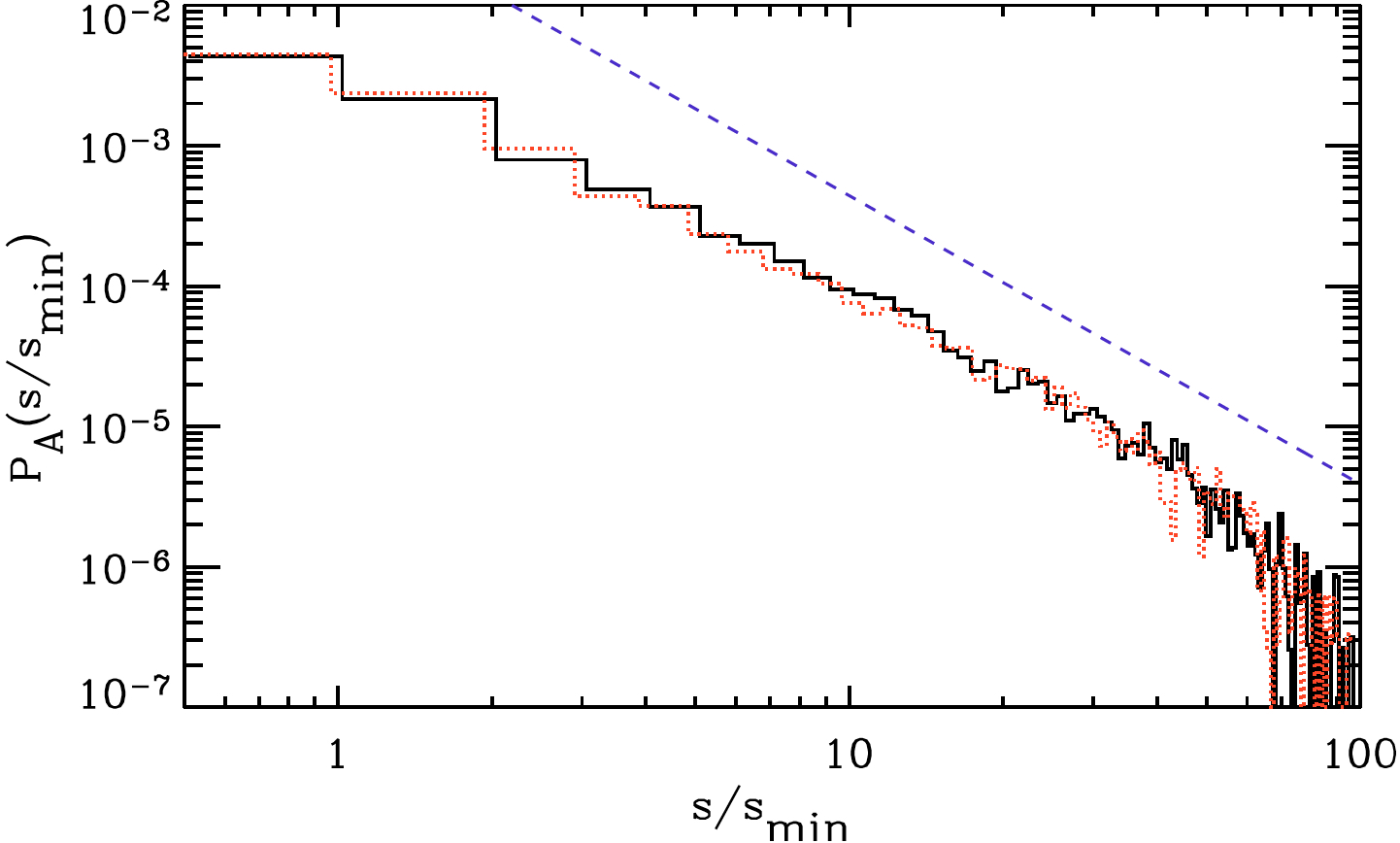}
  \begin{flushleft}
  \includegraphics[width=0.318\textwidth, valign=t]{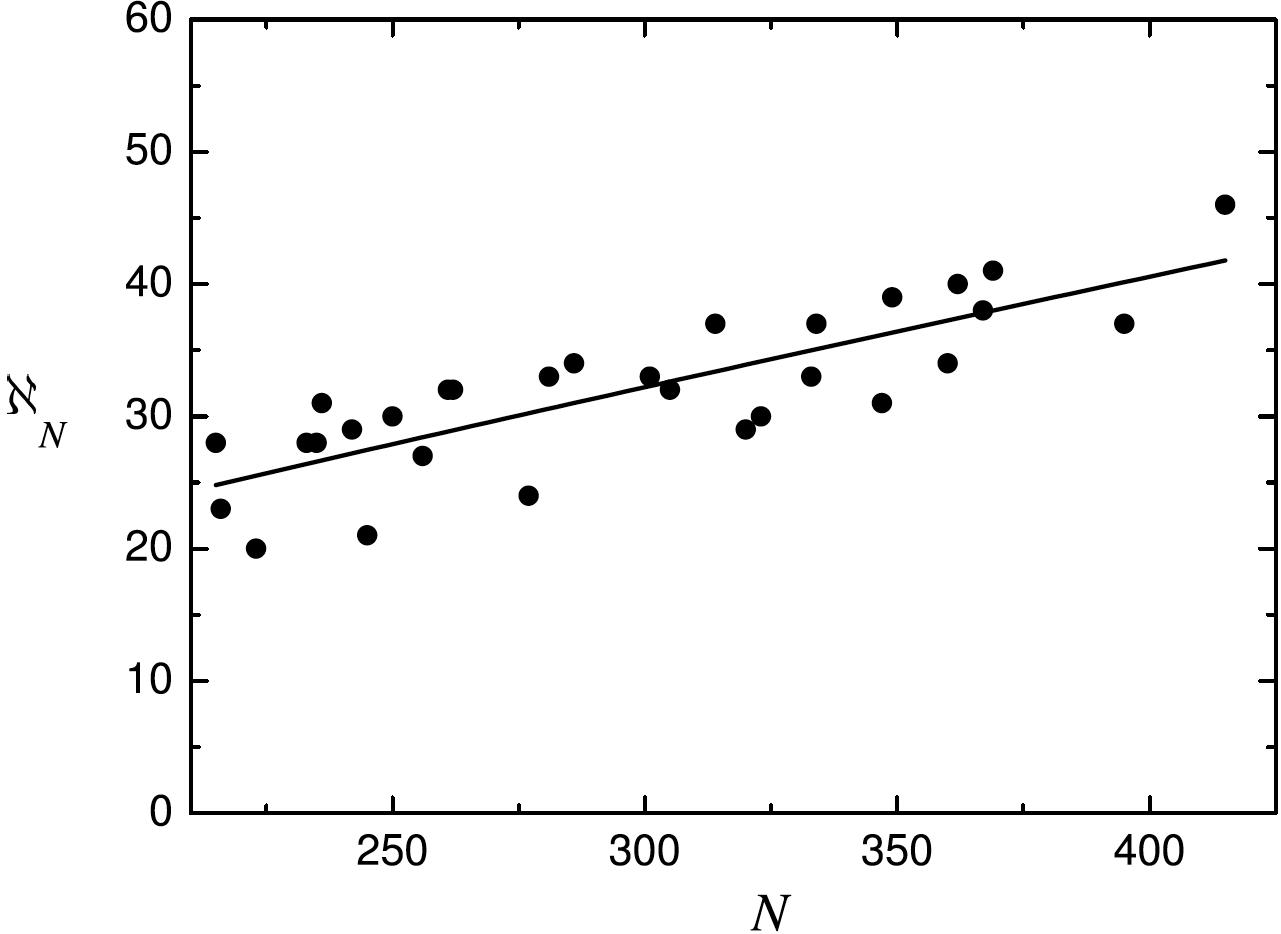}
  \includegraphics[trim={0.2cm 0 0 0.25cm}, clip, width=0.327\textwidth, height = 4.21cm, valign=t]{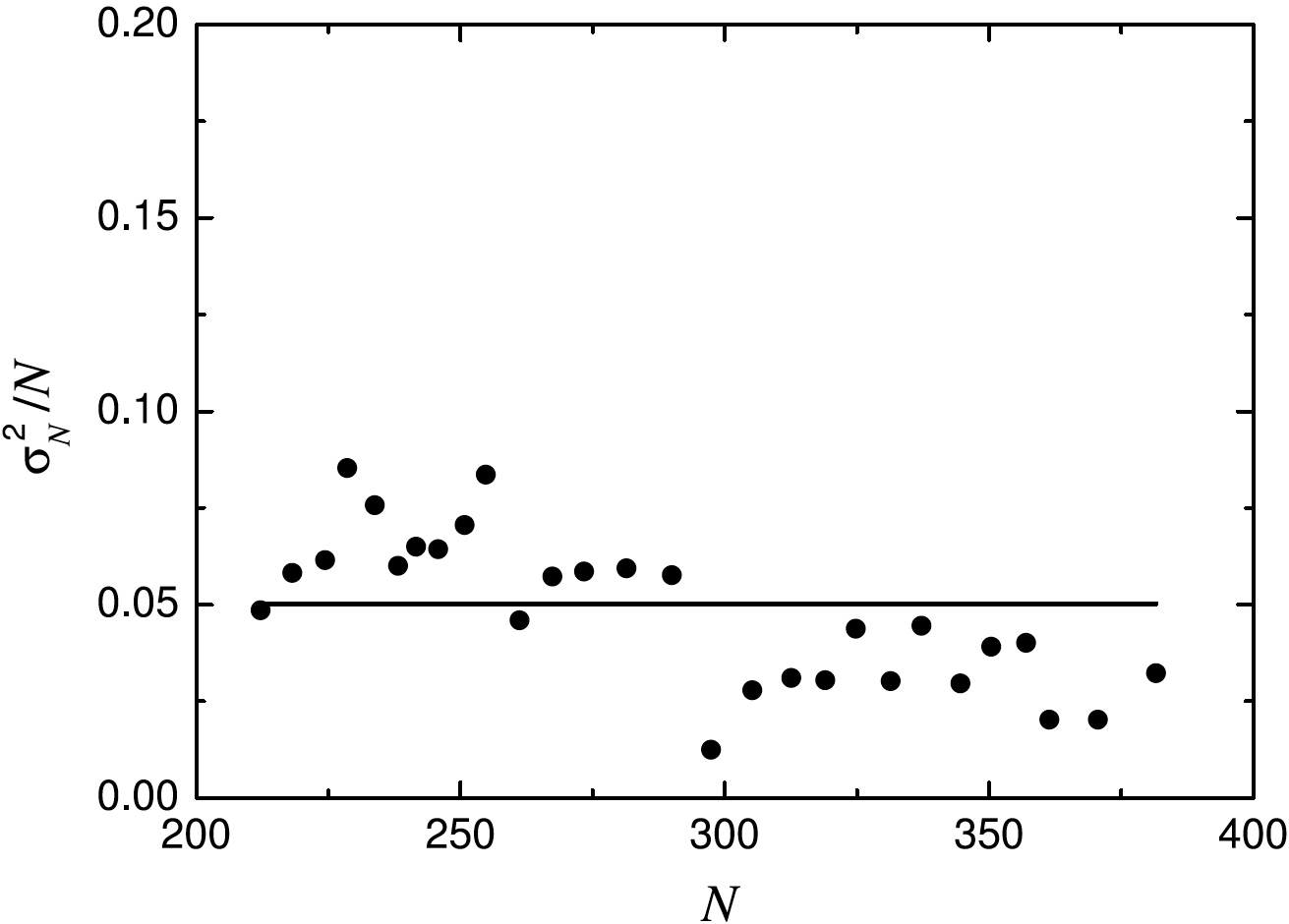}
  \begin{picture}(0,0)
  \put(-102,-58){\includegraphics[width=3cm]{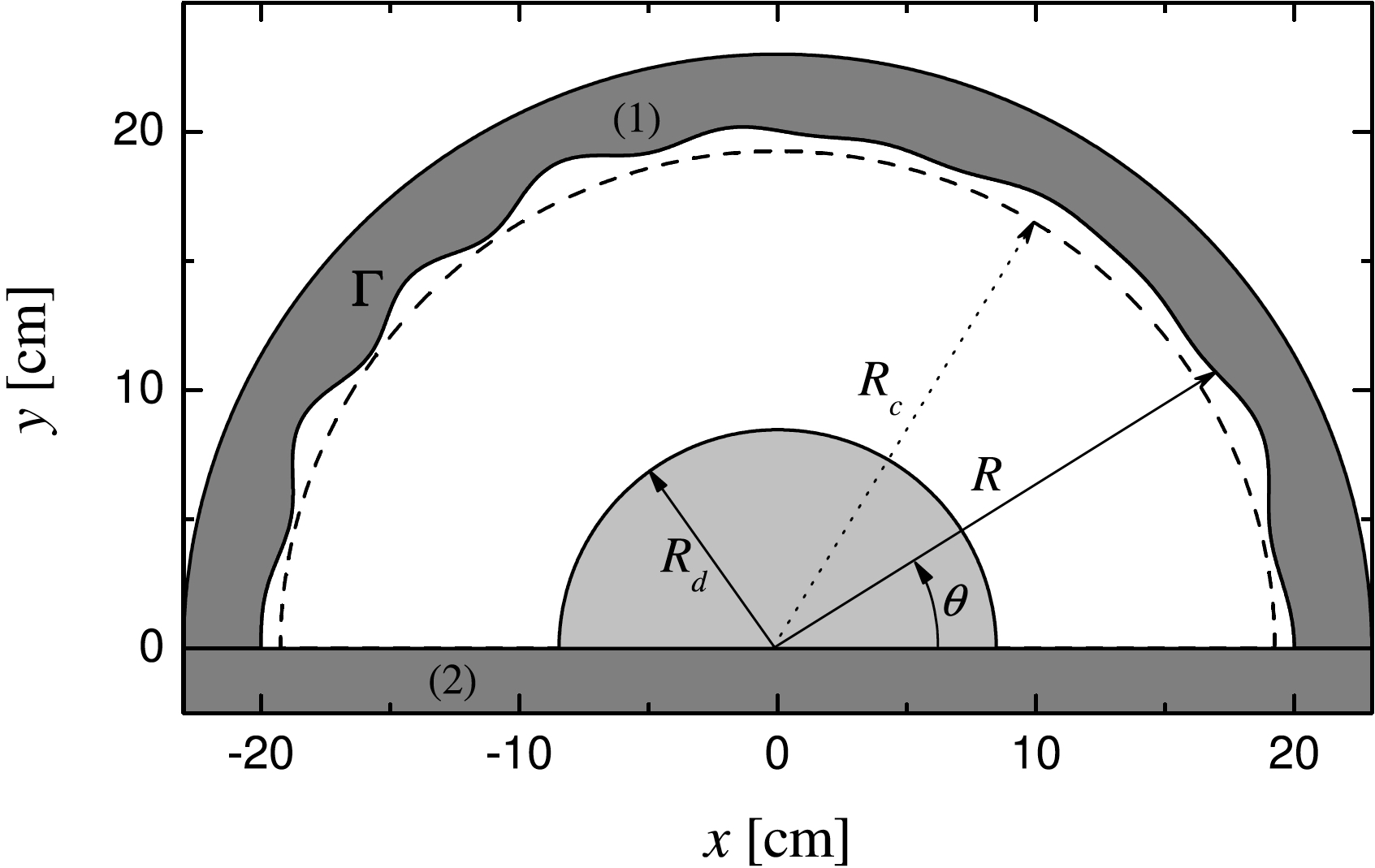}}
  \end{picture} 
  \includegraphics[trim={0.42cm 0 0 -0.4cm}, clip, width=0.318\textwidth, height = 4.32cm, valign=t]{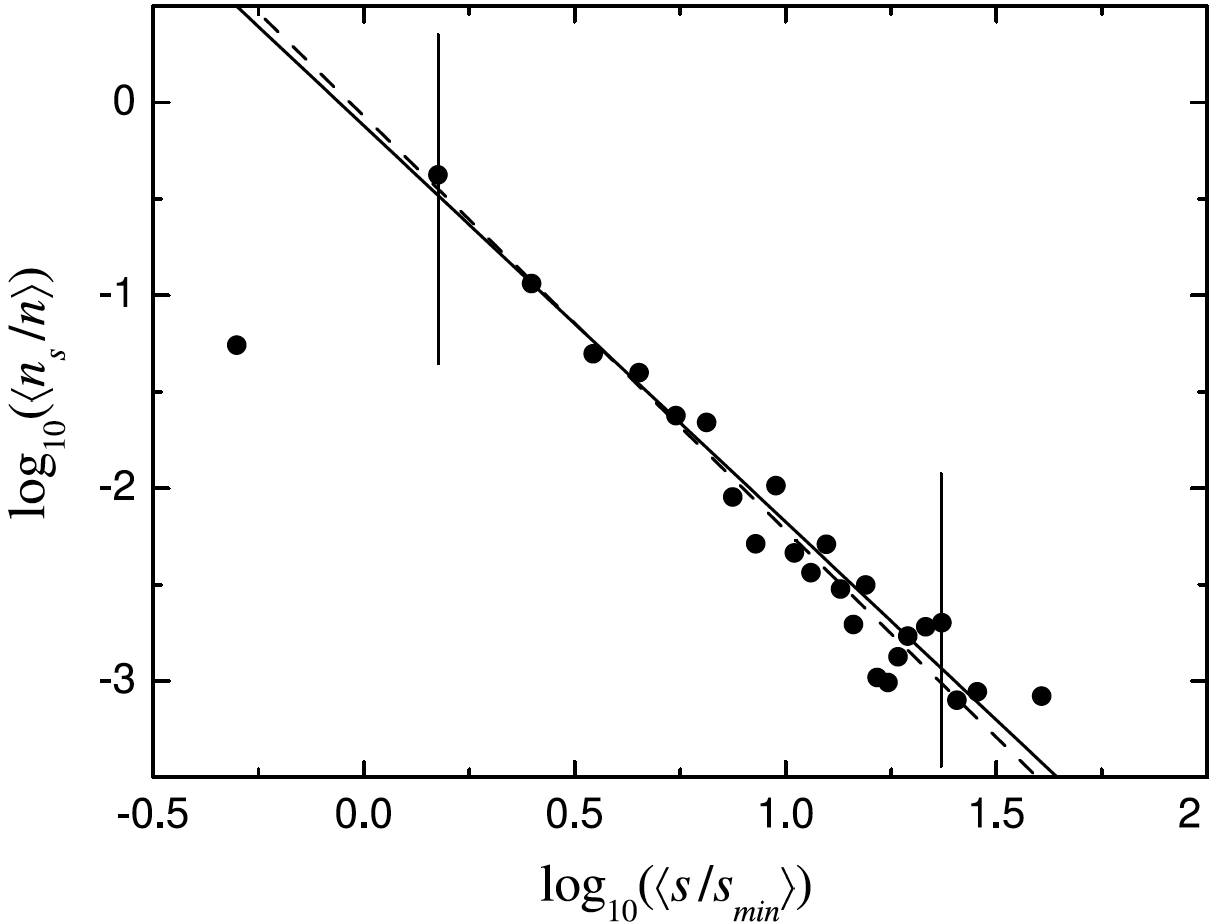}
  \end{flushleft}
  \caption{\label{fig:Exp}[Left to right]: The experimentally measured number of nodal domains, its scaled variance and the normalized area distribution for two different billiards (inset). The predictions of the Bogomolny-Schmit percolation model (or the linear fits thereto) are drawn as solid/dashed lines for comparison. [Top]: The billiard is a quantum dot-like structure of a rectangular shape with rounded corners and two attached leads. Data points for the real (imaginary) part are represented by crosses/solid histogram (diamonds/red dotted histogram). [Bottom]: The corresponding data for the chaotic half-circular microwave ray-splitting rough billiard, now plotted as solid circles. Adapted from \citet{kuhl2007nodal} and \citet{hul2005investigation}.}
\end{figure*}

This stratagem was employed to good effect by \citet{kuhl2007nodal} in investigating a certain class of open cylindrical microwave billiards. Stepping in increments of 1 MHz, they measured  the transmission $S_{12}$ from antenna 1 to 2, and the reflection $S_{11}$ at antenna 1, over a frequency range of 1--18.6 GHz. The experimental minutiae have been meticulously chronicled by \citet{kuhl2000mixing, veble2000experimental}. A particularly noteworthy detail is that with the aid of a vector network analyzer (VNA), the phase of the transmission $S_{12}$, $\phi$, can also be measured \cite{kuhl2005classical}. Since $\mathscr{E} (x,y)$ is a real quantity, this corresponds to determining both the real and imaginary parts of the complex wave function  $\psi\,(x,y) = \lvert\psi\,(x,y) \rvert \exp\,(\mathrm{i}\,\phi)$. More often than not, the real ($\psi_\textsc{r}$) and imaginary ($\psi_\textsc{i}$) parts are correlated due to unwelcome global phase shifts $\phi_{g}$, originating primarily from the leads and the antennae. This global phase can always be removed by an overall rotation \cite{ishio2001wave, saichev2002statistics},
\begin{equation}
\psi_\textsc{r} + \mathrm{i}\,\psi_\textsc{i} = \mathrm{e}^{-\mathrm{i}\,\phi_{g}}\, (\psi_\textsc{r}' + \mathrm{i}\,\psi_\textsc{i}').
\end{equation}
It does, however, have physical implications. Fig.~\ref{fig:phase} displays the nodal domains and their number for the real part of a wavefunction, at $n_\mathrm{Weyl} \approx 223$, as $\phi_g$ is varied. To be pedantic, the Weyl law is not strictly valid for open systems \cite{lu2003fractal, nonnenmacher2005fractal} in which the resonances are shifted to the complex plane or are, occasionally, even removed from the spectrum altogether \cite{lehmann1995chaotic}. Nonetheless, in the absence of a viable alternative\footnote{A popular candidate is the resonance counting function $n_r (k) = \sum_{i=1}^c \Phi_i (k) - c/2$, $ \Phi_i $ being the eigenphases of the S-matrix and $c$ the number of open channels \cite{doron1992chaotic, doron1992semiclassical}. Sadly, absorption disallows the calculation of $n_r$ from data.} at our disposal, it is still the best approximation---a one-eyed king in the land of the blind. The wavefunction under scrutiny is additionally characterized by a phase rigidity \cite{van1997fluctuating} that quantifies the extent to which the system is open \begin{equation*}
{\displaystyle \lvert \rho \rvert^2 = \bigg \lvert \frac{\int \mathrm{d}\,\mathbf{r}\,\, \psi\,(\mathbf{r})^2}{\int \mathrm{d}\,\mathbf{r}\,\,\lvert\psi\,(\mathbf{r}) \rvert^2} \bigg \rvert^2= \bigg \lvert \frac{\langle \psi_\textsc{r}^2\rangle_\mathcal{A} - \langle \psi_\textsc{i}^2\rangle_\mathcal{A}}{\langle \psi_\textsc{r}^2\rangle_\mathcal{A} + \langle \psi_\textsc{i}^2\rangle_\mathcal{A}} \bigg \rvert^2} \approx 0.81,
\end{equation*}
derived under the assumption that the real and imaginary parts, $\psi_\textsc{r}$ and $\psi_\textsc{i}$, are not correlated. Indeed, the two are uncorrelated for $\phi_g = 0$ and remain so at $\phi_g = \pi/2$, switching identities (and getting correlated) in between. While the phase is changing, the nodal lines are shifted and permanently dissolved and reconnected \cite{kuhl2007wave} and such rearrangements continually alter the number of nodal domains. 
Hence, it is of utmost importance that the effects of the phase are corrected for before any statistical calculations. Respecting this obligation, \citet{kuhl2007nodal} fit the experimentally observed number of nodal domains to $\nu_n = a\,n + b\,\sqrt{n}$ \cite{blum2002nodal},
taking $n = n_\mathrm{Weyl}$, with $a = 0.059\, (0.060)$ and $b = 1.23\, (1.30)$ for the real (imaginary) part of the wavefunction. The second term above takes boundary effects into account. These values are in accordance with the Bogomonly-Schmit prediction of $a = 0.0624$ (see Fig.~\ref{fig:Exp}). The variance and the area distribution also concur well with the percolation model, leading \citet{kuhl2007wave} to conclude that ``there is no difference between the nodal domains statistics of real and imaginary parts of complex wave functions in open billiards, and the corresponding statistics for real wave functions in closed systems.''

\subsection{The perturbing bead method}
\label{bead}

Another widely-used approach for the determination of wave functions is, for reasons self-explanatory, called the perturbing bead method. Successfully applied by \citet{sridhar1991experimental, sridhar1992physical, dembowski2000first, wu1998probability} to a number of differently shaped billiards, it involves, quite literally, introducing a small metallic bead or perturber into the cavity, which then alters its resonant frequencies $f_n$. The difference induced is proportional to the square of the field strengths (in the unperturbed cavity) at the location of the bead and is given by 
\begin{equation}
\label{eq:perturb}
\Delta \,f_n = f - f_n = f_n \,\left(A\,\mathscr{B}_n^2 - B\,\mathscr{E}_n^2  \right),
\end{equation}
where $A$ and $B$ are geometrical factors \cite{maier1952field}. By measuring the frequency shift as the bead's position is varied, we procure a mapping of the field distribution. This, of course, is easier said than done and actually reconstructing the wavefunction proves to be a formidable challenge. A systematic method to do so using ``trial functions'' was first developed by \citet{savytskyy2002}, which we outline in the context of experiments on chaotic rough billiards. These, and associated systems, have received considerable attention from motley quarters such as in relation to dynamic localization \cite{sirko2000observation}, localization in discontinuous quantum systems \cite{borgonovi1998localization}, microdisk lasers \cite{yamamoto1993, nockel1997ray} and ballistic electron transport in microstructures \cite{blanter1998quantum}. \citet{savytskyy2004} constructed a billiard of this type out of an aluminum microwave cavity in the shape of a rough half-circle. The rough segment is described by the radius function $R\,(\theta) = R_0 + \sum_{m=2}^M a_m\,\sin\left(m\, \theta + \phi_m \right)$ with a mean radius $R_0$ and the phases $\phi_m$ uniformly distributed in $[0, 2\pi]$. On loading a semicircular Teflon insert of radius $R_d < R_0$ \cite{hul2005investigation}, the dielectric constant (or the Sch\"{o}dinger potential) becomes discontinuous inside the resonator and the billiard turns ray-splitting \cite{blumel2001ray, couchman1992quantum}. The qualifier ``ray-splitting'' refers to a class of chaotic systems with non-Newtonian and nondeterministic classical dynamics (e.g., \citet{sirko1997experimental, savytskyy2001parametric, bauch1998signature, hlushchuk2000autocorrelation}) in which rays, as the name suggests, split upon reflection from sharp boundaries. Both these sets of experiments, with and without ray-splitting, allow one to encroach upon the regime of Shnirelman ergodicity in which the wave functions are expected to be homogeneously distributed on the energy surface, abiding by the quantum ergodicity theorem \cite{shnirel1974ergodic}. 

Let us work out the details for the simple rough billiard, for which the calculations are a little less involved. The cardinal premise is that the wave functions $\psi_n (r, \theta)$ can be determined from the electric field $\mathscr{E}_n (R_c, \theta)$, evaluated on a semicircle of fixed radius $R_c < R_0$. Since the perturbation of Eq.~\eqref{eq:perturb} only concerns itself with $\lvert \mathscr{E}_n \rvert^2$,\footnote{Of course, $\lvert \mathscr{B}_n \rvert^2$ can also come into play, depending on the mode excited, but its influence on $\Delta f_n$ can be minimized by using a small, vertically-positioned piece of a metallic pin as a perturber.} we still need to somehow recover the signs. To do this, first, we identify all the minima of  $\lvert \psi_n (R_c, \theta) \rvert$ that are close to zero. Then, we assign the signs ``minus'' and ``plus'' in alternating fashion to the region between consecutive minima, starting (arbitrarily) with the negative, generating our trial wavefunction $ \psi_n (R_c, \theta)$ in the process. It cannot be ascertained a priori that this assignment of the signs is correct. An a posteriori sanity check is that \textsl{if} it indeed is, the reconstructed wave function should automatically vanish on the boundary, i.e., $\psi_n \left(r_{\partial\,\mathcal{D}}, \theta_{\partial\,\mathcal{D}} \right) = 0$. Formally, the wavefunctions of the rough half-circular billiard can be expanded in a basis of circular waves (only odd states are considered here) as
\begin{equation}
\psi_n (r, \theta) = \sum_{s=1}^L a_s\,C_s\,J_s \left(k_n r \right) \sin (s \,\theta),
\end{equation}
where the number of basis functions $L = k_n\,r_{\max}$ grows with the maximum radius of the cavity $\,r_{\max}$. The accompanying coefficients are
\begin{alignat}{1}
C_s &= \left[ \frac{\pi}{2} \int_0^{r_{\max}} \left \lvert J_s (k_n r) \right \rvert^2 r \,\,\mathrm{d}\, r \right]^{-1/2},\\
\nonumber a_s &= \bigg[\frac{\pi}{2} C_s\, J_s \left( k_n R_c \right) \bigg]^{-1} \int_0^\pi \psi_n \left(R_c,\theta \right)\,\sin(s\,\theta)\, \mathrm{d}\,\theta.
\end{alignat}
The utility of the trial wavefunction is now patent. Putting this technique to use, \citet{savytskyy2004} were able to reconstruct 156 experimental wavefunctions for the rough microwave billiard; \citet{hul2005investigation} extracted 30 with ray-splitting. The computed statistics (Fig.~\ref{fig:Exp}) were consistent with the percolation model with Bogomolny-Schmit parameters $a = 0.058 \pm 0.006$, $\tau =1.99 \pm0.14$ and $a = 0.063 \pm 0.023$, $\tau = 2.14 \pm 0.12$, in the absence and presence of ray-splitting, respectively.

\subsection{Current and vortex statistics}
\label{sec:expVortex}

Over the years, another broad class of experiments on the nodal structure of quantum billiards have gained prominence. Picking up where we left off in Sec.~\ref{flow}, we now sketch some of these experimental studies probing the distributions and correlations of wavefunctions, currents flows, and vortices.

\begin{figure*}[htb]
\subfigure[]{\includegraphics[width= 0.25\linewidth, trim={0 -2.6cm 0 0}, valign=t]{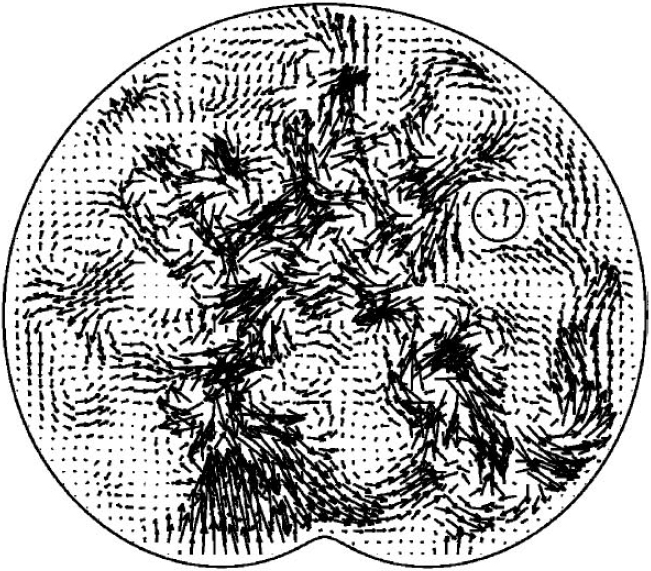}}
\subfigure[]{
\includegraphics[width= 0.35\linewidth, valign=t]{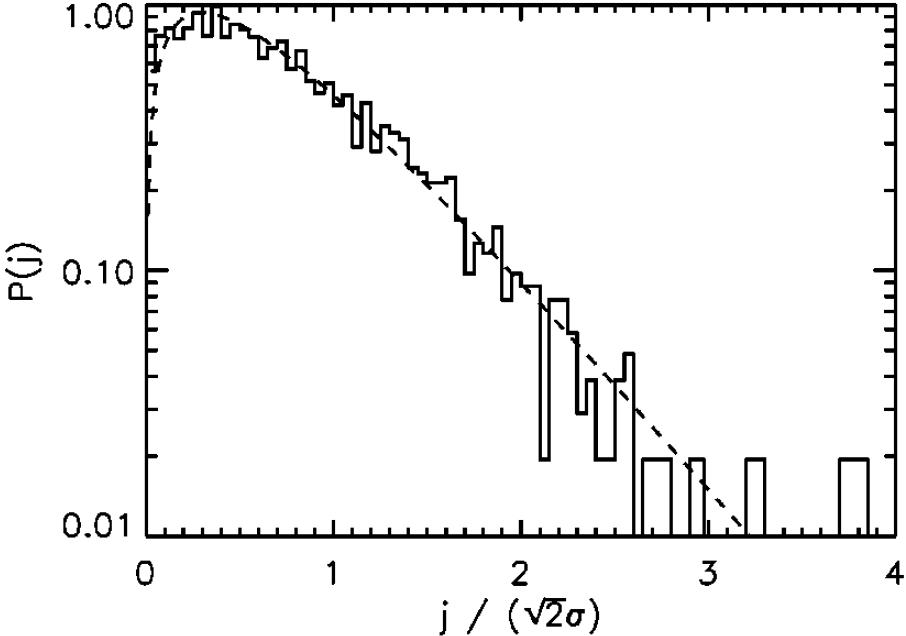}
\begin{picture}(0,0)
  \put(-78,-5){\includegraphics[width= 0.1375\linewidth, valign=t]{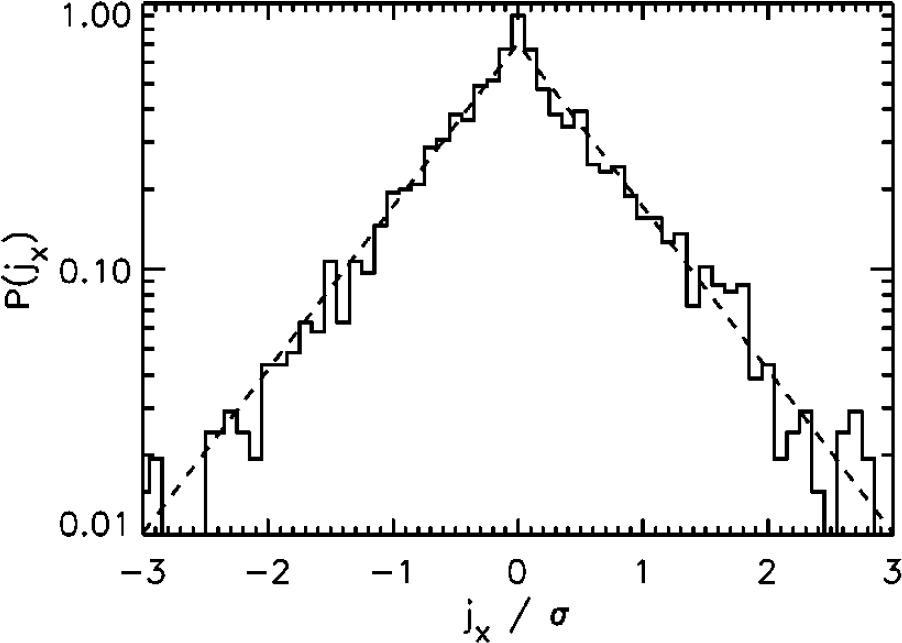}}
  \end{picture}  
}
\subfigure[]{\includegraphics[width= 0.35\linewidth, valign=t]{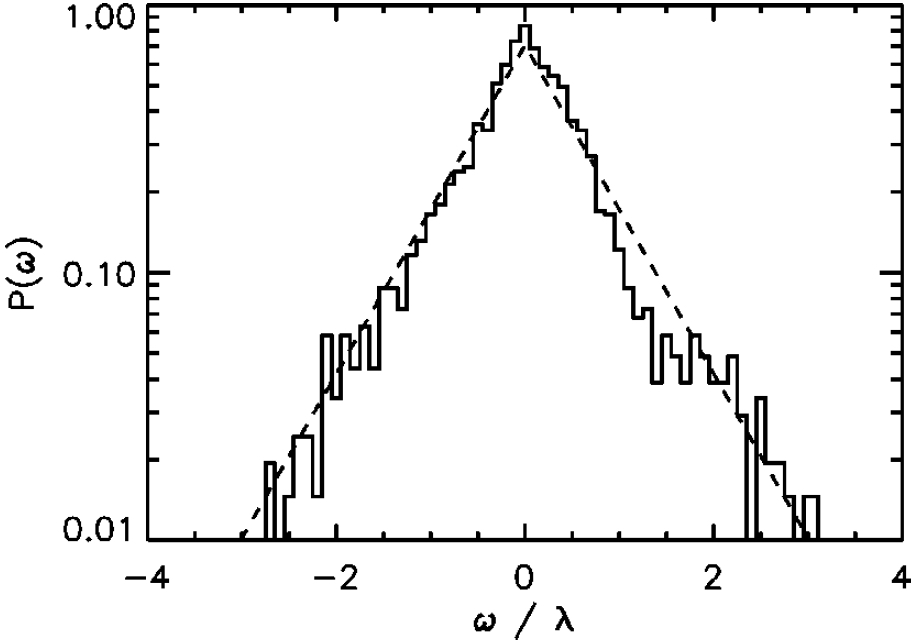}}
\caption{\label{fig:ferrite}(a) Map of the current in a ferrite billiard at 6.41 GHz. The lengths of the arrows correspond to the magnitude of the Poynting vector. (b) Distributions of $\lvert j \rvert$ (inset: $j_x$) and (c) $\Omega$, with $\sigma$ and $\lambda$ being shorthand for $\sqrt{\langle j^2_x\rangle}$ and $\langle \Omega^2 \rangle$, respectively. Dashed lines indicate the theoretical predictions of Eqs.~\eqref{eq:modj}--\eqref{eq:P(v)}. Adapted From \citet{barth2002current}.}
\end{figure*}

Following up on the correspondence between the Schr\"{o}dinger and Helmholtz equations, we can relate the Poynting vector \cite{seba1999experimental}
\begin{equation}
\mathbf{S} = \frac{c}{4\, \pi} \mathbf{E}\times \mathbf{H} =  \frac{c}{8\,\pi\,k} \im \left[ \mathbf{E}^* (\mathbf{r}) \,\nabla\, \mathbf{E}\,(\mathbf{r}) \right],
\end{equation}
in two dimensions, to the probability current density
\begin{equation}
\mathbf{j} \,(\mathbf{r}) \equiv \left(j_x,\,j_y \right) = \frac{\mathcal{A}}{k} \im \left[ \mathbf{\psi}^* (\mathbf{r}) \,\nabla \mathbf{\psi}\,(\mathbf{r}) \right].
\end{equation}
Using the random plane wave ansatz, it is not difficult to calculate the distributions of the current components \cite{saichev2002statistics}
\begin{alignat}{1}
\label{eq:modj}
P\,(\lvert j \rvert) &= \frac{4\,j}{\left \langle j^2 \right \rangle}\, K_0\, \bigg(\frac{2\,j}{\sqrt{\left \langle j^2 \right \rangle}} \bigg),\\
\label{eq:jx}
P\,(j_x) &= \frac{1}{\sqrt{2\left \langle j_x^2 \right \rangle}}\, \exp\, \bigg(-\sqrt{\frac{2}{\left \langle j_x^2 \right \rangle}} \left \lvert j_x \right \rvert \bigg),
\end{alignat}
where the parameter
\begin{equation}
\left \langle j_{x}^2 \right \rangle = \frac{1}{2} \left \langle j^2 \right \rangle = k^2 \left \langle \psi_R^2 \right \rangle\,\left \langle \psi_I^2 \right \rangle
\end{equation}
is directly accessible from experiments. Similarly, one can compute the corresponding distribution function for the vorticity (Eq.~\ref{eq:vorticity})
\begin{equation}
\label{eq:P(v)}
P\,(\Omega) = \frac{1}{\sqrt{2\left \langle \Omega^2 \right \rangle}}\, \exp\, \bigg(-\sqrt{\frac{2}{\left \langle \Omega^2 \right \rangle}} \left \lvert \Omega \right \rvert \bigg),
\end{equation}
where, once again,
\begin{equation}
\left \langle \Omega^2 \right \rangle = \frac{1}{2} \,k^4 \left \langle \psi_R^2 \right \rangle\,\left \langle \psi_I^2 \right \rangle
\end{equation}
can be taken straight from the experiment \cite{barth2002current}. The major stumbling block in experimentally determining the current distributions, however, is that the probe antenna, perforce, gives rise to a leakage current, which tampers with the statistical properties \cite{seba1999experimental}. The only way to guarantee that this influence is minimal is to either ensure a strong flow through the system or to choose frequencies such that the overall amplitudes are moderate \cite{barth2002current}. The statistical distributions at one such choice of frequency are shown in Fig.~\ref{fig:ferrite} for a lima{\c c}on billiard \cite{robnik1983classical, robnik1984quantising} with a TRS-breaking ferrite ring. In particular, the distributions for $j_x$ and $j_y$ are found to be identical but it need not always be so. For instance, in an open billiard the maximum of the $j_x$ distribution could be shifted significantly to negative (positive) values due to transport from the right (left) to the left (right) through the billiard \cite{barth2002current}.

Of related interest are the vortices of the current flow (the nodal points of the complex wavefunctions). In the plane, their positions are not independent but rather correlated. To characterize this, \citet{Berry2059} presented two types of vortex spatial autocorrelation functions. The first is the \textsl{pair correlation function} $ g\,(\mathbf{r})$, which quantifies the mean density of vortices at position $\mathbf{\bar{r} + r}$, given that there is a vortex at $\mathbf{\bar{r}}$. This is defined by
\begin{alignat}{1}
\label{eq:pair}
\nonumber g\, (r) =&\, g_0\, \big \langle \delta \left (\psi_R (\bar{\mathbf{r}} + \mathbf{r}) \right)\,\delta \left (\psi_I (\bar{\mathbf{r}} + \mathbf{r}) \right)\, \delta \left (\psi_R (\bar{\mathbf{r}}) \right) \, \delta \left (\psi_I (\bar{\mathbf{r}}) \right) \\
&\times \lvert \Omega \,(\bar{\mathbf{r}} + \mathbf{r}) \rvert \, \lvert \Omega\, (\bar{\mathbf{r}}) \rvert \big \rangle_{\bar{\mathbf{r}}},\\
\nonumber =&\, g_0\, \big \langle D_v (\bar{\mathbf{r}} + \mathbf{r}) \,D_v (\mathbf{r}) \big \rangle_{\bar{\mathbf{r}}},\\
=&\, \frac{2 \left(E^2 -  F_0\,(1-C^2) \right) }{\pi\,F_0\,(1- C^2)^2}\\
\nonumber&\times \int_0^\infty \mathrm{d}\,t\,\frac{3 - Z + 2Y + (3 + Z - 2Y)\,t^2 + 2\,Z\,t^4}{(1+t^2)^3 \sqrt{1 + (1+Z - Y)\,t^2 + Z\,t^4}},
\end{alignat}
where $g_0$ is a normalization factor such that $g\, (r) \rightarrow 1$ for $r \rightarrow \infty$; ordinarily, $g_0 = \rho^{-2}$ for a dislocation point density of $\rho$ (Eq.~\ref{eq:disdensity}). Let us go through the dictionary for the remaining symbols. $C$, as previously (see Eq.~\ref{eq:Berry}), is the non-local autocorrelation function
\begin{equation*}
C (r) = \left \langle \psi_{R,I}(1)\,\psi_{R,I}(2) \right \rangle =  \langle \langle J_0 (k\, r) \rangle \rangle = \bigg \langle \hspace*{-0.1cm} \bigg \langle \frac{\sin \,(K\,r)}{K\,r}\bigg \rangle \hspace*{-0.1cm} \bigg \rangle,
\end{equation*}
with $1$, $2$ denoting different positions and $k = \lvert (k_x, k_y ) \rvert$, $K = \lvert (k_x, k_y, k_z ) \rvert$. Building on this, we define
\begin{alignat*}{1}
E &\equiv C'\,(r); \, F \equiv - C''\,(r); \,H \equiv - C'\,(r)/r,\\
D_1 &\equiv \left[E^2 - (1 + C)\,(F_0 - F)\right]\,\left[E^2 - (1 - C)\,(F_0 + F)\right],\\
D_2 &\equiv F_0^2 - H^2,\quad \mbox{with} \quad F_0 \equiv - C''\,(0),
\end{alignat*}
and finally,
\begin{equation}
\nonumber Y \equiv \frac{H^2 \left(C E^2 -  F (1-C^2) \right)^2}{F_0^2 \left(E^2 - F_0 (1-C^2) \right)^2}; Z \equiv \frac{D_1\,D_2\,(1 - C^2)}{F_0^2 \left(E^2 - F_0 (1-C^2) \right)^2}.
\end{equation}
Knowing $g\,(r)$, one can calculate the nearest neighbor distribution of vortices. This distribution is dependent on whether the billiard is nominally either regular or irregular \cite{berggren1999signatures} and could thus potentially serve as yet another signature of quantum chaos. For ease of statement, we introduce the dimensionless pair correlation function
\begin{equation}
G (\ell) = g \left(  \frac{\ell}{\sqrt{\rho}}\right), 
\end{equation}
where $\rho = k^2 / 4 \pi$ is the bulk mean density of vortices for a homogeneous Gaussian field and $\ell = \sqrt{\rho}\,r$. Under the Poisson approximation, i.e., ignoring $n$-point correlations beyond $n = 2$, \citet{saichev2002statistics} calculated the distribution function for the nearest distances between the nodal points as
\begin{alignat}{1}
\label{eq:DFNDNP}
f\,(\ell) &\approx 2 \pi\, \ell\, \,G \,(\ell)\,\, \exp \left(- 2 \pi \int_0^\ell z\, G\,(z)\,\mathrm{d}\, z \right)\\
&\sim \frac{\pi}{2} \,\ell \hspace*{0.25cm} (\ell \rightarrow 0).
\end{alignat}
Although the Poisson approximation implicitly assumes that all nodal points around a given one are statistically independent, the end result, Eq.~\eqref{eq:DFNDNP}, is still an extremely useful reference point for experimental data nonetheless (e.g., \citet{kuhl2007wave, kim2003current}).

\begin{figure}[htb]
\includegraphics[width= \linewidth]{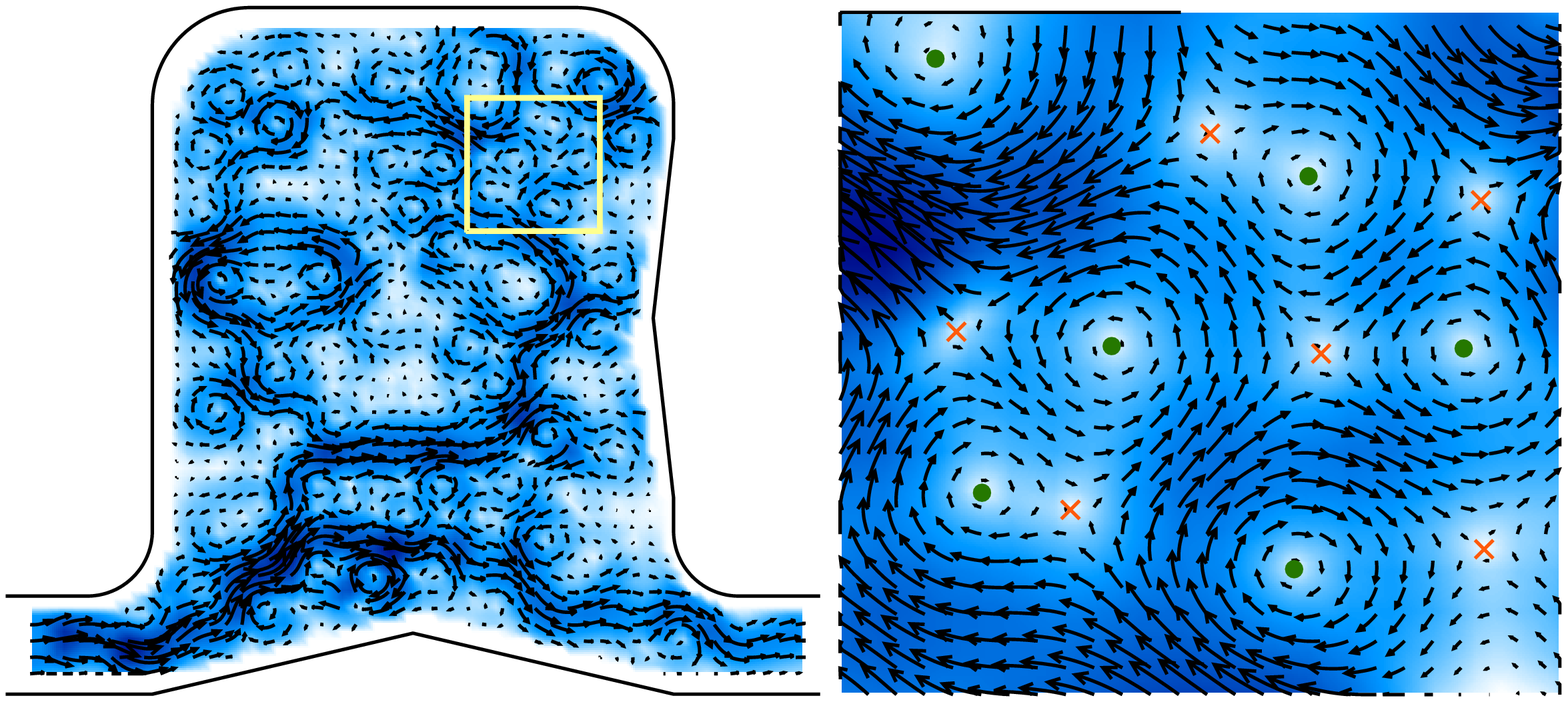}

\vspace{0.2cm}

\includegraphics[width= \linewidth, valign=t]{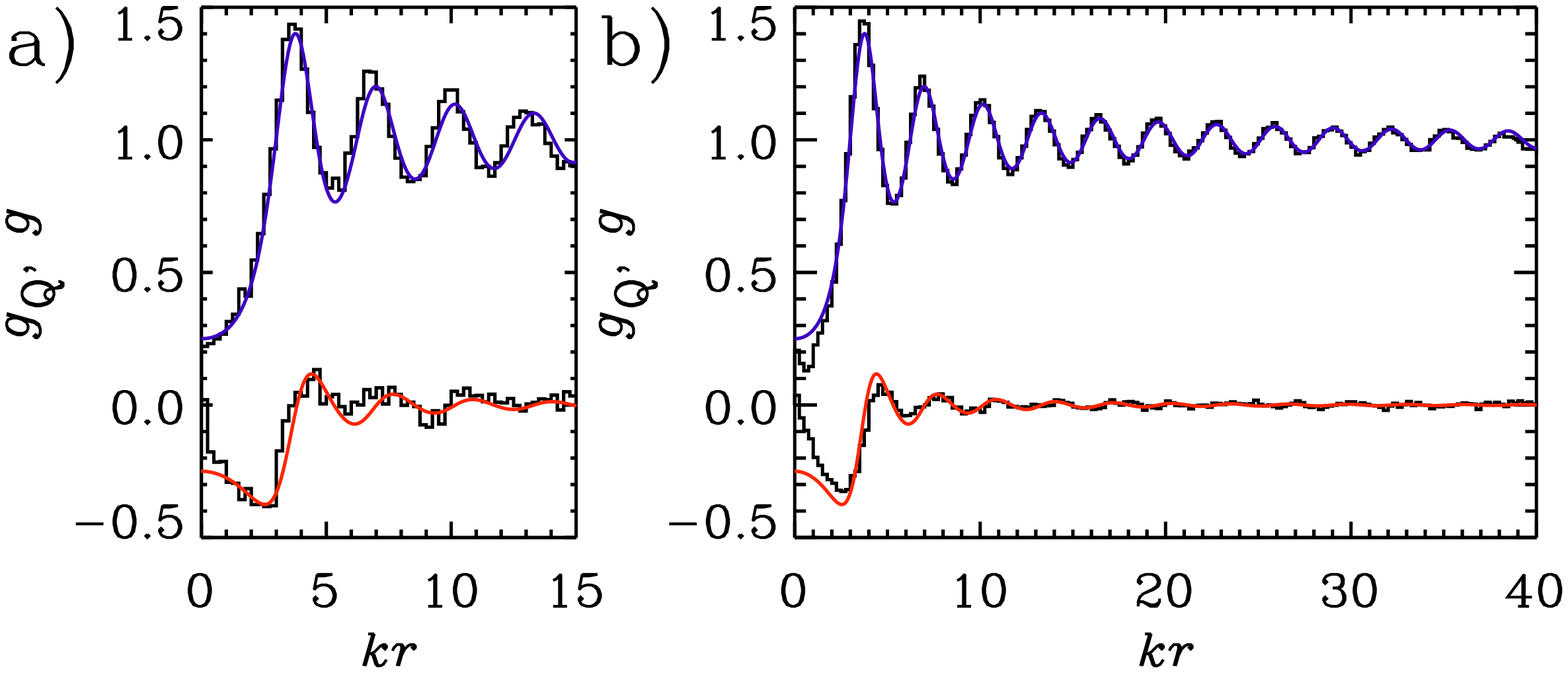}
\caption{\label{fig:current}[Top]: Probability current density $\mathbf{j}$ in an open microwave billiard. The
vortex (full disk) and saddle (crosses) structure is clearly observed in the enlarged area, showing clockwise and anticlockwise rotation, respectively. [Bottom]: Vortex pair correlation function $g$ and charge correlation function $g_Q$ for vortices taken from low (5--9 GHz, a) and high (15--18.6GHz, b) frequency regimes. The slight mismatch in the oscillation length is due to boundary effects in the low-frequency limit \cite{backer2002behaviour, hohmann2009density}, where the wavelength is no longer much smaller than the system size. From \citet{kuhl2007wave}. With kind permission of The European Physical Journal (EPJ).}
\end{figure}

A slightly simpler quantity to consider is the \textsl{charge correlation function} $g_Q \,(r)$ \cite{halperin1981}, which gives the normalized density of vortices separated by $r$, but now weighted by their strengths so that vortices of opposite sign (and sense of rotation) contribute antagonistically. Formally, we have
\begin{alignat}{1}
\label{eq:charge}
\nonumber g_Q (r) =&\, g_0\, \big \langle \delta \left (\psi_R (\bar{\mathbf{r}} + \mathbf{r}) \right)\,\delta \left (\psi_I (\bar{\mathbf{r}} + \mathbf{r}) \right)\, \delta \left (\psi_R (\bar{\mathbf{r}}) \right) \, \delta \left (\psi_I (\bar{\mathbf{r}}) \right) \\
&\times \Omega \,(\bar{\mathbf{r}} + \mathbf{r}) \, \Omega\, (\bar{\mathbf{r}})  \big \rangle_{\bar{\mathbf{r}}},\\
\nonumber =&\, g_0\, \big \langle D_v (\bar{\mathbf{r}} + \mathbf{r})\,\mathscr{S} (\bar{\mathbf{r}} + \mathbf{r}) \,D_v (\mathbf{r}) \,\mathscr{S} (\mathbf{r})\big \rangle_{\bar{\mathbf{r}}},\\
\nonumber=&\, \frac{2 E \left(C E^2 -  F (1-C^2) \right) }{r\,F_0^2\,(1- C^2)^2} = \frac{1}{F_0^2 \,r} \partial_r \bigg(\frac{E^2 (r)}{1- C^2 (r)} \bigg).
\end{alignat}
Eq.~\eqref{eq:charge} is essentially Eq.~\eqref{eq:pair} but without the modulus signs on the vorticity, wherefore $g_Q (r) \rightarrow 0$ as $r \rightarrow \infty$. At the origin, $g_Q$ and $g$ are related as
\begin{equation}
g \,(0) = - g_Q\,(0).
\end{equation}
Both the functions $g$ and $g_Q$ are plotted in Fig.~\ref{fig:current} for an open microwave billiard. Assuming isotropy, $g_Q (r)$ depends only on the scaled distance $R \equiv k\, \lvert \mathbf{r}_2 - \mathbf{r}_1 \rvert$ and can actually be kneaded into a surprisingly simple form \cite{freund1998critical, foltin2003signed, foltin2003distribution, dennis2003correlations, wilkinson2004screening}:
\begin{equation}
g_Q\,(R) = \frac{4}{R}\,\frac{\mathrm{d}}{\mathrm{d}\,R} \bigg[\frac{\mathrm{d}\,\arcsin\,\left(J_0 \,(R)\right)}{\mathrm{d}\,R} \bigg]^2.
\end{equation}
Moreover, since the distribution of the phases of the field $\psi$ is isotropic, $g_Q$ satisfies the ``topological charge screening relation'':
\begin{equation}
\frac{1}{2} \int_0^\infty \mathrm{d}\,R\, \,\,R\,g_Q \,(R) = -1,
\end{equation}
neglecting the self-interaction at $R=0$. Physically, this means that the integral of the topological charge over all $\mathbf{r}$ necessarily compensates the charge associated with the vortex at $\mathbf{r} = 0$ \cite{Berry2059}. This polyonymous local neutrality condition appears in many guises such as the Stillinger-Lovett sum rule \cite{stillinger1968ion, stillinger1968general} in the theory of ionic liquids, and ``critical-point screening'' in the context of dislocations \cite{freund1998critical}. 

For $R \gg 1$, $g_Q\,(R) \sim 8 \cos\, (2\, R) / \pi\,R^2$; the period of oscillation of $g_Q$ is thus twice that of the correlation function $C$. Contrarily, unsigned correlation functions, like those for saddle points, do not reduce to such simple forms. The asymptotic approximations to order $\mathcal{O} (R^{-1})$ of the unsigned RWM vortex-vortex, vortex-saddle, and saddle-saddle pair correlations were derived by \citet{hohmann2009density} to be
\begin{equation*}
g_{vv}\, (R),\, g_{ss}\, (R) \sim 1 + \frac{4 \sin\,2 R}{\pi\,R};\,\, g_{vs}\, (R) \sim 1 - \frac{4 \sin\,2 R}{\pi\,R}.
\end{equation*}
Finally, the most demanding asymptotics (short distances from the boundaries, for example) were addressed using supersymmetric techniques. \citet{klein2011critical} established that (unnormalized) pair correlations of both the unsigned density of critical points and the density of minima points of the Gaussian random field are given by 
\begin{equation}
\tilde{g}\,(r) = \begin{cases}
\langle D \rangle ^2 + \alpha_1 \nabla^4 C\,(r) + \alpha_2 \, \tr \left[ H_C (r)\right]^2; \hspace*{-0.2cm} &r \gg \sigma\\
\alpha_3\,r^{2-d + \mathcal{K}} & 0 < r \ll \sigma
\end{cases},
\end{equation}
with $\mathcal{K} = 0$ and $\mathcal{K} = 3$, respectively. The elements involved in computing this correlation function are $D$, the average density of critical points, $\alpha_i$ ,constants that depend on the type of critical-point density considered and the dimensionality of the system $d$, $[H_C (r)]_{ij} = \partial^2 C\,(r)/\partial r_i \,\partial r_j$, the Hessian matrix of the correlation function of the field, and $\sigma$, its typical length scale. 

A fitting end to this Section and effectively, the Review, is perhaps provided by Carlo Beenakker's Synopsis for the Seventh Annual Symposium on Frontiers of Science at Irvine, California, 1995. Addressing the then-nascent field, he remarks, ``Quantum billiards is a game played by physicists at a few academic and industrial laboratories in various parts of the world. It's a serious game: we are actually getting paid for it. It's also fun and exciting.'' We hope that the experiments described above---and our larger discussion on nodal portraits---have been successful in sharing some of this excitement.

\section{Concluding remarks}
\label{sec:conclude}

One of the key features of mathematical description---portability---allows us to reach out under one shade to nodal domains, percolating clusters, spin domains, electromagnetic modes, water waves, and others. From Chladni plates to gallium arsenide tables, the essence of the fundamental questions stays the same even if the details do not. What is clear by now is that nodal portraits distinguish between integrable and chaotic systems and afford details of their geometrical features. However, with a great number of new results comes an equally large number of new puzzles, many of which remain markedly unclear. It is to some of these that we turn now, before closing. 
\begin{enumerate}

\item The area distribution for the pseudointegrable barrier billiard is known \cite{dietz2008properties} to follow the same scaling as Eq.~\eqref{eq:fisher}, suggesting that the excited states are about as random as those for chaotic billiards. This is somewhat reminiscent of the time when ``linear level repulsion'' in spacing distributions was (fallaciously) believed to be an indicator of quantum chaos. A counterexample was the pseudointegrable rhombus billiard \cite{biswas1990quantum}, which belongs to a different universality class  altogether \cite{gremaud1998spacing,bogomolny1999models,auberson2001class}. Similarly, there might be some novel variant of a percolation model lurking here as well, awaiting discovery.

\item There is still a lot to be understood about the nodal statistical features of nonseparable but integrable, and quasi-integrable billiards, the wavefunctions of which have been spurned by both checkerboards and the random wave model. Some noteworthy results have been obtained for special systems \cite{prado2009superscars} but the evolution of the domains with quasi-integrable perturbations is an open problem. 

\item As we have seen, the statistics of nodal volumes are in tune with Yau's conjecture and thus, when appropriately scaled, are non-zero only over a finite interval. In light of this conjecture, it would certainly be interesting to seek exact limiting distributions for systems other than the cuboid and the RWM, which lie at opposite ends of the spectrum from order to chaos.

\item For the nonseparable, integrable billiards, it is fascinating to be able to set up difference equations for $\nu_{m,n}$ but there are important questions that this gives birth to. Firstly, how does one connect the topology of the eigenfunctions to the algebra of the difference equations? Moreover, even though one can count the domains, analytical forms for the limiting distributions $P\,(\xi)$ and the trace formulae portended by the statistics of nodal loops continue to elude us. In this vein, it might be worthwhile to study the sums of trigonometric products and explore the possibility of constructing difference equations for the domains of these functions. It is perhaps not too unreasonable to speculate that these sums of large number of trigonometric products, with random coefficients, might serve as good models for chaotic wavefunctions.

\item The idea of counting nodal domains of chaotic billiards with Potts spins is, by all means, innovative. However, at this stage, it seems far removed from actual quantitative counts and any developments in this direction would be a welcome addition.

\item Although there exist isospectral, convex, connected domains in dimensions larger than three \cite{gordon1994isospectral}, no known examples thereof have been found for planar billiards. Of course, absence of evidence is not evidence of absence and one would ideally hope to see this conjecture conclusively settled, once and for all.

\item Shortly after Kac's famous question, \citet{fisher1966hearing} posed the first instance of this query in the context of graphs. Half a century thence, the nodal domains of quantum graphs \cite{kuchment2008quantum} have today morphed into a subject of extensive investigation \cite{gnutzmann2004nodal, band2008nodal, schapotschnikow2006eigenvalue}, especially in the context of isospectrality \cite{band2006nodal, gutkin2001can, band2014nodal} and in its defiance of distinction by nodal counts \cite{oren2012isospectral}.
Borrowing and adapting some of these ideas---such as the fruits of representation theory \cite{band2009isospectral}---for counting the shape of a drum holds promise for the future. 

\end{enumerate}

\begin{acknowledgments}
RS is supported by the Purcell Fellowship. The authors thank Arnd B\"{a}cker and Arseni Goussev for their helpful comments on the manuscript.
\end{acknowledgments}

\appendix

\section{Isoperimetric inequalities}
\label{sec:isoperimetric}

The nomenclature ``isoperimetric inequality'' is somewhat of a misnomer as ``isoperimetric'' literally just means ``having the same perimeter".  The name is certainly appropriate for what may be regarded as the classical isoperimetric inequality---among all planar shapes with the \textsl{same perimeter} the circle has the largest area. Expressed differently, we have the relation
\begin{equation}
\mathcal{A} \le \frac{\mathcal{P}^2}{4\,\pi}
\end{equation}
between the area $\mathcal{A}$ enclosed by a planar closed curve and its perimeter $\mathcal{P}$, where the equality holds if and only if the curve is a circle. While this may have been the spirit of the first inquiries, it would be prudent to shed the restrictive connotations of isoperimetry that the label brings with it. In most general terms, isoperimetric inequalities address the question of which geometrical layout of some physical system maximizes or minimizes a certain quantity \cite{benguria2011isoperimetric}. Here, this quantity of interest is an eigenvalue of the Laplacian. The eigenvalues, in some sense, can be thought of as ``geometric objects'' in that they not only depend on the geometry of the domain but they also bear information about the underlying geometry thereof. This twin correspondence motivates the hunt for isoperimetric inequalities characterizing the eigenvalues of the Laplacian. Considering the number of excellent reviews on the subject already \cite{ashbaugh1999isoperimetric, ashbaugh2007isoperimetric, bandle1980isoperimetric, benguria2011isoperimetric, Hansen1994, hile1980inequalities, payne, polya1951isoperimetric}, here, we only enlist some of the best-known inequalities. 

\subsection{Rayleigh-Faber-Krahn and related inequalities}

On a Euclidean domain $\mathcal{D}$, we know that both the Dirichlet and Neumann eigenvalues scale as the square of the length inverse. It therefore seems intuitive, from simple dimensional analysis, that if one is to construct universal inequalities, the geometric quantity to compare the eigenvalue to must be the area $\mathcal{A}\,(\mathcal{D})$. Can one obtain such a universal (and preferably, sharp) bound on $\lambda_1$? This question was first pondered upon by Lord \citet{rayleigh1945theory} in his disquisition on the theory of sound wherein he conjectured that among all drums of the same area, tuned to the same tension, the circular membrane possesses the lowest fundamental frequency. Reworded more precisely as the Faber-Krahn inequality, this states that among all equiareal planar regions, the disk has the smallest first Dirichlet eigenvalue:
\begin{equation}
\label{eq:fk}
\lambda^\textsc{d}_1 \ge \frac{\pi}{\mathcal{A}} \left (\mathscr{J}_{\,0,1} \right)^2.
\end{equation}
This inequality, developed independently by \citet{faber1923beweis} and \citet{krahn1925rayleigh}, was generalized to $d$ dimensions by \citet{krahn1926minimaleigenschaften}
\begin{equation}
\lambda^\textsc{d}_1 \ge \left( \frac{\omega_d}{\mu_d\, (\mathcal{D})} \right)^{2/d} \left(\mathscr{J}_{\frac{d}{2}-1, 1} \right)^2
\end{equation}
and later, to regions inside circular sectors by \citet{payne1960faber}. It is possible to further improve this result if we now restrict the class of domains under consideration \cite{antunes2006new}. One such possibility is to consider the $n$-polygons, for which \citet{polya1951isoperimetric} conjectured that ``of all $n$-polygons with the same area, the regular $n$-polygon has the smallest first Dirichlet eigenvalue.'' Utilizing this hypothesis for triangles and quadrilaterals (see, for example, \citet{siudeja2007sharp, freitas2006upper, freitas2007precise}) yields
\begin{equation}
\lambda^\textsc{d}_1 (\triangle) \ge \frac{4 \sqrt{3} \,\pi^2}{3 \,\mathcal{A}}, \hspace*{0.25cm} \mbox{and} \hspace*{0.25cm} \lambda^\textsc{d}_1 (\square) \ge \frac{2\, \pi^2}{\mathcal{A}}.
\end{equation}

Analogously, the second Dirichlet eigenvalue $\lambda^\textsc{d}_2$ satisfies
\begin{equation}
\lambda^\textsc{d}_2 \ge 2 ^{2/d} \left(\frac{\omega_d}{\mu_d\,(\mathcal{D})} \right)^{2/d} \left(\mathscr{J}_{\frac{d}{2}-1, 1} \right)^2,
\end{equation}
which is minimized by the union of two identical disks \cite{krahn1926minimaleigenschaften, polya1955characteristic}. A related pair of bounds for $\lambda^\textsc{d}_1$ in a simply connected planar domain is given by
\begin{equation}
\frac{\alpha}{\mathcal{R}^2} \le \lambda^\textsc{d}_1 \le \frac{1}{\mathcal{R}^2} \, \left(\mathscr{J}_{\frac{d}{2}-1, 1} \right)^2,
\end{equation}
where $\mathcal{R}$ is the inradius of $\mathcal{D}$. The lower bound is due to \citet{makai1965lower} and \citet{hayman1978some}. Subsequent to several iterative refinements of their original estimates, the best value of the constant $\alpha$ known today is $\alpha = \pi^2/4 \approx 2.4674$ for convex domains \cite{hersch1960frequence} and $\alpha= 0.6197$ for a general $\mathcal{D}$ \cite{banuelos1994brownian}. The upper bound, on the other hand, follows directly from domain monotonicity and has been further tightened for planar \cite{polya1951isoperimetric} and higher-dimensional \cite{freitas2008sharp} star-shaped domains. The final result concerning $\lambda_1^{\textsc{d}}$ that deserves special mention here is Barta's inequality \cite{barta1937vibration}, which states that if $\phi$ is a positive, twice-continuously-differentiable function on $\mathbb{R}$, then
\begin{equation}
\inf_\mathbb{R} \left(-\frac{\Delta\,\phi}{\phi} \right) \le \lambda_1^{\textsc{d}} \le \sup_\mathbb{R} \left(-\frac{\Delta\,\phi}{\phi} \right).
\end{equation}

\subsection{Payne-P{\' o}lya-Weinberger inequality}

The preceding inequalities provide individual estimates for the first and second eigenvalues but stop short of relating them. The obvious question that follows is: how do the two compare? For an answer, we turn to the conjecture by \citet{payne1956ratio}, which posits that
\begin{equation}
\label{eq:polya}
\frac{\lambda_2^{\textsc{d}}}{\lambda_1^{\textsc{d}}} \le  \left (\frac{\mathscr{J}_{\frac{d}{2}, 1}}{\mathscr{J}_{\frac{d}{2}-1, 1}} \right)^2,
\end{equation}
and, more generally,
\begin{equation}
\lambda^{\textsc{d}}_{n+1} \le 3 \lambda^{\textsc{d}}_n.
\end{equation}
Although  \citet{payne1956ratio} originally proved Eq.~\eqref{eq:polya} for a weaker upper bound of $1 + 4 /d$ in $d= 2$ dimensions, it was not until much later that the rigorous proof of the inequality, as it stands today, was provided by \citet{ashbaugh1991proof, ashbaugh1992sharp, ashbaugh1993universal, ashbaugh1993more}. Alternatively, instead of looking at the ratios of the two eigenvalues, one can consider their differences. In this case, we know \cite{singer1985estimate}
\begin{equation}
\frac{\pi^2}{4\,\delta^2} \le \lambda_2^{\textsc{d}} - \lambda_1^{\textsc{d}} \le \frac{d\,\pi^2}{\mathcal{R}^2},
\end{equation}
where $\delta = \max \,\{ \lvert x - y \rvert; x, y \in \mathcal{D} \}$ is the diameter of $\mathcal{D} \in \mathbb{R}^d$. As suggested by \citet{donnelly2011spectral} and shown by \citet{andrews2011proof}, the lower bound can actually be sharpened to $3 \,\pi^2 / \delta^2$, thereby further narrowing the range.

\subsection{Szeg{\"o}-Weinberger inequality}

So far we have only talked about the properties of the Dirichlet eigenvalues---equally interesting are the  eigenvalues given Neumann boundary conditions. The first nontrivial Neumann eigenvalue $ \lambda^\textsc{n}_2$ (since $ \lambda^\textsc{n}_1 = 0$) is constrained as
\begin{equation}
\label{eq:B2}
\frac{\pi^2}{\delta^2} \le \lambda^\textsc{n}_2 \le \left(\frac{\omega_d}{\mu_d\,(\mathcal{D})} \right)^{2/d} \left(\tilde{\mathscr{J}}_{\frac{d}{2}, 1} \right)^2,
\end{equation}
where $\tilde{\mathscr{J}}_{\,\upsilon, 1}$ is the first positive zero of the function
\begin{equation*}
\frac{\mathrm{d}}{\mathrm{d}\,z} \left[z^{1-d/2} \,J_{\frac{d}{2}-1 + \upsilon} (z) \right] = {J}^{'}_\upsilon (z) \quad (\mbox{for } d = 2).
\end{equation*}
This relation actually consists of two independent isoperimetric inequalities compactified into one for ease of presentation. The supremum, proven for simply-connected, planar domains by \citet{szego1954inequalities} and \citet{weinberger1956isoperimetric}, is usually identified as the inequality bearing the name of its proponents. The other component of Eq.~\eqref{eq:B2}---the infimum---was obtained by \citet{payne1960optimal}. For Neumann eigenvalues in planar, bounded, regular domains (domains with a discrete Neumann eigenspectrum), \citet{polya1954mathematics} hypothesized that
\begin{equation}
\lambda_j^{\textsc{n}} \le \frac{4\,(j-1)\,\pi}{\mathcal{A}} \quad(j = 2, 3, 4,\ldots),
\end{equation}
which is, incontrovertibly, known to be true for any domain that tiles the plane \cite{polya1961eigenvalues}. Still, it remains only a conjecture for $d > 2$, the best (albeit weaker) proven estimate being
$\lambda_j^{\textsc{n}} \le 8\,\pi \,(j-1)$ \cite{kroger1992upper}. In particular, akin to Eq.~\eqref{eq:B2}, the third Neumann eigenvalue is bounded from above
\begin{equation}
\lambda^\textsc{n}_3 \le \frac{2 \pi \left(\tilde{\mathscr{J}}_{\,0, 1}\right)^2}{\mathcal{A}}
\end{equation}
for simply-connected, regular, planar domains \cite{girouard2009maximization} with the equality attained (in the limit) by a family of domains degenerating to a disjoint union of two identical disks \cite{Grebenkov2013}. Incidentally, it can be shown that the harmonic mean of the first two nontrivial Neumann eigenvalues is also minimized for a disk \cite{szego1954inequalities, weinberger1956isoperimetric}, i.e.,
\begin{equation}
\frac{1}{\lambda_2^{\textsc{n}}} + \frac{1}{\lambda_3^{\textsc{n}}} \ge \frac{2 \,\mathcal{A}}{\pi\,\left(\tilde{\mathscr{J}}_{\,1, 1} \right)^2},
\end{equation}
which permits generalization to the longer sequence \cite{ashbaugh1993universal}
\begin{equation}
\frac{1}{\lambda_2^{\textsc{n}}} + \cdots+ \frac{1}{\lambda_{d+1}^{\textsc{n}}} \ge \frac{d}{d+2} \left(\frac{\mu_d (\mathcal{D})}{\omega_d}\right)^{2/d}.
\end{equation}
To the reader perplexed as to the point of this compendium of abstruse relations (which might rightfully seem an archival exercise in mathematical stamp collecting), we can only offer solace in the reassurance that isoperimetric inequalities prove to be extremely useful in shape optimization problems, an overview of which is presented in Sec.~\ref{sec:inverse}.

\bibliographystyle{apsrmp4-1}
\bibliography{RMPRef.bib}

\begin{thebibliography}{631}%
\makeatletter
\providecommand \@ifxundefined [1]{%
 \@ifx{#1\undefined}
}%
\providecommand \@ifnum [1]{%
 \ifnum #1\expandafter \@firstoftwo
 \else \expandafter \@secondoftwo
 \fi
}%
\providecommand \@ifx [1]{%
 \ifx #1\expandafter \@firstoftwo
 \else \expandafter \@secondoftwo
 \fi
}%
\providecommand \natexlab [1]{#1}%
\providecommand \enquote  [1]{``#1''}%
\providecommand \bibnamefont  [1]{#1}%
\providecommand \bibfnamefont [1]{#1}%
\providecommand \citenamefont [1]{#1}%
\providecommand \href@noop [0]{\@secondoftwo}%
\providecommand \href [0]{\begingroup \@sanitize@url \@href}%
\providecommand \@href[1]{\@@startlink{#1}\@@href}%
\providecommand \@@href[1]{\endgroup#1\@@endlink}%
\providecommand \@sanitize@url [0]{\catcode `\\12\catcode `\$12\catcode
  `\&12\catcode `\#12\catcode `\^12\catcode `\_12\catcode `\%12\relax}%
\providecommand \@@startlink[1]{}%
\providecommand \@@endlink[0]{}%
\providecommand \url  [0]{\begingroup\@sanitize@url \@url }%
\providecommand \@url [1]{\endgroup\@href {#1}{\urlprefix }}%
\providecommand \urlprefix  [0]{URL }%
\providecommand \Eprint [0]{\href }%
\providecommand \doibase [0]{http://dx.doi.org/}%
\providecommand \selectlanguage [0]{\@gobble}%
\providecommand \bibinfo  [0]{\@secondoftwo}%
\providecommand \bibfield  [0]{\@secondoftwo}%
\providecommand \translation [1]{[#1]}%
\providecommand \BibitemOpen [0]{}%
\providecommand \bibitemStop [0]{}%
\providecommand \bibitemNoStop [0]{.\EOS\space}%
\providecommand \EOS [0]{\spacefactor3000\relax}%
\providecommand \BibitemShut  [1]{\csname bibitem#1\endcsname}%
\let\auto@bib@innerbib\@empty
\bibitem [{\citenamefont {{\AA}berg}\ \emph {et~al.}(2008)\citenamefont
  {{\AA}berg}, \citenamefont {Guhr}, \citenamefont {Miski-Oglu},\ and\
  \citenamefont {Richter}}]{aaberg2008superscars}%
  \BibitemOpen
  \bibfield  {author} {\bibinfo {author} {\bibnamefont {{\AA}berg},
  \bibfnamefont {S.}}, \bibinfo {author} {\bibfnamefont {T.}~\bibnamefont
  {Guhr}}, \bibinfo {author} {\bibfnamefont {M.}~\bibnamefont {Miski-Oglu}}, \
  and\ \bibinfo {author} {\bibfnamefont {A.}~\bibnamefont {Richter}}} (\bibinfo
  {year} {2008}),\ \href {\doibase 10.1103/PhysRevLett.100.204101} {\bibfield
  {journal} {\bibinfo  {journal} {Phys. Rev. Lett.}\ }\textbf {\bibinfo
  {volume} {100}}~(\bibinfo {number} {20}),\ \bibinfo {pages}
  {204101}}\BibitemShut {NoStop}%
\bibitem [{\citenamefont {Agam}\ and\ \citenamefont
  {Brenner}(1995)}]{agam1995semiclassical}%
  \BibitemOpen
  \bibfield  {author} {\bibinfo {author} {\bibnamefont {Agam}, \bibfnamefont
  {O.}}, \ and\ \bibinfo {author} {\bibfnamefont {N.}~\bibnamefont {Brenner}}}
  (\bibinfo {year} {1995}),\ \href {\doibase 10.1088/0305-4470/28/5/020}
  {\bibfield  {journal} {\bibinfo  {journal} {J. Phys. A: Math. Gen.}\ }\textbf
  {\bibinfo {volume} {28}}~(\bibinfo {number} {5}),\ \bibinfo {pages}
  {1345}}\BibitemShut {NoStop}%
\bibitem [{\citenamefont {Agam}\ and\ \citenamefont
  {Fishman}(1993)}]{agam1993quantum}%
  \BibitemOpen
  \bibfield  {author} {\bibinfo {author} {\bibnamefont {Agam}, \bibfnamefont
  {O.}}, \ and\ \bibinfo {author} {\bibfnamefont {S.}~\bibnamefont {Fishman}}}
  (\bibinfo {year} {1993}),\ \href {\doibase 10.1088/0305-4470/26/9/010}
  {\bibfield  {journal} {\bibinfo  {journal} {J. Phys. A: Math. Gen.}\ }\textbf
  {\bibinfo {volume} {26}}~(\bibinfo {number} {9}),\ \bibinfo {pages}
  {2113}}\BibitemShut {NoStop}%
\bibitem [{\citenamefont {Agam}\ and\ \citenamefont
  {Fishman}(1994)}]{agam1994semiclassical}%
  \BibitemOpen
  \bibfield  {author} {\bibinfo {author} {\bibnamefont {Agam}, \bibfnamefont
  {O.}}, \ and\ \bibinfo {author} {\bibfnamefont {S.}~\bibnamefont {Fishman}}}
  (\bibinfo {year} {1994}),\ \href {\doibase 10.1103/PhysRevLett.73.806}
  {\bibfield  {journal} {\bibinfo  {journal} {Phys. Rev. Lett.}\ }\textbf
  {\bibinfo {volume} {73}}~(\bibinfo {number} {6}),\ \bibinfo {pages}
  {806}}\BibitemShut {NoStop}%
\bibitem [{\citenamefont {Ahlfors}(2010)}]{ahlfors2010conformal}%
  \BibitemOpen
  \bibfield  {author} {\bibinfo {author} {\bibnamefont {Ahlfors}, \bibfnamefont
  {L.~V.}}} (\bibinfo {year} {2010}),\ \href@noop {} {\emph {\bibinfo {title}
  {Conformal invariants: topics in geometric function theory}}},\ Vol.\
  \bibinfo {volume} {371}\ (\bibinfo  {publisher} {American Mathematical
  Soc.},\ \bibinfo {address} {Providence, RI})\BibitemShut {NoStop}%
\bibitem [{\citenamefont {Alberts}(2008)}]{alberts2008slides}%
  \BibitemOpen
  \bibfield  {author} {\bibinfo {author} {\bibnamefont {Alberts}, \bibfnamefont
  {T.}}} (\bibinfo {year} {2008}),\ \href {https://www.math.utah.edu/~alberts}
  {\enquote {\bibinfo {title} {An {I}ntroduction to the {S}chramm-{L}oewner
  {E}volution},}\ }\bibinfo {howpublished} {Presentation (Courant Institute of
  Mathematical Sciences)}\BibitemShut {NoStop}%
\bibitem [{\citenamefont {Aleiner}\ \emph {et~al.}(2002)\citenamefont
  {Aleiner}, \citenamefont {Brouwer},\ and\ \citenamefont
  {Glazman}}]{aleiner2002quantum}%
  \BibitemOpen
  \bibfield  {author} {\bibinfo {author} {\bibnamefont {Aleiner}, \bibfnamefont
  {I.~L.}}, \bibinfo {author} {\bibfnamefont {P.~W.}\ \bibnamefont {Brouwer}},
  \ and\ \bibinfo {author} {\bibfnamefont {L.~I.}\ \bibnamefont {Glazman}}}
  (\bibinfo {year} {2002}),\ \href {\doibase 10.1016/S0370-1573(01)00063-1}
  {\bibfield  {journal} {\bibinfo  {journal} {Phys. Rep.}\ }\textbf {\bibinfo
  {volume} {358}}~(\bibinfo {number} {5}),\ \bibinfo {pages} {309}}\BibitemShut
  {NoStop}%
\bibitem [{\citenamefont {Alessandrini}(1994)}]{alessandrini1994nodal}%
  \BibitemOpen
  \bibfield  {author} {\bibinfo {author} {\bibnamefont {Alessandrini},
  \bibfnamefont {G.}}} (\bibinfo {year} {1994}),\ \href {\doibase
  10.1007/BF02564478} {\bibfield  {journal} {\bibinfo  {journal} {Comment.
  Math. Helv.}\ }\textbf {\bibinfo {volume} {69}}~(\bibinfo {number} {1}),\
  \bibinfo {pages} {142}}\BibitemShut {NoStop}%
\bibitem [{\citenamefont {Alessandrini}(1998)}]{alessandrini1998courant}%
  \BibitemOpen
  \bibfield  {author} {\bibinfo {author} {\bibnamefont {Alessandrini},
  \bibfnamefont {G.}}} (\bibinfo {year} {1998}),\ in\ \href {\doibase
  10.1515/form.10.5.521} {\emph {\bibinfo {booktitle} {Forum Math.}}},\
  Vol.~\bibinfo {volume} {10},\ pp.\ \bibinfo {pages} {521--532}\BibitemShut
  {NoStop}%
\bibitem [{\citenamefont {Alhassid}(2000)}]{alhassid2000statistical}%
  \BibitemOpen
  \bibfield  {author} {\bibinfo {author} {\bibnamefont {Alhassid},
  \bibfnamefont {Y.}}} (\bibinfo {year} {2000}),\ \href {\doibase
  10.1103/RevModPhys.72.895} {\bibfield  {journal} {\bibinfo  {journal} {Rev.
  Mod. Phys.}\ }\textbf {\bibinfo {volume} {72}}~(\bibinfo {number} {4}),\
  \bibinfo {pages} {895}}\BibitemShut {NoStop}%
\bibitem [{\citenamefont {Alhassid}\ and\ \citenamefont
  {Lewenkopf}(1997)}]{alhassid1997signatures}%
  \BibitemOpen
  \bibfield  {author} {\bibinfo {author} {\bibnamefont {Alhassid},
  \bibfnamefont {Y.}}, \ and\ \bibinfo {author} {\bibfnamefont {C.~H.}\
  \bibnamefont {Lewenkopf}}} (\bibinfo {year} {1997}),\ \href {\doibase
  10.1103/PhysRevB.55.7749} {\bibfield  {journal} {\bibinfo  {journal} {Phys.
  Rev. B}\ }\textbf {\bibinfo {volume} {55}}~(\bibinfo {number} {12}),\
  \bibinfo {pages} {7749}}\BibitemShut {NoStop}%
\bibitem [{\citenamefont {Allen}\ \emph {et~al.}(1992)\citenamefont {Allen},
  \citenamefont {Beijersbergen}, \citenamefont {Spreeuw},\ and\ \citenamefont
  {Woerdman}}]{allen1992orbital}%
  \BibitemOpen
  \bibfield  {author} {\bibinfo {author} {\bibnamefont {Allen}, \bibfnamefont
  {L.}}, \bibinfo {author} {\bibfnamefont {M.~W.}\ \bibnamefont
  {Beijersbergen}}, \bibinfo {author} {\bibfnamefont {R.~J.~C.}\ \bibnamefont
  {Spreeuw}}, \ and\ \bibinfo {author} {\bibfnamefont {J.~P.}\ \bibnamefont
  {Woerdman}}} (\bibinfo {year} {1992}),\ \href {\doibase
  10.1103/PhysRevA.45.8185} {\bibfield  {journal} {\bibinfo  {journal} {Phys.
  Rev. A}\ }\textbf {\bibinfo {volume} {45}}~(\bibinfo {number} {11}),\
  \bibinfo {pages} {8185}}\BibitemShut {NoStop}%
\bibitem [{\citenamefont {Allen}\ \emph {et~al.}(1999)\citenamefont {Allen},
  \citenamefont {Padgett},\ and\ \citenamefont {Babiker}}]{allen1999iv}%
  \BibitemOpen
  \bibfield  {author} {\bibinfo {author} {\bibnamefont {Allen}, \bibfnamefont
  {L.}}, \bibinfo {author} {\bibfnamefont {M.~J.}\ \bibnamefont {Padgett}}, \
  and\ \bibinfo {author} {\bibfnamefont {M.}~\bibnamefont {Babiker}}} (\bibinfo
  {year} {1999}),\ \href {\doibase 10.1016/S0079-6638(08)70391-3} {\bibfield
  {journal} {\bibinfo  {journal} {Prog. Opt.}\ }\textbf {\bibinfo {volume}
  {39}},\ \bibinfo {pages} {291}}\BibitemShut {NoStop}%
\bibitem [{\citenamefont {Alt}\ \emph {et~al.}(1997)\citenamefont {Alt},
  \citenamefont {Dembowski}, \citenamefont {Gr{\"a}f}, \citenamefont
  {Hofferbert}, \citenamefont {Rehfeld}, \citenamefont {Richter}, \citenamefont
  {Schuhmann},\ and\ \citenamefont {Weiland}}]{alt1997wave}%
  \BibitemOpen
  \bibfield  {author} {\bibinfo {author} {\bibnamefont {Alt}, \bibfnamefont
  {H.}}, \bibinfo {author} {\bibfnamefont {C.}~\bibnamefont {Dembowski}},
  \bibinfo {author} {\bibfnamefont {H.-D.}\ \bibnamefont {Gr{\"a}f}}, \bibinfo
  {author} {\bibfnamefont {R.}~\bibnamefont {Hofferbert}}, \bibinfo {author}
  {\bibfnamefont {H.}~\bibnamefont {Rehfeld}}, \bibinfo {author} {\bibfnamefont
  {A.}~\bibnamefont {Richter}}, \bibinfo {author} {\bibfnamefont
  {R.}~\bibnamefont {Schuhmann}}, \ and\ \bibinfo {author} {\bibfnamefont
  {T.}~\bibnamefont {Weiland}}} (\bibinfo {year} {1997}),\ \href {\doibase
  10.1103/PhysRevLett.79.1026} {\bibfield  {journal} {\bibinfo  {journal}
  {Phys. Rev. Lett.}\ }\textbf {\bibinfo {volume} {79}}~(\bibinfo {number}
  {6}),\ \bibinfo {pages} {1026}}\BibitemShut {NoStop}%
\bibitem [{\citenamefont {Alt}\ \emph {et~al.}(1995)\citenamefont {Alt},
  \citenamefont {Gr{\"a}f}, \citenamefont {Harney}, \citenamefont {Hofferbert},
  \citenamefont {Lengeler}, \citenamefont {Richter}, \citenamefont {Schardt},\
  and\ \citenamefont {Weidenm{\"u}ller}}]{alt1995gaussian}%
  \BibitemOpen
  \bibfield  {author} {\bibinfo {author} {\bibnamefont {Alt}, \bibfnamefont
  {H.}}, \bibinfo {author} {\bibfnamefont {H.-D.}\ \bibnamefont {Gr{\"a}f}},
  \bibinfo {author} {\bibfnamefont {H.~L.}\ \bibnamefont {Harney}}, \bibinfo
  {author} {\bibfnamefont {R.}~\bibnamefont {Hofferbert}}, \bibinfo {author}
  {\bibfnamefont {H.}~\bibnamefont {Lengeler}}, \bibinfo {author}
  {\bibfnamefont {A.}~\bibnamefont {Richter}}, \bibinfo {author} {\bibfnamefont
  {P.}~\bibnamefont {Schardt}}, \ and\ \bibinfo {author} {\bibfnamefont
  {H.~A.}\ \bibnamefont {Weidenm{\"u}ller}}} (\bibinfo {year} {1995}),\ \href
  {\doibase 10.1103/PhysRevLett.74.62} {\bibfield  {journal} {\bibinfo
  {journal} {Phys. Rev. Lett.}\ }\textbf {\bibinfo {volume} {74}}~(\bibinfo
  {number} {1}),\ \bibinfo {pages} {62}}\BibitemShut {NoStop}%
\bibitem [{\citenamefont {Alt}\ \emph {et~al.}(1996)\citenamefont {Alt},
  \citenamefont {Gr{\"a}f}, \citenamefont {Hofferbert}, \citenamefont
  {Rangacharyulu}, \citenamefont {Rehfeld}, \citenamefont {Richter},
  \citenamefont {Schardt},\ and\ \citenamefont {Wirzba}}]{alt1996chaotic}%
  \BibitemOpen
  \bibfield  {author} {\bibinfo {author} {\bibnamefont {Alt}, \bibfnamefont
  {H.}}, \bibinfo {author} {\bibfnamefont {H.-D.}\ \bibnamefont {Gr{\"a}f}},
  \bibinfo {author} {\bibfnamefont {R.}~\bibnamefont {Hofferbert}}, \bibinfo
  {author} {\bibfnamefont {C.}~\bibnamefont {Rangacharyulu}}, \bibinfo {author}
  {\bibfnamefont {H.}~\bibnamefont {Rehfeld}}, \bibinfo {author} {\bibfnamefont
  {A.}~\bibnamefont {Richter}}, \bibinfo {author} {\bibfnamefont
  {P.}~\bibnamefont {Schardt}}, \ and\ \bibinfo {author} {\bibfnamefont
  {A.}~\bibnamefont {Wirzba}}} (\bibinfo {year} {1996}),\ \href {\doibase
  10.1103/PhysRevE.54.2303} {\bibfield  {journal} {\bibinfo  {journal} {Phys.
  Rev. E}\ }\textbf {\bibinfo {volume} {54}}~(\bibinfo {number} {3}),\ \bibinfo
  {pages} {2303}}\BibitemShut {NoStop}%
\bibitem [{\citenamefont {Anantharaman}(2008)}]{anantharaman2008entropy}%
  \BibitemOpen
  \bibfield  {author} {\bibinfo {author} {\bibnamefont {Anantharaman},
  \bibfnamefont {N.}}} (\bibinfo {year} {2008}),\ \href {\doibase
  10.4007/annals.2008.168.435} {\bibfield  {journal} {\bibinfo  {journal} {Ann.
  Math.}\ }\textbf {\bibinfo {volume} {168}},\ \bibinfo {pages}
  {435}}\BibitemShut {NoStop}%
\bibitem [{\citenamefont {Anantharaman}\ and\ \citenamefont
  {Nonnenmacher}(2007)}]{Anantharaman2007}%
  \BibitemOpen
  \bibfield  {author} {\bibinfo {author} {\bibnamefont {Anantharaman},
  \bibfnamefont {N.}}, \ and\ \bibinfo {author} {\bibfnamefont
  {S.}~\bibnamefont {Nonnenmacher}}} (\bibinfo {year} {2007}),\ \href {\doibase
  10.5802/aif.2340} {\bibfield  {journal} {\bibinfo  {journal} {Ann. Inst.
  Fourier, Grenoble}\ }\textbf {\bibinfo {volume} {57}}~(\bibinfo {number}
  {7}),\ \bibinfo {pages} {2465}}\BibitemShut {NoStop}%
\bibitem [{\citenamefont {Ancona}\ \emph {et~al.}(2004)\citenamefont {Ancona},
  \citenamefont {Helffer},\ and\ \citenamefont
  {Hoffmann-Ostenhof}}]{ancona2004nodal}%
  \BibitemOpen
  \bibfield  {author} {\bibinfo {author} {\bibnamefont {Ancona}, \bibfnamefont
  {A.}}, \bibinfo {author} {\bibfnamefont {B.}~\bibnamefont {Helffer}}, \ and\
  \bibinfo {author} {\bibfnamefont {T.}~\bibnamefont {Hoffmann-Ostenhof}}}
  (\bibinfo {year} {2004}),\ \href
  {https://www.math.uni-bielefeld.de/documenta/vol-09/16.pdf} {\bibfield
  {journal} {\bibinfo  {journal} {Documenta Math.}\ }\textbf {\bibinfo {volume}
  {9}},\ \bibinfo {pages} {283}}\BibitemShut {NoStop}%
\bibitem [{\citenamefont {Andersen}\ \emph {et~al.}(2006)\citenamefont
  {Andersen}, \citenamefont {Kaplan}, \citenamefont {Gr{\"u}nzweig},\ and\
  \citenamefont {Davidson}}]{andersen2006decay}%
  \BibitemOpen
  \bibfield  {author} {\bibinfo {author} {\bibnamefont {Andersen},
  \bibfnamefont {M.~F.}}, \bibinfo {author} {\bibfnamefont {A.}~\bibnamefont
  {Kaplan}}, \bibinfo {author} {\bibfnamefont {T.}~\bibnamefont
  {Gr{\"u}nzweig}}, \ and\ \bibinfo {author} {\bibfnamefont {N.}~\bibnamefont
  {Davidson}}} (\bibinfo {year} {2006}),\ \href {\doibase
  10.1103/PhysRevLett.97.104102} {\bibfield  {journal} {\bibinfo  {journal}
  {Phys. Rev. Lett.}\ }\textbf {\bibinfo {volume} {97}}~(\bibinfo {number}
  {10}),\ \bibinfo {pages} {104102}}\BibitemShut {NoStop}%
\bibitem [{\citenamefont {Andrews}\ and\ \citenamefont
  {Clutterbuck}(2011)}]{andrews2011proof}%
  \BibitemOpen
  \bibfield  {author} {\bibinfo {author} {\bibnamefont {Andrews}, \bibfnamefont
  {B.}}, \ and\ \bibinfo {author} {\bibfnamefont {J.}~\bibnamefont
  {Clutterbuck}}} (\bibinfo {year} {2011}),\ \href {\doibase
  10.1090/S0894-0347-2011-00699-1} {\bibfield  {journal} {\bibinfo  {journal}
  {J. Amer. Math. Soc.}\ }\textbf {\bibinfo {volume} {24}}~(\bibinfo {number}
  {3}),\ \bibinfo {pages} {899}}\BibitemShut {NoStop}%
\bibitem [{\citenamefont {Antunes}\ and\ \citenamefont
  {Freitas}(2006)}]{antunes2006new}%
  \BibitemOpen
  \bibfield  {author} {\bibinfo {author} {\bibnamefont {Antunes}, \bibfnamefont
  {P.}}, \ and\ \bibinfo {author} {\bibfnamefont {P.}~\bibnamefont {Freitas}}}
  (\bibinfo {year} {2006}),\ \href {\doibase 10.1080/10586458.2006.10128966}
  {\bibfield  {journal} {\bibinfo  {journal} {Exper. Math.}\ }\textbf {\bibinfo
  {volume} {15}}~(\bibinfo {number} {3}),\ \bibinfo {pages} {333}}\BibitemShut
  {NoStop}%
\bibitem [{\citenamefont {Arendt}\ \emph {et~al.}(2009)\citenamefont {Arendt},
  \citenamefont {Nittka}, \citenamefont {Peter},\ and\ \citenamefont
  {Steiner}}]{arendt2009weyl}%
  \BibitemOpen
  \bibfield  {author} {\bibinfo {author} {\bibnamefont {Arendt}, \bibfnamefont
  {W.}}, \bibinfo {author} {\bibfnamefont {R.}~\bibnamefont {Nittka}}, \bibinfo
  {author} {\bibfnamefont {W.}~\bibnamefont {Peter}}, \ and\ \bibinfo {author}
  {\bibfnamefont {F.}~\bibnamefont {Steiner}}} (\bibinfo {year} {2009}),\
  \enquote {\bibinfo {title} {Weyl's law: Spectral properties of the
  {L}aplacian in mathematics and physics},}\ in\ \href@noop {} {\emph {\bibinfo
  {booktitle} {Mathematical Analysis of Evolution, Information, and
  Complexity}}},\ \bibinfo {editor} {edited by\ \bibinfo {editor}
  {\bibfnamefont {W.}~\bibnamefont {Arendt}}\ and\ \bibinfo {editor}
  {\bibfnamefont {W.~P.}\ \bibnamefont {Schleich}}}\ (\bibinfo  {publisher}
  {Wiley-VCH Verlag GmbH \& Co.},\ \bibinfo {address} {Weinheim})\ pp.\
  \bibinfo {pages} {1--71}\BibitemShut {NoStop}%
\bibitem [{\citenamefont {Arnold}(1963)}]{arnold1963proof}%
  \BibitemOpen
  \bibfield  {author} {\bibinfo {author} {\bibnamefont {Arnold}, \bibfnamefont
  {V.~I.}}} (\bibinfo {year} {1963}),\ \href {\doibase
  10.1070/rm1963v018n05abeh004130} {\bibfield  {journal} {\bibinfo  {journal}
  {Russ. Math. Surv.}\ }\textbf {\bibinfo {volume} {18}}~(\bibinfo {number}
  {5}),\ \bibinfo {pages} {9}}\BibitemShut {NoStop}%
\bibitem [{\citenamefont {Arnold}(1973)}]{arnold1973topology}%
  \BibitemOpen
  \bibfield  {author} {\bibinfo {author} {\bibnamefont {Arnold}, \bibfnamefont
  {V.~I.}}} (\bibinfo {year} {1973}),\ \href {\doibase
  10.1007/978-3-642-31031-7_27} {\bibfield  {journal} {\bibinfo  {journal}
  {Uspekhi Mat. Nauk}\ }\textbf {\bibinfo {volume} {28}}~(\bibinfo {number}
  {5}),\ \bibinfo {pages} {260}}\BibitemShut {NoStop}%
\bibitem [{\citenamefont {Arnold}(2011)}]{arnold2011topological}%
  \BibitemOpen
  \bibfield  {author} {\bibinfo {author} {\bibnamefont {Arnold}, \bibfnamefont
  {V.~I.}}} (\bibinfo {year} {2011}),\ \href {\doibase
  10.1134/S0081543811040031} {\bibfield  {journal} {\bibinfo  {journal} {Proc.
  Steklov Inst. Math.}\ }\textbf {\bibinfo {volume} {273}}~(\bibinfo {number}
  {1}),\ \bibinfo {pages} {25}}\BibitemShut {NoStop}%
\bibitem [{\citenamefont {Arnold}(2013)}]{arnol2013mathematical}%
  \BibitemOpen
  \bibfield  {author} {\bibinfo {author} {\bibnamefont {Arnold}, \bibfnamefont
  {V.~I.}}} (\bibinfo {year} {2013}),\ \href@noop {} {\emph {\bibinfo {title}
  {Mathematical Methods of Classical Mechanics}}},\ \bibinfo {edition} {2nd}\
  ed.,\ Vol.~\bibinfo {volume} {60}\ (\bibinfo  {publisher} {Springer},\
  \bibinfo {address} {New York})\BibitemShut {NoStop}%
\bibitem [{\citenamefont {Arnold}\ and\ \citenamefont
  {Avez}(1967)}]{arnold1967theorie}%
  \BibitemOpen
  \bibfield  {author} {\bibinfo {author} {\bibnamefont {Arnold}, \bibfnamefont
  {V.~I.}}, \ and\ \bibinfo {author} {\bibfnamefont {A.}~\bibnamefont {Avez}}}
  (\bibinfo {year} {1967}),\ \href@noop {} {\emph {\bibinfo {title} {Theorie
  ergodique des systemes dynamiques}}}\ (\bibinfo  {publisher}
  {Gauthier-Villars},\ \bibinfo {address} {Paris})\BibitemShut {NoStop}%
\bibitem [{\citenamefont {Aronovitch}\ \emph {et~al.}(2012)\citenamefont
  {Aronovitch}, \citenamefont {Band}, \citenamefont {Fajman},\ and\
  \citenamefont {Gnutzmann}}]{aronovitch2012nodal}%
  \BibitemOpen
  \bibfield  {author} {\bibinfo {author} {\bibnamefont {Aronovitch},
  \bibfnamefont {A.}}, \bibinfo {author} {\bibfnamefont {R.}~\bibnamefont
  {Band}}, \bibinfo {author} {\bibfnamefont {D.}~\bibnamefont {Fajman}}, \ and\
  \bibinfo {author} {\bibfnamefont {S.}~\bibnamefont {Gnutzmann}}} (\bibinfo
  {year} {2012}),\ \href {\doibase 10.1088/1751-8113/45/8/085209} {\bibfield
  {journal} {\bibinfo  {journal} {J. Phys. A: Math. Theor.}\ }\textbf {\bibinfo
  {volume} {45}}~(\bibinfo {number} {8}),\ \bibinfo {pages}
  {085209}}\BibitemShut {NoStop}%
\bibitem [{\citenamefont {Aronovitch}\ and\ \citenamefont
  {Smilansky}(2006)}]{aronovitch2006slides}%
  \BibitemOpen
  \bibfield  {author} {\bibinfo {author} {\bibnamefont {Aronovitch},
  \bibfnamefont {A.}}, \ and\ \bibinfo {author} {\bibfnamefont
  {U.}~\bibnamefont {Smilansky}}} (\bibinfo {year} {2006}),\ \href
  {https://webhome.weizmann.ac.il/home/feamit/nodalweek/a_aronovitch_nodalweek.pdf}
  {\enquote {\bibinfo {title} {Nodal intersections with a test curve},}\
  }\bibinfo {howpublished} {Presentation (Weizmann Institute of
  Science)}\BibitemShut {NoStop}%
\bibitem [{\citenamefont {Aronovitch}\ and\ \citenamefont
  {Smilansky}(2007)}]{aronovitch2007statistics}%
  \BibitemOpen
  \bibfield  {author} {\bibinfo {author} {\bibnamefont {Aronovitch},
  \bibfnamefont {A.}}, \ and\ \bibinfo {author} {\bibfnamefont
  {U.}~\bibnamefont {Smilansky}}} (\bibinfo {year} {2007}),\ \href {\doibase
  10.1088/1751-8113/40/32/003} {\bibfield  {journal} {\bibinfo  {journal} {J.
  Phys. A: Math. Theor.}\ }\textbf {\bibinfo {volume} {40}}~(\bibinfo {number}
  {32}),\ \bibinfo {pages} {9743}}\BibitemShut {NoStop}%
\bibitem [{\citenamefont {Ash}\ and\ \citenamefont
  {Dol{\'e}ans-Dade}(2000)}]{ash2000probability}%
  \BibitemOpen
  \bibfield  {author} {\bibinfo {author} {\bibnamefont {Ash}, \bibfnamefont
  {R.~B.}}, \ and\ \bibinfo {author} {\bibfnamefont {C.}~\bibnamefont
  {Dol{\'e}ans-Dade}}} (\bibinfo {year} {2000}),\ \href@noop {} {\emph
  {\bibinfo {title} {Probability and measure theory}}},\ \bibinfo {edition}
  {2nd}\ ed.\ (\bibinfo  {publisher} {Academic Press},\ \bibinfo {address} {San
  Diego})\BibitemShut {NoStop}%
\bibitem [{\citenamefont {Ashbaugh}(1999)}]{ashbaugh1999isoperimetric}%
  \BibitemOpen
  \bibfield  {author} {\bibinfo {author} {\bibnamefont {Ashbaugh},
  \bibfnamefont {M.~S.}}} (\bibinfo {year} {1999}),\ in\ \href@noop {} {\emph
  {\bibinfo {booktitle} {Spectral Theory and Geometry, London Math. Soc.
  Lecture Note Ser.}}},\ \bibinfo {editor} {edited by\ \bibinfo {editor}
  {\bibfnamefont {E.~B.}\ \bibnamefont {Davies}}\ and\ \bibinfo {editor}
  {\bibfnamefont {Y.}~\bibnamefont {Safarov}}}\ (\bibinfo  {publisher}
  {Cambridge University Press},\ \bibinfo {address} {Cambridge, UK})\ pp.\
  \bibinfo {pages} {95--139}\BibitemShut {NoStop}%
\bibitem [{\citenamefont {Ashbaugh}\ and\ \citenamefont
  {Benguria}(1991)}]{ashbaugh1991proof}%
  \BibitemOpen
  \bibfield  {author} {\bibinfo {author} {\bibnamefont {Ashbaugh},
  \bibfnamefont {M.~S.}}, \ and\ \bibinfo {author} {\bibfnamefont {R.~D.}\
  \bibnamefont {Benguria}}} (\bibinfo {year} {1991}),\ \href {\doibase
  10.1090/S0273-0979-1991-16016-7} {\bibfield  {journal} {\bibinfo  {journal}
  {Bull. Amer. Math. Soc.}\ }\textbf {\bibinfo {volume} {25}}~(\bibinfo
  {number} {1}),\ \bibinfo {pages} {19}}\BibitemShut {NoStop}%
\bibitem [{\citenamefont {Ashbaugh}\ and\ \citenamefont
  {Benguria}(1992)}]{ashbaugh1992sharp}%
  \BibitemOpen
  \bibfield  {author} {\bibinfo {author} {\bibnamefont {Ashbaugh},
  \bibfnamefont {M.~S.}}, \ and\ \bibinfo {author} {\bibfnamefont {R.~D.}\
  \bibnamefont {Benguria}}} (\bibinfo {year} {1992}),\ \href {\doibase
  10.2307/2946578} {\bibfield  {journal} {\bibinfo  {journal} {Ann. Math.}\
  }\textbf {\bibinfo {volume} {135}}~(\bibinfo {number} {3}),\ \bibinfo {pages}
  {601}}\BibitemShut {NoStop}%
\bibitem [{\citenamefont {Ashbaugh}\ and\ \citenamefont
  {Benguria}(1993{\natexlab{a}})}]{ashbaugh1993more}%
  \BibitemOpen
  \bibfield  {author} {\bibinfo {author} {\bibnamefont {Ashbaugh},
  \bibfnamefont {M.~S.}}, \ and\ \bibinfo {author} {\bibfnamefont {R.~D.}\
  \bibnamefont {Benguria}}} (\bibinfo {year} {1993}{\natexlab{a}}),\ \href
  {\doibase 10.1137/0524091} {\bibfield  {journal} {\bibinfo  {journal} {SIAM
  J. Math. Anal.}\ }\textbf {\bibinfo {volume} {24}}~(\bibinfo {number} {6}),\
  \bibinfo {pages} {1622}}\BibitemShut {NoStop}%
\bibitem [{\citenamefont {Ashbaugh}\ and\ \citenamefont
  {Benguria}(1993{\natexlab{b}})}]{ashbaugh1993universal}%
  \BibitemOpen
  \bibfield  {author} {\bibinfo {author} {\bibnamefont {Ashbaugh},
  \bibfnamefont {M.~S.}}, \ and\ \bibinfo {author} {\bibfnamefont {R.~D.}\
  \bibnamefont {Benguria}}} (\bibinfo {year} {1993}{\natexlab{b}}),\ \href
  {\doibase 10.1137/0524034} {\bibfield  {journal} {\bibinfo  {journal} {SIAM
  J. Math. Anal.}\ }\textbf {\bibinfo {volume} {24}}~(\bibinfo {number} {3}),\
  \bibinfo {pages} {557}}\BibitemShut {NoStop}%
\bibitem [{\citenamefont {Ashbaugh}\ and\ \citenamefont
  {Benguria}(2007)}]{ashbaugh2007isoperimetric}%
  \BibitemOpen
  \bibfield  {author} {\bibinfo {author} {\bibnamefont {Ashbaugh},
  \bibfnamefont {M.~S.}}, \ and\ \bibinfo {author} {\bibfnamefont {R.~D.}\
  \bibnamefont {Benguria}}} (\bibinfo {year} {2007}),\ \enquote {\bibinfo
  {title} {Isoperimetric inequalities for eigenvalues of the laplacian},}\ in\
  \href@noop {} {\emph {\bibinfo {booktitle} {Spectral Theory and Mathematical
  Physics}}},\ \bibinfo {series} {Proc. Symp. Pure Math.}, Vol.~\bibinfo
  {volume} {76}\ (\bibinfo  {publisher} {AMS},\ \bibinfo {address} {Providence,
  RI})\ pp.\ \bibinfo {pages} {105--139}\BibitemShut {NoStop}%
\bibitem [{\citenamefont {Auberson}\ \emph {et~al.}(2001)\citenamefont
  {Auberson}, \citenamefont {Jain},\ and\ \citenamefont
  {Khare}}]{auberson2001class}%
  \BibitemOpen
  \bibfield  {author} {\bibinfo {author} {\bibnamefont {Auberson},
  \bibfnamefont {G.}}, \bibinfo {author} {\bibfnamefont {S.~R.}\ \bibnamefont
  {Jain}}, \ and\ \bibinfo {author} {\bibfnamefont {A.}~\bibnamefont {Khare}}}
  (\bibinfo {year} {2001}),\ \href {\doibase 10.1088/0305-4470/34/4/302}
  {\bibfield  {journal} {\bibinfo  {journal} {J. Phys. A: Math. Gen.}\ }\textbf
  {\bibinfo {volume} {34}}~(\bibinfo {number} {4}),\ \bibinfo {pages}
  {695}}\BibitemShut {NoStop}%
\bibitem [{\citenamefont {Auerhammer}\ \emph {et~al.}(1993)\citenamefont
  {Auerhammer}, \citenamefont {Genz}, \citenamefont {Gr{\"a}f}, \citenamefont
  {Hahn}, \citenamefont {Hoffmann-Stascheck}, \citenamefont {L{\"u}ttge},
  \citenamefont {Nething}, \citenamefont {R{\"u}hl}, \citenamefont {Richter},\
  and\ \citenamefont {Rietdorf}}]{auerhammer1993}%
  \BibitemOpen
  \bibfield  {author} {\bibinfo {author} {\bibnamefont {Auerhammer},
  \bibfnamefont {J.}}, \bibinfo {author} {\bibfnamefont {H.}~\bibnamefont
  {Genz}}, \bibinfo {author} {\bibfnamefont {H.-D.}\ \bibnamefont {Gr{\"a}f}},
  \bibinfo {author} {\bibfnamefont {R.}~\bibnamefont {Hahn}}, \bibinfo {author}
  {\bibfnamefont {P.}~\bibnamefont {Hoffmann-Stascheck}}, \bibinfo {author}
  {\bibfnamefont {C.}~\bibnamefont {L{\"u}ttge}}, \bibinfo {author}
  {\bibfnamefont {U.}~\bibnamefont {Nething}}, \bibinfo {author} {\bibfnamefont
  {K.}~\bibnamefont {R{\"u}hl}}, \bibinfo {author} {\bibfnamefont
  {A.}~\bibnamefont {Richter}}, \ and\ \bibinfo {author} {\bibfnamefont
  {T.}~\bibnamefont {Rietdorf}}} (\bibinfo {year} {1993}),\ \href {\doibase
  10.1016/0375-9474(93)90708-6} {\bibfield  {journal} {\bibinfo  {journal}
  {Nucl. Phys. A}\ }\textbf {\bibinfo {volume} {553}},\ \bibinfo {pages}
  {841}}\BibitemShut {NoStop}%
\bibitem [{\citenamefont {Aurich}\ and\ \citenamefont
  {Steiner}(1988)}]{aurich1988periodic}%
  \BibitemOpen
  \bibfield  {author} {\bibinfo {author} {\bibnamefont {Aurich}, \bibfnamefont
  {R.}}, \ and\ \bibinfo {author} {\bibfnamefont {F.}~\bibnamefont {Steiner}}}
  (\bibinfo {year} {1988}),\ \href {\doibase 10.1016/0167-2789(88)90068-1}
  {\bibfield  {journal} {\bibinfo  {journal} {Physica D}\ }\textbf {\bibinfo
  {volume} {32}}~(\bibinfo {number} {3}),\ \bibinfo {pages} {451}}\BibitemShut
  {NoStop}%
\bibitem [{\citenamefont {Aurich}\ and\ \citenamefont
  {Steiner}(1993)}]{aurich1993statistical}%
  \BibitemOpen
  \bibfield  {author} {\bibinfo {author} {\bibnamefont {Aurich}, \bibfnamefont
  {R.}}, \ and\ \bibinfo {author} {\bibfnamefont {F.}~\bibnamefont {Steiner}}}
  (\bibinfo {year} {1993}),\ \href {\doibase 10.1016/0167-2789(93)90255-Y}
  {\bibfield  {journal} {\bibinfo  {journal} {Physica D}\ }\textbf {\bibinfo
  {volume} {64}}~(\bibinfo {number} {1}),\ \bibinfo {pages} {185}}\BibitemShut
  {NoStop}%
\bibitem [{\citenamefont {B{\"a}cker}(2007{\natexlab{a}})}]{arnd_thesis}%
  \BibitemOpen
  \bibfield  {author} {\bibinfo {author} {\bibnamefont {B{\"a}cker},
  \bibfnamefont {A.}}} (\bibinfo {year} {2007}{\natexlab{a}}),\ \emph {\bibinfo
  {title} {Eigenfunctions in chaotic quantum systems}},\ \href@noop {} {Ph.D.
  thesis}\ (\bibinfo  {school} {Technische Universit\"{a}t
  Dresden})\BibitemShut {NoStop}%
\bibitem [{\citenamefont {B{\"a}cker}(2007{\natexlab{b}})}]{backer2007random}%
  \BibitemOpen
  \bibfield  {author} {\bibinfo {author} {\bibnamefont {B{\"a}cker},
  \bibfnamefont {A.}}} (\bibinfo {year} {2007}{\natexlab{b}}),\ \href {\doibase
  10.1140/epjst/e2007-00153-4} {\bibfield  {journal} {\bibinfo  {journal} {Eur.
  Phys. J. Spec. Top}\ }\textbf {\bibinfo {volume} {145}}~(\bibinfo {number}
  {1}),\ \bibinfo {pages} {161}}\BibitemShut {NoStop}%
\bibitem [{\citenamefont {B{\"a}cker}\ \emph {et~al.}(2002)\citenamefont
  {B{\"a}cker}, \citenamefont {F{\"u}rstberger}, \citenamefont {Schubert},\
  and\ \citenamefont {Steiner}}]{backer2002behaviour}%
  \BibitemOpen
  \bibfield  {author} {\bibinfo {author} {\bibnamefont {B{\"a}cker},
  \bibfnamefont {A.}}, \bibinfo {author} {\bibfnamefont {S.}~\bibnamefont
  {F{\"u}rstberger}}, \bibinfo {author} {\bibfnamefont {R.}~\bibnamefont
  {Schubert}}, \ and\ \bibinfo {author} {\bibfnamefont {F.}~\bibnamefont
  {Steiner}}} (\bibinfo {year} {2002}),\ \href {\doibase
  10.1088/0305-4470/35/48/306} {\bibfield  {journal} {\bibinfo  {journal} {J.
  Phys. A: Math. Gen.}\ }\textbf {\bibinfo {volume} {35}}~(\bibinfo {number}
  {48}),\ \bibinfo {pages} {10293}}\BibitemShut {NoStop}%
\bibitem [{\citenamefont {B{\"a}cker}\ \emph {et~al.}(2005)\citenamefont
  {B{\"a}cker}, \citenamefont {Ketzmerick},\ and\ \citenamefont
  {Monastra}}]{backer2005flooding}%
  \BibitemOpen
  \bibfield  {author} {\bibinfo {author} {\bibnamefont {B{\"a}cker},
  \bibfnamefont {A.}}, \bibinfo {author} {\bibfnamefont {R.}~\bibnamefont
  {Ketzmerick}}, \ and\ \bibinfo {author} {\bibfnamefont {A.~G.}\ \bibnamefont
  {Monastra}}} (\bibinfo {year} {2005}),\ \href {\doibase
  10.1103/PhysRevLett.94.054102} {\bibfield  {journal} {\bibinfo  {journal}
  {Phys. Rev. Lett.}\ }\textbf {\bibinfo {volume} {94}}~(\bibinfo {number}
  {5}),\ \bibinfo {pages} {054102}}\BibitemShut {NoStop}%
\bibitem [{\citenamefont {B{\"a}cker}\ and\ \citenamefont
  {Schubert}(2002)}]{arnd2002amplitude}%
  \BibitemOpen
  \bibfield  {author} {\bibinfo {author} {\bibnamefont {B{\"a}cker},
  \bibfnamefont {A.}}, \ and\ \bibinfo {author} {\bibfnamefont
  {R.}~\bibnamefont {Schubert}}} (\bibinfo {year} {2002}),\ \href {\doibase
  10.1088/0305-4470/35/3/306} {\bibfield  {journal} {\bibinfo  {journal} {J
  Phys. A: Math. Gen.}\ }\textbf {\bibinfo {volume} {35}}~(\bibinfo {number}
  {3}),\ \bibinfo {pages} {527}}\BibitemShut {NoStop}%
\bibitem [{\citenamefont {B{\"a}cker}\ \emph
  {et~al.}(1998{\natexlab{a}})\citenamefont {B{\"a}cker}, \citenamefont
  {Schubert},\ and\ \citenamefont {Stifter}}]{PhysRevE.58.5192}%
  \BibitemOpen
  \bibfield  {author} {\bibinfo {author} {\bibnamefont {B{\"a}cker},
  \bibfnamefont {A.}}, \bibinfo {author} {\bibfnamefont {R.}~\bibnamefont
  {Schubert}}, \ and\ \bibinfo {author} {\bibfnamefont {P.}~\bibnamefont
  {Stifter}}} (\bibinfo {year} {1998}{\natexlab{a}}),\ \href {\doibase
  10.1103/PhysRevE.58.5192} {\bibfield  {journal} {\bibinfo  {journal} {Phys.
  Rev. E}\ }\textbf {\bibinfo {volume} {58}},\ \bibinfo {pages}
  {5192}}\BibitemShut {NoStop}%
\bibitem [{\citenamefont {B{\"a}cker}\ \emph
  {et~al.}(1998{\natexlab{b}})\citenamefont {B{\"a}cker}, \citenamefont
  {Schubert},\ and\ \citenamefont {Stifter}}]{backer1998rate}%
  \BibitemOpen
  \bibfield  {author} {\bibinfo {author} {\bibnamefont {B{\"a}cker},
  \bibfnamefont {A.}}, \bibinfo {author} {\bibfnamefont {R.}~\bibnamefont
  {Schubert}}, \ and\ \bibinfo {author} {\bibfnamefont {P.}~\bibnamefont
  {Stifter}}} (\bibinfo {year} {1998}{\natexlab{b}}),\ \href {\doibase
  10.1103/PhysRevE.57.5425} {\bibfield  {journal} {\bibinfo  {journal} {Phys.
  Rev. E}\ }\textbf {\bibinfo {volume} {57}}~(\bibinfo {number} {5}),\ \bibinfo
  {pages} {5425}}\BibitemShut {NoStop}%
\bibitem [{\citenamefont {Balian}\ and\ \citenamefont
  {Bloch}(1970)}]{balian1970}%
  \BibitemOpen
  \bibfield  {author} {\bibinfo {author} {\bibnamefont {Balian}, \bibfnamefont
  {R.}}, \ and\ \bibinfo {author} {\bibfnamefont {C.}~\bibnamefont {Bloch}}}
  (\bibinfo {year} {1970}),\ \href {\doibase 10.1016/0003-4916(70)90497-5}
  {\bibfield  {journal} {\bibinfo  {journal} {Ann. Phys.}\ }\textbf {\bibinfo
  {volume} {60}}~(\bibinfo {number} {2}),\ \bibinfo {pages} {401}}\BibitemShut
  {NoStop}%
\bibitem [{\citenamefont {Balian}\ and\ \citenamefont
  {Bloch}(1971)}]{balian1971}%
  \BibitemOpen
  \bibfield  {author} {\bibinfo {author} {\bibnamefont {Balian}, \bibfnamefont
  {R.}}, \ and\ \bibinfo {author} {\bibfnamefont {C.}~\bibnamefont {Bloch}}}
  (\bibinfo {year} {1971}),\ \href {\doibase 10.1016/0003-4916(71)90286-7}
  {\bibfield  {journal} {\bibinfo  {journal} {Ann. Phys.}\ }\textbf {\bibinfo
  {volume} {64}}~(\bibinfo {number} {1}),\ \bibinfo {pages} {271}}\BibitemShut
  {NoStop}%
\bibitem [{\citenamefont {Baltes}\ and\ \citenamefont
  {Hilf}(1976)}]{baltes1976spectra}%
  \BibitemOpen
  \bibfield  {author} {\bibinfo {author} {\bibnamefont {Baltes}, \bibfnamefont
  {H.~P.}}, \ and\ \bibinfo {author} {\bibfnamefont {E.~R.}\ \bibnamefont
  {Hilf}}} (\bibinfo {year} {1976}),\ \href@noop {} {\emph {\bibinfo {title}
  {Spectra of finite systems}}}\ (\bibinfo  {publisher}
  {BI-Wissenschaftsverlag},\ \bibinfo {address} {Mannheim})\BibitemShut
  {NoStop}%
\bibitem [{\citenamefont {Band}(2014)}]{band2014nodal}%
  \BibitemOpen
  \bibfield  {author} {\bibinfo {author} {\bibnamefont {Band}, \bibfnamefont
  {R.}}} (\bibinfo {year} {2014}),\ \href {\doibase 10.1098/rsta.2012.0504}
  {\bibfield  {journal} {\bibinfo  {journal} {Phil. Trans. R. Soc. A}\ }\textbf
  {\bibinfo {volume} {372}}~(\bibinfo {number} {2007}),\ \bibinfo {pages}
  {20120504}}\BibitemShut {NoStop}%
\bibitem [{\citenamefont {Band}\ \emph {et~al.}(2008)\citenamefont {Band},
  \citenamefont {Oren},\ and\ \citenamefont {Smilansky}}]{band2008nodal}%
  \BibitemOpen
  \bibfield  {author} {\bibinfo {author} {\bibnamefont {Band}, \bibfnamefont
  {R.}}, \bibinfo {author} {\bibfnamefont {I.}~\bibnamefont {Oren}}, \ and\
  \bibinfo {author} {\bibfnamefont {U.}~\bibnamefont {Smilansky}}} (\bibinfo
  {year} {2008}),\ in\ \href {\doibase 10.1090/pspum/077/2459876} {\emph
  {\bibinfo {booktitle} {Analysis on Graphs and Its Applications}}},\ \bibinfo
  {series} {Proc. Symp. Pure Math.}, Vol.~\bibinfo {volume} {77},\ \bibinfo
  {editor} {edited by\ \bibinfo {editor} {\bibfnamefont {P.}~\bibnamefont
  {Exner}}, \bibinfo {editor} {\bibfnamefont {J.~P.}\ \bibnamefont {Keating}},
  \bibinfo {editor} {\bibfnamefont {P.}~\bibnamefont {Kuchment}}, \bibinfo
  {editor} {\bibfnamefont {T.}~\bibnamefont {Sunada}}, \ and\ \bibinfo {editor}
  {\bibfnamefont {A.}~\bibnamefont {Teplyaev}}}\ (\bibinfo  {publisher}
  {American Mathematical Soc.},\ \bibinfo {address} {Providence, RI})\
  Chap.~\bibinfo {chapter} {1},\ pp.\ \bibinfo {pages} {5--27}\BibitemShut
  {NoStop}%
\bibitem [{\citenamefont {Band}\ \emph {et~al.}(2009)\citenamefont {Band},
  \citenamefont {Parzanchevski},\ and\ \citenamefont
  {Ben-Shach}}]{band2009isospectral}%
  \BibitemOpen
  \bibfield  {author} {\bibinfo {author} {\bibnamefont {Band}, \bibfnamefont
  {R.}}, \bibinfo {author} {\bibfnamefont {O.}~\bibnamefont {Parzanchevski}}, \
  and\ \bibinfo {author} {\bibfnamefont {G.}~\bibnamefont {Ben-Shach}}}
  (\bibinfo {year} {2009}),\ \href {\doibase 10.1088/1751-8113/42/17/175202}
  {\bibfield  {journal} {\bibinfo  {journal} {J. Phys. A: Math. Theor.}\
  }\textbf {\bibinfo {volume} {42}}~(\bibinfo {number} {17}),\ \bibinfo {pages}
  {175202}}\BibitemShut {NoStop}%
\bibitem [{\citenamefont {Band}\ \emph {et~al.}(2006)\citenamefont {Band},
  \citenamefont {Shapira},\ and\ \citenamefont {Smilansky}}]{band2006nodal}%
  \BibitemOpen
  \bibfield  {author} {\bibinfo {author} {\bibnamefont {Band}, \bibfnamefont
  {R.}}, \bibinfo {author} {\bibfnamefont {T.}~\bibnamefont {Shapira}}, \ and\
  \bibinfo {author} {\bibfnamefont {U.}~\bibnamefont {Smilansky}}} (\bibinfo
  {year} {2006}),\ \href {\doibase 10.1088/0305-4470/39/45/009} {\bibfield
  {journal} {\bibinfo  {journal} {J. Phys. A: Math. Gen.}\ }\textbf {\bibinfo
  {volume} {39}}~(\bibinfo {number} {45}),\ \bibinfo {pages}
  {13999}}\BibitemShut {NoStop}%
\bibitem [{\citenamefont {Bandle}(1980)}]{bandle1980isoperimetric}%
  \BibitemOpen
  \bibfield  {author} {\bibinfo {author} {\bibnamefont {Bandle}, \bibfnamefont
  {C.}}} (\bibinfo {year} {1980}),\ \href@noop {} {\emph {\bibinfo {title}
  {Isoperimetric Inequalities and Applications}}},\ Vol.~\bibinfo {volume} {7}\
  (\bibinfo  {publisher} {Pitman Publishing},\ \bibinfo {address}
  {London})\BibitemShut {NoStop}%
\bibitem [{\citenamefont {Ba{\~n}uelos}\ and\ \citenamefont
  {Carroll}(1994)}]{banuelos1994brownian}%
  \BibitemOpen
  \bibfield  {author} {\bibinfo {author} {\bibnamefont {Ba{\~n}uelos},
  \bibfnamefont {R.}}, \ and\ \bibinfo {author} {\bibfnamefont
  {T.}~\bibnamefont {Carroll}}} (\bibinfo {year} {1994}),\ \href {\doibase
  10.1215/s0012-7094-94-07517-0} {\bibfield  {journal} {\bibinfo  {journal}
  {Duke Math. J.}\ }\textbf {\bibinfo {volume} {75}}~(\bibinfo {number} {3}),\
  \bibinfo {pages} {575}}\BibitemShut {NoStop}%
\bibitem [{\citenamefont {Baranger}\ and\ \citenamefont
  {Mello}(1994)}]{baranger1994mesoscopic}%
  \BibitemOpen
  \bibfield  {author} {\bibinfo {author} {\bibnamefont {Baranger},
  \bibfnamefont {H.~U.}}, \ and\ \bibinfo {author} {\bibfnamefont {P.~A.}\
  \bibnamefont {Mello}}} (\bibinfo {year} {1994}),\ \href {\doibase
  10.1103/PhysRevLett.73.142} {\bibfield  {journal} {\bibinfo  {journal} {Phys.
  Rev. Lett.}\ }\textbf {\bibinfo {volume} {73}}~(\bibinfo {number} {1}),\
  \bibinfo {pages} {142}}\BibitemShut {NoStop}%
\bibitem [{\citenamefont {Baranova}\ \emph {et~al.}(1981)\citenamefont
  {Baranova}, \citenamefont {Zel'Dovich}, \citenamefont {Mamaev}, \citenamefont
  {Pilipetskii},\ and\ \citenamefont {Shkukov}}]{baranova1981dislocations}%
  \BibitemOpen
  \bibfield  {author} {\bibinfo {author} {\bibnamefont {Baranova},
  \bibfnamefont {N.~B.}}, \bibinfo {author} {\bibfnamefont {B.~Y.}\
  \bibnamefont {Zel'Dovich}}, \bibinfo {author} {\bibfnamefont {A.~V.}\
  \bibnamefont {Mamaev}}, \bibinfo {author} {\bibfnamefont {N.}~\bibnamefont
  {Pilipetskii}}, \ and\ \bibinfo {author} {\bibfnamefont {V.~V.}\ \bibnamefont
  {Shkukov}}} (\bibinfo {year} {1981}),\ \href@noop {} {\bibfield  {journal}
  {\bibinfo  {journal} {JETP Lett.}\ }\textbf {\bibinfo {volume}
  {33}}~(\bibinfo {number} {4}),\ \bibinfo {pages} {206}}\BibitemShut {NoStop}%
\bibitem [{\citenamefont {Barta}(1937)}]{barta1937vibration}%
  \BibitemOpen
  \bibfield  {author} {\bibinfo {author} {\bibnamefont {Barta}, \bibfnamefont
  {J.}}} (\bibinfo {year} {1937}),\ \href@noop {} {\bibfield  {journal}
  {\bibinfo  {journal} {C. R. Acad. Sci. Paris}\ }\textbf {\bibinfo {volume}
  {204}}~(\bibinfo {number} {7}),\ \bibinfo {pages} {472}}\BibitemShut
  {NoStop}%
\bibitem [{\citenamefont {Barth}\ and\ \citenamefont
  {St{\"o}ckmann}(2002)}]{barth2002current}%
  \BibitemOpen
  \bibfield  {author} {\bibinfo {author} {\bibnamefont {Barth}, \bibfnamefont
  {M.}}, \ and\ \bibinfo {author} {\bibfnamefont {H.-J.}\ \bibnamefont
  {St{\"o}ckmann}}} (\bibinfo {year} {2002}),\ \href {\doibase
  10.1103/PhysRevE.65.066208} {\bibfield  {journal} {\bibinfo  {journal} {Phys.
  Rev. E}\ }\textbf {\bibinfo {volume} {65}}~(\bibinfo {number} {6}),\ \bibinfo
  {pages} {066208}}\BibitemShut {NoStop}%
\bibitem [{\citenamefont {Bauch}\ \emph {et~al.}(1998)\citenamefont {Bauch},
  \citenamefont {B{\l}{\c e}dowski}, \citenamefont {Sirko}, \citenamefont
  {Koch},\ and\ \citenamefont {Bl{\"u}mel}}]{bauch1998signature}%
  \BibitemOpen
  \bibfield  {author} {\bibinfo {author} {\bibnamefont {Bauch}, \bibfnamefont
  {{\relax Sz}.}}, \bibinfo {author} {\bibfnamefont {A.}~\bibnamefont {B{\l}{\c
  e}dowski}}, \bibinfo {author} {\bibfnamefont {L.}~\bibnamefont {Sirko}},
  \bibinfo {author} {\bibfnamefont {P.~M.}\ \bibnamefont {Koch}}, \ and\
  \bibinfo {author} {\bibfnamefont {R.}~\bibnamefont {Bl{\"u}mel}}} (\bibinfo
  {year} {1998}),\ \href {\doibase 10.1103/PhysRevE.57.304} {\bibfield
  {journal} {\bibinfo  {journal} {Phys. Rev. E}\ }\textbf {\bibinfo {volume}
  {57}}~(\bibinfo {number} {1}),\ \bibinfo {pages} {304}}\BibitemShut {NoStop}%
\bibitem [{\citenamefont {Bauer}\ and\ \citenamefont
  {Bernard}(2002)}]{bauer2002sle}%
  \BibitemOpen
  \bibfield  {author} {\bibinfo {author} {\bibnamefont {Bauer}, \bibfnamefont
  {M.}}, \ and\ \bibinfo {author} {\bibfnamefont {D.}~\bibnamefont {Bernard}}}
  (\bibinfo {year} {2002}),\ \href {\doibase 10.1016/S0370-2693(02)02423-1}
  {\bibfield  {journal} {\bibinfo  {journal} {Phys. Lett. B}\ }\textbf
  {\bibinfo {volume} {543}}~(\bibinfo {number} {1}),\ \bibinfo {pages}
  {135}}\BibitemShut {NoStop}%
\bibitem [{\citenamefont {Bauer}\ and\ \citenamefont
  {Bernard}(2003)}]{bauer2003conformal}%
  \BibitemOpen
  \bibfield  {author} {\bibinfo {author} {\bibnamefont {Bauer}, \bibfnamefont
  {M.}}, \ and\ \bibinfo {author} {\bibfnamefont {D.}~\bibnamefont {Bernard}}}
  (\bibinfo {year} {2003}),\ \href {\doibase 10.1007/s00220-003-0881-x}
  {\bibfield  {journal} {\bibinfo  {journal} {Commun. Math. Phys.}\ }\textbf
  {\bibinfo {volume} {239}}~(\bibinfo {number} {3}),\ \bibinfo {pages}
  {493}}\BibitemShut {NoStop}%
\bibitem [{\citenamefont {Bauer}\ \emph {et~al.}(2005)\citenamefont {Bauer},
  \citenamefont {Bernard},\ and\ \citenamefont {Houdayer}}]{bauer2005dipolar}%
  \BibitemOpen
  \bibfield  {author} {\bibinfo {author} {\bibnamefont {Bauer}, \bibfnamefont
  {M.}}, \bibinfo {author} {\bibfnamefont {D.}~\bibnamefont {Bernard}}, \ and\
  \bibinfo {author} {\bibfnamefont {J.}~\bibnamefont {Houdayer}}} (\bibinfo
  {year} {2005}),\ \href {\doibase 10.1088/1742-5468/2005/03/P03001} {\bibfield
   {journal} {\bibinfo  {journal} {J. Stat. Mech.}\ }\textbf {\bibinfo {volume}
  {2005}}~(\bibinfo {number} {03}),\ \bibinfo {pages} {P03001}}\BibitemShut
  {NoStop}%
\bibitem [{\citenamefont {Bauer}(2003)}]{bauer2003discrete}%
  \BibitemOpen
  \bibfield  {author} {\bibinfo {author} {\bibnamefont {Bauer}, \bibfnamefont
  {R.~O.}}} (\bibinfo {year} {2003}),\ \href {\doibase 10.5802/afst.1056}
  {\bibfield  {journal} {\bibinfo  {journal} {Ann. Fac. Sci. Toulouse Math.
  (6)}\ }\textbf {\bibinfo {volume} {12}}~(\bibinfo {number} {4}),\ \bibinfo
  {pages} {433}}\BibitemShut {NoStop}%
\bibitem [{\citenamefont {Baxter}(2007)}]{baxter2007exactly}%
  \BibitemOpen
  \bibfield  {author} {\bibinfo {author} {\bibnamefont {Baxter}, \bibfnamefont
  {R.~J.}}} (\bibinfo {year} {2007}),\ \href@noop {} {\emph {\bibinfo {title}
  {Exactly solved models in statistical mechanics}}}\ (\bibinfo  {publisher}
  {Dover Publications},\ \bibinfo {address} {Mineola, New York})\BibitemShut
  {NoStop}%
\bibitem [{\citenamefont {Bayfield}\ and\ \citenamefont
  {Koch}(1974)}]{bayfield1974multiphoton}%
  \BibitemOpen
  \bibfield  {author} {\bibinfo {author} {\bibnamefont {Bayfield},
  \bibfnamefont {J.~E.}}, \ and\ \bibinfo {author} {\bibfnamefont {P.~M.}\
  \bibnamefont {Koch}}} (\bibinfo {year} {1974}),\ \href {\doibase
  10.1103/PhysRevLett.33.258} {\bibfield  {journal} {\bibinfo  {journal} {Phys.
  Rev. Lett.}\ }\textbf {\bibinfo {volume} {33}}~(\bibinfo {number} {5}),\
  \bibinfo {pages} {258}}\BibitemShut {NoStop}%
\bibitem [{\citenamefont {Beenakker}(1991)}]{beenakker1991theory}%
  \BibitemOpen
  \bibfield  {author} {\bibinfo {author} {\bibnamefont {Beenakker},
  \bibfnamefont {C.~W.~J.}}} (\bibinfo {year} {1991}),\ \href {\doibase
  10.1103/PhysRevB.44.1646} {\bibfield  {journal} {\bibinfo  {journal} {Phys.
  Rev. B}\ }\textbf {\bibinfo {volume} {44}}~(\bibinfo {number} {4}),\ \bibinfo
  {pages} {1646}}\BibitemShut {NoStop}%
\bibitem [{\citenamefont {Beenakker}(1997)}]{beenakker1997random}%
  \BibitemOpen
  \bibfield  {author} {\bibinfo {author} {\bibnamefont {Beenakker},
  \bibfnamefont {C.~W.~J.}}} (\bibinfo {year} {1997}),\ \href {\doibase
  10.1103/RevModPhys.69.731} {\bibfield  {journal} {\bibinfo  {journal} {Rev.
  Mod. Phys.}\ }\textbf {\bibinfo {volume} {69}}~(\bibinfo {number} {3}),\
  \bibinfo {pages} {731}}\BibitemShut {NoStop}%
\bibitem [{\citenamefont {Beffara}(2004)}]{beffara2004hausdorff}%
  \BibitemOpen
  \bibfield  {author} {\bibinfo {author} {\bibnamefont {Beffara}, \bibfnamefont
  {V.}}} (\bibinfo {year} {2004}),\ \href {\doibase 10.1214/009117904000000072}
  {\bibfield  {journal} {\bibinfo  {journal} {Ann. Probab.}\ }\textbf {\bibinfo
  {volume} {32}},\ \bibinfo {pages} {2606}}\BibitemShut {NoStop}%
\bibitem [{\citenamefont {Beffara}(2008)}]{beffara2008dimension}%
  \BibitemOpen
  \bibfield  {author} {\bibinfo {author} {\bibnamefont {Beffara}, \bibfnamefont
  {V.}}} (\bibinfo {year} {2008}),\ \href {\doibase 10.1214/07-AOP364}
  {\bibfield  {journal} {\bibinfo  {journal} {Ann. Probab.}\ }\textbf {\bibinfo
  {volume} {36}},\ \bibinfo {pages} {1421}}\BibitemShut {NoStop}%
\bibitem [{\citenamefont {Beijersbergen}(1996)}]{bm_thesis}%
  \BibitemOpen
  \bibfield  {author} {\bibinfo {author} {\bibnamefont {Beijersbergen},
  \bibfnamefont {M.}}} (\bibinfo {year} {1996}),\ \emph {\bibinfo {title}
  {Phase singularities in optical beams}},\ \href@noop {} {Ph.D. thesis}\
  (\bibinfo  {school} {Huygens Laboratory, Universiteit Leiden})\BibitemShut
  {NoStop}%
\bibitem [{\citenamefont {Belavin}\ \emph
  {et~al.}(1984{\natexlab{a}})\citenamefont {Belavin}, \citenamefont
  {Polyakov},\ and\ \citenamefont {Zamolodchikov}}]{belavin1984infinite2}%
  \BibitemOpen
  \bibfield  {author} {\bibinfo {author} {\bibnamefont {Belavin}, \bibfnamefont
  {A.~A.}}, \bibinfo {author} {\bibfnamefont {A.~M.}\ \bibnamefont {Polyakov}},
  \ and\ \bibinfo {author} {\bibfnamefont {A.~B.}\ \bibnamefont
  {Zamolodchikov}}} (\bibinfo {year} {1984}{\natexlab{a}}),\ \href {\doibase
  10.1016/0550-3213(84)90052-X} {\bibfield  {journal} {\bibinfo  {journal}
  {Nucl. Phys. B}\ }\textbf {\bibinfo {volume} {241}}~(\bibinfo {number} {2}),\
  \bibinfo {pages} {333}}\BibitemShut {NoStop}%
\bibitem [{\citenamefont {Belavin}\ \emph
  {et~al.}(1984{\natexlab{b}})\citenamefont {Belavin}, \citenamefont
  {Polyakov},\ and\ \citenamefont {Zamolodchikov}}]{belavin1984infinite}%
  \BibitemOpen
  \bibfield  {author} {\bibinfo {author} {\bibnamefont {Belavin}, \bibfnamefont
  {A.~A.}}, \bibinfo {author} {\bibfnamefont {A.~M.}\ \bibnamefont {Polyakov}},
  \ and\ \bibinfo {author} {\bibfnamefont {A.~B.}\ \bibnamefont
  {Zamolodchikov}}} (\bibinfo {year} {1984}{\natexlab{b}}),\ \href {\doibase
  10.1007/BF01009438} {\bibfield  {journal} {\bibinfo  {journal} {J. Stat.
  Phys.}\ }\textbf {\bibinfo {volume} {34}}~(\bibinfo {number} {5-6}),\
  \bibinfo {pages} {763}}\BibitemShut {NoStop}%
\bibitem [{\citenamefont {Beliaev}\ and\ \citenamefont
  {Kereta}(2013)}]{beliaev2013bogomolny}%
  \BibitemOpen
  \bibfield  {author} {\bibinfo {author} {\bibnamefont {Beliaev}, \bibfnamefont
  {D.}}, \ and\ \bibinfo {author} {\bibfnamefont {Z.}~\bibnamefont {Kereta}}}
  (\bibinfo {year} {2013}),\ \href {\doibase 10.1088/1751-8113/46/45/455003}
  {\bibfield  {journal} {\bibinfo  {journal} {J. Phys. A: Math. Theor.}\
  }\textbf {\bibinfo {volume} {46}}~(\bibinfo {number} {45}),\ \bibinfo {pages}
  {455003}}\BibitemShut {NoStop}%
\bibitem [{\citenamefont {Beliaev}\ and\ \citenamefont
  {Wigman}(2016)}]{beliaev2016volume}%
  \BibitemOpen
  \bibfield  {author} {\bibinfo {author} {\bibnamefont {Beliaev}, \bibfnamefont
  {D.}}, \ and\ \bibinfo {author} {\bibfnamefont {I.}~\bibnamefont {Wigman}}}
  (\bibinfo {year} {2016}),\ \href {http://www.arxiv.org/abs/1606.05766}
  {\enquote {\bibinfo {title} {Volume distribution of nodal domains of random
  band-limited functions},}\ }\bibinfo {note} {{a}rXiv preprint
  1606.05766}\BibitemShut {NoStop}%
\bibitem [{\citenamefont {Benguria}(2011)}]{benguria2011isoperimetric}%
  \BibitemOpen
  \bibfield  {author} {\bibinfo {author} {\bibnamefont {Benguria},
  \bibfnamefont {R.~D.}}} (\bibinfo {year} {2011}),\ \enquote {\bibinfo {title}
  {Isoperimetric {I}nequalities for {E}igenvalues of the {L}aplacian},}\ in\
  \href@noop {} {\emph {\bibinfo {booktitle} {Entropy and the Quantum II}}},\
  \bibinfo {series} {Contemporary Mathematics}, Vol.\ \bibinfo {volume} {552},\
  \bibinfo {editor} {edited by\ \bibinfo {editor} {\bibfnamefont
  {R.}~\bibnamefont {Sims}}\ and\ \bibinfo {editor} {\bibfnamefont
  {D.}~\bibnamefont {Ueltschi}}},\ pp.\ \bibinfo {pages} {21--60}\BibitemShut
  {NoStop}%
\bibitem [{\citenamefont {B{\'e}rard}(1985)}]{berardvolume}%
  \BibitemOpen
  \bibfield  {author} {\bibinfo {author} {\bibnamefont {B{\'e}rard},
  \bibfnamefont {P.}}} (\bibinfo {year} {1984--1985}),\ \href@noop {} {\enquote
  {\bibinfo {title} {Volume des ensembles nodaux des fonctions propres du
  laplacien},}\ }\bibinfo {howpublished} {Bony-Sjoestrand-Meyer seminar, Exp.
  No. 14}\BibitemShut {NoStop}%
\bibitem [{\citenamefont {B{\'e}rard}(1992)}]{berard1992transplantation}%
  \BibitemOpen
  \bibfield  {author} {\bibinfo {author} {\bibnamefont {B{\'e}rard},
  \bibfnamefont {P.}}} (\bibinfo {year} {1992}),\ \href {\doibase
  10.1007/BF01444635} {\bibfield  {journal} {\bibinfo  {journal} {Math. Ann.}\
  }\textbf {\bibinfo {volume} {292}}~(\bibinfo {number} {1}),\ \bibinfo {pages}
  {547}}\BibitemShut {NoStop}%
\bibitem [{\citenamefont {B{\'e}rard}\ and\ \citenamefont
  {Helffer}(2015)}]{Berard2015}%
  \BibitemOpen
  \bibfield  {author} {\bibinfo {author} {\bibnamefont {B{\'e}rard},
  \bibfnamefont {P.}}, \ and\ \bibinfo {author} {\bibfnamefont
  {B.}~\bibnamefont {Helffer}}} (\bibinfo {year} {2015}),\ \enquote {\bibinfo
  {title} {Dirichlet eigenfunctions of the square membrane: Courant's property,
  and a. stern's and {\aa}. pleijel's analyses},}\ in\ \href {\doibase
  10.1007/978-3-319-17443-3_6} {\emph {\bibinfo {booktitle} {Analysis and
  Geometry: MIMS-GGTM, Tunis, Tunisia, March 2014. In Honour of Mohammed Salah
  Baouendi}}},\ \bibinfo {editor} {edited by\ \bibinfo {editor} {\bibfnamefont
  {A.}~\bibnamefont {Baklouti}}, \bibinfo {editor} {\bibfnamefont
  {A.}~\bibnamefont {El~Kacimi}}, \bibinfo {editor} {\bibfnamefont
  {S.}~\bibnamefont {Kallel}}, \ and\ \bibinfo {editor} {\bibfnamefont
  {N.}~\bibnamefont {Mir}}}\ (\bibinfo  {publisher} {Springer},\ \bibinfo
  {address} {Cham})\ pp.\ \bibinfo {pages} {69--114}\BibitemShut {NoStop}%
\bibitem [{\citenamefont {B{\'e}rard}\ and\ \citenamefont
  {Meyer}(1982)}]{berard1982inegalites}%
  \BibitemOpen
  \bibfield  {author} {\bibinfo {author} {\bibnamefont {B{\'e}rard},
  \bibfnamefont {P.}}, \ and\ \bibinfo {author} {\bibfnamefont
  {D.}~\bibnamefont {Meyer}}} (\bibinfo {year} {1982}),\ \href
  {http://www.numdam.org/item?id=ASENS_1982_4_15_3_513_0} {\bibfield  {journal}
  {\bibinfo  {journal} {Ann. scient. {\'E}c. Norm. Sup.}\ }\textbf {\bibinfo
  {volume} {15}}~(\bibinfo {number} {3}),\ \bibinfo {pages} {513}}\BibitemShut
  {NoStop}%
\bibitem [{\citenamefont {Van~den Berg}\ and\ \citenamefont
  {Srisatkunarajah}(1988)}]{van1988heat}%
  \BibitemOpen
  \bibfield  {author} {\bibinfo {author} {\bibnamefont {Van~den Berg},
  \bibfnamefont {M.}}, \ and\ \bibinfo {author} {\bibfnamefont
  {S.}~\bibnamefont {Srisatkunarajah}}} (\bibinfo {year} {1988}),\ \href
  {\doibase 10.1112/jlms/s2-37.121.119} {\bibfield  {journal} {\bibinfo
  {journal} {J. London Math. Soc.}\ }\textbf {\bibinfo {volume} {37}}~(\bibinfo
  {number} {1}),\ \bibinfo {pages} {119}}\BibitemShut {NoStop}%
\bibitem [{\citenamefont {Berggren}\ \emph {et~al.}(1999)\citenamefont
  {Berggren}, \citenamefont {Pichugin}, \citenamefont {Sadreev},\ and\
  \citenamefont {Starikov}}]{berggren1999signatures}%
  \BibitemOpen
  \bibfield  {author} {\bibinfo {author} {\bibnamefont {Berggren},
  \bibfnamefont {K.-F.}}, \bibinfo {author} {\bibfnamefont {K.~N.}\
  \bibnamefont {Pichugin}}, \bibinfo {author} {\bibfnamefont {A.~F.}\
  \bibnamefont {Sadreev}}, \ and\ \bibinfo {author} {\bibfnamefont
  {A.}~\bibnamefont {Starikov}}} (\bibinfo {year} {1999}),\ \href {\doibase
  10.1134/1.568188} {\bibfield  {journal} {\bibinfo  {journal} {JETP Lett.}\
  }\textbf {\bibinfo {volume} {70}}~(\bibinfo {number} {6}),\ \bibinfo {pages}
  {403}}\BibitemShut {NoStop}%
\bibitem [{\citenamefont {Berggren}\ \emph {et~al.}(2002)\citenamefont
  {Berggren}, \citenamefont {Sadreev},\ and\ \citenamefont
  {Starikov}}]{berggren2002crossover}%
  \BibitemOpen
  \bibfield  {author} {\bibinfo {author} {\bibnamefont {Berggren},
  \bibfnamefont {K.-F.}}, \bibinfo {author} {\bibfnamefont {A.~F.}\
  \bibnamefont {Sadreev}}, \ and\ \bibinfo {author} {\bibfnamefont {A.~A.}\
  \bibnamefont {Starikov}}} (\bibinfo {year} {2002}),\ \href {\doibase
  10.1103/PhysRevE.66.016218} {\bibfield  {journal} {\bibinfo  {journal} {Phys.
  Rev. E}\ }\textbf {\bibinfo {volume} {66}}~(\bibinfo {number} {1}),\ \bibinfo
  {pages} {016218}}\BibitemShut {NoStop}%
\bibitem [{\citenamefont {Berk}(1987)}]{berk1987scattering}%
  \BibitemOpen
  \bibfield  {author} {\bibinfo {author} {\bibnamefont {Berk}, \bibfnamefont
  {N.~F.}}} (\bibinfo {year} {1987}),\ \href {\doibase
  10.1103/PhysRevLett.58.2718} {\bibfield  {journal} {\bibinfo  {journal}
  {Phys. Rev. Lett.}\ }\textbf {\bibinfo {volume} {58}}~(\bibinfo {number}
  {25}),\ \bibinfo {pages} {2718}}\BibitemShut {NoStop}%
\bibitem [{\citenamefont {Bernard}\ \emph {et~al.}(2006)\citenamefont
  {Bernard}, \citenamefont {Boffetta}, \citenamefont {Celani},\ and\
  \citenamefont {Falkovich}}]{bernard2006conformal}%
  \BibitemOpen
  \bibfield  {author} {\bibinfo {author} {\bibnamefont {Bernard}, \bibfnamefont
  {D.}}, \bibinfo {author} {\bibfnamefont {G.}~\bibnamefont {Boffetta}},
  \bibinfo {author} {\bibfnamefont {A.}~\bibnamefont {Celani}}, \ and\ \bibinfo
  {author} {\bibfnamefont {G.}~\bibnamefont {Falkovich}}} (\bibinfo {year}
  {2006}),\ \href {\doibase 10.1038/nphys217} {\bibfield  {journal} {\bibinfo
  {journal} {Nat. Phys.}\ }\textbf {\bibinfo {volume} {2}}~(\bibinfo {number}
  {2}),\ \bibinfo {pages} {124}}\BibitemShut {NoStop}%
\bibitem [{\citenamefont {Bernstein}(1950)}]{bernstein1950existence}%
  \BibitemOpen
  \bibfield  {author} {\bibinfo {author} {\bibnamefont {Bernstein},
  \bibfnamefont {D.~L.}}} (\bibinfo {year} {1950}),\ \href@noop {} {\emph
  {\bibinfo {title} {Existence Theorems in Partial Differential Equations}}},\
  \bibinfo {series} {Annals of Mathematics Studies}, Vol.~\bibinfo {volume}
  {23}\ (\bibinfo  {publisher} {Princeton University Press},\ \bibinfo
  {address} {Princeton, NJ})\BibitemShut {NoStop}%
\bibitem [{\citenamefont {Berry}\ \emph {et~al.}(1994)\citenamefont {Berry},
  \citenamefont {Katine}, \citenamefont {Westervelt},\ and\ \citenamefont
  {Gossard}}]{berry1994influence}%
  \BibitemOpen
  \bibfield  {author} {\bibinfo {author} {\bibnamefont {Berry}, \bibfnamefont
  {M.~J.}}, \bibinfo {author} {\bibfnamefont {J.~A.}\ \bibnamefont {Katine}},
  \bibinfo {author} {\bibfnamefont {R.~M.}\ \bibnamefont {Westervelt}}, \ and\
  \bibinfo {author} {\bibfnamefont {A.~C.}\ \bibnamefont {Gossard}}} (\bibinfo
  {year} {1994}),\ \href {\doibase 10.1103/PhysRevB.50.17721} {\bibfield
  {journal} {\bibinfo  {journal} {Phys. Rev. B}\ }\textbf {\bibinfo {volume}
  {50}}~(\bibinfo {number} {23}),\ \bibinfo {pages} {17721}}\BibitemShut
  {NoStop}%
\bibitem [{\citenamefont {Berry}(1977)}]{berry1977regular}%
  \BibitemOpen
  \bibfield  {author} {\bibinfo {author} {\bibnamefont {Berry}, \bibfnamefont
  {M.~V.}}} (\bibinfo {year} {1977}),\ \href {\doibase
  10.1088/0305-4470/10/12/016} {\bibfield  {journal} {\bibinfo  {journal} {J.
  Phys. A: Math. Gen.}\ }\textbf {\bibinfo {volume} {10}}~(\bibinfo {number}
  {12}),\ \bibinfo {pages} {2083}}\BibitemShut {NoStop}%
\bibitem [{\citenamefont {Berry}(1978)}]{berry1978disruption}%
  \BibitemOpen
  \bibfield  {author} {\bibinfo {author} {\bibnamefont {Berry}, \bibfnamefont
  {M.~V.}}} (\bibinfo {year} {1978}),\ \href {\doibase
  10.1088/0305-4470/11/1/007} {\bibfield  {journal} {\bibinfo  {journal} {J.
  Phys. A: Math. Gen.}\ }\textbf {\bibinfo {volume} {11}}~(\bibinfo {number}
  {1}),\ \bibinfo {pages} {27}}\BibitemShut {NoStop}%
\bibitem [{\citenamefont {Berry}(1979)}]{Berry1979}%
  \BibitemOpen
  \bibfield  {author} {\bibinfo {author} {\bibnamefont {Berry}, \bibfnamefont
  {M.~V.}}} (\bibinfo {year} {1979}),\ \enquote {\bibinfo {title} {Distribution
  of modes in fractal resonators},}\ in\ \href {\doibase
  10.1007/978-3-642-67363-4_7} {\emph {\bibinfo {booktitle} {Structural
  Stability in Physics}}},\ \bibinfo {editor} {edited by\ \bibinfo {editor}
  {\bibfnamefont {W.}~\bibnamefont {G{\"u}ttinger}}\ and\ \bibinfo {editor}
  {\bibfnamefont {H.}~\bibnamefont {Eikemeier}}}\ (\bibinfo  {publisher}
  {Springer},\ \bibinfo {address} {Berlin, Heidelberg})\ pp.\ \bibinfo {pages}
  {51--53}\BibitemShut {NoStop}%
\bibitem [{\citenamefont {Berry}(1980)}]{berry1980wavefront}%
  \BibitemOpen
  \bibfield  {author} {\bibinfo {author} {\bibnamefont {Berry}, \bibfnamefont
  {M.~V.}}} (\bibinfo {year} {1980}),\ \bibfield  {booktitle} {\emph {\bibinfo
  {booktitle} {Geometry of the Laplace operator}},\ }\href@noop {} {\ \bibinfo
  {series} {Proc. Symp. Pure Math},\ \textbf {\bibinfo {volume} {36}},\
  \bibinfo {pages} {13}}\BibitemShut {NoStop}%
\bibitem [{\citenamefont {Berry}(1981{\natexlab{a}})}]{berry1981regularity}%
  \BibitemOpen
  \bibfield  {author} {\bibinfo {author} {\bibnamefont {Berry}, \bibfnamefont
  {M.~V.}}} (\bibinfo {year} {1981}{\natexlab{a}}),\ \href {\doibase
  10.1088/0143-0807/2/2/006} {\bibfield  {journal} {\bibinfo  {journal} {Eur.
  J. Phys.}\ }\textbf {\bibinfo {volume} {2}}~(\bibinfo {number} {2}),\
  \bibinfo {pages} {91}}\BibitemShut {NoStop}%
\bibitem [{\citenamefont {Berry}(1981{\natexlab{b}})}]{berry1981singularities}%
  \BibitemOpen
  \bibfield  {author} {\bibinfo {author} {\bibnamefont {Berry}, \bibfnamefont
  {M.~V.}}} (\bibinfo {year} {1981}{\natexlab{b}}),\ in\ \href@noop {} {\emph
  {\bibinfo {booktitle} {Physics of Defects, Les Houches Lecture Series Session
  XXXV}}},\ Vol.~\bibinfo {volume} {35},\ \bibinfo {editor} {edited by\
  \bibinfo {editor} {\bibfnamefont {R.}~\bibnamefont {Balian}}, \bibinfo
  {editor} {\bibfnamefont {M.}~\bibnamefont {Kl{\'e}man}}, \ and\ \bibinfo
  {editor} {\bibfnamefont {J.-P.}\ \bibnamefont {Poirier}}}\ (\bibinfo
  {publisher} {Elsevier},\ \bibinfo {address} {North-Holland, Amsterdam})\ pp.\
  \bibinfo {pages} {453--543}\BibitemShut {NoStop}%
\bibitem [{\citenamefont {Berry}(1983)}]{berry1983semiclassical}%
  \BibitemOpen
  \bibfield  {author} {\bibinfo {author} {\bibnamefont {Berry}, \bibfnamefont
  {M.~V.}}} (\bibinfo {year} {1983}),\ in\ \href@noop {} {\emph {\bibinfo
  {booktitle} {Chaotic Behaviour of Deterministic Systems, Les Houches Lecture
  Series Session XXXVI}}},\ Vol.~\bibinfo {volume} {36},\ \bibinfo {editor}
  {edited by\ \bibinfo {editor} {\bibfnamefont {G.}~\bibnamefont {Iooss}},
  \bibinfo {editor} {\bibfnamefont {R.~H.~G.}\ \bibnamefont {Helleman}}, \ and\
  \bibinfo {editor} {\bibfnamefont {S.}~\bibnamefont {R}}}\ (\bibinfo
  {publisher} {North-Holland Amsterdam})\ pp.\ \bibinfo {pages}
  {171--271}\BibitemShut {NoStop}%
\bibitem [{\citenamefont {Berry}(1989)}]{berry1989quantum}%
  \BibitemOpen
  \bibfield  {author} {\bibinfo {author} {\bibnamefont {Berry}, \bibfnamefont
  {M.~V.}}} (\bibinfo {year} {1989}),\ \href {\doibase 10.1098/rspa.1989.0052}
  {\bibfield  {journal} {\bibinfo  {journal} {Proc. R. Soc. Lond. A}\ }\textbf
  {\bibinfo {volume} {423}}~(\bibinfo {number} {1864}),\ \bibinfo {pages}
  {219}}\BibitemShut {NoStop}%
\bibitem [{\citenamefont {Berry}(1991)}]{berry1991houches}%
  \BibitemOpen
  \bibfield  {author} {\bibinfo {author} {\bibnamefont {Berry}, \bibfnamefont
  {M.~V.}}} (\bibinfo {year} {1991}),\ in\ \href@noop {} {\emph {\bibinfo
  {booktitle} {Chaos and Quantum Physics, Les Houches Lecture Series LII
  (1989)}}},\ \bibinfo {editor} {edited by\ \bibinfo {editor} {\bibfnamefont
  {M.~J.}\ \bibnamefont {Giannoni}}, \bibinfo {editor} {\bibfnamefont
  {A.}~\bibnamefont {Voros}}, \ and\ \bibinfo {editor} {\bibfnamefont
  {J.}~\bibnamefont {Zinn-Justin}}}\ (\bibinfo  {publisher} {Elsevier},\
  \bibinfo {address} {North-Holland, Amsterdam})\ pp.\ \bibinfo {pages}
  {251--305}\BibitemShut {NoStop}%
\bibitem [{\citenamefont {Berry}(1998)}]{berry1998much}%
  \BibitemOpen
  \bibfield  {author} {\bibinfo {author} {\bibnamefont {Berry}, \bibfnamefont
  {M.~V.}}} (\bibinfo {year} {1998}),\ in\ \href {\doibase 10.1117/12.317693}
  {\emph {\bibinfo {booktitle} {Proc. SPIE 3487, International Conference on
  Singular Optics}}},\ \bibinfo {editor} {edited by\ \bibinfo {editor}
  {\bibfnamefont {M.~S.}\ \bibnamefont {Soskin}}},\ p.~\bibinfo {pages}
  {1}\BibitemShut {NoStop}%
\bibitem [{\citenamefont {Berry}(2002)}]{berry2002statistics}%
  \BibitemOpen
  \bibfield  {author} {\bibinfo {author} {\bibnamefont {Berry}, \bibfnamefont
  {M.~V.}}} (\bibinfo {year} {2002}),\ \href {\doibase
  10.1088/0305-4470/35/13/301} {\bibfield  {journal} {\bibinfo  {journal} {J.
  Phys. A: Math. Gen.}\ }\textbf {\bibinfo {volume} {35}}~(\bibinfo {number}
  {13}),\ \bibinfo {pages} {3025}}\BibitemShut {NoStop}%
\bibitem [{\citenamefont {Berry}\ and\ \citenamefont
  {Dennis}(2000)}]{Berry2059}%
  \BibitemOpen
  \bibfield  {author} {\bibinfo {author} {\bibnamefont {Berry}, \bibfnamefont
  {M.~V.}}, \ and\ \bibinfo {author} {\bibfnamefont {M.~R.}\ \bibnamefont
  {Dennis}}} (\bibinfo {year} {2000}),\ \href {\doibase 10.1098/rspa.2000.0602}
  {\bibfield  {journal} {\bibinfo  {journal} {Proc. R. Soc. Lond. A}\ }\textbf
  {\bibinfo {volume} {456}}~(\bibinfo {number} {2001}),\ \bibinfo {pages}
  {2059}}\BibitemShut {NoStop}%
\bibitem [{\citenamefont {Berry}\ and\ \citenamefont
  {Ishio}(2002)}]{berry2002nodal}%
  \BibitemOpen
  \bibfield  {author} {\bibinfo {author} {\bibnamefont {Berry}, \bibfnamefont
  {M.~V.}}, \ and\ \bibinfo {author} {\bibfnamefont {H.}~\bibnamefont {Ishio}}}
  (\bibinfo {year} {2002}),\ \href {\doibase 10.1088/0305-4470/35/29/302}
  {\bibfield  {journal} {\bibinfo  {journal} {J. Phys. A: Math. Gen.}\ }\textbf
  {\bibinfo {volume} {35}}~(\bibinfo {number} {29}),\ \bibinfo {pages}
  {5961}}\BibitemShut {NoStop}%
\bibitem [{\citenamefont {Berry}\ and\ \citenamefont
  {Robnik}(1984)}]{berry1984semiclassical}%
  \BibitemOpen
  \bibfield  {author} {\bibinfo {author} {\bibnamefont {Berry}, \bibfnamefont
  {M.~V.}}, \ and\ \bibinfo {author} {\bibfnamefont {M.}~\bibnamefont
  {Robnik}}} (\bibinfo {year} {1984}),\ \href {\doibase
  10.1088/0305-4470/17/12/013} {\bibfield  {journal} {\bibinfo  {journal} {J.
  Phys. A: Math. Gen.}\ }\textbf {\bibinfo {volume} {17}}~(\bibinfo {number}
  {12}),\ \bibinfo {pages} {2413}}\BibitemShut {NoStop}%
\bibitem [{\citenamefont {Berry}\ and\ \citenamefont
  {Robnik}(1986)}]{berry1986quantum}%
  \BibitemOpen
  \bibfield  {author} {\bibinfo {author} {\bibnamefont {Berry}, \bibfnamefont
  {M.~V.}}, \ and\ \bibinfo {author} {\bibfnamefont {M.}~\bibnamefont
  {Robnik}}} (\bibinfo {year} {1986}),\ \href {\doibase
  10.1088/0305-4470/19/8/018} {\bibfield  {journal} {\bibinfo  {journal} {J.
  Phys. A: Math. Gen.}\ }\textbf {\bibinfo {volume} {19}}~(\bibinfo {number}
  {8}),\ \bibinfo {pages} {1365}}\BibitemShut {NoStop}%
\bibitem [{\citenamefont {Berry}\ and\ \citenamefont
  {Tabor}(1976)}]{berry1976closed}%
  \BibitemOpen
  \bibfield  {author} {\bibinfo {author} {\bibnamefont {Berry}, \bibfnamefont
  {M.~V.}}, \ and\ \bibinfo {author} {\bibfnamefont {M.}~\bibnamefont {Tabor}}}
  (\bibinfo {year} {1976}),\ \href {\doibase 10.1098/rspa.1976.0062} {\bibfield
   {journal} {\bibinfo  {journal} {Proc. R. Soc. Lond. A}\ }\textbf {\bibinfo
  {volume} {349}}~(\bibinfo {number} {1656}),\ \bibinfo {pages}
  {101}}\BibitemShut {NoStop}%
\bibitem [{\citenamefont {Berry}\ and\ \citenamefont
  {Tabor}(1977{\natexlab{a}})}]{berry1977calculating}%
  \BibitemOpen
  \bibfield  {author} {\bibinfo {author} {\bibnamefont {Berry}, \bibfnamefont
  {M.~V.}}, \ and\ \bibinfo {author} {\bibfnamefont {M.}~\bibnamefont {Tabor}}}
  (\bibinfo {year} {1977}{\natexlab{a}}),\ \href {\doibase
  10.1088/0305-4470/10/3/009} {\bibfield  {journal} {\bibinfo  {journal} {J.
  Phys. A: Math. Gen.}\ }\textbf {\bibinfo {volume} {10}}~(\bibinfo {number}
  {3}),\ \bibinfo {pages} {371}}\BibitemShut {NoStop}%
\bibitem [{\citenamefont {Berry}\ and\ \citenamefont
  {Tabor}(1977{\natexlab{b}})}]{berry1977level}%
  \BibitemOpen
  \bibfield  {author} {\bibinfo {author} {\bibnamefont {Berry}, \bibfnamefont
  {M.~V.}}, \ and\ \bibinfo {author} {\bibfnamefont {M.}~\bibnamefont {Tabor}}}
  (\bibinfo {year} {1977}{\natexlab{b}}),\ \href {\doibase
  10.1098/rspa.1977.0140} {\bibfield  {journal} {\bibinfo  {journal} {Proc. R.
  Soc. Lond. A}\ }\textbf {\bibinfo {volume} {356}}~(\bibinfo {number}
  {1686}),\ \bibinfo {pages} {375}}\BibitemShut {NoStop}%
\bibitem [{\citenamefont {Bers}(1955)}]{bers1955local}%
  \BibitemOpen
  \bibfield  {author} {\bibinfo {author} {\bibnamefont {Bers}, \bibfnamefont
  {L.}}} (\bibinfo {year} {1955}),\ \href {\doibase 10.1002/cpa.3160080404}
  {\bibfield  {journal} {\bibinfo  {journal} {Comm. Pure and Appl. Math.}\
  }\textbf {\bibinfo {volume} {8}}~(\bibinfo {number} {4}),\ \bibinfo {pages}
  {473}}\BibitemShut {NoStop}%
\bibitem [{\citenamefont {Bieberbach}(1916)}]{bieberbach1916}%
  \BibitemOpen
  \bibfield  {author} {\bibinfo {author} {\bibnamefont {Bieberbach},
  \bibfnamefont {L.}}} (\bibinfo {year} {1916}),\ \href@noop {} {\bibfield
  {journal} {\bibinfo  {journal} {Preuss. Akad. Wiss. Sitzungsb.}\ }\textbf
  {\bibinfo {volume} {138}},\ \bibinfo {pages} {940}}\BibitemShut {NoStop}%
\bibitem [{\citenamefont {Bies}\ and\ \citenamefont
  {Heller}(2002)}]{bies2002nodal}%
  \BibitemOpen
  \bibfield  {author} {\bibinfo {author} {\bibnamefont {Bies}, \bibfnamefont
  {W.~E.}}, \ and\ \bibinfo {author} {\bibfnamefont {E.~J.}\ \bibnamefont
  {Heller}}} (\bibinfo {year} {2002}),\ \href {\doibase
  10.1088/0305-4470/35/27/309} {\bibfield  {journal} {\bibinfo  {journal} {J.
  Phys. A: Math. Gen.}\ }\textbf {\bibinfo {volume} {35}}~(\bibinfo {number}
  {27}),\ \bibinfo {pages} {5673}}\BibitemShut {NoStop}%
\bibitem [{\citenamefont {Bies}\ \emph {et~al.}(2003)\citenamefont {Bies},
  \citenamefont {Lepore},\ and\ \citenamefont {Heller}}]{bies2003}%
  \BibitemOpen
  \bibfield  {author} {\bibinfo {author} {\bibnamefont {Bies}, \bibfnamefont
  {W.~E.}}, \bibinfo {author} {\bibfnamefont {N.}~\bibnamefont {Lepore}}, \
  and\ \bibinfo {author} {\bibfnamefont {E.~J.}\ \bibnamefont {Heller}}}
  (\bibinfo {year} {2003}),\ \href {\doibase 10.1088/0305-4470/36/6/306}
  {\bibfield  {journal} {\bibinfo  {journal} {J. Phys. A: Math. Gen.}\ }\textbf
  {\bibinfo {volume} {36}}~(\bibinfo {number} {6}),\ \bibinfo {pages}
  {1605}}\BibitemShut {NoStop}%
\bibitem [{\citenamefont {Binney}\ \emph {et~al.}(1992)\citenamefont {Binney},
  \citenamefont {Dowrick}, \citenamefont {Fisher},\ and\ \citenamefont
  {Newman}}]{binney1992theory}%
  \BibitemOpen
  \bibfield  {author} {\bibinfo {author} {\bibnamefont {Binney}, \bibfnamefont
  {J.~J.}}, \bibinfo {author} {\bibfnamefont {N.~J.}\ \bibnamefont {Dowrick}},
  \bibinfo {author} {\bibfnamefont {A.~J.}\ \bibnamefont {Fisher}}, \ and\
  \bibinfo {author} {\bibfnamefont {M.}~\bibnamefont {Newman}}} (\bibinfo
  {year} {1992}),\ \href@noop {} {\emph {\bibinfo {title} {The Theory of
  Critical Phenomena: An Introduction to the Renormalization Group}}}\
  (\bibinfo  {publisher} {Oxford University Press},\ \bibinfo {address} {New
  York})\BibitemShut {NoStop}%
\bibitem [{\citenamefont {Birkhoff}\ and\ \citenamefont
  {Fix}(1970)}]{birkhoff1970accurate}%
  \BibitemOpen
  \bibfield  {author} {\bibinfo {author} {\bibnamefont {Birkhoff},
  \bibfnamefont {G.}}, \ and\ \bibinfo {author} {\bibfnamefont
  {G.}~\bibnamefont {Fix}}} (\bibinfo {year} {1970}),\ in\ \href@noop {} {\emph
  {\bibinfo {booktitle} {Numerical Solution of Field Problems in Continuum
  Physics, SIAM-AMS Proceedings}}},\ Vol.~\bibinfo {volume} {2},\ \bibinfo
  {editor} {edited by\ \bibinfo {editor} {\bibfnamefont {G.}~\bibnamefont
  {Birkhoff}}\ and\ \bibinfo {editor} {\bibfnamefont {R.~S.}\ \bibnamefont
  {Varga}}}\ (\bibinfo  {publisher} {AMS},\ \bibinfo {address} {Providence,
  RI})\ pp.\ \bibinfo {pages} {111--151}\BibitemShut {NoStop}%
\bibitem [{\citenamefont {Biswas}\ and\ \citenamefont
  {Jain}(1990)}]{biswas1990quantum}%
  \BibitemOpen
  \bibfield  {author} {\bibinfo {author} {\bibnamefont {Biswas}, \bibfnamefont
  {D.}}, \ and\ \bibinfo {author} {\bibfnamefont {S.~R.}\ \bibnamefont {Jain}}}
  (\bibinfo {year} {1990}),\ \href {\doibase 10.1103/PhysRevA.42.3170}
  {\bibfield  {journal} {\bibinfo  {journal} {Phys. Rev. A}\ }\textbf {\bibinfo
  {volume} {42}}~(\bibinfo {number} {6}),\ \bibinfo {pages} {3170}}\BibitemShut
  {NoStop}%
\bibitem [{\citenamefont {Blanter}\ \emph {et~al.}(1998)\citenamefont
  {Blanter}, \citenamefont {Mirlin},\ and\ \citenamefont
  {Muzykantskii}}]{blanter1998quantum}%
  \BibitemOpen
  \bibfield  {author} {\bibinfo {author} {\bibnamefont {Blanter}, \bibfnamefont
  {{\relax Ya}.~M.}}, \bibinfo {author} {\bibfnamefont {A.~D.}\ \bibnamefont
  {Mirlin}}, \ and\ \bibinfo {author} {\bibfnamefont {B.~A.}\ \bibnamefont
  {Muzykantskii}}} (\bibinfo {year} {1998}),\ \href {\doibase
  10.1103/PhysRevLett.80.4161} {\bibfield  {journal} {\bibinfo  {journal}
  {Phys. Rev. Lett.}\ }\textbf {\bibinfo {volume} {80}}~(\bibinfo {number}
  {19}),\ \bibinfo {pages} {4161}}\BibitemShut {NoStop}%
\bibitem [{\citenamefont {Blatt}\ and\ \citenamefont
  {Weisskopf}(1979)}]{blatt1979theoretical}%
  \BibitemOpen
  \bibfield  {author} {\bibinfo {author} {\bibnamefont {Blatt}, \bibfnamefont
  {J.~M.}}, \ and\ \bibinfo {author} {\bibfnamefont {V.~F.}\ \bibnamefont
  {Weisskopf}}} (\bibinfo {year} {1979}),\ \href@noop {} {\emph {\bibinfo
  {title} {Theoretical nuclear physics}}}\ (\bibinfo  {publisher} {Springer},\
  \bibinfo {address} {New York})\BibitemShut {NoStop}%
\bibitem [{\citenamefont {Bleher}(1994)}]{bleher1994distribution}%
  \BibitemOpen
  \bibfield  {author} {\bibinfo {author} {\bibnamefont {Bleher}, \bibfnamefont
  {P.}}} (\bibinfo {year} {1994}),\ \href {\doibase
  10.1215/S0012-7094-94-07403-6} {\bibfield  {journal} {\bibinfo  {journal}
  {Duke Math. J.}\ }\textbf {\bibinfo {volume} {74}}~(\bibinfo {number} {1}),\
  \bibinfo {pages} {45}}\BibitemShut {NoStop}%
\bibitem [{\citenamefont {Bliokh}\ \emph {et~al.}(2017)\citenamefont {Bliokh},
  \citenamefont {Ivanov}, \citenamefont {Guzzinati}, \citenamefont {Clark},
  \citenamefont {Van~Boxem}, \citenamefont {B{\'e}ch{\'e}}, \citenamefont
  {Juchtmans}, \citenamefont {Alonso}, \citenamefont {Schattschneider},
  \citenamefont {Nori},\ and\ \citenamefont {Verbeek}}]{bliokh2017theory}%
  \BibitemOpen
  \bibfield  {author} {\bibinfo {author} {\bibnamefont {Bliokh}, \bibfnamefont
  {K.~Y.}}, \bibinfo {author} {\bibfnamefont {I.~P.}\ \bibnamefont {Ivanov}},
  \bibinfo {author} {\bibfnamefont {G.}~\bibnamefont {Guzzinati}}, \bibinfo
  {author} {\bibfnamefont {L.}~\bibnamefont {Clark}}, \bibinfo {author}
  {\bibfnamefont {R.}~\bibnamefont {Van~Boxem}}, \bibinfo {author}
  {\bibfnamefont {A.}~\bibnamefont {B{\'e}ch{\'e}}}, \bibinfo {author}
  {\bibfnamefont {R.}~\bibnamefont {Juchtmans}}, \bibinfo {author}
  {\bibfnamefont {M.~A.}\ \bibnamefont {Alonso}}, \bibinfo {author}
  {\bibfnamefont {P.}~\bibnamefont {Schattschneider}}, \bibinfo {author}
  {\bibfnamefont {F.}~\bibnamefont {Nori}}, \ and\ \bibinfo {author}
  {\bibfnamefont {J.}~\bibnamefont {Verbeek}}} (\bibinfo {year} {2017}),\ \href
  {http://www.arxiv.org/abs/1703.06879} {\enquote {\bibinfo {title} {Theory and
  applications of free-electron vortex states},}\ }\bibinfo {note} {{a}rXiv
  preprint 1703.06879}\BibitemShut {NoStop}%
\bibitem [{\citenamefont {Blum}\ \emph {et~al.}(2002)\citenamefont {Blum},
  \citenamefont {Gnutzmann},\ and\ \citenamefont {Smilansky}}]{blum2002nodal}%
  \BibitemOpen
  \bibfield  {author} {\bibinfo {author} {\bibnamefont {Blum}, \bibfnamefont
  {G.}}, \bibinfo {author} {\bibfnamefont {S.}~\bibnamefont {Gnutzmann}}, \
  and\ \bibinfo {author} {\bibfnamefont {U.}~\bibnamefont {Smilansky}}}
  (\bibinfo {year} {2002}),\ \href {\doibase 10.1103/PhysRevLett.88.114101}
  {\bibfield  {journal} {\bibinfo  {journal} {Phys. Rev. Lett.}\ }\textbf
  {\bibinfo {volume} {88}},\ \bibinfo {pages} {114101}}\BibitemShut {NoStop}%
\bibitem [{\citenamefont {Bl{\"u}mel}\ \emph {et~al.}(1992)\citenamefont
  {Bl{\"u}mel}, \citenamefont {Davidson}, \citenamefont {Reinhardt},
  \citenamefont {Lin},\ and\ \citenamefont {Sharnoff}}]{blumel1992quasilinear}%
  \BibitemOpen
  \bibfield  {author} {\bibinfo {author} {\bibnamefont {Bl{\"u}mel},
  \bibfnamefont {R.}}, \bibinfo {author} {\bibfnamefont {I.~H.}\ \bibnamefont
  {Davidson}}, \bibinfo {author} {\bibfnamefont {W.~P.}\ \bibnamefont
  {Reinhardt}}, \bibinfo {author} {\bibfnamefont {H.}~\bibnamefont {Lin}}, \
  and\ \bibinfo {author} {\bibfnamefont {M.}~\bibnamefont {Sharnoff}}}
  (\bibinfo {year} {1992}),\ \href {\doibase 10.1103/PhysRevA.45.2641}
  {\bibfield  {journal} {\bibinfo  {journal} {Phys. Rev. A}\ }\textbf {\bibinfo
  {volume} {45}}~(\bibinfo {number} {4}),\ \bibinfo {pages} {2641}}\BibitemShut
  {NoStop}%
\bibitem [{\citenamefont {Bl{\"u}mel}\ \emph {et~al.}(2001)\citenamefont
  {Bl{\"u}mel}, \citenamefont {Koch},\ and\ \citenamefont
  {Sirko}}]{blumel2001ray}%
  \BibitemOpen
  \bibfield  {author} {\bibinfo {author} {\bibnamefont {Bl{\"u}mel},
  \bibfnamefont {R.}}, \bibinfo {author} {\bibfnamefont {P.~M.}\ \bibnamefont
  {Koch}}, \ and\ \bibinfo {author} {\bibfnamefont {L.}~\bibnamefont {Sirko}}}
  (\bibinfo {year} {2001}),\ \href {\doibase 10.1023/A:1017590503566}
  {\bibfield  {journal} {\bibinfo  {journal} {Found. Phys.}\ }\textbf {\bibinfo
  {volume} {31}}~(\bibinfo {number} {2}),\ \bibinfo {pages} {269}}\BibitemShut
  {NoStop}%
\bibitem [{\citenamefont {B{\^ o}cher}(1898)}]{bocher1898}%
  \BibitemOpen
  \bibfield  {author} {\bibinfo {author} {\bibnamefont {B{\^ o}cher},
  \bibfnamefont {M.}}} (\bibinfo {year} {1898}),\ \href {\doibase
  10.1090/s0002-9904-1898-00500-1} {\bibfield  {journal} {\bibinfo  {journal}
  {Bull. Amer. Math. Soc.}\ }\textbf {\bibinfo {volume} {4}}~(\bibinfo {number}
  {7}),\ \bibinfo {pages} {295}}\BibitemShut {NoStop}%
\bibitem [{\citenamefont {Bogomolny}(1988)}]{bogomolny1988smoothed}%
  \BibitemOpen
  \bibfield  {author} {\bibinfo {author} {\bibnamefont {Bogomolny},
  \bibfnamefont {E.~B.}}} (\bibinfo {year} {1988}),\ \href {\doibase
  10.1016/0167-2789(88)90075-9} {\bibfield  {journal} {\bibinfo  {journal}
  {Physica D}\ }\textbf {\bibinfo {volume} {31}}~(\bibinfo {number} {2}),\
  \bibinfo {pages} {169}}\BibitemShut {NoStop}%
\bibitem [{\citenamefont {Bogomolny}\ \emph {et~al.}(1996)\citenamefont
  {Bogomolny}, \citenamefont {Bohigas},\ and\ \citenamefont
  {Leboeuf}}]{bogomolny1996quantum}%
  \BibitemOpen
  \bibfield  {author} {\bibinfo {author} {\bibnamefont {Bogomolny},
  \bibfnamefont {E.~B.}}, \bibinfo {author} {\bibfnamefont {O.}~\bibnamefont
  {Bohigas}}, \ and\ \bibinfo {author} {\bibfnamefont {P.}~\bibnamefont
  {Leboeuf}}} (\bibinfo {year} {1996}),\ \href {\doibase 10.1007/BF02199359}
  {\bibfield  {journal} {\bibinfo  {journal} {J. Stat. Phys.}\ }\textbf
  {\bibinfo {volume} {85}}~(\bibinfo {number} {5-6}),\ \bibinfo {pages}
  {639}}\BibitemShut {NoStop}%
\bibitem [{\citenamefont {Bogomolny}\ \emph
  {et~al.}(2006{\natexlab{a}})\citenamefont {Bogomolny}, \citenamefont {Dietz},
  \citenamefont {Friedrich}, \citenamefont {Miski-Oglu}, \citenamefont
  {Richter}, \citenamefont {Sch{\"a}fer},\ and\ \citenamefont
  {Schmit}}]{bogomolny2006first}%
  \BibitemOpen
  \bibfield  {author} {\bibinfo {author} {\bibnamefont {Bogomolny},
  \bibfnamefont {E.~B.}}, \bibinfo {author} {\bibfnamefont {B.}~\bibnamefont
  {Dietz}}, \bibinfo {author} {\bibfnamefont {T.}~\bibnamefont {Friedrich}},
  \bibinfo {author} {\bibfnamefont {M.}~\bibnamefont {Miski-Oglu}}, \bibinfo
  {author} {\bibfnamefont {A.}~\bibnamefont {Richter}}, \bibinfo {author}
  {\bibfnamefont {F.}~\bibnamefont {Sch{\"a}fer}}, \ and\ \bibinfo {author}
  {\bibfnamefont {C.}~\bibnamefont {Schmit}}} (\bibinfo {year}
  {2006}{\natexlab{a}}),\ \href {\doibase 10.1103/PhysRevLett.97.254102}
  {\bibfield  {journal} {\bibinfo  {journal} {Phys. Rev. Lett.}\ }\textbf
  {\bibinfo {volume} {97}}~(\bibinfo {number} {25}),\ \bibinfo {pages}
  {254102}}\BibitemShut {NoStop}%
\bibitem [{\citenamefont {Bogomolny}\ \emph
  {et~al.}(2006{\natexlab{b}})\citenamefont {Bogomolny}, \citenamefont
  {Dubertrand},\ and\ \citenamefont {Schmit}}]{bogomolny2006sle}%
  \BibitemOpen
  \bibfield  {author} {\bibinfo {author} {\bibnamefont {Bogomolny},
  \bibfnamefont {E.~B.}}, \bibinfo {author} {\bibfnamefont {R.}~\bibnamefont
  {Dubertrand}}, \ and\ \bibinfo {author} {\bibfnamefont {C.}~\bibnamefont
  {Schmit}}} (\bibinfo {year} {2006}{\natexlab{b}}),\ \href {\doibase
  10.1088/1751-8113/40/3/003} {\bibfield  {journal} {\bibinfo  {journal} {J.
  Phys. A: Math. Theor.}\ }\textbf {\bibinfo {volume} {40}}~(\bibinfo {number}
  {3}),\ \bibinfo {pages} {381}}\BibitemShut {NoStop}%
\bibitem [{\citenamefont {Bogomolny}\ \emph {et~al.}(1992)\citenamefont
  {Bogomolny}, \citenamefont {Georgeot}, \citenamefont {Giannoni},\ and\
  \citenamefont {Schmit}}]{bogomolny1992chaotic}%
  \BibitemOpen
  \bibfield  {author} {\bibinfo {author} {\bibnamefont {Bogomolny},
  \bibfnamefont {E.~B.}}, \bibinfo {author} {\bibfnamefont {B.}~\bibnamefont
  {Georgeot}}, \bibinfo {author} {\bibfnamefont {M.-J.}\ \bibnamefont
  {Giannoni}}, \ and\ \bibinfo {author} {\bibfnamefont {C.}~\bibnamefont
  {Schmit}}} (\bibinfo {year} {1992}),\ \href {\doibase
  10.1103/PhysRevLett.69.1477} {\bibfield  {journal} {\bibinfo  {journal}
  {Phys. Rev. Lett.}\ }\textbf {\bibinfo {volume} {69}}~(\bibinfo {number}
  {10}),\ \bibinfo {pages} {1477}}\BibitemShut {NoStop}%
\bibitem [{\citenamefont {Bogomolny}\ \emph {et~al.}(1999)\citenamefont
  {Bogomolny}, \citenamefont {Gerland},\ and\ \citenamefont
  {Schmit}}]{bogomolny1999models}%
  \BibitemOpen
  \bibfield  {author} {\bibinfo {author} {\bibnamefont {Bogomolny},
  \bibfnamefont {E.~B.}}, \bibinfo {author} {\bibfnamefont {U.}~\bibnamefont
  {Gerland}}, \ and\ \bibinfo {author} {\bibfnamefont {C.}~\bibnamefont
  {Schmit}}} (\bibinfo {year} {1999}),\ \href {\doibase
  10.1103/PhysRevE.59.R1315} {\bibfield  {journal} {\bibinfo  {journal} {Phys.
  Rev. E}\ }\textbf {\bibinfo {volume} {59}}~(\bibinfo {number} {2}),\ \bibinfo
  {pages} {R1315}}\BibitemShut {NoStop}%
\bibitem [{\citenamefont {Bogomolny}\ and\ \citenamefont
  {Schmit}(2002)}]{bogomolny2002percolation}%
  \BibitemOpen
  \bibfield  {author} {\bibinfo {author} {\bibnamefont {Bogomolny},
  \bibfnamefont {E.~B.}}, \ and\ \bibinfo {author} {\bibfnamefont
  {C.}~\bibnamefont {Schmit}}} (\bibinfo {year} {2002}),\ \href {\doibase
  10.1103/PhysRevLett.88.114102} {\bibfield  {journal} {\bibinfo  {journal}
  {Phys. Rev. Lett.}\ }\textbf {\bibinfo {volume} {88}}~(\bibinfo {number}
  {11}),\ \bibinfo {pages} {114102}}\BibitemShut {NoStop}%
\bibitem [{\citenamefont {Bogomolny}\ and\ \citenamefont
  {Schmit}(2004)}]{bogomolny2004structure}%
  \BibitemOpen
  \bibfield  {author} {\bibinfo {author} {\bibnamefont {Bogomolny},
  \bibfnamefont {E.~B.}}, \ and\ \bibinfo {author} {\bibfnamefont
  {C.}~\bibnamefont {Schmit}}} (\bibinfo {year} {2004}),\ \href {\doibase
  10.1103/PhysRevLett.92.244102} {\bibfield  {journal} {\bibinfo  {journal}
  {Phys. Rev. Lett.}\ }\textbf {\bibinfo {volume} {92}}~(\bibinfo {number}
  {24}),\ \bibinfo {pages} {244102}}\BibitemShut {NoStop}%
\bibitem [{\citenamefont {Bogomolny}\ and\ \citenamefont
  {Schmit}(2007)}]{bogomolnyRWM}%
  \BibitemOpen
  \bibfield  {author} {\bibinfo {author} {\bibnamefont {Bogomolny},
  \bibfnamefont {E.~B.}}, \ and\ \bibinfo {author} {\bibfnamefont
  {C.}~\bibnamefont {Schmit}}} (\bibinfo {year} {2007}),\ \href {\doibase
  10.1088/1751-8113/40/47/001} {\bibfield  {journal} {\bibinfo  {journal} {J.
  Phys. A: Math. Theor.}\ }\textbf {\bibinfo {volume} {40}}~(\bibinfo {number}
  {47}),\ \bibinfo {pages} {14033}}\BibitemShut {NoStop}%
\bibitem [{\citenamefont {Bohigas}\ \emph {et~al.}(1984)\citenamefont
  {Bohigas}, \citenamefont {Giannoni},\ and\ \citenamefont
  {Schmit}}]{bohigas1984characterization}%
  \BibitemOpen
  \bibfield  {author} {\bibinfo {author} {\bibnamefont {Bohigas}, \bibfnamefont
  {O.}}, \bibinfo {author} {\bibfnamefont {M.-J.}\ \bibnamefont {Giannoni}}, \
  and\ \bibinfo {author} {\bibfnamefont {C.}~\bibnamefont {Schmit}}} (\bibinfo
  {year} {1984}),\ \href {\doibase 10.1103/PhysRevLett.52.1} {\bibfield
  {journal} {\bibinfo  {journal} {Phys. Rev. Lett.}\ }\textbf {\bibinfo
  {volume} {52}}~(\bibinfo {number} {1}),\ \bibinfo {pages} {1}}\BibitemShut
  {NoStop}%
\bibitem [{\citenamefont {Bohr}\ and\ \citenamefont
  {Mottelson}(1998)}]{bohr1998nuclear}%
  \BibitemOpen
  \bibfield  {author} {\bibinfo {author} {\bibnamefont {Bohr}, \bibfnamefont
  {A.}}, \ and\ \bibinfo {author} {\bibfnamefont {B.~R.}\ \bibnamefont
  {Mottelson}}} (\bibinfo {year} {1998}),\ \href@noop {} {\emph {\bibinfo
  {title} {Nuclear structure}}},\ Vol.~\bibinfo {volume} {1}\ (\bibinfo
  {publisher} {World Scientific},\ \bibinfo {address} {Singapore})\BibitemShut
  {NoStop}%
\bibitem [{\citenamefont {Bolte}(1993)}]{bolte1993some}%
  \BibitemOpen
  \bibfield  {author} {\bibinfo {author} {\bibnamefont {Bolte}, \bibfnamefont
  {J.}}} (\bibinfo {year} {1993}),\ \href {\doibase 10.1142/S0217979293003759}
  {\bibfield  {journal} {\bibinfo  {journal} {Int. J. Mod. Phys. B}\ }\textbf
  {\bibinfo {volume} {7}}~(\bibinfo {number} {27}),\ \bibinfo {pages}
  {4451}}\BibitemShut {NoStop}%
\bibitem [{\citenamefont {Bolte}\ \emph {et~al.}(1992)\citenamefont {Bolte},
  \citenamefont {Steil},\ and\ \citenamefont
  {Steiner}}]{bolte1992arithmetical}%
  \BibitemOpen
  \bibfield  {author} {\bibinfo {author} {\bibnamefont {Bolte}, \bibfnamefont
  {J.}}, \bibinfo {author} {\bibfnamefont {G.}~\bibnamefont {Steil}}, \ and\
  \bibinfo {author} {\bibfnamefont {F.}~\bibnamefont {Steiner}}} (\bibinfo
  {year} {1992}),\ \href {\doibase 10.1103/PhysRevLett.69.2188} {\bibfield
  {journal} {\bibinfo  {journal} {Phys. Rev. Lett.}\ }\textbf {\bibinfo
  {volume} {69}}~(\bibinfo {number} {15}),\ \bibinfo {pages}
  {2188}}\BibitemShut {NoStop}%
\bibitem [{\citenamefont {Borell}(1975)}]{borell1975brunn}%
  \BibitemOpen
  \bibfield  {author} {\bibinfo {author} {\bibnamefont {Borell}, \bibfnamefont
  {C.}}} (\bibinfo {year} {1975}),\ \href {\doibase 10.1007/BF01425510}
  {\bibfield  {journal} {\bibinfo  {journal} {Invent. Math.}\ }\textbf
  {\bibinfo {volume} {30}}~(\bibinfo {number} {2}),\ \bibinfo {pages}
  {207}}\BibitemShut {NoStop}%
\bibitem [{\citenamefont {Borgonovi}(1998)}]{borgonovi1998localization}%
  \BibitemOpen
  \bibfield  {author} {\bibinfo {author} {\bibnamefont {Borgonovi},
  \bibfnamefont {F.}}} (\bibinfo {year} {1998}),\ \href {\doibase
  10.1103/PhysRevLett.80.4653} {\bibfield  {journal} {\bibinfo  {journal}
  {Phys. Rev. Lett.}\ }\textbf {\bibinfo {volume} {80}}~(\bibinfo {number}
  {21}),\ \bibinfo {pages} {4653}}\BibitemShut {NoStop}%
\bibitem [{\citenamefont {Brack}(1993)}]{brack1993physics}%
  \BibitemOpen
  \bibfield  {author} {\bibinfo {author} {\bibnamefont {Brack}, \bibfnamefont
  {M.}}} (\bibinfo {year} {1993}),\ \href {\doibase 10.1103/RevModPhys.65.677}
  {\bibfield  {journal} {\bibinfo  {journal} {Rev. Mod. Phys.}\ }\textbf
  {\bibinfo {volume} {65}}~(\bibinfo {number} {3}),\ \bibinfo {pages}
  {677}}\BibitemShut {NoStop}%
\bibitem [{\citenamefont {Brack}\ and\ \citenamefont
  {Bhaduri}(2003)}]{brack2003semiclassical}%
  \BibitemOpen
  \bibfield  {author} {\bibinfo {author} {\bibnamefont {Brack}, \bibfnamefont
  {M.}}, \ and\ \bibinfo {author} {\bibfnamefont {R.~K.}\ \bibnamefont
  {Bhaduri}}} (\bibinfo {year} {2003}),\ \href@noop {} {\emph {\bibinfo {title}
  {Semiclassical physics}}},\ Vol.~\bibinfo {volume} {96}\ (\bibinfo
  {publisher} {Westview Press},\ \bibinfo {address} {Boulder, CO})\BibitemShut
  {NoStop}%
\bibitem [{\citenamefont {Brink}\ and\ \citenamefont
  {Stephen}(1963)}]{brink1963widths}%
  \BibitemOpen
  \bibfield  {author} {\bibinfo {author} {\bibnamefont {Brink}, \bibfnamefont
  {D.~M.}}, \ and\ \bibinfo {author} {\bibfnamefont {R.~O.}\ \bibnamefont
  {Stephen}}} (\bibinfo {year} {1963}),\ \href {\doibase
  10.1016/S0375-9601(63)80037-7} {\bibfield  {journal} {\bibinfo  {journal}
  {Phys. Lett.}\ }\textbf {\bibinfo {volume} {5}}~(\bibinfo {number} {1}),\
  \bibinfo {pages} {77}}\BibitemShut {NoStop}%
\bibitem [{\citenamefont {Brossard}\ and\ \citenamefont
  {Carmona}(1986)}]{Brossard1986}%
  \BibitemOpen
  \bibfield  {author} {\bibinfo {author} {\bibnamefont {Brossard},
  \bibfnamefont {J.}}, \ and\ \bibinfo {author} {\bibfnamefont
  {R.}~\bibnamefont {Carmona}}} (\bibinfo {year} {1986}),\ \href {\doibase
  10.1007/BF01210795} {\bibfield  {journal} {\bibinfo  {journal} {Commun. Math.
  Phys.}\ }\textbf {\bibinfo {volume} {104}}~(\bibinfo {number} {1}),\ \bibinfo
  {pages} {103}}\BibitemShut {NoStop}%
\bibitem [{\citenamefont {Br{\"u}ning}(1978)}]{bruning1978knoten}%
  \BibitemOpen
  \bibfield  {author} {\bibinfo {author} {\bibnamefont {Br{\"u}ning},
  \bibfnamefont {J.}}} (\bibinfo {year} {1978}),\ \href {\doibase
  10.1007/BF01214561} {\bibfield  {journal} {\bibinfo  {journal} {Math. Z.}\
  }\textbf {\bibinfo {volume} {158}}~(\bibinfo {number} {1}),\ \bibinfo {pages}
  {15}}\BibitemShut {NoStop}%
\bibitem [{\citenamefont {Br{\"u}ning}\ and\ \citenamefont
  {Gromes}(1972)}]{bruning1972lange}%
  \BibitemOpen
  \bibfield  {author} {\bibinfo {author} {\bibnamefont {Br{\"u}ning},
  \bibfnamefont {J.}}, \ and\ \bibinfo {author} {\bibfnamefont
  {D.}~\bibnamefont {Gromes}}} (\bibinfo {year} {1972}),\ \href {\doibase
  10.1007/BF01142586} {\bibfield  {journal} {\bibinfo  {journal} {Math. Z.}\
  }\textbf {\bibinfo {volume} {124}}~(\bibinfo {number} {1}),\ \bibinfo {pages}
  {79}}\BibitemShut {NoStop}%
\bibitem [{\citenamefont {Buckley}\ and\ \citenamefont
  {Wigman}(2016)}]{buckley2016number}%
  \BibitemOpen
  \bibfield  {author} {\bibinfo {author} {\bibnamefont {Buckley}, \bibfnamefont
  {J.}}, \ and\ \bibinfo {author} {\bibfnamefont {I.}~\bibnamefont {Wigman}}}
  (\bibinfo {year} {2016}),\ \href {\doibase 10.1007/s00023-016-0476-7}
  {\bibfield  {journal} {\bibinfo  {journal} {Ann. Henri Poincar{\'e}}\
  }\textbf {\bibinfo {volume} {17}}~(\bibinfo {number} {11}),\ \bibinfo {pages}
  {3027}}\BibitemShut {NoStop}%
\bibitem [{\citenamefont {Bunimovich}(2001)}]{bunimovich2001mushrooms}%
  \BibitemOpen
  \bibfield  {author} {\bibinfo {author} {\bibnamefont {Bunimovich},
  \bibfnamefont {L.~A.}}} (\bibinfo {year} {2001}),\ \href {\doibase
  10.1063/1.1418763} {\bibfield  {journal} {\bibinfo  {journal} {Chaos}\
  }\textbf {\bibinfo {volume} {11}}~(\bibinfo {number} {4}),\ \bibinfo {pages}
  {802}}\BibitemShut {NoStop}%
\bibitem [{\citenamefont {Bunimovich}(2007)}]{bunimovich2007dynamical}%
  \BibitemOpen
  \bibfield  {author} {\bibinfo {author} {\bibnamefont {Bunimovich},
  \bibfnamefont {L.~A.}}} (\bibinfo {year} {2007}),\ \href {\doibase
  10.4249/scholarpedia.1813} {\bibfield  {journal} {\bibinfo  {journal}
  {Scholarpedia}\ }\textbf {\bibinfo {volume} {2}}~(\bibinfo {number} {8}),\
  \bibinfo {pages} {1813}}\BibitemShut {NoStop}%
\bibitem [{\citenamefont {Bunimovich}\ \emph {et~al.}(2013)\citenamefont
  {Bunimovich}, \citenamefont {Burago}, \citenamefont {Chernov}, \citenamefont
  {Cohen}, \citenamefont {Dettmann}, \citenamefont {Dorfman}, \citenamefont
  {Ferleger}, \citenamefont {Hirschl}, \citenamefont {Kononenko}, \citenamefont
  {Lebowitz}, \citenamefont {Liverani}, \citenamefont {Murphy}, \citenamefont
  {Piasecki}, \citenamefont {Posch}, \citenamefont {Simanyi}, \citenamefont
  {Sina{\u i}}, \citenamefont {Tel}, \citenamefont {van Beijeren},
  \citenamefont {van Zon}, \citenamefont {Vollmer},\ and\ \citenamefont
  {Young}}]{bunimovich2013hard}%
  \BibitemOpen
  \bibfield  {author} {\bibinfo {author} {\bibnamefont {Bunimovich},
  \bibfnamefont {L.~A.}}, \bibinfo {author} {\bibfnamefont {D.}~\bibnamefont
  {Burago}}, \bibinfo {author} {\bibfnamefont {N.}~\bibnamefont {Chernov}},
  \bibinfo {author} {\bibfnamefont {E.~G.~D.}\ \bibnamefont {Cohen}}, \bibinfo
  {author} {\bibfnamefont {C.~P.}\ \bibnamefont {Dettmann}}, \bibinfo {author}
  {\bibfnamefont {J.~R.}\ \bibnamefont {Dorfman}}, \bibinfo {author}
  {\bibfnamefont {S.}~\bibnamefont {Ferleger}}, \bibinfo {author}
  {\bibfnamefont {R.}~\bibnamefont {Hirschl}}, \bibinfo {author} {\bibfnamefont
  {A.}~\bibnamefont {Kononenko}}, \bibinfo {author} {\bibfnamefont {J.~L.}\
  \bibnamefont {Lebowitz}}, \bibinfo {author} {\bibfnamefont {C.}~\bibnamefont
  {Liverani}}, \bibinfo {author} {\bibfnamefont {T.~J.}\ \bibnamefont
  {Murphy}}, \bibinfo {author} {\bibfnamefont {J.}~\bibnamefont {Piasecki}},
  \bibinfo {author} {\bibfnamefont {H.~A.}\ \bibnamefont {Posch}}, \bibinfo
  {author} {\bibfnamefont {N.}~\bibnamefont {Simanyi}}, \bibinfo {author}
  {\bibfnamefont {{\relax Ya}.~G.}\ \bibnamefont {Sina{\u i}}}, \bibinfo
  {author} {\bibfnamefont {T.}~\bibnamefont {Tel}}, \bibinfo {author}
  {\bibfnamefont {H.}~\bibnamefont {van Beijeren}}, \bibinfo {author}
  {\bibfnamefont {R.}~\bibnamefont {van Zon}}, \bibinfo {author} {\bibfnamefont
  {J.}~\bibnamefont {Vollmer}}, \ and\ \bibinfo {author} {\bibfnamefont
  {L.~S.}\ \bibnamefont {Young}}} (\bibinfo {year} {2013}),\ \href {\doibase
  10.1007/978-3-662-04062-1} {\emph {\bibinfo {title} {Hard Ball Systems and
  the Lorentz Gas}}},\ edited by\ \bibinfo {editor} {\bibfnamefont
  {D.}~\bibnamefont {Sz{\' a}sz}},\ \bibinfo {series} {Encylopaedia of
  Mathematical Sciences}, Vol.\ \bibinfo {volume} {101}\ (\bibinfo  {publisher}
  {Springer},\ \bibinfo {address} {Heidelberg})\BibitemShut {NoStop}%
\bibitem [{\citenamefont {Bunimovich}\ and\ \citenamefont {Sina{\u
  i}}(1981)}]{bunimovich1981statistical}%
  \BibitemOpen
  \bibfield  {author} {\bibinfo {author} {\bibnamefont {Bunimovich},
  \bibfnamefont {L.~A.}}, \ and\ \bibinfo {author} {\bibfnamefont {{\relax
  Ya}.~G.}\ \bibnamefont {Sina{\u i}}}} (\bibinfo {year} {1981}),\ \href
  {\doibase 10.1007/BF02046760} {\bibfield  {journal} {\bibinfo  {journal}
  {Commun. Math. Phys.}\ }\textbf {\bibinfo {volume} {78}}~(\bibinfo {number}
  {4}),\ \bibinfo {pages} {479}}\BibitemShut {NoStop}%
\bibitem [{\citenamefont {Burenkov}\ and\ \citenamefont
  {Davies}(2002)}]{burenkov2002spectral}%
  \BibitemOpen
  \bibfield  {author} {\bibinfo {author} {\bibnamefont {Burenkov},
  \bibfnamefont {V.~I.}}, \ and\ \bibinfo {author} {\bibfnamefont {E.~B.}\
  \bibnamefont {Davies}}} (\bibinfo {year} {2002}),\ \href {\doibase
  10.1016/S0022-0396(02)00033-5} {\bibfield  {journal} {\bibinfo  {journal} {J.
  Differential Equations}\ }\textbf {\bibinfo {volume} {186}}~(\bibinfo
  {number} {2}),\ \bibinfo {pages} {485}}\BibitemShut {NoStop}%
\bibitem [{\citenamefont {Buser}\ \emph {et~al.}(1994)\citenamefont {Buser},
  \citenamefont {Conway}, \citenamefont {Doyle},\ and\ \citenamefont
  {Semmler}}]{buser1994some}%
  \BibitemOpen
  \bibfield  {author} {\bibinfo {author} {\bibnamefont {Buser}, \bibfnamefont
  {P.}}, \bibinfo {author} {\bibfnamefont {J.}~\bibnamefont {Conway}}, \bibinfo
  {author} {\bibfnamefont {P.}~\bibnamefont {Doyle}}, \ and\ \bibinfo {author}
  {\bibfnamefont {K.-D.}\ \bibnamefont {Semmler}}} (\bibinfo {year} {1994}),\
  \href {\doibase 10.1155/S1073792894000437} {\bibfield  {journal} {\bibinfo
  {journal} {Int. Math. Res. Not.}\ }\textbf {\bibinfo {volume}
  {1994}}~(\bibinfo {number} {9}),\ \bibinfo {pages} {391}}\BibitemShut
  {NoStop}%
\bibitem [{\citenamefont {Canzani}\ and\ \citenamefont
  {Sarnak}(2014)}]{canzani2014topology}%
  \BibitemOpen
  \bibfield  {author} {\bibinfo {author} {\bibnamefont {Canzani}, \bibfnamefont
  {Y.}}, \ and\ \bibinfo {author} {\bibfnamefont {P.}~\bibnamefont {Sarnak}}}
  (\bibinfo {year} {2014}),\ \href {http://www.arxiv.org/abs/1412.4437}
  {\enquote {\bibinfo {title} {On the topology of the zero sets of
  monochromatic random waves},}\ }\bibinfo {note} {{a}rXiv preprint
  1412.4437}\BibitemShut {NoStop}%
\bibitem [{\citenamefont {Canzani}\ and\ \citenamefont
  {Sarnak}(2016)}]{canzani2016topology}%
  \BibitemOpen
  \bibfield  {author} {\bibinfo {author} {\bibnamefont {Canzani}, \bibfnamefont
  {Y.}}, \ and\ \bibinfo {author} {\bibfnamefont {P.}~\bibnamefont {Sarnak}}}
  (\bibinfo {year} {2016}),\ \href {http://www.arxiv.org/abs/1701.00034}
  {\enquote {\bibinfo {title} {Topology and nesting of the zero set components
  of monochromatic random waves},}\ }\bibinfo {note} {{a}rXiv preprint
  1701.00034}\BibitemShut {NoStop}%
\bibitem [{\citenamefont {Cardy}(1984)}]{cardy1984conformal}%
  \BibitemOpen
  \bibfield  {author} {\bibinfo {author} {\bibnamefont {Cardy}, \bibfnamefont
  {J.~L.}}} (\bibinfo {year} {1984}),\ \href {\doibase
  10.1016/0550-3213(84)90241-4} {\bibfield  {journal} {\bibinfo  {journal}
  {Nucl. Phys. B}\ }\textbf {\bibinfo {volume} {240}}~(\bibinfo {number} {4}),\
  \bibinfo {pages} {514}}\BibitemShut {NoStop}%
\bibitem [{\citenamefont {Cardy}(1992)}]{cardy1992critical}%
  \BibitemOpen
  \bibfield  {author} {\bibinfo {author} {\bibnamefont {Cardy}, \bibfnamefont
  {J.~L.}}} (\bibinfo {year} {1992}),\ \href {\doibase
  10.1088/0305-4470/25/4/009} {\bibfield  {journal} {\bibinfo  {journal} {J.
  Phys. A: Math. Gen.}\ }\textbf {\bibinfo {volume} {25}}~(\bibinfo {number}
  {4}),\ \bibinfo {pages} {L201}}\BibitemShut {NoStop}%
\bibitem [{\citenamefont {Cardy}(2003)}]{cardy2003stochastic}%
  \BibitemOpen
  \bibfield  {author} {\bibinfo {author} {\bibnamefont {Cardy}, \bibfnamefont
  {J.~L.}}} (\bibinfo {year} {2003}),\ \href {\doibase
  10.1088/0305-4470/36/24/101} {\bibfield  {journal} {\bibinfo  {journal} {J.
  Phys. A: Math. Gen.}\ }\textbf {\bibinfo {volume} {36}}~(\bibinfo {number}
  {24}),\ \bibinfo {pages} {L379}}\BibitemShut {NoStop}%
\bibitem [{\citenamefont {Cardy}(2005)}]{cardy2005sle}%
  \BibitemOpen
  \bibfield  {author} {\bibinfo {author} {\bibnamefont {Cardy}, \bibfnamefont
  {J.~L.}}} (\bibinfo {year} {2005}),\ \href {\doibase
  10.1016/j.aop.2005.04.001} {\bibfield  {journal} {\bibinfo  {journal} {Ann.
  Phys.}\ }\textbf {\bibinfo {volume} {318}}~(\bibinfo {number} {1}),\ \bibinfo
  {pages} {81}}\BibitemShut {NoStop}%
\bibitem [{\citenamefont {Cardy}(2006)}]{cardy2006turbulence}%
  \BibitemOpen
  \bibfield  {author} {\bibinfo {author} {\bibnamefont {Cardy}, \bibfnamefont
  {J.~L.}}} (\bibinfo {year} {2006}),\ \href {\doibase 10.1038/nphys223}
  {\bibfield  {journal} {\bibinfo  {journal} {Nat. Phys.}\ }\textbf {\bibinfo
  {volume} {2}}~(\bibinfo {number} {2}),\ \bibinfo {pages} {67}}\BibitemShut
  {NoStop}%
\bibitem [{\citenamefont {Casati}\ and\ \citenamefont
  {Ford}(1979)}]{casati1979stochastic}%
  \BibitemOpen
  \bibfield  {author} {\bibinfo {author} {\bibnamefont {Casati}, \bibfnamefont
  {G.}}, \ and\ \bibinfo {author} {\bibfnamefont {J.}~\bibnamefont {Ford}}}
  (\bibinfo {year} {1979}),\ \href@noop {} {\emph {\bibinfo {title} {Stochastic
  Behavior in Classical and Quantum Hamiltonian systems}}},\ \bibinfo {series}
  {Lecture Notes in Physics}, Vol.~\bibinfo {volume} {93}\ (\bibinfo
  {publisher} {Springer},\ \bibinfo {address} {Berlin})\BibitemShut {NoStop}%
\bibitem [{\citenamefont {Chang}\ and\ \citenamefont
  {DeTurck}(1989)}]{chang1989hearing}%
  \BibitemOpen
  \bibfield  {author} {\bibinfo {author} {\bibnamefont {Chang}, \bibfnamefont
  {P.-K.}}, \ and\ \bibinfo {author} {\bibfnamefont {D.}~\bibnamefont
  {DeTurck}}} (\bibinfo {year} {1989}),\ \href {\doibase
  10.1090/S0002-9939-1989-0953738-7} {\bibfield  {journal} {\bibinfo  {journal}
  {Proc. Amer. Math. Soc.}\ }\textbf {\bibinfo {volume} {105}}~(\bibinfo
  {number} {4}),\ \bibinfo {pages} {1033}}\BibitemShut {NoStop}%
\bibitem [{\citenamefont {Chapman}(1995)}]{chapman1995drums}%
  \BibitemOpen
  \bibfield  {author} {\bibinfo {author} {\bibnamefont {Chapman}, \bibfnamefont
  {S.~J.}}} (\bibinfo {year} {1995}),\ \href {\doibase 10.2307/2975346}
  {\bibfield  {journal} {\bibinfo  {journal} {Amer. Math. Monthly}\ }\textbf
  {\bibinfo {volume} {102}}~(\bibinfo {number} {2}),\ \bibinfo {pages}
  {124}}\BibitemShut {NoStop}%
\bibitem [{\citenamefont {Chavel}(1984)}]{chavel1984eigenvalues}%
  \BibitemOpen
  \bibfield  {author} {\bibinfo {author} {\bibnamefont {Chavel}, \bibfnamefont
  {I.}}} (\bibinfo {year} {1984}),\ \href@noop {} {\emph {\bibinfo {title}
  {Eigenvalues in Riemannian geometry}}},\ Vol.\ \bibinfo {volume} {115}\
  (\bibinfo  {publisher} {Academic Press},\ \bibinfo {address} {Orlando,
  FL})\BibitemShut {NoStop}%
\bibitem [{\citenamefont {Chen}\ and\ \citenamefont
  {Huang}(2003{\natexlab{a}})}]{chen2003vortex}%
  \BibitemOpen
  \bibfield  {author} {\bibinfo {author} {\bibnamefont {Chen}, \bibfnamefont
  {Y.~F.}}, \ and\ \bibinfo {author} {\bibfnamefont {K.~F.}\ \bibnamefont
  {Huang}}} (\bibinfo {year} {2003}{\natexlab{a}}),\ \href {\doibase
  10.1103/PhysRevE.68.066207} {\bibfield  {journal} {\bibinfo  {journal} {Phys.
  Rev. E}\ }\textbf {\bibinfo {volume} {68}}~(\bibinfo {number} {6}),\ \bibinfo
  {pages} {066207}}\BibitemShut {NoStop}%
\bibitem [{\citenamefont {Chen}\ and\ \citenamefont
  {Huang}(2003{\natexlab{b}})}]{chen2003vortex2}%
  \BibitemOpen
  \bibfield  {author} {\bibinfo {author} {\bibnamefont {Chen}, \bibfnamefont
  {Y.~F.}}, \ and\ \bibinfo {author} {\bibfnamefont {K.~F.}\ \bibnamefont
  {Huang}}} (\bibinfo {year} {2003}{\natexlab{b}}),\ \href {\doibase
  10.1088/0305-4470/36/28/305} {\bibfield  {journal} {\bibinfo  {journal} {J.
  Phys. A: Math. Gen.}\ }\textbf {\bibinfo {volume} {36}}~(\bibinfo {number}
  {28}),\ \bibinfo {pages} {7751}}\BibitemShut {NoStop}%
\bibitem [{\citenamefont {Chen}\ \emph {et~al.}(2002)\citenamefont {Chen},
  \citenamefont {Huang},\ and\ \citenamefont {Lan}}]{chen2002quantum}%
  \BibitemOpen
  \bibfield  {author} {\bibinfo {author} {\bibnamefont {Chen}, \bibfnamefont
  {Y.~F.}}, \bibinfo {author} {\bibfnamefont {K.~F.}\ \bibnamefont {Huang}}, \
  and\ \bibinfo {author} {\bibfnamefont {Y.~P.}\ \bibnamefont {Lan}}} (\bibinfo
  {year} {2002}),\ \href {\doibase 10.1103/PhysRevE.66.066210} {\bibfield
  {journal} {\bibinfo  {journal} {Phys. Rev. E}\ }\textbf {\bibinfo {volume}
  {66}}~(\bibinfo {number} {6}),\ \bibinfo {pages} {066210}}\BibitemShut
  {NoStop}%
\bibitem [{\citenamefont {Chen}\ \emph {et~al.}(2012)\citenamefont {Chen},
  \citenamefont {Lin}, \citenamefont {Zhuang}, \citenamefont {Liang},
  \citenamefont {Su},\ and\ \citenamefont {Huang}}]{chen2012generation}%
  \BibitemOpen
  \bibfield  {author} {\bibinfo {author} {\bibnamefont {Chen}, \bibfnamefont
  {Y.~F.}}, \bibinfo {author} {\bibfnamefont {Y.~C.}\ \bibnamefont {Lin}},
  \bibinfo {author} {\bibfnamefont {W.~Z.}\ \bibnamefont {Zhuang}}, \bibinfo
  {author} {\bibfnamefont {H.~C.}\ \bibnamefont {Liang}}, \bibinfo {author}
  {\bibfnamefont {K.~W.}\ \bibnamefont {Su}}, \ and\ \bibinfo {author}
  {\bibfnamefont {K.~F.}\ \bibnamefont {Huang}}} (\bibinfo {year} {2012}),\
  \href {\doibase 10.1103/PhysRevA.85.043833} {\bibfield  {journal} {\bibinfo
  {journal} {Phys. Rev. A}\ }\textbf {\bibinfo {volume} {85}}~(\bibinfo
  {number} {4}),\ \bibinfo {pages} {043833}}\BibitemShut {NoStop}%
\bibitem [{\citenamefont {Chen}\ \emph {et~al.}(2006)\citenamefont {Chen},
  \citenamefont {Su}, \citenamefont {Lu},\ and\ \citenamefont
  {Huang}}]{chen2006manifestation}%
  \BibitemOpen
  \bibfield  {author} {\bibinfo {author} {\bibnamefont {Chen}, \bibfnamefont
  {Y.~F.}}, \bibinfo {author} {\bibfnamefont {K.~W.}\ \bibnamefont {Su}},
  \bibinfo {author} {\bibfnamefont {T.-H.}\ \bibnamefont {Lu}}, \ and\ \bibinfo
  {author} {\bibfnamefont {K.~F.}\ \bibnamefont {Huang}}} (\bibinfo {year}
  {2006}),\ \href {\doibase 10.1103/PhysRevLett.96.033905} {\bibfield
  {journal} {\bibinfo  {journal} {Phys. Rev. Lett.}\ }\textbf {\bibinfo
  {volume} {96}}~(\bibinfo {number} {3}),\ \bibinfo {pages}
  {033905}}\BibitemShut {NoStop}%
\bibitem [{\citenamefont {Cheng}(1976)}]{cheng1976eigenfunctions}%
  \BibitemOpen
  \bibfield  {author} {\bibinfo {author} {\bibnamefont {Cheng}, \bibfnamefont
  {S.-Y.}}} (\bibinfo {year} {1976}),\ \href {\doibase 10.1007/BF02568142}
  {\bibfield  {journal} {\bibinfo  {journal} {Comment. Math. Helv.}\ }\textbf
  {\bibinfo {volume} {51}}~(\bibinfo {number} {1}),\ \bibinfo {pages}
  {43}}\BibitemShut {NoStop}%
\bibitem [{\citenamefont {Chibotaru}\ \emph {et~al.}(2001)\citenamefont
  {Chibotaru}, \citenamefont {Ceulemans}, \citenamefont {Bruyndoncx},\ and\
  \citenamefont {Moshchalkov}}]{chibotaru2001vortex}%
  \BibitemOpen
  \bibfield  {author} {\bibinfo {author} {\bibnamefont {Chibotaru},
  \bibfnamefont {L.~F.}}, \bibinfo {author} {\bibfnamefont {A.}~\bibnamefont
  {Ceulemans}}, \bibinfo {author} {\bibfnamefont {V.}~\bibnamefont
  {Bruyndoncx}}, \ and\ \bibinfo {author} {\bibfnamefont {V.~V.}\ \bibnamefont
  {Moshchalkov}}} (\bibinfo {year} {2001}),\ \href {\doibase
  10.1103/PhysRevLett.86.1323} {\bibfield  {journal} {\bibinfo  {journal}
  {Phys. Rev. Lett.}\ }\textbf {\bibinfo {volume} {86}}~(\bibinfo {number}
  {7}),\ \bibinfo {pages} {1323}}\BibitemShut {NoStop}%
\bibitem [{\citenamefont {Chladni}(1787)}]{chladni1787entdeckungen}%
  \BibitemOpen
  \bibfield  {author} {\bibinfo {author} {\bibnamefont {Chladni}, \bibfnamefont
  {E.~F.~F.}}} (\bibinfo {year} {1787}),\ \href@noop {} {\emph {\bibinfo
  {title} {Entdeckungen {\"u}ber die Theorie des Klanges}}}\ (\bibinfo
  {publisher} {Zentralantiquariat der DDR},\ \bibinfo {address}
  {Leipzig})\BibitemShut {NoStop}%
\bibitem [{\citenamefont {Chladni}(1802)}]{chladni1802akustik}%
  \BibitemOpen
  \bibfield  {author} {\bibinfo {author} {\bibnamefont {Chladni}, \bibfnamefont
  {E.~F.~F.}}} (\bibinfo {year} {1802}),\ \href@noop {} {\emph {\bibinfo
  {title} {Die {A}kustik}}}\ (\bibinfo  {publisher} {Breitkopf und
  H{\"a}rtel},\ \bibinfo {address} {Leipzig})\BibitemShut {NoStop}%
\bibitem [{\citenamefont {Cipra}(1992)}]{cipra1992you}%
  \BibitemOpen
  \bibfield  {author} {\bibinfo {author} {\bibnamefont {Cipra}, \bibfnamefont
  {B.}}} (\bibinfo {year} {1992}),\ \href {\doibase
  0.1126/science.255.5052.1642} {\bibfield  {journal} {\bibinfo  {journal}
  {Science}\ }\textbf {\bibinfo {volume} {255}}~(\bibinfo {number} {5052}),\
  \bibinfo {pages} {1642}}\BibitemShut {NoStop}%
\bibitem [{\citenamefont {Coffman}(1982)}]{coffman1982structure}%
  \BibitemOpen
  \bibfield  {author} {\bibinfo {author} {\bibnamefont {Coffman}, \bibfnamefont
  {C.~V.}}} (\bibinfo {year} {1982}),\ \href {\doibase 10.1137/0513051}
  {\bibfield  {journal} {\bibinfo  {journal} {SIAM J. Math. Anal.}\ }\textbf
  {\bibinfo {volume} {13}}~(\bibinfo {number} {5}),\ \bibinfo {pages}
  {746}}\BibitemShut {NoStop}%
\bibitem [{\citenamefont {Coffman}\ and\ \citenamefont
  {Duffin}(1980)}]{coffman1980structure}%
  \BibitemOpen
  \bibfield  {author} {\bibinfo {author} {\bibnamefont {Coffman}, \bibfnamefont
  {C.~V.}}, \ and\ \bibinfo {author} {\bibfnamefont {R.~J.}\ \bibnamefont
  {Duffin}}} (\bibinfo {year} {1980}),\ \href {\doibase
  10.1016/0196-8858(80)90018-4} {\bibfield  {journal} {\bibinfo  {journal}
  {Adv. Appl. Math.}\ }\textbf {\bibinfo {volume} {1}}~(\bibinfo {number}
  {4}),\ \bibinfo {pages} {373}}\BibitemShut {NoStop}%
\bibitem [{\citenamefont {Collin}(1960)}]{collin1960field}%
  \BibitemOpen
  \bibfield  {author} {\bibinfo {author} {\bibnamefont {Collin}, \bibfnamefont
  {R.~E.}}} (\bibinfo {year} {1960}),\ \href@noop {} {\emph {\bibinfo {title}
  {Field theory of guided waves}}}\ (\bibinfo  {publisher} {McGraw-Hill},\
  \bibinfo {address} {New York})\BibitemShut {NoStop}%
\bibitem [{\citenamefont {Couchman}\ \emph {et~al.}(1992)\citenamefont
  {Couchman}, \citenamefont {Ott},\ and\ \citenamefont
  {Antonsen~Jr.}}]{couchman1992quantum}%
  \BibitemOpen
  \bibfield  {author} {\bibinfo {author} {\bibnamefont {Couchman},
  \bibfnamefont {L.}}, \bibinfo {author} {\bibfnamefont {E.}~\bibnamefont
  {Ott}}, \ and\ \bibinfo {author} {\bibfnamefont {T.~M.}\ \bibnamefont
  {Antonsen~Jr.}}} (\bibinfo {year} {1992}),\ \href {\doibase
  10.1103/PhysRevA.46.6193} {\bibfield  {journal} {\bibinfo  {journal} {Phys.
  Rev. A}\ }\textbf {\bibinfo {volume} {46}}~(\bibinfo {number} {10}),\
  \bibinfo {pages} {6193}}\BibitemShut {NoStop}%
\bibitem [{\citenamefont {Courant}(1923)}]{courant1923}%
  \BibitemOpen
  \bibfield  {author} {\bibinfo {author} {\bibnamefont {Courant}, \bibfnamefont
  {R.}}} (\bibinfo {year} {1923}),\ \href
  {http://www.digizeitschriften.de/dms/resolveppn/?PID=GDZPPN002506300}
  {\bibfield  {journal} {\bibinfo  {journal} {Nachr. Ges. Wiss. G{\" o}ttingen,
  Math.-Phys. Kl.}\ }\textbf {\bibinfo {volume} {1}},\ \bibinfo {pages}
  {81}}\BibitemShut {NoStop}%
\bibitem [{\citenamefont {Courant}\ and\ \citenamefont
  {Hilbert}(1953)}]{courant1953methods}%
  \BibitemOpen
  \bibfield  {author} {\bibinfo {author} {\bibnamefont {Courant}, \bibfnamefont
  {R.}}, \ and\ \bibinfo {author} {\bibfnamefont {D.}~\bibnamefont {Hilbert}}}
  (\bibinfo {year} {1953}),\ \href {\doibase 10.1002/9783527617210} {\emph
  {\bibinfo {title} {Methods of Mathematical Physics}}},\ Vol.~\bibinfo
  {volume} {1}\ (\bibinfo  {publisher} {Interscience Publishers Inc.},\
  \bibinfo {address} {New York})\BibitemShut {NoStop}%
\bibitem [{\citenamefont {Courtial}\ and\ \citenamefont
  {O'Holleran}(2007)}]{courtial2007experiments}%
  \BibitemOpen
  \bibfield  {author} {\bibinfo {author} {\bibnamefont {Courtial},
  \bibfnamefont {J.}}, \ and\ \bibinfo {author} {\bibfnamefont
  {K.}~\bibnamefont {O'Holleran}}} (\bibinfo {year} {2007}),\ \href {\doibase
  10.1140/epjst/e2007-00146-3} {\bibfield  {journal} {\bibinfo  {journal} {Eur.
  Phys. J. Spec. Top.}\ }\textbf {\bibinfo {volume} {145}}~(\bibinfo {number}
  {1}),\ \bibinfo {pages} {35}}\BibitemShut {NoStop}%
\bibitem [{\citenamefont {Cram{\'e}r}\ and\ \citenamefont
  {Leadbetter}(2013)}]{cramer2013stationary}%
  \BibitemOpen
  \bibfield  {author} {\bibinfo {author} {\bibnamefont {Cram{\'e}r},
  \bibfnamefont {H.}}, \ and\ \bibinfo {author} {\bibfnamefont {M.~R.}\
  \bibnamefont {Leadbetter}}} (\bibinfo {year} {2013}),\ \href@noop {} {\emph
  {\bibinfo {title} {Stationary and related stochastic processes: Sample
  function properties and their applications}}}\ (\bibinfo  {publisher} {Dover
  Publications},\ \bibinfo {address} {Mineola, New York})\BibitemShut {NoStop}%
\bibitem [{\citenamefont {Cuesta}\ \emph {et~al.}(2000)\citenamefont {Cuesta},
  \citenamefont {De~Figueiredo},\ and\ \citenamefont
  {Gossez}}]{cuesta2000nodal}%
  \BibitemOpen
  \bibfield  {author} {\bibinfo {author} {\bibnamefont {Cuesta}, \bibfnamefont
  {M.}}, \bibinfo {author} {\bibfnamefont {D.~G.}\ \bibnamefont
  {De~Figueiredo}}, \ and\ \bibinfo {author} {\bibfnamefont {J.-P.}\
  \bibnamefont {Gossez}}} (\bibinfo {year} {2000}),\ \href {\doibase
  10.1016/S0764-4442(00)00245-7} {\bibfield  {journal} {\bibinfo  {journal} {C.
  R. Acad. Sci. Paris S{\' e}r. I Math.}\ }\textbf {\bibinfo {volume}
  {330}}~(\bibinfo {number} {8}),\ \bibinfo {pages} {669}}\BibitemShut
  {NoStop}%
\bibitem [{\citenamefont {Davies}(1996)}]{davies1996spectral}%
  \BibitemOpen
  \bibfield  {author} {\bibinfo {author} {\bibnamefont {Davies}, \bibfnamefont
  {E.~B.}}} (\bibinfo {year} {1996}),\ \href@noop {} {\emph {\bibinfo {title}
  {Spectral theory and differential operators}}},\ Vol.~\bibinfo {volume} {42}\
  (\bibinfo  {publisher} {Cambridge University Press},\ \bibinfo {address}
  {Cambridge, UK})\BibitemShut {NoStop}%
\bibitem [{\citenamefont {De~Branges}(1985)}]{de1985proof}%
  \BibitemOpen
  \bibfield  {author} {\bibinfo {author} {\bibnamefont {De~Branges},
  \bibfnamefont {L.}}} (\bibinfo {year} {1985}),\ \href {\doibase
  10.1007/BF02392821} {\bibfield  {journal} {\bibinfo  {journal} {Acta Math.}\
  }\textbf {\bibinfo {volume} {154}}~(\bibinfo {number} {1}),\ \bibinfo {pages}
  {137}}\BibitemShut {NoStop}%
\bibitem [{\citenamefont {DeMarco}(2011)}]{demarco2011conformal}%
  \BibitemOpen
  \bibfield  {author} {\bibinfo {author} {\bibnamefont {DeMarco}, \bibfnamefont
  {L.}}} (\bibinfo {year} {2011}),\ \href {\doibase
  10.1090/S0273-0979-2010-01322-7} {\bibfield  {journal} {\bibinfo  {journal}
  {Bull. Amer. Math. Soc.}\ }\textbf {\bibinfo {volume} {48}}~(\bibinfo
  {number} {1}),\ \bibinfo {pages} {33}}\BibitemShut {NoStop}%
\bibitem [{\citenamefont {Dembowski}\ \emph {et~al.}(2003)\citenamefont
  {Dembowski}, \citenamefont {Dietz}, \citenamefont {Gr{\"a}f}, \citenamefont
  {Heine}, \citenamefont {Leyvraz}, \citenamefont {Miski-Oglu}, \citenamefont
  {Richter},\ and\ \citenamefont {Seligman}}]{dembowski2003phase}%
  \BibitemOpen
  \bibfield  {author} {\bibinfo {author} {\bibnamefont {Dembowski},
  \bibfnamefont {C.}}, \bibinfo {author} {\bibfnamefont {B.}~\bibnamefont
  {Dietz}}, \bibinfo {author} {\bibfnamefont {H.-D.}\ \bibnamefont {Gr{\"a}f}},
  \bibinfo {author} {\bibfnamefont {A.}~\bibnamefont {Heine}}, \bibinfo
  {author} {\bibfnamefont {F.}~\bibnamefont {Leyvraz}}, \bibinfo {author}
  {\bibfnamefont {M.}~\bibnamefont {Miski-Oglu}}, \bibinfo {author}
  {\bibfnamefont {A.}~\bibnamefont {Richter}}, \ and\ \bibinfo {author}
  {\bibfnamefont {T.~H.}\ \bibnamefont {Seligman}}} (\bibinfo {year} {2003}),\
  \href {\doibase 10.1103/PhysRevLett.90.014102} {\bibfield  {journal}
  {\bibinfo  {journal} {Phys. Rev. Lett.}\ }\textbf {\bibinfo {volume}
  {90}}~(\bibinfo {number} {1}),\ \bibinfo {pages} {014102}}\BibitemShut
  {NoStop}%
\bibitem [{\citenamefont {Dembowski}\ \emph {et~al.}(2000)\citenamefont
  {Dembowski}, \citenamefont {Gr{\"a}f}, \citenamefont {Heine}, \citenamefont
  {Hofferbert}, \citenamefont {Rehfeld},\ and\ \citenamefont
  {Richter}}]{dembowski2000first}%
  \BibitemOpen
  \bibfield  {author} {\bibinfo {author} {\bibnamefont {Dembowski},
  \bibfnamefont {C.}}, \bibinfo {author} {\bibfnamefont {H.-D.}\ \bibnamefont
  {Gr{\"a}f}}, \bibinfo {author} {\bibfnamefont {A.}~\bibnamefont {Heine}},
  \bibinfo {author} {\bibfnamefont {R.}~\bibnamefont {Hofferbert}}, \bibinfo
  {author} {\bibfnamefont {H.}~\bibnamefont {Rehfeld}}, \ and\ \bibinfo
  {author} {\bibfnamefont {A.}~\bibnamefont {Richter}}} (\bibinfo {year}
  {2000}),\ \href {\doibase 10.1103/PhysRevLett.84.867} {\bibfield  {journal}
  {\bibinfo  {journal} {Phys. Rev. Lett.}\ }\textbf {\bibinfo {volume}
  {84}}~(\bibinfo {number} {5}),\ \bibinfo {pages} {867}}\BibitemShut {NoStop}%
\bibitem [{\citenamefont {Dennis}(2001)}]{dennis2001phase}%
  \BibitemOpen
  \bibfield  {author} {\bibinfo {author} {\bibnamefont {Dennis}, \bibfnamefont
  {M.~R.}}} (\bibinfo {year} {2001}),\ \href {\doibase
  10.1088/0305-4470/34/20/102} {\bibfield  {journal} {\bibinfo  {journal} {J.
  Phys. A: Math. Gen.}\ }\textbf {\bibinfo {volume} {34}}~(\bibinfo {number}
  {20}),\ \bibinfo {pages} {L297}}\BibitemShut {NoStop}%
\bibitem [{\citenamefont {Dennis}(2003)}]{dennis2003correlations}%
  \BibitemOpen
  \bibfield  {author} {\bibinfo {author} {\bibnamefont {Dennis}, \bibfnamefont
  {M.~R.}}} (\bibinfo {year} {2003}),\ \href {\doibase
  10.1088/0305-4470/36/24/301} {\bibfield  {journal} {\bibinfo  {journal} {J.
  Phys. A: Math. Gen.}\ }\textbf {\bibinfo {volume} {36}}~(\bibinfo {number}
  {24}),\ \bibinfo {pages} {6611}}\BibitemShut {NoStop}%
\bibitem [{\citenamefont {Dennis}\ \emph {et~al.}(2009)\citenamefont {Dennis},
  \citenamefont {O'Holleran},\ and\ \citenamefont
  {Padgett}}]{dennis2009singular}%
  \BibitemOpen
  \bibfield  {author} {\bibinfo {author} {\bibnamefont {Dennis}, \bibfnamefont
  {M.~R.}}, \bibinfo {author} {\bibfnamefont {K.}~\bibnamefont {O'Holleran}}, \
  and\ \bibinfo {author} {\bibfnamefont {M.~J.}\ \bibnamefont {Padgett}}}
  (\bibinfo {year} {2009}),\ \href {\doibase 10.1016/S0079-6638(08)00205-9}
  {\bibfield  {journal} {\bibinfo  {journal} {Prog. Opt.}\ }\textbf {\bibinfo
  {volume} {53}},\ \bibinfo {pages} {293}}\BibitemShut {NoStop}%
\bibitem [{\citenamefont {Desjardins}(1998)}]{DESJARDINS1998257}%
  \BibitemOpen
  \bibfield  {author} {\bibinfo {author} {\bibnamefont {Desjardins},
  \bibfnamefont {S.}}} (\bibinfo {year} {1998}),\ \href {\doibase
  10.1016/S0926-2245(98)00011-4} {\bibfield  {journal} {\bibinfo  {journal}
  {Differential Geom. Appl.}\ }\textbf {\bibinfo {volume} {8}}~(\bibinfo
  {number} {3}),\ \bibinfo {pages} {257}}\BibitemShut {NoStop}%
\bibitem [{\citenamefont {Desjardins}\ and\ \citenamefont
  {Gilkey}(1994)}]{Desjardins1994}%
  \BibitemOpen
  \bibfield  {author} {\bibinfo {author} {\bibnamefont {Desjardins},
  \bibfnamefont {S.}}, \ and\ \bibinfo {author} {\bibfnamefont
  {P.}~\bibnamefont {Gilkey}}} (\bibinfo {year} {1994}),\ \href {\doibase
  10.1007/BF02571714} {\bibfield  {journal} {\bibinfo  {journal} {Mathematische
  Z.}\ }\textbf {\bibinfo {volume} {215}}~(\bibinfo {number} {1}),\ \bibinfo
  {pages} {251}}\BibitemShut {NoStop}%
\bibitem [{\citenamefont {Deus}\ \emph {et~al.}(1995)\citenamefont {Deus},
  \citenamefont {Koch},\ and\ \citenamefont {Sirko}}]{deus1995statistical}%
  \BibitemOpen
  \bibfield  {author} {\bibinfo {author} {\bibnamefont {Deus}, \bibfnamefont
  {S.}}, \bibinfo {author} {\bibfnamefont {P.~M.}\ \bibnamefont {Koch}}, \ and\
  \bibinfo {author} {\bibfnamefont {L.}~\bibnamefont {Sirko}}} (\bibinfo {year}
  {1995}),\ \href {\doibase 10.1103/PhysRevE.52.1146} {\bibfield  {journal}
  {\bibinfo  {journal} {Phys. Rev. E}\ }\textbf {\bibinfo {volume}
  {52}}~(\bibinfo {number} {1}),\ \bibinfo {pages} {1146}}\BibitemShut
  {NoStop}%
\bibitem [{\citenamefont {Dietz}\ \emph {et~al.}(2008)\citenamefont {Dietz},
  \citenamefont {Friedrich}, \citenamefont {Miski-Oglu}, \citenamefont
  {Richter},\ and\ \citenamefont {Sch{\"a}fer}}]{dietz2008properties}%
  \BibitemOpen
  \bibfield  {author} {\bibinfo {author} {\bibnamefont {Dietz}, \bibfnamefont
  {B.}}, \bibinfo {author} {\bibfnamefont {T.}~\bibnamefont {Friedrich}},
  \bibinfo {author} {\bibfnamefont {M.}~\bibnamefont {Miski-Oglu}}, \bibinfo
  {author} {\bibfnamefont {A.}~\bibnamefont {Richter}}, \ and\ \bibinfo
  {author} {\bibfnamefont {F.}~\bibnamefont {Sch{\"a}fer}}} (\bibinfo {year}
  {2008}),\ \href {\doibase 10.1103/PhysRevE.78.045201} {\bibfield  {journal}
  {\bibinfo  {journal} {Phys. Rev. E}\ }\textbf {\bibinfo {volume}
  {78}}~(\bibinfo {number} {4}),\ \bibinfo {pages} {045201}}\BibitemShut
  {NoStop}%
\bibitem [{\citenamefont {Dietz}\ \emph {et~al.}(2017)\citenamefont {Dietz},
  \citenamefont {Heusler}, \citenamefont {Maier}, \citenamefont {Richter},\
  and\ \citenamefont {Brown}}]{dietz2017chaos}%
  \BibitemOpen
  \bibfield  {author} {\bibinfo {author} {\bibnamefont {Dietz}, \bibfnamefont
  {B.}}, \bibinfo {author} {\bibfnamefont {A.}~\bibnamefont {Heusler}},
  \bibinfo {author} {\bibfnamefont {K.~H.}\ \bibnamefont {Maier}}, \bibinfo
  {author} {\bibfnamefont {A.}~\bibnamefont {Richter}}, \ and\ \bibinfo
  {author} {\bibfnamefont {B.~A.}\ \bibnamefont {Brown}}} (\bibinfo {year}
  {2017}),\ \href {\doibase 10.1103/PhysRevLett.118.012501} {\bibfield
  {journal} {\bibinfo  {journal} {Phys. Rev. Lett.}\ }\textbf {\bibinfo
  {volume} {118}}~(\bibinfo {number} {1}),\ \bibinfo {pages}
  {012501}}\BibitemShut {NoStop}%
\bibitem [{\citenamefont {Dietz}\ \emph {et~al.}(2015)\citenamefont {Dietz},
  \citenamefont {Richter},\ and\ \citenamefont {Samajdar}}]{dietz2015cross}%
  \BibitemOpen
  \bibfield  {author} {\bibinfo {author} {\bibnamefont {Dietz}, \bibfnamefont
  {B.}}, \bibinfo {author} {\bibfnamefont {A.}~\bibnamefont {Richter}}, \ and\
  \bibinfo {author} {\bibfnamefont {R.}~\bibnamefont {Samajdar}}} (\bibinfo
  {year} {2015}),\ \href {\doibase 10.1103/PhysRevE.92.022904} {\bibfield
  {journal} {\bibinfo  {journal} {Phys. Rev. E}\ }\textbf {\bibinfo {volume}
  {92}}~(\bibinfo {number} {2}),\ \bibinfo {pages} {022904}}\BibitemShut
  {NoStop}%
\bibitem [{\citenamefont {Dirac}(1931)}]{dirac1931}%
  \BibitemOpen
  \bibfield  {author} {\bibinfo {author} {\bibnamefont {Dirac}, \bibfnamefont
  {P.~A.~M.}}} (\bibinfo {year} {1931}),\ \href {\doibase
  10.1098/rspa.1931.0130} {\bibfield  {journal} {\bibinfo  {journal} {Proc. R.
  Soc. Lond. A}\ }\textbf {\bibinfo {volume} {133}},\ \bibinfo {pages}
  {60}}\BibitemShut {NoStop}%
\bibitem [{\citenamefont {Donnelly}(2011)}]{donnelly2011spectral}%
  \BibitemOpen
  \bibfield  {author} {\bibinfo {author} {\bibnamefont {Donnelly},
  \bibfnamefont {H.}}} (\bibinfo {year} {2011}),\ \href {\doibase
  10.1007/s00209-009-0629-1} {\bibfield  {journal} {\bibinfo  {journal} {Math.
  Z.}\ }\textbf {\bibinfo {volume} {269}}~(\bibinfo {number} {1}),\ \bibinfo
  {pages} {1}}\BibitemShut {NoStop}%
\bibitem [{\citenamefont {Donnelly}\ and\ \citenamefont
  {Fefferman}(1988)}]{donnelly1988nodal}%
  \BibitemOpen
  \bibfield  {author} {\bibinfo {author} {\bibnamefont {Donnelly},
  \bibfnamefont {H.}}, \ and\ \bibinfo {author} {\bibfnamefont
  {C.}~\bibnamefont {Fefferman}}} (\bibinfo {year} {1988}),\ \href {\doibase
  10.1007/BF01393691} {\bibfield  {journal} {\bibinfo  {journal} {Invent.
  Math.}\ }\textbf {\bibinfo {volume} {93}}~(\bibinfo {number} {1}),\ \bibinfo
  {pages} {161}}\BibitemShut {NoStop}%
\bibitem [{\citenamefont {Donnelly}\ and\ \citenamefont
  {Fefferman}(1990)}]{donnelly1990nodal}%
  \BibitemOpen
  \bibfield  {author} {\bibinfo {author} {\bibnamefont {Donnelly},
  \bibfnamefont {H.}}, \ and\ \bibinfo {author} {\bibfnamefont
  {C.}~\bibnamefont {Fefferman}}} (\bibinfo {year} {1990}),\ \enquote {\bibinfo
  {title} {Nodal sets of eigenfunctions: Riemannian manifolds with boundary},}\
  in\ \href {\doibase 10.1016/B978-0-12-574249-8.50017-1} {\emph {\bibinfo
  {booktitle} {Analysis, et Cetera}}}\ (\bibinfo  {publisher} {Academic
  Press},\ \bibinfo {address} {Boston, MA})\ pp.\ \bibinfo {pages}
  {251--262}\BibitemShut {NoStop}%
\bibitem [{\citenamefont {Doron}\ and\ \citenamefont
  {Smilansky}(1992{\natexlab{a}})}]{doron1992chaotic}%
  \BibitemOpen
  \bibfield  {author} {\bibinfo {author} {\bibnamefont {Doron}, \bibfnamefont
  {E.}}, \ and\ \bibinfo {author} {\bibfnamefont {U.}~\bibnamefont
  {Smilansky}}} (\bibinfo {year} {1992}{\natexlab{a}}),\ \href {\doibase
  10.1103/PhysRevLett.68.1255} {\bibfield  {journal} {\bibinfo  {journal}
  {Phys. Rev. Lett.}\ }\textbf {\bibinfo {volume} {68}}~(\bibinfo {number}
  {9}),\ \bibinfo {pages} {1255}}\BibitemShut {NoStop}%
\bibitem [{\citenamefont {Doron}\ and\ \citenamefont
  {Smilansky}(1992{\natexlab{b}})}]{doron1992semiclassical}%
  \BibitemOpen
  \bibfield  {author} {\bibinfo {author} {\bibnamefont {Doron}, \bibfnamefont
  {E.}}, \ and\ \bibinfo {author} {\bibfnamefont {U.}~\bibnamefont
  {Smilansky}}} (\bibinfo {year} {1992}{\natexlab{b}}),\ \href {\doibase
  10.1088/0951-7715/5/5/003} {\bibfield  {journal} {\bibinfo  {journal}
  {Nonlinearity}\ }\textbf {\bibinfo {volume} {5}}~(\bibinfo {number} {5}),\
  \bibinfo {pages} {1055}}\BibitemShut {NoStop}%
\bibitem [{\citenamefont {Doron}\ \emph {et~al.}(1990)\citenamefont {Doron},
  \citenamefont {Smilansky},\ and\ \citenamefont
  {Frenkel}}]{doron1990experimental}%
  \BibitemOpen
  \bibfield  {author} {\bibinfo {author} {\bibnamefont {Doron}, \bibfnamefont
  {E.}}, \bibinfo {author} {\bibfnamefont {U.}~\bibnamefont {Smilansky}}, \
  and\ \bibinfo {author} {\bibfnamefont {A.}~\bibnamefont {Frenkel}}} (\bibinfo
  {year} {1990}),\ \href {\doibase 10.1103/PhysRevLett.65.3072} {\bibfield
  {journal} {\bibinfo  {journal} {Phys. Rev. Lett.}\ }\textbf {\bibinfo
  {volume} {65}}~(\bibinfo {number} {25}),\ \bibinfo {pages}
  {3072}}\BibitemShut {NoStop}%
\bibitem [{\citenamefont {Dr{\'a}bek}\ and\ \citenamefont
  {Robinson}(2002)}]{drabek2002generalization}%
  \BibitemOpen
  \bibfield  {author} {\bibinfo {author} {\bibnamefont {Dr{\'a}bek},
  \bibfnamefont {P.}}, \ and\ \bibinfo {author} {\bibfnamefont {S.~B.}\
  \bibnamefont {Robinson}}} (\bibinfo {year} {2002}),\ \href {\doibase
  10.1006/jdeq.2001.4070} {\bibfield  {journal} {\bibinfo  {journal} {J.
  Differential Equations}\ }\textbf {\bibinfo {volume} {181}}~(\bibinfo
  {number} {1}),\ \bibinfo {pages} {58}}\BibitemShut {NoStop}%
\bibitem [{\citenamefont {Duffin}(1948)}]{duffin1948question}%
  \BibitemOpen
  \bibfield  {author} {\bibinfo {author} {\bibnamefont {Duffin}, \bibfnamefont
  {R.~J.}}} (\bibinfo {year} {1948}),\ \href {\doibase 10.1002/sapm1948271253}
  {\bibfield  {journal} {\bibinfo  {journal} {J. Math. \& Phys. [Stud. Appl.
  Math.]}\ }\textbf {\bibinfo {volume} {27}}~(\bibinfo {number} {1-4}),\
  \bibinfo {pages} {253}}\BibitemShut {NoStop}%
\bibitem [{\citenamefont {Duplantier}(2000)}]{duplantier2000conformally}%
  \BibitemOpen
  \bibfield  {author} {\bibinfo {author} {\bibnamefont {Duplantier},
  \bibfnamefont {B.}}} (\bibinfo {year} {2000}),\ \href {\doibase
  10.1103/PhysRevLett.84.1363} {\bibfield  {journal} {\bibinfo  {journal}
  {Phys. Rev. Lett.}\ }\textbf {\bibinfo {volume} {84}}~(\bibinfo {number}
  {7}),\ \bibinfo {pages} {1363}}\BibitemShut {NoStop}%
\bibitem [{\citenamefont {Durvasula}(1968)}]{durvasula1968natural}%
  \BibitemOpen
  \bibfield  {author} {\bibinfo {author} {\bibnamefont {Durvasula},
  \bibfnamefont {S.}}} (\bibinfo {year} {1968}),\ \href {\doibase
  10.1121/1.1911307} {\bibfield  {journal} {\bibinfo  {journal} {J. Acoust.
  Soc. Am}\ }\textbf {\bibinfo {volume} {44}}~(\bibinfo {number} {6}),\
  \bibinfo {pages} {1636}}\BibitemShut {NoStop}%
\bibitem [{\citenamefont {Eckhardt}\ \emph {et~al.}(1984)\citenamefont
  {Eckhardt}, \citenamefont {Ford},\ and\ \citenamefont
  {Vivaldi}}]{eckhardt1984analytically}%
  \BibitemOpen
  \bibfield  {author} {\bibinfo {author} {\bibnamefont {Eckhardt},
  \bibfnamefont {B.}}, \bibinfo {author} {\bibfnamefont {J.}~\bibnamefont
  {Ford}}, \ and\ \bibinfo {author} {\bibfnamefont {F.}~\bibnamefont
  {Vivaldi}}} (\bibinfo {year} {1984}),\ \href {\doibase
  10.1016/0167-2789(84)90135-0} {\bibfield  {journal} {\bibinfo  {journal}
  {Physica D}\ }\textbf {\bibinfo {volume} {13}}~(\bibinfo {number} {3}),\
  \bibinfo {pages} {339}}\BibitemShut {NoStop}%
\bibitem [{\citenamefont {Edmunds}\ and\ \citenamefont
  {Evans}(1987)}]{edmunds1987spectral}%
  \BibitemOpen
  \bibfield  {author} {\bibinfo {author} {\bibnamefont {Edmunds}, \bibfnamefont
  {D.~E.}}, \ and\ \bibinfo {author} {\bibfnamefont {W.~D.}\ \bibnamefont
  {Evans}}} (\bibinfo {year} {1987}),\ \href@noop {} {\emph {\bibinfo {title}
  {Spectral theory and differential operators}}},\ \bibinfo {series} {Oxford
  Mathematical Monographs}, Vol.~\bibinfo {volume} {15}\ (\bibinfo  {publisher}
  {Oxford University Press},\ \bibinfo {address} {Oxford, UK})\BibitemShut
  {NoStop}%
\bibitem [{\citenamefont {Ellegaard}\ \emph {et~al.}(1995)\citenamefont
  {Ellegaard}, \citenamefont {Guhr}, \citenamefont {Lindemann}, \citenamefont
  {Lorensen}, \citenamefont {Nyg{\aa}rd},\ and\ \citenamefont
  {Oxborrow}}]{ellegaard1995spectral}%
  \BibitemOpen
  \bibfield  {author} {\bibinfo {author} {\bibnamefont {Ellegaard},
  \bibfnamefont {C.}}, \bibinfo {author} {\bibfnamefont {T.}~\bibnamefont
  {Guhr}}, \bibinfo {author} {\bibfnamefont {K.}~\bibnamefont {Lindemann}},
  \bibinfo {author} {\bibfnamefont {H.~Q.}\ \bibnamefont {Lorensen}}, \bibinfo
  {author} {\bibfnamefont {J.}~\bibnamefont {Nyg{\aa}rd}}, \ and\ \bibinfo
  {author} {\bibfnamefont {M.}~\bibnamefont {Oxborrow}}} (\bibinfo {year}
  {1995}),\ \href {\doibase 10.1103/PhysRevLett.75.1546} {\bibfield  {journal}
  {\bibinfo  {journal} {Phys. Rev. Lett.}\ }\textbf {\bibinfo {volume}
  {75}}~(\bibinfo {number} {8}),\ \bibinfo {pages} {1546}}\BibitemShut
  {NoStop}%
\bibitem [{\citenamefont {Elon}\ \emph {et~al.}(2007)\citenamefont {Elon},
  \citenamefont {Gnutzmann}, \citenamefont {Joas},\ and\ \citenamefont
  {Smilansky}}]{elon2007geometric}%
  \BibitemOpen
  \bibfield  {author} {\bibinfo {author} {\bibnamefont {Elon}, \bibfnamefont
  {Y.}}, \bibinfo {author} {\bibfnamefont {S.}~\bibnamefont {Gnutzmann}},
  \bibinfo {author} {\bibfnamefont {C.}~\bibnamefont {Joas}}, \ and\ \bibinfo
  {author} {\bibfnamefont {U.}~\bibnamefont {Smilansky}}} (\bibinfo {year}
  {2007}),\ \href {\doibase 10.1088/1751-8113/40/11/007} {\bibfield  {journal}
  {\bibinfo  {journal} {J. Phys. A: Math. Theor.}\ }\textbf {\bibinfo {volume}
  {40}}~(\bibinfo {number} {11}),\ \bibinfo {pages} {2689}}\BibitemShut
  {NoStop}%
\bibitem [{\citenamefont {Faber}(1923)}]{faber1923beweis}%
  \BibitemOpen
  \bibfield  {author} {\bibinfo {author} {\bibnamefont {Faber}, \bibfnamefont
  {G.}}} (\bibinfo {year} {1923}),\ \href
  {http://publikationen.badw.de/003399311} {\bibfield  {journal} {\bibinfo
  {journal} {Sitz. Bayer. Akad. Wiss.}\ }\textbf {\bibinfo {volume} {8}},\
  \bibinfo {pages} {169}}\BibitemShut {NoStop}%
\bibitem [{\citenamefont {Falkovich}(2007)}]{falkovich2007nodal}%
  \BibitemOpen
  \bibfield  {author} {\bibinfo {author} {\bibnamefont {Falkovich},
  \bibfnamefont {G.}}} (\bibinfo {year} {2007}),\ \href {\doibase
  10.1140/epjst/e2007-00157-0} {\bibfield  {journal} {\bibinfo  {journal} {Eur.
  Phys. J. Spec. Top.}\ }\textbf {\bibinfo {volume} {145}}~(\bibinfo {number}
  {1}),\ \bibinfo {pages} {211}}\BibitemShut {NoStop}%
\bibitem [{\citenamefont {Faraday}(1859)}]{faraday1859experimental}%
  \BibitemOpen
  \bibfield  {author} {\bibinfo {author} {\bibnamefont {Faraday}, \bibfnamefont
  {M.}}} (\bibinfo {year} {1859}),\ \href@noop {} {\emph {\bibinfo {title}
  {Experimental researches in chemistry and physics}}}\ (\bibinfo  {publisher}
  {Taylor \& Francis},\ \bibinfo {address} {London})\BibitemShut {NoStop}%
\bibitem [{\citenamefont {Feder}(1988)}]{feder1988fractals}%
  \BibitemOpen
  \bibfield  {author} {\bibinfo {author} {\bibnamefont {Feder}, \bibfnamefont
  {J.}}} (\bibinfo {year} {1988}),\ \href@noop {} {\emph {\bibinfo {title}
  {Fractals}}},\ Physics of Solids and Liquids\ (\bibinfo  {publisher}
  {Springer},\ \bibinfo {address} {New York})\BibitemShut {NoStop}%
\bibitem [{\citenamefont {Fisher}(1966)}]{fisher1966hearing}%
  \BibitemOpen
  \bibfield  {author} {\bibinfo {author} {\bibnamefont {Fisher}, \bibfnamefont
  {M.~E.}}} (\bibinfo {year} {1966}),\ \href {\doibase
  10.1016/S0021-9800(66)80008-X} {\bibfield  {journal} {\bibinfo  {journal} {J.
  Combin. Theory}\ }\textbf {\bibinfo {volume} {1}}~(\bibinfo {number} {1}),\
  \bibinfo {pages} {105}}\BibitemShut {NoStop}%
\bibitem [{\citenamefont {Fisher}(1967)}]{fisher1967}%
  \BibitemOpen
  \bibfield  {author} {\bibinfo {author} {\bibnamefont {Fisher}, \bibfnamefont
  {M.~E.}}} (\bibinfo {year} {1967}),\ \href {\doibase
  10.1088/0034-4885/30/2/306} {\bibfield  {journal} {\bibinfo  {journal} {Rep.
  Prog. Phys.}\ }\textbf {\bibinfo {volume} {30}}~(\bibinfo {number} {2}),\
  \bibinfo {pages} {615}}\BibitemShut {NoStop}%
\bibitem [{\citenamefont {Fishman}\ \emph {et~al.}(1996)\citenamefont
  {Fishman}, \citenamefont {Georgeot},\ and\ \citenamefont
  {Prange}}]{fishman1996fredholm}%
  \BibitemOpen
  \bibfield  {author} {\bibinfo {author} {\bibnamefont {Fishman}, \bibfnamefont
  {S.}}, \bibinfo {author} {\bibfnamefont {B.}~\bibnamefont {Georgeot}}, \ and\
  \bibinfo {author} {\bibfnamefont {R.~E.}\ \bibnamefont {Prange}}} (\bibinfo
  {year} {1996}),\ \href {\doibase 10.1088/0305-4470/29/4/019} {\bibfield
  {journal} {\bibinfo  {journal} {J. Phys. A: Math. Gen.}\ }\textbf {\bibinfo
  {volume} {29}}~(\bibinfo {number} {4}),\ \bibinfo {pages} {919}}\BibitemShut
  {NoStop}%
\bibitem [{\citenamefont {Flores}(2007)}]{flores2007nodal}%
  \BibitemOpen
  \bibfield  {author} {\bibinfo {author} {\bibnamefont {Flores}, \bibfnamefont
  {J.}}} (\bibinfo {year} {2007}),\ \href {\doibase
  10.1140/epjst/e2007-00148-1} {\bibfield  {journal} {\bibinfo  {journal} {Eur.
  Phys. J. Spec. Top.}\ }\textbf {\bibinfo {volume} {145}}~(\bibinfo {number}
  {1}),\ \bibinfo {pages} {63}}\BibitemShut {NoStop}%
\bibitem [{\citenamefont {Fogedby}(2012)}]{fogedby2012stochastic}%
  \BibitemOpen
  \bibfield  {author} {\bibinfo {author} {\bibnamefont {Fogedby}, \bibfnamefont
  {H.~C.}}} (\bibinfo {year} {2012}),\ in\ \href@noop {} {\emph {\bibinfo
  {booktitle} {Computational Complexity}}},\ \bibinfo {editor} {edited by\
  \bibinfo {editor} {\bibfnamefont {R.~A.}\ \bibnamefont {Meyers}}}\ (\bibinfo
  {publisher} {Springer})\ pp.\ \bibinfo {pages} {3075--3096}\BibitemShut
  {NoStop}%
\bibitem [{\citenamefont {Foltin}(2003{\natexlab{a}})}]{foltin2003}%
  \BibitemOpen
  \bibfield  {author} {\bibinfo {author} {\bibnamefont {Foltin}, \bibfnamefont
  {G.}}} (\bibinfo {year} {2003}{\natexlab{a}}),\ \href
  {http://www.arxiv.org/abs/nlin/0302049} {\enquote {\bibinfo {title} {Counting
  nodal domains},}\ }\bibinfo {note} {{a}rXiv preprint
  nlin/0302049}\BibitemShut {NoStop}%
\bibitem [{\citenamefont
  {Foltin}(2003{\natexlab{b}})}]{foltin2003distribution}%
  \BibitemOpen
  \bibfield  {author} {\bibinfo {author} {\bibnamefont {Foltin}, \bibfnamefont
  {G.}}} (\bibinfo {year} {2003}{\natexlab{b}}),\ \href {\doibase
  10.1088/0305-4470/36/16/307} {\bibfield  {journal} {\bibinfo  {journal} {J.
  Phys. A: Math. Gen.}\ }\textbf {\bibinfo {volume} {36}}~(\bibinfo {number}
  {6}),\ \bibinfo {pages} {4561}}\BibitemShut {NoStop}%
\bibitem [{\citenamefont {Foltin}(2003{\natexlab{c}})}]{foltin2003signed}%
  \BibitemOpen
  \bibfield  {author} {\bibinfo {author} {\bibnamefont {Foltin}, \bibfnamefont
  {G.}}} (\bibinfo {year} {2003}{\natexlab{c}}),\ \href {\doibase
  10.1088/0305-4470/36/6/316} {\bibfield  {journal} {\bibinfo  {journal} {J.
  Phys. A: Math. Gen.}\ }\textbf {\bibinfo {volume} {36}}~(\bibinfo {number}
  {6}),\ \bibinfo {pages} {1729}}\BibitemShut {NoStop}%
\bibitem [{\citenamefont {Foltin}\ \emph
  {et~al.}(2004{\natexlab{a}})\citenamefont {Foltin}, \citenamefont
  {Gnutzmann},\ and\ \citenamefont {Smilansky}}]{foltin2004morphology}%
  \BibitemOpen
  \bibfield  {author} {\bibinfo {author} {\bibnamefont {Foltin}, \bibfnamefont
  {G.}}, \bibinfo {author} {\bibfnamefont {S.}~\bibnamefont {Gnutzmann}}, \
  and\ \bibinfo {author} {\bibfnamefont {U.}~\bibnamefont {Smilansky}}}
  (\bibinfo {year} {2004}{\natexlab{a}}),\ \href {\doibase
  10.1088/0305-4470/37/47/005} {\bibfield  {journal} {\bibinfo  {journal} {J.
  Phys. A: Math. Gen.}\ }\textbf {\bibinfo {volume} {37}}~(\bibinfo {number}
  {47}),\ \bibinfo {pages} {11363}}\BibitemShut {NoStop}%
\bibitem [{\citenamefont {Foltin}\ \emph
  {et~al.}(2004{\natexlab{b}})\citenamefont {Foltin}, \citenamefont
  {Gnutzmann},\ and\ \citenamefont {Smilansky}}]{foltin2004arxiv}%
  \BibitemOpen
  \bibfield  {author} {\bibinfo {author} {\bibnamefont {Foltin}, \bibfnamefont
  {G.}}, \bibinfo {author} {\bibfnamefont {S.}~\bibnamefont {Gnutzmann}}, \
  and\ \bibinfo {author} {\bibfnamefont {U.}~\bibnamefont {Smilansky}}}
  (\bibinfo {year} {2004}{\natexlab{b}}),\ \href
  {http://www.arxiv.org/abs/nlin/0407012} {\enquote {\bibinfo {title} {The
  morphology of nodal lines-random waves versus percolation},}\ }\bibinfo
  {note} {{a}rXiv preprint nlin/0407012}\BibitemShut {NoStop}%
\bibitem [{\citenamefont {Fournais}(2001)}]{fournais2001nodal}%
  \BibitemOpen
  \bibfield  {author} {\bibinfo {author} {\bibnamefont {Fournais},
  \bibfnamefont {S.}}} (\bibinfo {year} {2001}),\ \href {\doibase
  10.1006/jdeq.2000.3868} {\bibfield  {journal} {\bibinfo  {journal} {J.
  Differential Equations}\ }\textbf {\bibinfo {volume} {173}}~(\bibinfo
  {number} {1}),\ \bibinfo {pages} {145}}\BibitemShut {NoStop}%
\bibitem [{\citenamefont {Freitas}(2006)}]{freitas2006upper}%
  \BibitemOpen
  \bibfield  {author} {\bibinfo {author} {\bibnamefont {Freitas}, \bibfnamefont
  {P.}}} (\bibinfo {year} {2006}),\ \href {http://www.jstor.org/stable/4098239}
  {\bibinfo  {journal} {Proc. Amer. Math. Soc.}\ ,\ \bibinfo {pages}
  {2083}}\BibitemShut {NoStop}%
\bibitem [{\citenamefont {Freitas}(2007)}]{freitas2007precise}%
  \BibitemOpen
\bibfield  {journal} {  }\bibfield  {author} {\bibinfo {author} {\bibnamefont
  {Freitas}, \bibfnamefont {P.}}} (\bibinfo {year} {2007}),\ \href {\doibase
  10.1016/j.jfa.2007.04.012} {\bibfield  {journal} {\bibinfo  {journal} {J.
  Funct. Anal.}\ }\textbf {\bibinfo {volume} {251}}~(\bibinfo {number} {1}),\
  \bibinfo {pages} {376}}\BibitemShut {NoStop}%
\bibitem [{\citenamefont {Freitas}\ and\ \citenamefont {Krej{\v c}i{\v
  r}ik}(2008)}]{freitas2008sharp}%
  \BibitemOpen
  \bibfield  {author} {\bibinfo {author} {\bibnamefont {Freitas}, \bibfnamefont
  {P.}}, \ and\ \bibinfo {author} {\bibfnamefont {D.}~\bibnamefont {Krej{\v
  c}i{\v r}ik}}} (\bibinfo {year} {2008}),\ \href {\doibase
  10.1090/S0002-9939-08-09399-4} {\bibfield  {journal} {\bibinfo  {journal}
  {Proc. Amer. Math. Soc.}\ }\textbf {\bibinfo {volume} {136}}~(\bibinfo
  {number} {8}),\ \bibinfo {pages} {2997}}\BibitemShut {NoStop}%
\bibitem [{\citenamefont {Freund}(1994)}]{freund1994optical}%
  \BibitemOpen
  \bibfield  {author} {\bibinfo {author} {\bibnamefont {Freund}, \bibfnamefont
  {I.}}} (\bibinfo {year} {1994}),\ \href {\doibase 10.1364/JOSAA.11.001644}
  {\bibfield  {journal} {\bibinfo  {journal} {J. Opt. Soc. Am. A}\ }\textbf
  {\bibinfo {volume} {11}}~(\bibinfo {number} {5}),\ \bibinfo {pages}
  {1644}}\BibitemShut {NoStop}%
\bibitem [{\citenamefont {Freund}(1997)}]{freund1997critical}%
  \BibitemOpen
  \bibfield  {author} {\bibinfo {author} {\bibnamefont {Freund}, \bibfnamefont
  {I.}}} (\bibinfo {year} {1997}),\ \href {\doibase 10.1364/JOSAA.14.001911}
  {\bibfield  {journal} {\bibinfo  {journal} {J. Opt. Soc. Am. A}\ }\textbf
  {\bibinfo {volume} {14}}~(\bibinfo {number} {8}),\ \bibinfo {pages}
  {1911}}\BibitemShut {NoStop}%
\bibitem [{\citenamefont {Freund}(1999)}]{freund1999critical}%
  \BibitemOpen
  \bibfield  {author} {\bibinfo {author} {\bibnamefont {Freund}, \bibfnamefont
  {I.}}} (\bibinfo {year} {1999}),\ \href {\doibase
  10.1016/S0030-4018(98)00591-4} {\bibfield  {journal} {\bibinfo  {journal}
  {Opt. Commun.}\ }\textbf {\bibinfo {volume} {159}}~(\bibinfo {number} {1}),\
  \bibinfo {pages} {99}}\BibitemShut {NoStop}%
\bibitem [{\citenamefont {Freund}\ and\ \citenamefont
  {Freilikher}(1997)}]{freund1997parameterization}%
  \BibitemOpen
  \bibfield  {author} {\bibinfo {author} {\bibnamefont {Freund}, \bibfnamefont
  {I.}}, \ and\ \bibinfo {author} {\bibfnamefont {V.}~\bibnamefont
  {Freilikher}}} (\bibinfo {year} {1997}),\ \href {\doibase
  10.1364/JOSAA.14.001902} {\bibfield  {journal} {\bibinfo  {journal} {J. Opt.
  Soc. Am. A}\ }\textbf {\bibinfo {volume} {14}}~(\bibinfo {number} {8}),\
  \bibinfo {pages} {1902}}\BibitemShut {NoStop}%
\bibitem [{\citenamefont {Freund}\ and\ \citenamefont
  {Shvartsman}(1994)}]{freund1994wave}%
  \BibitemOpen
  \bibfield  {author} {\bibinfo {author} {\bibnamefont {Freund}, \bibfnamefont
  {I.}}, \ and\ \bibinfo {author} {\bibfnamefont {N.}~\bibnamefont
  {Shvartsman}}} (\bibinfo {year} {1994}),\ \href {\doibase
  10.1103/PhysRevA.50.5164} {\bibfield  {journal} {\bibinfo  {journal} {Phys.
  Rev. A}\ }\textbf {\bibinfo {volume} {50}}~(\bibinfo {number} {6}),\ \bibinfo
  {pages} {5164}}\BibitemShut {NoStop}%
\bibitem [{\citenamefont {Freund}\ \emph {et~al.}(1993)\citenamefont {Freund},
  \citenamefont {Shvartsman},\ and\ \citenamefont
  {Freilikher}}]{freund1993optical}%
  \BibitemOpen
  \bibfield  {author} {\bibinfo {author} {\bibnamefont {Freund}, \bibfnamefont
  {I.}}, \bibinfo {author} {\bibfnamefont {N.}~\bibnamefont {Shvartsman}}, \
  and\ \bibinfo {author} {\bibfnamefont {V.}~\bibnamefont {Freilikher}}}
  (\bibinfo {year} {1993}),\ \href {\doibase 10.1016/0030-4018(93)90375-F}
  {\bibfield  {journal} {\bibinfo  {journal} {Opt. Commun.}\ }\textbf {\bibinfo
  {volume} {101}}~(\bibinfo {number} {3-4}),\ \bibinfo {pages}
  {247}}\BibitemShut {NoStop}%
\bibitem [{\citenamefont {Freund}\ and\ \citenamefont
  {Wilkinson}(1998)}]{freund1998critical}%
  \BibitemOpen
  \bibfield  {author} {\bibinfo {author} {\bibnamefont {Freund}, \bibfnamefont
  {I.}}, \ and\ \bibinfo {author} {\bibfnamefont {M.}~\bibnamefont
  {Wilkinson}}} (\bibinfo {year} {1998}),\ \href {\doibase
  10.1364/JOSAA.15.002892} {\bibfield  {journal} {\bibinfo  {journal} {J. Opt.
  Soc. Am. A}\ }\textbf {\bibinfo {volume} {15}}~(\bibinfo {number} {11}),\
  \bibinfo {pages} {2892}}\BibitemShut {NoStop}%
\bibitem [{\citenamefont {Friedman}\ \emph {et~al.}(2001)\citenamefont
  {Friedman}, \citenamefont {Kaplan}, \citenamefont {Carasso},\ and\
  \citenamefont {Davidson}}]{friedman2001observation}%
  \BibitemOpen
  \bibfield  {author} {\bibinfo {author} {\bibnamefont {Friedman},
  \bibfnamefont {N.}}, \bibinfo {author} {\bibfnamefont {A.}~\bibnamefont
  {Kaplan}}, \bibinfo {author} {\bibfnamefont {D.}~\bibnamefont {Carasso}}, \
  and\ \bibinfo {author} {\bibfnamefont {N.}~\bibnamefont {Davidson}}}
  (\bibinfo {year} {2001}),\ \href {\doibase 10.1103/PhysRevLett.86.1518}
  {\bibfield  {journal} {\bibinfo  {journal} {Phys. Rev. Lett.}\ }\textbf
  {\bibinfo {volume} {86}}~(\bibinfo {number} {8}),\ \bibinfo {pages}
  {1518}}\BibitemShut {NoStop}%
\bibitem [{\citenamefont {Fyodorov}\ \emph {et~al.}(2005)\citenamefont
  {Fyodorov}, \citenamefont {Savin},\ and\ \citenamefont
  {Sommers}}]{fyodorov2005scattering}%
  \BibitemOpen
  \bibfield  {author} {\bibinfo {author} {\bibnamefont {Fyodorov},
  \bibfnamefont {{\relax Yu}.~V.}}, \bibinfo {author} {\bibfnamefont {D.~V.}\
  \bibnamefont {Savin}}, \ and\ \bibinfo {author} {\bibfnamefont {H.-J.}\
  \bibnamefont {Sommers}}} (\bibinfo {year} {2005}),\ \href {\doibase
  10.1088/0305-4470/38/49/017} {\bibfield  {journal} {\bibinfo  {journal} {J.
  Phys. A: Math. Gen.}\ }\textbf {\bibinfo {volume} {38}}~(\bibinfo {number}
  {49}),\ \bibinfo {pages} {10731}}\BibitemShut {NoStop}%
\bibitem [{\citenamefont {Garabedian}(1950)}]{garabedian1950partial}%
  \BibitemOpen
  \bibfield  {author} {\bibinfo {author} {\bibnamefont {Garabedian},
  \bibfnamefont {P.~R.}}} (\bibinfo {year} {1950}),\ \href {\doibase
  10.2140/pjm.1951.1.485} {\bibfield  {journal} {\bibinfo  {journal} {Pacific
  J. Math.}\ }\textbf {\bibinfo {volume} {1}}~(\bibinfo {number} {4}),\
  \bibinfo {pages} {485}}\BibitemShut {NoStop}%
\bibitem [{\citenamefont {Garc{\'e}s-Ch{\'a}vez}\ \emph
  {et~al.}(2003)\citenamefont {Garc{\'e}s-Ch{\'a}vez}, \citenamefont {McGloin},
  \citenamefont {Padgett}, \citenamefont {Dultz}, \citenamefont {Schmitzer},\
  and\ \citenamefont {Dholakia}}]{garces2003observation}%
  \BibitemOpen
  \bibfield  {author} {\bibinfo {author} {\bibnamefont {Garc{\'e}s-Ch{\'a}vez},
  \bibfnamefont {V.}}, \bibinfo {author} {\bibfnamefont {D.}~\bibnamefont
  {McGloin}}, \bibinfo {author} {\bibfnamefont {M.~J.}\ \bibnamefont
  {Padgett}}, \bibinfo {author} {\bibfnamefont {W.}~\bibnamefont {Dultz}},
  \bibinfo {author} {\bibfnamefont {H.}~\bibnamefont {Schmitzer}}, \ and\
  \bibinfo {author} {\bibfnamefont {K.}~\bibnamefont {Dholakia}}} (\bibinfo
  {year} {2003}),\ \href {\doibase 10.1103/PhysRevLett.91.093602} {\bibfield
  {journal} {\bibinfo  {journal} {Phys. Rev. Lett.}\ }\textbf {\bibinfo
  {volume} {91}}~(\bibinfo {number} {9}),\ \bibinfo {pages}
  {093602}}\BibitemShut {NoStop}%
\bibitem [{\citenamefont {Gayet}\ and\ \citenamefont
  {Welschinger}(2014{\natexlab{a}})}]{gayet2014betti}%
  \BibitemOpen
  \bibfield  {author} {\bibinfo {author} {\bibnamefont {Gayet}, \bibfnamefont
  {D.}}, \ and\ \bibinfo {author} {\bibfnamefont {J.-Y.}\ \bibnamefont
  {Welschinger}}} (\bibinfo {year} {2014}{\natexlab{a}}),\ \href
  {http://www.arxiv.org/abs/1406.0934} {\enquote {\bibinfo {title} {Betti
  numbers of random nodal sets of elliptic pseudo-differential operators},}\
  }\bibinfo {note} {{a}rXiv preprint 1406.0934}\BibitemShut {NoStop}%
\bibitem [{\citenamefont {Gayet}\ and\ \citenamefont
  {Welschinger}(2014{\natexlab{b}})}]{gayet2014lower}%
  \BibitemOpen
  \bibfield  {author} {\bibinfo {author} {\bibnamefont {Gayet}, \bibfnamefont
  {D.}}, \ and\ \bibinfo {author} {\bibfnamefont {J.-Y.}\ \bibnamefont
  {Welschinger}}} (\bibinfo {year} {2014}{\natexlab{b}}),\ \href {\doibase
  10.1112/jlms/jdu018} {\bibfield  {journal} {\bibinfo  {journal} {J. London
  Math. Soc.}\ }\textbf {\bibinfo {volume} {90}}~(\bibinfo {number} {1}),\
  \bibinfo {pages} {105}}\BibitemShut {NoStop}%
\bibitem [{\citenamefont {Gayet}\ and\ \citenamefont
  {Welschinger}(2015)}]{gayet2015expected}%
  \BibitemOpen
  \bibfield  {author} {\bibinfo {author} {\bibnamefont {Gayet}, \bibfnamefont
  {D.}}, \ and\ \bibinfo {author} {\bibfnamefont {J.-Y.}\ \bibnamefont
  {Welschinger}}} (\bibinfo {year} {2015}),\ \href {\doibase
  10.1017/S1474748014000115} {\bibfield  {journal} {\bibinfo  {journal} {J.
  Inst. Math. Jussieu}\ }\textbf {\bibinfo {volume} {14}}~(\bibinfo {number}
  {4}),\ \bibinfo {pages} {673}}\BibitemShut {NoStop}%
\bibitem [{\citenamefont {G{\'e}rard}\ and\ \citenamefont
  {Leichtnam}(1993)}]{gerard1993ergodic}%
  \BibitemOpen
  \bibfield  {author} {\bibinfo {author} {\bibnamefont {G{\'e}rard},
  \bibfnamefont {P.}}, \ and\ \bibinfo {author} {\bibfnamefont
  {E.}~\bibnamefont {Leichtnam}}} (\bibinfo {year} {1993}),\ \href {\doibase
  10.1215/S0012-7094-93-07122-0} {\bibfield  {journal} {\bibinfo  {journal}
  {Duke Math. J.}\ }\textbf {\bibinfo {volume} {71}}~(\bibinfo {number} {2}),\
  \bibinfo {pages} {559}}\BibitemShut {NoStop}%
\bibitem [{\citenamefont {Ghosh}\ \emph {et~al.}(2013)\citenamefont {Ghosh},
  \citenamefont {Reznikov},\ and\ \citenamefont {Sarnak}}]{ghosh2013nodal}%
  \BibitemOpen
  \bibfield  {author} {\bibinfo {author} {\bibnamefont {Ghosh}, \bibfnamefont
  {A.}}, \bibinfo {author} {\bibfnamefont {A.}~\bibnamefont {Reznikov}}, \ and\
  \bibinfo {author} {\bibfnamefont {P.}~\bibnamefont {Sarnak}}} (\bibinfo
  {year} {2013}),\ \href {\doibase 10.1007/s00039-013-0237-4} {\bibfield
  {journal} {\bibinfo  {journal} {Geom. Funct. Anal.}\ }\textbf {\bibinfo
  {volume} {23}}~(\bibinfo {number} {5}),\ \bibinfo {pages} {1515}}\BibitemShut
  {NoStop}%
\bibitem [{\citenamefont {Ghosh}\ \emph {et~al.}(2015)\citenamefont {Ghosh},
  \citenamefont {Reznikov},\ and\ \citenamefont {Sarnak}}]{ghosh2015nodal}%
  \BibitemOpen
  \bibfield  {author} {\bibinfo {author} {\bibnamefont {Ghosh}, \bibfnamefont
  {A.}}, \bibinfo {author} {\bibfnamefont {A.}~\bibnamefont {Reznikov}}, \ and\
  \bibinfo {author} {\bibfnamefont {P.}~\bibnamefont {Sarnak}}} (\bibinfo
  {year} {2015}),\ \href {http://www.arxiv.org/abs/1510.02963} {\enquote
  {\bibinfo {title} {Nodal domains of maass forms ii},}\ }\bibinfo {note}
  {{a}rXiv preprint 1510.02963}\BibitemShut {NoStop}%
\bibitem [{\citenamefont {Giraud}\ and\ \citenamefont
  {Thas}(2010)}]{giraud2010hearing}%
  \BibitemOpen
  \bibfield  {author} {\bibinfo {author} {\bibnamefont {Giraud}, \bibfnamefont
  {O.}}, \ and\ \bibinfo {author} {\bibfnamefont {K.}~\bibnamefont {Thas}}}
  (\bibinfo {year} {2010}),\ \href {\doibase 10.1103/RevModPhys.82.2213}
  {\bibfield  {journal} {\bibinfo  {journal} {Rev. Mod. Phys.}\ }\textbf
  {\bibinfo {volume} {82}}~(\bibinfo {number} {3}),\ \bibinfo {pages}
  {2213}}\BibitemShut {NoStop}%
\bibitem [{\citenamefont {Girouard}\ \emph {et~al.}(2009)\citenamefont
  {Girouard}, \citenamefont {Nadirashvili},\ and\ \citenamefont
  {Polterovich}}]{girouard2009maximization}%
  \BibitemOpen
  \bibfield  {author} {\bibinfo {author} {\bibnamefont {Girouard},
  \bibfnamefont {A.}}, \bibinfo {author} {\bibfnamefont {N.}~\bibnamefont
  {Nadirashvili}}, \ and\ \bibinfo {author} {\bibfnamefont {I.}~\bibnamefont
  {Polterovich}}} (\bibinfo {year} {2009}),\ \href {\doibase
  10.4310/jdg/1264601037} {\bibfield  {journal} {\bibinfo  {journal} {J.
  Differential Geom.}\ }\textbf {\bibinfo {volume} {83}}~(\bibinfo {number}
  {3}),\ \bibinfo {pages} {637}}\BibitemShut {NoStop}%
\bibitem [{\citenamefont {Glashow}\ and\ \citenamefont
  {Mittag}(1997)}]{glashow1997three}%
  \BibitemOpen
  \bibfield  {author} {\bibinfo {author} {\bibnamefont {Glashow}, \bibfnamefont
  {S.~L.}}, \ and\ \bibinfo {author} {\bibfnamefont {L.}~\bibnamefont
  {Mittag}}} (\bibinfo {year} {1997}),\ \href {\doibase 10.1007/BF02181254}
  {\bibfield  {journal} {\bibinfo  {journal} {J. Stat. Phys.}\ }\textbf
  {\bibinfo {volume} {87}}~(\bibinfo {number} {3-4}),\ \bibinfo {pages}
  {937}}\BibitemShut {NoStop}%
\bibitem [{\citenamefont {Gnutzmann}\ \emph {et~al.}(2006)\citenamefont
  {Gnutzmann}, \citenamefont {Karageorge},\ and\ \citenamefont
  {Smilansky}}]{gnutzmann2006can}%
  \BibitemOpen
  \bibfield  {author} {\bibinfo {author} {\bibnamefont {Gnutzmann},
  \bibfnamefont {S.}}, \bibinfo {author} {\bibfnamefont {P.~D.}\ \bibnamefont
  {Karageorge}}, \ and\ \bibinfo {author} {\bibfnamefont {U.}~\bibnamefont
  {Smilansky}}} (\bibinfo {year} {2006}),\ \href {\doibase
  10.1103/PhysRevLett.97.090201} {\bibfield  {journal} {\bibinfo  {journal}
  {Phys. Rev. Lett.}\ }\textbf {\bibinfo {volume} {97}}~(\bibinfo {number}
  {9}),\ \bibinfo {pages} {090201}}\BibitemShut {NoStop}%
\bibitem [{\citenamefont {Gnutzmann}\ \emph {et~al.}(2007)\citenamefont
  {Gnutzmann}, \citenamefont {Karageorge},\ and\ \citenamefont
  {Smilansky}}]{gnutzmann2007trace}%
  \BibitemOpen
  \bibfield  {author} {\bibinfo {author} {\bibnamefont {Gnutzmann},
  \bibfnamefont {S.}}, \bibinfo {author} {\bibfnamefont {P.~D.}\ \bibnamefont
  {Karageorge}}, \ and\ \bibinfo {author} {\bibfnamefont {U.}~\bibnamefont
  {Smilansky}}} (\bibinfo {year} {2007}),\ \href {\doibase
  10.1140/epjst/e2007-00158-y} {\bibfield  {journal} {\bibinfo  {journal} {Eur.
  Phys. J. Spec. Top.}\ }\textbf {\bibinfo {volume} {145}}~(\bibinfo {number}
  {1}),\ \bibinfo {pages} {217}}\BibitemShut {NoStop}%
\bibitem [{\citenamefont {Gnutzmann}\ and\ \citenamefont
  {Lois}(2014)}]{gnutzmann2014remarks}%
  \BibitemOpen
  \bibfield  {author} {\bibinfo {author} {\bibnamefont {Gnutzmann},
  \bibfnamefont {S.}}, \ and\ \bibinfo {author} {\bibfnamefont
  {S.}~\bibnamefont {Lois}}} (\bibinfo {year} {2014}),\ \href {\doibase
  10.1098/rsta.2012.0521} {\bibfield  {journal} {\bibinfo  {journal} {Phil.
  Trans. R. Soc. A}\ }\textbf {\bibinfo {volume} {372}}~(\bibinfo {number}
  {2007}),\ \bibinfo {pages} {20120521}}\BibitemShut {NoStop}%
\bibitem [{\citenamefont {Gnutzmann}\ \emph {et~al.}(2005)\citenamefont
  {Gnutzmann}, \citenamefont {Smilansky},\ and\ \citenamefont
  {Sondergaard}}]{gnutzmann2005resolving}%
  \BibitemOpen
  \bibfield  {author} {\bibinfo {author} {\bibnamefont {Gnutzmann},
  \bibfnamefont {S.}}, \bibinfo {author} {\bibfnamefont {U.}~\bibnamefont
  {Smilansky}}, \ and\ \bibinfo {author} {\bibfnamefont {N.}~\bibnamefont
  {Sondergaard}}} (\bibinfo {year} {2005}),\ \href {\doibase
  10.1088/0305-4470/38/41/006} {\bibfield  {journal} {\bibinfo  {journal} {J.
  Phys. A: Math. Gen.}\ }\textbf {\bibinfo {volume} {38}}~(\bibinfo {number}
  {41}),\ \bibinfo {pages} {8921}}\BibitemShut {NoStop}%
\bibitem [{\citenamefont {Gnutzmann}\ \emph {et~al.}(2004)\citenamefont
  {Gnutzmann}, \citenamefont {Smilansky},\ and\ \citenamefont
  {Weber}}]{gnutzmann2004nodal}%
  \BibitemOpen
  \bibfield  {author} {\bibinfo {author} {\bibnamefont {Gnutzmann},
  \bibfnamefont {S.}}, \bibinfo {author} {\bibfnamefont {U.}~\bibnamefont
  {Smilansky}}, \ and\ \bibinfo {author} {\bibfnamefont {J.}~\bibnamefont
  {Weber}}} (\bibinfo {year} {2004}),\ \href {\doibase
  10.1088/0959-7174/14/1/011} {\bibfield  {journal} {\bibinfo  {journal} {Wave.
  Random Media}\ }\textbf {\bibinfo {volume} {14}}~(\bibinfo {number} {1}),\
  \bibinfo {pages} {S61}}\BibitemShut {NoStop}%
\bibitem [{\citenamefont {Goldstein}\ \emph {et~al.}(2006)\citenamefont
  {Goldstein}, \citenamefont {Lebowitz}, \citenamefont {Tumulka},\ and\
  \citenamefont {Zanghi}}]{goldstein2006distribution}%
  \BibitemOpen
  \bibfield  {author} {\bibinfo {author} {\bibnamefont {Goldstein},
  \bibfnamefont {S.}}, \bibinfo {author} {\bibfnamefont {J.~L.}\ \bibnamefont
  {Lebowitz}}, \bibinfo {author} {\bibfnamefont {R.}~\bibnamefont {Tumulka}}, \
  and\ \bibinfo {author} {\bibfnamefont {N.}~\bibnamefont {Zanghi}}} (\bibinfo
  {year} {2006}),\ \href {\doibase 10.1007/s10955-006-9210-z} {\bibfield
  {journal} {\bibinfo  {journal} {J. Stat. Phys.}\ }\textbf {\bibinfo {volume}
  {125}}~(\bibinfo {number} {5-6}),\ \bibinfo {pages} {1193}}\BibitemShut
  {NoStop}%
\bibitem [{\citenamefont {Gomes}(2015)}]{gomes2015percival}%
  \BibitemOpen
  \bibfield  {author} {\bibinfo {author} {\bibnamefont {Gomes}, \bibfnamefont
  {S.}}} (\bibinfo {year} {2015}),\ \href {http://www.arxiv.org/abs/1504.07332}
  {\enquote {\bibinfo {title} {Percival's {C}onjecture for the {B}unimovich
  {M}ushroom {B}illiard},}\ }\bibinfo {note} {{a}rXiv preprint
  1504.07332}\BibitemShut {NoStop}%
\bibitem [{\citenamefont {Gong}(1999)}]{gong1999bieberbach}%
  \BibitemOpen
  \bibfield  {author} {\bibinfo {author} {\bibnamefont {Gong}, \bibfnamefont
  {S.}}} (\bibinfo {year} {1999}),\ \href@noop {} {\emph {\bibinfo {title} {The
  Bieberbach Conjecture}}},\ \bibinfo {series} {Studies in Advanced
  Mathematics}, Vol.~\bibinfo {volume} {12}\ (\bibinfo  {publisher} {American
  Mathematical Soc.},\ \bibinfo {address} {Providence, RI})\BibitemShut
  {NoStop}%
\bibitem [{\citenamefont {Goodman}(2015)}]{goodman2015}%
  \BibitemOpen
  \bibfield  {author} {\bibinfo {author} {\bibnamefont {Goodman}, \bibfnamefont
  {J.~W.}}} (\bibinfo {year} {2015}),\ \href@noop {} {\emph {\bibinfo {title}
  {Statistical Optics}}},\ \bibinfo {edition} {2nd}\ ed.\ (\bibinfo
  {publisher} {John Wiley and Sons},\ \bibinfo {address} {New
  York})\BibitemShut {NoStop}%
\bibitem [{\citenamefont {Gordon}\ and\ \citenamefont
  {Webb}(1994)}]{gordon1994isospectral}%
  \BibitemOpen
  \bibfield  {author} {\bibinfo {author} {\bibnamefont {Gordon}, \bibfnamefont
  {C.}}, \ and\ \bibinfo {author} {\bibfnamefont {D.}~\bibnamefont {Webb}}}
  (\bibinfo {year} {1994}),\ \href {\doibase 10.4310/mrl.1994.v1.n5.a2}
  {\bibfield  {journal} {\bibinfo  {journal} {Math. Res. Lett}\ }\textbf
  {\bibinfo {volume} {1}}~(\bibinfo {number} {5}),\ \bibinfo {pages}
  {539}}\BibitemShut {NoStop}%
\bibitem [{\citenamefont {Gordon}\ \emph
  {et~al.}(1992{\natexlab{a}})\citenamefont {Gordon}, \citenamefont {Webb},\
  and\ \citenamefont {Wolpert}}]{gordon1992isospectral}%
  \BibitemOpen
  \bibfield  {author} {\bibinfo {author} {\bibnamefont {Gordon}, \bibfnamefont
  {C.}}, \bibinfo {author} {\bibfnamefont {D.~L.}\ \bibnamefont {Webb}}, \ and\
  \bibinfo {author} {\bibfnamefont {S.}~\bibnamefont {Wolpert}}} (\bibinfo
  {year} {1992}{\natexlab{a}}),\ \href {\doibase 10.1007/bf01231320} {\bibfield
   {journal} {\bibinfo  {journal} {Invent. Math.}\ }\textbf {\bibinfo {volume}
  {110}}~(\bibinfo {number} {1}),\ \bibinfo {pages} {1}}\BibitemShut {NoStop}%
\bibitem [{\citenamefont {Gordon}\ \emph
  {et~al.}(1992{\natexlab{b}})\citenamefont {Gordon}, \citenamefont {Webb},\
  and\ \citenamefont {Wolpert}}]{gordon1992one}%
  \BibitemOpen
  \bibfield  {author} {\bibinfo {author} {\bibnamefont {Gordon}, \bibfnamefont
  {C.}}, \bibinfo {author} {\bibfnamefont {D.~L.}\ \bibnamefont {Webb}}, \ and\
  \bibinfo {author} {\bibfnamefont {S.}~\bibnamefont {Wolpert}}} (\bibinfo
  {year} {1992}{\natexlab{b}}),\ \href {\doibase
  10.1090/S0273-0979-1992-00289-6} {\bibfield  {journal} {\bibinfo  {journal}
  {Bull. Amer. Math. Soc.}\ }\textbf {\bibinfo {volume} {27}}~(\bibinfo
  {number} {1}),\ \bibinfo {pages} {134}}\BibitemShut {NoStop}%
\bibitem [{\citenamefont {Gornyi}\ and\ \citenamefont
  {Mirlin}(2002)}]{gornyi2002wave}%
  \BibitemOpen
  \bibfield  {author} {\bibinfo {author} {\bibnamefont {Gornyi}, \bibfnamefont
  {I.~V.}}, \ and\ \bibinfo {author} {\bibfnamefont {A.~D.}\ \bibnamefont
  {Mirlin}}} (\bibinfo {year} {2002}),\ \href {\doibase
  10.1016/S1386-9477(01)00436-2} {\bibfield  {journal} {\bibinfo  {journal}
  {Physica E}\ }\textbf {\bibinfo {volume} {12}}~(\bibinfo {number} {1}),\
  \bibinfo {pages} {845}}\BibitemShut {NoStop}%
\bibitem [{\citenamefont {Gough}(2007)}]{gough2007violin}%
  \BibitemOpen
  \bibfield  {author} {\bibinfo {author} {\bibnamefont {Gough}, \bibfnamefont
  {C.}}} (\bibinfo {year} {2007}),\ \href {\doibase
  10.1140/epjst/e2007-00149-0} {\bibfield  {journal} {\bibinfo  {journal} {Eur.
  Phys. J. Spec. Top.}\ }\textbf {\bibinfo {volume} {145}}~(\bibinfo {number}
  {1}),\ \bibinfo {pages} {77}}\BibitemShut {NoStop}%
\bibitem [{\citenamefont {Gr{\"a}f}\ \emph {et~al.}(1992)\citenamefont
  {Gr{\"a}f}, \citenamefont {Harney}, \citenamefont {Lengeler}, \citenamefont
  {Lewenkopf}, \citenamefont {Rangacharyulu}, \citenamefont {Richter},
  \citenamefont {Schardt},\ and\ \citenamefont
  {Weidenm{\"u}ller}}]{graf1992distribution}%
  \BibitemOpen
  \bibfield  {author} {\bibinfo {author} {\bibnamefont {Gr{\"a}f},
  \bibfnamefont {H.-D.}}, \bibinfo {author} {\bibfnamefont {H.~L.}\
  \bibnamefont {Harney}}, \bibinfo {author} {\bibfnamefont {H.}~\bibnamefont
  {Lengeler}}, \bibinfo {author} {\bibfnamefont {C.~H.}\ \bibnamefont
  {Lewenkopf}}, \bibinfo {author} {\bibfnamefont {C.}~\bibnamefont
  {Rangacharyulu}}, \bibinfo {author} {\bibfnamefont {A.}~\bibnamefont
  {Richter}}, \bibinfo {author} {\bibfnamefont {P.}~\bibnamefont {Schardt}}, \
  and\ \bibinfo {author} {\bibfnamefont {H.~A.}\ \bibnamefont
  {Weidenm{\"u}ller}}} (\bibinfo {year} {1992}),\ \href {\doibase
  10.1103/PhysRevLett.69.1296} {\bibfield  {journal} {\bibinfo  {journal}
  {Phys. Rev. Lett.}\ }\textbf {\bibinfo {volume} {69}}~(\bibinfo {number}
  {9}),\ \bibinfo {pages} {1296}}\BibitemShut {NoStop}%
\bibitem [{\citenamefont {Grandy~Jr.}(1987)}]{grandy1987}%
  \BibitemOpen
  \bibfield  {author} {\bibinfo {author} {\bibnamefont {Grandy~Jr.},
  \bibfnamefont {W.}}} (\bibinfo {year} {1987}),\ \href@noop {} {\emph
  {\bibinfo {title} {Foundations of Statistical Mechanics: Equilibrium
  Theory}}},\ Vol.~\bibinfo {volume} {1}\ (\bibinfo  {publisher} {D. Reidel
  Publishing Company},\ \bibinfo {address} {Dordrecht})\BibitemShut {NoStop}%
\bibitem [{\citenamefont {Grebenkov}\ and\ \citenamefont
  {Nguyen}(2013)}]{Grebenkov2013}%
  \BibitemOpen
  \bibfield  {author} {\bibinfo {author} {\bibnamefont {Grebenkov},
  \bibfnamefont {D.~S.}}, \ and\ \bibinfo {author} {\bibfnamefont {B.-T.}\
  \bibnamefont {Nguyen}}} (\bibinfo {year} {2013}),\ \href {\doibase
  10.1137/120880173} {\bibfield  {journal} {\bibinfo  {journal} {SIAM Rev.}\
  }\textbf {\bibinfo {volume} {55}}~(\bibinfo {number} {4}),\ \bibinfo {pages}
  {601}}\BibitemShut {NoStop}%
\bibitem [{\citenamefont {Gr{\'e}maud}\ and\ \citenamefont
  {Jain}(1998)}]{gremaud1998spacing}%
  \BibitemOpen
  \bibfield  {author} {\bibinfo {author} {\bibnamefont {Gr{\'e}maud},
  \bibfnamefont {B.}}, \ and\ \bibinfo {author} {\bibfnamefont {S.~R.}\
  \bibnamefont {Jain}}} (\bibinfo {year} {1998}),\ \href {\doibase
  10.1088/0305-4470/31/37/003} {\bibfield  {journal} {\bibinfo  {journal} {J.
  Phys. A: Math. Gen.}\ }\textbf {\bibinfo {volume} {31}}~(\bibinfo {number}
  {37}),\ \bibinfo {pages} {L637}}\BibitemShut {NoStop}%
\bibitem [{\citenamefont {Grieser}\ and\ \citenamefont
  {Jerison}(1996)}]{grieser1996asymptotics}%
  \BibitemOpen
  \bibfield  {author} {\bibinfo {author} {\bibnamefont {Grieser}, \bibfnamefont
  {D.}}, \ and\ \bibinfo {author} {\bibfnamefont {D.}~\bibnamefont {Jerison}}}
  (\bibinfo {year} {1996}),\ \href {\doibase 10.1007/s002220050073} {\bibfield
  {journal} {\bibinfo  {journal} {Invent. Math.}\ }\textbf {\bibinfo {volume}
  {125}}~(\bibinfo {number} {2}),\ \bibinfo {pages} {197}}\BibitemShut
  {NoStop}%
\bibitem [{\citenamefont {Grieser}\ and\ \citenamefont
  {Maronna}(2013)}]{grieser2013hearing}%
  \BibitemOpen
  \bibfield  {author} {\bibinfo {author} {\bibnamefont {Grieser}, \bibfnamefont
  {D.}}, \ and\ \bibinfo {author} {\bibfnamefont {S.}~\bibnamefont {Maronna}}}
  (\bibinfo {year} {2013}),\ \href {\doibase 10.1090/noti1063} {\bibfield
  {journal} {\bibinfo  {journal} {Notices Amer. Math. Soc.}\ }\textbf {\bibinfo
  {volume} {60}}~(\bibinfo {number} {11}),\ 10.1090/noti1063}\BibitemShut
  {NoStop}%
\bibitem [{\citenamefont {Gruzberg}\ and\ \citenamefont
  {Kadanoff}(2004)}]{gruzberg2004loewner}%
  \BibitemOpen
  \bibfield  {author} {\bibinfo {author} {\bibnamefont {Gruzberg},
  \bibfnamefont {I.~A.}}, \ and\ \bibinfo {author} {\bibfnamefont {L.~P.}\
  \bibnamefont {Kadanoff}}} (\bibinfo {year} {2004}),\ \href {\doibase
  10.1023/B:JOSS.0000013973.40984.3b} {\bibfield  {journal} {\bibinfo
  {journal} {J. Stat. Phys.}\ }\textbf {\bibinfo {volume} {114}}~(\bibinfo
  {number} {5-6}),\ \bibinfo {pages} {1183}}\BibitemShut {NoStop}%
\bibitem [{\citenamefont {Gutkin}\ and\ \citenamefont
  {Smilansky}(2001)}]{gutkin2001can}%
  \BibitemOpen
  \bibfield  {author} {\bibinfo {author} {\bibnamefont {Gutkin}, \bibfnamefont
  {B.}}, \ and\ \bibinfo {author} {\bibfnamefont {U.}~\bibnamefont
  {Smilansky}}} (\bibinfo {year} {2001}),\ \href {\doibase
  10.1088/0305-4470/34/31/301} {\bibfield  {journal} {\bibinfo  {journal} {J.
  Phys. A: Math. Gen.}\ }\textbf {\bibinfo {volume} {34}}~(\bibinfo {number}
  {31}),\ \bibinfo {pages} {6061}}\BibitemShut {NoStop}%
\bibitem [{\citenamefont {Gutzwiller}(2013)}]{gutzwiller1990chaos}%
  \BibitemOpen
  \bibfield  {author} {\bibinfo {author} {\bibnamefont {Gutzwiller},
  \bibfnamefont {M.~C.}}} (\bibinfo {year} {2013}),\ \href@noop {} {\emph
  {\bibinfo {title} {Chaos in Classical and Quantum Mechanics}}},\ \bibinfo
  {series} {Interdisciplinary Applied Mathematics}, Vol.~\bibinfo {volume} {1}\
  (\bibinfo  {publisher} {Springer},\ \bibinfo {address} {New
  York})\BibitemShut {NoStop}%
\bibitem [{\citenamefont {Haake}(2013)}]{haake2013quantum}%
  \BibitemOpen
  \bibfield  {author} {\bibinfo {author} {\bibnamefont {Haake}, \bibfnamefont
  {F.}}} (\bibinfo {year} {2013}),\ \href@noop {} {\emph {\bibinfo {title}
  {Quantum signatures of chaos}}},\ \bibinfo {edition} {2nd}\ ed.,\
  Vol.~\bibinfo {volume} {54}\ (\bibinfo  {publisher} {Springer},\ \bibinfo
  {address} {Berlin Heidelberg})\BibitemShut {NoStop}%
\bibitem [{\citenamefont {Hale}(2005)}]{hale2005eigenvalues}%
  \BibitemOpen
  \bibfield  {author} {\bibinfo {author} {\bibnamefont {Hale}, \bibfnamefont
  {J.~K.}}} (\bibinfo {year} {2005}),\ \enquote {\bibinfo {title} {Eigenvalues
  and perturbed domains},}\ in\ \href {\doibase
  10.1016/b978-044451861-3/50003-3} {\emph {\bibinfo {booktitle} {Ten
  Mathematical Essays on Approximation in Analysis and Topology}}},\ \bibinfo
  {editor} {edited by\ \bibinfo {editor} {\bibfnamefont {J.}~\bibnamefont
  {Ferrera}}, \bibinfo {editor} {\bibfnamefont {J.}~\bibnamefont
  {L{\'o}pez-G{\'o}mez}}, \ and\ \bibinfo {editor} {\bibfnamefont {F.~R.~R.}\
  \bibnamefont {del Portal}}}\ (\bibinfo  {publisher} {Elsevier},\ \bibinfo
  {address} {Amsterdam})\ pp.\ \bibinfo {pages} {95--123}\BibitemShut {NoStop}%
\bibitem [{\citenamefont {Halperin}(1981)}]{halperin1981}%
  \BibitemOpen
  \bibfield  {author} {\bibinfo {author} {\bibnamefont {Halperin},
  \bibfnamefont {B.~I.}}} (\bibinfo {year} {1981}),\ in\ \href@noop {} {\emph
  {\bibinfo {booktitle} {Physics of Defects, Les Houches lecture series}}},\
  Vol.~\bibinfo {volume} {35},\ \bibinfo {editor} {edited by\ \bibinfo {editor}
  {\bibfnamefont {R.}~\bibnamefont {Balian}}, \bibinfo {editor} {\bibfnamefont
  {M.}~\bibnamefont {Kl{\'e}man}}, \ and\ \bibinfo {editor} {\bibfnamefont
  {J.-P.}\ \bibnamefont {Poirier}}}\ (\bibinfo  {publisher} {Elsevier},\
  \bibinfo {address} {North-Holland, Amsterdam})\ pp.\ \bibinfo {pages}
  {814--857}\BibitemShut {NoStop}%
\bibitem [{\citenamefont {Hannay}\ and\ \citenamefont
  {Berry}(1980)}]{hannay1980quantization}%
  \BibitemOpen
  \bibfield  {author} {\bibinfo {author} {\bibnamefont {Hannay}, \bibfnamefont
  {J.~H.}}, \ and\ \bibinfo {author} {\bibfnamefont {M.~V.}\ \bibnamefont
  {Berry}}} (\bibinfo {year} {1980}),\ \href {\doibase
  10.1016/0167-2789(80)90026-3} {\bibfield  {journal} {\bibinfo  {journal}
  {Physica D}\ }\textbf {\bibinfo {volume} {1}}~(\bibinfo {number} {3}),\
  \bibinfo {pages} {267}}\BibitemShut {NoStop}%
\bibitem [{\citenamefont {Hansen}\ and\ \citenamefont
  {Nadirashvili}(1994)}]{Hansen1994}%
  \BibitemOpen
  \bibfield  {author} {\bibinfo {author} {\bibnamefont {Hansen}, \bibfnamefont
  {W.}}, \ and\ \bibinfo {author} {\bibfnamefont {N.}~\bibnamefont
  {Nadirashvili}}} (\bibinfo {year} {1994}),\ \enquote {\bibinfo {title}
  {Isoperimetric inequalities in potential theory},}\ in\ \href {\doibase
  10.1007/978-94-011-1118-8_1} {\emph {\bibinfo {booktitle} {ICPT '91:
  Proceedings from the International Conference on Potential Theory}}},\
  \bibinfo {editor} {edited by\ \bibinfo {editor} {\bibfnamefont
  {E.}~\bibnamefont {Bertin}}}\ (\bibinfo  {publisher} {Springer Netherlands},\
  \bibinfo {address} {Dordrecht})\ pp.\ \bibinfo {pages} {1--14}\BibitemShut
  {NoStop}%
\bibitem [{\citenamefont {Hardy}\ and\ \citenamefont
  {Wright}(1979)}]{hardy1979introduction}%
  \BibitemOpen
  \bibfield  {author} {\bibinfo {author} {\bibnamefont {Hardy}, \bibfnamefont
  {G.~H.}}, \ and\ \bibinfo {author} {\bibfnamefont {E.~M.}\ \bibnamefont
  {Wright}}} (\bibinfo {year} {1979}),\ \href@noop {} {\emph {\bibinfo {title}
  {An Introduction to the Theory of Numbers}}},\ \bibinfo {edition} {5th}\ ed.\
  (\bibinfo  {publisher} {Oxford University Press},\ \bibinfo {address}
  {London})\BibitemShut {NoStop}%
\bibitem [{\citenamefont {Harris}(1974)}]{harris1974effect}%
  \BibitemOpen
  \bibfield  {author} {\bibinfo {author} {\bibnamefont {Harris}, \bibfnamefont
  {A.~B.}}} (\bibinfo {year} {1974}),\ \href {\doibase
  10.1088/0022-3719/7/9/009} {\bibfield  {journal} {\bibinfo  {journal} {J.
  Phys. C: Solid State Phys.}\ }\textbf {\bibinfo {volume} {7}}~(\bibinfo
  {number} {9}),\ \bibinfo {pages} {1671}}\BibitemShut {NoStop}%
\bibitem [{\citenamefont {Hayman}(1978)}]{hayman1978some}%
  \BibitemOpen
  \bibfield  {author} {\bibinfo {author} {\bibnamefont {Hayman}, \bibfnamefont
  {W.~K.}}} (\bibinfo {year} {1978}),\ \href {\doibase
  10.1080/00036817808839195} {\bibfield  {journal} {\bibinfo  {journal} {Appl.
  Anal.}\ }\textbf {\bibinfo {volume} {7}}~(\bibinfo {number} {3}),\ \bibinfo
  {pages} {247}}\BibitemShut {NoStop}%
\bibitem [{\citenamefont {He}\ \emph {et~al.}(1995)\citenamefont {He},
  \citenamefont {Friese}, \citenamefont {Heckenberg},\ and\ \citenamefont
  {Rubinsztein-Dunlop}}]{he1995direct}%
  \BibitemOpen
  \bibfield  {author} {\bibinfo {author} {\bibnamefont {He}, \bibfnamefont
  {H.}}, \bibinfo {author} {\bibfnamefont {M.~E.~J.}\ \bibnamefont {Friese}},
  \bibinfo {author} {\bibfnamefont {N.~R.}\ \bibnamefont {Heckenberg}}, \ and\
  \bibinfo {author} {\bibfnamefont {H.}~\bibnamefont {Rubinsztein-Dunlop}}}
  (\bibinfo {year} {1995}),\ \href {\doibase 10.1103/PhysRevLett.75.826}
  {\bibfield  {journal} {\bibinfo  {journal} {Phys. Rev. Lett.}\ }\textbf
  {\bibinfo {volume} {75}}~(\bibinfo {number} {5}),\ \bibinfo {pages}
  {826}}\BibitemShut {NoStop}%
\bibitem [{\citenamefont {Helffer}\ \emph {et~al.}(1987)\citenamefont
  {Helffer}, \citenamefont {Martinez},\ and\ \citenamefont
  {Robert}}]{helffer1987ergodicite}%
  \BibitemOpen
  \bibfield  {author} {\bibinfo {author} {\bibnamefont {Helffer}, \bibfnamefont
  {B.}}, \bibinfo {author} {\bibfnamefont {A.}~\bibnamefont {Martinez}}, \ and\
  \bibinfo {author} {\bibfnamefont {D.}~\bibnamefont {Robert}}} (\bibinfo
  {year} {1987}),\ \href {\doibase 10.1007/bf01215225} {\bibfield  {journal}
  {\bibinfo  {journal} {Commun. Math. Phys.}\ }\textbf {\bibinfo {volume}
  {109}}~(\bibinfo {number} {2}),\ \bibinfo {pages} {313}}\BibitemShut
  {NoStop}%
\bibitem [{\citenamefont {Helffer}\ and\ \citenamefont
  {Sundqvist}(2015)}]{helffer2015nodal}%
  \BibitemOpen
  \bibfield  {author} {\bibinfo {author} {\bibnamefont {Helffer}, \bibfnamefont
  {B.}}, \ and\ \bibinfo {author} {\bibfnamefont {M.~P.}\ \bibnamefont
  {Sundqvist}}} (\bibinfo {year} {2015}),\ \href
  {http://mi.mathnet.ru/eng/mmj571} {\bibfield  {journal} {\bibinfo  {journal}
  {Mosc. Math. J}\ }\textbf {\bibinfo {volume} {15}}~(\bibinfo {number} {3}),\
  \bibinfo {pages} {455}}\BibitemShut {NoStop}%
\bibitem [{\citenamefont {Heller}(1984)}]{heller1984bound}%
  \BibitemOpen
  \bibfield  {author} {\bibinfo {author} {\bibnamefont {Heller}, \bibfnamefont
  {E.~J.}}} (\bibinfo {year} {1984}),\ \href {\doibase
  10.1103/PhysRevLett.53.1515} {\bibfield  {journal} {\bibinfo  {journal}
  {Phys. Rev. Lett.}\ }\textbf {\bibinfo {volume} {53}}~(\bibinfo {number}
  {16}),\ \bibinfo {pages} {1515}}\BibitemShut {NoStop}%
\bibitem [{\citenamefont {Heller}\ and\ \citenamefont
  {Landry}(2007)}]{heller2007statistical}%
  \BibitemOpen
  \bibfield  {author} {\bibinfo {author} {\bibnamefont {Heller}, \bibfnamefont
  {E.~J.}}, \ and\ \bibinfo {author} {\bibfnamefont {B.}~\bibnamefont
  {Landry}}} (\bibinfo {year} {2007}),\ \href {\doibase
  10.1140/epjst/e2007-00159-x} {\bibfield  {journal} {\bibinfo  {journal} {Eur.
  Phys. J. Spec. Top.}\ }\textbf {\bibinfo {volume} {145}}~(\bibinfo {number}
  {1}),\ \bibinfo {pages} {231}}\BibitemShut {NoStop}%
\bibitem [{\citenamefont {von Helmholtz}(1865)}]{von1865lehre}%
  \BibitemOpen
  \bibfield  {author} {\bibinfo {author} {\bibnamefont {von Helmholtz},
  \bibfnamefont {H.}}} (\bibinfo {year} {1865}),\ \href@noop {} {\emph
  {\bibinfo {title} {Die Lehre von den Tonempfindungen als physiologische
  Grundlage f{\"u}r die Theorie der Musik}}}\ (\bibinfo  {publisher} {Vieweg},\
  \bibinfo {address} {Braunschweig})\BibitemShut {NoStop}%
\bibitem [{\citenamefont {Hempel}\ \emph {et~al.}(1991)\citenamefont {Hempel},
  \citenamefont {Seco},\ and\ \citenamefont {Simon}}]{hempel1991essential}%
  \BibitemOpen
  \bibfield  {author} {\bibinfo {author} {\bibnamefont {Hempel}, \bibfnamefont
  {R.}}, \bibinfo {author} {\bibfnamefont {L.~A.}\ \bibnamefont {Seco}}, \ and\
  \bibinfo {author} {\bibfnamefont {B.}~\bibnamefont {Simon}}} (\bibinfo {year}
  {1991}),\ \href {\doibase 10.1016/0022-1236(91)90130-W} {\bibfield  {journal}
  {\bibinfo  {journal} {J. Funct. Anal.}\ }\textbf {\bibinfo {volume}
  {102}}~(\bibinfo {number} {2}),\ \bibinfo {pages} {448}}\BibitemShut
  {NoStop}%
\bibitem [{\citenamefont {Henkel}\ and\ \citenamefont
  {Karevski}(2012)}]{henkel2012conformal}%
  \BibitemOpen
  \bibinfo {editor} {\bibnamefont {Henkel}, \bibfnamefont {M.}}, \ and\
  \bibinfo {editor} {\bibfnamefont {D.}~\bibnamefont {Karevski}},\ Eds.
  (\bibinfo {year} {2012}),\ \href@noop {} {\emph {\bibinfo {title} {Conformal
  Invariance: An Introduction to Loops, Interfaces and Stochastic Loewner
  Evolution}}},\ \bibinfo {series} {Lecture Notes in Physics}, Vol.\ \bibinfo
  {volume} {853}\ (\bibinfo  {publisher} {Springer},\ \bibinfo {address}
  {Berlin Heidelberg})\BibitemShut {NoStop}%
\bibitem [{\citenamefont {Henrot}(2006)}]{henrot2006extremum}%
  \BibitemOpen
  \bibfield  {author} {\bibinfo {author} {\bibnamefont {Henrot}, \bibfnamefont
  {A.}}} (\bibinfo {year} {2006}),\ \href
  {http://link.springer.com/book/10.1007/3-7643-7706-2} {\emph {\bibinfo
  {title} {Extremum Problems for Eigenvalues of Elliptic Operators}}},\ Front.
  Math.\ (\bibinfo  {publisher} {Basel: Birkh{\"a}user Verlag})\BibitemShut
  {NoStop}%
\bibitem [{\citenamefont {Herrmann}(1932)}]{herrmann1932beitrage}%
  \BibitemOpen
  \bibfield  {author} {\bibinfo {author} {\bibnamefont {Herrmann},
  \bibfnamefont {H.}}} (\bibinfo {year} {1932}),\ \emph {\bibinfo {title}
  {Beitr{\"a}ge zur Theorie der Eigenwerte und Eigenfunktionen}},\ \href@noop
  {} {Ph.D. thesis}\ (\bibinfo  {school} {Georg-August-Universit{\" a}t G{\"
  o}ttingen})\BibitemShut {NoStop}%
\bibitem [{\citenamefont {Herrmann}(1936)}]{herrmann1936beziehungen}%
  \BibitemOpen
  \bibfield  {author} {\bibinfo {author} {\bibnamefont {Herrmann},
  \bibfnamefont {H.}}} (\bibinfo {year} {1936}),\ \href {\doibase
  10.1007/BF01218852} {\bibfield  {journal} {\bibinfo  {journal} {Math. Z.}\
  }\textbf {\bibinfo {volume} {40}}~(\bibinfo {number} {1}),\ \bibinfo {pages}
  {221}}\BibitemShut {NoStop}%
\bibitem [{\citenamefont {Hersch}(1960)}]{hersch1960frequence}%
  \BibitemOpen
  \bibfield  {author} {\bibinfo {author} {\bibnamefont {Hersch}, \bibfnamefont
  {J.}}} (\bibinfo {year} {1960}),\ \href {\doibase 10.1007/BF01604498}
  {\bibfield  {journal} {\bibinfo  {journal} {Z. Angew. Math. Phys.}\ }\textbf
  {\bibinfo {volume} {11}}~(\bibinfo {number} {5}),\ \bibinfo {pages}
  {387}}\BibitemShut {NoStop}%
\bibitem [{\citenamefont {Hile}\ and\ \citenamefont
  {Protter}(1980)}]{hile1980inequalities}%
  \BibitemOpen
  \bibfield  {author} {\bibinfo {author} {\bibnamefont {Hile}, \bibfnamefont
  {G.~N.}}, \ and\ \bibinfo {author} {\bibfnamefont {M.~H.}\ \bibnamefont
  {Protter}}} (\bibinfo {year} {1980}),\ \href {\doibase
  10.1512/iumj.1980.29.29040} {\bibfield  {journal} {\bibinfo  {journal}
  {Indiana Univ. Math. J.}\ }\textbf {\bibinfo {volume} {29}}~(\bibinfo
  {number} {4}),\ \bibinfo {pages} {523}}\BibitemShut {NoStop}%
\bibitem [{\citenamefont {Hlushchuk}\ \emph {et~al.}(2000)\citenamefont
  {Hlushchuk}, \citenamefont {Kohler}, \citenamefont {Bauch}, \citenamefont
  {Sirko}, \citenamefont {Bl{\"u}mel}, \citenamefont {Barth},\ and\
  \citenamefont {St{\"o}ckmann}}]{hlushchuk2000autocorrelation}%
  \BibitemOpen
  \bibfield  {author} {\bibinfo {author} {\bibnamefont {Hlushchuk},
  \bibfnamefont {Y.}}, \bibinfo {author} {\bibfnamefont {A.}~\bibnamefont
  {Kohler}}, \bibinfo {author} {\bibfnamefont {{\relax Sz}.}~\bibnamefont
  {Bauch}}, \bibinfo {author} {\bibfnamefont {L.}~\bibnamefont {Sirko}},
  \bibinfo {author} {\bibfnamefont {R.}~\bibnamefont {Bl{\"u}mel}}, \bibinfo
  {author} {\bibfnamefont {M.}~\bibnamefont {Barth}}, \ and\ \bibinfo {author}
  {\bibfnamefont {H.-J.}\ \bibnamefont {St{\"o}ckmann}}} (\bibinfo {year}
  {2000}),\ \href {\doibase 10.1103/PhysRevE.61.366} {\bibfield  {journal}
  {\bibinfo  {journal} {Phys. Rev. E}\ }\textbf {\bibinfo {volume}
  {61}}~(\bibinfo {number} {1}),\ \bibinfo {pages} {366}}\BibitemShut {NoStop}%
\bibitem [{\citenamefont {Hoffmann-Ostenhof}\ \emph {et~al.}(1997)\citenamefont
  {Hoffmann-Ostenhof}, \citenamefont {Hoffmann-Ostenhof},\ and\ \citenamefont
  {Nadirashvili}}]{hoffmann1997nodal}%
  \BibitemOpen
  \bibfield  {author} {\bibinfo {author} {\bibnamefont {Hoffmann-Ostenhof},
  \bibfnamefont {M.}}, \bibinfo {author} {\bibfnamefont {T.}~\bibnamefont
  {Hoffmann-Ostenhof}}, \ and\ \bibinfo {author} {\bibfnamefont
  {N.}~\bibnamefont {Nadirashvili}}} (\bibinfo {year} {1997}),\ \href {\doibase
  10.1215/S0012-7094-97-09017-7} {\bibfield  {journal} {\bibinfo  {journal}
  {Duke Math. J.}\ }\textbf {\bibinfo {volume} {90}}~(\bibinfo {number} {3}),\
  \bibinfo {pages} {631}}\BibitemShut {NoStop}%
\bibitem [{\citenamefont {Hoffmann-Ostenhof}\ \emph
  {et~al.}(1999{\natexlab{a}})\citenamefont {Hoffmann-Ostenhof}, \citenamefont
  {Hoffmann-Ostenhof},\ and\ \citenamefont
  {Nadirashvili}}]{hoffmann1999multiplicity}%
  \BibitemOpen
  \bibfield  {author} {\bibinfo {author} {\bibnamefont {Hoffmann-Ostenhof},
  \bibfnamefont {M.}}, \bibinfo {author} {\bibfnamefont {T.}~\bibnamefont
  {Hoffmann-Ostenhof}}, \ and\ \bibinfo {author} {\bibfnamefont
  {N.}~\bibnamefont {Nadirashvili}}} (\bibinfo {year} {1999}{\natexlab{a}}),\
  \href {\doibase 10.1023/A:1006595115793} {\bibfield  {journal} {\bibinfo
  {journal} {Ann. Global Anal. Geom.}\ }\textbf {\bibinfo {volume}
  {17}}~(\bibinfo {number} {1}),\ \bibinfo {pages} {43}}\BibitemShut {NoStop}%
\bibitem [{\citenamefont {Hoffmann-Ostenhof}\ \emph
  {et~al.}(1999{\natexlab{b}})\citenamefont {Hoffmann-Ostenhof}, \citenamefont
  {Michor},\ and\ \citenamefont {Nadirashvili}}]{hoffmann1999bounds}%
  \BibitemOpen
  \bibfield  {author} {\bibinfo {author} {\bibnamefont {Hoffmann-Ostenhof},
  \bibfnamefont {T.}}, \bibinfo {author} {\bibfnamefont {P.~W.}\ \bibnamefont
  {Michor}}, \ and\ \bibinfo {author} {\bibfnamefont {N.}~\bibnamefont
  {Nadirashvili}}} (\bibinfo {year} {1999}{\natexlab{b}}),\ \href {\doibase
  10.1007/s000390050111} {\bibfield  {journal} {\bibinfo  {journal} {Geom.
  Funct. Anal.}\ }\textbf {\bibinfo {volume} {9}}~(\bibinfo {number} {6}),\
  \bibinfo {pages} {1169}}\BibitemShut {NoStop}%
\bibitem [{\citenamefont {H{\"o}hmann}\ \emph {et~al.}(2009)\citenamefont
  {H{\"o}hmann}, \citenamefont {Kuhl}, \citenamefont {St{\"o}ckmann},
  \citenamefont {Urbina},\ and\ \citenamefont {Dennis}}]{hohmann2009density}%
  \BibitemOpen
  \bibfield  {author} {\bibinfo {author} {\bibnamefont {H{\"o}hmann},
  \bibfnamefont {R.}}, \bibinfo {author} {\bibfnamefont {U.}~\bibnamefont
  {Kuhl}}, \bibinfo {author} {\bibfnamefont {H.-J.}\ \bibnamefont
  {St{\"o}ckmann}}, \bibinfo {author} {\bibfnamefont {J.~D.}\ \bibnamefont
  {Urbina}}, \ and\ \bibinfo {author} {\bibfnamefont {M.~R.}\ \bibnamefont
  {Dennis}}} (\bibinfo {year} {2009}),\ \href {\doibase
  10.1103/PhysRevE.79.016203} {\bibfield  {journal} {\bibinfo  {journal} {Phys.
  Rev. E}\ }\textbf {\bibinfo {volume} {79}}~(\bibinfo {number} {1}),\ \bibinfo
  {pages} {016203}}\BibitemShut {NoStop}%
\bibitem [{\citenamefont {Holle}\ \emph {et~al.}(1986)\citenamefont {Holle},
  \citenamefont {Wiebusch}, \citenamefont {Main}, \citenamefont {Hager},
  \citenamefont {Rottke},\ and\ \citenamefont {Welge}}]{holle1986diamagnetism}%
  \BibitemOpen
  \bibfield  {author} {\bibinfo {author} {\bibnamefont {Holle}, \bibfnamefont
  {A.}}, \bibinfo {author} {\bibfnamefont {G.}~\bibnamefont {Wiebusch}},
  \bibinfo {author} {\bibfnamefont {J.}~\bibnamefont {Main}}, \bibinfo {author}
  {\bibfnamefont {B.}~\bibnamefont {Hager}}, \bibinfo {author} {\bibfnamefont
  {H.}~\bibnamefont {Rottke}}, \ and\ \bibinfo {author} {\bibfnamefont {K.~H.}\
  \bibnamefont {Welge}}} (\bibinfo {year} {1986}),\ \href {\doibase
  10.1103/PhysRevLett.56.2594} {\bibfield  {journal} {\bibinfo  {journal}
  {Phys. Rev. Lett.}\ }\textbf {\bibinfo {volume} {56}}~(\bibinfo {number}
  {24}),\ \bibinfo {pages} {2594}}\BibitemShut {NoStop}%
\bibitem [{\citenamefont {Hortikar}\ and\ \citenamefont
  {Srednicki}(1998)}]{hortikar1998}%
  \BibitemOpen
  \bibfield  {author} {\bibinfo {author} {\bibnamefont {Hortikar},
  \bibfnamefont {S.}}, \ and\ \bibinfo {author} {\bibfnamefont
  {M.}~\bibnamefont {Srednicki}}} (\bibinfo {year} {1998}),\ \href {\doibase
  10.1103/PhysRevLett.80.1646} {\bibfield  {journal} {\bibinfo  {journal}
  {Phys. Rev. Lett.}\ }\textbf {\bibinfo {volume} {80}}~(\bibinfo {number}
  {8}),\ \bibinfo {pages} {1646}}\BibitemShut {NoStop}%
\bibitem [{\citenamefont {Hoshen}\ and\ \citenamefont
  {Kopelman}(1976)}]{hoshen1976percolation}%
  \BibitemOpen
  \bibfield  {author} {\bibinfo {author} {\bibnamefont {Hoshen}, \bibfnamefont
  {J.}}, \ and\ \bibinfo {author} {\bibfnamefont {R.}~\bibnamefont {Kopelman}}}
  (\bibinfo {year} {1976}),\ \href {\doibase 10.1103/PhysRevB.14.3438}
  {\bibfield  {journal} {\bibinfo  {journal} {Phys. Rev. B}\ }\textbf {\bibinfo
  {volume} {14}}~(\bibinfo {number} {8}),\ \bibinfo {pages} {3438}}\BibitemShut
  {NoStop}%
\bibitem [{\citenamefont {Houston}\ \emph {et~al.}(2017)\citenamefont
  {Houston}, \citenamefont {Gradhand},\ and\ \citenamefont
  {Dennis}}]{houston2017random}%
  \BibitemOpen
  \bibfield  {author} {\bibinfo {author} {\bibnamefont {Houston}, \bibfnamefont
  {A.~J.~H.}}, \bibinfo {author} {\bibfnamefont {M.}~\bibnamefont {Gradhand}},
  \ and\ \bibinfo {author} {\bibfnamefont {M.~R.}\ \bibnamefont {Dennis}}}
  (\bibinfo {year} {2017}),\ \href {\doibase 10.1088/1751-8121/aa660f}
  {\bibfield  {journal} {\bibinfo  {journal} {J. Phys. A: Math. Theor.}\
  }\textbf {\bibinfo {volume} {50}}~(\bibinfo {number} {20}),\ \bibinfo {pages}
  {205101}}\BibitemShut {NoStop}%
\bibitem [{\citenamefont {Huang}\ \emph {et~al.}(2002)\citenamefont {Huang},
  \citenamefont {Chen}, \citenamefont {Lai},\ and\ \citenamefont
  {Lan}}]{huang2002observation}%
  \BibitemOpen
  \bibfield  {author} {\bibinfo {author} {\bibnamefont {Huang}, \bibfnamefont
  {K.~F.}}, \bibinfo {author} {\bibfnamefont {Y.~F.}\ \bibnamefont {Chen}},
  \bibinfo {author} {\bibfnamefont {H.~C.}\ \bibnamefont {Lai}}, \ and\
  \bibinfo {author} {\bibfnamefont {Y.~P.}\ \bibnamefont {Lan}}} (\bibinfo
  {year} {2002}),\ \href {\doibase 10.1103/PhysRevLett.89.224102} {\bibfield
  {journal} {\bibinfo  {journal} {Phys. Rev. Lett.}\ }\textbf {\bibinfo
  {volume} {89}}~(\bibinfo {number} {22}),\ \bibinfo {pages}
  {224102}}\BibitemShut {NoStop}%
\bibitem [{\citenamefont {Hul}\ \emph {et~al.}(2004)\citenamefont {Hul},
  \citenamefont {Bauch}, \citenamefont {Pako{\'n}ski}, \citenamefont
  {Savytskyy}, \citenamefont {{\.Z}yczkowski},\ and\ \citenamefont
  {Sirko}}]{hul2004experimental}%
  \BibitemOpen
  \bibfield  {author} {\bibinfo {author} {\bibnamefont {Hul}, \bibfnamefont
  {O.}}, \bibinfo {author} {\bibfnamefont {S.}~\bibnamefont {Bauch}}, \bibinfo
  {author} {\bibfnamefont {P.}~\bibnamefont {Pako{\'n}ski}}, \bibinfo {author}
  {\bibfnamefont {N.}~\bibnamefont {Savytskyy}}, \bibinfo {author}
  {\bibfnamefont {K.}~\bibnamefont {{\.Z}yczkowski}}, \ and\ \bibinfo {author}
  {\bibfnamefont {L.}~\bibnamefont {Sirko}}} (\bibinfo {year} {2004}),\
  \href@noop {} {\bibfield  {journal} {\bibinfo  {journal} {Phys. Rev. E}\
  }\textbf {\bibinfo {volume} {69}}~(\bibinfo {number} {5}),\ \bibinfo {pages}
  {056205}}\BibitemShut {NoStop}%
\bibitem [{\citenamefont {Hul}\ \emph {et~al.}(2012)\citenamefont {Hul},
  \citenamefont {{\L}awniczak}, \citenamefont {Bauch}, \citenamefont {Sawicki},
  \citenamefont {Ku{\'s}},\ and\ \citenamefont {Sirko}}]{hul2012scattering}%
  \BibitemOpen
  \bibfield  {author} {\bibinfo {author} {\bibnamefont {Hul}, \bibfnamefont
  {O.}}, \bibinfo {author} {\bibfnamefont {M.}~\bibnamefont {{\L}awniczak}},
  \bibinfo {author} {\bibfnamefont {S.}~\bibnamefont {Bauch}}, \bibinfo
  {author} {\bibfnamefont {A.}~\bibnamefont {Sawicki}}, \bibinfo {author}
  {\bibfnamefont {M.}~\bibnamefont {Ku{\'s}}}, \ and\ \bibinfo {author}
  {\bibfnamefont {L.}~\bibnamefont {Sirko}}} (\bibinfo {year} {2012}),\ \href
  {\doibase 10.1103/PhysRevLett.109.040402} {\bibfield  {journal} {\bibinfo
  {journal} {Phys. Rev. Lett.}\ }\textbf {\bibinfo {volume} {109}}~(\bibinfo
  {number} {4}),\ \bibinfo {pages} {040402}}\BibitemShut {NoStop}%
\bibitem [{\citenamefont {Hul}\ \emph {et~al.}(2005)\citenamefont {Hul},
  \citenamefont {Savytskyy}, \citenamefont {Tymoshchuk}, \citenamefont
  {Bauch},\ and\ \citenamefont {Sirko}}]{hul2005investigation}%
  \BibitemOpen
  \bibfield  {author} {\bibinfo {author} {\bibnamefont {Hul}, \bibfnamefont
  {O.}}, \bibinfo {author} {\bibfnamefont {N.}~\bibnamefont {Savytskyy}},
  \bibinfo {author} {\bibfnamefont {O.}~\bibnamefont {Tymoshchuk}}, \bibinfo
  {author} {\bibfnamefont {S.}~\bibnamefont {Bauch}}, \ and\ \bibinfo {author}
  {\bibfnamefont {L.}~\bibnamefont {Sirko}}} (\bibinfo {year} {2005}),\ \href
  {\doibase 10.1103/PhysRevE.72.066212} {\bibfield  {journal} {\bibinfo
  {journal} {Phys. Rev. E}\ }\textbf {\bibinfo {volume} {72}}~(\bibinfo
  {number} {6}),\ \bibinfo {pages} {066212}}\BibitemShut {NoStop}%
\bibitem [{\citenamefont {Ikeda}(1980)}]{ikeda1980lens}%
  \BibitemOpen
  \bibfield  {author} {\bibinfo {author} {\bibnamefont {Ikeda}, \bibfnamefont
  {A.}}} (\bibinfo {year} {1980}),\ \href
  {http://www.numdam.org/item?id=ASENS_1980_4_13_3_303_0} {\bibfield  {journal}
  {\bibinfo  {journal} {Ann. scient. {\'E}c. Norm. Sup.}\ }\textbf {\bibinfo
  {volume} {13}}~(\bibinfo {number} {3}),\ \bibinfo {pages} {303}}\BibitemShut
  {NoStop}%
\bibitem [{\citenamefont {Ishio}\ \emph {et~al.}(2001)\citenamefont {Ishio},
  \citenamefont {Saichev}, \citenamefont {Sadreev},\ and\ \citenamefont
  {Berggren}}]{ishio2001wave}%
  \BibitemOpen
  \bibfield  {author} {\bibinfo {author} {\bibnamefont {Ishio}, \bibfnamefont
  {H.}}, \bibinfo {author} {\bibfnamefont {A.~I.}\ \bibnamefont {Saichev}},
  \bibinfo {author} {\bibfnamefont {A.~F.}\ \bibnamefont {Sadreev}}, \ and\
  \bibinfo {author} {\bibfnamefont {K.-F.}\ \bibnamefont {Berggren}}} (\bibinfo
  {year} {2001}),\ \href {\doibase 10.1103/PhysRevE.64.056208} {\bibfield
  {journal} {\bibinfo  {journal} {Phys. Rev. E}\ }\textbf {\bibinfo {volume}
  {64}}~(\bibinfo {number} {5}),\ \bibinfo {pages} {056208}}\BibitemShut
  {NoStop}%
\bibitem [{\citenamefont {It{\^o}}(1944)}]{ito1944109}%
  \BibitemOpen
  \bibfield  {author} {\bibinfo {author} {\bibnamefont {It{\^o}}, \bibfnamefont
  {K.}}} (\bibinfo {year} {1944}),\ \href {\doibase 10.3792/pia/1195572786}
  {\bibfield  {journal} {\bibinfo  {journal} {Proc. Imp. Acad.}\ }\textbf
  {\bibinfo {volume} {20}}~(\bibinfo {number} {8}),\ \bibinfo {pages}
  {519}}\BibitemShut {NoStop}%
\bibitem [{\citenamefont {Itzykson}\ and\ \citenamefont
  {Luck}(1986)}]{itzykson1986arithmetical}%
  \BibitemOpen
  \bibfield  {author} {\bibinfo {author} {\bibnamefont {Itzykson},
  \bibfnamefont {C.}}, \ and\ \bibinfo {author} {\bibfnamefont
  {J.}~\bibnamefont {Luck}}} (\bibinfo {year} {1986}),\ \href {\doibase
  10.1088/0305-4470/19/2/017} {\bibfield  {journal} {\bibinfo  {journal} {J.
  Phys. A: Math. Gen.}\ }\textbf {\bibinfo {volume} {19}}~(\bibinfo {number}
  {2}),\ \bibinfo {pages} {211}}\BibitemShut {NoStop}%
\bibitem [{\citenamefont {Ivrii}(1980)}]{Ivrii1980}%
  \BibitemOpen
  \bibfield  {author} {\bibinfo {author} {\bibnamefont {Ivrii}, \bibfnamefont
  {V.~{\relax Ya.}.}}} (\bibinfo {year} {1980}),\ \href {\doibase
  10.1007/BF01086550} {\bibfield  {journal} {\bibinfo  {journal} {Funct. Anal.
  Appl.}\ }\textbf {\bibinfo {volume} {14}}~(\bibinfo {number} {2}),\ \bibinfo
  {pages} {98}}\BibitemShut {NoStop}%
\bibitem [{\citenamefont {Jackson}(1999)}]{jackson1999classical}%
  \BibitemOpen
  \bibfield  {author} {\bibinfo {author} {\bibnamefont {Jackson}, \bibfnamefont
  {J.~D.}}} (\bibinfo {year} {1999}),\ \href@noop {} {\emph {\bibinfo {title}
  {Classical electrodynamics}}},\ \bibinfo {edition} {3rd}\ ed.\ (\bibinfo
  {publisher} {John Wiley and Sons},\ \bibinfo {address} {New
  York})\BibitemShut {NoStop}%
\bibitem [{\citenamefont {Jain}(2009)}]{jain2009ground}%
  \BibitemOpen
  \bibfield  {author} {\bibinfo {author} {\bibnamefont {Jain}, \bibfnamefont
  {S.~R.}}} (\bibinfo {year} {2009}),\ \href {\doibase
  10.1007/s12043-009-0054-3} {\bibfield  {journal} {\bibinfo  {journal}
  {Pramana - J. Phys.}\ }\textbf {\bibinfo {volume} {72}}~(\bibinfo {number}
  {3}),\ \bibinfo {pages} {611}}\BibitemShut {NoStop}%
\bibitem [{\citenamefont {Jain}\ \emph {et~al.}(2002)\citenamefont {Jain},
  \citenamefont {Gr{\'e}maud},\ and\ \citenamefont {Khare}}]{jain2002quantum}%
  \BibitemOpen
  \bibfield  {author} {\bibinfo {author} {\bibnamefont {Jain}, \bibfnamefont
  {S.~R.}}, \bibinfo {author} {\bibfnamefont {B.}~\bibnamefont {Gr{\'e}maud}},
  \ and\ \bibinfo {author} {\bibfnamefont {A.}~\bibnamefont {Khare}}} (\bibinfo
  {year} {2002}),\ \href {\doibase 10.1103/PhysRevE.66.016216} {\bibfield
  {journal} {\bibinfo  {journal} {Phys. Rev. E}\ }\textbf {\bibinfo {volume}
  {66}}~(\bibinfo {number} {1}),\ \bibinfo {pages} {016216}}\BibitemShut
  {NoStop}%
\bibitem [{\citenamefont {Jakobson}\ \emph {et~al.}(2001)\citenamefont
  {Jakobson}, \citenamefont {Nadirashvili},\ and\ \citenamefont
  {Toth}}]{jakobson2001geometric}%
  \BibitemOpen
  \bibfield  {author} {\bibinfo {author} {\bibnamefont {Jakobson},
  \bibfnamefont {D.}}, \bibinfo {author} {\bibfnamefont {N.}~\bibnamefont
  {Nadirashvili}}, \ and\ \bibinfo {author} {\bibfnamefont {J.~A.}\
  \bibnamefont {Toth}}} (\bibinfo {year} {2001}),\ \href {\doibase
  10.1070/RM2001v056n06ABEH000453} {\bibfield  {journal} {\bibinfo  {journal}
  {Russ. Math. Surv.}\ }\textbf {\bibinfo {volume} {56}}~(\bibinfo {number}
  {6}),\ \bibinfo {pages} {1085}}\BibitemShut {NoStop}%
\bibitem [{\citenamefont {Jalabert}\ \emph {et~al.}(1994)\citenamefont
  {Jalabert}, \citenamefont {Pichard},\ and\ \citenamefont
  {Beenakker}}]{jalabert1994universal}%
  \BibitemOpen
  \bibfield  {author} {\bibinfo {author} {\bibnamefont {Jalabert},
  \bibfnamefont {R.~A.}}, \bibinfo {author} {\bibfnamefont {J.-L.}\
  \bibnamefont {Pichard}}, \ and\ \bibinfo {author} {\bibfnamefont {C.~W.~J.}\
  \bibnamefont {Beenakker}}} (\bibinfo {year} {1994}),\ \href {\doibase
  10.1209/0295-5075/27/4/001} {\bibfield  {journal} {\bibinfo  {journal} {EPL}\
  }\textbf {\bibinfo {volume} {27}}~(\bibinfo {number} {4}),\ \bibinfo {pages}
  {255}}\BibitemShut {NoStop}%
\bibitem [{\citenamefont {Jalabert}\ \emph {et~al.}(1992)\citenamefont
  {Jalabert}, \citenamefont {Stone},\ and\ \citenamefont
  {Alhassid}}]{jalabert1992statistical}%
  \BibitemOpen
  \bibfield  {author} {\bibinfo {author} {\bibnamefont {Jalabert},
  \bibfnamefont {R.~A.}}, \bibinfo {author} {\bibfnamefont {A.~D.}\
  \bibnamefont {Stone}}, \ and\ \bibinfo {author} {\bibfnamefont
  {Y.}~\bibnamefont {Alhassid}}} (\bibinfo {year} {1992}),\ \href {\doibase
  10.1103/PhysRevLett.68.3468} {\bibfield  {journal} {\bibinfo  {journal}
  {Phys. Rev. Lett.}\ }\textbf {\bibinfo {volume} {68}}~(\bibinfo {number}
  {23}),\ \bibinfo {pages} {3468}}\BibitemShut {NoStop}%
\bibitem [{\citenamefont {Jensen}(1955)}]{jensen1955production}%
  \BibitemOpen
  \bibfield  {author} {\bibinfo {author} {\bibnamefont {Jensen}, \bibfnamefont
  {H.~C.}}} (\bibinfo {year} {1955}),\ \href {\doibase 10.1119/1.1934080}
  {\bibfield  {journal} {\bibinfo  {journal} {Am. J. Phys.}\ }\textbf {\bibinfo
  {volume} {23}}~(\bibinfo {number} {8}),\ \bibinfo {pages} {503}}\BibitemShut
  {NoStop}%
\bibitem [{\citenamefont {Jerison}(1991)}]{jerison1991first}%
  \BibitemOpen
  \bibfield  {author} {\bibinfo {author} {\bibnamefont {Jerison}, \bibfnamefont
  {D.}}} (\bibinfo {year} {1991}),\ \href {\doibase 10.1155/S1073792891000016}
  {\bibfield  {journal} {\bibinfo  {journal} {Int. Math. Res. Notices}\
  }\textbf {\bibinfo {volume} {1991}}~(\bibinfo {number} {1}),\ \bibinfo
  {pages} {1}}\BibitemShut {NoStop}%
\bibitem [{\citenamefont {Jerison}(1995)}]{jerison1995diameter}%
  \BibitemOpen
  \bibfield  {author} {\bibinfo {author} {\bibnamefont {Jerison}, \bibfnamefont
  {D.}}} (\bibinfo {year} {1995}),\ \href {\doibase 10.2307/2118626} {\bibfield
   {journal} {\bibinfo  {journal} {Ann. Math.}\ }\textbf {\bibinfo {volume}
  {141}}~(\bibinfo {number} {1}),\ \bibinfo {pages} {1}}\BibitemShut {NoStop}%
\bibitem [{\citenamefont {Johansson}(2002)}]{johansson2002non}%
  \BibitemOpen
  \bibfield  {author} {\bibinfo {author} {\bibnamefont {Johansson},
  \bibfnamefont {K.}}} (\bibinfo {year} {2002}),\ \href {\doibase
  10.1007/s004400100187} {\bibfield  {journal} {\bibinfo  {journal} {Probab.
  Theory Relat. Fields}\ }\textbf {\bibinfo {volume} {123}}~(\bibinfo {number}
  {2}),\ \bibinfo {pages} {225}}\BibitemShut {NoStop}%
\bibitem [{\citenamefont {Jung}\ and\ \citenamefont
  {Zelditch}(2016)}]{jung2016number}%
  \BibitemOpen
  \bibfield  {author} {\bibinfo {author} {\bibnamefont {Jung}, \bibfnamefont
  {J.}}, \ and\ \bibinfo {author} {\bibfnamefont {S.}~\bibnamefont {Zelditch}}}
  (\bibinfo {year} {2016}),\ \href {\doibase 10.1007/s00208-015-1236-6}
  {\bibfield  {journal} {\bibinfo  {journal} {Math. Ann.}\ }\textbf {\bibinfo
  {volume} {364}}~(\bibinfo {number} {3-4}),\ \bibinfo {pages}
  {813}}\BibitemShut {NoStop}%
\bibitem [{\citenamefont {Kac}(1959)}]{kac1959probability}%
  \BibitemOpen
  \bibfield  {author} {\bibinfo {author} {\bibnamefont {Kac}, \bibfnamefont
  {M.}}} (\bibinfo {year} {1959}),\ \href@noop {} {\emph {\bibinfo {title}
  {Probability and related topics in physical sciences}}},\ Vol.~\bibinfo
  {volume} {1}\ (\bibinfo  {publisher} {American Mathematical Soc.},\ \bibinfo
  {address} {Providence, RI})\BibitemShut {NoStop}%
\bibitem [{\citenamefont {Kac}(1966)}]{kac1966can}%
  \BibitemOpen
  \bibfield  {author} {\bibinfo {author} {\bibnamefont {Kac}, \bibfnamefont
  {M.}}} (\bibinfo {year} {1966}),\ \href {\doibase 10.2307/2313748} {\bibfield
   {journal} {\bibinfo  {journal} {Amer. Math. Monthly}\ }\textbf {\bibinfo
  {volume} {73}}~(\bibinfo {number} {4}),\ \bibinfo {pages} {1}}\BibitemShut
  {NoStop}%
\bibitem [{\citenamefont {Kager}\ and\ \citenamefont
  {Nienhuis}(2004)}]{kager2004guide}%
  \BibitemOpen
  \bibfield  {author} {\bibinfo {author} {\bibnamefont {Kager}, \bibfnamefont
  {W.}}, \ and\ \bibinfo {author} {\bibfnamefont {B.}~\bibnamefont {Nienhuis}}}
  (\bibinfo {year} {2004}),\ \href {\doibase
  10.1023/B:JOSS.0000028058.87266.be} {\bibfield  {journal} {\bibinfo
  {journal} {J. Stat. Phys.}\ }\textbf {\bibinfo {volume} {115}}~(\bibinfo
  {number} {5--6}),\ \bibinfo {pages} {1149}}\BibitemShut {NoStop}%
\bibitem [{\citenamefont {Kager}\ \emph {et~al.}(2004)\citenamefont {Kager},
  \citenamefont {Nienhuis},\ and\ \citenamefont {Kadanoff}}]{kager2004exact}%
  \BibitemOpen
  \bibfield  {author} {\bibinfo {author} {\bibnamefont {Kager}, \bibfnamefont
  {W.}}, \bibinfo {author} {\bibfnamefont {B.}~\bibnamefont {Nienhuis}}, \ and\
  \bibinfo {author} {\bibfnamefont {L.~P.}\ \bibnamefont {Kadanoff}}} (\bibinfo
  {year} {2004}),\ \href {\doibase 10.1023/B:JOSS.0000022380.93241.24}
  {\bibfield  {journal} {\bibinfo  {journal} {J. Stat. Phys.}\ }\textbf
  {\bibinfo {volume} {115}}~(\bibinfo {number} {3-4}),\ \bibinfo {pages}
  {805}}\BibitemShut {NoStop}%
\bibitem [{\citenamefont {Kaplan}(1999)}]{kaplan1999scars}%
  \BibitemOpen
  \bibfield  {author} {\bibinfo {author} {\bibnamefont {Kaplan}, \bibfnamefont
  {L.}}} (\bibinfo {year} {1999}),\ \href {\doibase 10.1088/0951-7715/12/2/009}
  {\bibfield  {journal} {\bibinfo  {journal} {Nonlinearity}\ }\textbf {\bibinfo
  {volume} {12}}~(\bibinfo {number} {2}),\ \bibinfo {pages} {R1}}\BibitemShut
  {NoStop}%
\bibitem [{\citenamefont {Kaplan}\ and\ \citenamefont
  {Heller}(1998)}]{kaplan1998linear}%
  \BibitemOpen
  \bibfield  {author} {\bibinfo {author} {\bibnamefont {Kaplan}, \bibfnamefont
  {L.}}, \ and\ \bibinfo {author} {\bibfnamefont {E.~J.}\ \bibnamefont
  {Heller}}} (\bibinfo {year} {1998}),\ \href {\doibase 10.1006/aphy.1997.5773}
  {\bibfield  {journal} {\bibinfo  {journal} {Ann. Phys.}\ }\textbf {\bibinfo
  {volume} {264}}~(\bibinfo {number} {2}),\ \bibinfo {pages} {171}}\BibitemShut
  {NoStop}%
\bibitem [{\citenamefont {Kaplan}\ and\ \citenamefont
  {Heller}(1999)}]{kaplan1999measuring}%
  \BibitemOpen
  \bibfield  {author} {\bibinfo {author} {\bibnamefont {Kaplan}, \bibfnamefont
  {L.}}, \ and\ \bibinfo {author} {\bibfnamefont {E.~J.}\ \bibnamefont
  {Heller}}} (\bibinfo {year} {1999}),\ \href {\doibase
  10.1103/PhysRevE.59.6609} {\bibfield  {journal} {\bibinfo  {journal} {Phys.
  Rev. E}\ }\textbf {\bibinfo {volume} {59}}~(\bibinfo {number} {6}),\ \bibinfo
  {pages} {6609}}\BibitemShut {NoStop}%
\bibitem [{\citenamefont {Karageorge}\ and\ \citenamefont
  {Smilansky}(2008)}]{karageorge2008counting}%
  \BibitemOpen
  \bibfield  {author} {\bibinfo {author} {\bibnamefont {Karageorge},
  \bibfnamefont {P.~D.}}, \ and\ \bibinfo {author} {\bibfnamefont
  {U.}~\bibnamefont {Smilansky}}} (\bibinfo {year} {2008}),\ \href {\doibase
  10.1088/1751-8113/41/20/205102} {\bibfield  {journal} {\bibinfo  {journal}
  {J. Phys. A: Math. Theor.}\ }\textbf {\bibinfo {volume} {41}}~(\bibinfo
  {number} {20}),\ \bibinfo {pages} {205102}}\BibitemShut {NoStop}%
\bibitem [{\citenamefont {Karman}\ \emph {et~al.}(1997)\citenamefont {Karman},
  \citenamefont {Beijersbergen}, \citenamefont {Van~Duijl},\ and\ \citenamefont
  {Woerdman}}]{karman1997creation}%
  \BibitemOpen
  \bibfield  {author} {\bibinfo {author} {\bibnamefont {Karman}, \bibfnamefont
  {G.~P.}}, \bibinfo {author} {\bibfnamefont {M.~W.}\ \bibnamefont
  {Beijersbergen}}, \bibinfo {author} {\bibfnamefont {A.}~\bibnamefont
  {Van~Duijl}}, \ and\ \bibinfo {author} {\bibfnamefont {J.~P.}\ \bibnamefont
  {Woerdman}}} (\bibinfo {year} {1997}),\ \href {\doibase 10.1364/OL.22.001503}
  {\bibfield  {journal} {\bibinfo  {journal} {Opt. Lett.}\ }\textbf {\bibinfo
  {volume} {22}}~(\bibinfo {number} {19}),\ \bibinfo {pages}
  {1503}}\BibitemShut {NoStop}%
\bibitem [{\citenamefont {Keating}(1991)}]{keating1991cat}%
  \BibitemOpen
  \bibfield  {author} {\bibinfo {author} {\bibnamefont {Keating}, \bibfnamefont
  {J.~P.}}} (\bibinfo {year} {1991}),\ \href {\doibase
  10.1088/0951-7715/4/2/006} {\bibfield  {journal} {\bibinfo  {journal}
  {Nonlinearity}\ }\textbf {\bibinfo {volume} {4}}~(\bibinfo {number} {2}),\
  \bibinfo {pages} {309}}\BibitemShut {NoStop}%
\bibitem [{\citenamefont {Keating}\ \emph {et~al.}(2006)\citenamefont
  {Keating}, \citenamefont {Marklof},\ and\ \citenamefont
  {Williams}}]{keating2006nodal}%
  \BibitemOpen
  \bibfield  {author} {\bibinfo {author} {\bibnamefont {Keating}, \bibfnamefont
  {J.~P.}}, \bibinfo {author} {\bibfnamefont {J.}~\bibnamefont {Marklof}}, \
  and\ \bibinfo {author} {\bibfnamefont {I.~G.}\ \bibnamefont {Williams}}}
  (\bibinfo {year} {2006}),\ \href {\doibase 10.1103/PhysRevLett.97.034101}
  {\bibfield  {journal} {\bibinfo  {journal} {Phys. Rev. Lett.}\ }\textbf
  {\bibinfo {volume} {97}}~(\bibinfo {number} {3}),\ \bibinfo {pages}
  {034101}}\BibitemShut {NoStop}%
\bibitem [{\citenamefont {Keating}\ \emph {et~al.}(2008)\citenamefont
  {Keating}, \citenamefont {Marklof},\ and\ \citenamefont
  {Williams}}]{keating2008nodal}%
  \BibitemOpen
  \bibfield  {author} {\bibinfo {author} {\bibnamefont {Keating}, \bibfnamefont
  {J.~P.}}, \bibinfo {author} {\bibfnamefont {J.}~\bibnamefont {Marklof}}, \
  and\ \bibinfo {author} {\bibfnamefont {I.~G.}\ \bibnamefont {Williams}}}
  (\bibinfo {year} {2008}),\ \href {\doibase 10.1088/1367-2630/10/8/083023}
  {\bibfield  {journal} {\bibinfo  {journal} {New J. Phys.}\ }\textbf {\bibinfo
  {volume} {10}}~(\bibinfo {number} {8}),\ \bibinfo {pages}
  {083023}}\BibitemShut {NoStop}%
\bibitem [{\citenamefont {Keating}\ and\ \citenamefont
  {Prado}(2001)}]{keating2001orbit}%
  \BibitemOpen
  \bibfield  {author} {\bibinfo {author} {\bibnamefont {Keating}, \bibfnamefont
  {J.~P.}}, \ and\ \bibinfo {author} {\bibfnamefont {S.~D.}\ \bibnamefont
  {Prado}}} (\bibinfo {year} {2001}),\ \href {\doibase 10.1098/rspa.2001.0790}
  {\bibfield  {journal} {\bibinfo  {journal} {Proc. R. Soc. Lond. A}\ }\textbf
  {\bibinfo {volume} {457}}~(\bibinfo {number} {2012}),\ \bibinfo {pages}
  {1855}}\BibitemShut {NoStop}%
\bibitem [{\citenamefont {Kelvin}(1901)}]{kelvin1901nineteenth}%
  \BibitemOpen
  \bibfield  {author} {\bibinfo {author} {\bibnamefont {Kelvin}, \bibfnamefont
  {L.}}} (\bibinfo {year} {1901}),\ \href {\doibase 10.1080/14786440109462664}
  {\bibfield  {journal} {\bibinfo  {journal} {Phil. Mag. Ser. 6}\ }\textbf
  {\bibinfo {volume} {2}}~(\bibinfo {number} {7}),\ \bibinfo {pages}
  {1}}\BibitemShut {NoStop}%
\bibitem [{\citenamefont {Kennedy}(2011)}]{kennedy2011nodal}%
  \BibitemOpen
  \bibfield  {author} {\bibinfo {author} {\bibnamefont {Kennedy}, \bibfnamefont
  {J.~B.}}} (\bibinfo {year} {2011}),\ \href {\doibase
  10.1016/j.jde.2011.08.012} {\bibfield  {journal} {\bibinfo  {journal} {J.
  Differential Equations}\ }\textbf {\bibinfo {volume} {251}}~(\bibinfo
  {number} {12}),\ \bibinfo {pages} {3606}}\BibitemShut {NoStop}%
\bibitem [{\citenamefont {Kennedy}(2007)}]{kennedy2007fast}%
  \BibitemOpen
  \bibfield  {author} {\bibinfo {author} {\bibnamefont {Kennedy}, \bibfnamefont
  {T.}}} (\bibinfo {year} {2007}),\ \href {\doibase 10.1007/s10955-007-9358-1}
  {\bibfield  {journal} {\bibinfo  {journal} {J. Stat. Phys.}\ }\textbf
  {\bibinfo {volume} {128}}~(\bibinfo {number} {5}),\ \bibinfo {pages}
  {1125}}\BibitemShut {NoStop}%
\bibitem [{\citenamefont {Kenyon}(2000)}]{kenyon2000}%
  \BibitemOpen
  \bibfield  {author} {\bibinfo {author} {\bibnamefont {Kenyon}, \bibfnamefont
  {R.}}} (\bibinfo {year} {2000}),\ \href {\doibase 10.1007/BF02392811}
  {\bibfield  {journal} {\bibinfo  {journal} {Acta Math.}\ }\textbf {\bibinfo
  {volume} {185}},\ \bibinfo {pages} {239}}\BibitemShut {NoStop}%
\bibitem [{\citenamefont {Kereta}(2012)}]{kereta2012}%
  \BibitemOpen
  \bibfield  {author} {\bibinfo {author} {\bibnamefont {Kereta}, \bibfnamefont
  {Z.}}} (\bibinfo {year} {2012}),\ \emph {\bibinfo {title} {Numerical study
  into validity of the Bogomolny-Schmit conjecture}},\ \href@noop {} {Master's
  thesis}\ (\bibinfo  {school} {University of Oxford})\BibitemShut {NoStop}%
\bibitem [{\citenamefont {Kim}\ \emph {et~al.}(2003)\citenamefont {Kim},
  \citenamefont {Barth}, \citenamefont {Kuhl},\ and\ \citenamefont
  {St{\"o}ckmann}}]{kim2003current}%
  \BibitemOpen
  \bibfield  {author} {\bibinfo {author} {\bibnamefont {Kim}, \bibfnamefont
  {Y.-H.}}, \bibinfo {author} {\bibfnamefont {M.}~\bibnamefont {Barth}},
  \bibinfo {author} {\bibfnamefont {U.}~\bibnamefont {Kuhl}}, \ and\ \bibinfo
  {author} {\bibfnamefont {H.-J.}\ \bibnamefont {St{\"o}ckmann}}} (\bibinfo
  {year} {2003}),\ \href {\doibase 10.1143/PTPS.150.105} {\bibfield  {journal}
  {\bibinfo  {journal} {Prog. Theor. Phys. Supp.}\ }\textbf {\bibinfo {volume}
  {150}},\ \bibinfo {pages} {105}}\BibitemShut {NoStop}%
\bibitem [{\citenamefont {Klawonn}(2009)}]{klawonn2009inverse}%
  \BibitemOpen
  \bibfield  {author} {\bibinfo {author} {\bibnamefont {Klawonn}, \bibfnamefont
  {D.}}} (\bibinfo {year} {2009}),\ \href {\doibase
  10.1088/1751-8113/42/17/175209} {\bibfield  {journal} {\bibinfo  {journal}
  {J. Phys. A: Math. Theor.}\ }\textbf {\bibinfo {volume} {42}}~(\bibinfo
  {number} {17}),\ \bibinfo {pages} {175209}}\BibitemShut {NoStop}%
\bibitem [{\citenamefont {Klein}\ and\ \citenamefont
  {Agam}(2011)}]{klein2011critical}%
  \BibitemOpen
  \bibfield  {author} {\bibinfo {author} {\bibnamefont {Klein}, \bibfnamefont
  {A.}}, \ and\ \bibinfo {author} {\bibfnamefont {O.}~\bibnamefont {Agam}}}
  (\bibinfo {year} {2011}),\ \href {\doibase 10.1088/1751-8113/45/2/025001}
  {\bibfield  {journal} {\bibinfo  {journal} {J. Phys. A: Math. Theor.}\
  }\textbf {\bibinfo {volume} {45}}~(\bibinfo {number} {2}),\ \bibinfo {pages}
  {025001}}\BibitemShut {NoStop}%
\bibitem [{\citenamefont {Kollmann}\ \emph {et~al.}(1994)\citenamefont
  {Kollmann}, \citenamefont {Stein}, \citenamefont {Stoffregen}, \citenamefont
  {St{\"o}ckmann},\ and\ \citenamefont {Eckhardt}}]{kollmann1994periodic}%
  \BibitemOpen
  \bibfield  {author} {\bibinfo {author} {\bibnamefont {Kollmann},
  \bibfnamefont {M.}}, \bibinfo {author} {\bibfnamefont {J.}~\bibnamefont
  {Stein}}, \bibinfo {author} {\bibfnamefont {U.}~\bibnamefont {Stoffregen}},
  \bibinfo {author} {\bibfnamefont {H.-J.}\ \bibnamefont {St{\"o}ckmann}}, \
  and\ \bibinfo {author} {\bibfnamefont {B.}~\bibnamefont {Eckhardt}}}
  (\bibinfo {year} {1994}),\ \href {\doibase 10.1103/PhysRevE.49.R1} {\bibfield
   {journal} {\bibinfo  {journal} {Phys. Rev. E}\ }\textbf {\bibinfo {volume}
  {49}}~(\bibinfo {number} {1}),\ \bibinfo {pages} {R1}}\BibitemShut {NoStop}%
\bibitem [{\citenamefont {Kolmogorov}(1954)}]{kolmogorov1954preservation}%
  \BibitemOpen
  \bibfield  {author} {\bibinfo {author} {\bibnamefont {Kolmogorov},
  \bibfnamefont {A.~N.}}} (\bibinfo {year} {1954}),\ \href {\doibase
  10.1007/978-94-011-3030-1_52} {\bibfield  {journal} {\bibinfo  {journal}
  {Dokl. Akad. Nauk SSSR}\ }\textbf {\bibinfo {volume} {98}}~(\bibinfo {number}
  {4}),\ \bibinfo {pages} {527}}\BibitemShut {NoStop}%
\bibitem [{\citenamefont {Konrad}(2012)}]{konrad2012}%
  \BibitemOpen
  \bibfield  {author} {\bibinfo {author} {\bibnamefont {Konrad}, \bibfnamefont
  {K.}}} (\bibinfo {year} {2012}),\ \href@noop {} {\enquote {\bibinfo {title}
  {Asymptotic statistics of nodal domains of quantum chaotic billiards in the
  semiclassical limit},}\ }\bibinfo {howpublished} {Senior thesis (Dartmouth
  College)}\BibitemShut {NoStop}%
\bibitem [{\citenamefont {Kozlov}\ \emph {et~al.}(1990)\citenamefont {Kozlov},
  \citenamefont {Kondrat'ev},\ and\ \citenamefont {Maz'ya}}]{kozlov1990sign}%
  \BibitemOpen
  \bibfield  {author} {\bibinfo {author} {\bibnamefont {Kozlov}, \bibfnamefont
  {V.~A.}}, \bibinfo {author} {\bibfnamefont {V.~A.}\ \bibnamefont
  {Kondrat'ev}}, \ and\ \bibinfo {author} {\bibfnamefont {V.~G.}\ \bibnamefont
  {Maz'ya}}} (\bibinfo {year} {1990}),\ \href {\doibase
  10.1070/IM1990v034n02ABEH000649} {\bibfield  {journal} {\bibinfo  {journal}
  {Math. USSR Izvestiya}\ }\textbf {\bibinfo {volume} {34}}~(\bibinfo {number}
  {2}),\ \bibinfo {pages} {337}}\BibitemShut {NoStop}%
\bibitem [{\citenamefont {Krahn}(1925)}]{krahn1925rayleigh}%
  \BibitemOpen
  \bibfield  {author} {\bibinfo {author} {\bibnamefont {Krahn}, \bibfnamefont
  {E.}}} (\bibinfo {year} {1925}),\ \href {\doibase 10.1007/BF01208645}
  {\bibfield  {journal} {\bibinfo  {journal} {Math. Ann.}\ }\textbf {\bibinfo
  {volume} {94}}~(\bibinfo {number} {1}),\ \bibinfo {pages} {97}}\BibitemShut
  {NoStop}%
\bibitem [{\citenamefont {Krahn}(1926)}]{krahn1926minimaleigenschaften}%
  \BibitemOpen
  \bibfield  {author} {\bibinfo {author} {\bibnamefont {Krahn}, \bibfnamefont
  {E.}}} (\bibinfo {year} {1926}),\ \href@noop {} {\bibfield  {journal}
  {\bibinfo  {journal} {Acta Comm. Univ. Tartu (Dorpat)}\ }\textbf {\bibinfo
  {volume} {A9}},\ \bibinfo {pages} {1}}\BibitemShut {NoStop}%
\bibitem [{\citenamefont {Krishnamurthy}\ \emph {et~al.}(1982)\citenamefont
  {Krishnamurthy}, \citenamefont {Mani},\ and\ \citenamefont
  {Verma}}]{krishnamurthy1982exact}%
  \BibitemOpen
  \bibfield  {author} {\bibinfo {author} {\bibnamefont {Krishnamurthy},
  \bibfnamefont {H.~R.}}, \bibinfo {author} {\bibfnamefont {H.~S.}\
  \bibnamefont {Mani}}, \ and\ \bibinfo {author} {\bibfnamefont {H.~C.}\
  \bibnamefont {Verma}}} (\bibinfo {year} {1982}),\ \href {\doibase
  10.1088/0305-4470/15/7/024} {\bibfield  {journal} {\bibinfo  {journal} {J.
  Phys. A: Math. Gen.}\ }\textbf {\bibinfo {volume} {15}}~(\bibinfo {number}
  {7}),\ \bibinfo {pages} {2131}}\BibitemShut {NoStop}%
\bibitem [{\citenamefont {Krishnapur}\ \emph {et~al.}(2013)\citenamefont
  {Krishnapur}, \citenamefont {Kurlberg},\ and\ \citenamefont
  {Wigman}}]{krishnapur2013nodal}%
  \BibitemOpen
  \bibfield  {author} {\bibinfo {author} {\bibnamefont {Krishnapur},
  \bibfnamefont {M.}}, \bibinfo {author} {\bibfnamefont {P.}~\bibnamefont
  {Kurlberg}}, \ and\ \bibinfo {author} {\bibfnamefont {I.}~\bibnamefont
  {Wigman}}} (\bibinfo {year} {2013}),\ \href {\doibase
  10.4007/annals.2013.177.2.8} {\bibfield  {journal} {\bibinfo  {journal} {Ann.
  Math.}\ }\textbf {\bibinfo {volume} {177}}~(\bibinfo {number} {2}),\ \bibinfo
  {pages} {699}}\BibitemShut {NoStop}%
\bibitem [{\citenamefont {Kr{\"o}ger}(1992)}]{kroger1992upper}%
  \BibitemOpen
  \bibfield  {author} {\bibinfo {author} {\bibnamefont {Kr{\"o}ger},
  \bibfnamefont {P.}}} (\bibinfo {year} {1992}),\ \href {\doibase
  10.1016/0022-1236(92)90052-K} {\bibfield  {journal} {\bibinfo  {journal} {J.
  Funct. Anal.}\ }\textbf {\bibinfo {volume} {106}}~(\bibinfo {number} {2}),\
  \bibinfo {pages} {353}}\BibitemShut {NoStop}%
\bibitem [{\citenamefont {Krylov}(1979)}]{krylov1979works}%
  \BibitemOpen
  \bibfield  {author} {\bibinfo {author} {\bibnamefont {Krylov}, \bibfnamefont
  {N.~S.}}} (\bibinfo {year} {1979}),\ \href@noop {} {\emph {\bibinfo {title}
  {Works on the Foundations of Statistical Physics}}}\ (\bibinfo  {publisher}
  {Princeton University Press},\ \bibinfo {address} {Princeton,
  NJ})\BibitemShut {NoStop}%
\bibitem [{\citenamefont {Kuchment}(2008)}]{kuchment2008quantum}%
  \BibitemOpen
  \bibfield  {author} {\bibinfo {author} {\bibnamefont {Kuchment},
  \bibfnamefont {P.}}} (\bibinfo {year} {2008}),\ in\ \href {\doibase
  10.1090/pspum/077/2459876} {\emph {\bibinfo {booktitle} {Analysis on Graphs
  and Its Applications}}},\ \bibinfo {series} {Proc. Symp. Pure Math.},
  Vol.~\bibinfo {volume} {77},\ \bibinfo {editor} {edited by\ \bibinfo {editor}
  {\bibfnamefont {P.}~\bibnamefont {Exner}}, \bibinfo {editor} {\bibfnamefont
  {J.~P.}\ \bibnamefont {Keating}}, \bibinfo {editor} {\bibfnamefont
  {P.}~\bibnamefont {Kuchment}}, \bibinfo {editor} {\bibfnamefont
  {T.}~\bibnamefont {Sunada}}, \ and\ \bibinfo {editor} {\bibfnamefont
  {A.}~\bibnamefont {Teplyaev}}}\ (\bibinfo  {publisher} {American Mathematical
  Soc.},\ \bibinfo {address} {Providence, RI})\ Chap.~\bibinfo {chapter} {3},\
  pp.\ \bibinfo {pages} {291--314}\BibitemShut {NoStop}%
\bibitem [{\citenamefont {Kudrolli}\ \emph {et~al.}(1995)\citenamefont
  {Kudrolli}, \citenamefont {Kidambi},\ and\ \citenamefont
  {Sridhar}}]{kudrolli1995experimental}%
  \BibitemOpen
  \bibfield  {author} {\bibinfo {author} {\bibnamefont {Kudrolli},
  \bibfnamefont {A.}}, \bibinfo {author} {\bibfnamefont {V.}~\bibnamefont
  {Kidambi}}, \ and\ \bibinfo {author} {\bibfnamefont {S.}~\bibnamefont
  {Sridhar}}} (\bibinfo {year} {1995}),\ \href {\doibase
  10.1103/PhysRevLett.75.822} {\bibfield  {journal} {\bibinfo  {journal} {Phys.
  Rev. Lett.}\ }\textbf {\bibinfo {volume} {75}}~(\bibinfo {number} {5}),\
  \bibinfo {pages} {822}}\BibitemShut {NoStop}%
\bibitem [{\citenamefont {Kudrolli}\ \emph {et~al.}(1994)\citenamefont
  {Kudrolli}, \citenamefont {Sridhar}, \citenamefont {Pandey},\ and\
  \citenamefont {Ramaswamy}}]{kudrolli1994signatures}%
  \BibitemOpen
  \bibfield  {author} {\bibinfo {author} {\bibnamefont {Kudrolli},
  \bibfnamefont {A.}}, \bibinfo {author} {\bibfnamefont {S.}~\bibnamefont
  {Sridhar}}, \bibinfo {author} {\bibfnamefont {A.}~\bibnamefont {Pandey}}, \
  and\ \bibinfo {author} {\bibfnamefont {R.}~\bibnamefont {Ramaswamy}}}
  (\bibinfo {year} {1994}),\ \href {\doibase 10.1103/PhysRevE.49.R11}
  {\bibfield  {journal} {\bibinfo  {journal} {Phys. Rev. E}\ }\textbf {\bibinfo
  {volume} {49}}~(\bibinfo {number} {1}),\ \bibinfo {pages} {R11}}\BibitemShut
  {NoStop}%
\bibitem [{\citenamefont {Kuhl}(2007)}]{kuhl2007wave}%
  \BibitemOpen
  \bibfield  {author} {\bibinfo {author} {\bibnamefont {Kuhl}, \bibfnamefont
  {U.}}} (\bibinfo {year} {2007}),\ \href {\doibase
  10.1140/epjst/e2007-00150-7} {\bibfield  {journal} {\bibinfo  {journal} {Eur.
  Phys. J. Spec. Top.}\ }\textbf {\bibinfo {volume} {145}}~(\bibinfo {number}
  {1}),\ \bibinfo {pages} {103}}\BibitemShut {NoStop}%
\bibitem [{\citenamefont {Kuhl}\ \emph {et~al.}(2007)\citenamefont {Kuhl},
  \citenamefont {H{\"o}hmann}, \citenamefont {St{\"o}ckmann},\ and\
  \citenamefont {Gnutzmann}}]{kuhl2007nodal}%
  \BibitemOpen
  \bibfield  {author} {\bibinfo {author} {\bibnamefont {Kuhl}, \bibfnamefont
  {U.}}, \bibinfo {author} {\bibfnamefont {R.}~\bibnamefont {H{\"o}hmann}},
  \bibinfo {author} {\bibfnamefont {H.-J.}\ \bibnamefont {St{\"o}ckmann}}, \
  and\ \bibinfo {author} {\bibfnamefont {S.}~\bibnamefont {Gnutzmann}}}
  (\bibinfo {year} {2007}),\ \href {\doibase 10.1103/PhysRevE.75.036204}
  {\bibfield  {journal} {\bibinfo  {journal} {Phys. Rev. E}\ }\textbf {\bibinfo
  {volume} {75}}~(\bibinfo {number} {3}),\ \bibinfo {pages}
  {036204}}\BibitemShut {NoStop}%
\bibitem [{\citenamefont {Kuhl}\ \emph
  {et~al.}(2005{\natexlab{a}})\citenamefont {Kuhl}, \citenamefont
  {Martinez-Mares}, \citenamefont {M{\'e}ndez-S{\'a}nchez},\ and\ \citenamefont
  {St{\"o}ckmann}}]{kuhl2005direct}%
  \BibitemOpen
  \bibfield  {author} {\bibinfo {author} {\bibnamefont {Kuhl}, \bibfnamefont
  {U.}}, \bibinfo {author} {\bibfnamefont {M.}~\bibnamefont {Martinez-Mares}},
  \bibinfo {author} {\bibfnamefont {R.~A.}\ \bibnamefont
  {M{\'e}ndez-S{\'a}nchez}}, \ and\ \bibinfo {author} {\bibfnamefont {H.-J.}\
  \bibnamefont {St{\"o}ckmann}}} (\bibinfo {year} {2005}{\natexlab{a}}),\ \href
  {\doibase 10.1103/PhysRevLett.94.144101} {\bibfield  {journal} {\bibinfo
  {journal} {Phys. Rev. Lett.}\ }\textbf {\bibinfo {volume} {94}}~(\bibinfo
  {number} {14}),\ \bibinfo {pages} {144101}}\BibitemShut {NoStop}%
\bibitem [{\citenamefont {Kuhl}\ \emph {et~al.}(2000)\citenamefont {Kuhl},
  \citenamefont {Persson}, \citenamefont {Barth},\ and\ \citenamefont
  {St{\"o}ckmann}}]{kuhl2000mixing}%
  \BibitemOpen
  \bibfield  {author} {\bibinfo {author} {\bibnamefont {Kuhl}, \bibfnamefont
  {U.}}, \bibinfo {author} {\bibfnamefont {E.}~\bibnamefont {Persson}},
  \bibinfo {author} {\bibfnamefont {M.}~\bibnamefont {Barth}}, \ and\ \bibinfo
  {author} {\bibfnamefont {H.-J.}\ \bibnamefont {St{\"o}ckmann}}} (\bibinfo
  {year} {2000}),\ \href {\doibase 10.1007/s100510070139} {\bibfield  {journal}
  {\bibinfo  {journal} {Eur. Phys. J. B}\ }\textbf {\bibinfo {volume}
  {17}}~(\bibinfo {number} {2}),\ \bibinfo {pages} {253}}\BibitemShut {NoStop}%
\bibitem [{\citenamefont {Kuhl}\ \emph
  {et~al.}(2005{\natexlab{b}})\citenamefont {Kuhl}, \citenamefont
  {St{\"o}ckmann},\ and\ \citenamefont {Weaver}}]{kuhl2005classical}%
  \BibitemOpen
  \bibfield  {author} {\bibinfo {author} {\bibnamefont {Kuhl}, \bibfnamefont
  {U.}}, \bibinfo {author} {\bibfnamefont {H.-J.}\ \bibnamefont
  {St{\"o}ckmann}}, \ and\ \bibinfo {author} {\bibfnamefont {R.}~\bibnamefont
  {Weaver}}} (\bibinfo {year} {2005}{\natexlab{b}}),\ \href {\doibase
  10.1088/0305-4470/38/49/001} {\bibfield  {journal} {\bibinfo  {journal} {J.
  Phys. A: Math. Gen.}\ }\textbf {\bibinfo {volume} {38}}~(\bibinfo {number}
  {49}),\ \bibinfo {pages} {10433}}\BibitemShut {NoStop}%
\bibitem [{\citenamefont {Kurlberg}\ and\ \citenamefont
  {Wigman}(2015)}]{kurlberg2015non}%
  \BibitemOpen
  \bibfield  {author} {\bibinfo {author} {\bibnamefont {Kurlberg},
  \bibfnamefont {P.}}, \ and\ \bibinfo {author} {\bibfnamefont
  {I.}~\bibnamefont {Wigman}}} (\bibinfo {year} {2015}),\ \href {\doibase
  10.1016/j.crma.2014.09.026} {\bibfield  {journal} {\bibinfo  {journal} {C. R.
  Math.}\ }\textbf {\bibinfo {volume} {353}}~(\bibinfo {number} {2}),\ \bibinfo
  {pages} {101}}\BibitemShut {NoStop}%
\bibitem [{\citenamefont {Kurlberg}\ and\ \citenamefont
  {Wigman}(2016)}]{kurlberg2016probability}%
  \BibitemOpen
  \bibfield  {author} {\bibinfo {author} {\bibnamefont {Kurlberg},
  \bibfnamefont {P.}}, \ and\ \bibinfo {author} {\bibfnamefont
  {I.}~\bibnamefont {Wigman}}} (\bibinfo {year} {2016}),\ \href {\doibase
  10.1007/s00208-016-1411-4} {\bibfield  {journal} {\bibinfo  {journal} {Math.
  Ann.}\ }\textbf {\bibinfo {volume} {0}},\ \bibinfo {pages} {1}}\BibitemShut
  {NoStop}%
\bibitem [{\citenamefont {Kuttler}\ and\ \citenamefont
  {Sigillito}(1984)}]{kuttler1984eigenvalues}%
  \BibitemOpen
  \bibfield  {author} {\bibinfo {author} {\bibnamefont {Kuttler}, \bibfnamefont
  {J.~R.}}, \ and\ \bibinfo {author} {\bibfnamefont {V.~G.}\ \bibnamefont
  {Sigillito}}} (\bibinfo {year} {1984}),\ \href {\doibase 10.1137/1026033}
  {\bibfield  {journal} {\bibinfo  {journal} {SIAM Rev.}\ }\textbf {\bibinfo
  {volume} {26}}~(\bibinfo {number} {2}),\ \bibinfo {pages} {163}}\BibitemShut
  {NoStop}%
\bibitem [{\citenamefont {Kuznetsov}(2015)}]{kuznetsov2015delusive}%
  \BibitemOpen
  \bibfield  {author} {\bibinfo {author} {\bibnamefont {Kuznetsov},
  \bibfnamefont {N.}}} (\bibinfo {year} {2015}),\ \href
  {http://www.arxiv.org/abs/1502.00601} {\enquote {\bibinfo {title} {On
  {D}elusive {N}odes of {F}ree {O}scillations},}\ }\bibinfo {note} {{a}rXiv
  preprint 1502.00601}\BibitemShut {NoStop}%
\bibitem [{\citenamefont {Lai}\ \emph {et~al.}(2009)\citenamefont {Lai},
  \citenamefont {Shi}, \citenamefont {Dinov}, \citenamefont {Chan},\ and\
  \citenamefont {Toga}}]{lai2009laplace}%
  \BibitemOpen
  \bibfield  {author} {\bibinfo {author} {\bibnamefont {Lai}, \bibfnamefont
  {R.}}, \bibinfo {author} {\bibfnamefont {Y.}~\bibnamefont {Shi}}, \bibinfo
  {author} {\bibfnamefont {I.}~\bibnamefont {Dinov}}, \bibinfo {author}
  {\bibfnamefont {T.~F.}\ \bibnamefont {Chan}}, \ and\ \bibinfo {author}
  {\bibfnamefont {A.~W.}\ \bibnamefont {Toga}}} (\bibinfo {year} {2009}),\ in\
  \href {\doibase 10.1109/ISBI.2009.5193142} {\emph {\bibinfo {booktitle} {2009
  IEEE International Symposium on Biomedical Imaging: From Nano to Macro}}}\
  (\bibinfo {organization} {IEEE})\ pp.\ \bibinfo {pages}
  {694--697}\BibitemShut {NoStop}%
\bibitem [{\citenamefont {Lai}\ \emph {et~al.}(2010)\citenamefont {Lai},
  \citenamefont {Shi}, \citenamefont {Scheibel}, \citenamefont {Fears},
  \citenamefont {Woods}, \citenamefont {Toga},\ and\ \citenamefont
  {Chan}}]{lai2010metric}%
  \BibitemOpen
  \bibfield  {author} {\bibinfo {author} {\bibnamefont {Lai}, \bibfnamefont
  {R.}}, \bibinfo {author} {\bibfnamefont {Y.}~\bibnamefont {Shi}}, \bibinfo
  {author} {\bibfnamefont {K.}~\bibnamefont {Scheibel}}, \bibinfo {author}
  {\bibfnamefont {S.}~\bibnamefont {Fears}}, \bibinfo {author} {\bibfnamefont
  {R.}~\bibnamefont {Woods}}, \bibinfo {author} {\bibfnamefont {A.~W.}\
  \bibnamefont {Toga}}, \ and\ \bibinfo {author} {\bibfnamefont {T.~F.}\
  \bibnamefont {Chan}}} (\bibinfo {year} {2010}),\ in\ \href {\doibase
  10.1109/CVPR.2010.5540023} {\emph {\bibinfo {booktitle} {2010 IEEE Computer
  Society Conference on Computer Vision and Pattern Recognition}}}\ (\bibinfo
  {organization} {IEEE})\ pp.\ \bibinfo {pages} {2871--2878}\BibitemShut
  {NoStop}%
\bibitem [{\citenamefont {Lam{\'e}}(1866)}]{lame1866leccons}%
  \BibitemOpen
  \bibfield  {author} {\bibinfo {author} {\bibnamefont {Lam{\'e}},
  \bibfnamefont {G.}}} (\bibinfo {year} {1866}),\ \href@noop {} {\emph
  {\bibinfo {title} {Le{\c{c}}ons sur la th{\'e}orie math{\'e}matique de
  l'elasticit{\'e} des corps solides}}}\ (\bibinfo  {publisher}
  {Gauthier-Villars},\ \bibinfo {address} {Bachelier, Paris})\BibitemShut
  {NoStop}%
\bibitem [{\citenamefont {Landau}(1909)}]{landau1909}%
  \BibitemOpen
  \bibfield  {author} {\bibinfo {author} {\bibnamefont {Landau}, \bibfnamefont
  {E.}}} (\bibinfo {year} {1909}),\ \href@noop {} {\bibfield  {journal}
  {\bibinfo  {journal} {Arch. Math. Phys. (III)}\ }\textbf {\bibinfo {volume}
  {13}},\ \bibinfo {pages} {305}}\BibitemShut {NoStop}%
\bibitem [{\citenamefont {Landau}\ and\ \citenamefont
  {Lifshitz}(1959)}]{landau1959course}%
  \BibitemOpen
  \bibfield  {author} {\bibinfo {author} {\bibnamefont {Landau}, \bibfnamefont
  {L.~D.}}, \ and\ \bibinfo {author} {\bibfnamefont {E.~M.}\ \bibnamefont
  {Lifshitz}}} (\bibinfo {year} {1959}),\ \href@noop {} {\emph {\bibinfo
  {title} {Theory of Elasticity}}},\ \bibinfo {series} {Course of Theoretical
  Physics}, Vol.~\bibinfo {volume} {7}\ (\bibinfo  {publisher} {Pergamon
  Press},\ \bibinfo {address} {Oxford})\BibitemShut {NoStop}%
\bibitem [{\citenamefont {van Langen}\ \emph {et~al.}(1997)\citenamefont {van
  Langen}, \citenamefont {Brouwer},\ and\ \citenamefont
  {Beenakker}}]{van1997fluctuating}%
  \BibitemOpen
  \bibfield  {author} {\bibinfo {author} {\bibnamefont {van Langen},
  \bibfnamefont {S.~A.}}, \bibinfo {author} {\bibfnamefont {P.~W.}\
  \bibnamefont {Brouwer}}, \ and\ \bibinfo {author} {\bibfnamefont {C.~W.~J.}\
  \bibnamefont {Beenakker}}} (\bibinfo {year} {1997}),\ \href {\doibase
  10.1103/PhysRevE.55.R1} {\bibfield  {journal} {\bibinfo  {journal} {Phys.
  Rev. E}\ }\textbf {\bibinfo {volume} {55}}~(\bibinfo {number} {1}),\ \bibinfo
  {pages} {R1}}\BibitemShut {NoStop}%
\bibitem [{\citenamefont {Lapidus}(1991)}]{lapidus1991fractal}%
  \BibitemOpen
  \bibfield  {author} {\bibinfo {author} {\bibnamefont {Lapidus}, \bibfnamefont
  {M.~L.}}} (\bibinfo {year} {1991}),\ \href {\doibase
  10.1090/S0002-9947-1991-0994168-5} {\bibfield  {journal} {\bibinfo  {journal}
  {Trans. Amer. Math. Soc.}\ }\textbf {\bibinfo {volume} {325}}~(\bibinfo
  {number} {2}),\ \bibinfo {pages} {465}}\BibitemShut {NoStop}%
\bibitem [{\citenamefont {Lapidus}\ and\ \citenamefont
  {Pomerance}(1993)}]{lapidus1993riemann}%
  \BibitemOpen
  \bibfield  {author} {\bibinfo {author} {\bibnamefont {Lapidus}, \bibfnamefont
  {M.~L.}}, \ and\ \bibinfo {author} {\bibfnamefont {C.}~\bibnamefont
  {Pomerance}}} (\bibinfo {year} {1993}),\ \href {\doibase
  10.1112/plms/s3-66.1.41} {\bibfield  {journal} {\bibinfo  {journal} {Proc.
  London Math. Soc.}\ }\textbf {\bibinfo {volume} {3}}~(\bibinfo {number}
  {1}),\ \bibinfo {pages} {41}}\BibitemShut {NoStop}%
\bibitem [{\citenamefont {Lapidus}\ and\ \citenamefont
  {Pomerance}(1996)}]{lapidus1996counterexamples}%
  \BibitemOpen
  \bibfield  {author} {\bibinfo {author} {\bibnamefont {Lapidus}, \bibfnamefont
  {M.~L.}}, \ and\ \bibinfo {author} {\bibfnamefont {C.}~\bibnamefont
  {Pomerance}}} (\bibinfo {year} {1996}),\ \href {\doibase
  10.1017/S0305004100074053} {\bibfield  {journal} {\bibinfo  {journal} {Math.
  Proc. Cambridge Philos. Soc.}\ }\textbf {\bibinfo {volume} {119}}~(\bibinfo
  {number} {1}),\ \bibinfo {pages} {167}}\BibitemShut {NoStop}%
\bibitem [{\citenamefont {Lawler}(2001)}]{lawler2001vienna}%
  \BibitemOpen
  \bibfield  {author} {\bibinfo {author} {\bibnamefont {Lawler}, \bibfnamefont
  {G.~F.}}} (\bibinfo {year} {2001}),\ in\ \href@noop {} {\emph {\bibinfo
  {booktitle} {Proceedings of Conference on Random Walks}}}\ (\bibinfo
  {organization} {Erwin Schr{\"o}dinger Institute},\ \bibinfo {address}
  {Vienna})\BibitemShut {NoStop}%
\bibitem [{\citenamefont {Lawler}(2008)}]{lawler2008conformally}%
  \BibitemOpen
  \bibfield  {author} {\bibinfo {author} {\bibnamefont {Lawler}, \bibfnamefont
  {G.~F.}}} (\bibinfo {year} {2008}),\ \href@noop {} {\emph {\bibinfo {title}
  {Conformally invariant processes in the plane}}},\ \bibinfo {series}
  {Mathematical Surveys and Monographs}, Vol.\ \bibinfo {volume} {114}\
  (\bibinfo  {publisher} {American Mathematical Soc.},\ \bibinfo {address}
  {Providence, RI})\BibitemShut {NoStop}%
\bibitem [{\citenamefont {Lawler}(2009)}]{lawler2009conformal}%
  \BibitemOpen
  \bibfield  {author} {\bibinfo {author} {\bibnamefont {Lawler}, \bibfnamefont
  {G.~F.}}} (\bibinfo {year} {2009}),\ \href {\doibase
  10.1090/S0273-0979-08-01229-9} {\bibfield  {journal} {\bibinfo  {journal}
  {Bull. Amer. Math. Soc.}\ }\textbf {\bibinfo {volume} {46}}~(\bibinfo
  {number} {1}),\ \bibinfo {pages} {35}}\BibitemShut {NoStop}%
\bibitem [{\citenamefont {Lawler}\ \emph {et~al.}(2000)\citenamefont {Lawler},
  \citenamefont {Schramm},\ and\ \citenamefont {Werner}}]{lawler2000dimension}%
  \BibitemOpen
  \bibfield  {author} {\bibinfo {author} {\bibnamefont {Lawler}, \bibfnamefont
  {G.~F.}}, \bibinfo {author} {\bibfnamefont {O.}~\bibnamefont {Schramm}}, \
  and\ \bibinfo {author} {\bibfnamefont {W.}~\bibnamefont {Werner}}} (\bibinfo
  {year} {2000}),\ \href {\doibase 10.4310/MRL.2001.v8.n4.a1} {\bibfield
  {journal} {\bibinfo  {journal} {Math. Res. Lett.}\ }\textbf {\bibinfo
  {volume} {8}}~(\bibinfo {number} {4}),\ \bibinfo {pages} {401}}\BibitemShut
  {NoStop}%
\bibitem [{\citenamefont {Lawler}\ \emph
  {et~al.}(2001{\natexlab{a}})\citenamefont {Lawler}, \citenamefont {Schramm},\
  and\ \citenamefont {Werner}}]{lsw2001-1}%
  \BibitemOpen
  \bibfield  {author} {\bibinfo {author} {\bibnamefont {Lawler}, \bibfnamefont
  {G.~F.}}, \bibinfo {author} {\bibfnamefont {O.}~\bibnamefont {Schramm}}, \
  and\ \bibinfo {author} {\bibfnamefont {W.}~\bibnamefont {Werner}}} (\bibinfo
  {year} {2001}{\natexlab{a}}),\ \href {\doibase 10.1007/BF02392618} {\bibfield
   {journal} {\bibinfo  {journal} {Acta Math.}\ }\textbf {\bibinfo {volume}
  {187}},\ \bibinfo {pages} {237}}\BibitemShut {NoStop}%
\bibitem [{\citenamefont {Lawler}\ \emph
  {et~al.}(2001{\natexlab{b}})\citenamefont {Lawler}, \citenamefont {Schramm},\
  and\ \citenamefont {Werner}}]{lsw2001-2}%
  \BibitemOpen
  \bibfield  {author} {\bibinfo {author} {\bibnamefont {Lawler}, \bibfnamefont
  {G.~F.}}, \bibinfo {author} {\bibfnamefont {O.}~\bibnamefont {Schramm}}, \
  and\ \bibinfo {author} {\bibfnamefont {W.}~\bibnamefont {Werner}}} (\bibinfo
  {year} {2001}{\natexlab{b}}),\ \href {\doibase 10.1007/BF02392619} {\bibfield
   {journal} {\bibinfo  {journal} {Acta Math.}\ }\textbf {\bibinfo {volume}
  {187}},\ \bibinfo {pages} {275}}\BibitemShut {NoStop}%
\bibitem [{\citenamefont {Lawler}\ \emph {et~al.}(2002)\citenamefont {Lawler},
  \citenamefont {Schramm},\ and\ \citenamefont {Werner}}]{lsw2002}%
  \BibitemOpen
  \bibfield  {author} {\bibinfo {author} {\bibnamefont {Lawler}, \bibfnamefont
  {G.~F.}}, \bibinfo {author} {\bibfnamefont {O.}~\bibnamefont {Schramm}}, \
  and\ \bibinfo {author} {\bibfnamefont {W.}~\bibnamefont {Werner}}} (\bibinfo
  {year} {2002}),\ \href {\doibase 10.1016/S0246-0203(01)01089-5} {\bibfield
  {journal} {\bibinfo  {journal} {Ann. Inst. H. Poincar{\'e} Probab. Statist.}\
  }\textbf {\bibinfo {volume} {38}},\ \bibinfo {pages} {109}}\BibitemShut
  {NoStop}%
\bibitem [{\citenamefont {Lawler}\ \emph {et~al.}(2003)\citenamefont {Lawler},
  \citenamefont {Schramm},\ and\ \citenamefont {Werner}}]{lawler2003conformal}%
  \BibitemOpen
  \bibfield  {author} {\bibinfo {author} {\bibnamefont {Lawler}, \bibfnamefont
  {G.~F.}}, \bibinfo {author} {\bibfnamefont {O.}~\bibnamefont {Schramm}}, \
  and\ \bibinfo {author} {\bibfnamefont {W.}~\bibnamefont {Werner}}} (\bibinfo
  {year} {2003}),\ \href {\doibase 10.1090/S0894-0347-03-00430-2} {\bibfield
  {journal} {\bibinfo  {journal} {J. Amer. Math. Soc.}\ }\textbf {\bibinfo
  {volume} {16}}~(\bibinfo {number} {4}),\ \bibinfo {pages} {917}}\BibitemShut
  {NoStop}%
\bibitem [{\citenamefont {Lawler}\ \emph
  {et~al.}(2004{\natexlab{a}})\citenamefont {Lawler}, \citenamefont {Schramm},\
  and\ \citenamefont {Werner}}]{lawler2004conformal}%
  \BibitemOpen
  \bibfield  {author} {\bibinfo {author} {\bibnamefont {Lawler}, \bibfnamefont
  {G.~F.}}, \bibinfo {author} {\bibfnamefont {O.}~\bibnamefont {Schramm}}, \
  and\ \bibinfo {author} {\bibfnamefont {W.}~\bibnamefont {Werner}}} (\bibinfo
  {year} {2004}{\natexlab{a}}),\ \href {\doibase 10.1214/aop/1079021469}
  {\bibfield  {journal} {\bibinfo  {journal} {Ann. Probab.}\ }\textbf {\bibinfo
  {volume} {32}},\ \bibinfo {pages} {939}}\BibitemShut {NoStop}%
\bibitem [{\citenamefont {Lawler}\ \emph
  {et~al.}(2004{\natexlab{b}})\citenamefont {Lawler}, \citenamefont {Schramm},\
  and\ \citenamefont {Werner}}]{lawler2004scaling}%
  \BibitemOpen
  \bibfield  {author} {\bibinfo {author} {\bibnamefont {Lawler}, \bibfnamefont
  {G.~F.}}, \bibinfo {author} {\bibfnamefont {O.}~\bibnamefont {Schramm}}, \
  and\ \bibinfo {author} {\bibfnamefont {W.}~\bibnamefont {Werner}}} (\bibinfo
  {year} {2004}{\natexlab{b}}),\ in\ \href {\doibase
  10.1090/pspum/072.2/2112127} {\emph {\bibinfo {booktitle} {Fractal Geometry
  and Applications: A Jubilee of Benoit Mandelbrot}}},\ \bibinfo {series}
  {Proc. Symp. Pure Math.}, Vol.\ \bibinfo {volume} {72.2},\ \bibinfo {editor}
  {edited by\ \bibinfo {editor} {\bibfnamefont {M.~L.}\ \bibnamefont
  {Lapidus}}\ and\ \bibinfo {editor} {\bibfnamefont {M.}~\bibnamefont {van
  Frankenhuijsen}}}\ (\bibinfo  {publisher} {American Mathematical Soc.},\
  \bibinfo {address} {Providence, RI})\BibitemShut {NoStop}%
\bibitem [{\citenamefont {{\L}awniczak}\ \emph {et~al.}(2014)\citenamefont
  {{\L}awniczak}, \citenamefont {Sawicki}, \citenamefont {Bauch}, \citenamefont
  {Ku{\'s}},\ and\ \citenamefont {Sirko}}]{lawniczak2014resonances}%
  \BibitemOpen
  \bibfield  {author} {\bibinfo {author} {\bibnamefont {{\L}awniczak},
  \bibfnamefont {M.}}, \bibinfo {author} {\bibfnamefont {A.}~\bibnamefont
  {Sawicki}}, \bibinfo {author} {\bibfnamefont {S.}~\bibnamefont {Bauch}},
  \bibinfo {author} {\bibfnamefont {M.}~\bibnamefont {Ku{\'s}}}, \ and\
  \bibinfo {author} {\bibfnamefont {L.}~\bibnamefont {Sirko}}} (\bibinfo {year}
  {2014}),\ \href {\doibase 10.1103/PhysRevE.89.032911} {\bibfield  {journal}
  {\bibinfo  {journal} {Phys. Rev. E}\ }\textbf {\bibinfo {volume}
  {89}}~(\bibinfo {number} {3}),\ \bibinfo {pages} {032911}}\BibitemShut
  {NoStop}%
\bibitem [{\citenamefont {Lehmann}\ \emph {et~al.}(1995)\citenamefont
  {Lehmann}, \citenamefont {Saher}, \citenamefont {Sokolov},\ and\
  \citenamefont {Sommers}}]{lehmann1995chaotic}%
  \BibitemOpen
  \bibfield  {author} {\bibinfo {author} {\bibnamefont {Lehmann}, \bibfnamefont
  {N.}}, \bibinfo {author} {\bibfnamefont {D.}~\bibnamefont {Saher}}, \bibinfo
  {author} {\bibfnamefont {V.~V.}\ \bibnamefont {Sokolov}}, \ and\ \bibinfo
  {author} {\bibfnamefont {H.-J.}\ \bibnamefont {Sommers}}} (\bibinfo {year}
  {1995}),\ \href {\doibase 10.1016/0375-9474(94)00460-5} {\bibfield  {journal}
  {\bibinfo  {journal} {Nucl. Phys. A}\ }\textbf {\bibinfo {volume}
  {582}}~(\bibinfo {number} {1}),\ \bibinfo {pages} {223}}\BibitemShut
  {NoStop}%
\bibitem [{\citenamefont {Levitin}\ and\ \citenamefont
  {Vassiliev}(1996)}]{levitin1996spectral}%
  \BibitemOpen
  \bibfield  {author} {\bibinfo {author} {\bibnamefont {Levitin}, \bibfnamefont
  {M.}}, \ and\ \bibinfo {author} {\bibfnamefont {D.}~\bibnamefont
  {Vassiliev}}} (\bibinfo {year} {1996}),\ \href {\doibase
  10.1112/plms/s3-72.1.188} {\bibfield  {journal} {\bibinfo  {journal} {Proc.
  London Math. Soc.}\ }\textbf {\bibinfo {volume} {3}}~(\bibinfo {number}
  {1}),\ \bibinfo {pages} {188}}\BibitemShut {NoStop}%
\bibitem [{\citenamefont {L{\'e}vy}(1965)}]{levy1965}%
  \BibitemOpen
  \bibfield  {author} {\bibinfo {author} {\bibnamefont {L{\'e}vy},
  \bibfnamefont {P.}}} (\bibinfo {year} {1965}),\ \href@noop {} {\emph
  {\bibinfo {title} {Processus stochastiques et mouvement brownien}}}\
  (\bibinfo  {publisher} {Gauthier-Villars},\ \bibinfo {address}
  {Paris})\BibitemShut {NoStop}%
\bibitem [{\citenamefont {Lewenkopf}\ \emph {et~al.}(1992)\citenamefont
  {Lewenkopf}, \citenamefont {M{\"u}ller},\ and\ \citenamefont
  {Doron}}]{lewenkopf1992microwave}%
  \BibitemOpen
  \bibfield  {author} {\bibinfo {author} {\bibnamefont {Lewenkopf},
  \bibfnamefont {C.~H.}}, \bibinfo {author} {\bibfnamefont {A.}~\bibnamefont
  {M{\"u}ller}}, \ and\ \bibinfo {author} {\bibfnamefont {E.}~\bibnamefont
  {Doron}}} (\bibinfo {year} {1992}),\ \href {\doibase
  10.1103/PhysRevA.45.2635} {\bibfield  {journal} {\bibinfo  {journal} {Phys.
  Rev. A}\ }\textbf {\bibinfo {volume} {45}}~(\bibinfo {number} {4}),\ \bibinfo
  {pages} {2635}}\BibitemShut {NoStop}%
\bibitem [{\citenamefont {Lewy}(1977)}]{lewy1977mininum}%
  \BibitemOpen
  \bibfield  {author} {\bibinfo {author} {\bibnamefont {Lewy}, \bibfnamefont
  {H.}}} (\bibinfo {year} {1977}),\ \href {\doibase 10.1080/03605307708820059}
  {\bibfield  {journal} {\bibinfo  {journal} {Commun. Part. Diff. Eq.}\
  }\textbf {\bibinfo {volume} {2}}~(\bibinfo {number} {12}),\ \bibinfo {pages}
  {1233}}\BibitemShut {NoStop}%
\bibitem [{\citenamefont {Leydold}(1996)}]{leydold1996number}%
  \BibitemOpen
  \bibfield  {author} {\bibinfo {author} {\bibnamefont {Leydold}, \bibfnamefont
  {J.}}} (\bibinfo {year} {1996}),\ \href {\doibase
  10.1016/0040-9383(95)00028-3} {\bibfield  {journal} {\bibinfo  {journal}
  {Topology}\ }\textbf {\bibinfo {volume} {35}}~(\bibinfo {number} {2}),\
  \bibinfo {pages} {301}}\BibitemShut {NoStop}%
\bibitem [{\citenamefont {Li}\ and\ \citenamefont
  {Robnik}(1994)}]{li1994statistical}%
  \BibitemOpen
  \bibfield  {author} {\bibinfo {author} {\bibnamefont {Li}, \bibfnamefont
  {B.}}, \ and\ \bibinfo {author} {\bibfnamefont {M.}~\bibnamefont {Robnik}}}
  (\bibinfo {year} {1994}),\ \href {\doibase 10.1088/0305-4470/27/16/017}
  {\bibfield  {journal} {\bibinfo  {journal} {J. Phys. A: Math. Gen.}\ }\textbf
  {\bibinfo {volume} {27}}~(\bibinfo {number} {16}),\ \bibinfo {pages}
  {5509}}\BibitemShut {NoStop}%
\bibitem [{\citenamefont {Liboff}(1994)}]{liboff1994nodal}%
  \BibitemOpen
  \bibfield  {author} {\bibinfo {author} {\bibnamefont {Liboff}, \bibfnamefont
  {R.~L.}}} (\bibinfo {year} {1994}),\ \href {\doibase 10.1063/1.530453}
  {\bibfield  {journal} {\bibinfo  {journal} {J. Math. Phys.}\ }\textbf
  {\bibinfo {volume} {35}}~(\bibinfo {number} {8}),\ \bibinfo {pages}
  {3881}}\BibitemShut {NoStop}%
\bibitem [{\citenamefont {Lichtenberg}\ and\ \citenamefont
  {Lieberman}(2013)}]{lichtenberg2013regular}%
  \BibitemOpen
  \bibfield  {author} {\bibinfo {author} {\bibnamefont {Lichtenberg},
  \bibfnamefont {A.~J.}}, \ and\ \bibinfo {author} {\bibfnamefont {M.~A.}\
  \bibnamefont {Lieberman}}} (\bibinfo {year} {2013}),\ \href@noop {} {\emph
  {\bibinfo {title} {Regular and stochastic motion}}},\ \bibinfo {series}
  {Applied Mathematical Sciences}, Vol.~\bibinfo {volume} {38}\ (\bibinfo
  {publisher} {Springer},\ \bibinfo {address} {New York})\BibitemShut {NoStop}%
\bibitem [{\citenamefont {Lin}(1987)}]{lin1987second}%
  \BibitemOpen
  \bibfield  {author} {\bibinfo {author} {\bibnamefont {Lin}, \bibfnamefont
  {C.-S.}}} (\bibinfo {year} {1987}),\ \href {\doibase 10.1007/BF01217758}
  {\bibfield  {journal} {\bibinfo  {journal} {Commun. Math. Phys.}\ }\textbf
  {\bibinfo {volume} {111}}~(\bibinfo {number} {2}),\ \bibinfo {pages}
  {161}}\BibitemShut {NoStop}%
\bibitem [{\citenamefont {Lin}\ and\ \citenamefont
  {Ni}(1988)}]{lin1988counterexample}%
  \BibitemOpen
  \bibfield  {author} {\bibinfo {author} {\bibnamefont {Lin}, \bibfnamefont
  {C.~S.}}, \ and\ \bibinfo {author} {\bibfnamefont {W.-M.}\ \bibnamefont
  {Ni}}} (\bibinfo {year} {1988}),\ \href {\doibase
  10.1090/S0002-9939-1988-0920985-9} {\bibfield  {journal} {\bibinfo  {journal}
  {Proc. Amer. Math. Soc.}\ }\textbf {\bibinfo {volume} {102}}~(\bibinfo
  {number} {2}),\ \bibinfo {pages} {271}}\BibitemShut {NoStop}%
\bibitem [{\citenamefont {Loewner}(1953)}]{charles1953generation}%
  \BibitemOpen
  \bibfield  {author} {\bibinfo {author} {\bibnamefont {Loewner}, \bibfnamefont
  {{\relax Ch}.}}} (\bibinfo {year} {1953}),\ \href {\doibase
  10.2140/pjm.1953.3.417} {\bibfield  {journal} {\bibinfo  {journal} {Pacific
  J. Math}\ }\textbf {\bibinfo {volume} {3}},\ \bibinfo {pages}
  {417}}\BibitemShut {NoStop}%
\bibitem [{\citenamefont {Logunov}(2016{\natexlab{a}})}]{logunov2016nodal2}%
  \BibitemOpen
  \bibfield  {author} {\bibinfo {author} {\bibnamefont {Logunov}, \bibfnamefont
  {A.}}} (\bibinfo {year} {2016}{\natexlab{a}}),\ \href
  {http://www.arxiv.org/abs/1605.02587} {\enquote {\bibinfo {title} {Nodal sets
  of laplace eigenfunctions: polynomial upper estimates of the hausdorff
  measure},}\ }\bibinfo {note} {{a}rXiv preprint 1605.02587}\BibitemShut
  {NoStop}%
\bibitem [{\citenamefont {Logunov}(2016{\natexlab{b}})}]{logunov2016nodal1}%
  \BibitemOpen
  \bibfield  {author} {\bibinfo {author} {\bibnamefont {Logunov}, \bibfnamefont
  {A.}}} (\bibinfo {year} {2016}{\natexlab{b}}),\ \href
  {http://www.arxiv.org/abs/1605.02589} {\enquote {\bibinfo {title} {Nodal sets
  of laplace eigenfunctions: proof of nadirashvili's conjecture and of the
  lower bound in yau's conjecture},}\ }\bibinfo {note} {{a}rXiv preprint
  1605.02589}\BibitemShut {NoStop}%
\bibitem [{\citenamefont {Logunov}\ and\ \citenamefont
  {Malinnikova}(2016)}]{logunov2016nodal3}%
  \BibitemOpen
  \bibfield  {author} {\bibinfo {author} {\bibnamefont {Logunov}, \bibfnamefont
  {A.}}, \ and\ \bibinfo {author} {\bibfnamefont {E.}~\bibnamefont
  {Malinnikova}}} (\bibinfo {year} {2016}),\ \href
  {http://www.arxiv.org/abs/1605.02595} {\enquote {\bibinfo {title} {Nodal sets
  of laplace eigenfunctions: estimates of the hausdorff measure in dimension
  two and three},}\ }\bibinfo {note} {{a}rXiv preprint 1605.02595}\BibitemShut
  {NoStop}%
\bibitem [{\citenamefont {Longuet-Higgins}(1957)}]{longuet1957statistical}%
  \BibitemOpen
  \bibfield  {author} {\bibinfo {author} {\bibnamefont {Longuet-Higgins},
  \bibfnamefont {M.~S.}}} (\bibinfo {year} {1957}),\ \href@noop {} {\bibfield
  {journal} {\bibinfo  {journal} {Phil. Trans. R. Soc. A}\ }\textbf {\bibinfo
  {volume} {249}}~(\bibinfo {number} {966}),\ \bibinfo {pages}
  {321}}\BibitemShut {NoStop}%
\bibitem [{\citenamefont {L{\"o}wner}(1923)}]{lowner1923untersuchungen}%
  \BibitemOpen
  \bibfield  {author} {\bibinfo {author} {\bibnamefont {L{\"o}wner},
  \bibfnamefont {K.}}} (\bibinfo {year} {1923}),\ \href {\doibase
  10.1007/BF01448091} {\bibfield  {journal} {\bibinfo  {journal} {Math. Ann.}\
  }\textbf {\bibinfo {volume} {89}}~(\bibinfo {number} {1}),\ \bibinfo {pages}
  {103}}\BibitemShut {NoStop}%
\bibitem [{\citenamefont {Lu}\ \emph {et~al.}(2003)\citenamefont {Lu},
  \citenamefont {Sridhar},\ and\ \citenamefont {Zworski}}]{lu2003fractal}%
  \BibitemOpen
  \bibfield  {author} {\bibinfo {author} {\bibnamefont {Lu}, \bibfnamefont
  {W.~T.}}, \bibinfo {author} {\bibfnamefont {S.}~\bibnamefont {Sridhar}}, \
  and\ \bibinfo {author} {\bibfnamefont {M.}~\bibnamefont {Zworski}}} (\bibinfo
  {year} {2003}),\ \href {\doibase 10.1103/PhysRevLett.91.154101} {\bibfield
  {journal} {\bibinfo  {journal} {Phys. Rev. Lett.}\ }\textbf {\bibinfo
  {volume} {91}}~(\bibinfo {number} {15}),\ \bibinfo {pages}
  {154101}}\BibitemShut {NoStop}%
\bibitem [{\citenamefont {Lukaschuk}\ \emph {et~al.}(2007)\citenamefont
  {Lukaschuk}, \citenamefont {Denissenko},\ and\ \citenamefont
  {Falkovich}}]{lukaschuk2007nodal}%
  \BibitemOpen
  \bibfield  {author} {\bibinfo {author} {\bibnamefont {Lukaschuk},
  \bibfnamefont {S.}}, \bibinfo {author} {\bibfnamefont {P.}~\bibnamefont
  {Denissenko}}, \ and\ \bibinfo {author} {\bibfnamefont {G.}~\bibnamefont
  {Falkovich}}} (\bibinfo {year} {2007}),\ \href {\doibase
  10.1140/epjst/e2007-00151-6} {\bibfield  {journal} {\bibinfo  {journal} {Eur.
  Phys. J. Spec. Top.}\ }\textbf {\bibinfo {volume} {145}}~(\bibinfo {number}
  {1}),\ \bibinfo {pages} {125}}\BibitemShut {NoStop}%
\bibitem [{\citenamefont {Maffucci}(2016)}]{maffucci2016nodal}%
  \BibitemOpen
  \bibfield  {author} {\bibinfo {author} {\bibnamefont {Maffucci},
  \bibfnamefont {R.~W.}}} (\bibinfo {year} {2016}),\ \href {\doibase
  10.1007/s00605-016-1001-2} {\bibinfo  {journal} {Monatsh. Math.}\ ,\ \bibinfo
  {pages} {1}}\BibitemShut {NoStop}%
\bibitem [{\citenamefont {Maier~Jr}\ and\ \citenamefont
  {Slater}(1952)}]{maier1952field}%
  \BibitemOpen
\bibfield  {journal} {  }\bibfield  {author} {\bibinfo {author} {\bibnamefont
  {Maier~Jr}, \bibfnamefont {L.~C.}}, \ and\ \bibinfo {author} {\bibfnamefont
  {J.~C.}\ \bibnamefont {Slater}}} (\bibinfo {year} {1952}),\ \href {\doibase
  10.1063/1.1701980} {\bibfield  {journal} {\bibinfo  {journal} {J. Appl.
  Phys.}\ }\textbf {\bibinfo {volume} {23}}~(\bibinfo {number} {1}),\ \bibinfo
  {pages} {68}}\BibitemShut {NoStop}%
\bibitem [{\citenamefont {Main}\ \emph {et~al.}(1986)\citenamefont {Main},
  \citenamefont {Wiebusch}, \citenamefont {Holle},\ and\ \citenamefont
  {Welge}}]{main1986new}%
  \BibitemOpen
  \bibfield  {author} {\bibinfo {author} {\bibnamefont {Main}, \bibfnamefont
  {J.}}, \bibinfo {author} {\bibfnamefont {G.}~\bibnamefont {Wiebusch}},
  \bibinfo {author} {\bibfnamefont {A.}~\bibnamefont {Holle}}, \ and\ \bibinfo
  {author} {\bibfnamefont {K.~H.}\ \bibnamefont {Welge}}} (\bibinfo {year}
  {1986}),\ \href {\doibase 10.1103/PhysRevLett.57.2789} {\bibfield  {journal}
  {\bibinfo  {journal} {Phys. Rev. Lett.}\ }\textbf {\bibinfo {volume}
  {57}}~(\bibinfo {number} {22}),\ \bibinfo {pages} {2789}}\BibitemShut
  {NoStop}%
\bibitem [{\citenamefont {Makai}(1965)}]{makai1965lower}%
  \BibitemOpen
  \bibfield  {author} {\bibinfo {author} {\bibnamefont {Makai}, \bibfnamefont
  {E.}}} (\bibinfo {year} {1965}),\ \href {\doibase 10.1007/bf01904840}
  {\bibfield  {journal} {\bibinfo  {journal} {Acta Math. Hung.}\ }\textbf
  {\bibinfo {volume} {16}}~(\bibinfo {number} {3-4}),\ \bibinfo {pages}
  {319}}\BibitemShut {NoStop}%
\bibitem [{\citenamefont {Mandelbrot}(1983)}]{mandelbrot1983fractal}%
  \BibitemOpen
  \bibfield  {author} {\bibinfo {author} {\bibnamefont {Mandelbrot},
  \bibfnamefont {B.~B.}}} (\bibinfo {year} {1983}),\ \href@noop {} {\emph
  {\bibinfo {title} {The Fractal Geometry of Nature}}},\ Vol.\ \bibinfo
  {volume} {173}\ (\bibinfo  {publisher} {Macmillan})\BibitemShut {NoStop}%
\bibitem [{\citenamefont {Mandwal}\ and\ \citenamefont
  {Jain}(2017)}]{mandwal2017}%
  \BibitemOpen
  \bibfield  {author} {\bibinfo {author} {\bibnamefont {Mandwal}, \bibfnamefont
  {A.~K.}}, \ and\ \bibinfo {author} {\bibfnamefont {S.~R.}\ \bibnamefont
  {Jain}}} (\bibinfo {year} {2017}),\ \href {\doibase
  10.1007/s12043-017-1432-x} {\bibfield  {journal} {\bibinfo  {journal}
  {Pramana - J. Phys.}\ }\textbf {\bibinfo {volume} {89}},\ \bibinfo {pages}
  {35}}\BibitemShut {NoStop}%
\bibitem [{\citenamefont {Mangoubi}(2008)}]{mangoubi2008inner}%
  \BibitemOpen
  \bibfield  {author} {\bibinfo {author} {\bibnamefont {Mangoubi},
  \bibfnamefont {D.}}} (\bibinfo {year} {2008}),\ \href {\doibase
  10.4153/CMB-2008-026-2} {\bibfield  {journal} {\bibinfo  {journal} {Canad.
  Math. Bull.}\ }\textbf {\bibinfo {volume} {51}}~(\bibinfo {number} {2}),\
  \bibinfo {pages} {249}}\BibitemShut {NoStop}%
\bibitem [{\citenamefont {Manjunath}\ \emph {et~al.}(2016)\citenamefont
  {Manjunath}, \citenamefont {Samajdar},\ and\ \citenamefont
  {Jain}}]{manjunath2016difference}%
  \BibitemOpen
  \bibfield  {author} {\bibinfo {author} {\bibnamefont {Manjunath},
  \bibfnamefont {N.}}, \bibinfo {author} {\bibfnamefont {R.}~\bibnamefont
  {Samajdar}}, \ and\ \bibinfo {author} {\bibfnamefont {S.~R.}\ \bibnamefont
  {Jain}}} (\bibinfo {year} {2016}),\ \href {\doibase
  10.1016/j.aop.2016.04.014} {\bibfield  {journal} {\bibinfo  {journal} {Ann.
  Phys.}\ }\textbf {\bibinfo {volume} {372}},\ \bibinfo {pages}
  {68}}\BibitemShut {NoStop}%
\bibitem [{\citenamefont {Marinucci}\ \emph {et~al.}(2016)\citenamefont
  {Marinucci}, \citenamefont {Peccati}, \citenamefont {Rossi},\ and\
  \citenamefont {Wigman}}]{marinucci2016non}%
  \BibitemOpen
  \bibfield  {author} {\bibinfo {author} {\bibnamefont {Marinucci},
  \bibfnamefont {D.}}, \bibinfo {author} {\bibfnamefont {G.}~\bibnamefont
  {Peccati}}, \bibinfo {author} {\bibfnamefont {M.}~\bibnamefont {Rossi}}, \
  and\ \bibinfo {author} {\bibfnamefont {I.}~\bibnamefont {Wigman}}} (\bibinfo
  {year} {2016}),\ \href {\doibase 10.1007/s00039-016-0376-5} {\bibfield
  {journal} {\bibinfo  {journal} {Geom. Funct. Anal.}\ }\textbf {\bibinfo
  {volume} {26}},\ \bibinfo {pages} {926}}\BibitemShut {NoStop}%
\bibitem [{\citenamefont {Marshall}\ and\ \citenamefont
  {Rohde}(2006)}]{marshall2006convergence}%
  \BibitemOpen
  \bibfield  {author} {\bibinfo {author} {\bibnamefont {Marshall},
  \bibfnamefont {D.~E.}}, \ and\ \bibinfo {author} {\bibfnamefont
  {S.}~\bibnamefont {Rohde}}} (\bibinfo {year} {2006}),\ \href
  {http://www.arxiv.org/abs/math/0605532} {\enquote {\bibinfo {title}
  {Convergence of the {Z}ipper algorithm for conformal mapping},}\ }\bibinfo
  {note} {{a}rXiv preprint math/0605532}\BibitemShut {NoStop}%
\bibitem [{\citenamefont {McCartin}(2003)}]{mccartin2003eigenstructure}%
  \BibitemOpen
  \bibfield  {author} {\bibinfo {author} {\bibnamefont {McCartin},
  \bibfnamefont {B.~J.}}} (\bibinfo {year} {2003}),\ \href {\doibase
  10.1137/S003614450238720} {\bibfield  {journal} {\bibinfo  {journal} {SIAM
  Rev.}\ }\textbf {\bibinfo {volume} {45}}~(\bibinfo {number} {2}),\ \bibinfo
  {pages} {267}}\BibitemShut {NoStop}%
\bibitem [{\citenamefont {McDonald}\ and\ \citenamefont
  {Fulling}(2014)}]{mcdonald2014neumann}%
  \BibitemOpen
  \bibfield  {author} {\bibinfo {author} {\bibnamefont {McDonald},
  \bibfnamefont {R.~B.}}, \ and\ \bibinfo {author} {\bibfnamefont {S.~A.}\
  \bibnamefont {Fulling}}} (\bibinfo {year} {2014}),\ \href {\doibase
  10.1098/rsta.2012.0505} {\bibfield  {journal} {\bibinfo  {journal} {Phil.
  Trans. R. Soc. A}\ }\textbf {\bibinfo {volume} {372}}~(\bibinfo {number}
  {2007}),\ \bibinfo {pages} {20120505}}\BibitemShut {NoStop}%
\bibitem [{\citenamefont {McDonald}\ and\ \citenamefont
  {Kaufman}(1979)}]{mcdonald1979spectrum}%
  \BibitemOpen
  \bibfield  {author} {\bibinfo {author} {\bibnamefont {McDonald},
  \bibfnamefont {S.~W.}}, \ and\ \bibinfo {author} {\bibfnamefont {A.~N.}\
  \bibnamefont {Kaufman}}} (\bibinfo {year} {1979}),\ \href {\doibase
  10.1103/PhysRevLett.42.1189} {\bibfield  {journal} {\bibinfo  {journal}
  {Phys. Rev. Lett.}\ }\textbf {\bibinfo {volume} {42}}~(\bibinfo {number}
  {18}),\ \bibinfo {pages} {1189}}\BibitemShut {NoStop}%
\bibitem [{\citenamefont {McDonald}\ and\ \citenamefont
  {Kaufman}(1988)}]{mcdonald1988wave}%
  \BibitemOpen
  \bibfield  {author} {\bibinfo {author} {\bibnamefont {McDonald},
  \bibfnamefont {S.~W.}}, \ and\ \bibinfo {author} {\bibfnamefont {A.~N.}\
  \bibnamefont {Kaufman}}} (\bibinfo {year} {1988}),\ \href {\doibase
  10.1103/PhysRevA.37.3067} {\bibfield  {journal} {\bibinfo  {journal} {Phys.
  Rev. A}\ }\textbf {\bibinfo {volume} {37}}~(\bibinfo {number} {8}),\ \bibinfo
  {pages} {3067}}\BibitemShut {NoStop}%
\bibitem [{\citenamefont {McKean}\ and\ \citenamefont
  {Singer}(1967)}]{mckean1967curvature}%
  \BibitemOpen
  \bibfield  {author} {\bibinfo {author} {\bibnamefont {McKean}, \bibfnamefont
  {H.~P.}}, \ and\ \bibinfo {author} {\bibfnamefont {I.~M.}\ \bibnamefont
  {Singer}}} (\bibinfo {year} {1967}),\ \href {\doibase
  10.1007/978-3-319-22237-0_6} {\bibfield  {journal} {\bibinfo  {journal} {J.
  Differential Geometry}\ }\textbf {\bibinfo {volume} {1}}~(\bibinfo {number}
  {1}),\ \bibinfo {pages} {43}}\BibitemShut {NoStop}%
\bibitem [{\citenamefont {Mehta}(2004)}]{mehta2004random}%
  \BibitemOpen
  \bibfield  {author} {\bibinfo {author} {\bibnamefont {Mehta}, \bibfnamefont
  {M.~L.}}} (\bibinfo {year} {2004}),\ \href@noop {} {\emph {\bibinfo {title}
  {Random matrices}}},\ \bibinfo {edition} {3rd}\ ed.,\ Vol.\ \bibinfo {volume}
  {142}\ (\bibinfo  {publisher} {Academic Press},\ \bibinfo {address} {San
  Diego, CA})\BibitemShut {NoStop}%
\bibitem [{\citenamefont {Melas}(1992)}]{melas1992nodal}%
  \BibitemOpen
  \bibfield  {author} {\bibinfo {author} {\bibnamefont {Melas}, \bibfnamefont
  {A.~D.}}} (\bibinfo {year} {1992}),\ \href {\doibase 10.4310/jdg/1214447811}
  {\bibfield  {journal} {\bibinfo  {journal} {J. Differential Geom.}\ }\textbf
  {\bibinfo {volume} {35}}~(\bibinfo {number} {1}),\ \bibinfo {pages}
  {255}}\BibitemShut {NoStop}%
\bibitem [{\citenamefont {Melrose}(1980)}]{melrose1980weyl}%
  \BibitemOpen
  \bibfield  {author} {\bibinfo {author} {\bibnamefont {Melrose}, \bibfnamefont
  {R.~B.}}} (\bibinfo {year} {1980}),\ in\ \href@noop {} {\emph {\bibinfo
  {booktitle} {Geometry of the Laplace Operator}}},\ \bibinfo {series} {Proc.
  Symp. Pure Math.}, Vol.~\bibinfo {volume} {36},\ \bibinfo {editor} {edited
  by\ \bibinfo {editor} {\bibfnamefont {R.}~\bibnamefont {Osserman}}\ and\
  \bibinfo {editor} {\bibfnamefont {A.}~\bibnamefont {Weinstein}}}\ (\bibinfo
  {publisher} {AMS, Providence, RI})\ pp.\ \bibinfo {pages}
  {257--273}\BibitemShut {NoStop}%
\bibitem [{\citenamefont {Milner}\ \emph {et~al.}(2001)\citenamefont {Milner},
  \citenamefont {Hanssen}, \citenamefont {Campbell},\ and\ \citenamefont
  {Raizen}}]{milner2001optical}%
  \BibitemOpen
  \bibfield  {author} {\bibinfo {author} {\bibnamefont {Milner}, \bibfnamefont
  {V.}}, \bibinfo {author} {\bibfnamefont {J.~L.}\ \bibnamefont {Hanssen}},
  \bibinfo {author} {\bibfnamefont {W.~C.}\ \bibnamefont {Campbell}}, \ and\
  \bibinfo {author} {\bibfnamefont {M.~G.}\ \bibnamefont {Raizen}}} (\bibinfo
  {year} {2001}),\ \href {\doibase 10.1103/PhysRevLett.86.1514} {\bibfield
  {journal} {\bibinfo  {journal} {Phys. Rev. Lett.}\ }\textbf {\bibinfo
  {volume} {86}}~(\bibinfo {number} {8}),\ \bibinfo {pages} {1514}}\BibitemShut
  {NoStop}%
\bibitem [{\citenamefont {Milnor}(1964)}]{milnor1964eigenvalues}%
  \BibitemOpen
  \bibfield  {author} {\bibinfo {author} {\bibnamefont {Milnor}, \bibfnamefont
  {J.}}} (\bibinfo {year} {1964}),\ \href {\doibase 10.1073/pnas.51.4.542}
  {\bibfield  {journal} {\bibinfo  {journal} {Proc. Nat. Acad. Sci.}\ }\textbf
  {\bibinfo {volume} {51}}~(\bibinfo {number} {4}),\ \bibinfo {pages}
  {542}}\BibitemShut {NoStop}%
\bibitem [{\citenamefont {Mirlin}(2000)}]{mirlin2000statistics}%
  \BibitemOpen
  \bibfield  {author} {\bibinfo {author} {\bibnamefont {Mirlin}, \bibfnamefont
  {A.~D.}}} (\bibinfo {year} {2000}),\ \href {\doibase Statistics of energy
  levels and eigenfunctions in disordered systems} {\bibfield  {journal}
  {\bibinfo  {journal} {Phys. Rep.}\ }\textbf {\bibinfo {volume}
  {326}}~(\bibinfo {number} {5}),\ \bibinfo {pages} {259}}\BibitemShut
  {NoStop}%
\bibitem [{\citenamefont {Moler}(2012)}]{moler2012}%
  \BibitemOpen
  \bibfield  {author} {\bibinfo {author} {\bibnamefont {Moler}, \bibfnamefont
  {C.}}} (\bibinfo {year} {2012}),\ \href
  {http://blogs.mathworks.com/cleve/2012/08/13/can-one-hear-the-shape-of-a-drum-part-2-eigenfunctions/}
  {\enquote {\bibinfo {title} {Can {O}ne {H}ear the {S}hape of a {D}rum? {P}art
  2, {E}igenfunctions},}\ }\bibinfo {howpublished} {Cleve's Corner: Cleve Moler
  on Mathematics and Computing, MathWorks Blogs}\BibitemShut {NoStop}%
\bibitem [{\citenamefont {Molina}\ \emph {et~al.}(2012)\citenamefont {Molina},
  \citenamefont {Jalabert}, \citenamefont {Weinmann},\ and\ \citenamefont
  {Jacquod}}]{molina2012scattering}%
  \BibitemOpen
  \bibfield  {author} {\bibinfo {author} {\bibnamefont {Molina}, \bibfnamefont
  {R.~A.}}, \bibinfo {author} {\bibfnamefont {R.~A.}\ \bibnamefont {Jalabert}},
  \bibinfo {author} {\bibfnamefont {D.}~\bibnamefont {Weinmann}}, \ and\
  \bibinfo {author} {\bibfnamefont {P.}~\bibnamefont {Jacquod}}} (\bibinfo
  {year} {2012}),\ \href {\doibase 10.1103/PhysRevLett.108.076803} {\bibfield
  {journal} {\bibinfo  {journal} {Phys. Rev. Lett.}\ }\textbf {\bibinfo
  {volume} {108}}~(\bibinfo {number} {7}),\ \bibinfo {pages}
  {076803}}\BibitemShut {NoStop}%
\bibitem [{\citenamefont {Molina-Terriza}\ \emph {et~al.}(2001)\citenamefont
  {Molina-Terriza}, \citenamefont {Recolons}, \citenamefont {Torres},
  \citenamefont {Torner},\ and\ \citenamefont
  {Wright}}]{molina2001observation}%
  \BibitemOpen
  \bibfield  {author} {\bibinfo {author} {\bibnamefont {Molina-Terriza},
  \bibfnamefont {G.}}, \bibinfo {author} {\bibfnamefont {J.}~\bibnamefont
  {Recolons}}, \bibinfo {author} {\bibfnamefont {J.~P.}\ \bibnamefont
  {Torres}}, \bibinfo {author} {\bibfnamefont {L.}~\bibnamefont {Torner}}, \
  and\ \bibinfo {author} {\bibfnamefont {E.~M.}\ \bibnamefont {Wright}}}
  (\bibinfo {year} {2001}),\ \href {\doibase 10.1103/PhysRevLett.87.023902}
  {\bibfield  {journal} {\bibinfo  {journal} {Phys. Rev. Lett.}\ }\textbf
  {\bibinfo {volume} {87}}~(\bibinfo {number} {2}),\ \bibinfo {pages}
  {023902}}\BibitemShut {NoStop}%
\bibitem [{\citenamefont {Monastra}\ \emph {et~al.}(2002)\citenamefont
  {Monastra}, \citenamefont {Smilansky},\ and\ \citenamefont
  {Gnutzmann}}]{monastra2002avoided}%
  \BibitemOpen
  \bibfield  {author} {\bibinfo {author} {\bibnamefont {Monastra},
  \bibfnamefont {A.~G.}}, \bibinfo {author} {\bibfnamefont {U.}~\bibnamefont
  {Smilansky}}, \ and\ \bibinfo {author} {\bibfnamefont {S.}~\bibnamefont
  {Gnutzmann}}} (\bibinfo {year} {2002}),\ \href
  {http://www.arxiv.org/abs/nlin/0212006} {\enquote {\bibinfo {title} {Avoided
  intersections of nodal lines},}\ }\bibinfo {note} {{a}rXiv preprint
  nlin/0212006}\BibitemShut {NoStop}%
\bibitem [{\citenamefont {Monastra}\ \emph {et~al.}(2003)\citenamefont
  {Monastra}, \citenamefont {Smilansky},\ and\ \citenamefont
  {Gnutzmann}}]{monastra2003avoided}%
  \BibitemOpen
  \bibfield  {author} {\bibinfo {author} {\bibnamefont {Monastra},
  \bibfnamefont {A.~G.}}, \bibinfo {author} {\bibfnamefont {U.}~\bibnamefont
  {Smilansky}}, \ and\ \bibinfo {author} {\bibfnamefont {S.}~\bibnamefont
  {Gnutzmann}}} (\bibinfo {year} {2003}),\ \href {\doibase
  10.1088/0305-4470/36/7/304} {\bibfield  {journal} {\bibinfo  {journal} {J.
  Phys. A: Math. Gen.}\ }\textbf {\bibinfo {volume} {36}}~(\bibinfo {number}
  {7}),\ \bibinfo {pages} {1845}}\BibitemShut {NoStop}%
\bibitem [{\citenamefont {Mondragon}\ and\ \citenamefont
  {Berry}(1989)}]{mondragon1989quantum}%
  \BibitemOpen
  \bibfield  {author} {\bibinfo {author} {\bibnamefont {Mondragon},
  \bibfnamefont {R.~J.}}, \ and\ \bibinfo {author} {\bibfnamefont {M.~V.}\
  \bibnamefont {Berry}}} (\bibinfo {year} {1989}),\ \href {\doibase
  10.1098/rspa.1989.0081} {\bibfield  {journal} {\bibinfo  {journal} {Proc. R.
  Soc. Lond. A}\ }\textbf {\bibinfo {volume} {424}}~(\bibinfo {number}
  {1867}),\ \bibinfo {pages} {263}}\BibitemShut {NoStop}%
\bibitem [{\citenamefont {Montangero}\ \emph {et~al.}(2009)\citenamefont
  {Montangero}, \citenamefont {Frustaglia}, \citenamefont {Calarco},\ and\
  \citenamefont {Fazio}}]{montangero2009quantum}%
  \BibitemOpen
  \bibfield  {author} {\bibinfo {author} {\bibnamefont {Montangero},
  \bibfnamefont {S.}}, \bibinfo {author} {\bibfnamefont {D.}~\bibnamefont
  {Frustaglia}}, \bibinfo {author} {\bibfnamefont {T.}~\bibnamefont {Calarco}},
  \ and\ \bibinfo {author} {\bibfnamefont {R.}~\bibnamefont {Fazio}}} (\bibinfo
  {year} {2009}),\ \href {\doibase 10.1209/0295-5075/88/30006} {\bibfield
  {journal} {\bibinfo  {journal} {EPL}\ }\textbf {\bibinfo {volume}
  {88}}~(\bibinfo {number} {3}),\ \bibinfo {pages} {30006}}\BibitemShut
  {NoStop}%
\bibitem [{\citenamefont {Moser}(1962)}]{moser1962invariant}%
  \BibitemOpen
  \bibfield  {author} {\bibinfo {author} {\bibnamefont {Moser}, \bibfnamefont
  {J.}}} (\bibinfo {year} {1962}),\ \href@noop {} {\bibfield  {journal}
  {\bibinfo  {journal} {Nachr. Akad. Wiss. GÂšottingen}\ }\textbf {\bibinfo
  {volume} {1}},\ \bibinfo {pages} {1}}\BibitemShut {NoStop}%
\bibitem [{\citenamefont {Moudgalya}\ \emph {et~al.}(2015)\citenamefont
  {Moudgalya}, \citenamefont {Chandra},\ and\ \citenamefont
  {Jain}}]{moudgalya2015finite}%
  \BibitemOpen
  \bibfield  {author} {\bibinfo {author} {\bibnamefont {Moudgalya},
  \bibfnamefont {S.}}, \bibinfo {author} {\bibfnamefont {S.}~\bibnamefont
  {Chandra}}, \ and\ \bibinfo {author} {\bibfnamefont {S.~R.}\ \bibnamefont
  {Jain}}} (\bibinfo {year} {2015}),\ \href {\doibase
  10.1016/j.aop.2015.05.033} {\bibfield  {journal} {\bibinfo  {journal} {Ann.
  Phys.}\ }\textbf {\bibinfo {volume} {361}},\ \bibinfo {pages}
  {82}}\BibitemShut {NoStop}%
\bibitem [{\citenamefont {Nadirashvili}(1988)}]{0036-0279-43-4-L23}%
  \BibitemOpen
  \bibfield  {author} {\bibinfo {author} {\bibnamefont {Nadirashvili},
  \bibfnamefont {N.}}} (\bibinfo {year} {1988}),\ \href {\doibase
  10.1070/RM1988v043n04ABEH001905} {\bibfield  {journal} {\bibinfo  {journal}
  {Russ. Math. Surv.}\ }\textbf {\bibinfo {volume} {43}}~(\bibinfo {number}
  {4}),\ \bibinfo {pages} {227}}\BibitemShut {NoStop}%
\bibitem [{\citenamefont {Nakamura}\ and\ \citenamefont
  {Harayama}(2004)}]{nakamura2004quantum}%
  \BibitemOpen
  \bibfield  {author} {\bibinfo {author} {\bibnamefont {Nakamura},
  \bibfnamefont {K.}}, \ and\ \bibinfo {author} {\bibfnamefont
  {T.}~\bibnamefont {Harayama}}} (\bibinfo {year} {2004}),\ \href@noop {}
  {\emph {\bibinfo {title} {Quantum Chaos and Quantum Dots}}},\ Vol.~\bibinfo
  {volume} {3}\ (\bibinfo  {publisher} {Oxford University Press},\ \bibinfo
  {address} {Oxford, UK})\BibitemShut {NoStop}%
\bibitem [{\citenamefont {Nakayama}(2015)}]{nakayama2015scale}%
  \BibitemOpen
  \bibfield  {author} {\bibinfo {author} {\bibnamefont {Nakayama},
  \bibfnamefont {{\relax Yu}.}}} (\bibinfo {year} {2015}),\ \href {\doibase
  10.1016/j.physrep.2014.12.003} {\bibfield  {journal} {\bibinfo  {journal}
  {Phys. Rep.}\ }\textbf {\bibinfo {volume} {569}},\ \bibinfo {pages}
  {1}}\BibitemShut {NoStop}%
\bibitem [{\citenamefont {Narimanov}\ \emph {et~al.}(2001)\citenamefont
  {Narimanov}, \citenamefont {Baranger}, \citenamefont {Cerruti},\ and\
  \citenamefont {Tomsovic}}]{narimanov2001semiclassical}%
  \BibitemOpen
  \bibfield  {author} {\bibinfo {author} {\bibnamefont {Narimanov},
  \bibfnamefont {E.~E.}}, \bibinfo {author} {\bibfnamefont {H.~U.}\
  \bibnamefont {Baranger}}, \bibinfo {author} {\bibfnamefont {N.~R.}\
  \bibnamefont {Cerruti}}, \ and\ \bibinfo {author} {\bibfnamefont
  {S.}~\bibnamefont {Tomsovic}}} (\bibinfo {year} {2001}),\ \href {\doibase
  10.1103/PhysRevB.64.235329} {\bibfield  {journal} {\bibinfo  {journal} {Phys.
  Rev. B}\ }\textbf {\bibinfo {volume} {64}}~(\bibinfo {number} {23}),\
  \bibinfo {pages} {235329}}\BibitemShut {NoStop}%
\bibitem [{\citenamefont {Nastasescu}(2011)}]{nastasescu2011}%
  \BibitemOpen
  \bibfield  {author} {\bibinfo {author} {\bibnamefont {Nastasescu},
  \bibfnamefont {M.~M.}}} (\bibinfo {year} {2011}),\ \href@noop {} {\enquote
  {\bibinfo {title} {The number of ovals of a random real plane curve},}\
  }\bibinfo {howpublished} {Senior thesis (Princeton University)}\BibitemShut
  {NoStop}%
\bibitem [{\citenamefont {Nazarov}\ and\ \citenamefont
  {Sodin}(2009)}]{nazarov2009number}%
  \BibitemOpen
  \bibfield  {author} {\bibinfo {author} {\bibnamefont {Nazarov}, \bibfnamefont
  {F.}}, \ and\ \bibinfo {author} {\bibfnamefont {M.}~\bibnamefont {Sodin}}}
  (\bibinfo {year} {2009}),\ \href {\doibase 10.1353/ajm.0.0070} {\bibinfo
  {journal} {Am. J. Math.}\ ,\ \bibinfo {pages} {1337}}\BibitemShut {NoStop}%
\bibitem [{\citenamefont {Nazarov}\ and\ \citenamefont
  {Sodin}(2016)}]{nazarov2016asymptotic}%
  \BibitemOpen
\bibfield  {journal} {  }\bibfield  {author} {\bibinfo {author} {\bibnamefont
  {Nazarov}, \bibfnamefont {F.}}, \ and\ \bibinfo {author} {\bibfnamefont
  {M.}~\bibnamefont {Sodin}}} (\bibinfo {year} {2016}),\ \href {\doibase
  10.15407/mag12.03.205} {\bibfield  {journal} {\bibinfo  {journal} {Zh. Mat.
  Fiz. Anal. Geom.}\ }\textbf {\bibinfo {volume} {12}}~(\bibinfo {number}
  {3}),\ \bibinfo {pages} {205}}\BibitemShut {NoStop}%
\bibitem [{\citenamefont {Netrusov}\ and\ \citenamefont
  {Safarov}(2005)}]{netrusov2005weyl}%
  \BibitemOpen
  \bibfield  {author} {\bibinfo {author} {\bibnamefont {Netrusov},
  \bibfnamefont {Y.}}, \ and\ \bibinfo {author} {\bibfnamefont
  {Y.}~\bibnamefont {Safarov}}} (\bibinfo {year} {2005}),\ \href {\doibase
  10.1007/s00220-004-1158-8} {\bibfield  {journal} {\bibinfo  {journal}
  {Commun. Math. Phys.}\ }\textbf {\bibinfo {volume} {253}}~(\bibinfo {number}
  {2}),\ \bibinfo {pages} {481}}\BibitemShut {NoStop}%
\bibitem [{\citenamefont {Nienhuis}(1987)}]{nienhuis1987}%
  \BibitemOpen
  \bibfield  {author} {\bibinfo {author} {\bibnamefont {Nienhuis},
  \bibfnamefont {B.}}} (\bibinfo {year} {1987}),\ in\ \href@noop {} {\emph
  {\bibinfo {booktitle} {Phase transitions and critical phenomena}}},\
  Vol.~\bibinfo {volume} {11},\ \bibinfo {editor} {edited by\ \bibinfo {editor}
  {\bibfnamefont {C.}~\bibnamefont {Domb}}\ and\ \bibinfo {editor}
  {\bibfnamefont {J.~L.}\ \bibnamefont {Lebowitz}}}\ (\bibinfo  {publisher}
  {Academic Press},\ \bibinfo {address} {London})\BibitemShut {NoStop}%
\bibitem [{\citenamefont {N{\"o}ckel}\ and\ \citenamefont
  {Stone}(1997)}]{nockel1997ray}%
  \BibitemOpen
  \bibfield  {author} {\bibinfo {author} {\bibnamefont {N{\"o}ckel},
  \bibfnamefont {J.~U.}}, \ and\ \bibinfo {author} {\bibfnamefont {A.~D.}\
  \bibnamefont {Stone}}} (\bibinfo {year} {1997}),\ \href {\doibase
  10.1038/385045a0} {\bibfield  {journal} {\bibinfo  {journal} {Nature}\
  }\textbf {\bibinfo {volume} {385}},\ \bibinfo {pages} {45}}\BibitemShut
  {NoStop}%
\bibitem [{\citenamefont {Nonnenmacher}(2013)}]{nonnenmacher2013}%
  \BibitemOpen
  \bibfield  {author} {\bibinfo {author} {\bibnamefont {Nonnenmacher},
  \bibfnamefont {S.}}} (\bibinfo {year} {2013}),\ in\ \href {\doibase
  10.1007/978-3-0348-0697-8_6} {\emph {\bibinfo {booktitle} {Chaos:
  Poincar{\'e} Seminar 2010}}},\ \bibinfo {series and number} {Progress in
  Mathematical Physics}\ (\bibinfo  {publisher} {Springer},\ \bibinfo {address}
  {Basel})\ pp.\ \bibinfo {pages} {193--238}\BibitemShut {NoStop}%
\bibitem [{\citenamefont {Nonnenmacher}\ and\ \citenamefont
  {Zworski}(2005)}]{nonnenmacher2005fractal}%
  \BibitemOpen
  \bibfield  {author} {\bibinfo {author} {\bibnamefont {Nonnenmacher},
  \bibfnamefont {S.}}, \ and\ \bibinfo {author} {\bibfnamefont
  {M.}~\bibnamefont {Zworski}}} (\bibinfo {year} {2005}),\ \href {\doibase
  10.1088/0305-4470/38/49/014} {\bibfield  {journal} {\bibinfo  {journal} {J.
  Phys. A: Math. Gen.}\ }\textbf {\bibinfo {volume} {38}}~(\bibinfo {number}
  {49}),\ \bibinfo {pages} {10683}}\BibitemShut {NoStop}%
\bibitem [{\citenamefont {Nye}(1999)}]{nye1999natural}%
  \BibitemOpen
  \bibfield  {author} {\bibinfo {author} {\bibnamefont {Nye}, \bibfnamefont
  {J.~F.}}} (\bibinfo {year} {1999}),\ \href@noop {} {\emph {\bibinfo {title}
  {Natural focusing and fine structure of light: caustics and wave
  dislocations}}}\ (\bibinfo  {publisher} {Institute of Physics Publishing},\
  \bibinfo {address} {Bristol})\BibitemShut {NoStop}%
\bibitem [{\citenamefont {Nye}\ and\ \citenamefont
  {Berry}(1974)}]{nye1974dislocations}%
  \BibitemOpen
  \bibfield  {author} {\bibinfo {author} {\bibnamefont {Nye}, \bibfnamefont
  {J.~F.}}, \ and\ \bibinfo {author} {\bibfnamefont {M.~V.}\ \bibnamefont
  {Berry}}} (\bibinfo {year} {1974}),\ \href {\doibase 10.1098/rspa.1974.0012}
  {\bibfield  {journal} {\bibinfo  {journal} {Proc. R. Soc. Lond. A}\ }\textbf
  {\bibinfo {volume} {336}}~(\bibinfo {number} {1605}),\ \bibinfo {pages}
  {165}}\BibitemShut {NoStop}%
\bibitem [{\citenamefont {O'Connor}\ and\ \citenamefont
  {Heller}(1988)}]{o1988quantum}%
  \BibitemOpen
  \bibfield  {author} {\bibinfo {author} {\bibnamefont {O'Connor},
  \bibfnamefont {P.~W.}}, \ and\ \bibinfo {author} {\bibfnamefont {E.~J.}\
  \bibnamefont {Heller}}} (\bibinfo {year} {1988}),\ \href {\doibase
  10.1103/PhysRevLett.61.2288} {\bibfield  {journal} {\bibinfo  {journal}
  {Phys. Rev. Lett.}\ }\textbf {\bibinfo {volume} {61}}~(\bibinfo {number}
  {20}),\ \bibinfo {pages} {2288}}\BibitemShut {NoStop}%
\bibitem [{\citenamefont {O'Holleran}\ \emph
  {et~al.}(2006{\natexlab{a}})\citenamefont {O'Holleran}, \citenamefont
  {Dennis},\ and\ \citenamefont {Padgett}}]{o2006illustrations}%
  \BibitemOpen
  \bibfield  {author} {\bibinfo {author} {\bibnamefont {O'Holleran},
  \bibfnamefont {K.}}, \bibinfo {author} {\bibfnamefont {M.~R.}\ \bibnamefont
  {Dennis}}, \ and\ \bibinfo {author} {\bibfnamefont {M.~J.}\ \bibnamefont
  {Padgett}}} (\bibinfo {year} {2006}{\natexlab{a}}),\ \href {\doibase
  10.2971/jeos.2006.06008} {\bibfield  {journal} {\bibinfo  {journal} {J. Eur.
  Opt. Soc.-Rapid Publ.}\ }\textbf {\bibinfo {volume} {1}},\ \bibinfo {pages}
  {06008}}\BibitemShut {NoStop}%
\bibitem [{\citenamefont {O'Holleran}\ \emph
  {et~al.}(2006{\natexlab{b}})\citenamefont {O'Holleran}, \citenamefont
  {Padgett},\ and\ \citenamefont {Dennis}}]{o2006topology}%
  \BibitemOpen
  \bibfield  {author} {\bibinfo {author} {\bibnamefont {O'Holleran},
  \bibfnamefont {K.}}, \bibinfo {author} {\bibfnamefont {M.~J.}\ \bibnamefont
  {Padgett}}, \ and\ \bibinfo {author} {\bibfnamefont {M.~R.}\ \bibnamefont
  {Dennis}}} (\bibinfo {year} {2006}{\natexlab{b}}),\ \href {\doibase
  10.1364/OE.14.003039} {\bibfield  {journal} {\bibinfo  {journal} {Opt.
  Express}\ }\textbf {\bibinfo {volume} {14}}~(\bibinfo {number} {7}),\
  \bibinfo {pages} {3039}}\BibitemShut {NoStop}%
\bibitem [{\citenamefont {Olendski}\ and\ \citenamefont
  {Mikhailovska}(2003)}]{olendski2003localized}%
  \BibitemOpen
  \bibfield  {author} {\bibinfo {author} {\bibnamefont {Olendski},
  \bibfnamefont {O.}}, \ and\ \bibinfo {author} {\bibfnamefont
  {L.}~\bibnamefont {Mikhailovska}}} (\bibinfo {year} {2003}),\ \href {\doibase
  10.1103/PhysRevE.67.056625} {\bibfield  {journal} {\bibinfo  {journal} {Phys.
  Rev. E}\ }\textbf {\bibinfo {volume} {67}}~(\bibinfo {number} {5}),\ \bibinfo
  {pages} {056625}}\BibitemShut {NoStop}%
\bibitem [{\citenamefont {O'Neil}\ \emph {et~al.}(2002)\citenamefont {O'Neil},
  \citenamefont {MacVicar}, \citenamefont {Allen},\ and\ \citenamefont
  {Padgett}}]{o2002intrinsic}%
  \BibitemOpen
  \bibfield  {author} {\bibinfo {author} {\bibnamefont {O'Neil}, \bibfnamefont
  {A.~T.}}, \bibinfo {author} {\bibfnamefont {I.}~\bibnamefont {MacVicar}},
  \bibinfo {author} {\bibfnamefont {L.}~\bibnamefont {Allen}}, \ and\ \bibinfo
  {author} {\bibfnamefont {M.~J.}\ \bibnamefont {Padgett}}} (\bibinfo {year}
  {2002}),\ \href {\doibase 10.1103/PhysRevLett.88.053601} {\bibfield
  {journal} {\bibinfo  {journal} {Phys. Rev. Lett.}\ }\textbf {\bibinfo
  {volume} {88}}~(\bibinfo {number} {5}),\ \bibinfo {pages}
  {053601}}\BibitemShut {NoStop}%
\bibitem [{\citenamefont {Oren}\ and\ \citenamefont
  {Band}(2012)}]{oren2012isospectral}%
  \BibitemOpen
  \bibfield  {author} {\bibinfo {author} {\bibnamefont {Oren}, \bibfnamefont
  {I.}}, \ and\ \bibinfo {author} {\bibfnamefont {R.}~\bibnamefont {Band}}}
  (\bibinfo {year} {2012}),\ \href {\doibase 10.1088/1751-8113/45/13/135203}
  {\bibfield  {journal} {\bibinfo  {journal} {J. Phys. A: Math. Theor.}\
  }\textbf {\bibinfo {volume} {45}}~(\bibinfo {number} {13}),\ \bibinfo {pages}
  {135203}}\BibitemShut {NoStop}%
\bibitem [{\citenamefont {Ott}(2002)}]{ott2002chaos}%
  \BibitemOpen
  \bibfield  {author} {\bibinfo {author} {\bibnamefont {Ott}, \bibfnamefont
  {E.}}} (\bibinfo {year} {2002}),\ \href@noop {} {\emph {\bibinfo {title}
  {Chaos in Dynamical Systems}}},\ \bibinfo {edition} {2nd}\ ed.\ (\bibinfo
  {publisher} {Cambridge University Press},\ \bibinfo {address} {Cambridge,
  UK})\BibitemShut {NoStop}%
\bibitem [{\citenamefont {Patel}\ \emph {et~al.}(1998)\citenamefont {Patel},
  \citenamefont {Stewart}, \citenamefont {Marcus}, \citenamefont
  {G{\"o}k{\c{c}}eda{\u{g}}}, \citenamefont {Alhassid}, \citenamefont {Stone},
  \citenamefont {Duru{\"o}z},\ and\ \citenamefont
  {Harris~Jr.}}]{patel1998changing}%
  \BibitemOpen
  \bibfield  {author} {\bibinfo {author} {\bibnamefont {Patel}, \bibfnamefont
  {S.~R.}}, \bibinfo {author} {\bibfnamefont {D.~R.}\ \bibnamefont {Stewart}},
  \bibinfo {author} {\bibfnamefont {C.~M.}\ \bibnamefont {Marcus}}, \bibinfo
  {author} {\bibfnamefont {M.}~\bibnamefont {G{\"o}k{\c{c}}eda{\u{g}}}},
  \bibinfo {author} {\bibfnamefont {Y.}~\bibnamefont {Alhassid}}, \bibinfo
  {author} {\bibfnamefont {A.~D.}\ \bibnamefont {Stone}}, \bibinfo {author}
  {\bibfnamefont {r.}~\bibnamefont {Duru{\"o}z}}, \ and\ \bibinfo {author}
  {\bibfnamefont {J.~S.}\ \bibnamefont {Harris~Jr.}}} (\bibinfo {year}
  {1998}),\ \href {\doibase 10.1103/PhysRevLett.81.5900} {\bibfield  {journal}
  {\bibinfo  {journal} {Phys. Rev. Lett.}\ }\textbf {\bibinfo {volume}
  {81}}~(\bibinfo {number} {26}),\ \bibinfo {pages} {5900}}\BibitemShut
  {NoStop}%
\bibitem [{\citenamefont {Payne}(1967)}]{payne}%
  \BibitemOpen
  \bibfield  {author} {\bibinfo {author} {\bibnamefont {Payne}, \bibfnamefont
  {L.~E.}}} (\bibinfo {year} {1967}),\ \href {\doibase 10.1137/1009070}
  {\bibfield  {journal} {\bibinfo  {journal} {SIAM Rev.}\ }\textbf {\bibinfo
  {volume} {9}}~(\bibinfo {number} {3}),\ \bibinfo {pages} {453}}\BibitemShut
  {NoStop}%
\bibitem [{\citenamefont {Payne}(1973)}]{payne1973two}%
  \BibitemOpen
  \bibfield  {author} {\bibinfo {author} {\bibnamefont {Payne}, \bibfnamefont
  {L.~E.}}} (\bibinfo {year} {1973}),\ \href {\doibase 10.1007/BF01597076}
  {\bibfield  {journal} {\bibinfo  {journal} {Z. Angew. Math. Phys.}\ }\textbf
  {\bibinfo {volume} {24}}~(\bibinfo {number} {5}),\ \bibinfo {pages}
  {721}}\BibitemShut {NoStop}%
\bibitem [{\citenamefont {Payne}\ \emph {et~al.}(1956)\citenamefont {Payne},
  \citenamefont {P{\'o}lya},\ and\ \citenamefont
  {Weinberger}}]{payne1956ratio}%
  \BibitemOpen
  \bibfield  {author} {\bibinfo {author} {\bibnamefont {Payne}, \bibfnamefont
  {L.~E.}}, \bibinfo {author} {\bibfnamefont {G.}~\bibnamefont {P{\'o}lya}}, \
  and\ \bibinfo {author} {\bibfnamefont {H.~F.}\ \bibnamefont {Weinberger}}}
  (\bibinfo {year} {1956}),\ \href {\doibase 10.1002/sapm1956351289} {\bibfield
   {journal} {\bibinfo  {journal} {J. Math. \& Phys. [Stud. Appl. Math.]}\
  }\textbf {\bibinfo {volume} {35}}~(\bibinfo {number} {1-4}),\ \bibinfo
  {pages} {289}}\BibitemShut {NoStop}%
\bibitem [{\citenamefont {Payne}\ and\ \citenamefont
  {Weinberger}(1960{\natexlab{a}})}]{payne1960faber}%
  \BibitemOpen
  \bibfield  {author} {\bibinfo {author} {\bibnamefont {Payne}, \bibfnamefont
  {L.~E.}}, \ and\ \bibinfo {author} {\bibfnamefont {H.~F.}\ \bibnamefont
  {Weinberger}}} (\bibinfo {year} {1960}{\natexlab{a}}),\ \href {\doibase
  10.1002/sapm1960391182} {\bibfield  {journal} {\bibinfo  {journal} {J. Math.
  \& Phys. [Stud. Appl. Math.]}\ }\textbf {\bibinfo {volume} {39}}~(\bibinfo
  {number} {1-4}),\ \bibinfo {pages} {182}}\BibitemShut {NoStop}%
\bibitem [{\citenamefont {Payne}\ and\ \citenamefont
  {Weinberger}(1960{\natexlab{b}})}]{payne1960optimal}%
  \BibitemOpen
  \bibfield  {author} {\bibinfo {author} {\bibnamefont {Payne}, \bibfnamefont
  {L.~E.}}, \ and\ \bibinfo {author} {\bibfnamefont {H.~F.}\ \bibnamefont
  {Weinberger}}} (\bibinfo {year} {1960}{\natexlab{b}}),\ \href {\doibase
  10.1007/BF00252910} {\bibfield  {journal} {\bibinfo  {journal} {Arch.
  Rational Mech. Anal.}\ }\textbf {\bibinfo {volume} {5}}~(\bibinfo {number}
  {1}),\ \bibinfo {pages} {286}}\BibitemShut {NoStop}%
\bibitem [{\citenamefont {Peetre}(1957)}]{peetre1957generalization}%
  \BibitemOpen
  \bibfield  {author} {\bibinfo {author} {\bibnamefont {Peetre}, \bibfnamefont
  {J.}}} (\bibinfo {year} {1957}),\ \href {\doibase 10.7146/math.scand.a-10484}
  {\bibinfo  {journal} {Math. Scand.}\ ,\ \bibinfo {pages} {15}}\BibitemShut
  {NoStop}%
\bibitem [{\citenamefont {Percival}(1973)}]{percival1973regular}%
  \BibitemOpen
\bibfield  {journal} {  }\bibfield  {author} {\bibinfo {author} {\bibnamefont
  {Percival}, \bibfnamefont {I.~C.}}} (\bibinfo {year} {1973}),\ \href
  {\doibase 10.1088/0022-3700/6/9/002} {\bibfield  {journal} {\bibinfo
  {journal} {J. Phys. B: At. Mol. Phys.}\ }\textbf {\bibinfo {volume}
  {6}}~(\bibinfo {number} {9}),\ \bibinfo {pages} {L229}}\BibitemShut {NoStop}%
\bibitem [{\citenamefont {Percival}(1979)}]{percival1979variational}%
  \BibitemOpen
  \bibfield  {author} {\bibinfo {author} {\bibnamefont {Percival},
  \bibfnamefont {I.~C.}}} (\bibinfo {year} {1979}),\ \href {\doibase
  10.1088/0305-4470/12/3/001} {\bibfield  {journal} {\bibinfo  {journal} {J.
  Phys. A: Math. Gen.}\ }\textbf {\bibinfo {volume} {12}}~(\bibinfo {number}
  {3}),\ \bibinfo {pages} {L57}}\BibitemShut {NoStop}%
\bibitem [{\citenamefont {Pleijel}(1954)}]{pleijel1954study}%
  \BibitemOpen
  \bibfield  {author} {\bibinfo {author} {\bibnamefont {Pleijel}, \bibfnamefont
  {A.}}} (\bibinfo {year} {1954}),\ \href {\doibase 10.1007/BF02591229}
  {\bibfield  {journal} {\bibinfo  {journal} {Ark. Mat.}\ }\textbf {\bibinfo
  {volume} {2}}~(\bibinfo {number} {6}),\ \bibinfo {pages} {553}}\BibitemShut
  {NoStop}%
\bibitem [{\citenamefont {Pleijel}(1956)}]{pleijel1956remarks}%
  \BibitemOpen
  \bibfield  {author} {\bibinfo {author} {\bibnamefont {Pleijel}, \bibfnamefont
  {A.}}} (\bibinfo {year} {1956}),\ \href {\doibase 10.1002/cpa.3160090324}
  {\bibfield  {journal} {\bibinfo  {journal} {Comm. Pure Appl. Math.}\ }\textbf
  {\bibinfo {volume} {9}}~(\bibinfo {number} {3}),\ \bibinfo {pages}
  {543}}\BibitemShut {NoStop}%
\bibitem [{\citenamefont {Pockels}(1892)}]{pockels1892uber}%
  \BibitemOpen
  \bibfield  {author} {\bibinfo {author} {\bibnamefont {Pockels}, \bibfnamefont
  {F.}}} (\bibinfo {year} {1892}),\ \href@noop {} {\emph {\bibinfo {title}
  {{\"U}ber die partielle Differentialgleichung $\Delta u+ k^2\,u= 0$ and deren
  Auftreten in Mathematischen Physik}}},\ Historical Math. Monographs, Cornell
  University\ (\bibinfo  {publisher} {BG Teubner},\ \bibinfo {address}
  {Liepzig})\BibitemShut {NoStop}%
\bibitem [{\citenamefont {Poisson}(1829)}]{poisson1829memoire}%
  \BibitemOpen
  \bibfield  {author} {\bibinfo {author} {\bibnamefont {Poisson}, \bibfnamefont
  {S.-D.}}} (\bibinfo {year} {1829}),\ \href@noop {} {\emph {\bibinfo {title}
  {M{\'e}moire sur l'{\'e}quilibre et mouvement des corps {\'e}lastiques}}},\
  \bibinfo {series} {Mem. Acad. Sci. Paris}, Vol.~\bibinfo {volume} {9}\
  (\bibinfo  {publisher} {L'Acad{\'e}mie des sciences})\BibitemShut {NoStop}%
\bibitem [{\citenamefont {Polterovich}(2009)}]{polterovich2009pleijel}%
  \BibitemOpen
  \bibfield  {author} {\bibinfo {author} {\bibnamefont {Polterovich},
  \bibfnamefont {I.}}} (\bibinfo {year} {2009}),\ \href {\doibase
  10.1090/S0002-9939-08-09596-8} {\bibfield  {journal} {\bibinfo  {journal}
  {Proc. Amer. Math. Soc.}\ }\textbf {\bibinfo {volume} {137}}~(\bibinfo
  {number} {3}),\ \bibinfo {pages} {1021}}\BibitemShut {NoStop}%
\bibitem [{\citenamefont {P{\'o}lya}(1954)}]{polya1954mathematics}%
  \BibitemOpen
  \bibfield  {author} {\bibinfo {author} {\bibnamefont {P{\'o}lya},
  \bibfnamefont {G.}}} (\bibinfo {year} {1954}),\ \href@noop {} {\emph
  {\bibinfo {title} {Induction and analogy in mathematics}}},\ \bibinfo
  {series} {Mathematics and plausible reasoning}, Vol.~\bibinfo {volume} {1}\
  (\bibinfo  {publisher} {Princeton University Press},\ \bibinfo {address}
  {Princeton, NJ})\BibitemShut {NoStop}%
\bibitem [{\citenamefont {P{\'o}lya}(1955)}]{polya1955characteristic}%
  \BibitemOpen
  \bibfield  {author} {\bibinfo {author} {\bibnamefont {P{\'o}lya},
  \bibfnamefont {G.}}} (\bibinfo {year} {1955}),\ \href {\doibase
  10.1007/BF01187944} {\bibfield  {journal} {\bibinfo  {journal} {Math. Z.}\
  }\textbf {\bibinfo {volume} {63}}~(\bibinfo {number} {1}),\ \bibinfo {pages}
  {331}}\BibitemShut {NoStop}%
\bibitem [{\citenamefont {P{\'o}lya}(1961)}]{polya1961eigenvalues}%
  \BibitemOpen
  \bibfield  {author} {\bibinfo {author} {\bibnamefont {P{\'o}lya},
  \bibfnamefont {G.}}} (\bibinfo {year} {1961}),\ \href {\doibase
  10.1112/plms/s3-11.1.419} {\bibfield  {journal} {\bibinfo  {journal} {Proc.
  London Math. Soc.}\ }\textbf {\bibinfo {volume} {3}}~(\bibinfo {number}
  {1}),\ \bibinfo {pages} {419}}\BibitemShut {NoStop}%
\bibitem [{\citenamefont {P{\'o}lya}\ and\ \citenamefont {Szeg{\"
  o}}(1951)}]{polya1951isoperimetric}%
  \BibitemOpen
  \bibfield  {author} {\bibinfo {author} {\bibnamefont {P{\'o}lya},
  \bibfnamefont {G.}}, \ and\ \bibinfo {author} {\bibfnamefont {A.~M.~S.}\
  \bibnamefont {Szeg{\" o}}, \bibfnamefont {G{\'a}bor}}} (\bibinfo {year}
  {1951}),\ \href@noop {} {\emph {\bibinfo {title} {Isoperimetric inequalities
  in mathematical physics}}},\ \bibinfo {series} {Annals of Mathematics
  Studies}\ No.~\bibinfo {number} {27}\ (\bibinfo  {publisher} {Princeton
  University Press},\ \bibinfo {address} {Princeton, NJ})\BibitemShut {NoStop}%
\bibitem [{\citenamefont {Porter}(1965)}]{porter1965statistical}%
  \BibitemOpen
  \bibfield  {author} {\bibinfo {author} {\bibnamefont {Porter}, \bibfnamefont
  {C.~E.}}} (\bibinfo {year} {1965}),\ \href@noop {} {\emph {\bibinfo {title}
  {Statistical Theory of Spectra: Fluctuations}}}\ (\bibinfo  {publisher}
  {Academic Press},\ \bibinfo {address} {New York})\BibitemShut {NoStop}%
\bibitem [{\citenamefont {Prado}\ \emph {et~al.}(2009)\citenamefont {Prado},
  \citenamefont {Vergini}, \citenamefont {Benito},\ and\ \citenamefont
  {Borondo}}]{prado2009superscars}%
  \BibitemOpen
  \bibfield  {author} {\bibinfo {author} {\bibnamefont {Prado}, \bibfnamefont
  {S.~D.}}, \bibinfo {author} {\bibfnamefont {E.}~\bibnamefont {Vergini}},
  \bibinfo {author} {\bibfnamefont {R.~M.}\ \bibnamefont {Benito}}, \ and\
  \bibinfo {author} {\bibfnamefont {F.}~\bibnamefont {Borondo}}} (\bibinfo
  {year} {2009}),\ \href {\doibase 10.1209/0295-5075/88/40003} {\bibfield
  {journal} {\bibinfo  {journal} {EPL}\ }\textbf {\bibinfo {volume}
  {88}}~(\bibinfo {number} {4}),\ \bibinfo {pages} {40003}}\BibitemShut
  {NoStop}%
\bibitem [{\citenamefont {Protter}(1987)}]{protter1987can}%
  \BibitemOpen
  \bibfield  {author} {\bibinfo {author} {\bibnamefont {Protter}, \bibfnamefont
  {M.~H.}}} (\bibinfo {year} {1987}),\ \href {\doibase 10.1137/1029041}
  {\bibfield  {journal} {\bibinfo  {journal} {SIAM Rev.}\ }\textbf {\bibinfo
  {volume} {29}}~(\bibinfo {number} {2}),\ \bibinfo {pages} {185}}\BibitemShut
  {NoStop}%
\bibitem [{\citenamefont {Rabouw}\ and\ \citenamefont
  {Ruijgrok}(1981)}]{rabouw1981three}%
  \BibitemOpen
  \bibfield  {author} {\bibinfo {author} {\bibnamefont {Rabouw}, \bibfnamefont
  {F.}}, \ and\ \bibinfo {author} {\bibfnamefont {{\relax Th}.~W.}\
  \bibnamefont {Ruijgrok}}} (\bibinfo {year} {1981}),\ \href {\doibase
  10.1016/0378-4371(81)90008-X} {\bibfield  {journal} {\bibinfo  {journal}
  {Physica A}\ }\textbf {\bibinfo {volume} {109}}~(\bibinfo {number} {3}),\
  \bibinfo {pages} {500}}\BibitemShut {NoStop}%
\bibitem [{\citenamefont {Rayleigh}(1945)}]{rayleigh1945theory}%
  \BibitemOpen
  \bibfield  {author} {\bibinfo {author} {\bibnamefont {Rayleigh},
  \bibfnamefont {J.~W.~S.}}} (\bibinfo {year} {1945}),\ \href@noop {} {\emph
  {\bibinfo {title} {The Theory of Sound, Volumes I and II}}}\ (\bibinfo
  {publisher} {Dover Publications},\ \bibinfo {address} {New York})\BibitemShut
  {NoStop}%
\bibitem [{\citenamefont {Reichl}(1992)}]{reichltransition}%
  \BibitemOpen
  \bibfield  {author} {\bibinfo {author} {\bibnamefont {Reichl}, \bibfnamefont
  {L.~E.}}} (\bibinfo {year} {1992}),\ \href@noop {} {\emph {\bibinfo {title}
  {The Transition to Chaos In Conservative Classical Systems: Quantum
  Manifestations}}}\ (\bibinfo  {publisher} {Springer},\ \bibinfo {address}
  {New York})\BibitemShut {NoStop}%
\bibitem [{\citenamefont {Reichl}(2016)}]{reichl2016modern}%
  \BibitemOpen
  \bibfield  {author} {\bibinfo {author} {\bibnamefont {Reichl}, \bibfnamefont
  {L.~E.}}} (\bibinfo {year} {2016}),\ \href@noop {} {\emph {\bibinfo {title}
  {A modern course in statistical physics}}},\ \bibinfo {edition} {4th}\ ed.\
  (\bibinfo  {publisher} {Wiley-VCH},\ \bibinfo {address} {New
  York})\BibitemShut {NoStop}%
\bibitem [{\citenamefont {Reuter}\ \emph {et~al.}(2006)\citenamefont {Reuter},
  \citenamefont {Wolter},\ and\ \citenamefont {Peinecke}}]{reuter2006laplace}%
  \BibitemOpen
  \bibfield  {author} {\bibinfo {author} {\bibnamefont {Reuter}, \bibfnamefont
  {M.}}, \bibinfo {author} {\bibfnamefont {F.-E.}\ \bibnamefont {Wolter}}, \
  and\ \bibinfo {author} {\bibfnamefont {N.}~\bibnamefont {Peinecke}}}
  (\bibinfo {year} {2006}),\ \href {\doibase 10.1016/j.cad.2005.10.011}
  {\bibfield  {journal} {\bibinfo  {journal} {Comput. Aided Des.}\ }\textbf
  {\bibinfo {volume} {38}}~(\bibinfo {number} {4}),\ \bibinfo {pages}
  {342}}\BibitemShut {NoStop}%
\bibitem [{\citenamefont {Richens}\ and\ \citenamefont
  {Berry}(1981)}]{richens1981pseudointegrable}%
  \BibitemOpen
  \bibfield  {author} {\bibinfo {author} {\bibnamefont {Richens}, \bibfnamefont
  {P.~J.}}, \ and\ \bibinfo {author} {\bibfnamefont {M.~V.}\ \bibnamefont
  {Berry}}} (\bibinfo {year} {1981}),\ \href {\doibase
  10.1016/0167-2789(81)90024-5} {\bibfield  {journal} {\bibinfo  {journal}
  {Physica D}\ }\textbf {\bibinfo {volume} {2}}~(\bibinfo {number} {3}),\
  \bibinfo {pages} {495}}\BibitemShut {NoStop}%
\bibitem [{\citenamefont {Richter}(1999)}]{richter1999playing}%
  \BibitemOpen
  \bibfield  {author} {\bibinfo {author} {\bibnamefont {Richter}, \bibfnamefont
  {A.}}} (\bibinfo {year} {1999}),\ in\ \href {\doibase
  10.1007/978-1-4612-1544-8_20} {\emph {\bibinfo {booktitle} {Emerging
  Applications of Number Theory}}},\ \bibinfo {series} {The IMA Volumes in
  Mathematics and its Applications}, Vol.\ \bibinfo {volume} {109},\ \bibinfo
  {editor} {edited by\ \bibinfo {editor} {\bibfnamefont {D.~A.}\ \bibnamefont
  {Hejhal}}, \bibinfo {editor} {\bibfnamefont {J.}~\bibnamefont {Friedman}},
  \bibinfo {editor} {\bibfnamefont {M.~C.}\ \bibnamefont {Gutzwiller}}, \ and\
  \bibinfo {editor} {\bibfnamefont {A.~M.}\ \bibnamefont {Odlyzko}}}\ (\bibinfo
   {publisher} {Springer},\ \bibinfo {address} {New York})\ pp.\ \bibinfo
  {pages} {479--523}\BibitemShut {NoStop}%
\bibitem [{\citenamefont {Richter}(2008)}]{richter2008superscars}%
  \BibitemOpen
  \bibfield  {author} {\bibinfo {author} {\bibnamefont {Richter}, \bibfnamefont
  {A.}}} (\bibinfo {year} {2008}),\ \href {\doibase 10.1063/1.2915603}
  {\bibfield  {journal} {\bibinfo  {journal} {AIP Conf. Proc.}\ }\textbf
  {\bibinfo {volume} {995}}~(\bibinfo {number} {1}),\ \bibinfo {pages}
  {202}}\BibitemShut {NoStop}%
\bibitem [{\citenamefont {Robnik}(1983)}]{robnik1983classical}%
  \BibitemOpen
  \bibfield  {author} {\bibinfo {author} {\bibnamefont {Robnik}, \bibfnamefont
  {M.}}} (\bibinfo {year} {1983}),\ \href {\doibase
  10.1088/0305-4470/16/17/014} {\bibfield  {journal} {\bibinfo  {journal} {J.
  Phys. A: Math. Gen.}\ }\textbf {\bibinfo {volume} {16}}~(\bibinfo {number}
  {17}),\ \bibinfo {pages} {3971}}\BibitemShut {NoStop}%
\bibitem [{\citenamefont {Robnik}(1984)}]{robnik1984quantising}%
  \BibitemOpen
  \bibfield  {author} {\bibinfo {author} {\bibnamefont {Robnik}, \bibfnamefont
  {M.}}} (\bibinfo {year} {1984}),\ \href {\doibase 10.1088/0305-4470/17/5/027}
  {\bibfield  {journal} {\bibinfo  {journal} {J. Phys. A: Math. Gen.}\ }\textbf
  {\bibinfo {volume} {17}}~(\bibinfo {number} {5}),\ \bibinfo {pages}
  {1049}}\BibitemShut {NoStop}%
\bibitem [{\citenamefont {Rohde}\ and\ \citenamefont
  {Schramm}(2005)}]{schramm2005basic}%
  \BibitemOpen
  \bibfield  {author} {\bibinfo {author} {\bibnamefont {Rohde}, \bibfnamefont
  {S.}}, \ and\ \bibinfo {author} {\bibfnamefont {O.}~\bibnamefont {Schramm}}}
  (\bibinfo {year} {2005}),\ \href {\doibase 10.4007/annals.2005.161.883}
  {\bibfield  {journal} {\bibinfo  {journal} {Ann. Math.}\ }\textbf {\bibinfo
  {volume} {161}}~(\bibinfo {number} {2}),\ \bibinfo {pages} {883}}\BibitemShut
  {NoStop}%
\bibitem [{\citenamefont {Rossiter}(1871)}]{rossiter1871}%
  \BibitemOpen
  \bibfield  {author} {\bibinfo {author} {\bibnamefont {Rossiter},
  \bibfnamefont {W.}}} (\bibinfo {year} {1871}),\ \href@noop {} {\emph
  {\bibinfo {title} {An Elementary Handbook of Physics}}}\ (\bibinfo
  {publisher} {William Blackwood and Sons},\ \bibinfo {address} {Edinburgh and
  London})\BibitemShut {NoStop}%
\bibitem [{\citenamefont {Rozenshein}(2016)}]{rozenshein2016number}%
  \BibitemOpen
  \bibfield  {author} {\bibinfo {author} {\bibnamefont {Rozenshein},
  \bibfnamefont {Y.}}} (\bibinfo {year} {2016}),\ \href
  {http://www.arxiv.org/abs/1604.00638} {\enquote {\bibinfo {title} {The
  {N}umber of {N}odal {C}omponents of {A}rithmetic {R}andom {W}aves},}\
  }\bibinfo {note} {{a}rXiv preprint 1604.00638}\BibitemShut {NoStop}%
\bibitem [{\citenamefont {Rudnick}\ and\ \citenamefont
  {Sarnak}(1994)}]{rudnick1994behaviour}%
  \BibitemOpen
  \bibfield  {author} {\bibinfo {author} {\bibnamefont {Rudnick}, \bibfnamefont
  {Z.}}, \ and\ \bibinfo {author} {\bibfnamefont {P.}~\bibnamefont {Sarnak}}}
  (\bibinfo {year} {1994}),\ \href {\doibase 10.1007/BF02099418} {\bibfield
  {journal} {\bibinfo  {journal} {Commun.Math. Phys.}\ }\textbf {\bibinfo
  {volume} {161}}~(\bibinfo {number} {1}),\ \bibinfo {pages} {195}}\BibitemShut
  {NoStop}%
\bibitem [{\citenamefont {Rudnick}\ and\ \citenamefont
  {Wigman}(2008)}]{rudnick2008volume}%
  \BibitemOpen
  \bibfield  {author} {\bibinfo {author} {\bibnamefont {Rudnick}, \bibfnamefont
  {Z.}}, \ and\ \bibinfo {author} {\bibfnamefont {I.}~\bibnamefont {Wigman}}}
  (\bibinfo {year} {2008}),\ \href {\doibase 10.1007/s00023-007-0352-6}
  {\bibfield  {journal} {\bibinfo  {journal} {Ann. Henri Poincar{\' e}}\
  }\textbf {\bibinfo {volume} {9}}~(\bibinfo {number} {1}),\ \bibinfo {pages}
  {109}}\BibitemShut {NoStop}%
\bibitem [{\citenamefont {Rudnick}\ and\ \citenamefont
  {Wigman}(2016)}]{rudnick2016nodal}%
  \BibitemOpen
  \bibfield  {author} {\bibinfo {author} {\bibnamefont {Rudnick}, \bibfnamefont
  {Z.}}, \ and\ \bibinfo {author} {\bibfnamefont {I.}~\bibnamefont {Wigman}}}
  (\bibinfo {year} {2016}),\ \href {\doibase 10.1353/ajm.2016.0048} {\bibfield
  {journal} {\bibinfo  {journal} {Am. J. Math.}\ }\textbf {\bibinfo {volume}
  {138}}~(\bibinfo {number} {6}),\ \bibinfo {pages} {1605}}\BibitemShut
  {NoStop}%
\bibitem [{\citenamefont {Rudnick}\ \emph {et~al.}(2015)\citenamefont
  {Rudnick}, \citenamefont {Wigman},\ and\ \citenamefont
  {Yesha}}]{rudnick2015nodal}%
  \BibitemOpen
  \bibfield  {author} {\bibinfo {author} {\bibnamefont {Rudnick}, \bibfnamefont
  {Z.}}, \bibinfo {author} {\bibfnamefont {I.}~\bibnamefont {Wigman}}, \ and\
  \bibinfo {author} {\bibfnamefont {N.}~\bibnamefont {Yesha}}} (\bibinfo {year}
  {2015}),\ \href {http://www.arxiv.org/abs/1501.07410} {\enquote {\bibinfo
  {title} {Nodal intersections for random waves on the 3-dimensional torus},}\
  }\bibinfo {note} {{a}rXiv preprint 1501.07410}\BibitemShut {NoStop}%
\bibitem [{\citenamefont {Sadreev}\ and\ \citenamefont
  {Berggren}(2004)}]{sadreev2004current}%
  \BibitemOpen
  \bibfield  {author} {\bibinfo {author} {\bibnamefont {Sadreev}, \bibfnamefont
  {A.~F.}}, \ and\ \bibinfo {author} {\bibfnamefont {K.-F.}\ \bibnamefont
  {Berggren}}} (\bibinfo {year} {2004}),\ \href {\doibase
  10.1103/PhysRevE.70.026201} {\bibfield  {journal} {\bibinfo  {journal} {Phys.
  Rev. E}\ }\textbf {\bibinfo {volume} {70}}~(\bibinfo {number} {2}),\ \bibinfo
  {pages} {026201}}\BibitemShut {NoStop}%
\bibitem [{\citenamefont {Saichev}\ \emph {et~al.}(2001)\citenamefont
  {Saichev}, \citenamefont {Berggren},\ and\ \citenamefont
  {Sadreev}}]{saichev2001distribution}%
  \BibitemOpen
  \bibfield  {author} {\bibinfo {author} {\bibnamefont {Saichev}, \bibfnamefont
  {A.~I.}}, \bibinfo {author} {\bibfnamefont {K.-F.}\ \bibnamefont {Berggren}},
  \ and\ \bibinfo {author} {\bibfnamefont {A.~F.}\ \bibnamefont {Sadreev}}}
  (\bibinfo {year} {2001}),\ \href {\doibase 10.1103/PhysRevE.64.036222}
  {\bibfield  {journal} {\bibinfo  {journal} {Phys. Rev. E}\ }\textbf {\bibinfo
  {volume} {64}}~(\bibinfo {number} {3}),\ \bibinfo {pages}
  {036222}}\BibitemShut {NoStop}%
\bibitem [{\citenamefont {Saichev}\ \emph {et~al.}(2002)\citenamefont
  {Saichev}, \citenamefont {Ishio}, \citenamefont {Sadreev},\ and\
  \citenamefont {Berggren}}]{saichev2002statistics}%
  \BibitemOpen
  \bibfield  {author} {\bibinfo {author} {\bibnamefont {Saichev}, \bibfnamefont
  {A.~I.}}, \bibinfo {author} {\bibfnamefont {H.}~\bibnamefont {Ishio}},
  \bibinfo {author} {\bibfnamefont {A.~F.}\ \bibnamefont {Sadreev}}, \ and\
  \bibinfo {author} {\bibfnamefont {K.-F.}\ \bibnamefont {Berggren}}} (\bibinfo
  {year} {2002}),\ \href {\doibase 10.1088/0305-4470/35/7/103} {\bibfield
  {journal} {\bibinfo  {journal} {J. Phys. A: Math. Gen.}\ }\textbf {\bibinfo
  {volume} {35}}~(\bibinfo {number} {7}),\ \bibinfo {pages} {L87}}\BibitemShut
  {NoStop}%
\bibitem [{\citenamefont {Saito}(2007)}]{saito2007}%
  \BibitemOpen
  \bibfield  {author} {\bibinfo {author} {\bibnamefont {Saito}, \bibfnamefont
  {N.}}} (\bibinfo {year} {2007}),\ \href
  {https://www.math.ucdavis.edu/~saito/courses/LapEig/lecpdf/lecture7.pdf}
  {\enquote {\bibinfo {title} {Lecture 7: Nodal sets},}\ }\bibinfo
  {howpublished} {MAT 280 Laplacian Eigenfunctions: Theory, Applications, and
  Computations},\ \bibinfo {note} {{U}niversity of California,
  Davis}\BibitemShut {NoStop}%
\bibitem [{\citenamefont {Saito}\ and\ \citenamefont
  {Woei}(2009)}]{saito2009analysis}%
  \BibitemOpen
  \bibfield  {author} {\bibinfo {author} {\bibnamefont {Saito}, \bibfnamefont
  {N.}}, \ and\ \bibinfo {author} {\bibfnamefont {E.}~\bibnamefont {Woei}}}
  (\bibinfo {year} {2009}),\ \href {\doibase 10.14495/jsiaml.1.13} {\bibfield
  {journal} {\bibinfo  {journal} {JSIAM Lett.}\ }\textbf {\bibinfo {volume}
  {1}}~(\bibinfo {number} {0}),\ \bibinfo {pages} {13}}\BibitemShut {NoStop}%
\bibitem [{\citenamefont {Samajdar}\ and\ \citenamefont
  {Jain}(2014{\natexlab{a}})}]{samajdar2014nodal}%
  \BibitemOpen
  \bibfield  {author} {\bibinfo {author} {\bibnamefont {Samajdar},
  \bibfnamefont {R.}}, \ and\ \bibinfo {author} {\bibfnamefont {S.~R.}\
  \bibnamefont {Jain}}} (\bibinfo {year} {2014}{\natexlab{a}}),\ \href
  {\doibase 10.1016/j.aop.2014.08.010} {\bibfield  {journal} {\bibinfo
  {journal} {Ann. Phys.}\ }\textbf {\bibinfo {volume} {351}},\ \bibinfo {pages}
  {1}}\BibitemShut {NoStop}%
\bibitem [{\citenamefont {Samajdar}\ and\ \citenamefont
  {Jain}(2014{\natexlab{b}})}]{samajdar2014JPA}%
  \BibitemOpen
  \bibfield  {author} {\bibinfo {author} {\bibnamefont {Samajdar},
  \bibfnamefont {R.}}, \ and\ \bibinfo {author} {\bibfnamefont {S.~R.}\
  \bibnamefont {Jain}}} (\bibinfo {year} {2014}{\natexlab{b}}),\ \href
  {\doibase 10.1088/1751-8113/47/19/195101} {\bibfield  {journal} {\bibinfo
  {journal} {J. Phys. A: Math. Theor.}\ }\textbf {\bibinfo {volume}
  {47}}~(\bibinfo {number} {19}),\ \bibinfo {pages} {195101}}\BibitemShut
  {NoStop}%
\bibitem [{\citenamefont {Sapoval}\ and\ \citenamefont
  {Gobron}(1993)}]{sapoval1993vibrations}%
  \BibitemOpen
  \bibfield  {author} {\bibinfo {author} {\bibnamefont {Sapoval}, \bibfnamefont
  {B.}}, \ and\ \bibinfo {author} {\bibfnamefont {{\relax Th}.}~\bibnamefont
  {Gobron}}} (\bibinfo {year} {1993}),\ \href {\doibase
  10.1103/PhysRevE.47.3013} {\bibfield  {journal} {\bibinfo  {journal} {Phys.
  Rev. E}\ }\textbf {\bibinfo {volume} {47}}~(\bibinfo {number} {5}),\ \bibinfo
  {pages} {3013}}\BibitemShut {NoStop}%
\bibitem [{\citenamefont {Sapoval}\ \emph {et~al.}(1991)\citenamefont
  {Sapoval}, \citenamefont {Gobron},\ and\ \citenamefont
  {Margolina}}]{sapoval1991vibrations}%
  \BibitemOpen
  \bibfield  {author} {\bibinfo {author} {\bibnamefont {Sapoval}, \bibfnamefont
  {B.}}, \bibinfo {author} {\bibfnamefont {{\relax Th}.}~\bibnamefont
  {Gobron}}, \ and\ \bibinfo {author} {\bibfnamefont {A.}~\bibnamefont
  {Margolina}}} (\bibinfo {year} {1991}),\ \href {\doibase
  10.1103/PhysRevLett.67.2974} {\bibfield  {journal} {\bibinfo  {journal}
  {Phys. Rev. Lett.}\ }\textbf {\bibinfo {volume} {67}}~(\bibinfo {number}
  {21}),\ \bibinfo {pages} {2974}}\BibitemShut {NoStop}%
\bibitem [{\citenamefont {Sapoval}\ \emph {et~al.}(1997)\citenamefont
  {Sapoval}, \citenamefont {Haeberl{\'e}},\ and\ \citenamefont
  {Russ}}]{sapoval1997acoustical}%
  \BibitemOpen
  \bibfield  {author} {\bibinfo {author} {\bibnamefont {Sapoval}, \bibfnamefont
  {B.}}, \bibinfo {author} {\bibfnamefont {O.}~\bibnamefont {Haeberl{\'e}}}, \
  and\ \bibinfo {author} {\bibfnamefont {S.}~\bibnamefont {Russ}}} (\bibinfo
  {year} {1997}),\ \href {\doibase 10.1121/1.419653} {\bibfield  {journal}
  {\bibinfo  {journal} {J. Acoust. Soc. Am.}\ }\textbf {\bibinfo {volume}
  {102}}~(\bibinfo {number} {4}),\ \bibinfo {pages} {2014}}\BibitemShut
  {NoStop}%
\bibitem [{\citenamefont {Sarnak}(1995)}]{sarnak1995arithmetic}%
  \BibitemOpen
  \bibfield  {author} {\bibinfo {author} {\bibnamefont {Sarnak}, \bibfnamefont
  {P.}}} (\bibinfo {year} {1995}),\ in\ \href@noop {} {\emph {\bibinfo
  {booktitle} {Israel Math. Conf. Proc.}}},\ Vol.~\bibinfo {volume} {8}\
  (\bibinfo  {publisher} {The Schur Lectures (1992)},\ \bibinfo {address} {Tel
  Aviv})\ pp.\ \bibinfo {pages} {183--236}\BibitemShut {NoStop}%
\bibitem [{\citenamefont {Savo}(2000)}]{savo2000eigenvalue}%
  \BibitemOpen
  \bibfield  {author} {\bibinfo {author} {\bibnamefont {Savo}, \bibfnamefont
  {A.}}} (\bibinfo {year} {2000}),\ in\ \href
  {http://emis.ams.org/proceedings/CDGD2000/pdf/K_Savo.pdf} {\emph {\bibinfo
  {booktitle} {Steps in differential geometry}}},\ \bibinfo {series and number}
  {Proceedings of the Colloquium on Differential Geometry},\ \bibinfo {editor}
  {edited by\ \bibinfo {editor} {\bibfnamefont {L.}~\bibnamefont {Kozma}},
  \bibinfo {editor} {\bibfnamefont {P.~T.}\ \bibnamefont {Nagy}}, \ and\
  \bibinfo {editor} {\bibfnamefont {L.}~\bibnamefont {Tam{\'a}ssy}}},\ pp.\
  \bibinfo {pages} {295--301}\BibitemShut {NoStop}%
\bibitem [{\citenamefont {Savytskyy}\ \emph {et~al.}(2004)\citenamefont
  {Savytskyy}, \citenamefont {Hul},\ and\ \citenamefont
  {Sirko}}]{savytskyy2004}%
  \BibitemOpen
  \bibfield  {author} {\bibinfo {author} {\bibnamefont {Savytskyy},
  \bibfnamefont {N.}}, \bibinfo {author} {\bibfnamefont {O.}~\bibnamefont
  {Hul}}, \ and\ \bibinfo {author} {\bibfnamefont {L.}~\bibnamefont {Sirko}}}
  (\bibinfo {year} {2004}),\ \href {\doibase 10.1103/PhysRevE.70.056209}
  {\bibfield  {journal} {\bibinfo  {journal} {Phys. Rev. E}\ }\textbf {\bibinfo
  {volume} {70}}~(\bibinfo {number} {5}),\ \bibinfo {pages}
  {056209}}\BibitemShut {NoStop}%
\bibitem [{\citenamefont {Savytskyy}\ \emph {et~al.}(2001)\citenamefont
  {Savytskyy}, \citenamefont {Kohler}, \citenamefont {Bauch}, \citenamefont
  {Bl{\"u}mel},\ and\ \citenamefont {Sirko}}]{savytskyy2001parametric}%
  \BibitemOpen
  \bibfield  {author} {\bibinfo {author} {\bibnamefont {Savytskyy},
  \bibfnamefont {N.}}, \bibinfo {author} {\bibfnamefont {A.}~\bibnamefont
  {Kohler}}, \bibinfo {author} {\bibfnamefont {{\relax Sz}.}~\bibnamefont
  {Bauch}}, \bibinfo {author} {\bibfnamefont {R.}~\bibnamefont {Bl{\"u}mel}}, \
  and\ \bibinfo {author} {\bibfnamefont {L.}~\bibnamefont {Sirko}}} (\bibinfo
  {year} {2001}),\ \href {\doibase 10.1103/PhysRevE.64.036211} {\bibfield
  {journal} {\bibinfo  {journal} {Phys. Rev. E}\ }\textbf {\bibinfo {volume}
  {64}}~(\bibinfo {number} {3}),\ \bibinfo {pages} {036211}}\BibitemShut
  {NoStop}%
\bibitem [{\citenamefont {Savytskyy}\ and\ \citenamefont
  {Sirko}(2002)}]{savytskyy2002}%
  \BibitemOpen
  \bibfield  {author} {\bibinfo {author} {\bibnamefont {Savytskyy},
  \bibfnamefont {N.}}, \ and\ \bibinfo {author} {\bibfnamefont
  {L.}~\bibnamefont {Sirko}}} (\bibinfo {year} {2002}),\ \href {\doibase
  10.1103/PhysRevE.65.066202} {\bibfield  {journal} {\bibinfo  {journal} {Phys.
  Rev. E}\ }\textbf {\bibinfo {volume} {65}}~(\bibinfo {number} {6}),\ \bibinfo
  {pages} {066202}}\BibitemShut {NoStop}%
\bibitem [{\citenamefont {Sch{\"a}fer}\ \emph {et~al.}(2003)\citenamefont
  {Sch{\"a}fer}, \citenamefont {Gorin}, \citenamefont {Seligman},\ and\
  \citenamefont {St{\"o}ckmann}}]{schafer2003correlation}%
  \BibitemOpen
  \bibfield  {author} {\bibinfo {author} {\bibnamefont {Sch{\"a}fer},
  \bibfnamefont {R.}}, \bibinfo {author} {\bibfnamefont {T.}~\bibnamefont
  {Gorin}}, \bibinfo {author} {\bibfnamefont {T.~H.}\ \bibnamefont {Seligman}},
  \ and\ \bibinfo {author} {\bibfnamefont {H.-J.}\ \bibnamefont
  {St{\"o}ckmann}}} (\bibinfo {year} {2003}),\ \href {\doibase
  10.1088/0305-4470/36/12/325} {\bibfield  {journal} {\bibinfo  {journal} {J.
  Phys. A: Math. Gen.}\ }\textbf {\bibinfo {volume} {36}}~(\bibinfo {number}
  {12}),\ \bibinfo {pages} {3289}}\BibitemShut {NoStop}%
\bibitem [{\citenamefont
  {Schapotschnikow}(2006)}]{schapotschnikow2006eigenvalue}%
  \BibitemOpen
  \bibfield  {author} {\bibinfo {author} {\bibnamefont {Schapotschnikow},
  \bibfnamefont {P.}}} (\bibinfo {year} {2006}),\ \href {\doibase
  10.1080/17455030600702535} {\bibfield  {journal} {\bibinfo  {journal} {Wave.
  Random Complex}\ }\textbf {\bibinfo {volume} {16}}~(\bibinfo {number} {3}),\
  \bibinfo {pages} {167}}\BibitemShut {NoStop}%
\bibitem [{\citenamefont {Schnabel}\ \emph {et~al.}(2007)\citenamefont
  {Schnabel}, \citenamefont {Kaschube}, \citenamefont {L{\"o}wel},\ and\
  \citenamefont {Wolf}}]{schnabel2007random}%
  \BibitemOpen
  \bibfield  {author} {\bibinfo {author} {\bibnamefont {Schnabel},
  \bibfnamefont {M.}}, \bibinfo {author} {\bibfnamefont {M.}~\bibnamefont
  {Kaschube}}, \bibinfo {author} {\bibfnamefont {S.}~\bibnamefont {L{\"o}wel}},
  \ and\ \bibinfo {author} {\bibfnamefont {F.}~\bibnamefont {Wolf}}} (\bibinfo
  {year} {2007}),\ \href {\doibase 10.1140/epjst/e2007-00152-5} {\bibfield
  {journal} {\bibinfo  {journal} {Eur. Phys. J. Spec. Top.}\ }\textbf {\bibinfo
  {volume} {145}}~(\bibinfo {number} {1}),\ \bibinfo {pages} {137}}\BibitemShut
  {NoStop}%
\bibitem [{\citenamefont {Schramm}(2000)}]{schramm2000scaling}%
  \BibitemOpen
  \bibfield  {author} {\bibinfo {author} {\bibnamefont {Schramm}, \bibfnamefont
  {O.}}} (\bibinfo {year} {2000}),\ \href {\doibase 10.1007/BF02803524}
  {\bibfield  {journal} {\bibinfo  {journal} {Isr. J. Math.}\ }\textbf
  {\bibinfo {volume} {118}}~(\bibinfo {number} {1}),\ \bibinfo {pages}
  {221}}\BibitemShut {NoStop}%
\bibitem [{\citenamefont {Schroeder}(1987)}]{schroeder1987normal}%
  \BibitemOpen
  \bibfield  {author} {\bibinfo {author} {\bibnamefont {Schroeder},
  \bibfnamefont {M.~R.}}} (\bibinfo {year} {1987}),\ \href
  {http://www.aes.org/e-lib/browse.cfm?elib=5207} {\bibfield  {journal}
  {\bibinfo  {journal} {J. Audio. Eng. Soc.}\ }\textbf {\bibinfo {volume}
  {35}}~(\bibinfo {number} {5}),\ \bibinfo {pages} {307}}\BibitemShut {NoStop}%
\bibitem [{\citenamefont {{\v S}eba}\ \emph {et~al.}(1999)\citenamefont {{\v
  S}eba}, \citenamefont {Kuhl}, \citenamefont {Barth},\ and\ \citenamefont
  {St{\"o}ckmann}}]{seba1999experimental}%
  \BibitemOpen
  \bibfield  {author} {\bibinfo {author} {\bibnamefont {{\v S}eba},
  \bibfnamefont {P.}}, \bibinfo {author} {\bibfnamefont {U.}~\bibnamefont
  {Kuhl}}, \bibinfo {author} {\bibfnamefont {M.}~\bibnamefont {Barth}}, \ and\
  \bibinfo {author} {\bibfnamefont {H.-J.}\ \bibnamefont {St{\"o}ckmann}}}
  (\bibinfo {year} {1999}),\ \href {\doibase 10.1088/0305-4470/32/47/302}
  {\bibfield  {journal} {\bibinfo  {journal} {J. Phys. A: Math. Gen.}\ }\textbf
  {\bibinfo {volume} {32}}~(\bibinfo {number} {47}),\ \bibinfo {pages}
  {8225}}\BibitemShut {NoStop}%
\bibitem [{\citenamefont {Sen}(1996)}]{sen1996multispecies}%
  \BibitemOpen
  \bibfield  {author} {\bibinfo {author} {\bibnamefont {Sen}, \bibfnamefont
  {D.}}} (\bibinfo {year} {1996}),\ \href {\doibase
  10.1016/0550-3213(96)00420-8} {\bibfield  {journal} {\bibinfo  {journal}
  {Nucl. Phys. B}\ }\textbf {\bibinfo {volume} {479}}~(\bibinfo {number} {3}),\
  \bibinfo {pages} {554}}\BibitemShut {NoStop}%
\bibitem [{\citenamefont {Shapiro}\ and\ \citenamefont
  {Tegmark}(1994)}]{shapiro1994elementary}%
  \BibitemOpen
  \bibfield  {author} {\bibinfo {author} {\bibnamefont {Shapiro}, \bibfnamefont
  {H.~S.}}, \ and\ \bibinfo {author} {\bibfnamefont {M.}~\bibnamefont
  {Tegmark}}} (\bibinfo {year} {1994}),\ \href {\doibase 10.1137/1036005}
  {\bibfield  {journal} {\bibinfo  {journal} {SIAM Rev.}\ }\textbf {\bibinfo
  {volume} {36}}~(\bibinfo {number} {1}),\ \bibinfo {pages} {99}}\BibitemShut
  {NoStop}%
\bibitem [{\citenamefont {Shi}\ \emph {et~al.}(2017)\citenamefont {Shi},
  \citenamefont {Zhang}, \citenamefont {Tang}, \citenamefont {Caselli},\ and\
  \citenamefont {Wang}}]{shi2017conformal}%
  \BibitemOpen
  \bibfield  {author} {\bibinfo {author} {\bibnamefont {Shi}, \bibfnamefont
  {J.}}, \bibinfo {author} {\bibfnamefont {W.}~\bibnamefont {Zhang}}, \bibinfo
  {author} {\bibfnamefont {M.}~\bibnamefont {Tang}}, \bibinfo {author}
  {\bibfnamefont {R.~J.}\ \bibnamefont {Caselli}}, \ and\ \bibinfo {author}
  {\bibfnamefont {Y.}~\bibnamefont {Wang}}} (\bibinfo {year} {2017}),\ \href
  {\doibase 10.1016/j.media.2016.09.001} {\bibfield  {journal} {\bibinfo
  {journal} {Med. Image Anal.}\ }\textbf {\bibinfo {volume} {35}},\ \bibinfo
  {pages} {517}}\BibitemShut {NoStop}%
\bibitem [{\citenamefont {Shnirel'man}(1974)}]{shnirel1974ergodic}%
  \BibitemOpen
  \bibfield  {author} {\bibinfo {author} {\bibnamefont {Shnirel'man},
  \bibfnamefont {A.~I.}}} (\bibinfo {year} {1974}),\ \href
  {http://mi.mathnet.ru/eng/umn4463} {\bibfield  {journal} {\bibinfo  {journal}
  {Usp. Mat. Nauk}\ }\textbf {\bibinfo {volume} {29}}~(\bibinfo {number} {6}),\
  \bibinfo {pages} {181}}\BibitemShut {NoStop}%
\bibitem [{\citenamefont {Shvartsman}\ and\ \citenamefont
  {Freund}(1994{\natexlab{a}})}]{shvartsman1994vortices}%
  \BibitemOpen
  \bibfield  {author} {\bibinfo {author} {\bibnamefont {Shvartsman},
  \bibfnamefont {N.}}, \ and\ \bibinfo {author} {\bibfnamefont
  {I.}~\bibnamefont {Freund}}} (\bibinfo {year} {1994}{\natexlab{a}}),\ \href
  {\doibase 10.1103/PhysRevLett.72.1008} {\bibfield  {journal} {\bibinfo
  {journal} {Phys. Rev. Lett.}\ }\textbf {\bibinfo {volume} {72}}~(\bibinfo
  {number} {7}),\ \bibinfo {pages} {1008}}\BibitemShut {NoStop}%
\bibitem [{\citenamefont {Shvartsman}\ and\ \citenamefont
  {Freund}(1994{\natexlab{b}})}]{shvartsman1994wave}%
  \BibitemOpen
  \bibfield  {author} {\bibinfo {author} {\bibnamefont {Shvartsman},
  \bibfnamefont {N.}}, \ and\ \bibinfo {author} {\bibfnamefont
  {I.}~\bibnamefont {Freund}}} (\bibinfo {year} {1994}{\natexlab{b}}),\ \href
  {\doibase 10.1364/JOSAA.11.002710} {\bibfield  {journal} {\bibinfo  {journal}
  {J. Opt. Soc. Am. A}\ }\textbf {\bibinfo {volume} {11}}~(\bibinfo {number}
  {10}),\ \bibinfo {pages} {2710}}\BibitemShut {NoStop}%
\bibitem [{\citenamefont {Siegel}(1941)}]{siegel1941ann}%
  \BibitemOpen
  \bibfield  {author} {\bibinfo {author} {\bibnamefont {Siegel}, \bibfnamefont
  {C.~L.}}} (\bibinfo {year} {1941}),\ \href@noop {} {\bibfield  {journal}
  {\bibinfo  {journal} {Ann. Math.}\ }\textbf {\bibinfo {volume}
  {42}}~(\bibinfo {number} {3}),\ \bibinfo {pages} {806}}\BibitemShut {NoStop}%
\bibitem [{\citenamefont {Siegel}\ \emph {et~al.}(1969)\citenamefont {Siegel},
  \citenamefont {Shenitzer},\ and\ \citenamefont {Solitar}}]{siegel1969topics}%
  \BibitemOpen
  \bibfield  {author} {\bibinfo {author} {\bibnamefont {Siegel}, \bibfnamefont
  {C.~L.}}, \bibinfo {author} {\bibfnamefont {A.}~\bibnamefont {Shenitzer}}, \
  and\ \bibinfo {author} {\bibfnamefont {D.}~\bibnamefont {Solitar}}} (\bibinfo
  {year} {1969}),\ \href@noop {} {\emph {\bibinfo {title} {Topics in complex
  function theory: Vol. 1, Elliptic functions and uniformization theory}}}\
  (\bibinfo  {publisher} {Wiley Interscience})\BibitemShut {NoStop}%
\bibitem [{\citenamefont {Simmel}\ and\ \citenamefont
  {Eckert}(1996)}]{simmel1996statistical}%
  \BibitemOpen
  \bibfield  {author} {\bibinfo {author} {\bibnamefont {Simmel}, \bibfnamefont
  {F.}}, \ and\ \bibinfo {author} {\bibfnamefont {M.}~\bibnamefont {Eckert}}}
  (\bibinfo {year} {1996}),\ \href {\doibase 10.1016/0167-2789(96)00040-1}
  {\bibfield  {journal} {\bibinfo  {journal} {Physica D}\ }\textbf {\bibinfo
  {volume} {97}}~(\bibinfo {number} {4}),\ \bibinfo {pages} {517}}\BibitemShut
  {NoStop}%
\bibitem [{\citenamefont {Simon}(2005)}]{simon2005sturm}%
  \BibitemOpen
  \bibfield  {author} {\bibinfo {author} {\bibnamefont {Simon}, \bibfnamefont
  {B.}}} (\bibinfo {year} {2005}),\ \enquote {\bibinfo {title} {Sturm
  {O}scillation and {C}omparison {T}heorems},}\ in\ \href {\doibase
  10.1007/3-7643-7359-8_2} {\emph {\bibinfo {booktitle} {Sturm-Liouville
  Theory: Past and Present}}},\ \bibinfo {editor} {edited by\ \bibinfo {editor}
  {\bibfnamefont {W.~O.}\ \bibnamefont {Amrein}}, \bibinfo {editor}
  {\bibfnamefont {A.~M.}\ \bibnamefont {Hinz}}, \ and\ \bibinfo {editor}
  {\bibfnamefont {D.~P.}\ \bibnamefont {Pearson}}},\ Chap.~\bibinfo {chapter}
  {2}\ (\bibinfo  {publisher} {Birkh{\"a}user Basel},\ \bibinfo {address}
  {Basel})\ pp.\ \bibinfo {pages} {29--43}\BibitemShut {NoStop}%
\bibitem [{\citenamefont {Simon}\ and\ \citenamefont
  {Reed}(1979)}]{simon1979methods}%
  \BibitemOpen
  \bibfield  {author} {\bibinfo {author} {\bibnamefont {Simon}, \bibfnamefont
  {B.}}, \ and\ \bibinfo {author} {\bibfnamefont {M.~C.}\ \bibnamefont {Reed}}}
  (\bibinfo {year} {1979}),\ \href@noop {} {\emph {\bibinfo {title} {Methods of
  Modern Mathematical Physics}}}\ (\bibinfo  {publisher} {Academic Press},\
  \bibinfo {address} {New York})\BibitemShut {NoStop}%
\bibitem [{\citenamefont {Simpson}\ \emph {et~al.}(1996)\citenamefont
  {Simpson}, \citenamefont {Allen},\ and\ \citenamefont
  {Padgett}}]{simpson1996optical}%
  \BibitemOpen
  \bibfield  {author} {\bibinfo {author} {\bibnamefont {Simpson}, \bibfnamefont
  {N.~B.}}, \bibinfo {author} {\bibfnamefont {L.}~\bibnamefont {Allen}}, \ and\
  \bibinfo {author} {\bibfnamefont {M.~J.}\ \bibnamefont {Padgett}}} (\bibinfo
  {year} {1996}),\ \href {\doibase 10.1080/09500349608230675} {\bibfield
  {journal} {\bibinfo  {journal} {J. Mod. Opt.}\ }\textbf {\bibinfo {volume}
  {43}}~(\bibinfo {number} {12}),\ \bibinfo {pages} {2485}}\BibitemShut
  {NoStop}%
\bibitem [{\citenamefont {Sina{\u i}}(1970)}]{sinai1970dynamical}%
  \BibitemOpen
  \bibfield  {author} {\bibinfo {author} {\bibnamefont {Sina{\u i}},
  \bibfnamefont {{\relax Ya}.~G.}}} (\bibinfo {year} {1970}),\ \href {\doibase
  10.1070/RM1970v025n02ABEH003794} {\bibfield  {journal} {\bibinfo  {journal}
  {Russ. Math. Surv.}\ }\textbf {\bibinfo {volume} {25}}~(\bibinfo {number}
  {2}),\ \bibinfo {pages} {137}}\BibitemShut {NoStop}%
\bibitem [{\citenamefont {Sina{\u i}}(1976)}]{sinaui1976introduction}%
  \BibitemOpen
  \bibfield  {author} {\bibinfo {author} {\bibnamefont {Sina{\u i}},
  \bibfnamefont {{\relax Ya}.~G.}}} (\bibinfo {year} {1976}),\ \href@noop {}
  {\emph {\bibinfo {title} {Introduction to Ergodic Theory}}},\ Vol.~\bibinfo
  {volume} {18}\ (\bibinfo  {publisher} {Princeton University Press},\ \bibinfo
  {address} {Princeton, NJ})\BibitemShut {NoStop}%
\bibitem [{\citenamefont {Singer}\ \emph {et~al.}(1985)\citenamefont {Singer},
  \citenamefont {Wong}, \citenamefont {Yau},\ and\ \citenamefont
  {Yau}}]{singer1985estimate}%
  \BibitemOpen
  \bibfield  {author} {\bibinfo {author} {\bibnamefont {Singer}, \bibfnamefont
  {I.~M.}}, \bibinfo {author} {\bibfnamefont {B.}~\bibnamefont {Wong}},
  \bibinfo {author} {\bibfnamefont {S.-T.}\ \bibnamefont {Yau}}, \ and\
  \bibinfo {author} {\bibfnamefont {S.~S.-T.}\ \bibnamefont {Yau}}} (\bibinfo
  {year} {1985}),\ \href
  {http://www.numdam.org/item?id=ASNSP_1985_4_12_2_319_0} {\bibfield  {journal}
  {\bibinfo  {journal} {Ann. Scuola Norm. Sup. Pisa}\ }\textbf {\bibinfo
  {volume} {12}}~(\bibinfo {number} {2}),\ \bibinfo {pages} {319}}\BibitemShut
  {NoStop}%
\bibitem [{\citenamefont {Sirko}\ \emph {et~al.}(2000)\citenamefont {Sirko},
  \citenamefont {Bauch}, \citenamefont {Hlushchuk}, \citenamefont {Koch},
  \citenamefont {Bl{\"u}mel}, \citenamefont {Barth}, \citenamefont {Kuhl},\
  and\ \citenamefont {St{\"o}ckmann}}]{sirko2000observation}%
  \BibitemOpen
  \bibfield  {author} {\bibinfo {author} {\bibnamefont {Sirko}, \bibfnamefont
  {L.}}, \bibinfo {author} {\bibfnamefont {S.}~\bibnamefont {Bauch}}, \bibinfo
  {author} {\bibfnamefont {Y.}~\bibnamefont {Hlushchuk}}, \bibinfo {author}
  {\bibfnamefont {P.~M.}\ \bibnamefont {Koch}}, \bibinfo {author}
  {\bibfnamefont {R.}~\bibnamefont {Bl{\"u}mel}}, \bibinfo {author}
  {\bibfnamefont {M.}~\bibnamefont {Barth}}, \bibinfo {author} {\bibfnamefont
  {U.}~\bibnamefont {Kuhl}}, \ and\ \bibinfo {author} {\bibfnamefont {H.-J.}\
  \bibnamefont {St{\"o}ckmann}}} (\bibinfo {year} {2000}),\ \href {\doibase
  10.1016/S0375-9601(00)00052-9} {\bibfield  {journal} {\bibinfo  {journal}
  {Phys. Lett. A}\ }\textbf {\bibinfo {volume} {266}}~(\bibinfo {number} {4}),\
  \bibinfo {pages} {331}}\BibitemShut {NoStop}%
\bibitem [{\citenamefont {Sirko}\ \emph {et~al.}(1997)\citenamefont {Sirko},
  \citenamefont {Koch},\ and\ \citenamefont
  {Bl{\"u}mel}}]{sirko1997experimental}%
  \BibitemOpen
  \bibfield  {author} {\bibinfo {author} {\bibnamefont {Sirko}, \bibfnamefont
  {L.}}, \bibinfo {author} {\bibfnamefont {P.~M.}\ \bibnamefont {Koch}}, \ and\
  \bibinfo {author} {\bibfnamefont {R.}~\bibnamefont {Bl{\"u}mel}}} (\bibinfo
  {year} {1997}),\ \href {\doibase 10.1103/PhysRevLett.78.2940} {\bibfield
  {journal} {\bibinfo  {journal} {Phys. Rev. Lett.}\ }\textbf {\bibinfo
  {volume} {78}}~(\bibinfo {number} {15}),\ \bibinfo {pages}
  {2940}}\BibitemShut {NoStop}%
\bibitem [{\citenamefont {Siudeja}(2007)}]{siudeja2007sharp}%
  \BibitemOpen
  \bibfield  {author} {\bibinfo {author} {\bibnamefont {Siudeja}, \bibfnamefont
  {B.}}} (\bibinfo {year} {2007}),\ \href {\doibase 10.1307/mmj/1187646992}
  {\bibfield  {journal} {\bibinfo  {journal} {Michigan Math. J.}\ }\textbf
  {\bibinfo {volume} {55}}~(\bibinfo {number} {2}),\ \bibinfo {pages}
  {243}}\BibitemShut {NoStop}%
\bibitem [{\citenamefont {Smilansky}\ and\ \citenamefont
  {Sankaranarayanan}(2005)}]{smilansky2005nodal}%
  \BibitemOpen
  \bibfield  {author} {\bibinfo {author} {\bibnamefont {Smilansky},
  \bibfnamefont {U.}}, \ and\ \bibinfo {author} {\bibfnamefont
  {R.}~\bibnamefont {Sankaranarayanan}}} (\bibinfo {year} {2005}),\ \href
  {http://www.arxiv.org/abs/nlin/0503002} {\enquote {\bibinfo {title} {Nodal
  domain distribution of rectangular drums},}\ }\bibinfo {note} {{a}rXiv
  preprint nlin/0503002}\BibitemShut {NoStop}%
\bibitem [{\citenamefont {Smirnov}(2001)}]{smirnov2001critical}%
  \BibitemOpen
  \bibfield  {author} {\bibinfo {author} {\bibnamefont {Smirnov}, \bibfnamefont
  {S.}}} (\bibinfo {year} {2001}),\ \href {\doibase
  10.1016/S0764-4442(01)01991-7} {\bibfield  {journal} {\bibinfo  {journal} {C.
  R. Acad. Sci. Paris S{\'e}r. I Math.}\ }\textbf {\bibinfo {volume}
  {333}}~(\bibinfo {number} {3}),\ \bibinfo {pages} {239}}\BibitemShut
  {NoStop}%
\bibitem [{\citenamefont {So}\ \emph {et~al.}(1995)\citenamefont {So},
  \citenamefont {Anlage}, \citenamefont {Ott},\ and\ \citenamefont
  {Oerter}}]{so1995wave}%
  \BibitemOpen
  \bibfield  {author} {\bibinfo {author} {\bibnamefont {So}, \bibfnamefont
  {P.}}, \bibinfo {author} {\bibfnamefont {S.~M.}\ \bibnamefont {Anlage}},
  \bibinfo {author} {\bibfnamefont {E.}~\bibnamefont {Ott}}, \ and\ \bibinfo
  {author} {\bibfnamefont {R.~N.}\ \bibnamefont {Oerter}}} (\bibinfo {year}
  {1995}),\ \href {\doibase 10.1103/PhysRevLett.74.2662} {\bibfield  {journal}
  {\bibinfo  {journal} {Phys. Rev. Lett.}\ }\textbf {\bibinfo {volume}
  {74}}~(\bibinfo {number} {14}),\ \bibinfo {pages} {2662}}\BibitemShut
  {NoStop}%
\bibitem [{\citenamefont {Sodin}(2016)}]{sodin2016lectures}%
  \BibitemOpen
  \bibfield  {author} {\bibinfo {author} {\bibnamefont {Sodin}, \bibfnamefont
  {M.}}} (\bibinfo {year} {2016}),\ in\ \href@noop {} {\emph {\bibinfo
  {booktitle} {Probability and {S}tatistical Physics in {S}t. {P}etersburg}}},\
  \bibinfo {series} {Proc. Symp. Pure Math.}, Vol.~\bibinfo {volume} {91},\
  \bibinfo {editor} {edited by\ \bibinfo {editor} {\bibfnamefont
  {V.}~\bibnamefont {Sidoravicius}}\ and\ \bibinfo {editor} {\bibfnamefont
  {S.}~\bibnamefont {Smirnov}}}\ (\bibinfo  {publisher} {American Mathematical
  Soc.})\ p.\ \bibinfo {pages} {395}\BibitemShut {NoStop}%
\bibitem [{\citenamefont {Sokolov}\ and\ \citenamefont
  {Zelevinsky}(1997)}]{sokolov1997simple}%
  \BibitemOpen
  \bibfield  {author} {\bibinfo {author} {\bibnamefont {Sokolov}, \bibfnamefont
  {V.~V.}}, \ and\ \bibinfo {author} {\bibfnamefont {V.}~\bibnamefont
  {Zelevinsky}}} (\bibinfo {year} {1997}),\ \href {\doibase
  10.1103/PhysRevC.56.311} {\bibfield  {journal} {\bibinfo  {journal} {Phys.
  Rev. C}\ }\textbf {\bibinfo {volume} {56}}~(\bibinfo {number} {1}),\ \bibinfo
  {pages} {311}}\BibitemShut {NoStop}%
\bibitem [{\citenamefont {Sridhar}(1991)}]{sridhar1991experimental}%
  \BibitemOpen
  \bibfield  {author} {\bibinfo {author} {\bibnamefont {Sridhar}, \bibfnamefont
  {S.}}} (\bibinfo {year} {1991}),\ \href {\doibase 10.1103/PhysRevLett.67.785}
  {\bibfield  {journal} {\bibinfo  {journal} {Phys. Rev. Lett.}\ }\textbf
  {\bibinfo {volume} {67}}~(\bibinfo {number} {7}),\ \bibinfo {pages}
  {785}}\BibitemShut {NoStop}%
\bibitem [{\citenamefont {Sridhar}\ and\ \citenamefont
  {Heller}(1992)}]{sridhar1992physical}%
  \BibitemOpen
  \bibfield  {author} {\bibinfo {author} {\bibnamefont {Sridhar}, \bibfnamefont
  {S.}}, \ and\ \bibinfo {author} {\bibfnamefont {E.~J.}\ \bibnamefont
  {Heller}}} (\bibinfo {year} {1992}),\ \href {\doibase
  10.1103/PhysRevA.46.R1728} {\bibfield  {journal} {\bibinfo  {journal} {Phys.
  Rev. A}\ }\textbf {\bibinfo {volume} {46}}~(\bibinfo {number} {4}),\ \bibinfo
  {pages} {R1728}}\BibitemShut {NoStop}%
\bibitem [{\citenamefont {Sridhar}\ and\ \citenamefont
  {Kudrolli}(1994)}]{sridhar1994experiments}%
  \BibitemOpen
  \bibfield  {author} {\bibinfo {author} {\bibnamefont {Sridhar}, \bibfnamefont
  {S.}}, \ and\ \bibinfo {author} {\bibfnamefont {A.}~\bibnamefont {Kudrolli}}}
  (\bibinfo {year} {1994}),\ \href {\doibase 10.1103/PhysRevLett.72.2175}
  {\bibfield  {journal} {\bibinfo  {journal} {Phys. Rev. Lett.}\ }\textbf
  {\bibinfo {volume} {72}}~(\bibinfo {number} {14}),\ \bibinfo {pages}
  {2175}}\BibitemShut {NoStop}%
\bibitem [{\citenamefont {Stanley}(1987)}]{stanley1987introduction}%
  \BibitemOpen
  \bibfield  {author} {\bibinfo {author} {\bibnamefont {Stanley}, \bibfnamefont
  {H.~E.}}} (\bibinfo {year} {1987}),\ \href@noop {} {\emph {\bibinfo {title}
  {Introduction to Phase Transitions and Critical Phenomena}}}\ (\bibinfo
  {publisher} {Oxford University Press},\ \bibinfo {address} {New
  York})\BibitemShut {NoStop}%
\bibitem [{\citenamefont {Stauffer}\ and\ \citenamefont
  {Aharony}(1994)}]{stauffer1994introduction}%
  \BibitemOpen
  \bibfield  {author} {\bibinfo {author} {\bibnamefont {Stauffer},
  \bibfnamefont {D.}}, \ and\ \bibinfo {author} {\bibfnamefont
  {A.}~\bibnamefont {Aharony}}} (\bibinfo {year} {1994}),\ \href@noop {} {\emph
  {\bibinfo {title} {Introduction to percolation theory}}}\ (\bibinfo
  {publisher} {Taylor \& Francis},\ \bibinfo {address} {London})\BibitemShut
  {NoStop}%
\bibitem [{\citenamefont {Stein}\ and\ \citenamefont
  {St{\"o}ckmann}(1992)}]{stein1992experimental}%
  \BibitemOpen
  \bibfield  {author} {\bibinfo {author} {\bibnamefont {Stein}, \bibfnamefont
  {J.}}, \ and\ \bibinfo {author} {\bibfnamefont {H.-J.}\ \bibnamefont
  {St{\"o}ckmann}}} (\bibinfo {year} {1992}),\ \href {\doibase
  10.1103/PhysRevLett.68.2867} {\bibfield  {journal} {\bibinfo  {journal}
  {Phys. Rev. Lett.}\ }\textbf {\bibinfo {volume} {68}}~(\bibinfo {number}
  {19}),\ \bibinfo {pages} {2867}}\BibitemShut {NoStop}%
\bibitem [{\citenamefont {Stein}\ \emph {et~al.}(1995)\citenamefont {Stein},
  \citenamefont {St{\"o}ckmann},\ and\ \citenamefont
  {Stoffregen}}]{stein1995microwave}%
  \BibitemOpen
  \bibfield  {author} {\bibinfo {author} {\bibnamefont {Stein}, \bibfnamefont
  {J.}}, \bibinfo {author} {\bibfnamefont {H.-J.}\ \bibnamefont
  {St{\"o}ckmann}}, \ and\ \bibinfo {author} {\bibfnamefont {U.}~\bibnamefont
  {Stoffregen}}} (\bibinfo {year} {1995}),\ \href {\doibase
  10.1103/PhysRevLett.75.53} {\bibfield  {journal} {\bibinfo  {journal} {Phys.
  Rev. Lett.}\ }\textbf {\bibinfo {volume} {75}}~(\bibinfo {number} {1}),\
  \bibinfo {pages} {53}}\BibitemShut {NoStop}%
\bibitem [{\citenamefont {Stern}(1924)}]{stern1924}%
  \BibitemOpen
  \bibfield  {author} {\bibinfo {author} {\bibnamefont {Stern}, \bibfnamefont
  {A.}}} (\bibinfo {year} {1924}),\ \emph {\bibinfo {title} {Bemerkungen {\"
  u}ber asymptotisches Verhalten von Eigenwerten und Eigenfunktionen}},\
  \href@noop {} {Ph.D. thesis}\ (\bibinfo  {school} {Georg-August-Universit{\"
  a}t G{\" o}ttingen})\BibitemShut {NoStop}%
\bibitem [{\citenamefont {Stillinger~Jr}\ and\ \citenamefont
  {Lovett}(1968{\natexlab{a}})}]{stillinger1968general}%
  \BibitemOpen
  \bibfield  {author} {\bibinfo {author} {\bibnamefont {Stillinger~Jr},
  \bibfnamefont {F.~H.}}, \ and\ \bibinfo {author} {\bibfnamefont
  {R.}~\bibnamefont {Lovett}}} (\bibinfo {year} {1968}{\natexlab{a}}),\ \href
  {\doibase 10.1063/1.1670358} {\bibfield  {journal} {\bibinfo  {journal} {J.
  Chem. Phys.}\ }\textbf {\bibinfo {volume} {49}}~(\bibinfo {number} {5}),\
  \bibinfo {pages} {1991}}\BibitemShut {NoStop}%
\bibitem [{\citenamefont {Stillinger~Jr}\ and\ \citenamefont
  {Lovett}(1968{\natexlab{b}})}]{stillinger1968ion}%
  \BibitemOpen
  \bibfield  {author} {\bibinfo {author} {\bibnamefont {Stillinger~Jr},
  \bibfnamefont {F.~H.}}, \ and\ \bibinfo {author} {\bibfnamefont
  {R.}~\bibnamefont {Lovett}}} (\bibinfo {year} {1968}{\natexlab{b}}),\ \href
  {\doibase 10.1063/1.1669709} {\bibfield  {journal} {\bibinfo  {journal} {J.
  Chem Phys.}\ }\textbf {\bibinfo {volume} {48}}~(\bibinfo {number} {9}),\
  \bibinfo {pages} {3858}}\BibitemShut {NoStop}%
\bibitem [{\citenamefont {St{\"o}ckmann}(2006)}]{stockmann2006quantum}%
  \BibitemOpen
  \bibfield  {author} {\bibinfo {author} {\bibnamefont {St{\"o}ckmann},
  \bibfnamefont {H.-J.}}} (\bibinfo {year} {2006}),\ \href@noop {} {\emph
  {\bibinfo {title} {Quantum chaos: an introduction}}}\ (\bibinfo  {publisher}
  {Cambridge University Press},\ \bibinfo {address} {New York})\BibitemShut
  {NoStop}%
\bibitem [{\citenamefont {St{\"o}ckmann}(2007)}]{stockmann2007chladni}%
  \BibitemOpen
  \bibfield  {author} {\bibinfo {author} {\bibnamefont {St{\"o}ckmann},
  \bibfnamefont {H.-J.}}} (\bibinfo {year} {2007}),\ \href {\doibase
  10.1140/epjst/e2007-00144-5} {\bibfield  {journal} {\bibinfo  {journal} {Eur.
  Phys. J. Spec. Top.}\ }\textbf {\bibinfo {volume} {145}}~(\bibinfo {number}
  {1}),\ \bibinfo {pages} {15}}\BibitemShut {NoStop}%
\bibitem [{\citenamefont {St{\"o}ckmann}\ and\ \citenamefont
  {Stein}(1990)}]{stockmann1990quantum}%
  \BibitemOpen
  \bibfield  {author} {\bibinfo {author} {\bibnamefont {St{\"o}ckmann},
  \bibfnamefont {H.-J.}}, \ and\ \bibinfo {author} {\bibfnamefont
  {J.}~\bibnamefont {Stein}}} (\bibinfo {year} {1990}),\ \href {\doibase
  10.1103/PhysRevLett.64.2215} {\bibfield  {journal} {\bibinfo  {journal}
  {Phys. Rev. Lett.}\ }\textbf {\bibinfo {volume} {64}}~(\bibinfo {number}
  {19}),\ \bibinfo {pages} {2215}}\BibitemShut {NoStop}%
\bibitem [{\citenamefont {Strauss}(1992)}]{strausspartial}%
  \BibitemOpen
  \bibfield  {author} {\bibinfo {author} {\bibnamefont {Strauss}, \bibfnamefont
  {W.~A.}}} (\bibinfo {year} {1992}),\ \href@noop {} {\bibinfo  {journal} {John
  Wiley \& Sons}\ }\BibitemShut {NoStop}%
\bibitem [{\citenamefont {Struik}(2012)}]{struik2012lectures}%
  \BibitemOpen
\bibfield  {journal} {  }\bibfield  {author} {\bibinfo {author} {\bibnamefont
  {Struik}, \bibfnamefont {D.~J.}}} (\bibinfo {year} {2012}),\ \href@noop {}
  {\emph {\bibinfo {title} {Lectures on classical differential geometry}}},\
  \bibinfo {edition} {2nd}\ ed.\ (\bibinfo  {publisher} {Dover Publications},\
  \bibinfo {address} {New York})\BibitemShut {NoStop}%
\bibitem [{\citenamefont {Sturm}(1836{\natexlab{a}})}]{sturm1836a}%
  \BibitemOpen
  \bibfield  {author} {\bibinfo {author} {\bibnamefont {Sturm}, \bibfnamefont
  {C.~F.}}} (\bibinfo {year} {1836}{\natexlab{a}}),\ \href {\doibase
  10.1007/978-3-7643-7990-2_30} {\bibfield  {journal} {\bibinfo  {journal} {J.
  Math. Pures Appl.}\ }\textbf {\bibinfo {volume} {1}},\ \bibinfo {pages}
  {106}}\BibitemShut {NoStop}%
\bibitem [{\citenamefont {Sturm}(1836{\natexlab{b}})}]{sturm1836b}%
  \BibitemOpen
  \bibfield  {author} {\bibinfo {author} {\bibnamefont {Sturm}, \bibfnamefont
  {C.~F.}}} (\bibinfo {year} {1836}{\natexlab{b}}),\ \href {\doibase
  10.1007/978-3-7643-7990-2_33} {\bibfield  {journal} {\bibinfo  {journal} {J.
  Math. Pures Appl.}\ }\textbf {\bibinfo {volume} {1}},\ \bibinfo {pages}
  {373}}\BibitemShut {NoStop}%
\bibitem [{\citenamefont {Sudakov}\ and\ \citenamefont
  {Tsirel'son}(1978)}]{sudakov1978extremal}%
  \BibitemOpen
  \bibfield  {author} {\bibinfo {author} {\bibnamefont {Sudakov}, \bibfnamefont
  {V.~N.}}, \ and\ \bibinfo {author} {\bibfnamefont {B.~S.}\ \bibnamefont
  {Tsirel'son}}} (\bibinfo {year} {1978}),\ \href {\doibase 10.1007/BF01086099}
  {\bibfield  {journal} {\bibinfo  {journal} {J. Math. Sci.}\ }\textbf
  {\bibinfo {volume} {9}}~(\bibinfo {number} {1}),\ \bibinfo {pages}
  {9}}\BibitemShut {NoStop}%
\bibitem [{\citenamefont {Sunada}(1985)}]{sunada1985riemannian}%
  \BibitemOpen
  \bibfield  {author} {\bibinfo {author} {\bibnamefont {Sunada}, \bibfnamefont
  {T.}}} (\bibinfo {year} {1985}),\ \href {\doibase /doi.org/10.2307/1971195}
  {\bibfield  {journal} {\bibinfo  {journal} {Ann. Math.}\ }\textbf {\bibinfo
  {volume} {121}}~(\bibinfo {number} {1}),\ \bibinfo {pages} {169}}\BibitemShut
  {NoStop}%
\bibitem [{\citenamefont {Sweers}(2001)}]{sweers2001first}%
  \BibitemOpen
  \bibfield  {author} {\bibinfo {author} {\bibnamefont {Sweers}, \bibfnamefont
  {G.}}} (\bibinfo {year} {2001}),\ in\ \href
  {http://ejde.math.txstate.edu/conf-proc/06/s3/sweers.pdf} {\emph {\bibinfo
  {booktitle} {Electron. J. Diff. Eqns., Conf. 06,}}}\ (\bibinfo {organization}
  {USA-Chile Workshop on Nonlinear Analysis})\ pp.\ \bibinfo {pages}
  {285--296}\BibitemShut {NoStop}%
\bibitem [{\citenamefont {Szeg{\"o}}(1950)}]{szego1950membranes}%
  \BibitemOpen
  \bibfield  {author} {\bibinfo {author} {\bibnamefont {Szeg{\"o}},
  \bibfnamefont {G.}}} (\bibinfo {year} {1950}),\ \href {\doibase
  10.1073/pnas.36.3.210} {\bibfield  {journal} {\bibinfo  {journal} {Proc. Nat.
  Acad. Sci.}\ }\textbf {\bibinfo {volume} {36}}~(\bibinfo {number} {3}),\
  \bibinfo {pages} {210}}\BibitemShut {NoStop}%
\bibitem [{\citenamefont {Szeg{\" o}}(1954)}]{szego1954inequalities}%
  \BibitemOpen
  \bibfield  {author} {\bibinfo {author} {\bibnamefont {Szeg{\" o}},
  \bibfnamefont {G.}}} (\bibinfo {year} {1954}),\ \href {\doibase
  10.1512/iumj.1954.3.53017} {\bibfield  {journal} {\bibinfo  {journal} {J.
  Rational Mech. Anal.}\ }\textbf {\bibinfo {volume} {3}}~(\bibinfo {number}
  {4}),\ \bibinfo {pages} {343}}\BibitemShut {NoStop}%
\bibitem [{\citenamefont {Tanner}\ and\ \citenamefont
  {S{\o}ndergaard}(2007)}]{tanner2007wave}%
  \BibitemOpen
  \bibfield  {author} {\bibinfo {author} {\bibnamefont {Tanner}, \bibfnamefont
  {G.}}, \ and\ \bibinfo {author} {\bibfnamefont {N.}~\bibnamefont
  {S{\o}ndergaard}}} (\bibinfo {year} {2007}),\ \href {\doibase
  10.1088/1751-8113/40/50/R01} {\bibfield  {journal} {\bibinfo  {journal} {J.
  Phys. A: Math. Theor.}\ }\textbf {\bibinfo {volume} {40}}~(\bibinfo {number}
  {50}),\ \bibinfo {pages} {R443}}\BibitemShut {NoStop}%
\bibitem [{\citenamefont {Taylor}\ and\ \citenamefont
  {Dennis}(2014)}]{taylor2014geometry}%
  \BibitemOpen
  \bibfield  {author} {\bibinfo {author} {\bibnamefont {Taylor}, \bibfnamefont
  {A.~J.}}, \ and\ \bibinfo {author} {\bibfnamefont {M.~R.}\ \bibnamefont
  {Dennis}}} (\bibinfo {year} {2014}),\ \href {\doibase
  10.1088/1751-8113/47/46/465101} {\bibfield  {journal} {\bibinfo  {journal}
  {J. Phys. A: Math. Theor.}\ }\textbf {\bibinfo {volume} {47}}~(\bibinfo
  {number} {46}),\ \bibinfo {pages} {465101}}\BibitemShut {NoStop}%
\bibitem [{\citenamefont {Taylor}\ and\ \citenamefont
  {Dennis}(2016)}]{taylor2016vortex}%
  \BibitemOpen
  \bibfield  {author} {\bibinfo {author} {\bibnamefont {Taylor}, \bibfnamefont
  {A.~J.}}, \ and\ \bibinfo {author} {\bibfnamefont {M.~R.}\ \bibnamefont
  {Dennis}}} (\bibinfo {year} {2016}),\ \href {\doibase 10.1038/ncomms12346}
  {\bibfield  {journal} {\bibinfo  {journal} {Nat. Commun.}\ }\textbf {\bibinfo
  {volume} {7}},\ \bibinfo {pages} {12346}}\BibitemShut {NoStop}%
\bibitem [{\citenamefont {Toth}\ and\ \citenamefont
  {Wigman}(2009)}]{toth2009counting}%
  \BibitemOpen
  \bibfield  {author} {\bibinfo {author} {\bibnamefont {Toth}, \bibfnamefont
  {J.~A.}}, \ and\ \bibinfo {author} {\bibfnamefont {I.}~\bibnamefont
  {Wigman}}} (\bibinfo {year} {2009}),\ \href {\doibase 10.1093/imrn/rnp052}
  {\bibfield  {journal} {\bibinfo  {journal} {Int. Math. Res. Notices}\
  }\textbf {\bibinfo {volume} {18}},\ \bibinfo {pages} {3337}}\BibitemShut
  {NoStop}%
\bibitem [{\citenamefont {Toth}\ and\ \citenamefont
  {Zelditch}(2009)}]{toth2009}%
  \BibitemOpen
  \bibfield  {author} {\bibinfo {author} {\bibnamefont {Toth}, \bibfnamefont
  {J.~A.}}, \ and\ \bibinfo {author} {\bibfnamefont {S.}~\bibnamefont
  {Zelditch}}} (\bibinfo {year} {2009}),\ \href {\doibase
  10.4310/jdg/1236604347} {\bibfield  {journal} {\bibinfo  {journal} {J.
  Differential Geom.}\ }\textbf {\bibinfo {volume} {81}}~(\bibinfo {number}
  {3}),\ \bibinfo {pages} {649}}\BibitemShut {NoStop}%
\bibitem [{\citenamefont {Tuan}\ \emph {et~al.}(2015)\citenamefont {Tuan},
  \citenamefont {Tung}, \citenamefont {Liang}, \citenamefont {Chiang},
  \citenamefont {Huang},\ and\ \citenamefont {Chen}}]{tuan2015resolving}%
  \BibitemOpen
  \bibfield  {author} {\bibinfo {author} {\bibnamefont {Tuan}, \bibfnamefont
  {P.~H.}}, \bibinfo {author} {\bibfnamefont {J.~C.}\ \bibnamefont {Tung}},
  \bibinfo {author} {\bibfnamefont {H.~C.}\ \bibnamefont {Liang}}, \bibinfo
  {author} {\bibfnamefont {P.~Y.}\ \bibnamefont {Chiang}}, \bibinfo {author}
  {\bibfnamefont {K.~F.}\ \bibnamefont {Huang}}, \ and\ \bibinfo {author}
  {\bibfnamefont {Y.~F.}\ \bibnamefont {Chen}}} (\bibinfo {year} {2015}),\
  \href {\doibase 10.1209/0295-5075/111/64004} {\bibfield  {journal} {\bibinfo
  {journal} {EPL}\ }\textbf {\bibinfo {volume} {111}}~(\bibinfo {number} {6}),\
  \bibinfo {pages} {64004}}\BibitemShut {NoStop}%
\bibitem [{\citenamefont {Turnbull}\ \emph {et~al.}(1996)\citenamefont
  {Turnbull}, \citenamefont {Robertson}, \citenamefont {Smith}, \citenamefont
  {Allen},\ and\ \citenamefont {Padgett}}]{turnbull1996generation}%
  \BibitemOpen
  \bibfield  {author} {\bibinfo {author} {\bibnamefont {Turnbull},
  \bibfnamefont {G.~A.}}, \bibinfo {author} {\bibfnamefont {D.~A.}\
  \bibnamefont {Robertson}}, \bibinfo {author} {\bibfnamefont {G.~M.}\
  \bibnamefont {Smith}}, \bibinfo {author} {\bibfnamefont {L.}~\bibnamefont
  {Allen}}, \ and\ \bibinfo {author} {\bibfnamefont {M.~J.}\ \bibnamefont
  {Padgett}}} (\bibinfo {year} {1996}),\ \href {\doibase
  10.1016/0030-4018(96)00070-3} {\bibfield  {journal} {\bibinfo  {journal}
  {Opt. Commun.}\ }\textbf {\bibinfo {volume} {127}}~(\bibinfo {number}
  {4-6}),\ \bibinfo {pages} {183}}\BibitemShut {NoStop}%
\bibitem [{\citenamefont {Turner}(1984)}]{turner1984quantum}%
  \BibitemOpen
  \bibfield  {author} {\bibinfo {author} {\bibnamefont {Turner}, \bibfnamefont
  {J.~W.}}} (\bibinfo {year} {1984}),\ \href {\doibase
  10.1088/0305-4470/17/14/022} {\bibfield  {journal} {\bibinfo  {journal} {.
  Phys. A: Math. Gen.}\ }\textbf {\bibinfo {volume} {17}}~(\bibinfo {number}
  {14}),\ \bibinfo {pages} {2791}}\BibitemShut {NoStop}%
\bibitem [{\citenamefont {Uhlenbeck}(1972)}]{uhlenbeck1972eigenfunctions}%
  \BibitemOpen
  \bibfield  {author} {\bibinfo {author} {\bibnamefont {Uhlenbeck},
  \bibfnamefont {K.}}} (\bibinfo {year} {1972}),\ \href {\doibase
  10.1090/s0002-9904-1972-13117-3} {\bibfield  {journal} {\bibinfo  {journal}
  {Bull. Am. Math. Soc.}\ }\textbf {\bibinfo {volume} {78}}~(\bibinfo {number}
  {6}),\ 10.1090/s0002-9904-1972-13117-3}\BibitemShut {NoStop}%
\bibitem [{\citenamefont {Uhlenbeck}(1976)}]{uhlenbeck1976generic}%
  \BibitemOpen
  \bibfield  {author} {\bibinfo {author} {\bibnamefont {Uhlenbeck},
  \bibfnamefont {K.}}} (\bibinfo {year} {1976}),\ \href {\doibase
  10.2307/2374041} {\bibfield  {journal} {\bibinfo  {journal} {Am. J. Math.}\
  }\textbf {\bibinfo {volume} {98}}~(\bibinfo {number} {4}),\ \bibinfo {pages}
  {1059}}\BibitemShut {NoStop}%
\bibitem [{\citenamefont {Ullmann}(2007)}]{ullmann2007life}%
  \BibitemOpen
  \bibfield  {author} {\bibinfo {author} {\bibnamefont {Ullmann}, \bibfnamefont
  {D.}}} (\bibinfo {year} {2007}),\ \href {\doibase
  10.1140/epjst/e2007-00145-4} {\bibfield  {journal} {\bibinfo  {journal} {Eur.
  Phys. J. Spec. Top.}\ }\textbf {\bibinfo {volume} {145}}~(\bibinfo {number}
  {1}),\ \bibinfo {pages} {25}}\BibitemShut {NoStop}%
\bibitem [{\citenamefont {Ullmo}(2008)}]{ullmo2008many}%
  \BibitemOpen
  \bibfield  {author} {\bibinfo {author} {\bibnamefont {Ullmo}, \bibfnamefont
  {D.}}} (\bibinfo {year} {2008}),\ \href {\doibase
  10.1088/0034-4885/71/2/026001} {\bibfield  {journal} {\bibinfo  {journal}
  {Rep. Prog. Phys.}\ }\textbf {\bibinfo {volume} {71}}~(\bibinfo {number}
  {2}),\ \bibinfo {pages} {026001}}\BibitemShut {NoStop}%
\bibitem [{\citenamefont {Ullmo}\ and\ \citenamefont
  {Baranger}(2001)}]{ullmo2001interactions}%
  \BibitemOpen
  \bibfield  {author} {\bibinfo {author} {\bibnamefont {Ullmo}, \bibfnamefont
  {D.}}, \ and\ \bibinfo {author} {\bibfnamefont {H.~U.}\ \bibnamefont
  {Baranger}}} (\bibinfo {year} {2001}),\ \href {\doibase
  10.1103/PhysRevB.64.245324} {\bibfield  {journal} {\bibinfo  {journal} {Phys.
  Rev. B}\ }\textbf {\bibinfo {volume} {64}}~(\bibinfo {number} {24}),\
  \bibinfo {pages} {245324}}\BibitemShut {NoStop}%
\bibitem [{\citenamefont {Urakawa}(1982)}]{urakawa1982bounded}%
  \BibitemOpen
  \bibfield  {author} {\bibinfo {author} {\bibnamefont {Urakawa}, \bibfnamefont
  {H.}}} (\bibinfo {year} {1982}),\ \href
  {http://www.numdam.org/item?id=ASENS_1982_4_15_3_441_0} {\bibfield  {journal}
  {\bibinfo  {journal} {Ann. scient. {\'E}c. Norm. Sup.}\ }\textbf {\bibinfo
  {volume} {15}}~(\bibinfo {number} {3}),\ \bibinfo {pages} {441}}\BibitemShut
  {NoStop}%
\bibitem [{\citenamefont {Urbina}\ and\ \citenamefont
  {Richter}(2003)}]{urbina2003supporting}%
  \BibitemOpen
  \bibfield  {author} {\bibinfo {author} {\bibnamefont {Urbina}, \bibfnamefont
  {J.~D.}}, \ and\ \bibinfo {author} {\bibfnamefont {K.}~\bibnamefont
  {Richter}}} (\bibinfo {year} {2003}),\ \href {\doibase
  10.1088/0305-4470/36/38/102} {\bibfield  {journal} {\bibinfo  {journal} {J.
  Phys. A: Math. Gen.}\ }\textbf {\bibinfo {volume} {36}}~(\bibinfo {number}
  {38}),\ \bibinfo {pages} {L495}}\BibitemShut {NoStop}%
\bibitem [{\citenamefont {Urbina}\ and\ \citenamefont
  {Richter}(2004)}]{urbina2004semiclassical}%
  \BibitemOpen
  \bibfield  {author} {\bibinfo {author} {\bibnamefont {Urbina}, \bibfnamefont
  {J.~D.}}, \ and\ \bibinfo {author} {\bibfnamefont {K.}~\bibnamefont
  {Richter}}} (\bibinfo {year} {2004}),\ \href {\doibase
  10.1103/PhysRevE.70.015201} {\bibfield  {journal} {\bibinfo  {journal} {Phys.
  Rev. E}\ }\textbf {\bibinfo {volume} {70}}~(\bibinfo {number} {1}),\ \bibinfo
  {pages} {015201 (R)}}\BibitemShut {NoStop}%
\bibitem [{\citenamefont {Urbina}\ and\ \citenamefont
  {Richter}(2006)}]{urbina2006statistical}%
  \BibitemOpen
  \bibfield  {author} {\bibinfo {author} {\bibnamefont {Urbina}, \bibfnamefont
  {J.~D.}}, \ and\ \bibinfo {author} {\bibfnamefont {K.}~\bibnamefont
  {Richter}}} (\bibinfo {year} {2006}),\ \href {\doibase
  10.1103/PhysRevLett.97.214101} {\bibfield  {journal} {\bibinfo  {journal}
  {Phys. Rev. Lett.}\ }\textbf {\bibinfo {volume} {97}}~(\bibinfo {number}
  {21}),\ \bibinfo {pages} {214101}}\BibitemShut {NoStop}%
\bibitem [{\citenamefont {Urbina}\ and\ \citenamefont
  {Richter}(2007)}]{urbina2007random}%
  \BibitemOpen
  \bibfield  {author} {\bibinfo {author} {\bibnamefont {Urbina}, \bibfnamefont
  {J.~D.}}, \ and\ \bibinfo {author} {\bibfnamefont {K.}~\bibnamefont
  {Richter}}} (\bibinfo {year} {2007}),\ \href {\doibase
  10.1140/epjst/e2007-00161-4} {\bibfield  {journal} {\bibinfo  {journal} {Eur.
  Phys. J. Spec. Top.}\ }\textbf {\bibinfo {volume} {145}}~(\bibinfo {number}
  {1}),\ \bibinfo {pages} {255}}\BibitemShut {NoStop}%
\bibitem [{\citenamefont {Urbina}\ and\ \citenamefont
  {Richter}(2013)}]{urbina2013random}%
  \BibitemOpen
  \bibfield  {author} {\bibinfo {author} {\bibnamefont {Urbina}, \bibfnamefont
  {J.~D.}}, \ and\ \bibinfo {author} {\bibfnamefont {K.}~\bibnamefont
  {Richter}}} (\bibinfo {year} {2013}),\ \href {\doibase
  10.1080/00018732.2013.860277} {\bibfield  {journal} {\bibinfo  {journal}
  {Adv. Phys.}\ }\textbf {\bibinfo {volume} {62}}~(\bibinfo {number} {4-6}),\
  \bibinfo {pages} {363}}\BibitemShut {NoStop}%
\bibitem [{\citenamefont {Urbina}\ \emph {et~al.}(2013)\citenamefont {Urbina},
  \citenamefont {Wimmer}, \citenamefont {Bauernfeind}, \citenamefont {Espitia},
  \citenamefont {Adagideli},\ and\ \citenamefont
  {Richter}}]{urbina2013universal}%
  \BibitemOpen
  \bibfield  {author} {\bibinfo {author} {\bibnamefont {Urbina}, \bibfnamefont
  {J.~D.}}, \bibinfo {author} {\bibfnamefont {M.}~\bibnamefont {Wimmer}},
  \bibinfo {author} {\bibfnamefont {D.}~\bibnamefont {Bauernfeind}}, \bibinfo
  {author} {\bibfnamefont {D.}~\bibnamefont {Espitia}}, \bibinfo {author}
  {\bibfnamefont {I.}~\bibnamefont {Adagideli}}, \ and\ \bibinfo {author}
  {\bibfnamefont {K.}~\bibnamefont {Richter}}} (\bibinfo {year} {2013}),\ \href
  {\doibase 10.1103/PhysRevE.87.042115} {\bibfield  {journal} {\bibinfo
  {journal} {Phys. Rev. E}\ }\textbf {\bibinfo {volume} {87}}~(\bibinfo
  {number} {4}),\ \bibinfo {pages} {042115}}\BibitemShut {NoStop}%
\bibitem [{\citenamefont {Vaa}\ \emph {et~al.}(2005)\citenamefont {Vaa},
  \citenamefont {Koch},\ and\ \citenamefont {Bl{\"u}mel}}]{vaa2005weyl}%
  \BibitemOpen
  \bibfield  {author} {\bibinfo {author} {\bibnamefont {Vaa}, \bibfnamefont
  {C.}}, \bibinfo {author} {\bibfnamefont {P.~M.}\ \bibnamefont {Koch}}, \ and\
  \bibinfo {author} {\bibfnamefont {R.}~\bibnamefont {Bl{\"u}mel}}} (\bibinfo
  {year} {2005}),\ \href {\doibase 10.1103/PhysRevE.72.056211} {\bibfield
  {journal} {\bibinfo  {journal} {Phys. Rev. E}\ }\textbf {\bibinfo {volume}
  {72}}~(\bibinfo {number} {5}),\ \bibinfo {pages} {056211}}\BibitemShut
  {NoStop}%
\bibitem [{\citenamefont {Veble}\ \emph {et~al.}(2000)\citenamefont {Veble},
  \citenamefont {Kuhl}, \citenamefont {Robnik}, \citenamefont {St{\"o}ckmann},
  \citenamefont {Liu},\ and\ \citenamefont {Barth}}]{veble2000experimental}%
  \BibitemOpen
  \bibfield  {author} {\bibinfo {author} {\bibnamefont {Veble}, \bibfnamefont
  {G.}}, \bibinfo {author} {\bibfnamefont {U.}~\bibnamefont {Kuhl}}, \bibinfo
  {author} {\bibfnamefont {M.}~\bibnamefont {Robnik}}, \bibinfo {author}
  {\bibfnamefont {H.-J.}\ \bibnamefont {St{\"o}ckmann}}, \bibinfo {author}
  {\bibfnamefont {J.}~\bibnamefont {Liu}}, \ and\ \bibinfo {author}
  {\bibfnamefont {M.}~\bibnamefont {Barth}}} (\bibinfo {year} {2000}),\ \href
  {\doibase 10.1143/PTPS.139.283} {\bibfield  {journal} {\bibinfo  {journal}
  {Prog. Theor. Phys. Supp.}\ }\textbf {\bibinfo {volume} {139}},\ \bibinfo
  {pages} {283}}\BibitemShut {NoStop}%
\bibitem [{\citenamefont {Colin~de Verdi{\`e}re}(1985)}]{colin1985ergodicite}%
  \BibitemOpen
  \bibfield  {author} {\bibinfo {author} {\bibnamefont {Colin~de Verdi{\`e}re},
  \bibfnamefont {Y.}}} (\bibinfo {year} {1985}),\ \href {\doibase
  10.1007/bf01209296} {\bibfield  {journal} {\bibinfo  {journal} {Commun. Math.
  Phys.}\ }\textbf {\bibinfo {volume} {102}}~(\bibinfo {number} {3}),\ \bibinfo
  {pages} {497}}\BibitemShut {NoStop}%
\bibitem [{\citenamefont {Vign{\'e}ras}(1980)}]{vigneras1980varietes}%
  \BibitemOpen
  \bibfield  {author} {\bibinfo {author} {\bibnamefont {Vign{\'e}ras},
  \bibfnamefont {M.-F.}}} (\bibinfo {year} {1980}),\ \href {\doibase
  10.2307/1971319} {\bibfield  {journal} {\bibinfo  {journal} {Ann. Math.}\
  }\textbf {\bibinfo {volume} {112}}~(\bibinfo {number} {1}),\ \bibinfo {pages}
  {21}}\BibitemShut {NoStop}%
\bibitem [{\citenamefont {Viro}(1979)}]{viro1979construction}%
  \BibitemOpen
  \bibfield  {author} {\bibinfo {author} {\bibnamefont {Viro}, \bibfnamefont
  {O.~J.}}} (\bibinfo {year} {1979}),\ \href@noop {} {\bibfield  {journal}
  {\bibinfo  {journal} {Doklady Akad. nauk SSSR}\ }\textbf {\bibinfo {volume}
  {248}}~(\bibinfo {number} {2}),\ \bibinfo {pages} {279}},\ \bibinfo {note}
  {[\textsl{Soviet Math. Dokl.} \textbf{20}, 991--995]}\BibitemShut {NoStop}%
\bibitem [{\citenamefont {Voros}(1976)}]{voros1976semi}%
  \BibitemOpen
  \bibfield  {author} {\bibinfo {author} {\bibnamefont {Voros}, \bibfnamefont
  {A.}}} (\bibinfo {year} {1976}),\ \href
  {http://www.numdam.org/item?id=AIHPA_1976__24_1_31_0} {\bibfield  {journal}
  {\bibinfo  {journal} {Ann. Inst. Henri Poincar{\'e} Sect. A}\ }\textbf
  {\bibinfo {volume} {24}},\ \bibinfo {pages} {31}}\BibitemShut {NoStop}%
\bibitem [{\citenamefont {Voros}(1977)}]{voros1977asymptotic}%
  \BibitemOpen
  \bibfield  {author} {\bibinfo {author} {\bibnamefont {Voros}, \bibfnamefont
  {A.}}} (\bibinfo {year} {1977}),\ \href
  {http://www.numdam.org/item?id=AIHPA_1977__26_4_343_0} {\bibfield  {journal}
  {\bibinfo  {journal} {Ann. Inst. H. Poincar{\'e} Sect. A}\ }\textbf {\bibinfo
  {volume} {26}},\ \bibinfo {pages} {343}}\BibitemShut {NoStop}%
\bibitem [{\citenamefont {Voros}(1979)}]{voros1979semi}%
  \BibitemOpen
  \bibfield  {author} {\bibinfo {author} {\bibnamefont {Voros}, \bibfnamefont
  {A.}}} (\bibinfo {year} {1979}),\ in\ \href {\doibase 10.1007/BFb0021756}
  {\emph {\bibinfo {booktitle} {Stochastic behavior in classical and quantum
  Hamiltonian systems}}},\ \bibinfo {series} {Lecture Notes in Physics},
  Vol.~\bibinfo {volume} {93},\ \bibinfo {editor} {edited by\ \bibinfo {editor}
  {\bibfnamefont {G.}~\bibnamefont {Casati}}\ and\ \bibinfo {editor}
  {\bibfnamefont {J.}~\bibnamefont {Ford}}}\ (\bibinfo  {publisher}
  {Springer})\ pp.\ \bibinfo {pages} {326--333}\BibitemShut {NoStop}%
\bibitem [{\citenamefont {Waller}(1937)}]{waller1937production}%
  \BibitemOpen
  \bibfield  {author} {\bibinfo {author} {\bibnamefont {Waller}, \bibfnamefont
  {M.~D.}}} (\bibinfo {year} {1937}),\ \href {\doibase
  10.1088/0959-5309/49/5/308} {\bibfield  {journal} {\bibinfo  {journal} {Proc.
  Phys. Soc.}\ }\textbf {\bibinfo {volume} {49}}~(\bibinfo {number} {5}),\
  \bibinfo {pages} {522}}\BibitemShut {NoStop}%
\bibitem [{\citenamefont {Waller}(1938)}]{waller1938vibrations}%
  \BibitemOpen
  \bibfield  {author} {\bibinfo {author} {\bibnamefont {Waller}, \bibfnamefont
  {M.~D.}}} (\bibinfo {year} {1938}),\ \href {\doibase
  10.1088/0959-5309/50/1/307} {\bibfield  {journal} {\bibinfo  {journal} {Proc.
  Phys. Soc.}\ }\textbf {\bibinfo {volume} {50}}~(\bibinfo {number} {1}),\
  \bibinfo {pages} {77}}\BibitemShut {NoStop}%
\bibitem [{\citenamefont {Waller}(1940)}]{waller1940vibrations}%
  \BibitemOpen
  \bibfield  {author} {\bibinfo {author} {\bibnamefont {Waller}, \bibfnamefont
  {M.~D.}}} (\bibinfo {year} {1940}),\ \href {\doibase
  10.1088/0959-5309/52/4/304} {\bibfield  {journal} {\bibinfo  {journal} {Proc.
  Phys. Soc.}\ }\textbf {\bibinfo {volume} {52}}~(\bibinfo {number} {4}),\
  \bibinfo {pages} {452}}\BibitemShut {NoStop}%
\bibitem [{\citenamefont {Waller}(1952)}]{waller1952vibrations}%
  \BibitemOpen
  \bibfield  {author} {\bibinfo {author} {\bibnamefont {Waller}, \bibfnamefont
  {M.~D.}}} (\bibinfo {year} {1952}),\ \href {\doibase 10.1098/rspa.1952.0038}
  {\bibfield  {journal} {\bibinfo  {journal} {Proc. R. Soc. Lond. A}\ }\textbf
  {\bibinfo {volume} {211}}~(\bibinfo {number} {1105}),\ \bibinfo {pages}
  {265}}\BibitemShut {NoStop}%
\bibitem [{\citenamefont {Waller}(1954)}]{waller1954symmetry}%
  \BibitemOpen
  \bibfield  {author} {\bibinfo {author} {\bibnamefont {Waller}, \bibfnamefont
  {M.~D.}}} (\bibinfo {year} {1954}),\ \href {\doibase
  10.1088/0370-1301/67/12/406} {\bibfield  {journal} {\bibinfo  {journal}
  {Proc. Phys. Soc. B}\ }\textbf {\bibinfo {volume} {67}}~(\bibinfo {number}
  {12}),\ \bibinfo {pages} {895}}\BibitemShut {NoStop}%
\bibitem [{\citenamefont {Waller}(1957)}]{waller1957interpreting}%
  \BibitemOpen
  \bibfield  {author} {\bibinfo {author} {\bibnamefont {Waller}, \bibfnamefont
  {M.~D.}}} (\bibinfo {year} {1957}),\ \href {\doibase 10.1119/1.1934385}
  {\bibfield  {journal} {\bibinfo  {journal} {Am. J. Phys.}\ }\textbf {\bibinfo
  {volume} {25}}~(\bibinfo {number} {3}),\ \bibinfo {pages} {157}}\BibitemShut
  {NoStop}%
\bibitem [{\citenamefont {Wang}\ and\ \citenamefont
  {Wang}(2017)}]{wang2017towards}%
  \BibitemOpen
  \bibfield  {author} {\bibinfo {author} {\bibnamefont {Wang}, \bibfnamefont
  {G.}}, \ and\ \bibinfo {author} {\bibfnamefont {Y.}~\bibnamefont {Wang}}}
  (\bibinfo {year} {2017}),\ \href {\doibase 10.1016/j.neuroimage.2016.12.014}
  {\bibfield  {journal} {\bibinfo  {journal} {NeuroImage}\ }\textbf {\bibinfo
  {volume} {147}},\ \bibinfo {pages} {360}}\BibitemShut {NoStop}%
\bibitem [{\citenamefont {Watson}(1995)}]{watson1995treatise}%
  \BibitemOpen
  \bibfield  {author} {\bibinfo {author} {\bibnamefont {Watson}, \bibfnamefont
  {G.~N.}}} (\bibinfo {year} {1995}),\ \href@noop {} {\emph {\bibinfo {title}
  {A treatise on the theory of Bessel functions}}}\ (\bibinfo  {publisher}
  {Cambridge University Press},\ \bibinfo {address} {Cambridge,
  UK})\BibitemShut {NoStop}%
\bibitem [{\citenamefont {Watts}(1996)}]{watts1996crossing}%
  \BibitemOpen
  \bibfield  {author} {\bibinfo {author} {\bibnamefont {Watts}, \bibfnamefont
  {G.~M.~T.}}} (\bibinfo {year} {1996}),\ \href {\doibase
  10.1088/0305-4470/29/14/002} {\bibfield  {journal} {\bibinfo  {journal} {J.
  Phys. A: Math. Gen.}\ }\textbf {\bibinfo {volume} {29}}~(\bibinfo {number}
  {14}),\ \bibinfo {pages} {L363}}\BibitemShut {NoStop}%
\bibitem [{\citenamefont {Weaver}(1989)}]{weaver1989spectral}%
  \BibitemOpen
  \bibfield  {author} {\bibinfo {author} {\bibnamefont {Weaver}, \bibfnamefont
  {R.~L.}}} (\bibinfo {year} {1989}),\ \href {\doibase 10.1121/1.397484}
  {\bibfield  {journal} {\bibinfo  {journal} {J. Acoust. Soc. Am.}\ }\textbf
  {\bibinfo {volume} {85}}~(\bibinfo {number} {3}),\ \bibinfo {pages}
  {1005}}\BibitemShut {NoStop}%
\bibitem [{\citenamefont {Weinberger}(1956)}]{weinberger1956isoperimetric}%
  \BibitemOpen
  \bibfield  {author} {\bibinfo {author} {\bibnamefont {Weinberger},
  \bibfnamefont {H.~F.}}} (\bibinfo {year} {1956}),\ \href {\doibase
  10.1512/iumj.1956.5.55021} {\bibfield  {journal} {\bibinfo  {journal} {J.
  Rational Mech. Anal.}\ }\textbf {\bibinfo {volume} {5}}~(\bibinfo {number}
  {4}),\ \bibinfo {pages} {633}}\BibitemShut {NoStop}%
\bibitem [{\citenamefont {Weinrib}(1982)}]{weinrib1982percolation}%
  \BibitemOpen
  \bibfield  {author} {\bibinfo {author} {\bibnamefont {Weinrib}, \bibfnamefont
  {A.}}} (\bibinfo {year} {1982}),\ \href {\doibase 10.1103/PhysRevB.26.1352}
  {\bibfield  {journal} {\bibinfo  {journal} {Phys. Rev. B}\ }\textbf {\bibinfo
  {volume} {26}}~(\bibinfo {number} {3}),\ \bibinfo {pages} {1352}}\BibitemShut
  {NoStop}%
\bibitem [{\citenamefont {Weinrib}(1984)}]{weinrib1984long}%
  \BibitemOpen
  \bibfield  {author} {\bibinfo {author} {\bibnamefont {Weinrib}, \bibfnamefont
  {A.}}} (\bibinfo {year} {1984}),\ \href {\doibase 10.1103/PhysRevB.29.387}
  {\bibfield  {journal} {\bibinfo  {journal} {Phys. Rev. B}\ }\textbf {\bibinfo
  {volume} {29}}~(\bibinfo {number} {1}),\ \bibinfo {pages} {387}}\BibitemShut
  {NoStop}%
\bibitem [{\citenamefont {Weinstein}(1966)}]{weinstein1966some}%
  \BibitemOpen
  \bibfield  {author} {\bibinfo {author} {\bibnamefont {Weinstein},
  \bibfnamefont {A.}}} (\bibinfo {year} {1966}),\ in\ \href@noop {} {\emph
  {\bibinfo {booktitle} {Numerical solution of partial differential
  equations}}},\ \bibinfo {editor} {edited by\ \bibinfo {editor} {\bibfnamefont
  {J.~H.}\ \bibnamefont {Bramble}}}\ (\bibinfo  {publisher} {Academic Press,
  New York})\ pp.\ \bibinfo {pages} {167--191}\BibitemShut {NoStop}%
\bibitem [{\citenamefont {Werner}(2004)}]{werner2004random}%
  \BibitemOpen
  \bibfield  {author} {\bibinfo {author} {\bibnamefont {Werner}, \bibfnamefont
  {W.}}} (\bibinfo {year} {2004}),\ in\ \href@noop {} {\emph {\bibinfo
  {booktitle} {Lectures on Probability Theory and Statistics}}},\ Vol.\
  \bibinfo {volume} {1840},\ \bibinfo {editor} {edited by\ \bibinfo {editor}
  {\bibfnamefont {J.}~\bibnamefont {Picard}}}\ (\bibinfo  {publisher}
  {Springer},\ \bibinfo {address} {Berlin Heidelberg})\ p.\ \bibinfo {pages}
  {113},\ \bibinfo {note} {{E}cole d'Et{\'e} de Probabilit{\'e}s de Saint-Flour
  XXXII (2002)}\BibitemShut {NoStop}%
\bibitem [{\citenamefont {Weyl}(1911)}]{weyl1911asymptotische}%
  \BibitemOpen
  \bibfield  {author} {\bibinfo {author} {\bibnamefont {Weyl}, \bibfnamefont
  {H.}}} (\bibinfo {year} {1911}),\ \href
  {http://resolver.sub.uni-goettingen.de/purl?PPN252457811_1911} {\bibfield
  {journal} {\bibinfo  {journal} {Gott. Nach.}\ }\textbf {\bibinfo {volume}
  {1911}},\ \bibinfo {pages} {110}}\BibitemShut {NoStop}%
\bibitem [{\citenamefont {Weyl}(1912)}]{Weyl1912}%
  \BibitemOpen
  \bibfield  {author} {\bibinfo {author} {\bibnamefont {Weyl}, \bibfnamefont
  {H.}}} (\bibinfo {year} {1912}),\ \href {\doibase 10.1007/BF01456804}
  {\bibfield  {journal} {\bibinfo  {journal} {Math. Ann.}\ }\textbf {\bibinfo
  {volume} {71}}~(\bibinfo {number} {4}),\ \bibinfo {pages} {441}}\BibitemShut
  {NoStop}%
\bibitem [{\citenamefont {Wheeler}(2005)}]{wheeler2005curved}%
  \BibitemOpen
  \bibfield  {author} {\bibinfo {author} {\bibnamefont {Wheeler}, \bibfnamefont
  {C.}}} (\bibinfo {year} {2005}),\ \href {\doibase 10.1088/0305-4470/38/7/006}
  {\bibfield  {journal} {\bibinfo  {journal} {J. Phys. A: Math. Gen.}\ }\textbf
  {\bibinfo {volume} {38}}~(\bibinfo {number} {7}),\ \bibinfo {pages}
  {1491}}\BibitemShut {NoStop}%
\bibitem [{\citenamefont {Whittaker}\ and\ \citenamefont
  {Watson}(1996)}]{whittaker1996course}%
  \BibitemOpen
  \bibfield  {author} {\bibinfo {author} {\bibnamefont {Whittaker},
  \bibfnamefont {E.~T.}}, \ and\ \bibinfo {author} {\bibfnamefont {G.~N.}\
  \bibnamefont {Watson}}} (\bibinfo {year} {1996}),\ \href@noop {} {\emph
  {\bibinfo {title} {A course of modern analysis}}}\ (\bibinfo  {publisher}
  {Cambridge University Press},\ \bibinfo {address} {Cambridge,
  UK})\BibitemShut {NoStop}%
\bibitem [{\citenamefont {Wiersig}(2002)}]{wiersig2002spectral}%
  \BibitemOpen
  \bibfield  {author} {\bibinfo {author} {\bibnamefont {Wiersig}, \bibfnamefont
  {J.}}} (\bibinfo {year} {2002}),\ \href {\doibase 10.1103/PhysRevE.65.046217}
  {\bibfield  {journal} {\bibinfo  {journal} {Phys. Rev. E}\ }\textbf {\bibinfo
  {volume} {65}}~(\bibinfo {number} {4}),\ \bibinfo {pages}
  {046217}}\BibitemShut {NoStop}%
\bibitem [{\citenamefont {Wigman}(2009)}]{wigman2009}%
  \BibitemOpen
  \bibfield  {author} {\bibinfo {author} {\bibnamefont {Wigman}, \bibfnamefont
  {I.}}} (\bibinfo {year} {2009}),\ \href {\doibase 10.1063/1.3056589}
  {\bibfield  {journal} {\bibinfo  {journal} {J. Math. Phys.}\ }\textbf
  {\bibinfo {volume} {50}}~(\bibinfo {number} {1}),\ \bibinfo {pages}
  {013521}}\BibitemShut {NoStop}%
\bibitem [{\citenamefont {Wigman}(2010)}]{wigman2010fluctuations}%
  \BibitemOpen
  \bibfield  {author} {\bibinfo {author} {\bibnamefont {Wigman}, \bibfnamefont
  {I.}}} (\bibinfo {year} {2010}),\ \href {\doibase 10.1007/s00220-010-1078-8}
  {\bibfield  {journal} {\bibinfo  {journal} {Commun. Math. Phys.}\ }\textbf
  {\bibinfo {volume} {298}}~(\bibinfo {number} {3}),\ \bibinfo {pages}
  {787}}\BibitemShut {NoStop}%
\bibitem [{\citenamefont {Wigman}(2012{\natexlab{a}})}]{wigman2012erratum}%
  \BibitemOpen
  \bibfield  {author} {\bibinfo {author} {\bibnamefont {Wigman}, \bibfnamefont
  {I.}}} (\bibinfo {year} {2012}{\natexlab{a}}),\ \href {\doibase
  10.1007/s00220-011-1367-x} {\bibfield  {journal} {\bibinfo  {journal}
  {Commun. Math. Phys.}\ }\textbf {\bibinfo {volume} {309}}~(\bibinfo {number}
  {1}),\ \bibinfo {pages} {293}}\BibitemShut {NoStop}%
\bibitem [{\citenamefont {Wigman}(2012{\natexlab{b}})}]{wigman2012nodal}%
  \BibitemOpen
  \bibfield  {author} {\bibinfo {author} {\bibnamefont {Wigman}, \bibfnamefont
  {I.}}} (\bibinfo {year} {2012}{\natexlab{b}}),\ in\ \href {\doibase
  10.1090/pspum/084} {\emph {\bibinfo {booktitle} {Spectral Geometry}}},\
  \bibinfo {series} {Proc. Symp. Pure Math.}, Vol.~\bibinfo {volume} {84},\
  \bibinfo {editor} {edited by\ \bibinfo {editor} {\bibfnamefont {A.~H.}\
  \bibnamefont {Barnett}}, \bibinfo {editor} {\bibfnamefont {C.~S.}\
  \bibnamefont {Gordon}}, \bibinfo {editor} {\bibfnamefont {P.~A.}\
  \bibnamefont {Perry}}, \ and\ \bibinfo {editor} {\bibfnamefont
  {A.}~\bibnamefont {Uribe}}}\ (\bibinfo  {publisher} {American Mathematical
  Soc.},\ \bibinfo {address} {Providence, RI})\ pp.\ \bibinfo {pages}
  {285--298}\BibitemShut {NoStop}%
\bibitem [{\citenamefont {Wigner}(1932)}]{wigner1932quantum}%
  \BibitemOpen
  \bibfield  {author} {\bibinfo {author} {\bibnamefont {Wigner}, \bibfnamefont
  {E.}}} (\bibinfo {year} {1932}),\ \href {\doibase 10.1103/PhysRev.40.749}
  {\bibfield  {journal} {\bibinfo  {journal} {Phys. Rev.}\ }\textbf {\bibinfo
  {volume} {40}}~(\bibinfo {number} {5}),\ \bibinfo {pages} {749}}\BibitemShut
  {NoStop}%
\bibitem [{\citenamefont {Wilkinson}(2004)}]{wilkinson2004screening}%
  \BibitemOpen
  \bibfield  {author} {\bibinfo {author} {\bibnamefont {Wilkinson},
  \bibfnamefont {M.}}} (\bibinfo {year} {2004}),\ \href {\doibase
  10.1088/0305-4470/37/26/012} {\bibfield  {journal} {\bibinfo  {journal} {J.
  Phys. A: Math. Gen.}\ }\textbf {\bibinfo {volume} {37}}~(\bibinfo {number}
  {26}),\ \bibinfo {pages} {6763}}\BibitemShut {NoStop}%
\bibitem [{\citenamefont {Wilson}\ and\ \citenamefont
  {Kogut}(1974)}]{wilson1974renormalization}%
  \BibitemOpen
  \bibfield  {author} {\bibinfo {author} {\bibnamefont {Wilson}, \bibfnamefont
  {K.~G.}}, \ and\ \bibinfo {author} {\bibfnamefont {J.}~\bibnamefont {Kogut}}}
  (\bibinfo {year} {1974}),\ \href@noop {} {\bibfield  {journal} {\bibinfo
  {journal} {Phys. Rep.}\ }\textbf {\bibinfo {volume} {12}}~(\bibinfo {number}
  {2}),\ \bibinfo {pages} {75}}\BibitemShut {NoStop}%
\bibitem [{\citenamefont {Wu}\ \emph {et~al.}(1998)\citenamefont {Wu},
  \citenamefont {Bridgewater}, \citenamefont {Gokirmak},\ and\ \citenamefont
  {Anlage}}]{wu1998probability}%
  \BibitemOpen
  \bibfield  {author} {\bibinfo {author} {\bibnamefont {Wu}, \bibfnamefont
  {D.~H.}}, \bibinfo {author} {\bibfnamefont {J.~S.~A.}\ \bibnamefont
  {Bridgewater}}, \bibinfo {author} {\bibfnamefont {A.}~\bibnamefont
  {Gokirmak}}, \ and\ \bibinfo {author} {\bibfnamefont {S.~M.}\ \bibnamefont
  {Anlage}}} (\bibinfo {year} {1998}),\ \href {\doibase
  10.1103/PhysRevLett.81.2890} {\bibfield  {journal} {\bibinfo  {journal}
  {Phys. Rev. Lett.}\ }\textbf {\bibinfo {volume} {81}}~(\bibinfo {number}
  {14}),\ \bibinfo {pages} {2890}}\BibitemShut {NoStop}%
\bibitem [{\citenamefont {Wu}(1982)}]{wu1982potts}%
  \BibitemOpen
  \bibfield  {author} {\bibinfo {author} {\bibnamefont {Wu}, \bibfnamefont
  {F.-Y.}}} (\bibinfo {year} {1982}),\ \href {\doibase
  10.1103/RevModPhys.54.235} {\bibfield  {journal} {\bibinfo  {journal} {Rev.
  Mod. Phys.}\ }\textbf {\bibinfo {volume} {54}}~(\bibinfo {number} {1}),\
  \bibinfo {pages} {235}}\BibitemShut {NoStop}%
\bibitem [{\citenamefont {Yamamoto}\ and\ \citenamefont
  {Slusher}(1993)}]{yamamoto1993}%
  \BibitemOpen
  \bibfield  {author} {\bibinfo {author} {\bibnamefont {Yamamoto},
  \bibfnamefont {Y.}}, \ and\ \bibinfo {author} {\bibfnamefont {R.~E.}\
  \bibnamefont {Slusher}}} (\bibinfo {year} {1993}),\ \href {\doibase
  10.1063/1.881356} {\bibfield  {journal} {\bibinfo  {journal} {Phys. Today}\
  }\textbf {\bibinfo {volume} {46}}~(\bibinfo {number} {6}),\ \bibinfo {pages}
  {66}}\BibitemShut {NoStop}%
\bibitem [{\citenamefont {Yang}\ and\ \citenamefont
  {Guo}(2013)}]{yang2013nodal}%
  \BibitemOpen
  \bibfield  {author} {\bibinfo {author} {\bibnamefont {Yang}, \bibfnamefont
  {D.-H.}}, \ and\ \bibinfo {author} {\bibfnamefont {B.-Z.}\ \bibnamefont
  {Guo}}} (\bibinfo {year} {2013}),\ \href {\doibase 10.1007/s11424-013-1175-9}
  {\bibfield  {journal} {\bibinfo  {journal} {J. Syst. Sci. Complex.}\ }\textbf
  {\bibinfo {volume} {26}}~(\bibinfo {number} {3}),\ \bibinfo {pages}
  {483}}\BibitemShut {NoStop}%
\bibitem [{\citenamefont {Yau}(1982)}]{yau1982survey}%
  \BibitemOpen
  \bibinfo {editor} {\bibnamefont {Yau}, \bibfnamefont {S.-T.}},\ Ed. (\bibinfo
  {year} {1982}),\ \href@noop {} {\emph {\bibinfo {title} {Seminar on
  Differential Geometry}}},\ \bibinfo {series} {Annals of Mathematics Studies},
  Vol.\ \bibinfo {volume} {102}\ (\bibinfo  {publisher} {Princeton University
  Press},\ \bibinfo {address} {Princeton, NJ})\BibitemShut {NoStop}%
\bibitem [{\citenamefont {Yau}(1993)}]{yau1993differential}%
  \BibitemOpen
  \bibfield  {author} {\bibinfo {author} {\bibnamefont {Yau}, \bibfnamefont
  {S.-T.}}} (\bibinfo {year} {1993}),\ in\ \href {\doibase 10.1090/pspum/054.1}
  {\emph {\bibinfo {booktitle} {Open problems in geometry}}},\ \bibinfo
  {series} {Proc. Symp. Pure Math.}, Vol.~\bibinfo {volume} {54}\ (\bibinfo
  {publisher} {American Mathematical Soc.},\ \bibinfo {address} {Providence,
  RI})\BibitemShut {NoStop}%
\bibitem [{\citenamefont {Zallen}\ and\ \citenamefont
  {Scher}(1971)}]{zallen1971percolation}%
  \BibitemOpen
  \bibfield  {author} {\bibinfo {author} {\bibnamefont {Zallen}, \bibfnamefont
  {R.}}, \ and\ \bibinfo {author} {\bibfnamefont {H.}~\bibnamefont {Scher}}}
  (\bibinfo {year} {1971}),\ \href {\doibase 10.1103/PhysRevB.4.4471}
  {\bibfield  {journal} {\bibinfo  {journal} {Phys. Rev. B}\ }\textbf {\bibinfo
  {volume} {4}}~(\bibinfo {number} {12}),\ \bibinfo {pages} {4471}}\BibitemShut
  {NoStop}%
\bibitem [{\citenamefont {Zelditch}(1987)}]{zelditch1987uniform}%
  \BibitemOpen
  \bibfield  {author} {\bibinfo {author} {\bibnamefont {Zelditch},
  \bibfnamefont {S.}}} (\bibinfo {year} {1987}),\ \href {\doibase
  10.1215/S0012-7094-87-05546-3} {\bibfield  {journal} {\bibinfo  {journal}
  {Duke Math. J}\ }\textbf {\bibinfo {volume} {55}}~(\bibinfo {number} {4}),\
  \bibinfo {pages} {919}}\BibitemShut {NoStop}%
\bibitem [{\citenamefont {Zelditch}(1990)}]{zelditch1990quantum}%
  \BibitemOpen
  \bibfield  {author} {\bibinfo {author} {\bibnamefont {Zelditch},
  \bibfnamefont {S.}}} (\bibinfo {year} {1990}),\ \href {\doibase
  10.1016/0022-1236(90)90021-C} {\bibfield  {journal} {\bibinfo  {journal} {J.
  Funct. Anal.}\ }\textbf {\bibinfo {volume} {94}}~(\bibinfo {number} {2}),\
  \bibinfo {pages} {415}}\BibitemShut {NoStop}%
\bibitem [{\citenamefont {Zelditch}(2000)}]{zelditch2000spectral}%
  \BibitemOpen
  \bibfield  {author} {\bibinfo {author} {\bibnamefont {Zelditch},
  \bibfnamefont {S.}}} (\bibinfo {year} {2000}),\ \href {\doibase
  10.1007/PL00001633} {\bibfield  {journal} {\bibinfo  {journal} {Geom. Funct.
  Anal.}\ }\textbf {\bibinfo {volume} {10}}~(\bibinfo {number} {3}),\ \bibinfo
  {pages} {628}}\BibitemShut {NoStop}%
\bibitem [{\citenamefont {Zelditch}(2004)}]{zelditch2004inverse}%
  \BibitemOpen
  \bibfield  {author} {\bibinfo {author} {\bibnamefont {Zelditch},
  \bibfnamefont {S.}}} (\bibinfo {year} {2004}),\ \href {\doibase
  10.1007/s00220-004-1074-y} {\bibfield  {journal} {\bibinfo  {journal}
  {Commun. Math. Phys.}\ }\textbf {\bibinfo {volume} {248}}~(\bibinfo {number}
  {2}),\ \bibinfo {pages} {357}}\BibitemShut {NoStop}%
\bibitem [{\citenamefont {Zelditch}(2009{\natexlab{a}})}]{zelditch2009inverse}%
  \BibitemOpen
  \bibfield  {author} {\bibinfo {author} {\bibnamefont {Zelditch},
  \bibfnamefont {S.}}} (\bibinfo {year} {2009}{\natexlab{a}}),\ \href {\doibase
  10.4007/annals.2009.170.205} {\bibfield  {journal} {\bibinfo  {journal} {Ann.
  Math.}\ }\textbf {\bibinfo {volume} {170}}~(\bibinfo {number} {1}),\ \bibinfo
  {pages} {205}}\BibitemShut {NoStop}%
\bibitem [{\citenamefont {Zelditch}(2009{\natexlab{b}})}]{zelditch2009real}%
  \BibitemOpen
  \bibfield  {author} {\bibinfo {author} {\bibnamefont {Zelditch},
  \bibfnamefont {S.}}} (\bibinfo {year} {2009}{\natexlab{b}}),\ in\ \href
  {\doibase 10.1090/conm/484/09482} {\emph {\bibinfo {booktitle} {Spectral
  Analysis in Geometry and Number Theory}}},\ \bibinfo {series} {Contemporary
  Mathematics}, Vol.\ \bibinfo {volume} {484},\ \bibinfo {editor} {edited by\
  \bibinfo {editor} {\bibfnamefont {M.}~\bibnamefont {Kotani}}, \bibinfo
  {editor} {\bibfnamefont {H.}~\bibnamefont {Naito}}, \ and\ \bibinfo {editor}
  {\bibfnamefont {T.}~\bibnamefont {Tate}}}\ (\bibinfo  {publisher} {American
  Mathematical Soc.},\ \bibinfo {address} {Providence, RI})\ p.\ \bibinfo
  {pages} {321}\BibitemShut {NoStop}%
\bibitem [{\citenamefont {Zelditch}(2013)}]{Zelditch_2013}%
  \BibitemOpen
  \bibfield  {author} {\bibinfo {author} {\bibnamefont {Zelditch},
  \bibfnamefont {S.}}} (\bibinfo {year} {2013}),\ \href {\doibase
  10.4310/sdg.2013.v18.n1.a7} {\bibfield  {journal} {\bibinfo  {journal} {Surv.
  Diff. Geom.}\ }\textbf {\bibinfo {volume} {18}}~(\bibinfo {number} {1}),\
  \bibinfo {pages} {237}}\BibitemShut {NoStop}%
\bibitem [{\citenamefont {Zelditch}\ and\ \citenamefont
  {Zworski}(1996)}]{zelditch1996ergodicity}%
  \BibitemOpen
  \bibfield  {author} {\bibinfo {author} {\bibnamefont {Zelditch},
  \bibfnamefont {S.}}, \ and\ \bibinfo {author} {\bibfnamefont
  {M.}~\bibnamefont {Zworski}}} (\bibinfo {year} {1996}),\ \href {\doibase
  10.1007/bf02099513} {\bibfield  {journal} {\bibinfo  {journal} {Commun. Math.
  Phys.}\ }\textbf {\bibinfo {volume} {175}}~(\bibinfo {number} {3}),\ \bibinfo
  {pages} {673}}\BibitemShut {NoStop}%
\bibitem [{\citenamefont {Zemlyakov}\ and\ \citenamefont
  {Katok}(1975)}]{zemlyakov1975topological}%
  \BibitemOpen
  \bibfield  {author} {\bibinfo {author} {\bibnamefont {Zemlyakov},
  \bibfnamefont {A.~N.}}, \ and\ \bibinfo {author} {\bibfnamefont {A.~B.}\
  \bibnamefont {Katok}}} (\bibinfo {year} {1975}),\ \href {\doibase
  10.1007/BF01818045} {\bibfield  {journal} {\bibinfo  {journal} {Math. Notes}\
  }\textbf {\bibinfo {volume} {18}}~(\bibinfo {number} {2}),\ \bibinfo {pages}
  {760}}\BibitemShut {NoStop}%
\bibitem [{\citenamefont {Ziff}(1986)}]{ziff1986test}%
  \BibitemOpen
  \bibfield  {author} {\bibinfo {author} {\bibnamefont {Ziff}, \bibfnamefont
  {R.~M.}}} (\bibinfo {year} {1986}),\ \href {\doibase
  10.1103/PhysRevLett.56.545} {\bibfield  {journal} {\bibinfo  {journal} {Phys.
  Rev. Lett.}\ }\textbf {\bibinfo {volume} {56}}~(\bibinfo {number} {6}),\
  \bibinfo {pages} {545}}\BibitemShut {NoStop}%
\bibitem [{\citenamefont {Ziff}\ \emph {et~al.}(1997)\citenamefont {Ziff},
  \citenamefont {Finch},\ and\ \citenamefont
  {Adamchik}}]{ziff1997universality}%
  \BibitemOpen
  \bibfield  {author} {\bibinfo {author} {\bibnamefont {Ziff}, \bibfnamefont
  {R.~M.}}, \bibinfo {author} {\bibfnamefont {S.~R.}\ \bibnamefont {Finch}}, \
  and\ \bibinfo {author} {\bibfnamefont {V.~S.}\ \bibnamefont {Adamchik}}}
  (\bibinfo {year} {1997}),\ \href {\doibase 10.1103/PhysRevLett.79.3447}
  {\bibfield  {journal} {\bibinfo  {journal} {Phys. Rev. Lett.}\ }\textbf
  {\bibinfo {volume} {79}}~(\bibinfo {number} {18}),\ \bibinfo {pages}
  {3447}}\BibitemShut {NoStop}%
\end{thebibliography}%

\end{document}